\documentclass[onecolumn,  numberedappendix]{openjournal}

\usepackage{amsmath, verbatim}
\usepackage[multidot]{grffile}
\usepackage{lineno}

\setcitestyle{square,numbers}

\RequirePackage{amsmath}
\RequirePackage{amssymb}
\RequirePackage{epsfig}
\RequirePackage{graphicx}

\RequirePackage[colorlinks=true
  ,urlcolor=blue
  ,anchorcolor=blue
  ,citecolor=blue
  ,filecolor=blue
  ,linkcolor=blue
  ,menucolor=blue
  ,linktocpage=true
  ,pdfproducer=medialab
  ,pdfa=true]{hyperref}

\usepackage{tablefootnote}
\usepackage[T1]{fontenc} 

\usepackage[normal]{caption}

\newcommand*{\rtensor}[1]{\bar{\bar{#1}}}
\def\pmb#1{\setbox0=\hbox{#1}%
    \kern-.025em\copy0\kern-\wd0
    \kern.05em\copy0\kern-\wd0
    \kern-.025em\raise.0433em\box0}
\def\ltsima{$\; \buildrel < \over \sim \;$}
\def\gtsima{$\; \buildrel > \over \sim \;$}
\def\simlt{\lower.5ex\hbox{\ltsima}}
\def\simgt{\lower.5ex\hbox{\gtsima}}
\def\muk{$(\mu{\rm K})^2$ }
\def\mask1{mask$\_$1}
\def\mask2{mask$\_$2}
\def\mask3{mask$\_$3}
\def\mask4{mask$\_$4}
\def\mask5{mask$\_$5}
\def\mask70{mask$\_$70}
\def\mask3pol{mask$\_$pol$\_$3}
\def\mask4pol{mask$\_$pol$\_$4}

\def\p2Y{\;_2Y}
\def\m2Y{\;_{-2}Y}
\newcommand{\isdraft}[1]{#1}
\newcommand{\GE}[1]{{\isdraft{\color{black}  #1}}}

\numberwithin{equation}{section}
\numberwithin{figure}{section}

\def\WMAP{\textit{WMAP}}

\newcommand{\healpix}{{\tt HEALpix}}
\newcommand{\camspec}{{\tt CamSpec}}
\newcommand{\commander}{{\tt Commander}}
\newcommand{\plik}{{\tt Plik}}
\newcommand{\simall}{{\tt SimAll}}
\newcommand{\SROLL}{{\tt SRoll}}
\newcommand{\FEBECOP}{{\tt FEBeCoP}}
\newcommand{\Quickpol}{{\tt QuickPol}}
\newcommand{\ILC}{{\tt ILC}}
\newcommand{\SMICA}{{\tt SMICA}}
\newcommand{\NILC}{{\tt NILC}}
\newcommand{\SEVEM}{{\tt SEVEM}}
\newcommand{\LOW}{{\tt LOW}}
\newcommand{\HIGH}{{\tt HIGH}}
\newcommand{\FULL}{{\tt FULL}}

\newcommand{\lcdm}{$\Lambda$CDM}

\newcommand{\Alens}{A_{\rm L}}

\newcommand{\omegak}{\Omega_K}

\newcommand{\adjust}{\hspace{-0.11 truein}}

\providecommand{\Planck}{\textit{Planck}}
\providecommand{\Plancks}{\textit{Planck }}
\providecommand{\planck}{\Planck}

\providecommand{\text}[1]{\rm{#1}}

\newcommand{\Mpc}{\text{Mpc}}

\newcommand{\Hunit}{~\text{km}~\text{s}^{-1} \Mpc^{-1}}

\providecommand{\muK}{\mu\rm{K}}

\providecommand{\COSMOMC}{{\tt CosmoMC}}

\providecommand{\LCDM}{{$\rm{\Lambda CDM}$}}

\newcommand{\begm}{\begin{pmatrix}}
\newcommand{\enm}{\end{pmatrix}}

\newcommand\ba{\begin{eqnarray}}
\newcommand\ea{\end{eqnarray}}
\newcommand\bea{\begin{eqnarray}}
\newcommand\eea{\end{eqnarray}}

\newcommand\be{\begin{equation}}
\newcommand\ee{\end{equation}}

\def\pmb#1{\setbox0=\hbox{#1}%
    \kern-.025em\copy0\kern-\wd0
    \kern.05em\copy0\kern-\wd0
    \kern-.025em\raise.0433em\box0}
\def\ltsima{$\; \buildrel < \over \sim \;$}
\def\gtsima{$\; \buildrel > \over \sim \;$}
\def\simlt{\lower.5ex\hbox{\ltsima}}
\def\simgt{\lower.5ex\hbox{\gtsima}}
\def\muk{$(\mu{\rm K})^2$ }

\def\p2Y{\;_2Y}
\def\m2Y{\;_{-2}Y}
\newcounter{parentequation1}

\def\apjl{ApJ}%
\def\apjs{ApJS}%
\def\ao{Appl.~Opt.}%
\def\aap{A\&A}%
\begin{document}

\title{A Detailed Description of the CamSpec Likelihood Pipeline and
a Reanalysis of the Planck  High Frequency Maps}
\author{G. Efstathiou  and S. Gratton}
\email{gpe@ast.cam.ac.uk}
\affiliation{Kavli Institute for Cosmology Cambridge, 
Madingley Road, Cambridge,  CB3 OHA, UK.}
\begin{abstract}
This paper presents a detailed description of the \camspec\ likelihood
which has been used to analyse \Plancks temperature and polarization
maps of the cosmic microwave background since the first \Plancks data
release.  The goal of the CamSpec pipeline has been to extract an
accurate likelihood based on the TT, TE and EE spectra from Planck
which can be used to test cosmological models.  \Plancks is an
important legacy dataset which is likely to be reanalysed by many
researchers for many years to come.  Our aim in this paper is to
present, in a single source, a comprehensive analysis of our
methodology including what we have learned about: (a) the CMB sky and
associated foregrounds at the \Plancks high frequencies ($\nu \ge 100$
GHz); (b) the consistency of the \Plancks data in temperature and
polarization; (c) experimental systematics in the \Plancks data which
need to be corrected when building a likelihood.  For this paper we
have created a number of temperature and polarization likelihoods
using a range of Galactic sky masks and different methods of
temperature foreground cleaning. Our most powerful likelihood uses
80\% of the sky in temperature and polarization at \GE{143 and 217 GHz,}
increasing the effective sky coverage compared to the likelihoods used
in the 2018 \Plancks data release. Our results show that the base
six-parameter \LCDM\ cosmology provides an excellent fit to the
\planck\ data. There is no evidence for statistically significant
internal tensions in the \Planck\ TT, TE and EE spectra computed for
different frequency combinations.  The cosmological parameters of the
base \LCDM\ model are entirely consistent with those reported by the
\Planck\ collaboration in \citep{PCP18} and earlier \Planck\ papers,
though our most powerful likelihood tightens up the statistical
uncertainties and reduces the residuals of the TT, TE and EE spectra
relative to the best fit model. We present evidence that the
tendencies for the \Planck\ temperature power spectra to favour a
lensing amplitude $A_L>1$ and positive spatial curvature $\Omega_k <
0$ reported in \citep{PCP18} are caused by statistical fluctuations in
the temperature power spectra in the multipole range $800 \simlt \ell
\simlt 1600$, which are repeatable between detectors and
frequencies. Using our statistically most powerful likelihood,
combined with \ the 2018 \Planck\ low \ multipole \ likelihoods for
$\ell < 30$, we find that the $A_L$ parameter determined from the
\Planck\ power spectra alone differs from unity at no more than the
$2.2 \sigma$ level. We find no evidence for anomalous shifts of
cosmological parameters with multipole range. In fact, we show that
the combined TTTEEE \camspec\ likelihood over the restricted multipole
range $2 \le \ell \le 800$ gives cosmological parameters for the base
\LCDM\ cosmology that are very close to those derived from the full
multipole range $2 \le \ell \le 2500$. We present revised constraints
on a few extensions of the base \LCDM\ cosmology, focussing on the sum
of neutrino masses, the number of relativistic species and the
tensor-scalar ratio. The results presented here show that the
\Planck\ data are remarkably consistent between detector-sets,
frequencies and sky area. We find no evidence in our analysis that
cosmological parameters determined from the \camspec\ likelihood are
affected to any significant degree by systematic errors in the
\Planck\ data.

\end{abstract}
\makeatletter
\let\start@align@nopar\start@align
\let\start@gather@nopar\start@gather
\let\start@multline@nopar\start@multline
\long\def\start@align{\par\start@align@nopar}
\long\def\start@gather{\par\start@gather@nopar}
\long\def\start@multline{\par\start@multline@nopar}
\makeatother

\section{Introduction}
\label{sec:intro}

Since the discovery of the cosmic microwave background radiation (CMB)
in 1965 \citep{Penzias:1965}, observations of the CMB have provided a
wealth of new information on the early and late time Universe. The
\Planck\ satellite\footnote{\Planck\ (http://www.esa.int/Planck) is a
  project of the European Space Agency (ESA) with instruments provided
  by two scientific consortia funded by ESA member states (in
  particular the lead countries France and Italy), with contributions
  from NASA (USA), and telescope reflectors provided by a
  collaboration between ESA and a scientific consortium led and funded
  by Denmark} \citep{Planck_mission:2014, Planck_mission:2015} is the
third space mission dedicated to measuring anisotropies in the CMB,
following COBE \citep{Smoot:1992} and WMAP \citep{Bennett:2003,
  Bennett:2013}.  The first cosmological results from the
\Planck\ nominal mission temperature data were presented in
\citep{PCP13} and results for the full mission\footnote{The nominal
  mission comprises the first 15.5 months of data from \Planck. The
  full mission uses 29 months of data for the \Planck\ High Frequency
  Instrument (HFI) and 48 months for the Low Frequency Instrument
  (LFI).}, including polarization data, have been reported in
\citep{PCP15} and \citep{PCP18}. To extract cosmological information
from CMB data requires the construction of a likelihood. The
likelihoods used in the \Planck\ analysis are described in abbreviated
form in \citep{PPL13,PPL15, PPL18}\footnote{This paper will refer
  extensively to the \Planck\ 2013, 2015 and 2018 cosmological
  parameters papers \citep{PCP13, PCP15, PCP18} , which will
  henceforth be referred to as PCP13, PCP15 and PCP18. The
  corresponding likelihood papers will be referred to as PPL13, PPL15
  and PPL18.}.

\begin{figure}[t]
\centering
\includegraphics[width=150mm,angle=0]{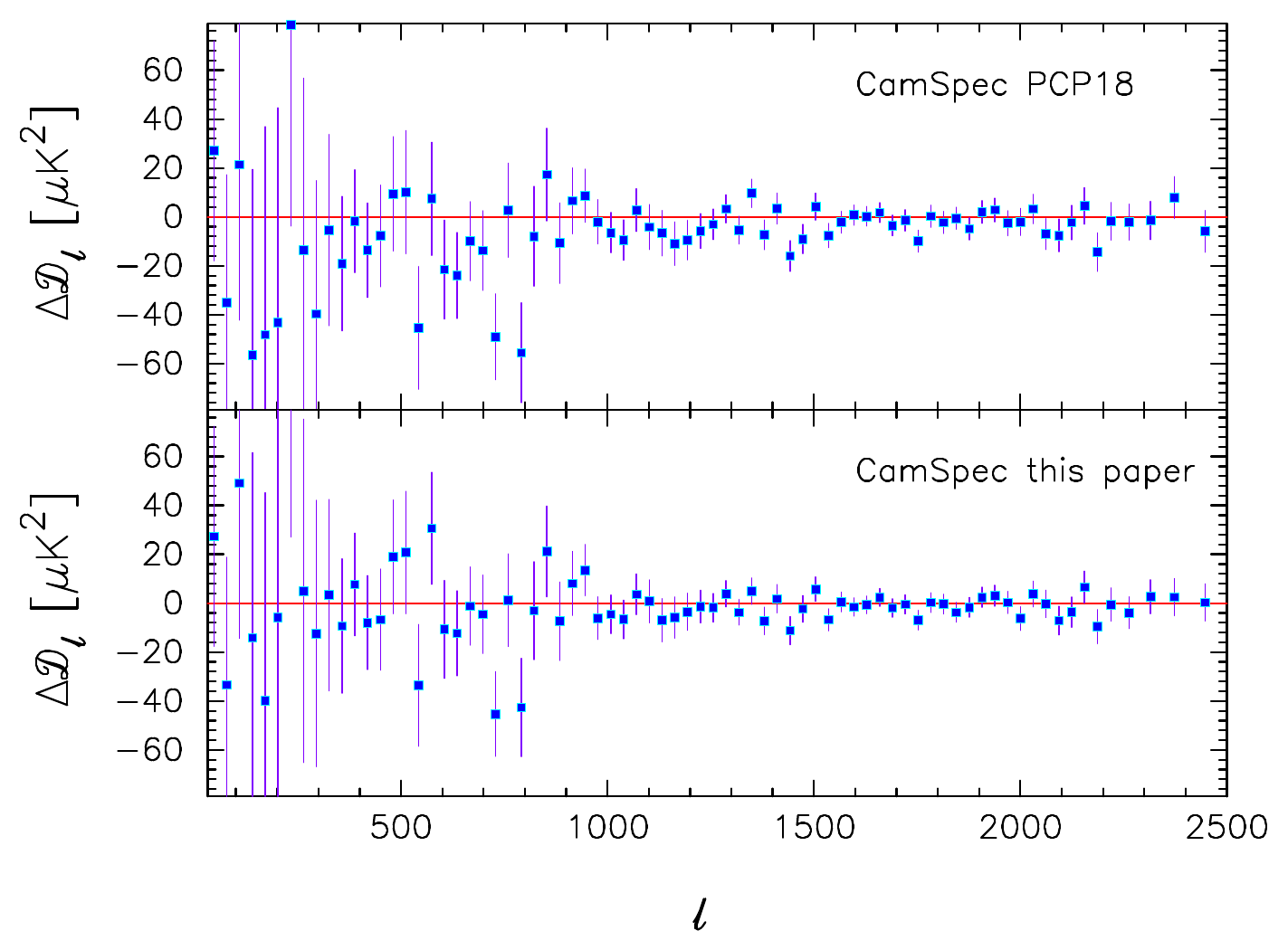} 
\caption {TT power spectrum residuals for the CamSpec likelihood as used in PCP18 (upper figure)
and for the most powerful likelihood (12.5HMcl) produced for this paper (lower figure). The 
residuals are computed with respect to the best-fit base \LCDM\ cosmology and foreground model 
fitted to the TT spectra at $\ell\ge 30$  in combination with the low multipole temperature
and polarization likelihoods at $\ell < 30$ (as discussed in Sect. \ref{sec:Likelihood}).}

\label{fig:fig1}

\vskip 0.3 truein

\end{figure}

In this paper we present a detailed description of the
\camspec\ likelihood that we developed and applied to \Planck\ in each
of the three \Planck\ data releases.  The \camspec\ likelihood has
been described in short form in the \Planck\ collaboration likelihood
papers PPL13, PPL15 and PPL18 and has been compared with the
\plik\ likelihood in PCP15 and PCP18.  \Planck\ is an important legacy
dataset and is likely to be analysed by many other researchers in the
future. We believe that it will be useful, particularly for
experimentalists who wish to combine \Plancks\ with ground based CMB
data, to present in a single source a detailed description of exactly
what we have done in constructing likelihoods for the \Plancks
collaboration. A second, and perhaps more significant, motivation for
this paper has been to address the consistency and fidelity of the
\Planck\ data. The \Planck\ data are consistent with the CMB
fluctuations predicted for a spatially flat Universe with a power-law
spectrum of scalar Gaussian adiabatic fluctuations. This model, which
we will refer to as the base \LCDM\ cosmology is described by six
parameters. The values of some of these parameters are not in perfect
agreement with some other data, for example direct measurements of the
Hubble constant (as will be discussed in
Sect.\ \ref{subsec:tensions}). It is therefore important to
demonstrate the consistency of the \Planck\ results. In developing
\camspec\ we focussed extensively on the fidelity of the
\Planck\ power spectra, testing for consistency between power spectra
determined from individual detector-sets and frequency combinations.
Such consistency checks are more direct than tests based on
consistency of cosmological parameters.  The third motivation for this
paper is to investigate a number of peculiar results reported in
PCP18.  These include the tendency of the \Planck\ temperature power
spectra to favour a lensing amplitude $A_L$ greater than
unity\footnote{See PCP18 for a definition of this parameter.} and to
favour closed universes. Neither of these results has been reported at
a high statistical significance (at $\simlt 3 \sigma$), but since they
could be signs of new physics, internal tensions within the
\Planck\ data, or small inconsistencies in the construction of the
\Plancks likelihoods, we felt that a closer investigation was
merited. We have therefore created statistically more powerful
\camspec\ likelihoods than those used in PPL18 and PCP18, primarily by
extending sky coverage\footnote{The 12.1HMcl likelihood is available
  as a \COSMOMC\ module at the following web site:
  \tt{https://people.ast.cam.ac.uk/$\sim$stg20/camspec/index.html}.}.
To motivate the reader, the upper panel in Fig. \ref{fig:fig1} shows
the residuals of the PCP18 \camspec\ coadded TT
spectrum\footnote{Which is almost identical to the coadded PCP18
  \plik\ TT spectrum.} with respect to the best-fit base \LCDM\ and
foreground model.  The lower panel in Fig. \ref{fig:fig1} shows
residuals for the 12.5HMcl likelihood discussed in this paper which
increases the sky area at 143 and 217 GHz compared to the likelihoods
used in the \Planck\ legacy papers. The base \LCDM\ cosmology fitted
to this likelihood is almost identical to the best fit base
\LCDM\ cosmology presented in PCP18 (see Sect. \ref{sec:base_lcdm})
but the residuals are {\it visibly} smaller.  In other words, by
constructing a more powerful likelihood, {\it the CMB power spectrum
  tightens up around the predictions of the base \LCDM\ cosmology.}
This is strong evidence in support of the \LCDM\ model as will be
discussed in detail in Sect. \ref{sec:base_lcdm}.

  A short summary of the mathematical framework underlying
  \camspec\ for both temperature and polarization is presented in
  Sect.\ \ref{sec:spectra}, with details relegated to  Appendix \ref{sec:appendix}.  The
  application of this theoretical framework to real CMB data requires:
  (a) an accurate model of unresolved foregrounds, including Galactic
  contamination; (b) accurate models of beams and instrumental noise;
  (c) control of instrumental systematics. Since (a)-(c) can lead to
  biases in the cosmology, a large part of this paper is devoted to
  these aspects of the analysis. Various choices need to be made to
  construct a likelihood, for example, the choices of sky cuts,
  multipole ranges and methods of foreground removal.  We present the
  rationale behind these choices and investigate the robustness of the
  \camspec\ results to these choices.

The rest of this paper is divided into four distinct blocks:

\smallskip 

\noindent
[1] {\it Sections \ref{sec:foreground_masks} - \ref{sec:beams}: Preliminaries and instrumental effects.}

Section \ref{sec:foreground_masks} summarizes the Galactic temperature
and polarization masks used in this paper. Section
\ref{sec:cross_spectra} discusses the maps that we have used to
estimate cross-spectra. PCP13 used nominal mission detector-set maps
whereas the likelihoods used in PCP15 and PCP18 used half mission
cross spectra. In this paper, we compare half mission spectra with
spectra constructed from the full mission detector-set maps.
Section \ref{sec:noise} discusses ways of estimating detector noise
and analyses correlated noise between detectors. Beam corrections and
polarization efficiencies are described in Sect. \ref{sec:beams}. This
section presents a detailed intercomparison of the temperature spectra
power spectra measured by different detectors and presents a
cross-check of the temperature to polarization leakage corrections
applied to the polarization spectra.

\smallskip 

\noindent
[2] {\it Sections \ref{sec:dust_temp} - \ref{sec:nuisance}: Galactic dust emission in temperature and polarization, extragalactic 
foregrounds and nuisance parameters.}

Section \ref{sec:dust_temp} presents an analysis of Galactic dust
emission in temperature. In early versions of \camspec\ (see PCP13) we
corrected for Galactic dust emission in temperature by constructing
and fitting dust power spectrum templates together with template power
spectra for extragalactic foregrounds. The construction of such
templates is discussed in  Sect.\ \ref{sec:dust_temp} 
together with an analysis of
the universality of the dust power spectrum as a function of frequency
and sky coverage. We analyse CMB-dust correlations, which introduce
substantial additional scatter to the power spectrum estimates,
especially at 217 GHz, limiting the sky area that can be used reliably
at this frequency. This motivates an alternative method of removing
Galactic dust by subtracting high frequency maps (as described by
\cite{Lueker:2010,Spergel:2015} and PPL15). We demonstrate that `cleaning' 
the $143\times 143$, $143 \times 217$ and $217 \times 217$ spectra\footnote{The notation $143 \times 143$ denotes the cross spectrum of two 143 GHz maps. In later sections we will use more specific notation, for example $143 {\rm HM1} \times143 {\rm HM2}$ denotes the cross spectrum of a first half mission 143 GHz map with
a second half mission 143 GHz map.}
with higher \Planck\ frequencies removes Galactic dust very accurately
leaving residual power-spectrum contributions from extragalactic
foregrounds that are well described by power-laws.  In this paper we construct
likelihoods using the standard power-spectrum template based foreground model, as
described in previous \Planck\ papers, and we also construct high frequency   `cleaned' likelihoods using
a much simpler foreground model. Comparison of these likelihoods gives
an indication of residual uncertainties in 
 the cosmological results associated with temperature
foreground modelling. We present, for the first time,  a detailed analysis to
demonstrate that  high frequency cleaning can be used 
to extend the sky coverage at $143$ and $217$ GHz reliably to 80\% of
the sky. Extragalactic foregrounds in polarization are well below the sensitivity level
of \Planck. All of our likelihoods use $353$ GHz maps to subtract polarizated dust emission,
as discussed in Sect. \ref{sec:dust_polarization}. Instrumental nuisance parameters and
the extragalactic foreground templates are discussed in Sect. \ref{sec:nuisance}, together
with the priors adopted in the likelihood analysis. We have made minor changes to the 
foreground/nuisance model used in PCP18. Here we fix the relative calibrations of the cross-spectra,
rather than carrying them as nuisance parameters, since they can be determined to high accuracy as described
in Sect. \ref{subsubsec:inter_frequency}; we also allow the amplitude of the Cosmic Infrared Background (CIB) contribution to the
 $143 \times 217$ spectrum to vary independently of the amplitude in the
 $217 \times 217$ spectrum. These changes have relatively little impact on cosmological parameters.

\smallskip

\noindent
[3] {\it Sections \ref{sec:Likelihood} - \ref{sec:inter_frequency_pol}: Likelihoods and inter-frequency comparisons of power spectra.}

Comparison of temperature power spectra at different frequencies requires a
likelihood analysis to determine the foreground parameters. In this
part of the paper, we adopt the six parameter base
\LCDM\ model. Section \ref{sec:inter_frequency} compares the
consistency of the temperature spectra in the half mission, cleaned
and full mission likelihoods and analyses spectrum residuals as a
function of sky coverage.  Section \ref{sec:inter_frequency_pol}
presents a similar analysis for the TE and EE
spectra cleaned with $353$ GHz. We also compare our spectra with the spectra used to form the
low multipole ($\ell < 30$) temperature and polarization likelihoods.

\smallskip 

\noindent
[4] {\it Sections \ref{sec:base_lcdm} - \ref{sec:extensions_lcdm}: Science results.}

Section \ref{sec:base_lcdm} discusses cosmological parameters for the
base \LCDM\ model,  demonstrating the consistency of the cosmological
parameters determined from various  sky areas, temperature
and polarization combinations, and different methods of
temperature foreground cleaning. As highlighted in Fig.\ \ref{fig:fig1},
by extending the sky coverage in temperature and polarization
we have created more powerful likelihoods than those discussed in the 2018 \Plancks legacy papers. This allows 
more sensitive tests of consistency with, and deviations from,  the base \LCDM\ cosmology.
  We therefore revisit the consistency of
cosmological parameters determined from different multipole ranges, the significance of oscillatory 
features in the temperature power spectra,  and
possible tensions of the base \LCDM\ model with other astrophysical data. One-parameter 
extensions to the base \LCDM\ model are
discussed in Sect. \ref{sec:extensions_lcdm}. Since, for the most part, our results are
consistent with those given in PCP18, we do not present a 
comprehensive analysis of extended models, but instead focus mainly on the
parameters $A_L$ and $\omegak$ for which there were hints of
anomalies in PCP15 and PCP18 and for which the \Planck\ papers
reported  differences between the \camspec\ and \plik\ likelihoods.
A comparison of our best fit \LCDM\ cosmology with the power 
spectra measured from ground based polarization experiments \citep{Louis:2017, Henning:2018, Choi:2020, Aiola:2020, Dutcher:2021}
is presented in Appendix \ref{sec:appendix2}.

\GE{
To further guide the reader we note the main differences between the \camspec\ analyses presented in PCP18
and in this paper:
\medskip

\noindent
$\bullet$ PCP18 presented two TT likelihoods using approximately
$60\%$, $70\%$ and $80\%$ of sky at 217, 143 and 100 GHz respectively
(mask60, mask70, mask80, see Fig. \ref{fig:tempmasks} and Table
\ref{tab:sky_fractions}). The default TT likelihood used the standard
multi-parametric foreground model described in PCP18 and earlier
\Planck\ papers (see also Sect.~\ref{sec:nuisance}).  However, we
also produced a 545 GHz `cleaned' TT likelihood which allowed the use of  a much
simpler heuristic model for the TT foreground contributions to the
143$\times$143, 143$\times$217 and 217$\times$217 spectra. In PCP18 we showed that the
spectra and cosmological parameters from the default and cleaned
\camspec\ likelihoods were in very good agreement. In this paper, we
increase the sky coverage in temperature to $80\%$ of the sky at 143 and
217 GHz. This can only be done reliably by constructing `cleaned'
likelihoods (to  eliminate noise arising from
CMB-foreground cross correlations). Thus a large part of this paper is
devoted to demonstrating that the residual cleaned foreground
contributions to the TT power spectra at 143 and 217 GHz are
statistically isotropic on the sky and therefore of extragalactic
orgin as sky coverage is increased (see
e.g. Fig. \ref{fig:inter_frequency143v217cleaned}).

\smallskip

\noindent
$\bullet$ We demonstrate that over $80\%$ of the sky, $143$ GHz maps dust-cleaned with $545$ GHz are `cleaner' than $100$ GHz maps cleaned with
$545$ GHz. This is because  at large angular scales $100$ GHz maps  contain low levels of
synchrotron and CO emission (even though we apply a CO mask at 100 GHz
see Fig. \ref{fig:pgcalib}). At small angular scales, the $100$ GHz maps are noisy
and are contaminated by extragalactic radio sources and thermal Sunyaev-Zeldovich
fluctuations. In  this paper we discard the $100 \times 100$ spectra from the cleaned TT
likelihood, significantly reducing the number of parameters required
to model TT foregrounds at negligible loss to the science of interest.

\smallskip

\noindent
$\bullet$ In PCP18 we allowed relative effective calibrations of the TT spectra
with (over-)generous priors to vary along with the foreground
parameters. We show in this paper that using 545 GHz cleaning, the
relative calibrations of the TT spectra can be determined to such high
accuracy that they do not need to be carried as nuisance parameters
(see Sect. \ref{sec:nuisance}).

\smallskip

\noindent
$\bullet$ The polarization analysis in PCP18 used approximately $60\%$ of sky at all
frequencies with a specially constructed polarization mask
(maskpol60, see Fig.~\ref{Polmasks}). In this paper, we construct
likelihoods using up to $80\%$ of sky in polarization at all
frequencies.

\medskip 

It is also worth summarizing the main differences between \plik,
used as the baseline likelihood in PCP18, and the \camspec\ likelihoods 
discussed in this paper:

\medskip

\noindent
$\bullet$ The `uncleaned' \camspec\ and \plik\   TT likelihoods are essentially identical
and give almost identical cosmological parameters as discussed in PCP18. In this paper,
we use 545 GHz cleaning to extend the sky coverage in temperature.
 
\medskip

\noindent
$\bullet$ In \camspec\ we remove polarized Galactic dust emission at
low multipoles ($\ell < 150$) by subtracting 353 GHz polarization
maps. At higher multipoles (where polarized foregrounds are sub-dominant to the
primordial fluctuations) we remove the contribution at Galactic
dust emission at the power spectrum level by subtracting power-law
dust models fitted to low multipoles. Since there are no other
detectable foreground contributions in polarization, we compress the
dust corrected polarization spectra to form frequency averaged TE and
EE spectra. The combined TTTEEE \camspec\ likelihoods are compact and
there is therefore computational requirement to band average the spectra. Since dust
emission has already been removed from the TE and EE spectra to high
accuracy, there is no need to carry nuisance parameters in the likelihood associated
with polarized dust emission. In contrast, the
\plik\ likelihood retains all distinct spectral combinations in the TE
and EE blocks (note that in \plik\ the TE and ET spectra are averaged).
Twelve additional nuisance parameters are then required to
characterise Galactic dust emission in TE and EE. In addition, to make
the \plik\ TTTEEE likelihood compact, the default \plik\ likelihood is
band averaged as described in Sect. 3.2.5 of PPL15.

\medskip

\noindent
$\bullet$ In \camspec\ we  calibrate each individual TE and EE
spectrum assuming a fiducial cosmological model to determine effective
(`spectrum-based') polarization efficiencies. The \plik\ team chose to use `map-based'
effective polarization efficiencies computed from the EE spectra to
correct the EE {\it and} TE spectra.  This procedure results in poor
$\chi^2$ values for the \plik\ TE spectra (cf Table 20 for PPL18).

\medskip

The  differences between \camspec\ and \plik\ are therefore mainly
in the TE and EE blocks of the likelihoods. Nevertheless, for almost
all cosmological models, the two likelihoods give closely similar
results. As shown in PCP18, the main science conclusions of that paper
would have been unaltered had \camspec\ been used as the baseline in
place of \plik. For some extended models, particularly $A_L$ and
$\Omega_k$, the \camspec\ TTTEEE likelihoods used in PCP18 (and in this
paper)  give results that are more consistent with the base \LCDM\ model
than those from \plik. These differences come almost entirely from the TE blocks
of the likelihood, since for \Planck\ EE is so noisy that the EE block contributes
little statistical weight to the TTTEEE likelihood. The main differences between the 
\camspec\ and \plik\ TTTEEE likelihoods can be traced  to the polarization 
efficiency corrections; \plik\ TTTEEE using spectrum-based polarization efficiencies
comes into closer agreement with \camspec\ (see Sect. 2.2.1 of PCP18).  In our view,
the \camspec\ TE efficiencies are internally self-consistent, since they lead to
acceptable $\chi^2$ values for  \LCDM-like models for the statistically dominant TT and TE  blocks 
of the TTTEEE likelihood. Readers who are unpersuaded by this argument should at the very least
treat the differences between \camspec\ and \plik\ TTTEEE as indicative of errors arising from 
inaccuracies in the calibrations of polarization  efficiencies and angles of the  \Planck\ HFI detectors.
}

\medskip

Since most of this paper is technical in nature, readers interested
only in the final cosmological results can skip to
Sect. \ref{sec:base_lcdm}, though the earlier sections are essential
reading if one wants to acquire an appreciation of the fidelity of the results. Our conclusions are summarized in
Sect. \ref{sec:conclusions}. Throughout this paper, we use the same
notation and definitions of cosmological parameters as in PCP18.

\section{Spectra and covariance matrices}
\label{sec:spectra}

This section presents a summary of the mathematical
framework developed for the \camspec\ pipeline. Analytic expressions for the covariance matrices
have been presented in \citep{Efstathiou:2004, Challinor:2005, Efstathiou:2006, Hamimeche:2008}
and are summarized in Appendix \ref{subsec:covariance_matrices}. These are based
on a number of idealised assumptions which do not apply exactly to the real \Planck\ data. We discuss
the mismatch between the theoretical framework  and the real data in this
section.

\subsection{Pseudo-cross spectra}
\label{subsec:PCL}

The \camspec\ likelihood uses pseudo-cross spectra computed on masked
skies. Since the masks are apodised (see
Sect.\ \ref{sec:foreground_masks}), they are described by weight
functions $w^T_p$ for temperature and $w^P_p$ for $Q$ and $U$
polarization maps at each map pixel $p$. Note that we always apply
identical weight functions to $Q$ and $U$ maps.\footnote{\GE{To achieve close to minimum variance (see the discussion in 
\cite{Efstathiou:2006}), 
the power spectra should be computed by assigning equal weight to each pixel in the signal dominated regime
(a good approximation for the TT spectra over most of the multipole range covered by \Planck) and inverse-noise
variance weighting for noise dominated spectra (a good approximation for the EE spectra over most of the multipole range covered by \Planck). To keep the computations of the covariance matrices simple, we chose to use the apodised weight functions $w_p^T$ and $w_p^P$ (i.e. equal weight per pixel over most of the sky) paying a small penalty in the statistical power
 of the polarization spectra.}}.

For a particular  weighting scheme we compute the following pseudo-spectra from maps $i$ and $j$ expressed as a vector:
\begin{equation}
 {\bf \rtensor C}^{ij}  
 =  (\rtensor C^{T_iT_j}_\ell,  \rtensor C^{T_iE_j}_\ell, \rtensor C^{E_iT_j}_\ell, 
\rtensor C^{E_iE_j}_\ell , 
 \rtensor C^{B_iB_j}_\ell, \rtensor C^{E_iB_j}_\ell, \rtensor C^{B_iE_j}_\ell,  
\rtensor C^{T_iB_j}_\ell,  
         \rtensor C^{B_iT_j}_\ell)^T.  \label{equ:C1}
\end{equation}

The pseudo-spectra of equation (\ref{equ:C1}) are constructed from the following transforms:
\begin{subequations}
\begin{equation}
 \rtensor a^{T_i}_{\ell m} = \sum_p  (T_i)_p w^{T_i}_p 
\Omega_p Y^*_{\ell m}({\pmb{$\theta$}}_p),    \label{equ:C2a}
\end{equation}
\begin{equation}
 \rtensor a^{E_i}_{\ell m} = -{1 \over 2}\sum_p (Q_i+iU_i)_p w^{P_i}_p \Omega_p \p2Y^*_{\ell m}({\pmb{$\theta$}}_p)
+ (Q_i-iU_i)_p w^{P_i}_p \Omega_p \m2Y^*_{\ell m}({\pmb{$\theta$}}_p),    \label{equ:C2b}
\end{equation}
\begin{equation}
 \rtensor a^{B_i}_{\ell m} = -{1 \over 2}\sum_p (U_i-iQ_i) w^{P_i}_p \Omega_p \p2Y^*_{\ell m}({\pmb{$\theta$}}_p)
+ (U_i+iQ_i)_p w^{P_i}_p \Omega_p \m2Y^*_{\ell m}({\pmb{$\theta$}}_p),    \label{equ:C2c}
\end{equation}
\end{subequations}
where the sums extend over the number of map pixels each of solid angle $\Omega_p$. The pseudo-power spectra are then computed in the usual way, for example,
\begin{equation}
\rtensor C^{T_iT_j}_\ell = {1 \over (2 \ell + 1)} \sum_m \rtensor  a^{T_i}_{\ell m}
\rtensor  a^{*T_j}_{\ell m}. \label{equ:C3}
\end{equation}

For the majority of this paper, we use the  \Planck\ 2018 HFI half-mission frequency 
maps in Healpix format \citep{Gorski:2005} at a resolution NSIDE=2048 available from the \Planck\ Legacy Archive\footnote{https://pla.esac.int.} (hereafter PLA). The only preprocessing applied to these maps before
applying the transforms \ref{equ:C2a}-\ref{equ:C2c} is to remove the means within
the unmasked area of sky. For example, for map $T_i$ we subtract the mean
\begin{equation}
 (T^i)_{\rm mean}  =  \sum_p w^{T_i}_p (T_i)_p / \sum_p w^{T_i}_p, \label{equ:C4}
\end{equation}
(and similarly for the $Q_i$ and $U_i$ maps) \GE{to eliminate  Galactic emission at low lattitudes leaking to higher multipoles}. We make no other corrections to the maps prior to transformation.

The expectation values of the pseudo-spectral estimates are related to the beam convolved theoretical spectra ${\bf \bar C}^{ij}$ via a coupling matrix ${\bf K}^{ij}$ \citep{Hivon:2002, Kogut:2003}. The components of the coupling matrix are given in Sect.\ \ref{subsec:coupling_matrices}.  At this stage, for clarity we recap on the notation used for various power spectra:

\noindent
$\bullet$ $\rtensor{C}_\ell$ is the beam convolved spectrum computed on the incomplete sky.

\noindent
$\bullet$  $\tilde C_\ell$ is a beam corrected spectrum computed on incomplete sky.

\noindent
$\bullet$ $\hat C_\ell$ is a beam corrected spectrum deconvolved for the sky mask.

\noindent
$\bullet$ $\bar C_\ell$ is a beam convolved theoretical spectrum

\noindent
$\bullet$ $C_\ell$ is the theoretical spectrum.

The power spectrum estimates need to be deconvolved for the effects of
the sky mask (described by the coupling matrix ${\bf K}^{ij}$), the
\Planck\ instrumental beams and the effects of the finite size of the
sky pixels, which we assume are described by functions that depend
only on multipole $\ell$.  To simplify the discussion, we will discuss
estimates of the temperature power spectrum. The power spectra
measured on the incomplete sky are related to the theoretical power
spectrum as \
\begin{subequations}
\begin{eqnarray}
  \langle \rtensor{C}^{TT}_\ell \rangle &=& \sum_{\ \ell^\prime} K^{TT}_{\ell\ell^\prime} \bar C^{TT}_{\ell^\prime} , \label{equ:C5a} \\
       & = &   \sum_{ \ell^\prime} K^{TT}_{\ell\ell^\prime}  C^{TT}_{\ell^\prime} W^{TTTT}_{\ell^\prime} \pi^2_{\ell^\prime},  \label{equ:C5b}
\end{eqnarray}
\end{subequations}
where $W^{TTTT}_{\ell}$ is the TTTT beam transfer function for the
particular pair of maps $(i, j)$ used to compute $\rtensor {C^{TT}}$
and $\pi_\ell$ is the correction for finite pixel size returned by the
\healpix\ routine {\tt pixel\_window} (which we have verified, by
simulation, accurately describes the effects of the finite pixel size
for the sky masks used in this paper). The beam transfer functions vary slightly
depending on how much sky is excluded by the masks; in our analysis we
use beam transfer functions computed \GE{for the smoothed apodised masks shown  Fig. \ref{fig:tempmasks}
  used to estimate the power spectra, ignoring  small differences arising from missing pixels,
  masks for point source holes, extended sources
and CO emission (which account for such a small sky area that they have negligible impact on the
isotropised beam transfer functions)}.  We then form the spectra
$\tilde C^{TT}_\ell$ and $\hat C^{TT}_\ell$: 
\begin{subequations}
\begin{eqnarray}
    \tilde C^{TT}_\ell &=& \rtensor{C}^{TT}_\ell/(W^{TTTT}_\ell \pi^2_\ell), \label{equ:C6a} \\
    \hat C^{TT}_\ell & = & \left (\sum_{ \ell^\prime} (K^{TT})^{-1}_{\ell \ell^\prime}\rtensor{C}^{TT}_{\ell^\prime}\right)/(W^{TTTT}_\ell \pi^2_\ell). \label{equ:C6b} 
\end{eqnarray}
\end{subequations}
and take $\hat C_\ell$ as our estimate of the theory spectrum $C_\ell$. Note that 
Eq. \ref{equ:C6b} assumes that the product $W^{TTTT}_\ell\pi^2_\ell$ is much broader than the 
width of the mask coupling matrix $(K^{TT})^{-1}_{\ell \ell^\prime}$. The generalization of these equations to 
polarized beams is discussed in Sect.\ \ref{subsec:planck_beams}.

In almost all of this paper, we show plots of the mask-deconvolved, beam- and pixel-window corrected spectra
\begin{equation}
\hat D_\ell \equiv   {\ell (\ell+1) \over 2 \pi} \hat C_\ell, \label{equ:C7}
\end{equation} 
  usually omitting
the circumflex accent. We will, however, apply accents rigorously if we display other spectra.

\subsection{Covariance Matrices}
\label{subsec:camspec_covariance_matrices}

\camspec\ uses analytic approximations to the covariance matrices of
the pseudo-spectra derived under the assumptions of narrow window
functions and uncorrelated, but anisotropic,  pixel noise
($(\sigma^T_i)^2$, $(\sigma^Q_i)^2$, $(\sigma^U_i)^2$)
\citep{Efstathiou:2004, Challinor:2005, Efstathiou:2006,
  Hamimeche:2008}. The components of the covariance matrix ${\bf M}$
are given in Sect.\ \ref{subsec:covariance_matrices}. Idealised
simulations \citep{Efstathiou:2004, Efstathiou:2006} show that these
expressions are accurate to typically percent level precision at high
multipoles for typical Galactic sky masks for TT, TE and EE spectra,
but are only accurate at the $\sim 10 \%$ level for spectra involving
B-modes.

In \camspec, we compute all of the spectra in Eq.\ \ref{equ:C1} and all
of the mask coupling matrices for all detector combinations.  In the
absence of parity violating physics in the early universe, the
primordial $C^{TB}$ and $C^{EB}$ spectra should be identically
zero. In the absence of tensor modes, the BB spectra should also be
zero apart from a small contribution from gravitational
lensing. Although we compute these spectra, they are used primarily as
diagnostic tools to test for systematics in the data.  The current version
of the \camspec\ temperature-polarization likelihood uses {\it only}
the TT, TE, ET and EE spectra and associated covariance matrices.

The \camspec\ covariance matrices are based on a number of approximations:

\smallskip 

\noindent
[1] For realistic experiments such as \Planck\ the noise is
non-white. As described in PPL13, in \camspec\ we adopt a heuristic
prescription for dealing with non-white noise by multiplying noise
weight functions (see Eqs.\ \ref{equ:A3b}-\ref{equ:A3d}) with
$\ell$-dependent functions, $\psi_\ell$, fitted to the noise power
spectra (see Sect.\ \ref{subsec:covariance_matrices}).  In
Sect.\ \ref{sec:noise}, we discuss ways of estimating noise directly
from the maps.  Inaccuracies in the noise model are the main source of
error in the \camspec\ polarization covariance matrices.  Correlated
noise between maps is demonstrably small (see
Sect. \ref{subsec:correlatednoise}) and is ignored in the covariance
matrices.

\smallskip

\noindent
[2] Foregrounds can make a significant contribution to the spectra and
are included in the covariance matrices by adding a best-fit
foreground model to the fiducial primordial CMB model. This assumes
that the foregrounds are well approximated as isotropic Gaussian
random fields. This is a good approximation at high multipoles, where
the dominant foreground contributions are extragalactic, but clearly
fails at multipoles $\ell \simlt 500$ where Galactic dust emission,
which is anisotropic on the sky, becomes the dominant foreground
contribution. A technique for incorporating anisotropic dust emission
into the covariance matrices is described in \citep{Mak:2017}, but
since Galactic dust emission in temperature is small compared to the
primordial CMB in the frequency range $100-217$ GHz we do not
implement the prescription of reference \citep{Mak:2017} in the
\camspec\ covariance matrices. For `uncleaned' temperature
likelihoods, designed for the standard temperature foreground model,
the covariance matrices include a \GE{Galactic} dust contribution under the assumption of
Gaussianity and isotropy, and so 
 underestimate\footnote{\GE {As the model of \citep{Mak:2017} shows, the effective 
Galactic dust amplitude relevant for CMB-Galactic dust correlations is weighted towards the edges of the mask and so is higher than the mean Galactic dust amplitude measured in the isotropised power spectra.}} the amplitude of CMB-\GE{Galactic dust}
correlations.
For high frequency `cleaned' temperature likelihoods,
CMB-dust correlations are strongly suppressed (see
Sect. \ref{subsec:spec_clean}) so dust is ignored in computing the
the covariance matrices. \GE{Also, the consistency between cleaned and uncleaned likelihoods,
which have very different levels of extragalactic foreground at 143 and 217 GHz,
demonstrates that complexities such as non-Gaussianity of the extragalactic foregrounds have no significant impact 
for \Planck.}
 All of the \camspec\ polarization spectra are
cleaned for polarized Galactic dust emission at low multipoles using
353 GHz polarization maps, as described in
Sect. \ref{sec:dust_polarization}.

\smallskip

\noindent
[3] Point source holes and missing pixels increase the effective widths of the window functions
introducing `leakage' of large-scale power to smaller scales. This leakage can introduce errors
in the analytic covariance matrices. However, experimentation with numerical simulations and
high pass filtered maps have shown that these errors have negligible effect on cosmological parameters
and so they are ignored. 

The covariance matrices require a fiducial theoretical power spectrum. In this paper, we use the \camspec\
TT base \LCDM\ best fit power spectrum from PCP18. 
The number of covariance matrices required to form a likelihood scales
as $N_{\rm map}^4$ and becomes prohibitively expensive as the number
of spectra becomes large. To reduce the number of operations, we
(usually) adopt the same polarization mask at all frequencies, though
this may differ from the masks applied to the temperature maps. The \Planck\ maps contain `missing' pixels, defined as
pixels which are either not scanned by \Planck, or for which the
map-making algorithm cannot return a reliable solution for T,Q and U
(see \citep{DataProcessing:2018}). The number of missing pixels is
small for half mission and full mission coadded frequency maps, but
can become significant for individual detector set (hereafter detset) maps (see Sect. \ref{subsec:maps}).  In \camspec,
we therefore compute coupling matrices ${\bf K}$ for each spectrum and
map combination including missing pixels, but ignore differences in
missing pixels when we compute the covariance matrices.  This
dramatically reduces the computational cost of computing covariance
matrices for all detector combinations for very little loss in
computational accuracy.

Following these computations, we end up with covariance matrices
for the TT, TE, ET and EE components of the data vectors ${\bf \tilde C^{ij}_\ell}$,
${\bf \hat C^{ij}_\ell}$ and also all cross-covariances. We can then easily compute covariance matrices for any linear
combination of these spectra.

\subsection{Data compression}

Since we have a relatively large number of maps (especially if we are analysing detset maps), 
the full cross-spectrum data vector 
and associated covariance matrix would be very large. To  make the computation of a high-$\ell$ 
likelihood fast enough for parameter estimation without any band-averaging  of the spectra\footnote{Avoiding band averaging
is important if one wants to test models that predict high frequency oscillatory power spectra, for example
axion monodromy  \cite{Flauger:2017}.}  we compress the 
the data vector. We therefore discard all of the spectra involving B modes, retaining only the TT, TE, ET and EE spectra.

In principle all of the mask/beam deconvolved temperature
cross-spectra {\it within} a given frequency combination should be
identical to within the levels set by instrument noise. These can then
be compressed into a single power spectrum estimate with negligible loss
of information. However,  we do not average across frequency combinations since
the unresolved temperature foregrounds depend on frequency. Further compression
can be accomplished only after unresolved temperature foreground parameters have
been determined via likelihood analysis\footnote{For example, as has
  been done to create the {\tt plik$\_$lite} likelihoods described in
  PPL15 and PPL18. }.

In temperature, we  form a linear combination of individual cross-spectra, 
\begin{equation}
  \hat C^{kT}_\ell =  \sum_{ij\subset k,  i\ne j} \alpha^{TT_{ij}}_\ell c_i c_j{\hat C^{T_{ij}}_\ell}.  \label{equ:C8}
\end{equation}
Here the index $k$ denotes each distinct  frequency cross-spectrum combination retained in the likelihood
({\it e.g.} $100\times100$, $143\times 217$, $\dots$) and the coefficients $c_i$
denote the relative calibration factors for each map.
The coefficients $\alpha_{ij}$ are normalized so that 
\begin{equation}
\sum_{ij\subset k,  i\ne j} \alpha^{TT_{ij}}_\ell  = 1, \qquad  \alpha^{TT_{ii}}_\ell = 0.  \label{equ:C9}
\end{equation}

To determine the coefficients $\alpha_{ij}$ we adopt another simplifying assumption: an optimal 
linear combination,  $\hat X^k_\ell$,  is given by solving 
\begin{equation}
\sum_{pq\subset k, p\ne q} \hat{\cal{M}}^{-1}_{pq} \hat X^k_{\ell} = \sum_{pq\subset k, p\ne q} \hat{\cal{M}}^{-1}_{pq}
\hat X^{pq}_\ell, \label{equ:C10}
\end{equation}
where $\hat{\cal{M}}^{-1}_{pq}$ is the block of the inverse covariance matrix appropriate to the spectrum combination $k$. If the covariance matrix $\hat{\cal{M}}$ accurately describes the data, the solution of Eq.\ (\ref{equ:C10}) properly accounts
for the correlations between the cross-spectra. However, solving Eq.\ (\ref{equ:C10}) requires the inversion of a
very large matrix, and so we adopt a simpler solution by  weighting each cross spectum  estimate 
by the diagonal component of the relevant covariance matrix, {\it i.e.}
\begin{equation}
\alpha^{TT_{ij}}_\ell  \propto  1 / {\rm Cov} (\hat
    C^{T_{ij}}_\ell \hat C^{T_{ij}}_{\ell}).  \label{equ:C11}
\end{equation}
This has the effect of assigning each cross-spectrum equal weight in
the signal dominated regime and inverse variance weighting in the
noise dominated regime, which is qualitatively correct but will be slightly
sub-optimal compared to solving Eq. (\ref{equ:C10}).  Thus, in temperature, we compress all
cross-spectra within a particular frequency combination into a single
cross-spectrum. This means that it is straightforward to compare, for
example, coadded full mission with half mission cross spectra. It is,
however, important to test the consistency of spectra within each
frequency combination prior to coaddition (see
Sect. \ref{subsec:intra_frequency}).

For TE and EE spectra, we adopt a different approach. For TE and EE,
the only frequency dependent foreground contribution detected in the
Planck spectra is polarized Galactic dust emission. This affects the
polarization spectra at multipoles $\simlt 500$ (see
Sect. \ref{sec:dust_polarization}). There is no evidence for a 
frequency dependent contribution from polarized point sources at high
multipoles at the \Planck\ sensitivity levels, consistent with the
results of high resolution ground-based CMB experiments \citep{Louis:2017,
  Henning:2018}. We therefore `clean' each of the TE and EE spectra
using 353 GHz maps as dust templates as described in
Sect. \ref{sec:dust_polarization}. With Galactic dust emission cleaned
at low multipoles, we coadd all frequency combinations of TE and EE
spectra using inverse diagonal weighting, analogous to
Eq. \ref{equ:C11}, to form a single coadded TE and a signle coadded EE spectrum {\it with
no further free parameters to model polarized foregrounds}.

The compressed data vector in the standard version of
\camspec\ consists of four TT spectra involving the frequency
combinations $100 \times 100$, $143 \times 143$, $143 \times 217$ and
$217 \times 217$ spectra, together with TE and EE spectra coadded over
all polarized HFI frequencies with the exception of $353$ GHz, which
is used as a Galactic dust template. We do not retain the $100 \times
143$ and $100 \times 217$ TT spectra since they add very little new
information on the primordial CMB, but would require carrying
additional foreground nuisance parameters. (See PCP13 for a
  discussion, and plots,  of the \Planck\ $100 \times 143$ and $100 \times 217$ TT
  spectra.)  We also produce `cleaned' likelihoods, in which we
subtract Galactic dust emission in temperature using 545 GHz
maps. As discussed in Sect. \ref{sec:inter_frequency}, for these
  cleaned likelihoods  we retain only the $143\times143$, $143
  \times 217$ and $217 \times 217$ TT spectra.

\section{Foreground masks}
\label{sec:foreground_masks}

\subsection{Temperature masks}
\label{subsec:temperature_masks}

\begin{figure}
\centering

\includegraphics[width=160mm,angle=0]{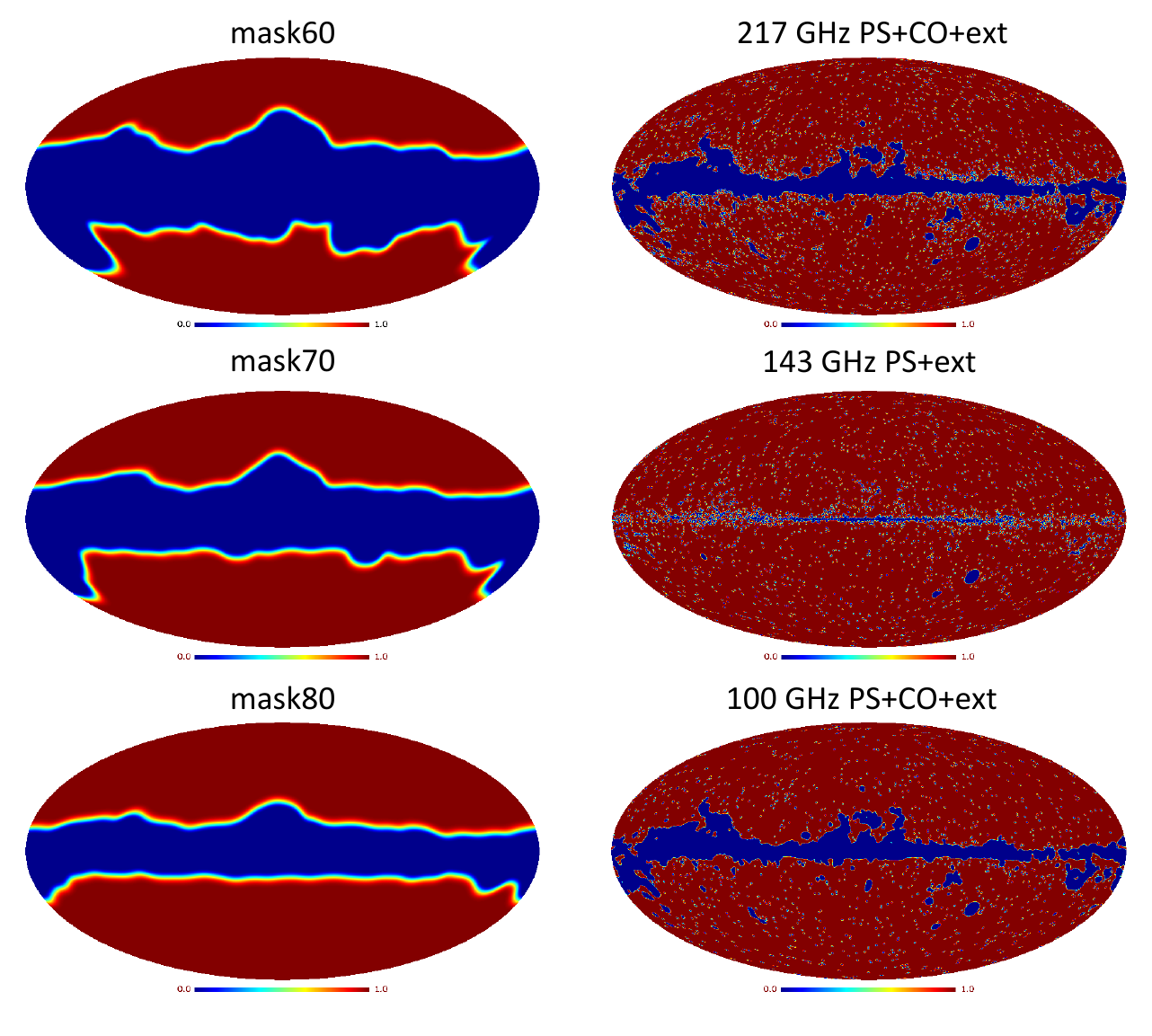} 

\caption {Figures to the left show the sequence of apodised diffuse foreground temperature masks  
(from top to bottom) applied to the 217, 143 and 100 GHz maps used to form our  12.1HM likelihood.
Figures to the right show the point source+CO+extended object masks that we apply to the 217, 143 and 100 GHz
maps.} 

\medskip

\label{fig:tempmasks}

\end{figure}

In this paper, we use the same family of temperature masks as used
in PCP15 and PCP18.  These masks are described in PPL15 and form a
sequence with unapodised sky fractions increasing in increments of 5\%
in sky area. The masks are apodised with a Gaussian window function of
width $\sigma = 2^\circ$.  Examples of the temperature masks used
frequently in this paper are shown in Fig.\ \ref{fig:tempmasks}.  We
will use the simple nomenclature mask25, mask60, mask70 {\it etc.} to
delineate these masks, where the numbers refer to the unapodised sky
areas retained after applying the masks.

The sky fraction over which an unapodised mask is non-zero is denoted $f_{\rm sky}$. Apodisation (and any additional
masking, for example, point source holes,  CO masks) reduces the effective sky area. We therefore define
a  weighted sky fraction $f^W_{\rm sky}$
\begin{equation}
      f^W_{\rm sky} = {1 \over 4 \pi} \sum_i w^2_i \Omega_i.  \label{mask1}
\end{equation}
Values for $f_{\rm sky}$ and $f^W_{\rm sky}$ are given in Table \ref{tab:sky_fractions}. \GE{Note that the CO mask
used here is based on  a multi-line CO  map produced as part of the 2013 \Planck\ data release \citep{Planck_CO13}, smoothed 
with a $2^\circ$ Gaussian and thresholded at a CO line brightness of $1 {\rm K}_{\rm RJ} \ {\rm km} \ {\rm s}^{-1}$. The CO mask
is plotted in Fig. A.2 of PPL13.}

In addition to the diffuse masks, we mask point sources, extended
objects (such as the Large Magellanic Cloud) and for 100 and 217 GHz maps we also mask out
areas of sky with strong CO line emission. These masks are identical
to those used in PCP18 and are described in PPL15. The point
source+CO+extended object masks are shown in
Fig.\ \ref{fig:tempmasks}. To avoid cumbersome nomenclature, we loosely
refer to these masks as `point source' masks in the rest of this
paper.

\begin{table}[h]
{\centering
\caption{\small{Sky fractions retained by the diffuse temperature and polarization masks}}
\label{tab:sky_fractions}
\begin{center}
\begin{tabular}{|c|c|c|c|c|c|} \hline
Temperature Mask  & $f_{\rm sky}$  (\%) & $f^W_{\rm sky}$ (\%) & Polarization Mask & $f_{\rm sky}$ (\%) & $f^W_{\rm sky}$ (\%) \\  \hline
mask25 &  24.68 & 18.55 & -- & -- & --\\
mask50 &  49.27 & 41.07 & maskpol50  &  50.26  & 38.94 \\
mask60 &  59.10 & 50.01 & maskpol60  &  59.57 &  48.81\\
mask70 &  69.40 & 60.19 & --  & --    & -- \\
mask80 &  79.13 & 70.15 & -- & --  & -- \\ \hline
\end{tabular}
\end{center}}
\end{table}

For this paper we have produced a series of likelihoods labelled
12.1-12.5 that use a range of different sky masks. These likelihoods are described
in Sect.\ \ref{sec:Likelihood}. The temperature masks shown in
Fig.\ \ref{fig:tempmasks} have been used to form the 12.1HM
likelihood. This is a half mission likelihood that is similar
(differing in minor ways that will be discussed in
Sect.\ \ref{sec:base_lcdm})  to the \camspec\ likelihood used in PCP18.

\subsection{Polarization masks}

\begin{figure}

\vspace{-0.2truein}
\centering

\includegraphics[width=160mm,angle=0]{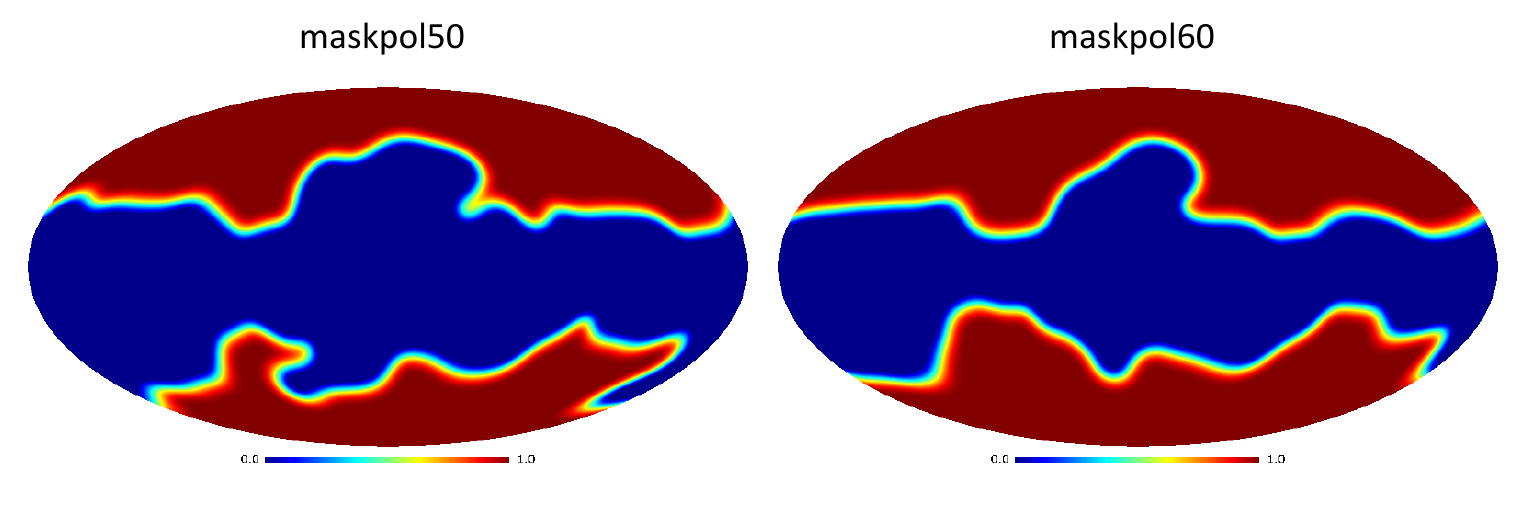}

\caption {Polarization masks constructed from a degraded resolution full mission  353 GHz $P = (Q^2 + U^2)^{1/2}$ map. \vspace{0.1 truein}}

\label{Polmasks}
\end{figure}

The large-scale features in \Planck\ Q and U maps are dominated by
Galactic dust emission at all HFI frequencies (see
Fig.\ \ref{fig:pol_cleaned_maps} of Sect.\ \ref{sec:dust_polarization}).
When we first started analysis of \Planck\ polarization, we created
diffuse polarization masks from the 353 GHz Q and U maps. We first
subtracted 143 GHz Q and U maps to remove the primordial CMB signal and
then smoothed the maps with a Gaussian of FWHM  of $10^\circ$. We
then applied a threshold in $P = (Q^2+U^2)^{1/2}$. The thresholded
mask was then apodised by smoothing with a Gaussian of FWHM of
$5^\circ$. To avoid isolated `islands' in the resulting polarization
masks, we iterated the thresholding and smoothing operations four
times. Two examples of the polarized masks constructed in this way are
shown in Fig.\ \ref{Polmasks}.  These polarization masks were created in the early stages of
our analysis of \Planck\ data. However subsequent experimentation showed that for fixed sky area,  the precise
shape of the polarization mask is unimportant. In this paper, we have therefore applied the
temperature masks of Fig.\ \ref{fig:tempmasks} to the Q and U maps to extend the sky area in polarization
beyond that of maskpol60.

Although we have found no strong evidence to suggest that bright
polarized point sources (e.g.\ bright AGN) influence the TE and EE
polarization spectra, for some likelihoods we have applied the $143$
GHz point source mask shown in Fig.\ \ref{fig:tempmasks}. Our
statistically most powerful likelihood (12.5HMcl, see
Sect.\ \ref{sec:Likelihood}) uses mask80 together with the point
source masks shown in Fig.\ \ref{fig:tempmasks} for temperature and
mask80 with the 143 GHz point source mask from
Fig.\ \ref{fig:tempmasks} applied to all polarization maps.

\section{Input maps and multipole ranges}
\label{sec:cross_spectra}

\subsection{Detector set and half mission maps}
\label{subsec:maps}

The Planck focal plane contains a mixture of spider web bolometers
(SWB) and polarization sensitive bolometers (PSB). The SWBs can be
processed individually to produce temperature maps and the PSBs can be
processed in pairs (i.e.\ 4 bolometers) to produce T, Q and U Stokes
parameter polarization maps. We refer to `detset' maps as the set of
$13$ maps constructed from the detector combinations listed in Table
\ref{tab:detsets}. A detset cross-spectrum
analysis involves cross-correlation analysis of all 13 detsets.
Excluding auto spectra, this leads to 78 TT, 72 TE/ET and 15 EE detset
cross spectra. This large number of spectra allows cross-checks of
potential instrumental systematics, as will be described in subsequent
sections.

We use the 2018 HFI maps which are described in
\cite{DataProcessing:2018}. Note that the 2018 HFI maps do not include
scanning rings 26050 to 27005 (which were included in the 2015 HFI
maps). This selection removes data from the end of the HFI cryogenic
phase of the \Planck\ mission which showed increased thermal
fluctuations in the HFI focal plane. As a consequence, the noise
levels of the 2018 HFI maps are slightly higher than those of the 2015
maps, though there is no other 
noticeable impact on the power spectra at high multipoles.  
In
addition to the full mission detset maps, we analyse frequency
averaged half mission (HM) maps which can be downloaded from the PLA\footnote{The 2018 detset maps are not available on
  the PLA. We are indebted to the \Plancks collaboration for
  permission to present results based on these maps in this
  paper. Apart from slightly increased noise levels (caused by the
  omission of some data at the end of mission survey 5), the 2018 
  detset maps at high multipoles ($\ell \simgt 30$) are almost
  identical to the 2015 detset maps, which are available on the
  PLA. The 2018 HFI DPC paper \cite{DataProcessing:2018} uses the
  detset maps for several internal consistency tests of the noise and
  calibration characteristics of these maps.}. The first half mission (HM1) maps are
constructed from scanning rings 240 to 13471 and the second
half mission (HM2) maps are constructed from scanning rings 13472 to
26050 (as summarized in the PLA).

\begin{table}[h]
{\centering \caption{\small{Detector combinations used in this analysis}}

\label{tab:detsets}
\begin{center}

\smallskip

\begin{tabular}{|c|c|l|c c c|} \hline
freq. (GHz) & detector & type & $\rtensor N^T$ & $\rtensor N^Q$ & $\rtensor N^U$ \\  \hline
100 & 1+4 (ds1) & PSB  & 1.708E-4 & 2.763E-4 & 2.575E-4 \\
100 & 2+3 (ds2) & PSB  & 7.340E-5 & 1.164E-4 & 1.172E-4 \\
143 & 5 & SWB  &   6.464E-5 & -- & --\\
143 & 6 & SWB  &   7.171E-5 & -- & --\\
143 & 7 & SWB  &   5.307E-5 & -- & --\\
143 & 1+3 (ds1) & PSB  & 3.230E-5 & 6.941E-5 & 6.989E-5\\
143 & 2+4 (ds2)  & PSB & 2.922E-5 & 5.687E-5 & 5.689E-5\\
217 & 1 & SWB &9.815E-5 & -- & -- \\
217 & 2 & SWB &1.179E-4& -- & -- \\
217 & 3 & SWB &1.038E-4& -- & -- \\
217 & 4 & SWB & 9.422E-5& -- & -- \\
217 & 5+7 (ds1) & PSB & 5.985E-5 & 1.306E-4 & 1.292E-4\\
217 & 6+8 (ds2) & PSB & 7.337E-5 & 1.616E-4 & 1.652E-4 \\ \hline
\end{tabular}
\end{center}}
\end{table}

The last three columns in Table \ref{tab:detsets} give the effective white-noise power
level computed from Eq.\ \ref{equ:Noise1} below\GE{, in units of $(\muK)^2$,}  over the default temperature+point source
masks shown in Fig.\  \ref{fig:tempmasks} (i.e. mask60 for 217 GHz, mask70 for 143 GHz and mask80 for 217 GHz)
and maskpol60 in polarization.
One can see from these entries that there are some signficant differences in the noise levels
of detsets within a frequency band. For example, 100ds1 maps are considerably noisier than 
100ds2 maps. The weighting that we apply to form coadded cross spectra for a given frequency 
combination (Eq.\ \ref{equ:C11}) downweights the noiser spectrum at high  multipoles where the
spectra are noise dominated.

\subsection{Multipole ranges}
\label{subsec:multipole_ranges}

\begin{table}[b]
\medskip\medskip\medskip
{\centering  \caption{\small{Multipole ranges used in the 12.1HM \camspec\ likelihood}}
\label{tab:multipoleranges}
\begin{center}
\smallskip

\begin{tabular}{|c |c  c| c c| c c|} \hline
 & \multicolumn{2}{c|}{TT} &  \multicolumn{2}{c|}{TE} & \multicolumn{2}{c|}{EE} \\ 
 \hline
spectrum      & $\ell_{\rm min}$ & $\ell_{\rm max}$       & $\ell_{\rm min}$ & $\ell_{\rm max}$       & $\ell_{\rm min}$ & $\ell_{\rm max}$ \\ \hline
$100\times100$  &  30 & 1400 & 30 & 1200 & 200 & 1200 \\ \hline
$100\times143$  &  -- &  -- & 30 & 1500 & 30 & 1500 \\ \hline
$100\times217$  &  -- & -- & 200 & 1500 & 200 & 1200 \\ \hline
$143\times143$  &  30 & 2000 & 30 & 2000 & 200 & 2000 \\ \hline
$143\times217$  &  500 & 2500 & 200 & 2000 & 300 & 2000 \\ \hline
$217\times217$  &  500 & 2500 & 500 & 2000 & 500 & 2000 \\ \hline
\end{tabular}
\end{center}}
\end{table}

We impose lower ($\ell_{\rm min}$) and upper ($\ell_{\rm max}$)
multipole cuts for each spectrum. At low multipoles, the pseudo-power
spectra are statistically sub-optimal \citep{Efstathiou:2004}. The
\Planck\ likelihoods are therefore hybrids \citep{Efstathiou:2004,
  Efstathiou:2006}, using more optimal likelihoods over the multipole
range $2 \le \ell \le 29$ (see
Sect.\ \ref{subsec:low_multipole_likelihoods}) patched to the
\Planck\ high multipole likelihoods.  The default values of the
multipole ranges used in \camspec\ are listed in Table
\ref{tab:multipoleranges}.  The rational for these choices is as
follows:

\noindent
$\bullet$ For each spectrum,  $\ell_{\rm max}$ is chosen so that we
do not use spectra at multipoles well into the inner beams (corresponding to angular scales much smaller than the
FWHM of the beams) where the 
beam transfer functions become small and issues such as beam errors and
correlated noise between detsets become significant.  At such high multipoles,
the noise in the beam corrected power spectra increases exponentially and so
little information is lost by truncating the spectra.

\noindent
$\bullet$ As discussed in the previous section,  we do not include the
$100\times143$ and $100\times217$ temperature  spectra in the likelihood,
since they add little cosmological information compared to the $100 \times 100$
and $143 \times 143$ spectra.

\noindent
$\bullet$ We impose $\ell_{\rm min}$ cuts on the
$143\times217$ and $217\times217$ temperature spectra for which Galactic dust
has a higher amplitude than in the $100 \times 100$ and $143 \times 143$
spectra. Since instrumental noise 
is negligible in temperature at multipoles $\ell \simlt 500$, no signficant
cosmological information is lost by truncating these spectra. However,
by eliminating the low multipoles in the $143\times 217$  and $217 \times 217$ spectra, 
we reduce the sensitivity of the temperature
likelihood to inaccurate dust subtraction and also supress the impact
of CMB-dust correlations on the likelihood (see Sect.\ \ref{sec:dust_temp}).
 
\noindent
$\bullet$ For TE and EE, the Planck spectra are noisy and so we coadd all
frequency combinations including the $100 \times 143$ and $100\times 217$ 
spectra to improve the signal-to-noise of the coadded spectra.

\noindent
$\bullet$ EE spectra involving $217$ GHz maps are strongly
contaminated by dust at low multipoles (see
Sect.\ \ref{sec:dust_polarization}). To avoid biases associated with
inaccurate dust subtraction we impose lower multipole cuts at the
expense of a reduction of signal-to-noise in the coadded spectra.  In
addition, at multipoles $\ell \simlt 50$ we find clear evidence of
systematics in EE spectra involving 217 GHz (see
Fig.\ \ref{fig:lowl_EE}). Therefore, we include only the $100
\times 143$ EE spectra at $\ell \le 200$ which gives consistency with the low
multipole EE likelihood at  $\ell < 30$. If we include the $100
\times 217$, $143 \times 217$ and $217 \times 217$ EE at $\ell < 200$
in the \camspec\ likelihoods, we find negligible impact on
cosmological parameters for \LCDM-like cosmologies. However, we then
see inconsistencies between the TT and EE solutions for models with
low frequency oscillatory features in the primordial power spectrum
\citep{Lemos:2018}.

\section{Estimating noise}
\label{sec:noise}

\subsection{Noise power spectra}
\label{subsec:noise+power_spectrum}

Accurate covariance matrices for a cross spectrum likelihood require 
accurate noise estimates.  The map making stage returns an estimate of the
noise at each pixel, $(\sigma^T)^2_i$, for unpolarized  maps and
a $3\times3$ noise matrix with diagonal components, $(\sigma^T)^2_i$,
$(\sigma^Q)^2_i$, $(\sigma^U)^2_i$ for polarized maps. If the noise is
uncorrelated between pixels, the noise power spectra of  maps computed with weighting
$w^X_i$ is
\begin{equation}
   \rtensor N^X  = {1 \over 4 \pi} \sum_i (\sigma^X)^2_i(w^X)^2_i\Omega^2_i,  \label{equ:Noise1}
\end{equation}
where $X=(T, Q, U)$. These are independent of multipole (i.e.\ the noise power spectra
are white).  However,  the \Plancks noise  differs
significantly  from white noise. In the time-ordered-data (TOI), the
\Plancks noise from each bolometer can be decomposed (roughly) into
three components: a white noise component at high frequencies; a
$1/f^\alpha$ component with $\alpha \sim 1$ at intermediate
frequencies (defining an effective `knee' frequency) coming from the
bolometer electronics; a $1/f^\beta$ component with $\beta \sim 2$ at
low frequencies from thermal noise (see \citep{DataProcessing:2018}). Low frequency noise at
the map level is substantially suppressed (but cannot be completely removed)  by destriping during the map
making stage. In addition,  as will be discussed in Sect.\ \ref{subsec:correlatednoise},  aspects of the  TOI
processing (such as deglitching thermal fluctuations caused by cosmic ray hits and removal of 4K cooler/bolometer 
interference lines)  introduce correlated noise at high multipoles in cotemporal
cross spectra.

The noise power spectra of \Plancks maps are therefore complex and
strongly dependent on the details of the TOI data processing, bolometer time constant corrections,  and
map-making.  The question then arises as to how to accurately
determine noise spectra. Ever since our first analyses of the \Plancks
data, we have preferred to use empirical noise estimates rather than
to rely on simulations (which, even in the elaborate end-to-end form
described in DPC18, do not include deglitching, direct injection of 4K lines
 and a
number of other key aspects of the data processing, as described in
their Appendix A4).  To determine the noise, we use differences of maps
constructed from each detector, or combination of detectors:

\smallskip

\noindent
{Half-ring differences (HRD): }
Half-ring (HR) maps are created from the first and second half of each \healpix\  pointing 
ring. Noise estimates can then be formed by differencing these maps. We call these half-ring 
difference (HRD) maps.

\smallskip

\noindent
{Odd-even differences (OED): }
Maps can be created from the odd and even numbered
\healpix\ pointing rings. The differences of these (OE) maps are called 
odd-even difference (OED)  maps.

\smallskip

Before showing results, it is extremely important to comment on
corrections for `missing' pixels. As a result of the \Plancks scanning
strategy, the sky coverage is highly inhomogeneous. Regions close to the
ecliptic poles are well sampled, but the coverage becomes sparser
towards the ecliptic plane. In addition, to construct polarization
maps, a threshold is applied to the $3\times3$ pixel noise covariance matrices
returned by the map-making algorithm. If the condition  number of
the $3 \times 3$ matrix is high at a particular pixel, then the pixel is
flagged as `missing'. For combined frequency maps, e.g.\ half mission
maps, the number of missing pixels is small. However, the number of
missing pixels is higher for detset maps,  since these are created
either from a single SWB or a PSB detset. The number of missing pixels
is even greater in HR and OE maps. In particular, in odd or even maps,
several million pixels (or several percent of sky) can be classified
as `missing'. Furthermore, there is very little overlap between
the missing pixels in odd and even splits.  Simply ignoring missing pixels 
leads to very large (up to 30\%) biases in the OED noise
power spectra for individual SWB maps.

\begin{figure}
\centering

\includegraphics[width=185mm,angle=0]{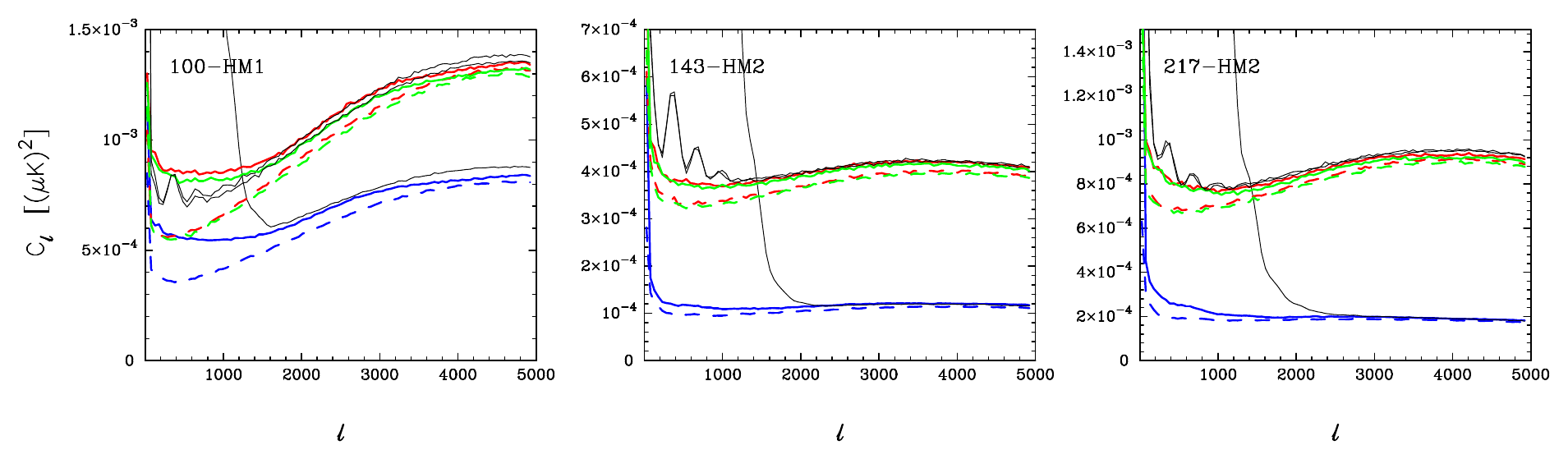}

\caption {Estimates of undeconvolved (i.e.\ uncorrected for missing sky and beam transfer functions) noise spectra for three
  half mission maps: 100 GHz HM1, 143 GHz HM2, 217 GHz HM2 computed for
  the masks used in the 12.1HM likelihood. The solid lines show noise
  estimates derived from OED maps and the dotted lines show noise
  estimates derived from HRD maps.  T, Q, and U noise spectra are
  shown by the blue, red and \GE{green} lines respectively. The solid
  black lines show the auto-spectra of the T, Q and U maps.}

\medskip
\medskip

\label{fig:noisespec}
\end{figure}
 
Consider two maps $M_1$ and $M_2$ (we drop the superscripts denoting
T,Q or U, and ignore noise correlations between these quantities) with noise
variances given by the relevant diagonal components of the $3\times 3$
covariance matrices  $\sigma^2_1$
and $\sigma^2_2$. The minimum variance combined map is 
\begin{subequations}
\begin{equation}
 M = {\sigma^2_1 \sigma^2_2 \over (\sigma^2_1 + \sigma^2_2)}\left ({M_1 \over \sigma^2_1} + {M_2 \over \sigma^2_2} \right ),  \label{equ:Noise2}
\end{equation}
with noise level
\begin{equation}
  \sigma^2  =  {\sigma^2_1 \sigma^2_2 \over (\sigma^2_1 +  \sigma^2_2) }.  \label{equ:Noise3}
\end{equation}
\end{subequations}
To match this noise level, we need to weight the difference map as follows:
\begin{equation}
  D  =  {\sigma_1 \sigma_2 \over (\sigma^2_1 +  \sigma^2_2)} (M_1 - M_2).  \label{equ:Noise4}
\end{equation}
However, in our application $\sigma_1$ and $\sigma_2$ are not
determined for missing pixels.  We therefore `infill' the missing
pixels in each of the maps $M_1$ and $M_2$ by replacing the missing
pixel and its noise estimate
with that of a pixel drawn at random from the nearest 100 pixels
within a disc centred on the missing pixel.

Typical noise estimates derived from OED and HRD maps are shown in Fig.\ \ref{fig:noisespec}. The noise
power spectra are clearly non-white and this needs to be acounted for in computing the power spectrum
covariance matrices (via the heuristic $\psi_\ell$ factors discussed in Sect.\ \ref{subsec:covariance_matrices}). The noise spectra can be  fitted accurately by the following functional form:
\begin{equation}
N_\ell  = A\left ({100 \over \ell} \right) ^\alpha + B {(\ell/1000)^\beta \over (1 + (\ell/\ell_c)^\gamma)^\delta}, \label{equ:Noise5}
\end{equation}
with $A$, $\alpha$, $B$, $\beta$, $\ell_c$, $\gamma$ and $\delta$ as
free parameters (see Fig. A.3 of PLP13). The second term in (\ref{equ:Noise5}) describes the non-white behaviour of the noise spectra at high multipoles, which is caused mainly by bolometer time-constant deconvolution. The first term describes the non-white noise behaviour at low multipoles which arises primarily from  bolometer `$1/f$' noise
and residual low-frequency thermal fluctuations in the focal plane.

An important result from this analysis is that the HRD maps
 systematically  {\it underestimate}  the noise power 
compared to the OED maps. Similar results are reported in PPL18
and in  Section 5.4 of \citep{DataProcessing:2018} and are
 caused by the fact that deglitching is performed
on the full set of \healpix\ pointing rings leading to correlated errors in half-rings which
cancel in the HRD maps.  In addition, the OED spectra fall off less
steeply than the HRD spectra at low multipoles. This is particularly
noticeable in the 100 GHz noise spectra.  The black lines in
Fig.\ \ref{fig:noisespec} show the (undeconvolved) auto spectra of each map
(with Q and U maps treated as scalar maps). \GE{One can see  that the
temperature auto spectra are signal dominated over most of the
multipole range used for cosmology and so errors in the noise spectra
are not particularly critical for cosmology derived from the TT spectra. However, this not true
for the polarization spectra which are noise dominated over most of
the multipole range. Errors in the noise modelling can therefore
affect the $\chi^2$ values of the EE power spectra, particularly for
100 GHz, for which the polarization noise power spectra are very significantly 
non-white.} Fig.\ \ref{fig:noisespec} shows that the auto spectra in both temperature
and polarization match well with the OED spectra at high multipoles for both temperature
and polarization. For 100 GHz, the OED polarization spectra sit high compared to the auto spectra
at multipoles $\simlt 1000$ which suggests that the OED spectra have some correlated 
component in addition to instrument noise. 

In the \Plancks 2013 and 2015 analyses, we used HRD maps to estimate noise 
in forming the \camspec\ likelihoods and so the noise was underestimated.
This is the main reason for high $\chi^2$ values for the \camspec\ 
EE spectra used  in PPL15
and PCP15,  rather than systematics in the polarization data\footnote{These papers suggested that temperature-polarization
leakage may have been responsible for the high EE $\chi^2$ values for both the 
\camspec\ and \plik\ likelihoods. In fact,
temperature-to-polarization leakage is a small effect for EE and the high
$\chi^2$ values were caused primarily by underestimated noise
and inaccurate polarization efficiencies (see  Sect.\ \ref{subsec:pol_cal}).}. In this paper, we use the OED 
noise estimates (as illustrated in Fig.\ \ref{fig:noisespec}) though our analysis  suggests that they
overestimate the noise contribution to the 100 GHz and possibly the 143 GHz EE spectra.

As pointed out in PCP13, the larger the data vector, the more accurately one needs to know the noise levels to avoid a significant bias in $\chi^2$. As a rough rule of thumb, for a data vector of length $N$ the noise estimates  need to satisfy
\begin{equation}
  {\Delta \sigma^2 \over \sigma^2} \simlt \sqrt{2 \over N},  \label{equ:Noise6}
\end{equation}
to give an accurate  $\chi^2$.
For the full \camspec\ TTTEEE likelihoods, $N\sim 12000$, so we require  covariance matrices accurate to $\sim 1\%$ if the value of $\chi^2$ is to be used as a simple `goodness-of-fit' criterion.  In reality,
the covariance matrices are accurate to only a few percent and this needs to be borne in mind when interpreting
$\chi^2$ values for the full likelihood. As we show in Table \ref{tab:chi_squared}, by adopting the OED noise
estimates, the full TTTEEE likelihood fitted to a base \LCDM\ cosmology gives an  acceptable $\chi^2$ (to within $\sim 2.2 \sigma$), though the $\chi^2$ values for individual  EE spectra are consistently (though not alarmingly)
low (see Table \ref{tab:chi_squared_pol}). \GE{Noise estimation remains a significant issue for \Planck\ EE analysis,
though we are currently investigating whether the techniques described in \cite{deBelsunce:2021} based on end-to-end simulations can be adapted to give improved noise estimates.}

\subsection{Correlated noise}
\label{subsec:correlatednoise}

As pointed out by Spergel et. al. \cite{Spergel:2015}, there is clear evidence for
correlated noise between cotemporal HFI temperature maps. For
this reason, in the 2015 and 2018 \Plancks analyses the \Plancks collaboration
used half mission
cross spectra to form Planck likelihoods rather than using full mission detset
spectra, sacrificing signal-to-noise in favour of reducing systematics
from correlated noise\footnote{PCP13 used nominal mission detset
  spectra. The coadded $217 \times 217$ detset spectrum showed a `dip' at $\ell
  \sim 1800$, which was traced to incomplete removal of $4 \ {\rm K}$ lines in
  the TOD data, which primarily affected survey 1. The systematic origin of this feature was demonstrated
  convincingly in \cite{Spergel:2015}.  Fortunately, the
  $217\times 217$  feature was not strong enough to significantly affect
  cosmological parameters.}. There are two simple ways of testing for
correlated noise: (i) we can compute 
the differences between full mission coadded detset (DS) spectra and half mission (HM) cross
spectra; (ii) we can cross-correlate the OED detset maps. 
The DS and HM cross spectra may differ because
of correlated noise, errors in the effective beams etc.The OED DS cross
spectra therefore give a direct measure of correlated noise, though we
use both tests in the results presented below.

\begin{figure}
\centering
\includegraphics[width=75mm,angle=0]{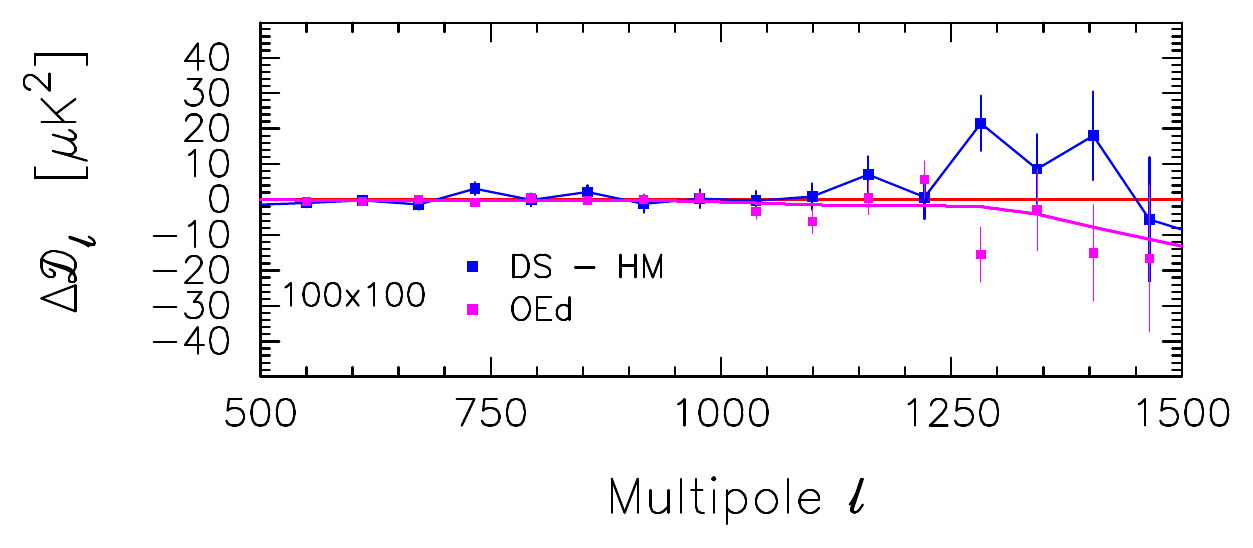} \includegraphics[width=75mm,angle=0]{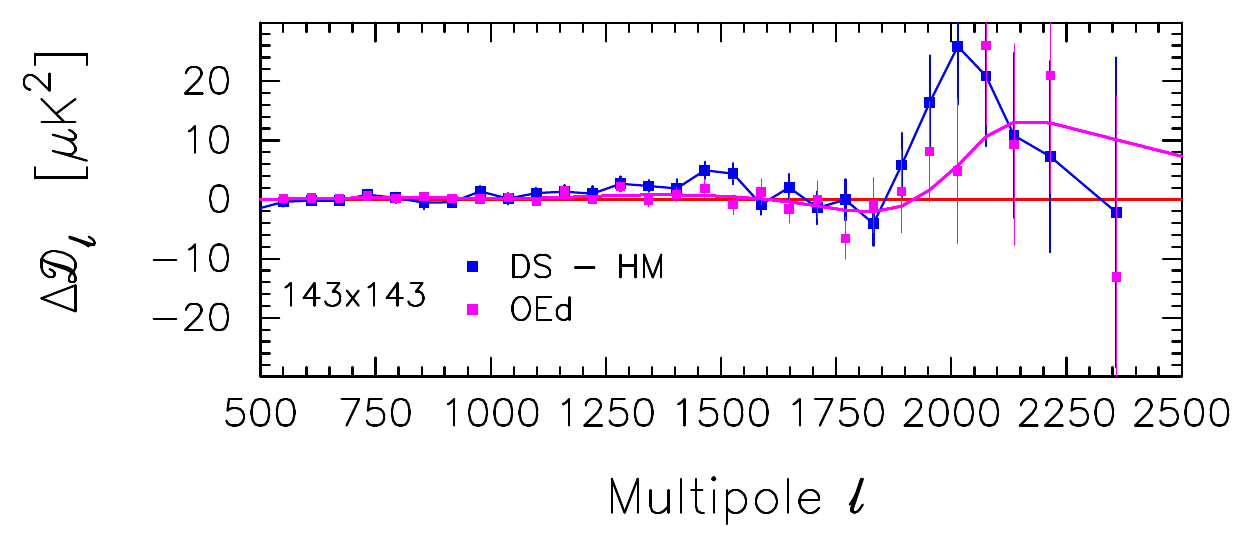} \\
\includegraphics[width=75mm,angle=0]{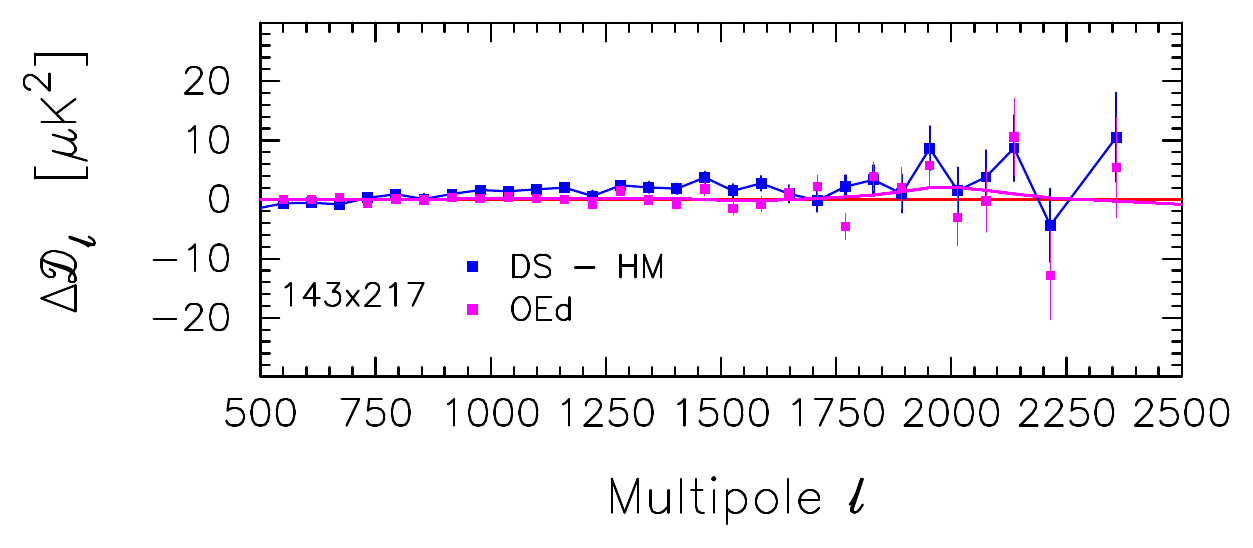}
\includegraphics[width=75mm,angle=0]{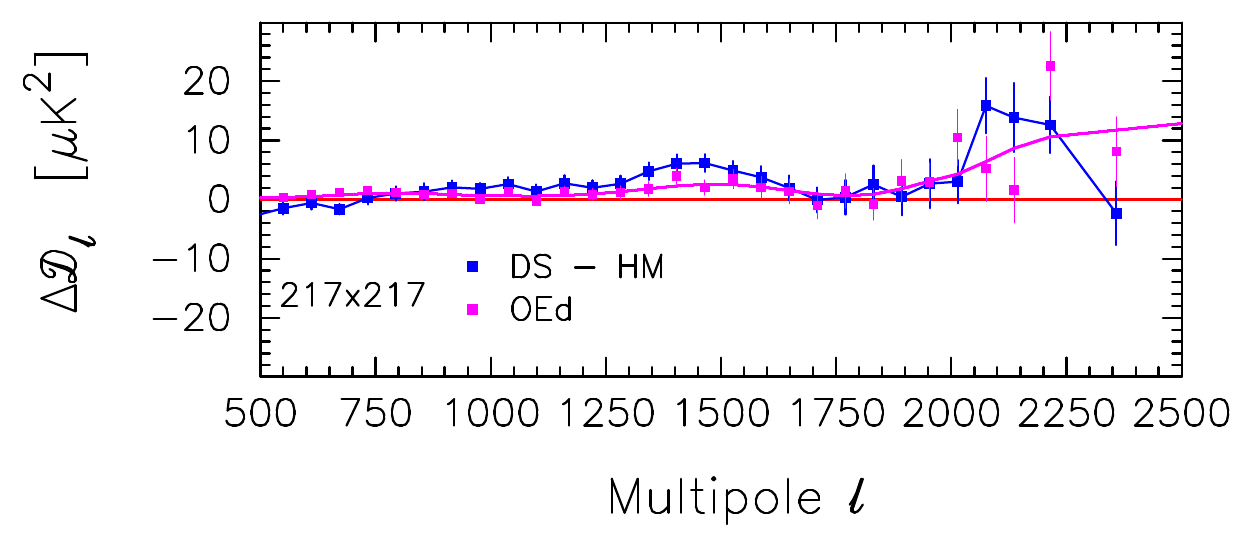} \\
\includegraphics[width=75mm,angle=0]{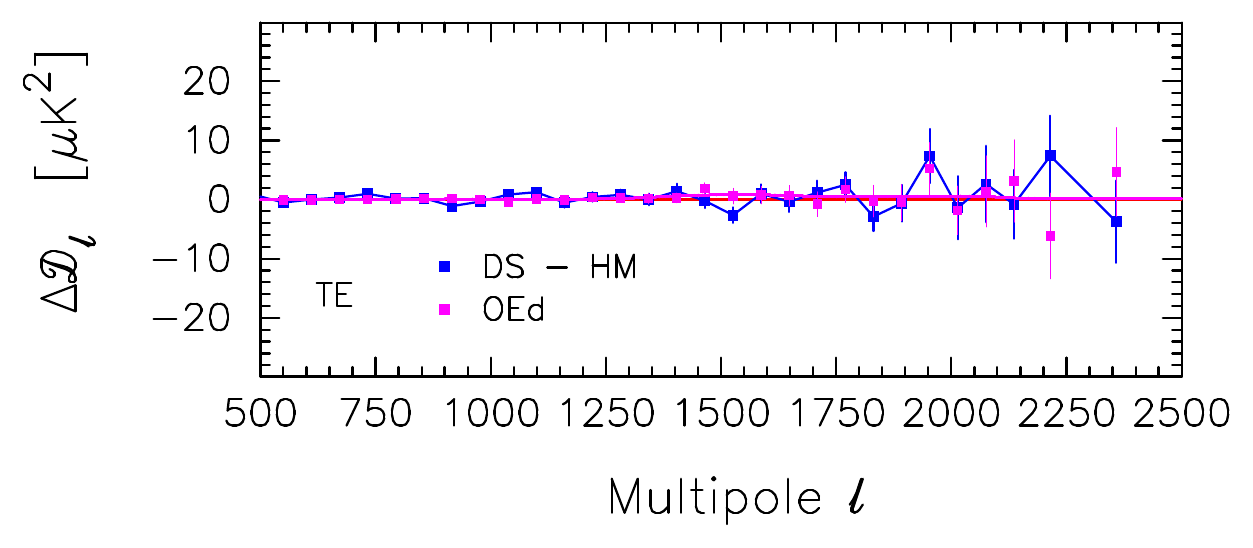}
\includegraphics[width=75mm,angle=0]{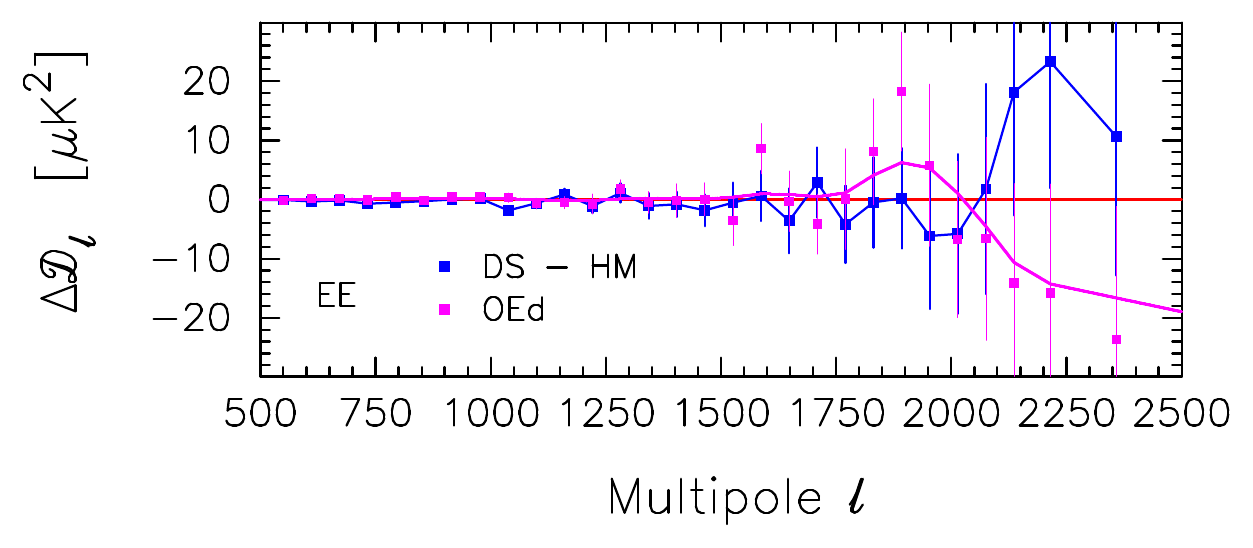} \\

\caption {Top four figures show $100\times100$, $143\times143$,
  $143\times217$ and $217\times217$ TT spectra.  The blue points show
  the differences between coadded DS TT spectra and HM spectra in
  bands of width $\Delta \ell=61$.  The purple points show the DS OED
  cross spectra, coadded using the same weights (Eq. \ref{equ:C11}) as
  those for the TT spectra. The lines show fits to the OED
  spectra. The error bars show the $1\sigma$ scatter of points within
  each bandpower.  The lower two figures show the equivalent plots for
  coadded TE and EE spectra.}
\label{fig:corrnoise}

\medskip
\medskip

\end{figure}

  Results are shown in Fig.\ \ref{fig:corrnoise}. The blue points show the differences between coadded DS and HM spectra
together with error bars reflecting variance caused by instrumental noise. The purple points show coadded DS OED spectra using the DS TT weights  (Eq.\ \ref{equ:C11}).
The purple lines show fits to the OED spectra computed by applying a three-point filter to the purple points. From these
plots we conclude the following:

\noindent
$\bullet$ There is evidence of correlated noise in the $143\times143$ and $217 \times 217$ OED TT spectra at high multipoles
 that matches reasonably well with the differences between DS and HM cross spectra. The level of correlated noise is comparable
to the $\pm 1\sigma$ errors from noise. 

\noindent
$\bullet$ The OED spectra show no evidence for correlated noise in the $100\times 100$ and $143 \times 217$ TT spectra
or in the coadded TE OED spectra at a level that could bias cosmological parameters in the high multipole likelihoods.

\noindent
$\bullet$ The EE OED spectra are very noisy at multipoles $\ell \simgt 2000$. At lower multipoles, there is no evidence
for correlated noise in either the OED spectra or the (DS-HM) difference
spectra at a level that could bias cosmological parameters in the high multipole
likelihoods. 

\noindent
$\bullet$ There is an indication of a small excess in the DS-HM $217\times 217$ TT spectrum compared to the OED spectrum
in the  multipole range $\ell \sim 1450$, which is qualitatively reproduced in the OED spectrum.

\noindent
$\bullet$ Although we have presented results in Fig.\ \ref{fig:corrnoise}
for OED detset spectra, we find similar results for OED HM  cross spectra.

With the choices of $\ell_{\rm max}$ in Table
\ref{tab:multipoleranges}, correlated noise will have little impact on
the DS spectra except for the $217 \times 217$ DS spectrum at $\ell
\simgt 2000$, where correlated noise appears to bias the DS spectrum
high by about $1 \sigma$. In forming a DS likelihood, we therefore
subtract the fits to the OED spectra (purple lines) from the coadded
DS TT spectra.

\newpage

\section{Beams, calibrations and polarization efficiencies}
\label{sec:beams}

\subsection{Planck effective beams}
\label{subsec:planck_beams}

As discussed in \cite{DataProcessing:2018}, the absolute calibration of Planck HFI maps is
based on the orbital dipole over the frequency range $100-353$ GHz and
on Uranus and Neptune at $545$ and $857$ GHz. As far as this paper is
concerned, the absolute calibration of the $100-217$ GHz DS or HM maps
at $\ell = 1$ is assumed to be exact, and any differences between TT cross spectra
is ascribed to errors in the effective beam transfer
functions. Analysis of the Solar dipole at 545 GHz reported in DPC18
shows that the absolute calibration at $545$ GHz agrees to within
$0.2\%$ of the calibrations at lower frequencies. Since we use
$353$, $545$ and $857$ GHz only as Galactic dust templates in this paper, any
calibration errors relative to lower frequencies are absorbed in the
cleaning coefficients (see Sects.\ \ref{sec:dust_temp} and \ref{sec:dust_polarization}).

In simplified form, the power absorbed by a detector at time $t$ on the sky is
\begin{equation}
P(t) = G[I + \rho(Q\cos 2(\psi(t) + \psi_0)+U \sin (2(\psi(t) + \psi_0))] + n(t), \label{equ:B0}
\end{equation}
where $I$, $Q$, $U$ are the beam convolved Stokes parameters seen by
the detector at time $t$, $G$ is the effective gain (setting the
absolute calibration), $\rho$ is the detector polarization
efficiency, $\psi(t)$ is the roll angle of the satellite, $\psi_0$ is
the detector polarization angle and $n(t)$ is the noise. For a perfect
PSB, $\rho=1$, while for a perfect SWB, $\rho = 0$. The polarization
efficiencies and polarization angles for the HFI bolometers were
measured on the ground and are reported by Rosset et
al.\ \cite{Rosset:2010}. For PSB detectors, the ground based
measurements of polarization angles were measured to an accuracy of
$\sim 1^\circ$ and the polarization efficiencies to an accuracy of
$\sim 0.1-0.3 \%$. The \SROLL\ map making algorithm used to produce the 2018 HFI maps (described in
\citep{DataProcessing:2018})
assumes the ground based measurements of polarization angles and efficiencies.
Within the  \SROLL\ formalism, polarization angles and efficiencies
 are degenerate and cannot be disentangled from each other.  Errors in
the polarization angles induce leakage from $E$ to $B$ modes while
errors in the polarization efficiencies induce leakage from
temperature to polarization. As discussed in \citep{DataProcessing:2018}, analysis of the Planck TB and EB spectra
(which should be zero in the absence of parity violating physics)
suggest errors in the polarization angles of $\simlt 0.5^\circ$,
within the errors reported by Rosset et al. \GE{with an error in the overall orientation of the focal
plane of $0.31^\circ \pm 0.28^\circ$ \citep{Planck_parity:2016}. Simulations including errors of this order suggest that polarization angle errors contribute systematic errors of less than $10\%$ of cosmic variance in the HFI EE spectra in the multipole  range 2-1000 (see Appendix A6 of \cite{SROLL:2016}).} However, by inter-comparing spectra,
\citep{DataProcessing:2018} concluded that systematic errors in the
polarization efficiencies are significantly larger than the
statistical errors reported by Rosset et al. (perhaps at the 1\%
level).  This conclusion is supported by the results presented in this
paper. As we will show in Sect.\ \ref{subsec:pol_cal}, the
polarization efficiencies and temperature-to-polarization leakage can
be constrained accurately by comparing different frequency
combinations of TE and EE spectra. In this paper, \GE{we assume that
polarization efficiencies dominate over errors in the polarization
angles and determine multiplicative polarization efficiencies
empirically, recognising that in reality, polarization angle errors
and polarization efficiencies are intertwined.}

We use the definitions of \cite{Planck_beams:2014},  distinguishing
between scanning and effective beams. The scanning beam is defined as
the coupled response of the optical system, deconvolved time response
function and software low pass filter applied to the TOD. The
effective beam describes the beam in the map domain, after
combining the TOD in the map making process. The effective beam will
vary from pixel to pixel in any given map. For some applications, {\it
  e.g.} analysis of individual sources, it is useful to have estimates
of the effective beams at each pixel, as provided by the \FEBECOP\footnote{Fast Effective Beam Convolution in Pixel Space.}
software \citep{Mitra:2011}. However, for power spectrum estimation it
is more useful to have an `isotropised' beam transfer function which
can be used to correct the power spectra as in
Eqs.\ \ref{equ:C6a} and \ref{equ:C6b}. Such isotropized beam transfer functions are
provided by the \Quickpol\  software \citep{Hivon:2017},  which can be
tuned to return beam transfer functions for the exact masks used in a
cosmological analysis. (In practice, we use beams computed on {\it almost} the same
sky maps used to compute spectra, differing in point-source holes, missing pixels, CO masks etc.) Discrete sampling of the sky can lead to a
small additive (rather than multiplicative) sub-pixel contribution to
the beam convolved power spectra with an amplitude that depends on the
temperature gradient within each pixel. These sub-pixel effects can be
computed in \Quickpol\  assuming fiducial spectra (including any foreground
contributions). Sub-pixel effects have been quantified in detail in PPL15 and PPL18, and have been shown to be small. We have  therefore neglected sub-pixel
effects in creating \camspec\ likelihoods.

The determination of Planck HFI beams is complex and is described in detail in \cite{Planck_beams:2014}. We summarize some 
of the main details here. The `main beam' is defined as the scanning beam out to $100^\prime$ from the beam axis. Smearing of
the main beam (caused by the time dependent response of a bolometer to a signal) is reduced by deconvolving the TOD. This deconvolution amplifies the noise at high frequencies, which is why the noise power spectra are non-white at high multipoles
(cf.\  Fig.\ \ref{fig:noisespec}). The main beams of the $100-353$ GHz detectors are determined by calibrating against scans of Saturn and Jupiter; at $545$ and $857$ GHz, Mars observations are used to calibrate the peak of the main beam since Saturn and Jupiter
saturate the detectors in the inner parts of the main beam. At beam radii that are larger than those set by the  noise levels of the Jupiter scans, the main beam is patched to a power law ($ \propto \theta^{-3}$, where $\theta$ is the angular distance from the main beam axis) where the  exponent is
 derived from GRASP\footnote{See {\tt https://www.ticra.com/software/grasp}.} physical optics models. The main beam calibrations do not correct for the filtering of the sky from far side-lobes (FSL)  arising from reflector and baffle spillover. 
The FSL for each detector is defined as the beam response  at $\theta > 5^\circ$ and is computed from GRASP models. These computations are used to generate FSL convolutions of the dipole and Galaxy, which are removed from the TOD during the \SROLL\ map making stage. Since the FSL contributions project onto the sky in different ways for odd and even sky surveys, odd-even survey null tests can be used to test for residual effects arising from FSL (and also Zodiacal emission) as described in Section 3 of \citep{DataProcessing:2018}.

The polarization maps are constructed from pairs of PSB
detectors. Mismatch of the beams for individual bolometers within each
PSB pair will introduce couplings between the temperature and
polarization maps. The \Quickpol\ formalism computes a beam matrix
relating the expectation values of the beam convolved spectra measured
on the sky ($\rtensor C_\ell^{TT}$, $\rtensor C_\ell^{TE}$, $\rtensor
C_\ell^{EE}$, $\rtensor C_\ell^{BB}$) to the beam uncorrected spectra ($\tilde C_\ell^{TT}$,
$\tilde C_\ell^{TE}$, $\tilde C_\ell^{EE}$, $\tilde C_\ell^{BB}$):
\begin{equation}
\left ( \begin{array}{c} \rtensor C^{TT}_\ell \\ \rtensor C^{TE}_\ell \\ \rtensor C^{EE}_\ell \\ \rtensor   C^{BB}_\ell   \end{array} \right ) =    \left ( \begin{array}{cccc} W^{TTTT}_\ell& W^{TTTE}_\ell &W^{TTEE}_\ell & W^{TTBB}_\ell \\ 
W^{TETT}_\ell& W^{TETE}_\ell& W^{TEEE}_\ell    &  W^{TEBB}_\ell \\
W^{EETT}_\ell & W^{EETE}_\ell & W^{EEEE}_\ell & W^{EEBB}_\ell \\
W^{BBTT}_\ell & W^{BBTE}_\ell & W^{BBEE}_\ell & W^{BBBB}_\ell \end{array} \right )
\left ( \begin{array}{c} \tilde C^{TT}_\ell \\  \tilde C^{TE}_\ell \\ \tilde C^{EE}_\ell \\  \tilde C^{BB}_\ell   \end{array} \right ) . \label{equ:B1}
\end{equation}
This  generalises the discussion given in Sect.\  \ref{subsec:PCL}  to polarized beams.
The diagonal components of this matrix are the dominant terms, and we (loosely)
refer to these diagonal components as `scalar beams'.

In \camspec, we ignore the off-diagonal terms in the full beam matrix for the TT spectra. However, for the TE and EE spectra we retain the most important off-diagonal terms
\begin{subequations}
\begin{eqnarray}
\rtensor{C}^{TE}_\ell &=&   W^{TETE}_\ell \tilde C^{TE}_\ell + W^{TETT}_\ell \tilde C^{TT}_\ell , \label{equ:B2a} \\
\rtensor{C}^{EE}_\ell &=&   W^{EEEE}_\ell \tilde C^{EE}_\ell + W^{EETT}_\ell \tilde C^{TT}_\ell , \label{equ:B2b}
\end{eqnarray}
\end{subequations}
describing beam-induced temperature-to-polarization (TP) leakage. For these,  we make corrections to the measured spectra
assuming the best fit base $\Lambda$CDM theoretical TT spectrum. These corrections have a small impact on the \Planck\ TE
spectra and are negligible for the EE spectra. Section \ref{subsec:pol_leak} presents tests of the TP leakage model
of Eqs. \ref{equ:B2a} and \ref{equ:B2b}.

A useful model for TP leakage has been proposed by Hivon et al. \citep{Hivon:2015}. Here it is assumed that 
TP leakage modifies the measured $a_{\ell m}$ coeffients and power spectra as follows:
\begin{subequations}
\begin{eqnarray}
a^T_{\ell m} &\rightarrow&  a^T_{\ell m},  \label{equ:B3a}\\
a^E_{\ell m} &\rightarrow& a^E_{\ell m} + \epsilon_{\ell} a^T_{\ell m}, \label{equ:B3b}
\end{eqnarray}
\end{subequations}
causing perturbations to the power spectra of
\begin{subequations}
\begin{eqnarray}
\Delta C^{TT}_{\ell } & = &  0,  \label{equ:B4a}\\
\Delta C^{TE}_{\ell } & = &  \epsilon_\ell C^{TT}_\ell,  \label{equ:B4b} \\
\Delta C^{EE}_{\ell } & = &  \epsilon^2_\ell C^{TT}_\ell + 2\epsilon_\ell C^{TE}_\ell.  \label{equ:B4c}
\end{eqnarray}
\end{subequations}
\GE{As noted by \cite{Hivon:2015},} if it is assumed that the main effect of beam mismatch arises from  $m=2$ and $m=4$ beam modes describing beam ellipticity
(note that beam modes with odd values of $m$ are small as a result  of the Planck scanning strategy), then the coefficients
$\epsilon_\ell$ should vary approximately as powers $\ell^m$. In this highly simplified  model,
\begin{equation}
\epsilon_\ell \approx \epsilon_2 \ell^2 + \epsilon_4 \ell^4    , \label{equ:B5}
\end{equation}
and so the coefficients $\epsilon_2$ and $\epsilon_4$, together with
Eqs. \ref{equ:B4b}-\ref{equ:B4c} describe TP leakage. In the Planck 2015 analyses, we did not at that
time have estimates of the full polarized beam matrices of
Eq. \ref{equ:B1} and so we used this simplified model to roughly
characterise TP leakage. We will revisit this
model in Sect.\ \ref{subsec:pol_leak}.

\subsection{Intra-frequency residuals in temperature}
\label{subsec:intra_frequency}

The detset spectra provide a good test of the accuracy of beams,
calibrations and other instrumental systematics. As discussed in Sect.\ 4.4 of \citep{DataProcessing:2018}
based on the consistency of the Solar dipole solutions, errors in the absolute calibration of \Planck\ over the $100-217$ GHz 
frequency range are extremely small ($\simlt 3 \times 10^{-4}$) and have negligible impact on the power spectra. The formal
errors on the main beam calibrations are also small (see Fig.\ 21 of PPL15) and are neglected in this paper.
However errors in characterising the beams beyond the main beams (at $\theta
\simgt 100^\prime$)  introduce beam transfer function errors
at multipoles $\ell \simlt 100$. At the power spectrum level, such
errors will appear to be nearly degenerate with multiplicative
(i.e.\ effective calibration) errors in the power spectra. To test for
such errors, we investigate intra-frequency residuals using the
detsets. Excluding auto spectra, the detsets provide $N_s=10$
$143\times143$ cross spectra, $15$ $217\times 217$ and $30$ $143\times
217$ cross spectra. Since the foregrounds are the same in each of
these frequency groups, the cross spectra within  each  group should be identical
provided beam, calibration, band-pass mismatch and other systematics are negligible.

To test this, we minimise
\begin{equation}
 \chi^2 = {\large \sum_{ij\subset k}}{\large \sum_{\ell_{\rm min}}^{\ell_{\rm max}}} \left (\psi_{ij}{\hat D}^{ij}_\ell - {1/N_s} \sum_{pq\subset k} \psi_{pq} {\hat D}^{pq}_\ell, \right )^2, \label{equ:B6}
\end{equation}
with respect to the coefficients $\psi_{ij}$ for all $N_s$ TT detset cross spectra within frequency group $k$. We impose the constraint that $\psi_{ij}=1$ for the first cross spectrum within the frequency group (with detsets ordered as in Table \ref{tab:detsets}) computed using our default masks.
(For example,   we fix $\psi_{56}=1$ for the $143{\rm-}5\times 143{\rm -}6$ cross spectrum).  The $\chi^2$ in Eq. \ref{equ:B6} is unnormalised, since the dominant source of variance at multipoles $\ell \simlt 1000$ comes from systematic errors in the 
far-field beams. Note that this differs from our analysis of intra-frequency residuals in PLP13, where we minimised to fit detset calibration coefficients, $c_i$,  at the map level. At that stage in the Planck analysis, the beam transfer functions beyond the main beams had not been  characterised accurately, which led to relatively large $1{\rm -}2\%$ effective calibration differences which were easily detectable in the power spectra. The tests described in this section are extremely sensitive and can
easily detect effective calibration differences at the $0.1\%$ level.

\begin{figure}[tp]
	\centering
	\includegraphics[width=73mm, angle=0]{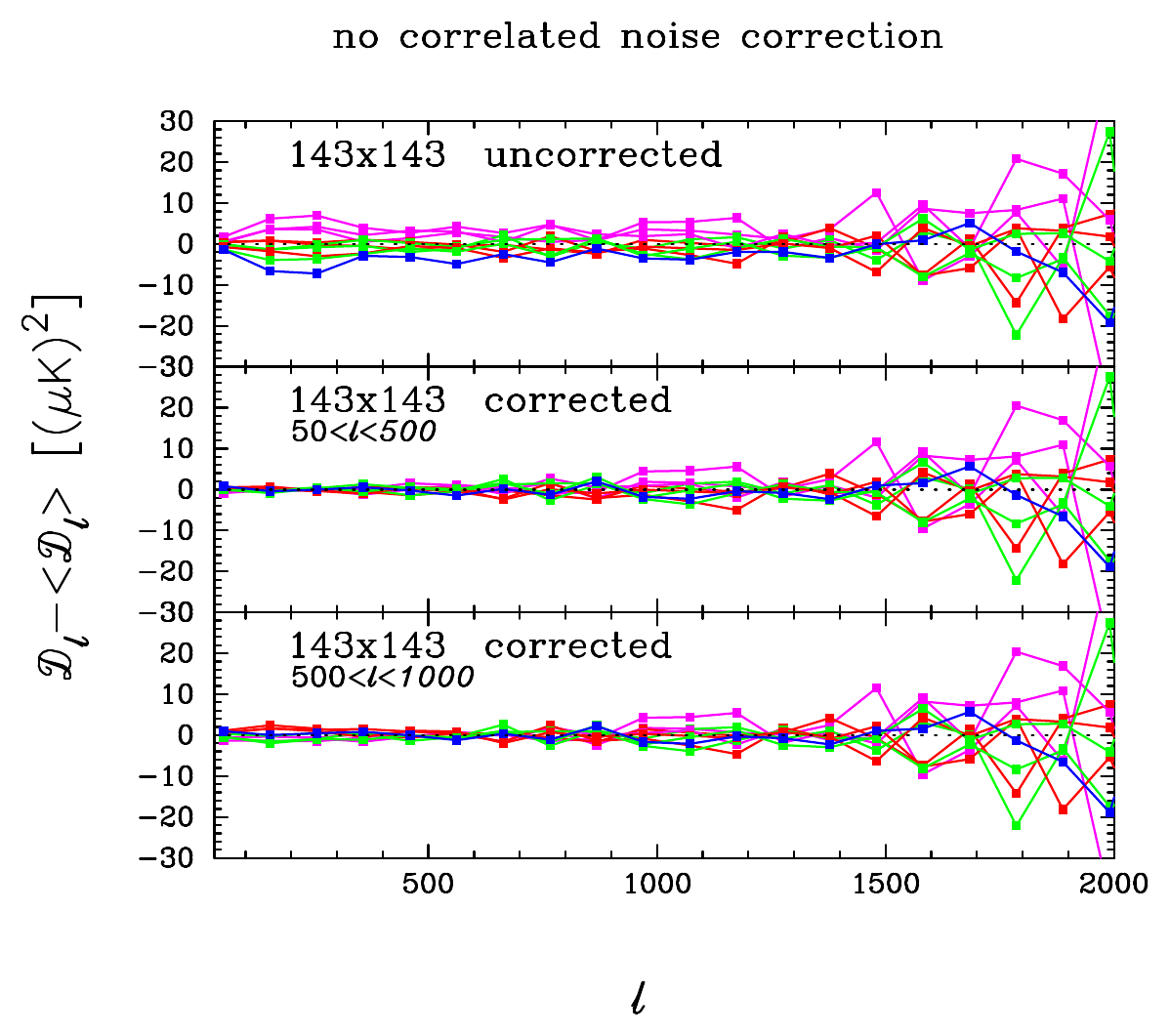} 	\includegraphics[width=73mm, angle=0]{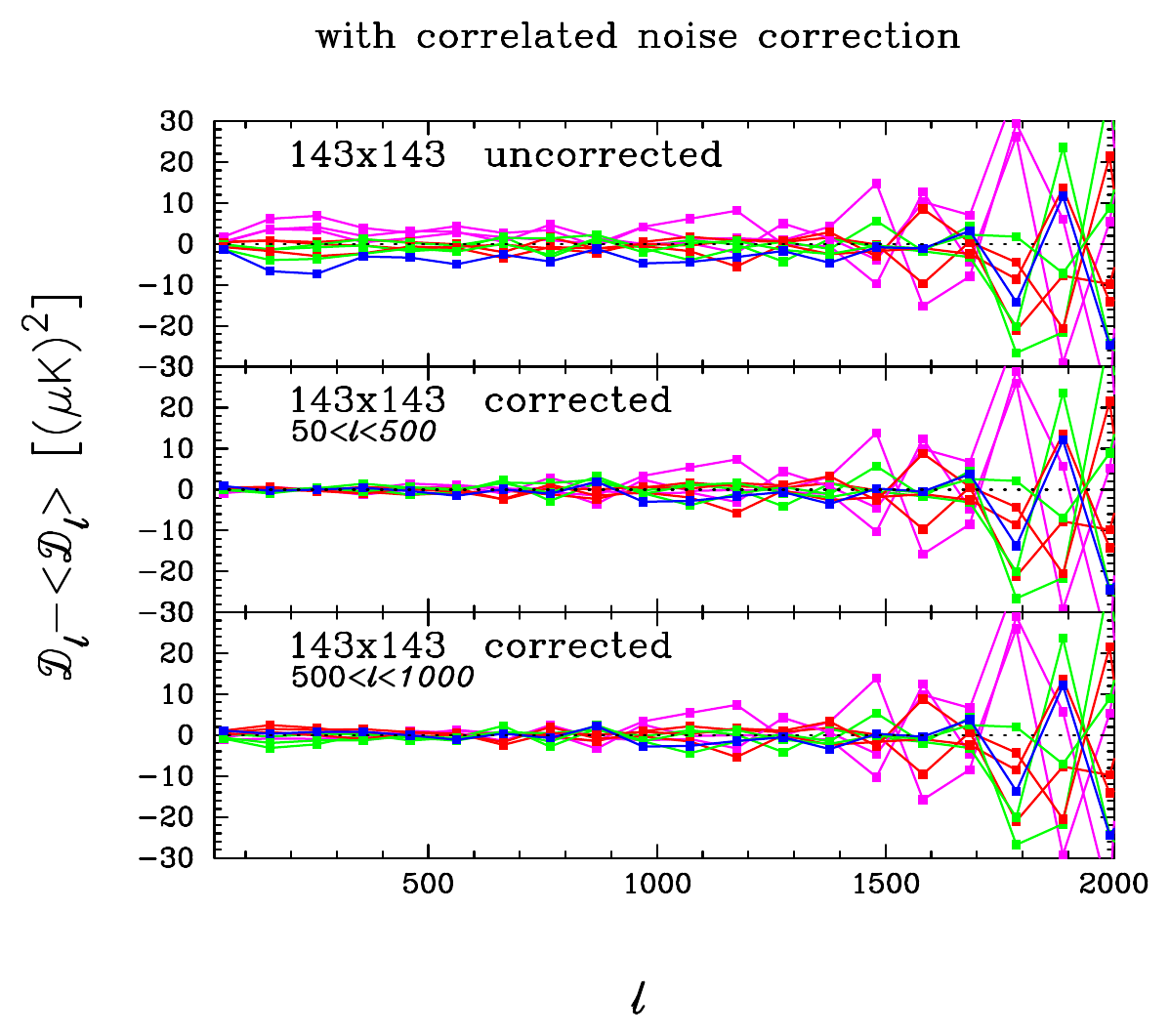} \\
	\includegraphics[width=73mm, angle=0]{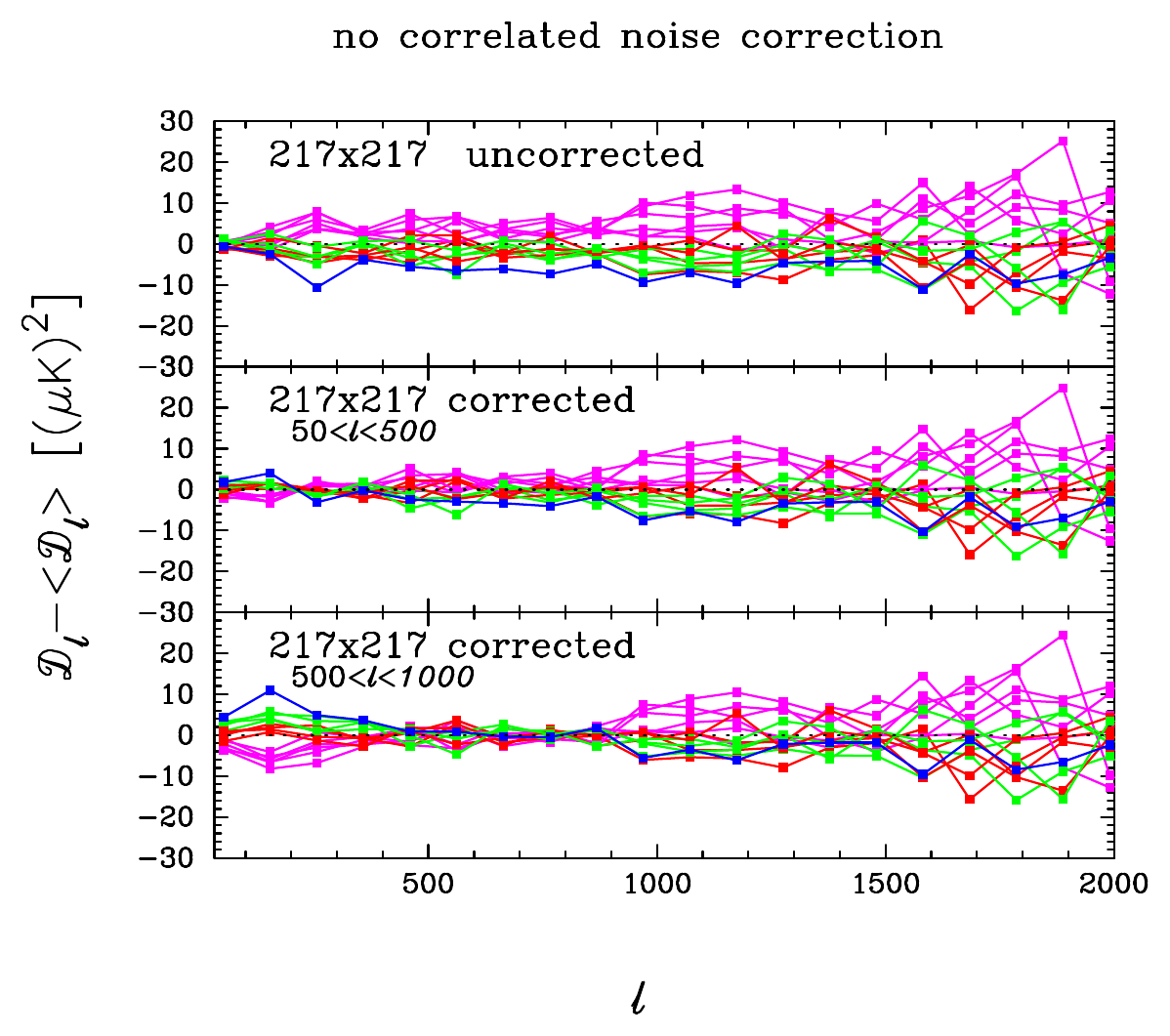} 	\includegraphics[width=73mm, angle=0]{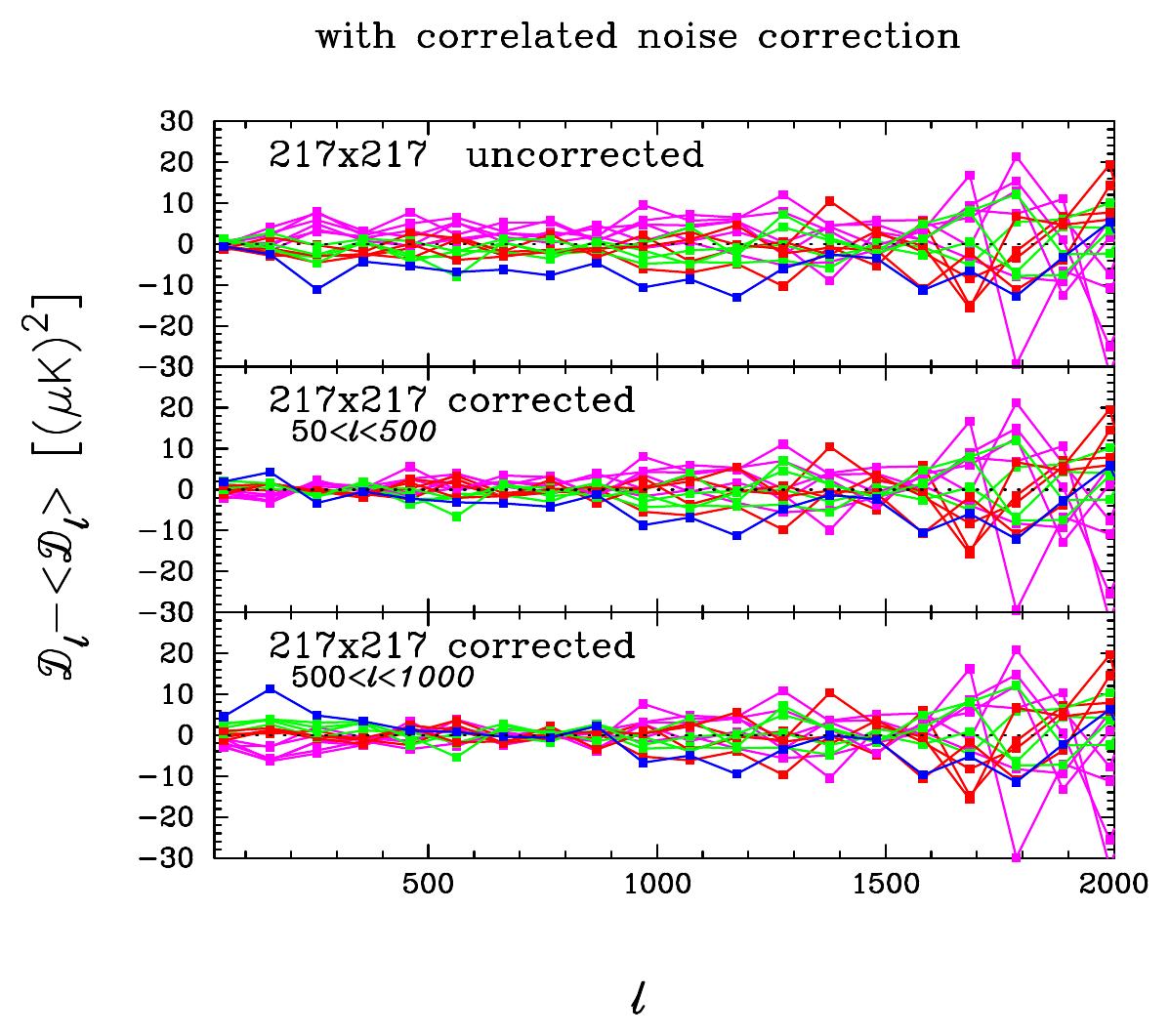} \\
	\includegraphics[width=73mm, angle=0]{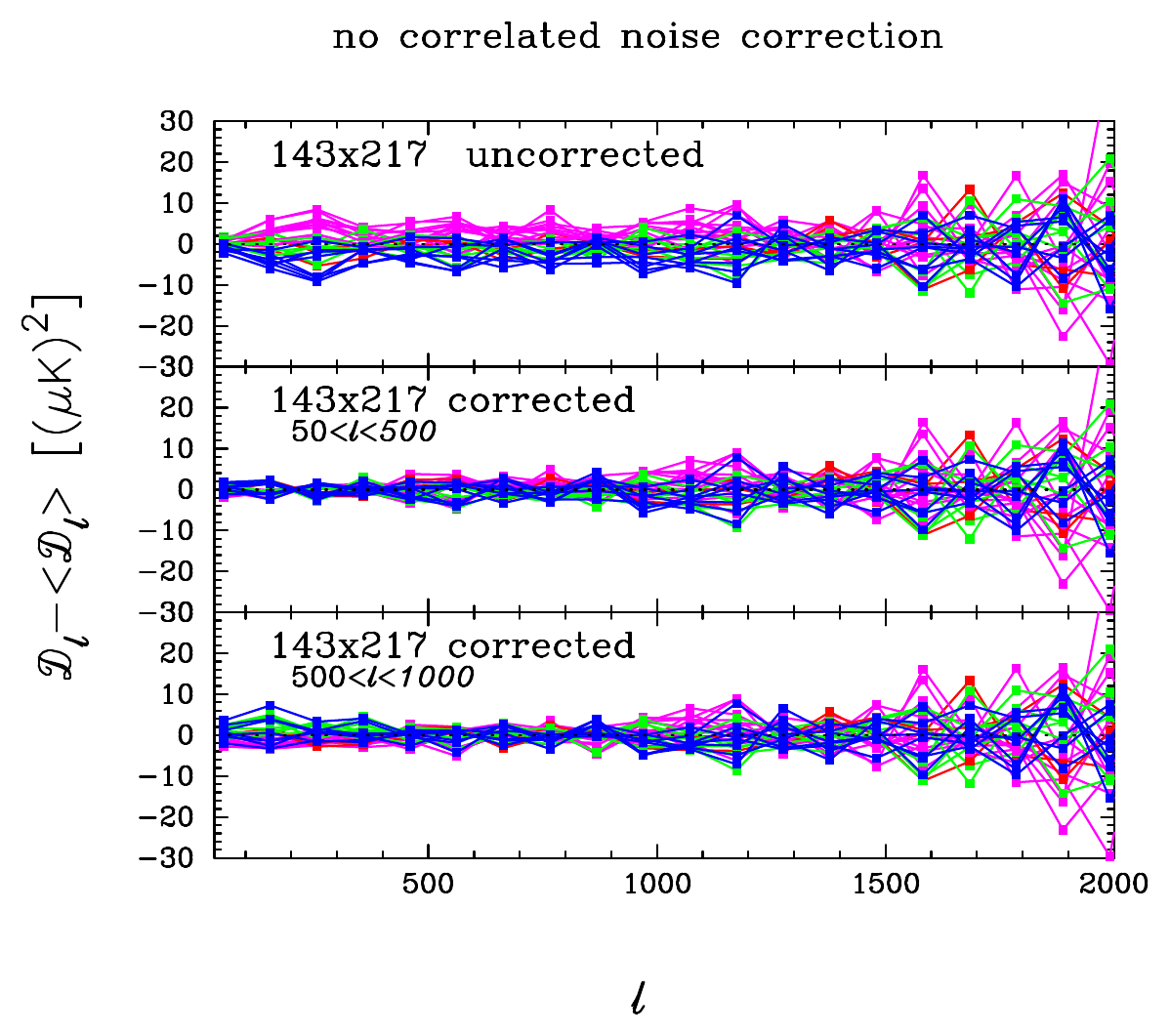} 	\includegraphics[width=73mm, angle=0]{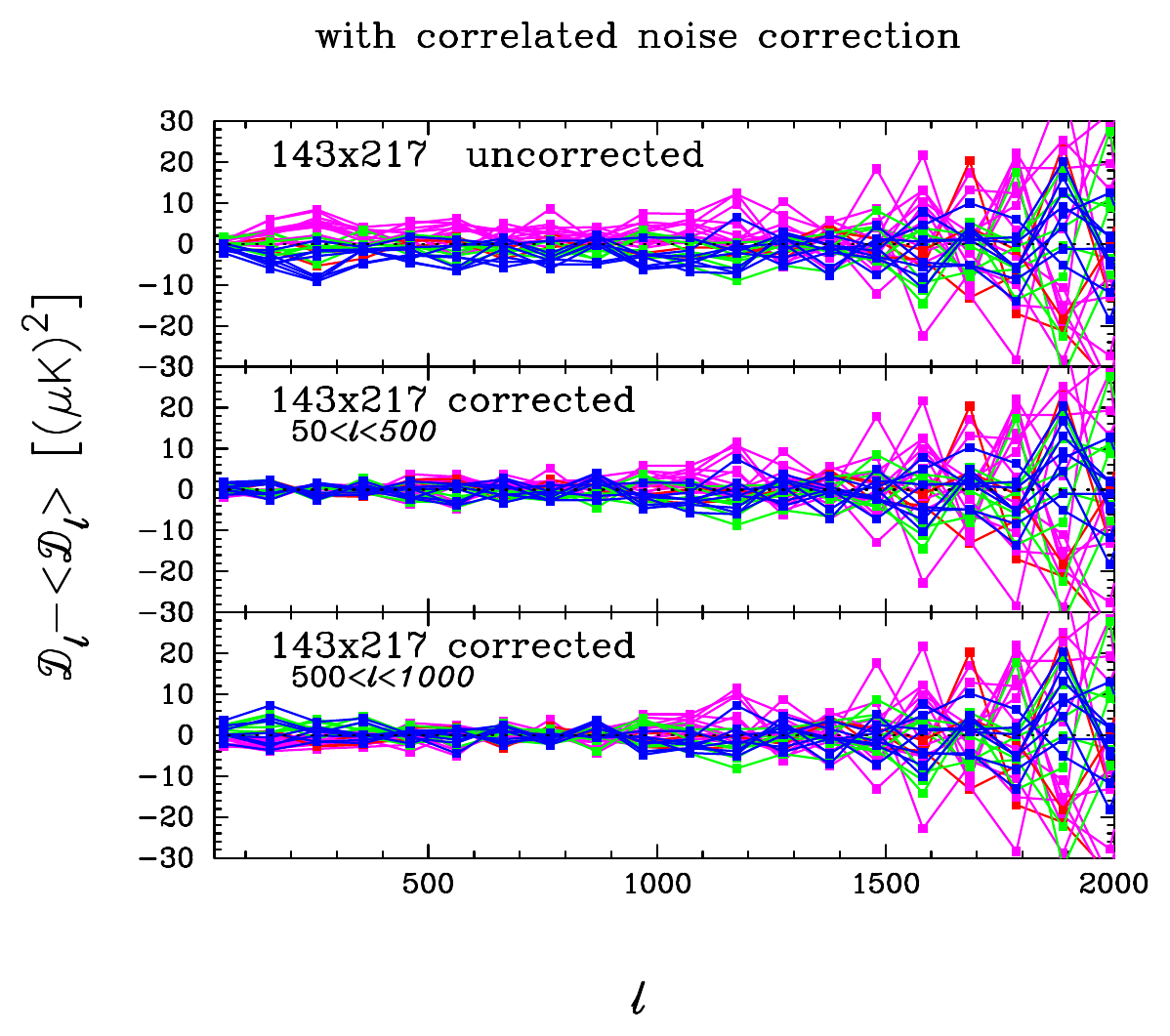} \\

	\caption{Intra-frequency residuals for the detset $143\times
          143$, $217\times217$ and $143\times 217$ TT spectra.  The
          spectra are colour coded as follows, SWB$\times$SWB spectra
          are in purple, SWB$\times$PSB in green and PSB$\times$PSB in
          blue. $\langle D_\ell \rangle$ is the mean of the spectra
          within each frequency group.  In  each panel, the top
          figure shows the raw spectra, the middle figure shows the
          corrected spectra with multiplicative corrections
          $\psi_{ij}$ determined by minimising Eq. \ref{equ:B6} over
          the multipole range $50 \le \ell \le 500$, the lower
          figure shows results of minimisation over the multipole range $500 \le
          \ell \le 1000$. The panels to the left show the cross
          spectra as measured from the maps with no correction for
          correlated noise.  The panels to the right show what
          happens if we subtract the odd-even difference spectra
          detset-by-detset as an indicator of correlated noise.}

	\label{fig:ifresiduals}

\end{figure}

Results are  shown in Fig.\ \ref{fig:ifresiduals}. The upper panel in each figure shows the residuals of the  beam corrected cross spectra
as measured from the detset maps relative to the mean spectrum $\langle D_\ell \rangle$ within each frequency group\footnote{We show residuals relative to the mean spectrum rather than to a theoretical model to eliminate residuals from cosmic variance and foregrounds.}. The middle and lower  
plots in each panel show the corrected spectra $\psi_{ij}{\hat D}^{ij}_\ell$
minimising Eq. \ref{equ:B6} over the multipole ranges $50 \le \ell \le 500$ and $500 \le \ell \le 1000$ respectively. We draw the following conclusions:

\smallskip 

\noindent
$\bullet$ For the $143\times143$ and $143\times217$ spectra, small calibration changes significantly reduce the scatter between the detset spectra with no systematic difference between SWB$\times$SWB, SWB$\times$PSB and PSB$\times$PSB spectra.  The corrections, $\psi_{ij}$, are stable with respect to multipole range and are almost identical for the two multipole ranges shown.

\smallskip

\noindent
$\bullet$  For the $217\times217$ cross spectra, we see a systematic separation between the SWB$\times$SWB, SWB$\times$PSB and PSB$\times$PSB spectra that remains after minimisation of Eq. \ref{equ:B6}. The figures to the right show what happens
if we subtract the odd-even difference spectra (see  Sect.\ \ref{subsec:correlatednoise})  detset-by-detset 
as a proxy for correlated noise. As can be seen, subtracting these corrections increases the noise levels at high multipoles. Nevertheless,
most of the difference between SWB$\times$SWB and SWB$\times$PSB in the 217$\times$217 spectra is
removed with this correction. The remaining differences between the spectra are
largely removed after minimisation of Eq. \ref{equ:B6}.

\smallskip

\noindent
$\bullet$ Correlated noise has no detectable effect on the 143$\times$143 and 143$\times$217  cross spectra.

\smallskip

The multiplicative factors $\psi_{ij}$ determined from Eq. \ref{equ:B6}
are all extremely close to unity. The means and standard deviations
about the mean are as follows: $\overline{\psi_{ij}}  = 1.0014$,
$\sigma_\psi = 7.77\times10^{-4}$ for $143 \times 143$; $\overline{
\psi_{ij}}  = 1.00057$, $\sigma_\psi = 7.59\times10^{-4}$ for
$217\times 217$; $\overline{ \psi_{ij}} = 1.00074$, $\sigma_\psi =
8.1\times10^{-4}$ for $143 \times 217$. These numbers are for minimising
Eq. \ref{equ:B6} over the multipole range $50 \le \ell \le 500$ with no
corrections of the spectra for correlated noise (and are very similar for
all of the fits shown in Fig.\ \ref{fig:ifresiduals}).

These results show that in temperature the \Quickpol\ beam transfer
functions have small residual errors at low multipoles which are
largely absorbed by  multiplicative calibration factors of order
$0.1\%$ in the power spectra. Correlated noise is responsible for part
of the mismatch in the $217 \times 217$ spectra shown in
Fig.\ \ref{fig:ifresiduals}. After correction for correlated noise and
multiplicative corrections, there is perhaps a hint of a transfer
function difference between the $217 \times 217$ PSB$\times$PSB spectrum
and the other detset spectra, but any difference is small and
unimportant for cosmology. These results demonstrate that the
intra-frequency spectra are consistent to extremely high accuracy.

\subsection{Relative calibrations of polarization spectra: effective polarization efficiencies}
\label{subsec:pol_cal}

Since there is a degeneracy between polarization efficiencies and
polarization angles (cf.\ Eq.\ \ref{equ:B0}), the \SROLL\ mapmaking
algorithm assumes the ground based calibrations of
\cite{Rosset:2010}. The combined systematic and statistical errors in
the polarization efficiencies are uncertain and may be as high as a
few percent. Errors in the polarization efficiencies will show up as
multiplicative calibration factors in the TE and EE spectra. However,
by intercomparing power spectra, we measure {\it effective
  polarization efficiencies}, because of the degeneracy between
polarization angles and polarization efficiencies in the map making
stage and because of transfer functions caused by errors in modelling
the beams beyond the main beam.  The relative calibrations discussed
here should be interpreted in this light and should not be interpreted
as bolometer polarization efficiencies.

Our analysis of polarization efficiencies differs from the TT
calibration analysis described in the previous section. Since we have
many fewer detset EE spectra than TT spectra within a frequency
combination (one only for each of $100 \times 100$, $143 \times 143$
and $217 \times 217$), we cannot minimise intra-frequency residuals
between detset spectra to fix calibration coefficients. However,
apart from Galactic dust emission, there are no other foreground
contributions detectable in the \Planck\ TE and EE spectra. We can
therefore perform {\it inter-frequency} comparisons of TE and EE spectra to determine effective polarization efficiencies with polarized dust emission removed using $353$ GHz maps as templates.
 The dust subtraction is discussed in detail in
Sect.\ \ref{sec:dust_polarization}. In addition, the TE and EE spectra
are corrected for TP leakage using Eqs
\ref{equ:B2a}-\ref{equ:B2b} with the \Quickpol\ polarized beams
assuming the best fit base \LCDM\ TT, TE and EE spectra for the
12.1HM TT likelihood (which we will henceforth refer to as the
fiducial theoretical model). Tests of the TP
leakage model are discussed in Sect.\ \ref{subsec:pol_leak}.  The
analysis described in this section has been done for both half mission
and detset spectra, though for reasons of economy we present only the
half mission results.

Since the TE and EE spectra are noisy, we determine calibration factors $c^{TE}_k$ and $c^{EE}_k$ for each TE and EE spectrum by minimising the residuals with respect to the theoretical  TE and EE spectra of the fiducial cosmology.
For either TE or EE we therefore minimise 
\begin{equation}
\chi^2 = \sum_{\ell_1\ell_2} ({\hat C}^k_{\ell_1} - c_k C^{\rm theory}_{\ell_1})(M^k)^{-1}_{{\ell_1}{\ell_2}}({\hat C}^k_{\ell_2} - 
c_k C^{\rm theory}_{\ell_2}), \label{equ:chi1}
\end{equation}
with respect to $c_k$, where the index $k$ identifies the spectrum, $M^k$ is the covariance matrix for spectrum $k$
and $C^{\rm theory}_\ell$ is the relevant  theoretical spectrum. This gives:
\begin{equation}
c_k = \sum_{\ell_1\ell_2} C^{\rm theory}_{\ell_1}(M^k)^{-1}_{{\ell_1}{\ell_2}}{\hat C}^k_{\ell_2}/
\sum_{\ell_1\ell_2} C^{\rm theory}_{\ell_1}(M^k)^{-1}_{{\ell_1}{\ell_2}}C^{\rm theory}_{\ell_2}.  \label{equ:chi2}
\end{equation}
The sums in Eqs. \ref{equ:chi1} and \ref{equ:chi2} extend over the
multipole ranges $\ell_{\rm min } \le \ell_1 \le \ell_{\rm max}$,
$\ell_{\rm min } \le \ell_2 \le \ell_{\rm max}$.  Note that with this
definition of $c_k$,   only theory terms enter in
the denominator in Eq. \ref{equ:chi2} giving reasonably stable
estimates of $c_k$ from the noisy \planck\ polarization spectra. Since
the $\chi^2$ in Eq. \ref{equ:chi1} is correctly normalized, we can calculate 
error estimates on the coefficients $c_k$.

We apply this scheme first to the half mission EE spectra. The HM maps
are labelled (1)-(6) in the order 100HM1, 100HM2, 143HM1, 143HM2,
217HM1, 217HM2, where HM1 refers to maps from the first half mission
and HM2 to the second half mission. We exclude auto spectra and any
other cotemporal spectra (e.g.\ 100HM1$\times$143HM1,
100HM2$\times$143HM2). This leaves nine EE spectra with map indices as
given in the first column of Table \ref{tab:EE_cal}. The next three
columns give the best fit calibration factors and 1$\sigma$ errors for
fits over three multipole ranges. The numbers are stable with respect
to multipole ranges and differ from unity by up to a few percent. In
two cases, the best fit calibrations exceed unity (which is possible
given that \SROLL\ assumes the  polarization
efficiencies of reference \cite{Rosset:2010}). Our results are consistent with the conclusions of
\citep{DataProcessing:2018} and PPL18, namely that systematic errors in the HFI
effective polarization calibrations are at the level of $1\%$ or more,
i.e.\ several times larger than the statistical errors on the detector
polarization efficiencies quoted in \cite{Rosset:2010}\footnote{Note
  that at the power spectrum level, the errors are doubled relative to the map level.}.

\begin{table}[h]
{       \centering
	\caption{\small{Relative calibrations of half mission EE spectra}}
	\label{tab:EE_cal}
       \begin{center}
	\begin{tabular}{ccccc} 
		\hline
        {\rm Spectrum}&{\rm EE index} &   $200-1000$ &  $200-1500$  &  $500-1000$ \\
 100HM1x100HM2 &(1,2)    & $0.978 \pm 0.010$ & $0.979 \pm 0.011$ & $0.978 \pm 0.019$ \\
 100HM1x143HM2 &(1,4)    & $1.010 \pm 0.008$ & $1.010 \pm 0.008$ & $1.013 \pm 0.014$ \\
 100HM1x217HM2 &(1,6)    & $0.958 \pm 0.010$ & $0.960 \pm 0.010$ & $0.949 \pm 0.016$ \\
 100HM2x143HM1 &(2,3)    & $0.998 \pm 0.008$ & $0.998 \pm 0.008$ & $1.011 \pm 0.013$ \\
 100HM2x217HM1 &(2,5)    & $0.956 \pm 0.010$ & $0.954 \pm 0.010$ & $0.960 \pm 0.016$ \\
 143HM1x143HM2 &(3,4)    & $1.034 \pm 0.006$ & $1.037 \pm 0.006$ & $1.043 \pm 0.010$  \\
 143HM1x217HM2 &(3,6)    & $0.982 \pm 0.008$ & $0.985 \pm 0.008$ & $0.966 \pm 0.011$ \\
 143HM2x217HM1 &(4,5)    & $0.985 \pm 0.008$ & $0.985 \pm 0.008$ & $0.984 \pm 0.012$ \\
 217HM1x217HM2 &(5,6)    & $0.959 \pm 0.010$ & $0.960 \pm 0.009$ & $0.930 \pm 0.014$ \\
		\hline
              \end{tabular}
\end{center}}

\end{table}

\begin{table}[h]
{
        \centering
	\caption{\small{Relative calibrations of half mission TE spectra}}
	\label{tab:TE_cal}
       \begin{center}
	\begin{tabular}{ccccc} 
		\hline
        {\rm Spectrum}  &{\rm TE index}   &  $200-1000$ &  $200-1500$  & $500-1000$ \\
 100HM1x100HM2 & (1,2)    & $0.990 \pm 0.013$ & $0.986 \pm 0.012$ &   $0.984 \pm 0.022$  \\
 100HM1x143HM2 & (1,4)    & $1.005 \pm 0.010$ & $1.004 \pm 0.010$ &   $0.980 \pm 0.015$ \\
 100HM1x217HM2 & (1,6)    & $0.987 \pm 0.012$ & $0.988 \pm 0.011$ &   $0.971 \pm 0.015$ \\
 100HM2x143HM1 & (2,3)    & $1.010 \pm 0.011$ & $1.010 \pm 0.010$ &   $1.010 \pm 0.016$ \\
 100HM2x217HM1 & (2,5)    & $0.969 \pm 0.013$ & $0.974 \pm 0.012$ &   $0.967 \pm 0.020$ \\
 143HM1x143HM2 & (3,4)    & $1.002 \pm 0.010$ & $1.002 \pm 0.009$ &   $0.977 \pm 0.015$ \\
 143HM1x217HM2 & (3,6)    & $0.991 \pm 0.012$ & $0.988 \pm 0.011$ &   $0.980 \pm 0.018$ \\
 143HM2x217HM1 & (4,5)    & $0.971 \pm 0.013$ & $0.972 \pm 0.012$ &   $0.968  \pm 0.019$ \\
 217HM1x217HM2 & (5,6)    & $0.995 \pm 0.012$ & $0.992 \pm 0.012$ &   $0.990 \pm 0.019$ \\
 100HM1x100HM2 & (2,1)    & $0.982 \pm 0.014$ & $0.984 \pm 0.013$ &   $0.966 \pm 0.023$ \\
 100HM1x143HM2 & (4,1)    & $0.978 \pm 0.014$ & $0.980 \pm 0.013$ &   $0.957 \pm 0.023$ \\
 100HM1x217HM2 & (6,1)    & $0.980 \pm 0.014$ & $0.982 \pm 0.014$ &   $0.964 \pm 0.024$ \\
 100HM2x143HM1 & (3,2)    & $0.979 \pm 0.013$ & $0.983 \pm 0.013$ &   $0.966 \pm 0.022$ \\
 100HM2x217HM1 & (5,2)    & $0.985 \pm 0.014$ & $0.989 \pm 0.013$ &   $0.966 \pm 0.023$ \\
 143HM1x143HM2 & (4,3)    & $1.012 \pm 0.011$ & $1.013 \pm 0.010$ &   $1.012 \pm 0.016$ \\
 143HM1x217HM2 & (6,3)    & $1.008 \pm 0.011$ & $1.010 \pm 0.010$ &   $1.007 \pm 0.017$ \\
 143HM2x217HM1 & (5,4)    & $1.007 \pm 0.011$ & $1.008 \pm 0.009$ &   $0.981 \pm 0.016$ \\
 217HM1x217HM2 & (6,5)    & $0.973 \pm 0.014$ & $0.974 \pm 0.012$ &   $0.970 \pm 0.020$ \\

		\hline
              \end{tabular}
\end{center}}

\end{table}

Table \ref{tab:TE_cal} summarizes results on the effective calibrations of the half mission TE spectra. As with the
EE spectra, the calibration coefficients are close to unity to within $2-3\%$ and are relatively insensitive to the multipole
ranges used in the fits. There is, however, a tendency for smaller effective calibrations if multipoles $\ell < 500$ are
excluded from the fits. This suggests that treating the TE efficiency factors as purely multiplicative may be an oversimplification. (The 143HM1$\times$143HM2 TE spectrum is the worst offender, though most of the spectra show a similar
trend.)

We now ask whether we can relate the numbers in Tables \ref{tab:EE_cal} and \ref{tab:TE_cal}. Section \ref{subsec:intra_frequency} demonstrates that any effective calibration errors in the TT maps are negligible compared to the polarization calibrations
given in Tables \ref{tab:EE_cal} and \ref{tab:TE_cal}. We can therefore assume that the temperature maps are perfect and that any deviations from unity  in the TE calibrations listed in Table \ref{tab:TE_cal}
are caused by  effective polarization efficiencies, $\rho_i$.
With this assumption, we can group the TE spectra into triplets giving an effective polarization efficiency 
 for each map. The results are summarized  in Table \ref{tab:map_pol_cal} (for which we use the calibrations from Table
\ref{tab:TE_cal} computed over the multipole range $200 \le \ell \le 1000$).

\begin{table}[h]
{       \centering
	\caption{\small{Effective polarization efficiencies}}
	\label{tab:map_pol_cal}
\begin{center}
	\begin{tabular}{cccccc} 
		\hline
        {\rm Map} &k      &   &   &  &  $\bar \rho_k$ \\
 100HM1  & 1  & (2,1) $0.982$ & (4,1) $0.978$ & (6,1)  $0.980$ & $0.980$ \\
 100HM2  & 2  & (1,2) $0.990$ & (3,2) $0.979$ & (5,2)  $0.985$ & $0.984$ \\
 143HM1  & 3  & (2,3) $1.010$ & (4,3) $1.012$ & (6,3)  $1.008$ & $1.007$ \\
 143HM2  & 4  & (1,4) $1.005$ & (3,4) $1.002$ & (5,4)  $1.007$ & $1.005$ \\
 217HM1  & 5  & (2,5) $0.969$ & (4,5) $0.971$ & (6,5)  $0.973$ & $0.971$\\
 217HM2  & 6  & (1,6) $0.987$ & (3,6) $0.991$ & (5,6)  $0.995$ & $0.991$ \\
		\hline
              \end{tabular}
\end{center}}

\end{table}

Notice the good agreement between the numbers within each triplet,
which are consistent to $\simlt 0.005$ (i.e.\ significantly better
than the errors of $0.010-0.013$ given in Table \ref{tab:TE_cal}). The
last column in Table \ref{tab:map_pol_cal} gives the average value of
the calibrations for each triplet, $\bar \rho_k$, which we take as our
estimate of the effective polarization efficiency for each half
mission map. Note also that these polarization efficiencies pair up by
frequency. The ground based polarization efficiencies measured by
\cite{Rosset:2010} are strongly correlated within a frequency band,
and so our results suggest that systematic biases in the ground based
calibrations are similar for all detectors within a frequency band.

\begin{figure}
	\centering
	\includegraphics[width=73mm, angle=0]{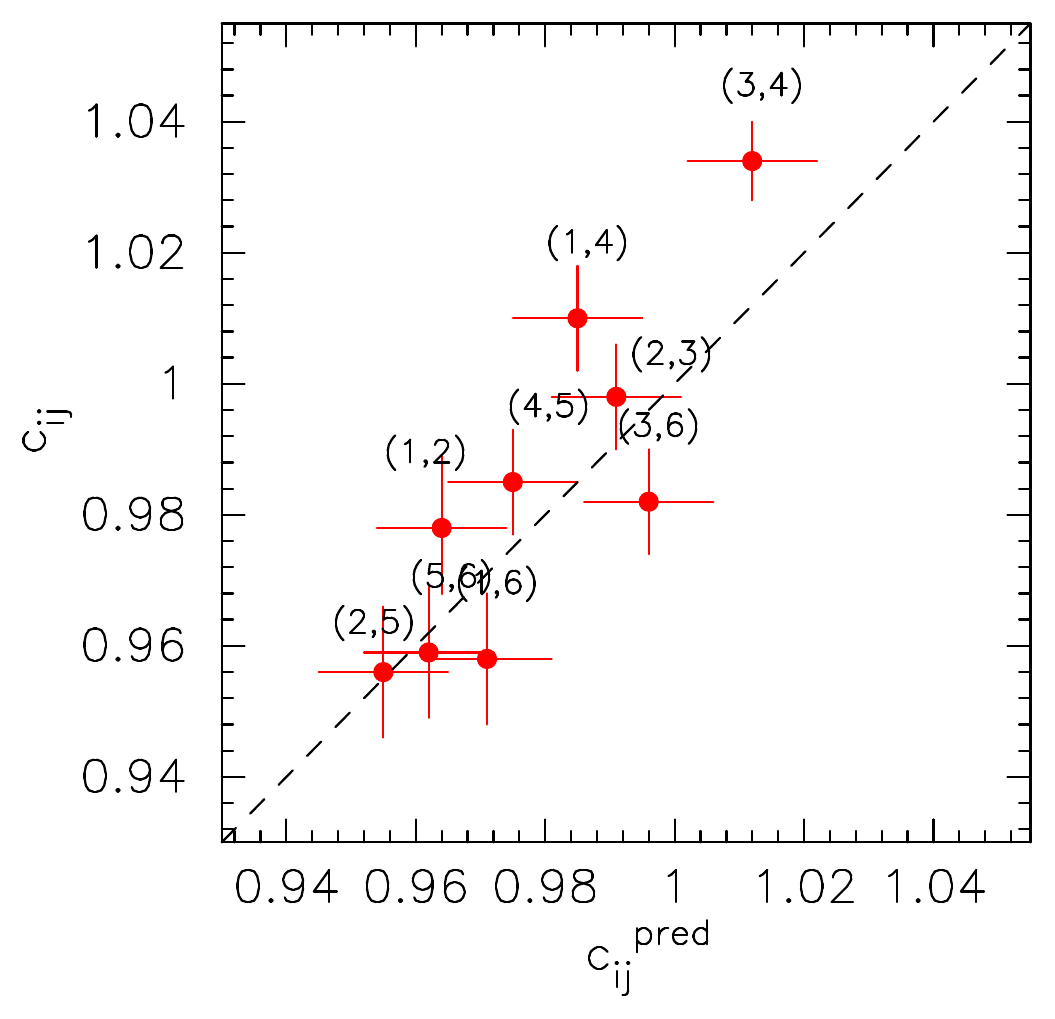}
	\caption{Calibrations of the EE spectra from Table \ref{tab:EE_cal} with $1\sigma$ errors plotted against the
predicted coefficients from Eq. \ref{equ:chi3}. We have assigned a nominal $1\sigma$ error of $\pm 0.01$ on $c^{\rm pred}_{ij}$.
Each point is labelled with the map indices $i,j$ as in Table \ref{tab:EE_cal}.}
	\label{fig:pol_cal}

\end{figure}

From these effective polarization efficiencies, we can try to predict the calibrations $c_{ij}$ for each EE spectrum listed in 
Table \ref{tab:EE_cal}:
\begin{equation}
c^{\rm pred}_{ij} = \bar \rho_i \bar \rho_j.  \label{equ:chi3}
\end{equation}
The results are summarized in Fig.\ \ref{fig:pol_cal}.

There is a strong correlation between the measured calibration coefficients $c_{ij}$ for the EE spectra and those
predicted from Eq. \ref{equ:chi3}. This is consistent with the idea that the effective calibrations measured from the EE
and TE spectra come mainly  from systematic errors in the ground based polarization efficiency calibrations. The correlation
in Fig.\  \ref{fig:pol_cal} is not perfect, however, and is not expected to be perfect since errors in polarization efficiencies
are strongly coupled to errors in polarization angles and far-field polarized beams.

In the \camspec\ half mission likelihoods, we apply the calibration
factors from Tables \ref{tab:EE_cal} and \ref{tab:TE_cal} to the EE
and TE half mission spectra prior to coaddition. For the full mission
detset likelihoods, we apply corrections determined from a similar
analysis using all detset polarization spectra. Since these calibrations are computed with
respect to a theoretical model determined from the TT likelihood,
amplitude parameters determined from the TE and  EE likelihoods are
not strictly independent of those determined from TT. Furthermore, as can be
seen from Tables \ref{tab:EE_cal}, \ref{tab:TE_cal} and
Fig.\ \ref{fig:pol_cal}, the calibration factors applied are accurate
to no better than about a percent.  We therefore include calibration
factors $c^{EE}$ and $c^{TE}$ for the coadded EE and TE spectra which
are treated as nuisance parameters in the likelihood, with priors as
given in Table \ref{tab:priors}.

The scheme discussed above for estimating polarization efficiencies in
\camspec\ differs from that used in the \plik\ likelihood (
described in PPL18). The \GE{publicly} distributed \plik\ likelihood code with its
default settings uses
effective polarization efficiencies determined from the  EE
spectra which are then applied to the TE spectra\footnote{Which in
  \plik\ are symmetrized, i.e.  T$^i$E$^j$ and
  E$^i$T$^j$ are averaged.}  on the assumption that the efficiencies are `map
based'. With this procedure, TE spectra involving $143$ GHz receive a
large (and uncertain) polarization efficiency correction (see Table \ref{tab:EE_cal}
and Fig.\ \ref{fig:pol_cal}) which is not wanted by the TE spectra. In
producing a TTTEEE likelihood it is essential that polarization
efficiencies are corrected accurately in the TE spectra; errors in
EE polarization efficiency corrections are of less importance,  since
the EE spectra are noisy and carry little statistical weight in a
TTTEEE likelihood. In producing \camspec\ we therefore corrected each
TE/ET spectrum with the polarization efficiency corrections given in
the third column of Table \ref{tab:TE_cal}. We can also see from Table
\ref{tab:map_pol_cal} that these efficiency corrections are consistent
with map based coefficients $\overline \rho_k$. In fact, for each
frequency we have six estimates of a map based effective polarization
efficiency which are consistent to typically $\pm 0.005$. For
\camspec\ it would therefore make no difference to the TE likelihood
had we adopted `map based' polarization efficiencies rather than
`spectrum based' efficiencies. The results of this section show that our
procedure for determining and correcting polarization efficiencies is 
strongly supported by the data. The differences in correcting polarization
efficiencies are largely responsible for the differences between the \camspec\
and \plik\ TTTEEE likelihoods for some cosmologies\footnote{As discussed in the revised version of PCP18.}, in particular when the
parameters $A_L$ and $\Omega_{\rm K}$ are allowed as extensions to the base \LCDM\
cosmology (see Sect. \ref{sec:extensions_lcdm} for further discussion). For these
cosmologies, the \camspec\ TTTEEE likelihood is more reliable than the \plik\
likelihood with its default settings.

\subsection{Tests of the temperature-polarization leakage model}

\label{subsec:pol_leak}

Beam mismatch introduces temperature-to-polarization (TP) leakage in the
\Planck\ maps, which can be characterised by the polarized beam matrix
of Eq.\ \ref{equ:B1}. For the TE spectra, TP leakage has a small but
non-negligible effect on cosmology. For the EE spectra, TP leakage is
small compared to the EE noise and can be safely neglected. In this
section, we test the \Planck\ polarized beam model for the TE
spectra. As in the previous section, we assume that the temperature
maps are perfect. We can then arrange the TE HM spectra, uncorrected
for beam leakage, into triplets as in Table \ref{tab:map_pol_cal} and
search for correlated residuals. These residuals should be caused by
temperature leakage into the (Q,U) maps, which are identical within each
triplet. We therefore expect that these residuals should match up with
the residuals computed from the polarized beam matrix.

\begin{figure}[tp]
	\centering
	\includegraphics[width=73mm, angle=0]{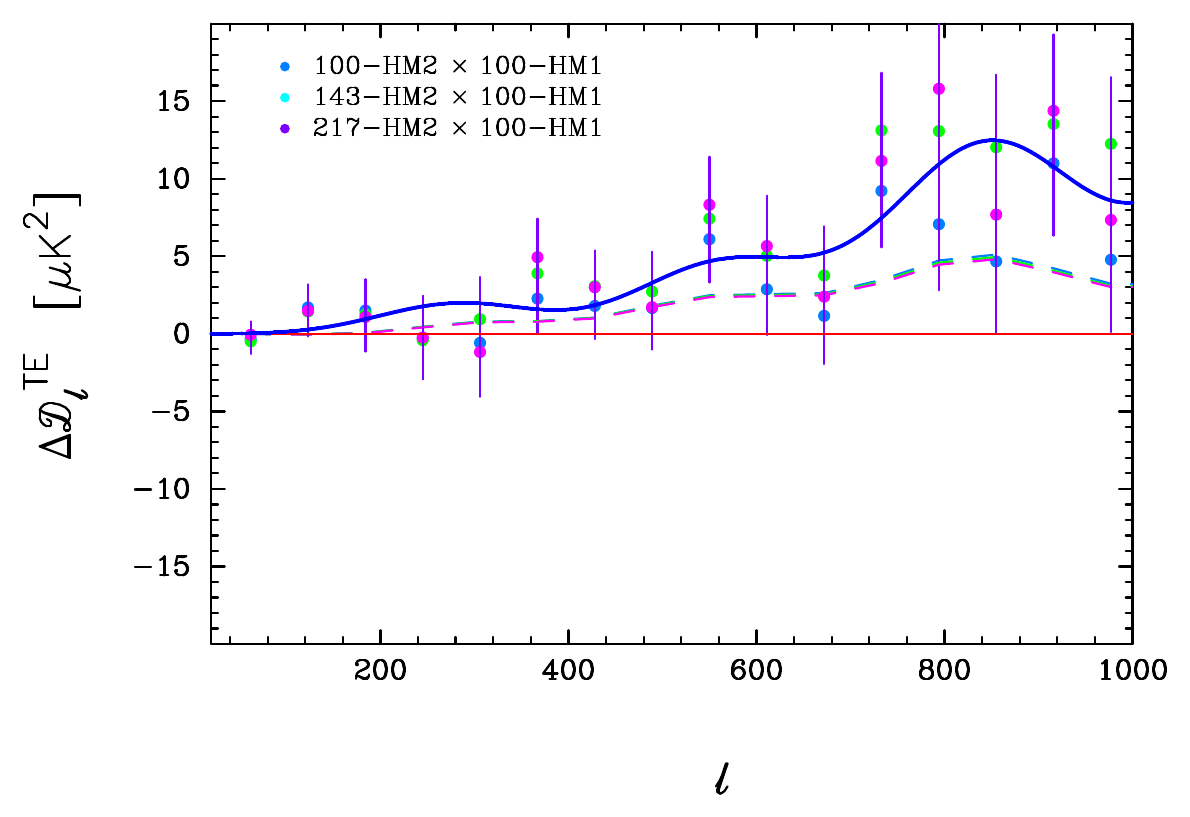} 	\includegraphics[width=73mm, angle=0]{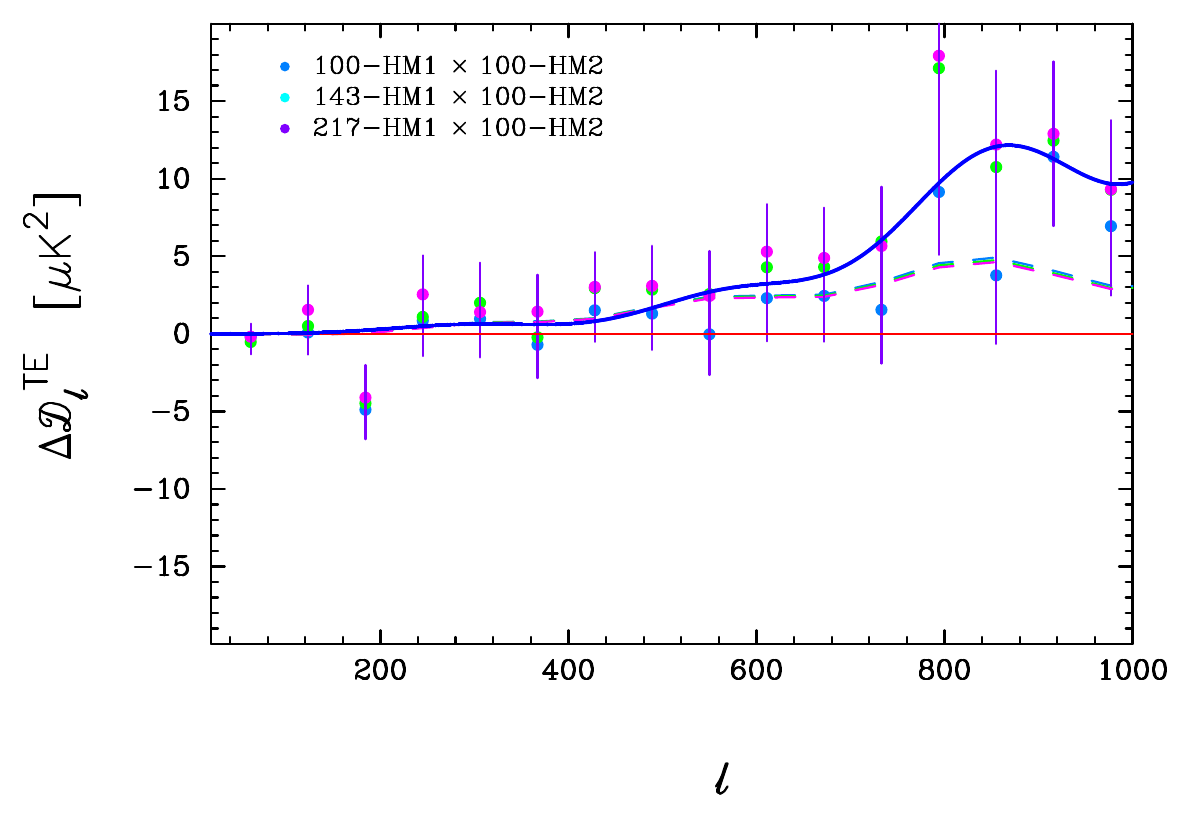} \\
	\includegraphics[width=73mm, angle=0]{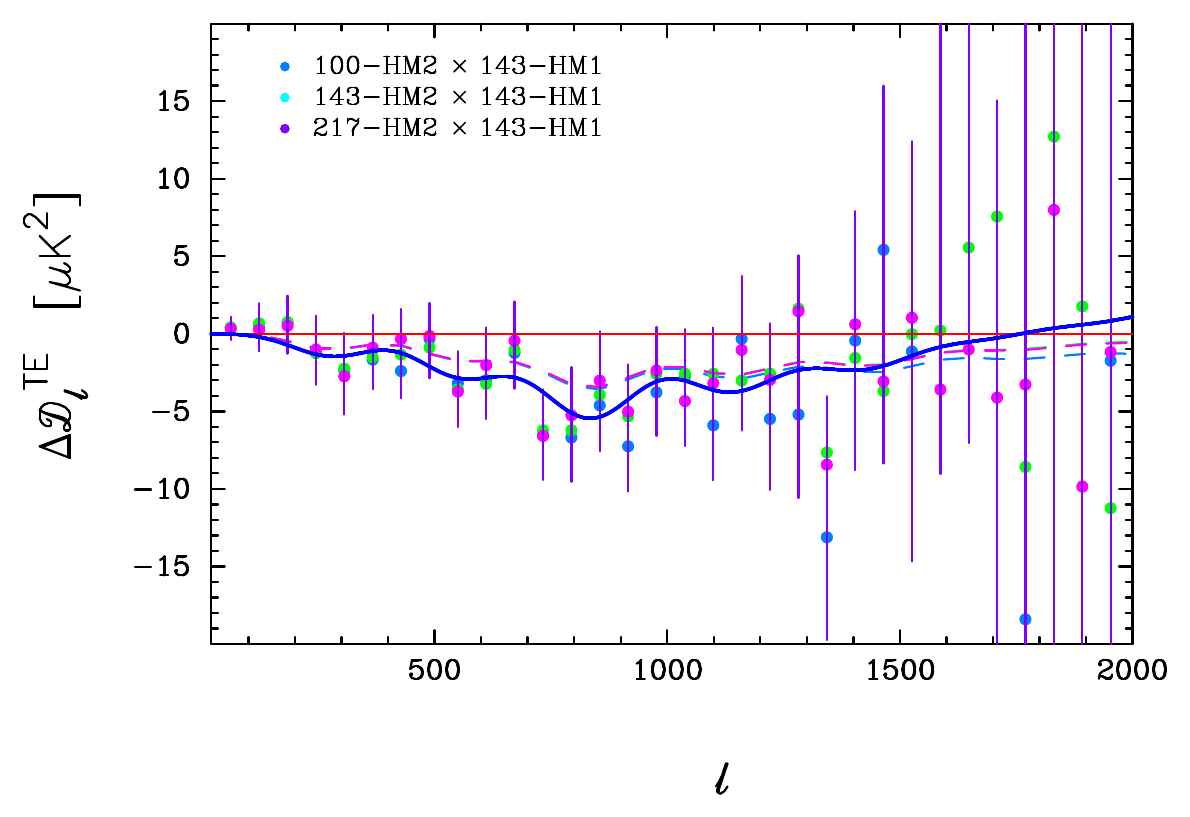} 	\includegraphics[width=73mm, angle=0]{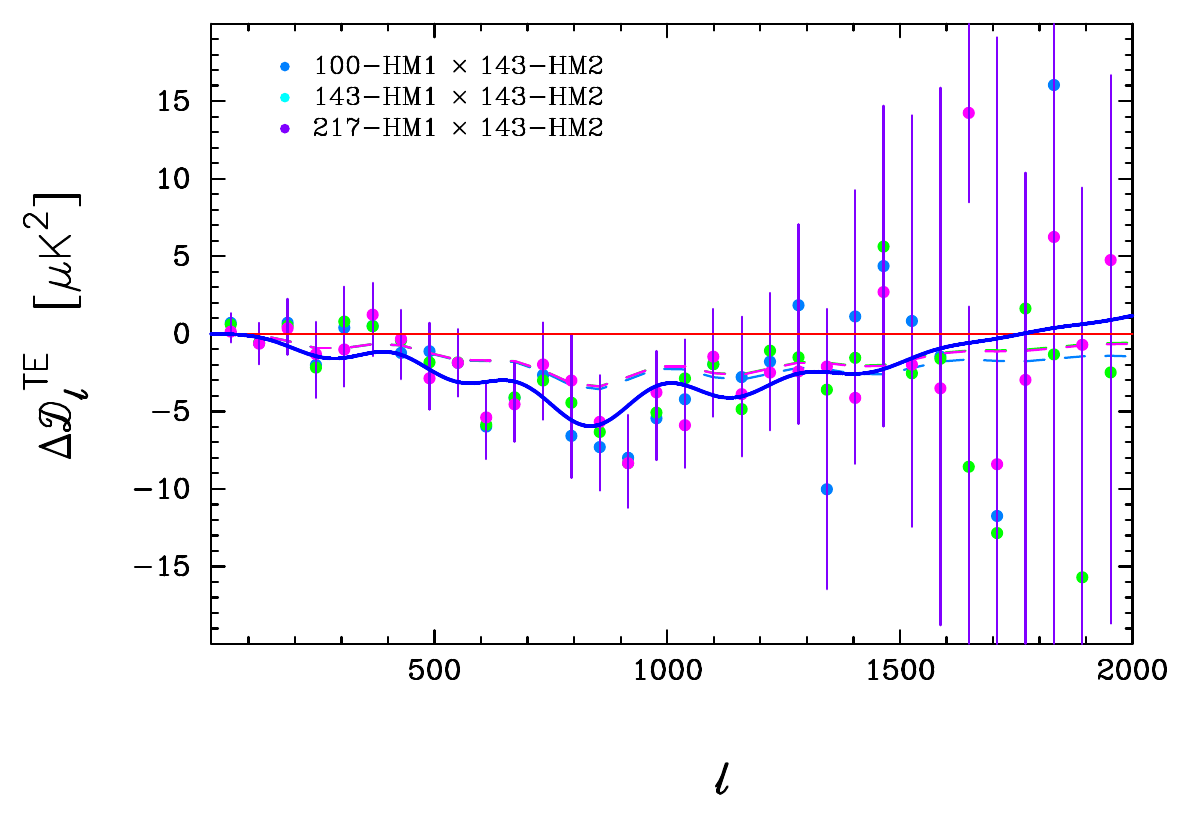} \\
	\includegraphics[width=73mm, angle=0]{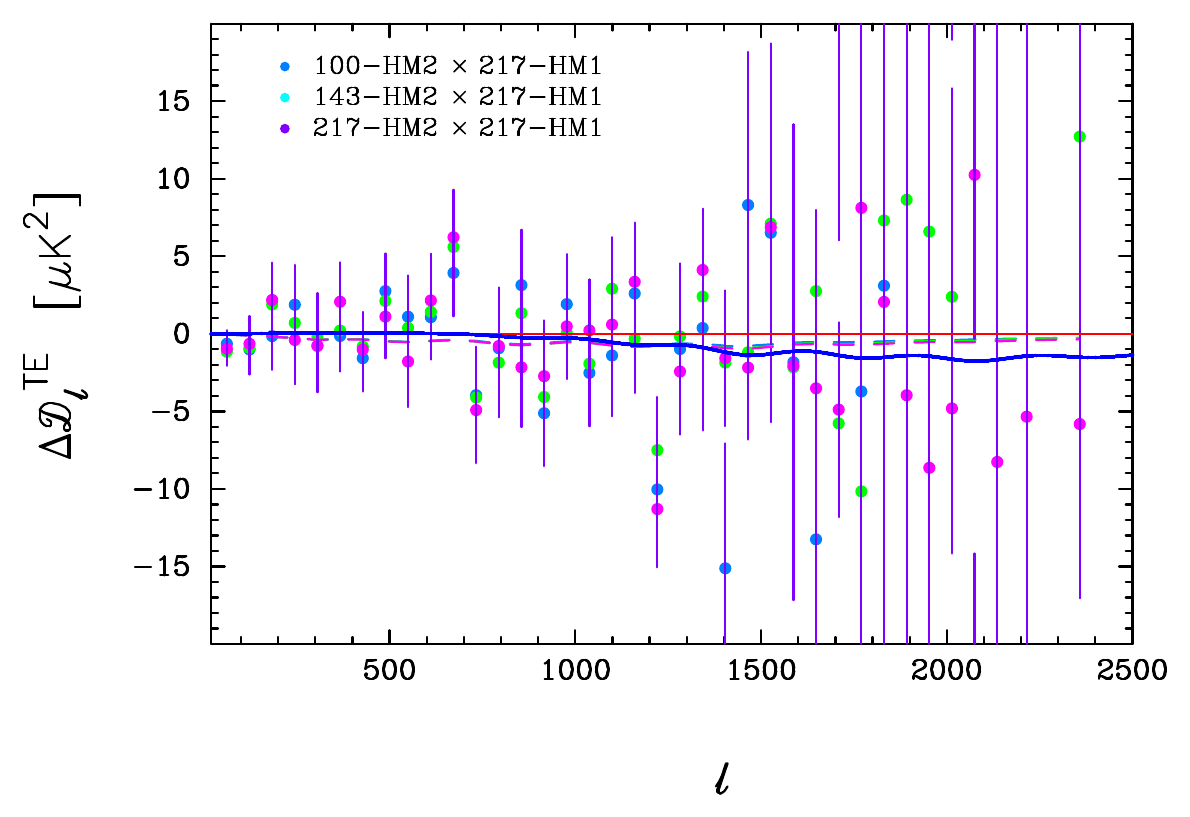} 	\includegraphics[width=73mm, angle=0]{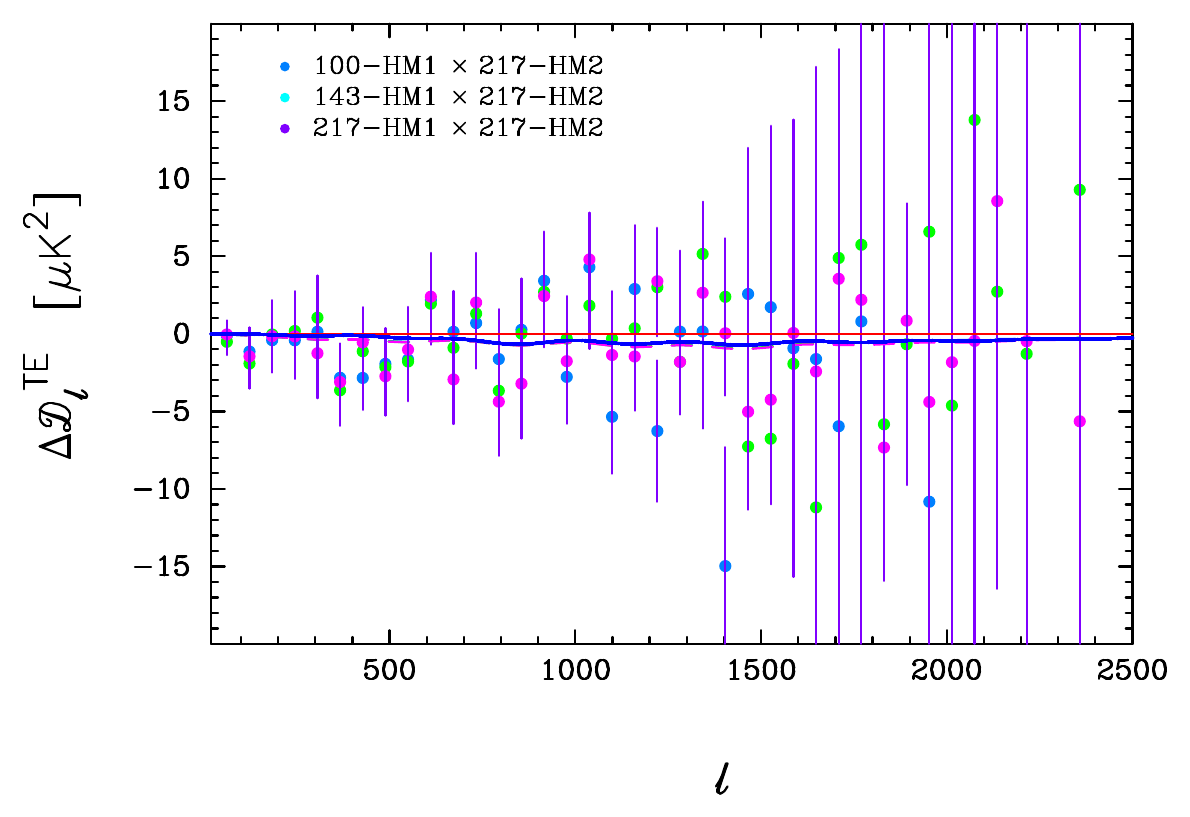} \\

	\caption{TE half mission spectrum residuals (mask and beam
          deconvolved and corrected for effective polarization
          efficiencies) organized into triplets. Each triplet
          corresponds to a distinct polarization half mission map
          (e.g. top left is 100HM1 and bottom right is 217HM2). The
          dotted lines show the TP leakage computed from the polarized
          beams as described in the text.  The solid lines show a fit
          to the leakage based on Eqs. \ref{equ:B4b} and \ref{equ:B5}.
          The error bars show $\pm 1\sigma$ errors computed from the
          \camspec\ covariance matrices.}
	\label{fig:leak}

\end{figure}

This test is illustrated in Fig.\ \ref{fig:leak}. The TE spectra shown in this figure are 
divided by the calibration coefficients given  in Table \ref{tab:TE_cal}. In each panel, we show the residuals of the TE
spectra in each triplet relative to the mean TE spectrum computed from all $18$ cross spectra (weighted by the
diagonals of the covariance matrices as in Eq. \ref{equ:C11}). Subtracting the mean TE spectrum reduces scatter from 
cosmic variance in these plots. The residuals within each panel match up extremely well. Furthermore, the patterns
of these residuals are very similar for polarization maps with the same frequency,  as expected from the \Quickpol\ polarized
beam matrix. The dotted lines in Fig.\ \ref{fig:leak}
show the residuals computed from the polarized beams  assuming the fiducial  base \LCDM\ cosmology.
(These dotted lines are the TP corrections applied to the  TE spectra in the \camspec\ half mission likelihoods used in this paper.)
The TP corrections computed from the polarized beams provide a good match to the TE residuals and have nearly identical
shapes within each triplet. The solid lines show fits of the  model of Eqs. \ref{equ:B4b} and \ref{equ:B5} to the data
points with $\epsilon_2$ and $\epsilon_4$ as free parameters. This simple model provides a good match to the polarized beam
TP leakage model, except at high multipoles where the TE spectra become noise dominated.

The tests shown in Fig.\ \ref{fig:leak} provide strong evidence that
the polarized beams accurately account for TP leakage in the TE
spectra. If we coadd the 18 half mission TE spectra, the TP leakage
corrections partially cancel so that the net effect of TP leakage is
relatively small. If we ignore TP leakage entirely in the \camspec\
TTTEEE likelihoods we find shifts of up to $1\sigma$ in base \LCDM\
parameters in agreement with the \plik\ results reported in PCP18 and
PPL18. Given the tests shown in Fig.\ \ref{fig:leak} we can be confident
that the TP corrections applied to the \camspec\ spectra are reasonably
accurate and that any errors in these corrections have significantly less
that $1\sigma$  impact on cosmological
parameters.  The
beam-derived TP leakage corrections for the EE spectra are negligible
and have no impact on cosmology.

In summary, the results of this section show that the main systematic
in the \planck\ polarization spectra is caused caused by errors in the
polarization efficiencies assumed in \SROLL\ and errors in the
far-field \Planck\ beams.  These effects can be accurately modelled by
multiplicative calibration factors (effective polarization
efficiencies) applied to each of the TE and EE spectra. TP leakage is
a subdominant effect and we have demonstrated via internal
consistency tests that TP leakage is described
accurately by the \Quickpol\ polarized beam matrices.

\section{Galactic dust emission in temperature}
\label{sec:dust_temp}

To extract cosmology from CMB experiments, Galactic and extragalactic
foregrounds need to be removed to high accuracy to reveal the primary
CMB anisotropies. There is a large literature on foreground cleaning
methods which will not be reviewed here.  Various techniques have been
applied to the \Planck\ data, and are described in
\citep{Planck_components:2014,
  Planck_component2015,Planck_foregrounds_2018} (to which we refer the
reader for details, including references to earlier work). Broadly
speaking, the methods can be divided into two classes: methods such as
\commander\ \citep{Eriksen:2006, Eriksen:2008}, which fit a parametric
model of foreground spectral energy distributions to a set of maps at
different frequencies, and template fitting methods which are based on
linear combinations of pixelised maps (or `needlets' in the multipole
domain) at different frequencies (e.g. \ILC, \SMICA, \SEVEM,
\NILC)\footnote{\ILC: Internal Linear Combination \cite{Bennett:2003};
  \SMICA: Spectral Matching Independent Component Analysis
  \cite{Cardoso:2008}; \NILC: Needlet Internal Linear Combination
  \cite{Delabrouille:2009}; \SEVEM: Spectral Estimation Via
  Expectation Maximisation \cite{Leach:2008, SEVEM}.}. In contrast to
these techniques, in the \camspec\ likelihood we remove foregrounds by
fitting parametric foreground models to the power spectra over a range
of frequencies.  By using power spectra, it is possible to fit
foreground components such as the cosmic infrared background (CIB)
which decorrelate with frequency (see \cite{Planckdust:2014b}). Such
foregrounds cannot be cleaned by conventional map based techniques. In
addition, we can make use of the fact that Galactic dust emission is
anisotropic on the sky, whereas extragalactic foregrounds are
isotropically distributed. It is then straightforward to use power
spectra computed on different areas of sky to separate Galactic dust
emission from the CIB (see \cite{Mak:2017}). A further advantage is
that a parametric foreground model is `controllable', in the sense
that we can investigate power spectrum residuals for each frequency
combination to assess whether a physically reasonable foreground model
(compatible with external data) provides an acceptable fit to the
measured power spectra.

At all HFI frequencies, Galactic dust emission is the dominant
foreground at low multipoles ($\ell \simlt 500$). In this section,
which focusses on temperature (polarized foregrounds are discussed in
Sect.\ \ref{sec:dust_polarization}), we investigate the properties of Galactic dust emission
in more detail to assess: (a) the `universality' of dust emission,
i.e.\ the accuracy with which a \Planck\ high frequency map can be used
as a template {\it with a single `cleaning' coefficient} to remove Galactic
dust emission at a lower frequency; (b) the amplitude of dust emission
at 100, 143 and 217 GHz; (c) the sensitivity of Galactic dust emission
to point source masking; (d) the creation of power spectrum templates
for Galactic dust emission; (e)
impact of template cleaning on the 100, 143 and 217 GHz spectra.

\subsection{Map-based cleaning coefficients}
\label{subsec:map_clean}

The simplest way of removing Galactic dust from the $100$-$217$ GHz maps is to use one of the
higher frequency maps as a dust template. The goal then is to determine an appropriate template
cleaning coefficient.
We have experimented with various different ways of determining cleaning coefficients.
Two of these methods are based on minimising map residuals:
\begin{subequations}
\begin{eqnarray}
\sigma^2 &=&  \sum_i((1+\alpha^{T_\nu}_m ) M^\nu_{i} - \alpha^{T_\nu}_m (M^T_{i} +c))^2,  \label{equ:Dust1a}\\
\sigma^{\prime2} &=&  \sum_i( (M^\nu_{i} - M^{\rm SMICA}_i) - 
\alpha^{T_\nu}_{m^\prime} (M^T_{i} - M^{\rm SMICA}_i +c^\prime))^2, \label{equ:Dust1b}
\end{eqnarray} 
\end{subequations}
where the sums extend over all unmasked pixels.  Here $M^\nu_i$ is the low frequency 
map,  $M^T_{i}$ is the high frequency `template' map and $M^{\rm SMICA}_i$ is an estimate of the primordial CMB
from the \SMICA\ component separation algorithm. The masks used in these summations
are the unapodised sequence of masks defined in Sect.\ \ref{subsec:temperature_masks}, except that we {\it always  exclude} the
high Galactic latitude regions of sky defined by mask25\footnote{Interestingly, mask25 roughly delineates the `shore line'
where the CIB dominates over Galactic dust emission at 217 GHz.}. In other words, the summations in
Eqs.\ \ref{equ:Dust1b} and \ref{equ:Dust1b} are over {\it annuli} on the sky. This is done to reduce the
impact of the CIB anisotropies on the scaling coefficients $\alpha^{T_\nu}_m$ and $\alpha^{T_\nu}_{m^\prime}$. 
(The relative importance of CIB anisotropies compared to Galactic dust emission 
is discussed in Sect.\ \ref{subsec:cleaned_TT}.)
We remove bright point sources, extended objects and CO emission by  
multiplying the three `point source' masks at $100$, $143$ and
$217$ GHz shown in Fig.\ \ref{fig:tempmasks}. By using a concatenated point source mask, the masks are identical 
for all frequencies. Note that the point source masks have a small effect on the shapes of the dust power
spectra, as discussed in Sect.\ \ref{subsec:dust_power}, but have little effect on the template coefficients
derived from Eqs.\ \ref{equ:Dust1a} and \ref{equ:Dust1b}.
The constants $c$ and $c^\prime$ are included in Eqs.\ \ref{equ:Dust1a} and \ref{equ:Dust1b}
to model the uncertainties in the absolute zero levels of the high frequency maps. Note further that the 
Planck 857 and 545 GHz maps are calibrated in ${\rm MJy}\ {\rm sr}^{-1}$. Throughout this paper, we convert these high 
frequency maps into units of thermodynamic temperature in $\muK$ by dividing the 857 and 545 GHz maps by factors
of $2.269$ and $57.98$ respectively as described in \cite{Planckdust:2014b}.

\begin{table}[tp]
{\centering

\caption{\small{Template cleaning coefficients: The first column gives the
  sky area used to compute the cleaning coefficients.  The map
  residuals are minimised within `annuli' in which the high Galactic
  latitude sky defined by mask25 is excluded. The second and third
  columns list map-based cleaning coefficients computed by minimising
  Eqs.\ \ref{equ:Dust1a} and \ref{equ:Dust1b}.  The fourth column lists
  the spectrum-based cleaning coefficients computed by minimising Eq.
  \ref{equ:MC3} over the mulipole range $100 \le \ell \le 500$, where
  Galactic dust emission dominates over the CIB. The coefficients
  listed in boldface were used to generate the `fake' maps shown in
  Fig.\ \ref{fig:fakemaps}.}}

\label{tab:cleaning_coeffs}
\begin{center}
\begin{small}
\begin{tabular}{|c|c|c|c|c|} \hline
$217  \ {\rm cleaned}\ {\rm with} \ 857$  & $\alpha^{T}_m/(1+\alpha^{T}_m)$ &    $\alpha^T_{m^\prime}$ & &  $\alpha^{T}_s/(1+\alpha^T_s)$  \\  \hline
mask50-mask25  &  $9.45 \times 10^{-5}$  & $9.46 \times 10^{-5}$  & mask50 & $1.02 \times 10^{-4}$   \\
mask60-mask25 &  $9.85 \times 10^{-5}$  & $9.64 \times 10^{-5}$ &  mask60 & $1.02 \times 10^{-4}$ \\
mask70-mask25 &  $1.06 \times 10^{-4}$   & ${\mathbf{1.00 \times 10^{-4}}}$ & mask70 &  $1.03 \times 10^{-4}$ \\
mask80-mask25 &  $1.04 \times 10^{-4}$   & $1.02 \times 10^{-4}$ &  mask80 & $1.03 \times 10^{-4}$  \\ \hline
$217  \ {\rm cleaned}\ {\rm with} \ 545 $  & $\alpha^{T}_m/(1+\alpha^{T}_m)$   & $\alpha^T_{m^\prime}$ &  & $\alpha^{T}_s/(1+\alpha^T_s)$  \\  \hline
mask50-mask25 &  $7.33 \times 10^{-3}$  & $7.47 \times 10^{-3}$  & mask50 &$7.83 \times 10^{-3}$   \\
mask60-mask25 &  $7.67 \times 10^{-3}$  & $7.55 \times 10^{-3}$  & mask60 & $7.86 \times 10^{-3}$ \\
mask70-mask25 &  $8.15 \times 10^{-3}$   & $\mathbf{7.67 \times 10^{-3}}$ & mask70 & $7.91 \times 10^{-3}$ \\
mask80-mask25 &  $7.95 \times 10^{-3}$   & $7.81 \times 10^{-3}$ &  mask80 & $7.77 \times 10^{-3}$  \\ \hline
$217  \ {\rm cleaned}\ {\rm with} \ 353$  & $\alpha^{T}_m/(1+\alpha^{T}_m)$   &$\alpha^T_{m^\prime}$ &  & $\alpha^{T}_s/(1+\alpha^T_s)$ \\  \hline
mask50-mask25 &  $1.27 \times 10^{-1}$  & $1.30 \times 10^{-1}$  & mask50 &$1.32 \times 10^{-1}$   \\
mask60-mask25 &  $1.31 \times 10^{-1}$  & $1.30 \times 10^{-1}$  & mask60 &  $1.32 \times 10^{-1}$ \\
mask70-mask25 &  $1.38 \times 10^{-1}$  & $\mathbf{1.31 \times 10^{-1}}$  & mask70 & $1.32 \times 10^{-1}$ \\
mask80-mask25 &  $1.36 \times 10^{-1}$  & $1.33 \times 10^{-1}$  & mask80 & $1.32 \times 10^{-1}$  \\ \hline
$143 \ {\rm cleaned}\ {\rm with} \ 857$  & $\alpha^{T}_m/(1+\alpha^{T}_m)$&   $\alpha^T_{m^\prime}$ &  & $\alpha^{T}_s/(1+\alpha^T_s)$  \\  \hline
mask50-mask25 &  $2.83 \times 10^{-5}$  & $2.68 \times 10^{-5}$  &  mask50 &$2.55 \times 10^{-5}$   \\
mask60-mask25 &  $3.09 \times 10^{-5}$  & $2.77 \times 10^{-5}$  &  mask60 &$2.27 \times 10^{-5}$   \\
mask70-mask25 &  $3.41 \times 10^{-5}$  & $\mathbf{2.91 \times 10^{-5}}$  &  mask70 &$2.52 \times 10^{-5}$   \\
mask80-mask25 &  $3.07 \times 10^{-5}$  & $3.02 \times 10^{-5}$  &  mask80 &$2.83 \times 10^{-5}$   \\ \hline
$143 \ {\rm cleaned}\ {\rm with} \  545$  & $\alpha^{T}_m/(1+\alpha^{T}_m)$ &  $\alpha^T_{m^\prime}$ & &  $\alpha^{T}_s/(1+\alpha^T_s)$   \\  \hline
mask50-mask25 &  $2.11 \times 10^{-3}$  & $2.12 \times 10^{-3}$  &  mask50 &$1.97 \times 10^{-3}$   \\
mask60-mask25 &  $2.39 \times 10^{-3}$  & $2.18 \times 10^{-3}$  &  mask60 & $1.76 \times 10^{-3}$   \\
mask70-mask25 &  $2.60 \times 10^{-3}$  & $\mathbf{2.23 \times 10^{-3}}$  &  mask70 & $1.92 \times 10^{-3}$  \\
mask80-mask25 &  $2.35 \times 10^{-3}$  & $2.30 \times 10^{-3}$  &  mask80 & $2.11 \times 10^{-3}$   \\ \hline
$143 \ {\rm cleaned}\ {\rm with} \  353$  & $\alpha^{T}_m/(1+\alpha^{T}_m)$ &     $\alpha^T_{m^\prime}$ &  & $\alpha^{T}_s/(1+\alpha^T_s)$  \\  \hline
mask50-mask25 &  $3.57\times 10^{-2}$  & $3.71 \times 10^{-2}$  &   mask50 & $3.36 \times 10^{-2}$  \\
mask60-mask25 &  $4.05 \times 10^{-2}$  & $3.78 \times 10^{-2}$  &  mask60 & $3.01 \times 10^{-2}$    \\
mask70-mask25 &  $4.40 \times 10^{-2}$  & $\mathbf{3.81 \times 10^{-2}}$  &  mask70 &  $3.26 \times 10^{-2}$   \\
mask80-mask25 &  $3.98 \times 10^{-2}$  & $3.91 \times 10^{-2}$  &   mask80 & $3.54 \times 10^{-2}$    \\ \hline
$100 \ {\rm cleaned}\ {\rm with} \  857$  & $\alpha^{T}_m/(1+\alpha^{T}_m)$ &  $\alpha^T_{m^\prime}$ & &  $\alpha^{T}_s/(1+\alpha^T_s)$  \\  \hline
mask50-mask25 &  $1.78 \times 10^{-5}$  &  $1.61 \times 10^{-5}$  &  mask50 & $1.79 \times 10^{-5}$   \\
mask60-mask25 &  $1.91 \times 10^{-5}$  &  $1.59 \times 10^{-5}$  &  mask60 & $1.60 \times 10^{-5}$   \\
mask70-mask25 &  $2.30 \times 10^{-5}$  &  $\mathbf{1.80 \times 10^{-5}}$  & mask70 &   $1.67 \times 10^{-5}$   \\
mask80-mask25 &  $2.21 \times 10^{-5}$  &  $2.17 \times 10^{-5}$  &  mask80 & $1.88 \times 10^{-5}$   \\ \hline
$100 \ {\rm cleaned}\ {\rm with} \ 545$  & $\alpha^{T}_m/(1+\alpha^{T}_m)$ &  $\alpha^T_{m^\prime}$ &  & $\alpha^{T}_s/(1+\alpha^T_s)$ \\  \hline
mask50-mask25 &  $1.26 \times 10^{-3}$  &   $1.26 \times 10^{-3}$  &  mask50 & $1.33 \times 10^{-3}$   \\
mask60-mask25 &  $1.45 \times 10^{-3}$  &  $1.25\times 10^{-3}$  &   mask60 & $1.20 \times 10^{-3}$  \\
mask70-mask25 &  $1.76 \times 10^{-3}$  &  $\mathbf{1.38 \times 10^{-3}}$  & mask70 &  $1.26 \times 10^{-3}$   \\
mask80-mask25 &  $1.68 \times 10^{-3}$  &  $1.64 \times 10^{-3}$  & mask80 &  $1.40 \times 10^{-3}$   \\ \hline
$100 \ {\rm cleaned}\ {\rm with} \  353$  & $\alpha^{T}_m/(1+\alpha^{T}_m)$  &  $\alpha^T_{m^\prime}$ & &  $\alpha^{T}_s/(1+\alpha^T_s)$  \\  \hline
mask50-mask25 &  $2.05 \times 10^{-2}$   &   $2.21 \times 10^{-2}$  &    mask50 &$2.23 \times 10^{-2}$   \\
mask60-mask25 &  $2.43 \times 10^{-2}$   &   $2.16 \times 10^{-2}$  &    mask60 &$2.00 \times 10^{-2}$   \\
mask70-mask25 &  $2.96 \times 10^{-2}$   &   $\mathbf{2.37 \times 10^{-2}}$  & mask70 &  $2.11 \times 10^{-2}$     \\
mask80-mask25 &  $2.86 \times 10^{-2}$   &   $2.79 \times 10^{-2}$  &   mask80 &$2.35 \times 10^{-2}$    \\ \hline

\end{tabular}
\end{small}
\end{center}}
\end{table}

In Eq. \ref{equ:Dust1a}, the maps are uncorrected for CMB
anisotropies. In Eq. \ref{equ:Dust1b} we subtract the
\Planck\ \SMICA\ component separated CMB map
\citep{Planck_component2015} from the low frequency maps to remove 
primordial CMB anisotropies\footnote{For the purposes of this section,
  it makes no difference whether we use the \SMICA\ component
  separated map, or any of the other component separated maps
  discussed in \citep{Planck_component2015}.}.  To reduce sensitivity
to noise and foreground contributions at high multipoles, we first
smooth all of the (unmasked) maps to a common resolution with a
Gaussian of FWHM of one degree. This smoothing has almost no effect on
the cleaning coefficients determined using $545$ and $857$ GHz as
templates (since these maps are effectively noise-free), but gives more stable results if $353$ GHz is used as a
template. The form of Eq. \ref{equ:Dust1a} is chosen to reduce the
sensitivity of the CMB component on the recovered cleaning
coefficients. However, Eq. \ref{equ:Dust1a} will lead to  biased
coefficients.  If we write 
\begin{subequations}
\begin{eqnarray}
  M^{\nu}   & = & S + \beta F,  \label{equ:Dust2a}\\
  M^{\nu_T}  & = & S + F,  \label{equ:Dust2b}
\end{eqnarray}
\end{subequations}
where $S$ is the CMB signal and $F$ is the foreground, then it is straightforward to show that if CMB-foreground
cross-correlations are negligible, then minimising Eq. \ref{equ:Dust1a} leads to a biased cleaning coefficient
with 
\begin{equation}
\alpha^{T_\nu}_m = {\beta \over (1 - \beta)}.
\end{equation}
Since an estimate of the CMB is subtracted from the maps in Eq. \ref{equ:Dust1b}, the cleaning coefficients
$\alpha^{T_\nu}_{m^\prime}$ give an unbiased estimate of $\beta$. We therefore compare $\alpha^{T_\nu}_{m^\prime}$ with
$\alpha^{T_\nu}_m/(1+ \alpha^{T_\nu}_m)$.

Results are listed in Table \ref{tab:cleaning_coeffs} for various
masks. For all frequencies and templates we see a trend for the
cleaning coefficients to increase with increasing sky area. We
postpone a discussion of whether this trend indicates a departure of
the dust properties from universality until
Sect.\ \ref{subsec:universality}. For $217$ GHz, which is the most
heavily dust-contaminated channel in the \camspec\ likelihood, the
cleaning coefficients vary by a few percent as the sky used
changes from (${\rm mask50}-{\rm mask25}$) to (${\rm mask80}-{\rm
  mask25}$). At $100$ GHz, the level of dust contamination is so low
that it is difficult to derive an accurate cleaning coefficient using
any of the high frequency templates.  The cleaning coefficients
derived from Eqs.\ \ref{equ:Dust1a} and \ref{equ:Dust1b} agree
reasonably well, but we expect \ref{equ:Dust1b} to be more reliable
since the cleaning coefficients are not biased by the CMB and
CMB-foreground correlations.

\subsection{Power spectrum of Galactic dust emission}
\label{subsec:dust_power}

\begin{figure*}
\centering
\includegraphics[width=105mm,angle=0]{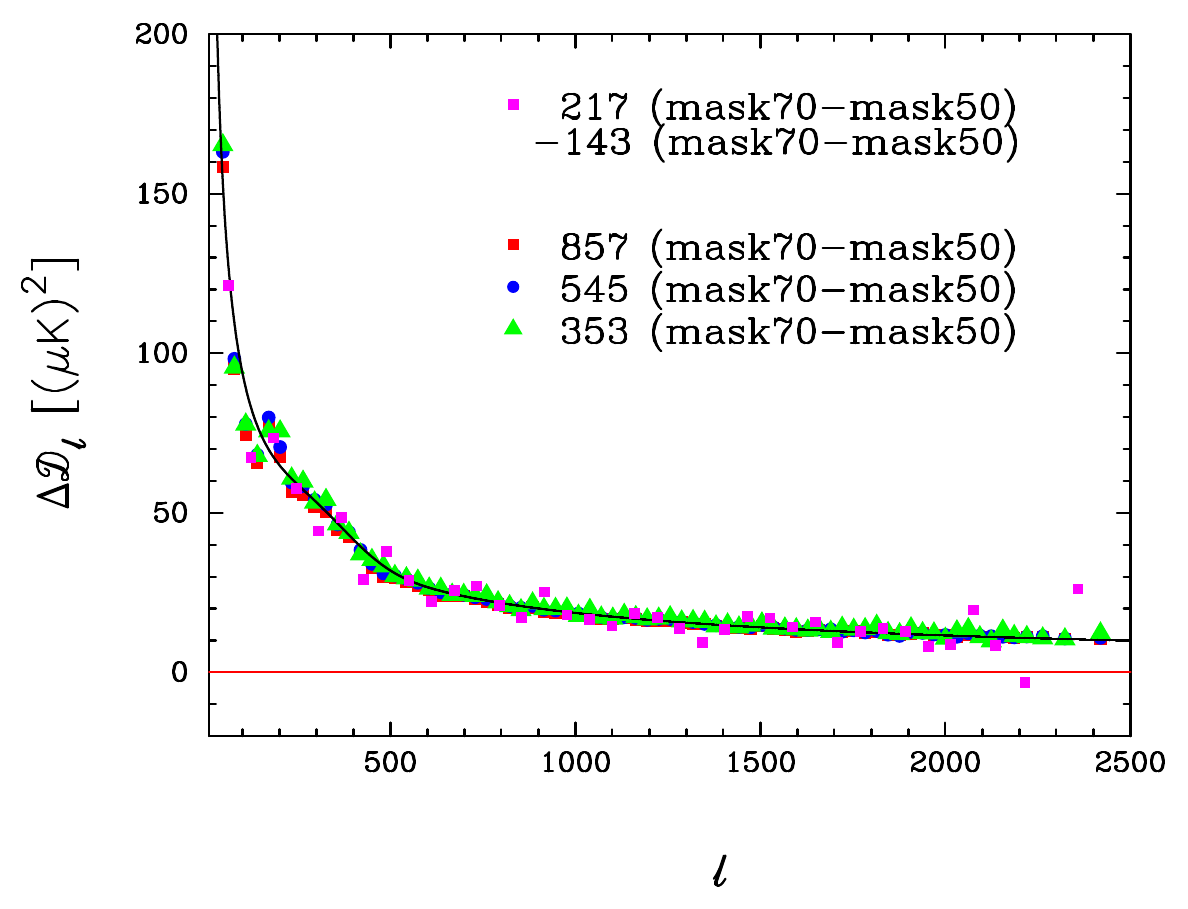}
\caption {Differences of power spectra computed on mask70 and mask50
  at 857, 545 and 353 GHz scaled to match the foreground emission at
  217 GHz using the cleaning coefficients given by the bold-faced
  entries in Table \ref{tab:cleaning_coeffs}. The pink points show the
  `double-differenced' $217\times 217 - 143\times 143$ spectrum
  (renormalized to the dust amplitude of the $217 \times 217$
  spectrum).  The solid line shows a fit of the 545 GHz spectrum to
  the expression given in Eq. \ref{equ:Noise5}.}
     
\label{fig:doublediff}

 \end{figure*}

We can eliminate all isotropic components, including the CMB,  CIB, and extragalactic point sources by differencing the power
spectra computed on different masks. Fig.\ \ref{fig:doublediff}  is similar to Fig.\ 3 of PPL13. This
figure  shows the mask-differenced HM1$\times$HM2 spectra at 857, 545 and 353 GHz
scaled to the amplitude of the foreground emission at  217 GHz using the  coefficients
listed in boldface in  Table \ref{tab:cleaning_coeffs}. We use the concatenated $100-217$ GHz point source mask
for the spectra in Fig.\ \ref{fig:doublediff}, consistent with the cleaning coefficients listed
in Table  \ref{tab:cleaning_coeffs}. Since we are plotting mask-differenced spectra, the
points plotted in Fig.\ \ref{fig:doublediff} reflect the properties of {\it Galactic
dust emission alone}. As can be seen, the rescaled $857$, $545$ and $353$ GHz spectra match to high accuracy
and are barely distinguishable in Fig.\ \ref{fig:doublediff}.

 The solid line in Fig.\ \ref{fig:doublediff} shows a fit of the 545
 GHz points to a simple analytic fitting function. We use the same
 parametric form,  Eq. \ref{equ:Noise5},   that was used to fit  the \Planck\ noise spectra.
  The second term in Eq. \ref{equ:Noise5} fits
 the small `bump' in the $545$ GHz spectrum at multipoles $\ell \sim
 300$. However, the dominant term is the power-law component $D^D_\ell
 = A(100/\ell)^\alpha$.  The best fit parameters are $A=90.661$ \muk,
 $\alpha = 0.6873$, $B=14.402$ \muk, \ $\beta=1.646$, $\ell_c=
 100.73$, $\gamma = 3.283$, $\delta=33.26$.  Note that at high
 multipoles, the dust power spectrum falls off with a slope of $C_\ell
 \propto \ell^{-2.69}$, consistent with the power-law slope of
 approximately $-2.7$ to $-2.8$ inferred from a very different
 analysis \citep{Planckdust:2014b} of the \Planck\ data.

The pink points in Fig.\ \ref{fig:doublediff} show the double-differenced
  $217\times 217 - 143\times 143$ power spectrum. The reason
for plotting the double-difference is to suppress the large cosmic
variance fluctuations in the primordial CMB contribution, which
dominate over foreground fluctuations at frequencies $\le 217$
GHz. The double-differenced spectrum is corrected by a scaling factor
to acount for the small amplitude of dust emission in the
143$\times$143 mask-differenced spectrum. As with the higher frequency
spectra, the mask-differencing isolates Galactic dust emission from
isotropic foregrounds. Evidently, Galactic dust emission at low
frequencies is extremely well approximated by the model of
Eq. \ref{equ:Noise5}. Over the same sky region, therefore, we have
demonstrated that the power spectrum of Galactic dust emission has the
same shape, to very high accuracy, over the wide frequency range $217-857$
GHz.

\begin{figure}
\centering
\includegraphics[width=75mm,angle=0]{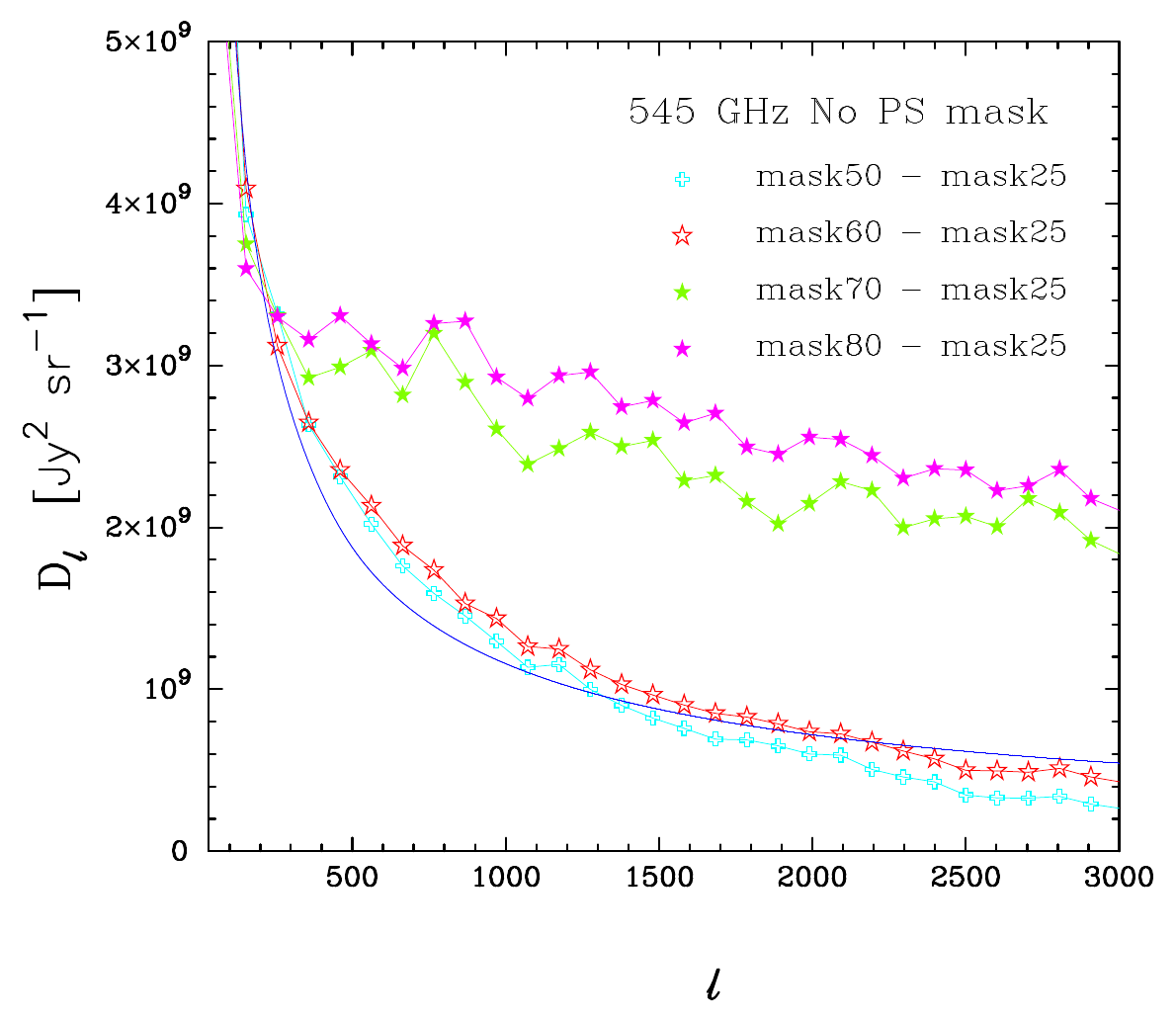} \includegraphics[width=75mm,angle=0]{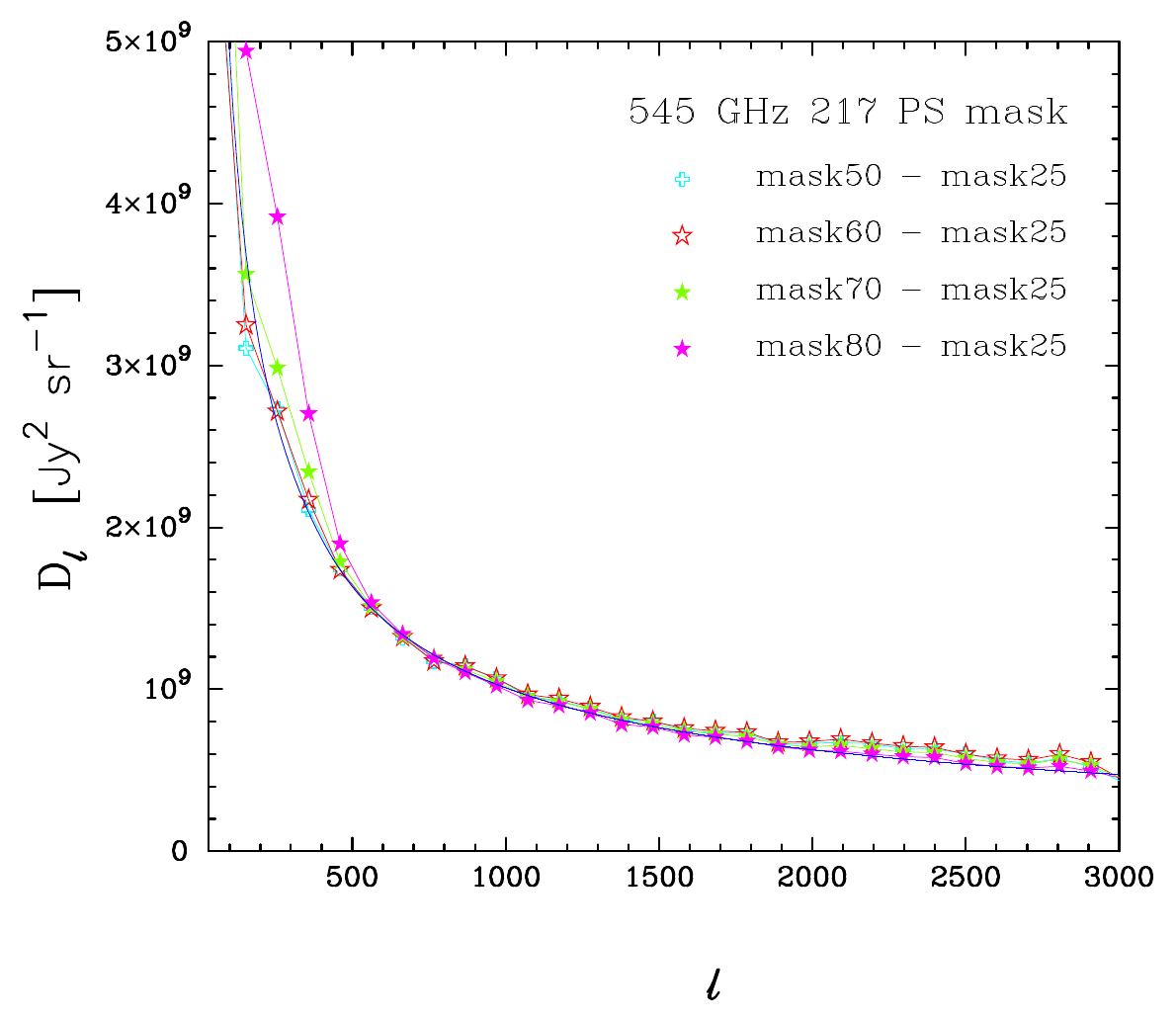} 
\\
\includegraphics[width=75mm,angle=0]{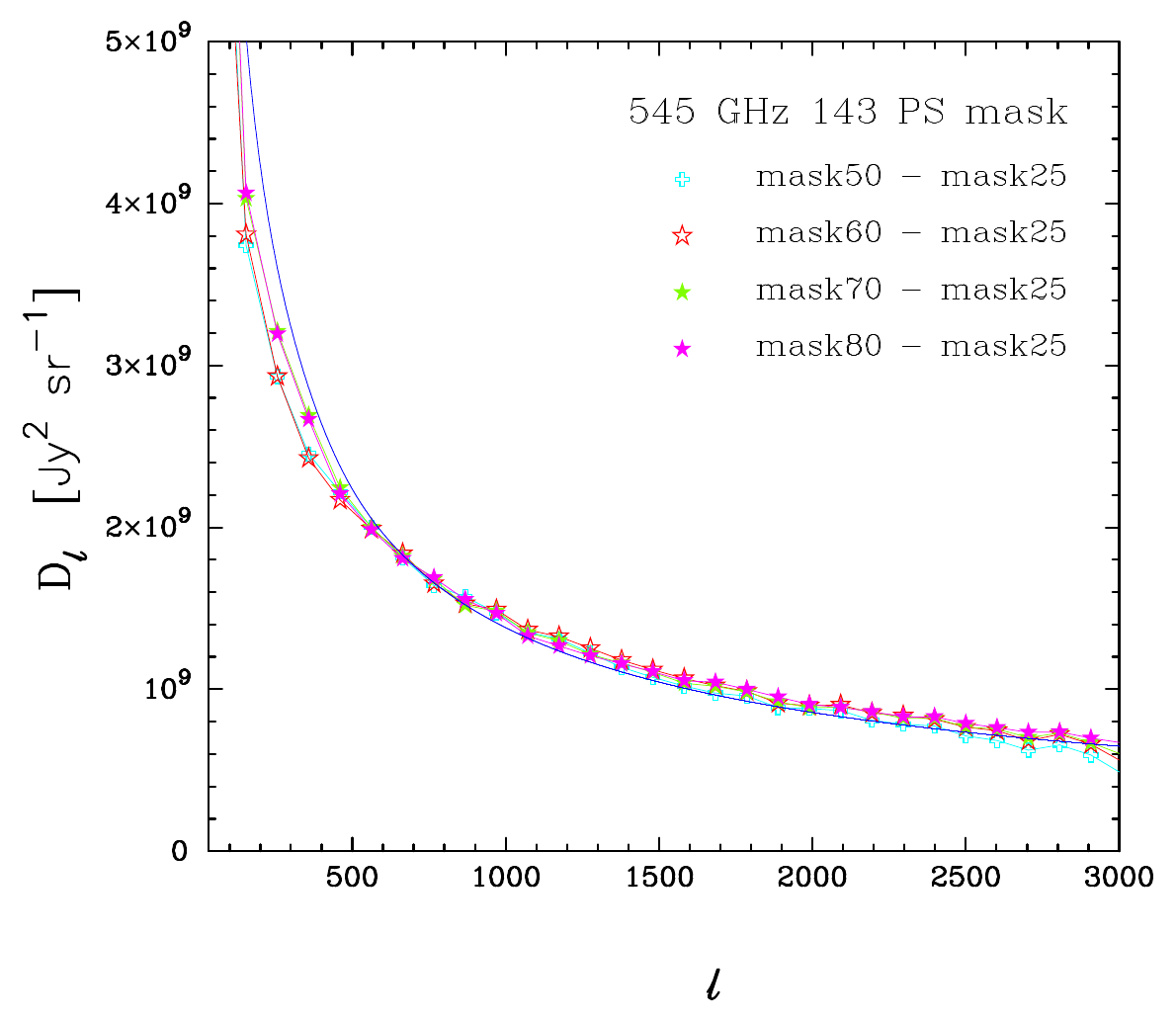} \includegraphics[width=75mm,angle=0]{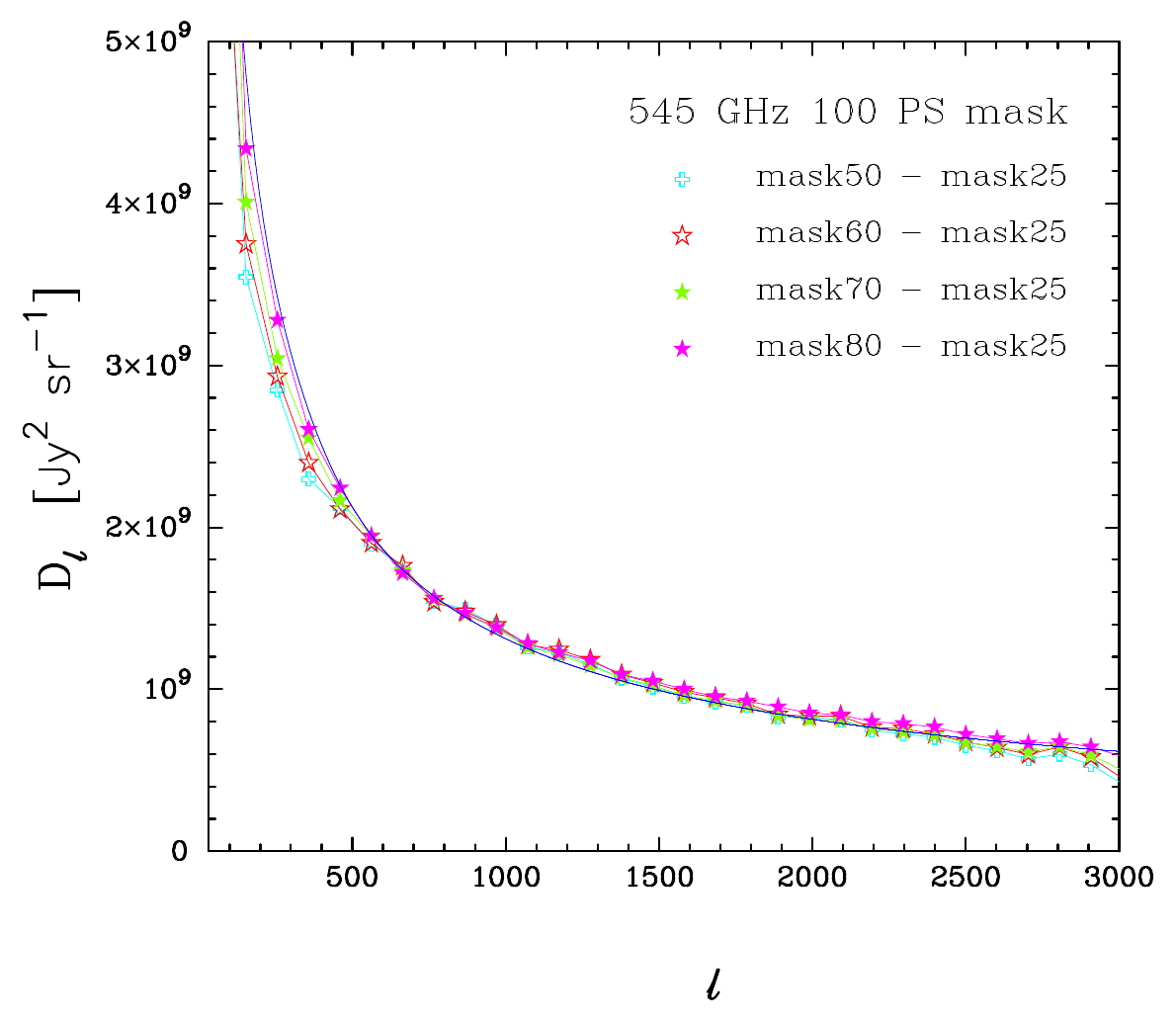}

   \caption {Differenced 545 GHz half mission spectra for varous sky
     masks. For the figure at the top left no point source mask was
     applied. The other figures show spectra for the 217, 143 and 100
     GHz point source masks as shown in Fig.\ \ref{fig:tempmasks}. The
     blue-lines show the best-fit dust spectrum from
     Fig.\ \ref{fig:doublediff}.}

\label{fig:dust545}
\end{figure}

We now investigate variations of the shape of the dust power spectra
with changes in the sky area and point source
mask. Fig.\ \ref{fig:dust545} shows mask-differenced power spectra for
the 545 GHz half mission maps as a function of sky area for the $217$,
$143$ and $100$ point source masks used in the cosmological
analysis. The amplitudes of the spectra have been matched to the
amplitude of the mask50-mask25 spectrum over the multipole range $500
\le \ell \le 1000$.  The blue lines show the best-fit dust spectrum
from Fig.\ \ref{fig:doublediff}. The 217 GHz point source mask is very
similar to the concatenated point source mask used in
Fig.\ \ref{fig:doublediff}, thus the fit provides a very good match to
the 217 GHz spectra for sky areas up to mask70. There are, however,
small differences in the shape of the dust spectrum computed on mask80
at lower frequencies. However, it would be incorrect to conclude that
this is caused by variations in diffuse Galactic dust emission over
the sky. One can see that the 545 GHz spectra using the $143$ and
$100$ points source masks are nearly identical over 50-80\% of
sky. The $217$ GHz point sources are distributed anisotropically over
the sky (see Fig.\ \ref{fig:tempmasks}) with a surface density that
increases strongly at low Galactic latitudes. At low Galactic
latitudes, where the dust emission is high and the background level
varies strongly, dust knots become identified as point sources. The
point source masks remove some of the dust emission from the unmasked
sky causing the estimated dust spectra to steepen if one uses  sky
areas extending to low Galactic latitudes.  This effect is less
important for the $143$ and $100$ point source masks, because at these
frequencies there is less contamination of the point source catalogues
by knots of dust emission. However, the dust spectra computed using
these point source masks are slightly shallower than the spectra
measured using the $217$ GHz point source masks. For this reason, we
tailor the Galactic dust template spectra used in the
\camspec\ likelihoods to the identical point source masks used to
compute the spectra (see Sect.\ \ref{subsec:galactic_dust_templates}).

The plot at the top left in Fig.\ \ref{fig:dust545} shows what happens
if we apply no point source mask. These spectra show a dramatic
departure from universality for mask70 and mask80. The excesses seen
in these spectra arise from a small number  ($\simlt 100$) of extremely bright sources
(such as Centaurus A and compact knots of dust emission). These bright
sources contaminate the \Planck\ temperature power spectra over the
entire HFI frequency range $100-857$ GHz and must be removed for any
meaningful science analysis.

\begin{figure}
\centering


\includegraphics[width=150mm,angle=0]{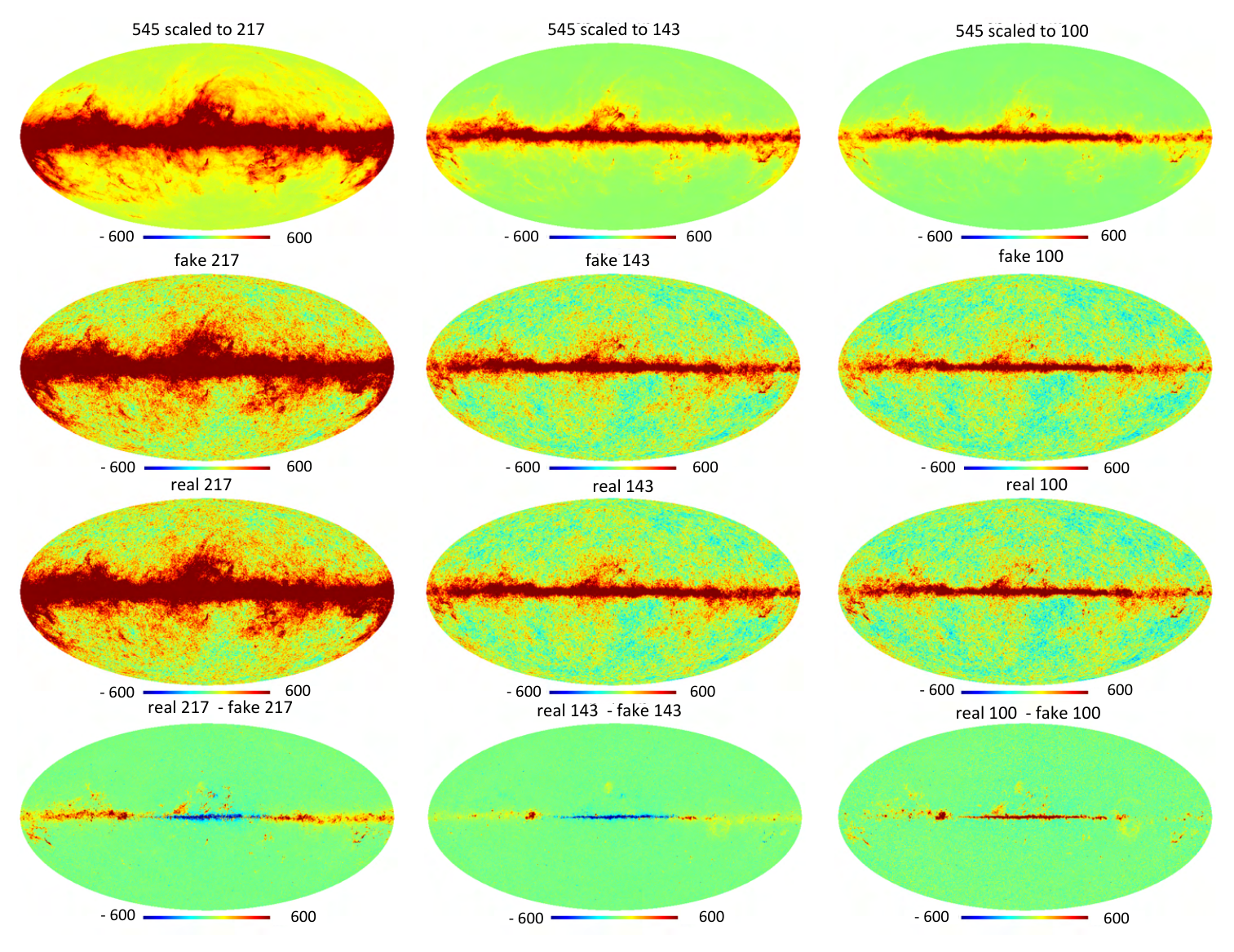} 
   \caption {The top row shows the 545 GHz maps scaled to match the
     dust emission at 217, 143 and 100 GHz (left to right) using the
     boldface dust cleaning coefficients listed in Table
     \ref{tab:cleaning_coeffs}.  The second row shows the addition of
     the \SMICA\ component separated CMB map to the scaled 545 GHz
     maps creating `fake' maps at 217, 143 and 100 GHz. The third row
     shows the real maps at these frequencies.  The bottom row shows
     the difference between the real and fake maps. The color scales
are in units of $\mu{\rm K}$.}
\label{fig:fakemaps}
\end{figure}

Even though we have plotted a double-differenced spectrum at low
frequencies in Fig.\ \ref{fig:doublediff}, the pink points show scatter
that is much higher than the scatter of the higher frequency spectra. This
scatter is also much higher than that expected from instrument noise. The
excess scatter is caused by `chance' CMB-foreground
cross-correlations.  Consider the simple model where the signal at
frequency $1$ is a sum of primordial CMB (denoted S) plus a
contribution from a foreground $F$. We assume that frequency $2$ is
dominated by the foreground $F$:
\begin{subequations}
\begin{eqnarray}
  M_1  & = & S + \alpha F, \label{equ:CMBxF1a} \\
  M_2  & = & F.  \label{equ:CMBxF1b}
\end{eqnarray}
\end{subequations}
Schematically, the power spectra of these two maps are
\begin{subequations}
\begin{eqnarray}
  C_1  & = & S *S + 2\alpha S*F +\alpha^2F*F, \label{equ:CMBxF2a} \\
  C_2  & = & F*F,  \label{equ:CMBxF2b}
\end{eqnarray}
\end{subequations}
and so the power spectrum of the low frequency map
will contain a CMB-foreground cross-term.  The excess scatter in the
$217\times 217 -143\times 143$ spectrum shown in Fig.\ \ref{fig:doublediff} compared to the
spectra at higher frequencies is caused by the CMB-foreground
cross-terms, not by any intrinsic variation of the dust emission
between low and high frequencies. We can demonstrate this conclusively
by  analysing `fake' maps at $217$, $143$ and $100$ GHz constructed by adding 
appropriately scaled $545$ GHz maps to the \SMICA\ half-mission CMB maps.

The top row of Fig.\ \ref{fig:fakemaps} shows the 545 GHz maps scaled to 217, 143 and 100 GHz respectively
(with the best fit constant, $c^\prime$, subtracted from each map). The second row shows the scaled
545 maps added to the \SMICA\ component separated map to produce `fake' 217, 143 and 100 GHz maps.
The real maps are shown in the third row. The fourth row shows the differences between the real
and the fake maps.  The broad \Planck\ frequency bands centred at 100 GHz and 217 GHz are contaminated 
by CO rotational transitions \citep{Planck_CO13} ($J=1 \rightarrow 0$ at 115 GHz and $J=2 \rightarrow 1$ at 230 GHz). CO emission contaminates the $100$ and $217$ GHz maps at low Galactic latitudes and accounts
for much of the residual emission seen in Fig.\ \ref{fig:fakemaps} at these frequencies. It is for this reason that
we apply CO masks at 100 and 217 GHz in the cosmological analysis (see Fig.\ \ref{fig:tempmasks}). At $143$ GHz, 
the residuals are at low levels except within a few degrees of the Galactic plane. There is 
some excess emission at $100-217$ GHz in the Ophiucus region, which 
has a higher dust temperature than the bulk of the Galactic dust \citep{Planck_dust}.

\begin{figure*}

\centering
\includegraphics[width=135mm,angle=0]{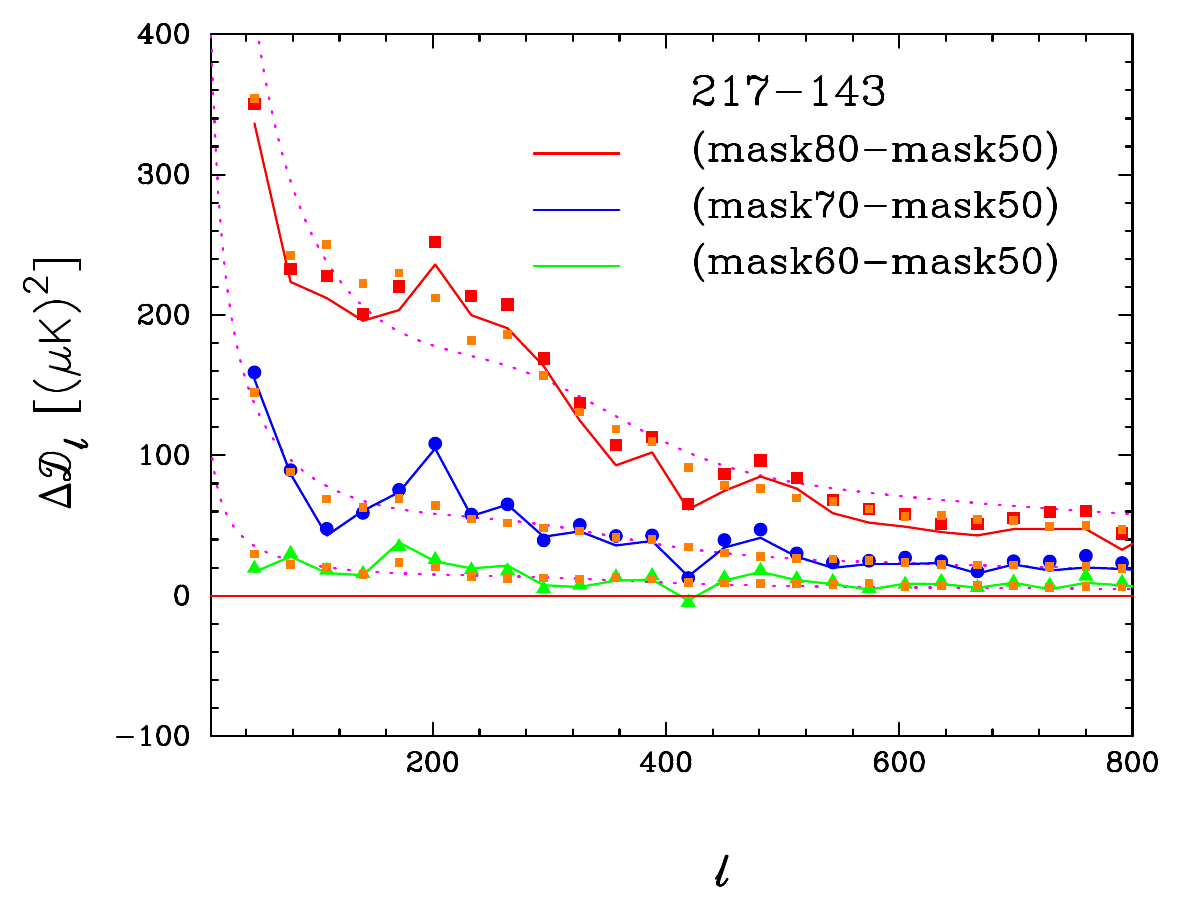}
\caption {The filled green, blue and red points show the double-differenced $217\times217 - 143\times143$
half mission power spectra  for various masks. These double-differences cancel the CMB and isotropic foreground components leaving only Galactic
dust components. The orange points show the half mission  power spectra computed from the 545 GHz maps, scaled to match the amplitudes of the $217\times 217-143\times143$ spectra.
 The green, blue and red solid lines show the double-differenced power spectra
computed from the `fake' 217 and 143 GHz maps shown in Fig.\ \ref{fig:fakemaps}.
The dotted lines show the best-fit dust templates from Eq. \ref{equ:Noise5} 
with amplitudes scaled to match the $217\times 217-143\times143$ double-differenced spectra. }
     
\label{fig:217-143sim}

\end{figure*}

The green, blue and red points in Fig.\ \ref{fig:217-143sim} show the
217-143 double-differenced power spectra for the real maps. The solid
lines show the double-differenced power spectra computed from the fake
545 GHz  + \SMICA\ maps described above.  One can see that the spectra for
the fake maps are in very good agreement with those for the real data,
tracking the fluctuations to high accuracy. The orange points in the
figure show the mask differenced power spectra for the 545 GHz
half mission maps, scaled to the lower frequencies. The dotted line
shows the best fitting model of Eq. \ref{equ:Noise5} scaled to match
the various mask sizes. Evidently, the increased scatter in the
$217 \times 217 -143 \times 143$ spectra compared to the $545\times 545$ 
spectra comes from chance CMB-dust
cross correlations and the spectra are almost perfectly matched by the
`fake' 545 GHz + \SMICA\ maps.

The additional variance arising from chance CMB-dust correlations
should be included in the covariance matrices if the 217 GHz spectra
are to be used at low multipoles in forming a likelihood. In the
\camspec\ likelihoods, we add the best fit foreground power spectrum
(which varies with frequency and sky-coverage) to the fiducial
theoretical model. In other words, we treat the foregrounds as an
additional statistically isotropic contribution to the primordial CMB
signal when we compute covariance matrices.  This is a good
approximation for the extragalactic foreground contributions, but as
mentioned in Sect.\ \ref{subsec:camspec_covariance_matrices} it is a poor
approximation for Galactic dust, which is statistically anisotropic on
the sky. This issue is discussed in detail in  \citep{Mak:2017}, which
presents a more accurate model based on the assumption that Galactic
dust can be approximated as a small-scale isotropic component
superimposed on a smooth field with large-scale gradients. In the
\camspec\ likelihood, we simply exclude the $217\times 217$ and $143
\times 217$ spectra at low mutipoles and so we have not adopted the
prescription of \citep{Mak:2017}.

\subsection{Spectrum-based cleaning coefficients}
\label{subsec:spec_clean}

 Another way to determine
cleaning coefficients, explicitly tuning to a selected range of multipoles, is to minimise
power spectrum residuals \citep{Lueker:2010, Spergel:2015}. Consider  the `cleaned' maps,
\begin{equation}
M^{T_\nu {\rm clean}} =   (1 + \alpha^{T_\nu}_s) M^{T_\nu} - \alpha^{T_\nu}_s M^{T_{\nu_T}}, \label{equ:MC1}
\end{equation} 
where $\nu_T$ is the frequency of the template map\footnote{To avoid cumbersome notation, we write
the cleaning coefficients as $\alpha^{T_\nu}_s$ rather than $\alpha^{T_\nu T_{\nu_T}}_s$.}.
The cross power spectrum of cleaned maps at frequencies $\nu_1$ and $\nu_2$ is:
\begin{eqnarray}
\hat C^{T_{\nu_1}T_{\nu_2} {\rm  clean}} & = & (1 + \alpha^{T_{\nu_1}}_s)(1 + \alpha^{T_{\nu_2}}_s) \hat C^{T_{\nu_1} T_{\nu_2}} - 
(1 + \alpha^{T_{\nu_1}}_s) \alpha^{T_{\nu_2}}_s \hat C^{T_{\nu_1}T_{\nu_T}}  \nonumber\\ 
& - & (1 + \alpha^{T_{\nu_2}}_s) \alpha^{T_{\nu_1}}_s \hat C^{T_{\nu_2}T_{\nu_T}} + \alpha^{T_{\nu_1}}_s \alpha^{T_{\nu_2}}_s \hat C^{T_{\nu_T}T_{\nu_T}},  \label{equ:MC2}
\end{eqnarray}
where  $\hat C^{T_{\nu_1}T_{\nu_2}}$  etc.\ are the mask-deconvolved beam corrected power spectra.
An  advantage  of working with power spectra rather than maps is that it is straightforward to 
correct for differences in the beams of the low frequency and high frequency template maps.
 The coefficients $\alpha^{T_{\nu_1}}_S$ are determined by minimising
\begin{equation} 
\Psi_{TT} = \sum_{\ell_{\rm min}}^{\ell_{\rm max}} \hat C^{T_{\nu_1}T_{\nu_1} {\rm  clean}}_\ell.  \label{equ:MC3}
\end{equation}
We minimise Eq. \ref{equ:MC3} instead of the usual $\chi^2$, which leads to 
 biased cleaning coefficients when $\hat C^{T_{\nu_1}T_{\nu_2} {\rm  clean}}$ becomes noise dominated.

As in the discussion of map based template fitting based on Eq. \ref{equ:Dust1a}, 
 minimisation of Eq. \ref{equ:MC3} leads to biased cleaning coefficients. If we adopt the simplified model of Eqs.\ \ref{equ:Dust2a} and \ref{equ:Dust2b}, then Eq. \ref{equ:MC3} is minimized for
\begin{equation}
  \alpha^{T_\nu}_s =  {\beta \over (1- \beta)} \left[ 1 + { S*F  \over F*F } \right ] \approx  {\beta \over (1- \beta)}, 
\end{equation}
(assuming $S*F \ll F*F$) and so gives a biased estimate of the true cleaning 
coefficient $\beta$, though the `cleaned' power spectrum from Eq. \ref{equ:MC2} is
unbiased, i.e.
\begin{equation}
\langle \hat C^{T_{\nu_1}T_{\nu_2} {\rm  clean}} \rangle  =  S*S.
\end{equation}
Evidently, for values of $\alpha^{T_\nu}_s \ll 1$, the bias is
negligible, but for larger values, e.g.\ the 
coefficient appropriate for cleaning   $217$ GHz with $353$ GHz, the bias can become significant. We therefore list
values of $\alpha^T_s/(1+\alpha^T_s)$ in the final column of Table
\ref{tab:cleaning_coeffs},  determined by minimising Eq. \ref{equ:MC3} over
the multipole range $100 \le \ell \le 500$. Over this multipole range,
Galactic dust emission dominates over the CIB.  The spectrum-based
cleaning coefficients listed in Table \ref{tab:cleaning_coeffs} are
generally in very good agreement with the map-based coefficients
determined by minimising Eq. \ref{equ:Dust1b} (i.e.\ with an
estimate of the CMB removed from the maps). However, we find less good agreement
with the map based coefficients based on Eq. \ref{equ:Dust1a} at low
frequencies and for small sky fractions where Galactic dust
emission does not dominate strongly compared to the CMB. For those cases,
the CMB can strongly bias the cleaning coefficients. The cleaning coefficients
listed in  columns 3 and 5 of Table \ref{tab:cleaning_coeffs} provide
our best estimates of the contribution of Galactic dust emission over the
frequency range $100-217$ GHz.

\begin{figure}
\centering
\includegraphics[width=100mm,angle=0]{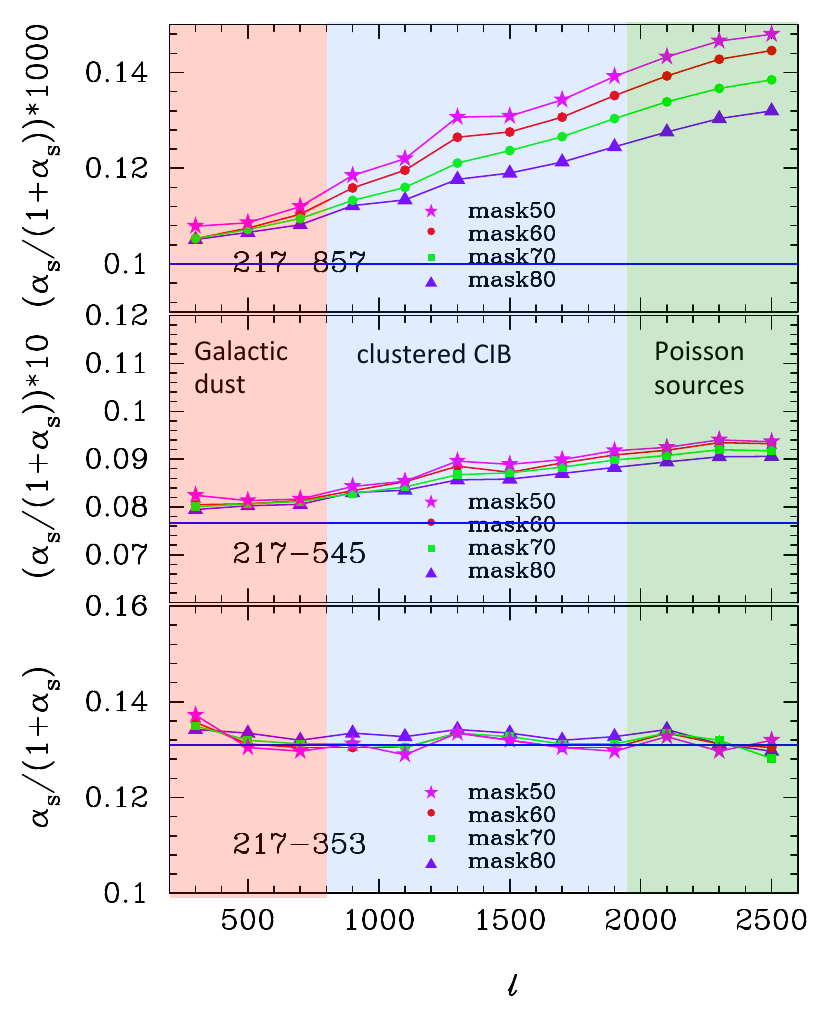}

   \caption {Cleaning coefficients derived by minimising
     Eq. \ref{equ:MC3} in bands of multipole of width $\Delta \ell =
     200$ for 217 GHz cleaned with 857 GHz (top), 545 GHz (middle) and
     353 GHz (bottom). Results are shown for four different masks.
     The horizontal lines show the map-based cleaning coefficients
     $\alpha^T_{m^\prime}$ listed in bold face in Table
     \ref{tab:cleaning_coeffs}.  The shaded areas delineate the
     multipole ranges where on mask60 the $217\times217$ spectrum is
     dominated by Galactic dust emission, clustered CIB and Poisson
     point sources (determined from the Monte-Carlo Markov Chain  fits to the 12.1HM
     \camspec\ likelihood). }

\label{fig:cleaning}
\end{figure}

Fig.\ \ref{fig:cleaning} shows the 217 GHz cleaning coefficients computed
by minimising Eq.\  \ref{equ:MC3} in multipole bands of width $\Delta
\ell = 200$ for four different masks using $353$, $545$ and $857$ GHz
maps as templates.  The shaded areas in Fig.\ \ref{fig:cleaning} show
the multipole ranges over which various foreground components dominate
on mask60: Galactic dust emission ($\ell \simlt 500$), clustered CIB
($500 \simlt \ell \simlt 2000$) and Poisson point sources ($\ell
\simgt 2000)$. (These ranges are computed from  the
fits of the combined CMB + foreground model to the 12.1HM likelihood
described in Sect.\ \ref{sec:Likelihood}.) The horizontal lines show
the map-based cleaning coefficients listed in bold face in Table
\ref{tab:cleaning_coeffs}.  Using 353 GHz as a template, we find
cleaning coefficients that are remarkably flat and insensitive to the
sky mask. Using 545 GHz as a template, we see a gradual drift in
$\alpha_S$ from $\approx 7.9 \times 10^{-3}$ at multipoles $\ell
\simlt 500$ to $9.3 \times 10^{-3}$ at $\ell \sim 2500$. This drift is
a consequence of the CIB having a slightly different spectral energy
distribution (SED) compared to Galactic dust. We see a stronger drift
if we use $857$ GHz as a template. The differences in the SEDs of
Galactic dust emission and the CIB can be exploited to construct 857
and 545 GHz difference maps that are dominated by the CIB and so can
be used as a tracer of large-scale structure and lensing of the CMB
\citep{Larsen:2016}.

In fact, we can use the cleaning coefficients in the last column of
Table \ref{tab:cleaning_coeffs} to estimate the SED of Galactic dust
emission. We use the central frequencies and conversions from
temperature to units of ${\rm MJy\ sr}^{-1}$ for a dust-like spectrum
from \citep{PlanckIX:2014}.  We estimate a rough error on each
cleaning coefficient from the scatter in the cleaning
coefficients measured for the four masks listed in Table
\ref{tab:cleaning_coeffs} and we add a $7\%$ absolute calibration
error for the $545$ and $857$ GHz maps \citep{Planck_calib:2016}. We
normalize the SED to unity at the $217$ GHz channel and fit the points
to a modified black-body (MBB): 
\begin{subequations}
\begin{equation}
I_\nu = \tau_{\rm obs} \left ( {\nu \over \nu_0} \right )^{\beta_{\rm obs}}
 {B_\nu(T_{\rm obs}) \over B_{\nu_0}(T_{\rm obs}) }, \label{equ:Dust4}
\end{equation}
where $B_\nu(T)$ is the Planck function
\begin{equation}
B_\nu (T) \propto  {\nu^3 \over [{\rm exp}(h\nu/kT) - 1 ]}.  \label{equ:Dust5}
\end{equation}
\end{subequations}

\begin{figure}
\centering
\includegraphics[width=100mm,angle=0]{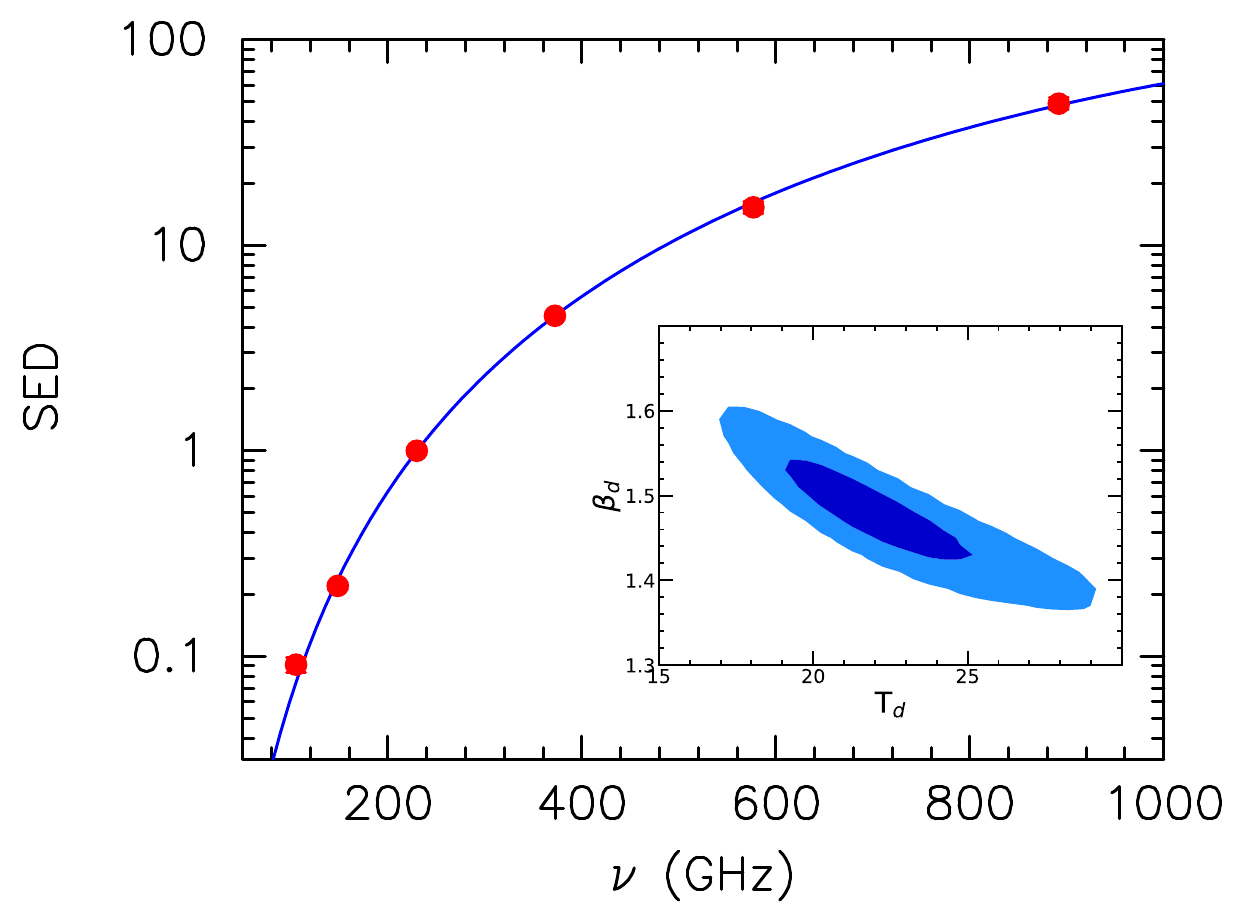}

\caption {Spectral energy distribution of dust emission based on the spectrum-based cleaning
coefficients listed in column 5 of Table \ref{tab:cleaning_coeffs}. The SED is normalized to unity at 229 GHz
(the effective central frequency for a dust-like SED in the 217 GHz band). The solid line
shows the best-fit modified black body distribution of Eqs.\ \ref{equ:Dust4} and \ref{equ:Dust5}.
68\% and 95\% confidence contours for the parameters $\beta_{\rm obs}$ and $T_{\rm obs}$ are 
plotted in the inset.}

\label{fig:dust_sed}

\vspace{0.20truein}
\end{figure}

 The SED is plotted in Fig.\ \ref{fig:dust_sed}. The inset in this
 figures shows the marginalized posteriors of $\beta_{\rm obs}$ and
 $T_{\rm obs}$ determined from an MCMC fit to the data points. We find
 $\beta_{\rm obs} = 1.49 \pm 0.05$, $T_{\rm obs} = 22.7 \pm 2.8\ {\rm
   K}$, in reasonable agreement with the map-based analysis of
 \citep{Planck_dust}, which finds $\beta_{\rm obs} = 1.59 \pm 0.12 $,
 $T_{\rm obs}=20.3 \pm 1.3\ {\rm K}$ by fitting to the \Planck\ 353,
 545, 857 GHz and IRAS $100 \ \mu{\rm m}$ data over the sky at $\vert b \vert >
 15^\circ$.  This single component MBB is an acceptable fit to the
 observed dust spectrum over the \Planck\ frequency range. The
 question of whether the Galactic dust spectrum is better fitted by a two component
  model (see e.g.\ \cite{Finkbeiner1999,Meisner2015}) is beyond the
 scope of this paper.

Note that Eq.\ \ref{equ:MC3} has a very broad minimum. If we make an error in the cleaning
coefficient of  $\alpha^{T_\nu}_s =  \beta/ (1- \beta) + \delta \alpha^{T_\nu}_s$, then assuming
our simplified model of a perfect foreground match between frequencies  (Eqs.\ \ref{equ:Dust2a} and \ref{equ:Dust2b}),
the leading contribution to the cleaned spectrum varies as
\begin{equation}
\hat C^{T_{\nu_1}T_{\nu_2} {\rm  clean}}  \approx  S*S +  (\delta \alpha^{T_\nu}_s)^2 (1 - \beta)^2 F*F,  \label{equ:MC4}
\end{equation}
 i.e.\ the bias varies as the square of the error in the cleaning coefficient. In practice, the
biases in the cleaned spectra are dominated by template mismatch, which cannot be removed via a
single cleaning coefficient.

\begin{figure}
\centering
\includegraphics[width=95mm,angle=0]{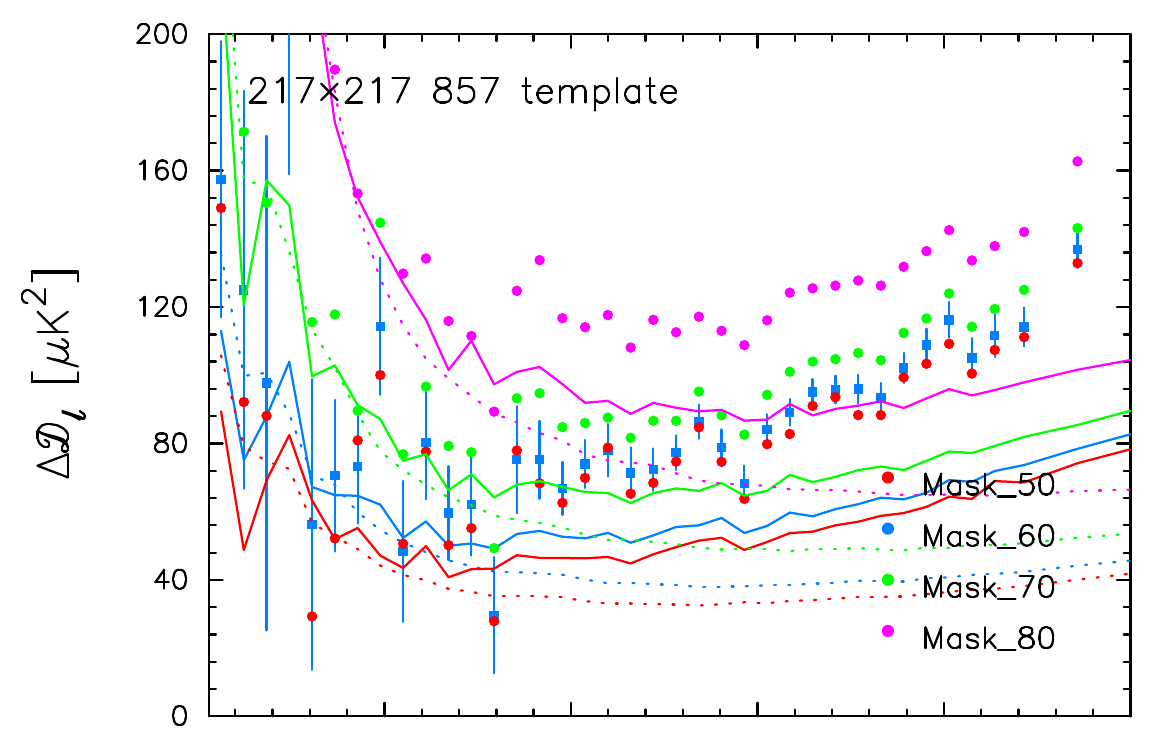} \\
\vskip -3.5mm 
\includegraphics[width=95mm,angle=0]{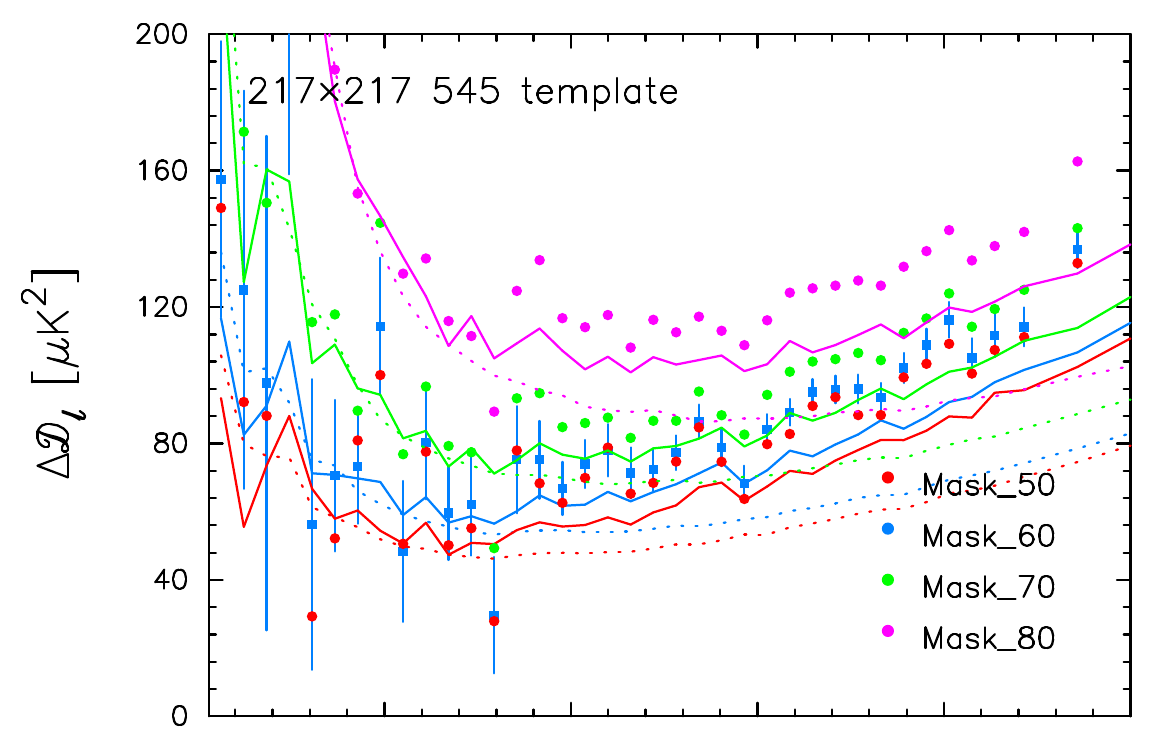} \\
\vskip -3.5mm 
\hskip 3mm \includegraphics[width=98.5mm,angle=0]{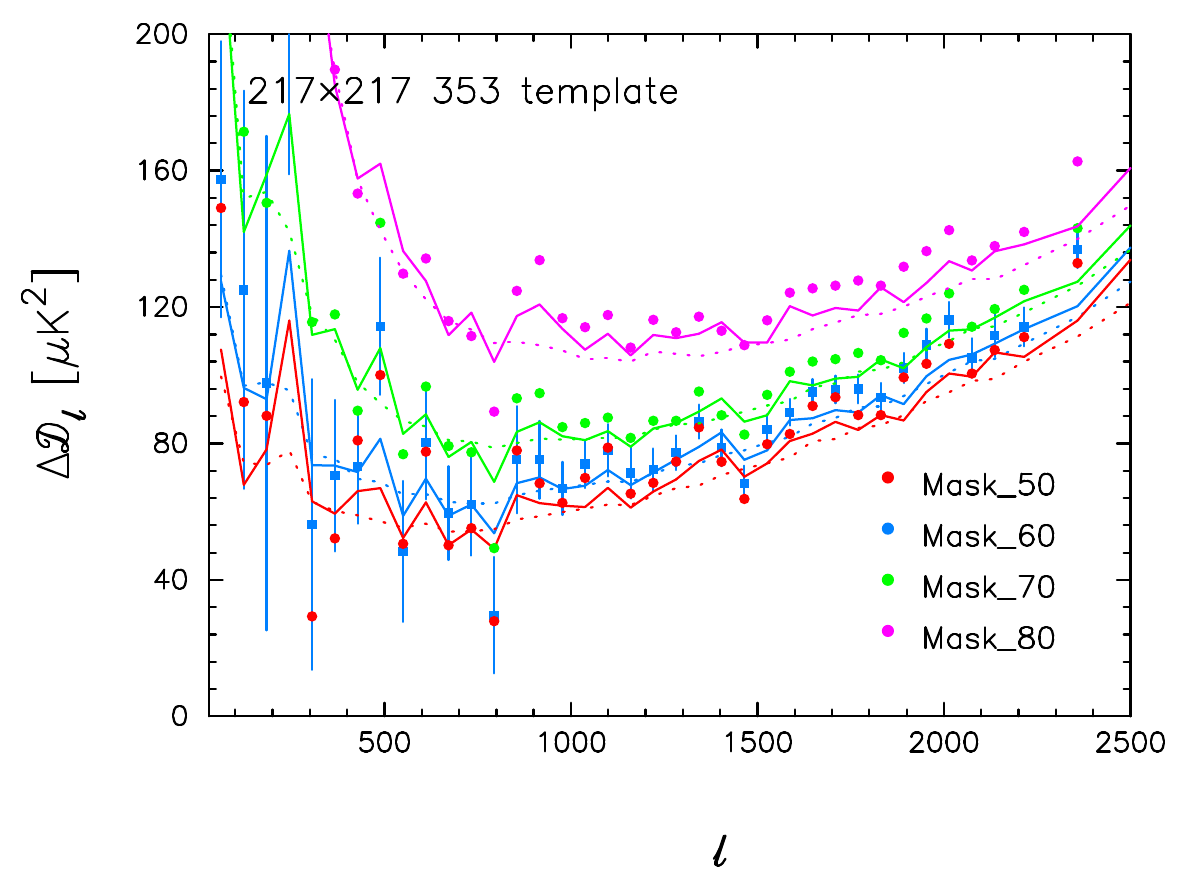}

\caption{The filled points show band averages of the $217 \times 217$ half mission
  cross spectra,  ${\hat D}^{T_{217}T_{217}}_\ell$  minus the fiducial base  \LCDM\ spectrum,
  $D^{\rm fid}_\ell$, for various masks. We plot the errors on the mask60
  spectrum, computed from the covariance matrix used in the 12.1HM \camspec\ likelihood. 
The dashed lines show  $\beta^2_{\nu_T} ({\hat D}^{T_{\nu_T}T_{\nu_T}}_\ell -   D^{\rm  fid}_\ell)$  
for the three template frequencies, illustrating the degree of template mismatch. 
The solid lines show $(1-\beta^2_{\nu_T}) [
2\alpha^{T_{\nu_T}}(1+\alpha^{T_{\nu_T}})({\hat D}^{T_{217}T_{\nu_T}}_\ell + {\hat D}^{T_{\nu_T}T_{217}}_\ell)  
 - (\alpha^{T_{\nu_T}})^2{\hat D}^{T_{\nu_T}T_{\nu_T}}_\ell + 2\alpha^{T_{\nu_T}}(2+\alpha^{T_{\nu_T}})D^{\rm fid}_\ell]$, which includes
signal-foreground correlations and partially compensates for template mismatch. }

\label{fig:217_template_res}
\end{figure}

The degree of template mismatch is shown clearly in
Fig.\ \ref{fig:217_template_res}. The filled points show the residuals
of the $217 \times 217$ half mission cross spectrum with respect to
the base \LCDM\ fiducial spectrum, $\Delta D_\ell = {\hat
  D}^{T_{217}T_{217}}_\ell - D^{\rm fid}_\ell$ for four masks.  (These
points are therefore the same in each panel.)  The dashed lines show
$\beta^2_{\nu_T} ({\hat D}^{T_{\nu_T}T_{\nu_T}}_\ell - D^{\rm
  fid}_\ell)$ for each of the three template frequencies. One can see
that these lines match the filled points quite well at low multipoles
for each of the templates, but for 545 and 857 GHz they fail to match
at multipoles $\simgt 500$.  This is consistent with the results shown
in Fig.\ \ref{fig:dust_sed} and occurs because the CIB
decorrelates between $217$ and $857$ GHz \citep{Planckdust:2014b}. One
can see from the constancy of the cleaning coefficients for $353$ GHz
in Fig.\ \ref{fig:cleaning} that
Galactic dust emission {\it and most of the CIB} can be removed  from $217$ GHz
 using $353$ GHz. However, since the CIB at 217 GHz
progressively decorrelates with the CIB at 545 and 857 GHz, template
cleaning with these frequencies removes Galactic dust emission
accurately at low multipoles (even for large sky masks), but leaves
residual CIB excesses at high multipoles,  which cannot be removed by
template cleaning for any value of the cleaning coefficient. The solid
lines in Fig.\ \ref{fig:217_template_res} show:
\begin{equation}
\Delta D_\ell =  (1-\beta^2_{\nu_T})  [ 2\alpha^{T_{\nu_T}}(1+\alpha^{T_{\nu_T}})({\hat D}^{T_{217}T_{\nu_T}}_\ell + {\hat D}^{T_{\nu_T}T_{217}}_\ell)   - (\alpha^{T_{\nu_T}})^2{\hat D}^{T_{\nu_T}T_{\nu_T}}_\ell   + 2\alpha^{T_{\nu_T}}(2+\alpha^{T_{\nu_T}})D^{\rm fid}_\ell  ], \label{equ:MC5}
\end{equation}

which accounts for signal-foreground correlations and partially
compensates for template mismatch. The differences between the filled
points and the solid lines give an accurate indication of the
remaining foreground residuals in the template cleaned spectra
(Eq. \ref{equ:MC2}). Notice that the solid lines show quite large
fluctuations from chance CMB-foreground correlations, even at
relatively high multipoles. We draw attention to the `dip' at $\ell
\approx 1500$, which is particularly prominent using $353$ and $545$
GHz as templates. This feature has a bearing on the parameters
$\Alens$ and $\Omega_{\rm K}$, as will be discussed in
Sects.\ \ref{sec:inter_frequency} and \ref{sec:extensions_lcdm}.

In PCP15 and PCP18, we formed `cleaned' \camspec\ likelihoods by cleaning
the $217\times217$, $143\times217$ and $143\times143$ TT spectra using
$545$ GHz as a template. As will be discussed in the next section, any
of $353$, $545$ or $857$ GHz could be used as an accurate tracer of
Galactic dust emission. However, the $545$ GHz maps are effectively
noise free and remove more of the CIB compared to using $857$ GHz as a
template.  Since the $353$ GHz maps are noisy (and the cleaning
coefficient relative to $217$ GHz is large), one pays a significant
signal-to-noise penalty if $353$ GHz is used as a template. For these
reasons, in this paper we use $545$ GHz as a template to form  cleaned temperature
likelihoods.

\subsection{Universality of Galactic dust emission}
\label{subsec:universality}

The results presented so far suggest that Galactic dust emission is
remarkably universal over most of the sky, i.e.\ a dust template
rescaled with a single template coefficient describes dust emission to
high accuracy over the entire \Planck\ frequency range of $100$ -- $857$
GHz.  This is illustrated by Fig.\ \ref{fig:Cleaned_maps}, which shows
the 217, 143 and 100 GHz maps cleaned with $353$, $545$ and $857$
GHz maps. The differences between the cleaned $143$ GHz maps and the
\SMICA\ CMB map  are shown in the bottom row of this figure on an expanded
scale. The large scale features visible in these plots reflect the
inhomogeneous noise levels in the \Planck\ maps. We do, however, see
residuals above the Galactic plane that are clearly physical. As in
Fig.\ \ref{fig:fakemaps} we see a prominent residual coincident with the
Ophiuchus molecular cloud complex, which has a higher temperature than
the diffuse Galactic dust emission. We also see residuals in the 100
GHz and 217 GHz maps in the region of the Aquila Rift, Perseus and the
Gum Nebula. These residuals correlate strongly with the
\Planck\ Galactic CO emission maps \citep{Planck_CO13}.  Evidently at
low Galactic latitudes, CO emission makes a significant contribution
to the residuals in the $100$ and $217$ GHz maps.  The universality of
Galactic dust emission over most of the sky leads to the following
(and somewhat unorthodox) conclusions concerning foregrounds at low
multipoles:

\begin{figure}
\centering

\includegraphics[width=150mm,angle=0]{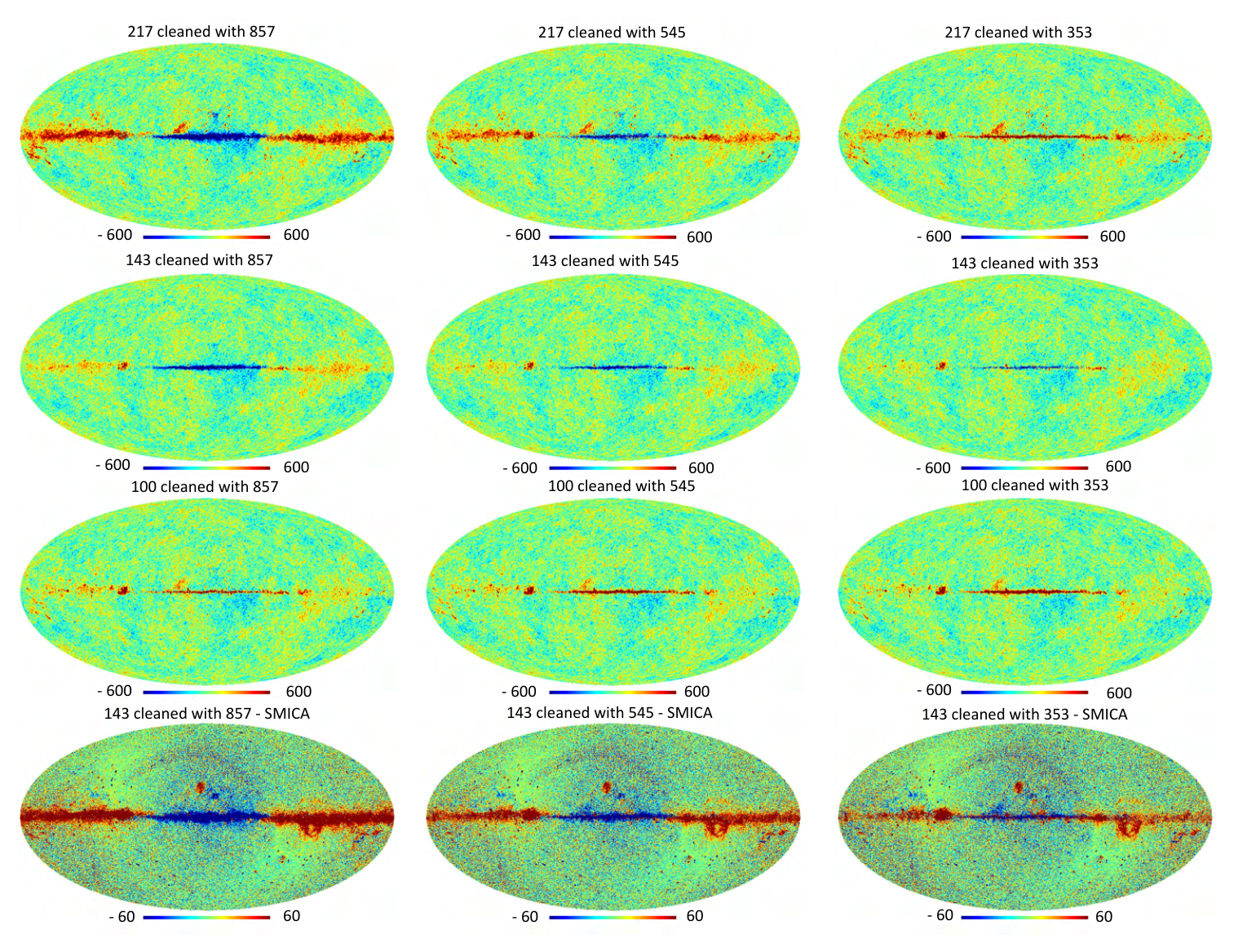}\\

   \caption {The top three rows show 217, 143 and 100 GHz  full mission maps cleaned with 857, 545 and 353 GHz full mission maps
(left to right)
using the bold-faced cleaning coefficients listed in Table \ref{tab:cleaning_coeffs}. The differences between the cleaned
143 GHz maps and the \SMICA\ map are shown in the bottom row on an expanded temperature scale. The color scales
are in units of $\mu{\rm K}$.}

\label{fig:Cleaned_maps}

\vspace{0.1truein}
\end{figure}

\begin{figure}
\centering
\includegraphics[width=145mm,angle=0]{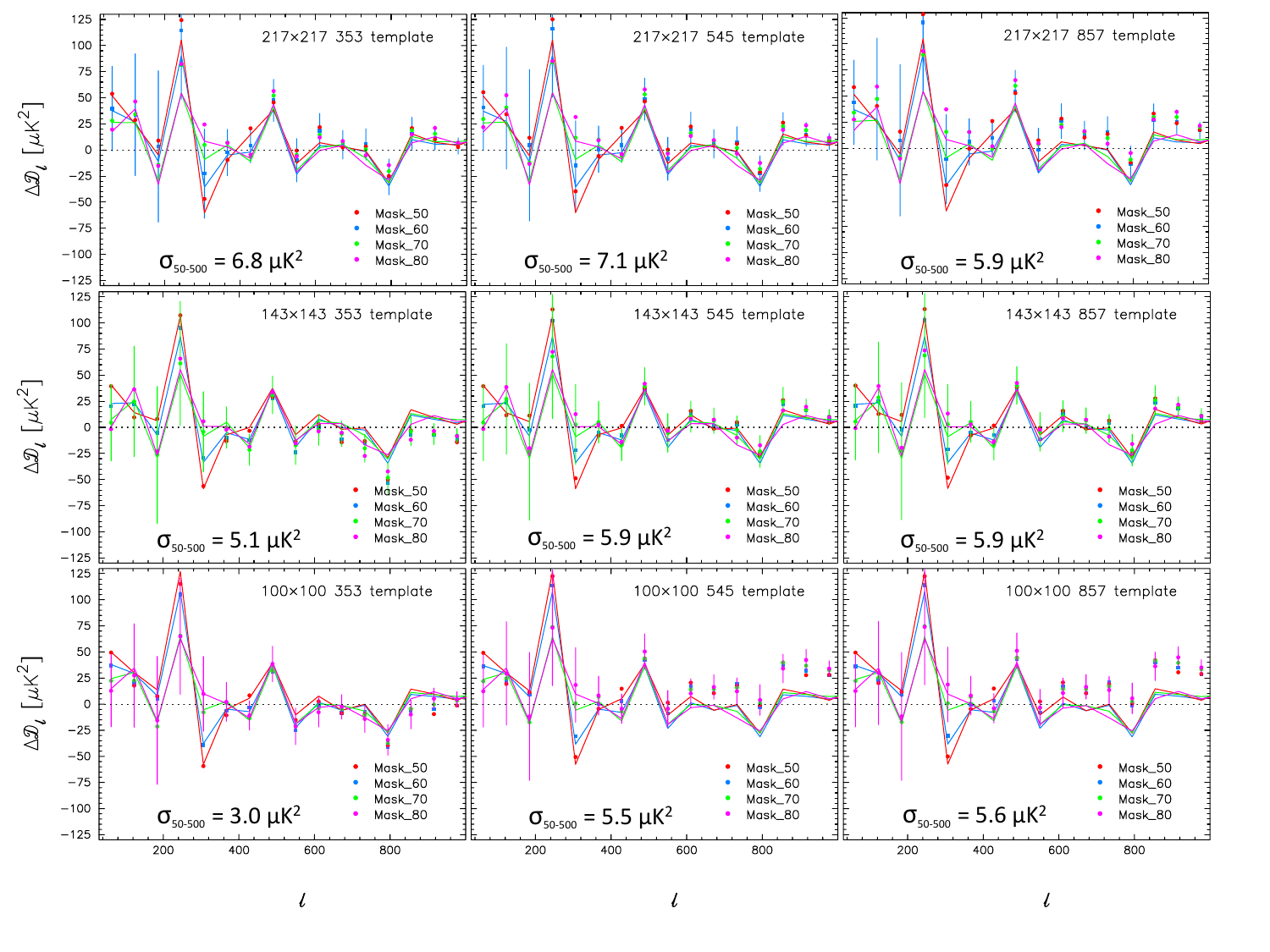} 

\caption{Cross-power spectra of template cleaned half mission  maps using the same cleaning coefficients as the full mission maps shown  in Fig.\ \ref{fig:Cleaned_maps}.  The filled points show the residuals of the power spectra with respect to the fiducial base \LCDM\ power spectrum  for four Galactic masks.
 The masks include the point source, extended object and (for $100$ and $217$ GHz) CO masks  appropriate to each frequency (as plotted in Fig.\ \ref{fig:tempmasks}). We show the \camspec\ errors on the set of points corresponding to the masks used in the 12.1HM  \camspec\ likelihood (mask80 for $100$ 
GHz, mask70 for $143$ GHz and mask60 for $217$ GHz). The solid lines show the power spectrum residuals for half mission SMICA maps. The numbers give the dispersion in the differences between the template cleaned and SMICA band-averaged  power spectra over the multipole range $50 \le \ell \le 500$ for the 12.1HM masks.}

\vspace{0.1truein}

\label{fig:mapclean}
\end{figure}

\noindent
$\bullet$  The `cleanest' \Planck\ channel is at $143$ GHz. 
Even though the net amplitude of foreground emission is
lower at $100$ and $70$ GHz, the dominant foreground at $143$ GHz is Galactic dust emission which can be subtracted
to high accuracy even at low Galactic latitudes using the higher frequency \Planck\ channels.

\noindent
$\bullet$  At $100$ GHz, CO emission is a  significant contaminant (and there is a small contribution from synchrotron emission \footnote{Typically, we use $\sim 80\%$ of the total sky area, excluding the Galactic plane,  to compute  100 GHz power spectra. Over such a sky area, the contribution of synchrotron emission to the $100\times100$ temperature 
spectrum  is $\simlt 10 (\mu{\rm K})^2$ at $\ell \simlt 400$ (see Figs. \ref{fig:pgcalib} and \ref{fig:inter_frequency100v143}) and is subdominant compared to CO emission if no CO mask is applied.}
 which will not be discussed here). Regions of intense CO emission (which coincide with molecular clouds) can be masked without paying a significant  penalty in loss of sky area.

\noindent
$\bullet$  Apart from a narrow region centered on the Galactic plane, the template subtracted $143$ GHz maps are almost indistiguishable from the \Planck\ component separated maps. (We compare with the \SMICA\ maps in  Fig.\  \ref{fig:Cleaned_maps}, but the results are almost identical if we compare with the \Planck\ 
\NILC\  or \SEVEM\ maps.) In fact, applying various statistical tests (see,  for example,
Fig.\ \ref{fig:mapclean}), we find that  differences between the cleaned 143 GHz maps and the SMICA maps are
of the same order as the differences between the various \Planck\ component separated maps. Over at least 80\% of sky, sophisticated component separation methods are not required to produce science quality maps of the CMB
free of Galactic dust emission; simple template subtraction of higher frequency maps is perfectly adequate.

\noindent
$\bullet$ The foregrounds at lower frequencies are more complicated
that at the HFI frequencies. At frequencies $\simlt 100$ GHz,
synchrotron, free-free and anomalous microwave emission become the
dominant foregrounds and vary significantly over the sky (see
\cite{Planck_foregrounds_2018}). To produce an accurate
foreground-cleaned CMB map, one needs to make a trade-off between the
net amplitude of the foregrounds and the complexity of the
foregrounds. This trade-off depends critically on the universality of
the foregrounds.  The results shown in Fig.\ \ref{fig:Cleaned_maps}
show that dust cleaning of the $143$ GHz \Planck\ maps produces
accurate maps of the CMB over almost all of the sky. Adding
information from low frequencies produces little additional gain in
the fidelity of the cleaned CMB maps and could, potentially, introduce
errors if the model adopted for the low frequency foregrounds differs
from reality. At very low Galactic latitudes (within about $\pm
5^\circ$ of the Galactic plane) Galactic emission becomes so
complicated that all component separation methods fail\footnote{To
  give some perspective, at $143$ GHz, the Galactic emission in the
  Galactic plane has a typical amplitude of $\sim 1 \times 10^5 {\mu
    {\rm K}}$, requiring foreground removal to an accuracy $\simlt 0.1
  \%$ to achieve an accuracy of $\sim 100 \ { \mu {\rm K}}$ on the
  primordial CMB.}. Our main conclusion (applicable even more strongly
to polarization, for which polarized Galactic emission is a major
problem in the search for B-modes) is that it is a better strategy to concentrate
science measurements at frequencies where the foregrounds are {\it
  simplest}. These may differ from the frequencies at which the
foregrounds have their lowest amplitude.

The power spectra of half mission template cleaned temperature maps
are plotted in Fig.\ \ref{fig:mapclean} and compared to those of half
mission SMICA maps for various Galactic masks. According to the
discussion of the previous subsection, at multipoles $\simlt 500$, the
template cleaned spectra should look almost identical and in close
agreement with the SMICA spectra, reflecting the universality of
Galactic dust emission. At higher multipoles, we should see excesses
that are independent of the sky area caused by extragalactic
foregrounds (primarily point sources at $100$ GHz and  CIB at $217$
GHz). This is what is observed in Fig.\ \ref{fig:mapclean}. The numbers in
each panel give the dispersion in the differences between the
template-cleaned and SMICA band-averaged power spectra over the
multipole range $50 \le \ell \le 500$ for the temperature masks used
in the 12.1HM \camspec\ likelihood. These are all within a few $(\mu
{\rm K})^2$.  This figure also shows that the 12.1HM masks are
conservative and that it is possible to use larger areas of sky in
forming temperature likelihood. This is explored further in
Sect.\ \ref{sec:inter_frequency}.

\section{Galactic dust emission in polarization}
\label{sec:dust_polarization}

The results of the previous section show that Galactic dust emission
in temperature can be subtracted to high accuracy from the
\Planck\ spectra. However, dust subtraction leaves residuals at high
multipoles in temperature which arise from frequency dependent
extragalactic foregrounds.  This necessarily requires that we fit for
frequency dependent extragalactic foregrounds described by `nuisance'
parameters.  For polarization,
the situation is different since extragalactic foregrounds are
expected to be very weakly polarized.  The high multipole polarization
measurements from ACTpol and SPTpol \citep{Louis:2017, Henning:2018}
provide strong evidence that extragalactic infrared sources should be
undetectable in the \Planck\ polarization power spectra (consistent
with what we see from the \Planck\ data). In polarization, Galactic dust emission is
the only foreground that needs to be removed from the
\Planck\  maps.

\newpage

\subsection{Spectrum-based cleaning coefficients}
\label{subsec:spec_clean_pol}

Since the $545$ and $857$ GHz bands are unpolarized, \Planck\ has a
limited frequency range over which to monitor polarized Galactic dust
emission ($100-353$ GHz). In addition, the polarization maps are noisy
and strongly `contaminated' by the primordial E-mode signal, even at
$353$ GHz.  We therefore do not use map-based cleaning coefficients in
polarization but instead use spectrum-based cleaning coefficients,
determined by minimising  residual functions $\Psi_{EE}$ an $\Psi_{BB}$ 
and generalising Eqs.\ \ref{equ:MC2} and \ref{equ:MC3} to the polarization
spectra.

Table \ref{tab:pol_cleaning_coeffs} lists cleaning coefficients, using
353 GHz maps as templates determined by minimising $\Psi_{EE}$ and
$\Psi_{BB}$ for the $217\times 217$, $143\times 143$ and $100\times
100$ EE and BB half mission spectra over the multipole range $30 \le
\ell \le 300$ using the 12.1HM \camspec\ polarization mask. For
completeness, we also list the TT cleaning coefficient computed over
this multipole range, which can be compared with the entries in Table
\ref{tab:cleaning_coeffs}. We recover almost identical polarization
cleaning coefficients using the EE and BB spectra. In addition, we
find that these coefficients are insensitive to the size of the
polarization mask. As discussed in Sect.\ \ref{subsec:spec_clean}, the
statistics $\Psi_{TT}$, $\Psi_{EE}$, $\Psi_{BB}$ have broad
minima and so the cleaned polarization spectra are insensitive to the
precise values of the cleaning coefficients. We adopt the $\alpha^T_s$
and $\alpha^E_s$ cleaning coefficients listed in Table
\ref{tab:pol_cleaning_coeffs} to clean the TE, ET and EE
polarization spectra used in the \camspec\ likelihoods at low
multipoles. Note that the polarization cleaning coefficients are quite
close to the temperature cleaning coefficients. Over the limited range
of frequencies probed by \Planck, the SED of polarized Galactic
emission is very close to the SED in intensity (see
Fig.\ \ref{fig:dust_sed}), in agreement with the results of
\cite{Planckdust:2015}.

\begin{table}[b]

\vspace{0.25 truein}

{\centering

\caption{\small{Cleaning coefficients using $353$ GHz half mission T,Q,U maps as templates. 
The cleaning coefficients $\alpha^E_s$ and $\alpha^B_s$ are determined
by separately minimising cleaning residuals in the $E$- and $B$-mode spectra over the multipole range  $30 \le \ell \le 300$, using the 12.1HM \camspec\ polarization mask. For reference, we also list the
temperature cleaning coefficient for the 12.1HM \camspec\ temperature masks,
computed over the same multipole range.} }

\label{tab:pol_cleaning_coeffs}
\begin{center}

\large{
\begin{tabular}{|c|c|c|c|} \hline

spectrum  & $\alpha^E_s$ &  $\alpha^B_s$ &  $\alpha^T_s$  \\  \hline
$217\times 217$  &  $0.141$  & $0.146$  & $0.143$   \\
$143\times 143$ &  $0.0392$  & $0.0381$ &  $0.0341$ \\
$100 \times 100$ &  $0.0192$   & $0.0171$ &  $0.0208$ \\
 \hline

\end{tabular}}
\end{center}}
\end{table}

\vspace{0.1truein}

\begin{figure*}
 \centering
\includegraphics[width=160mm,angle=0]{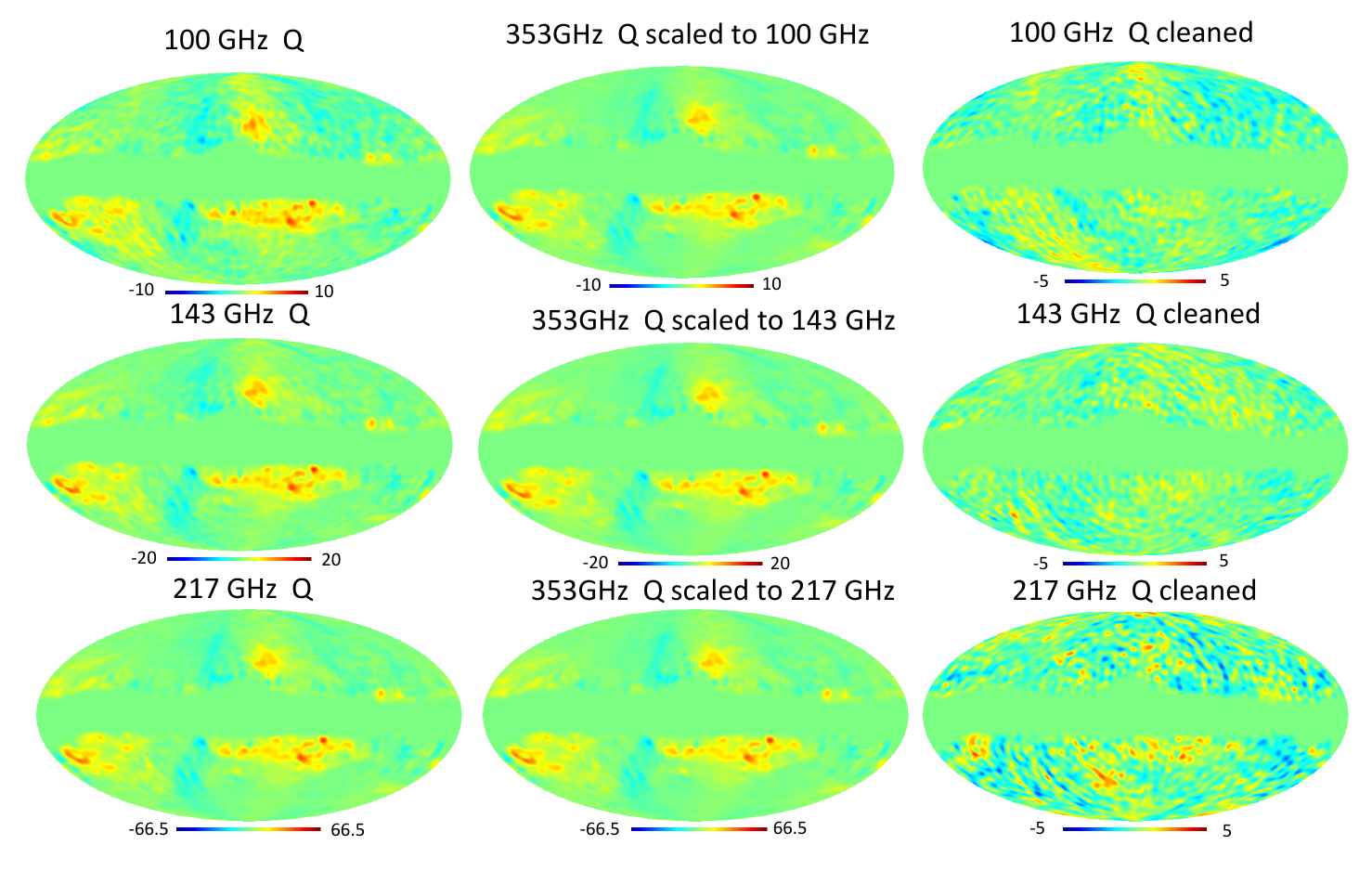}  \\
\includegraphics[width=160mm,angle=0]{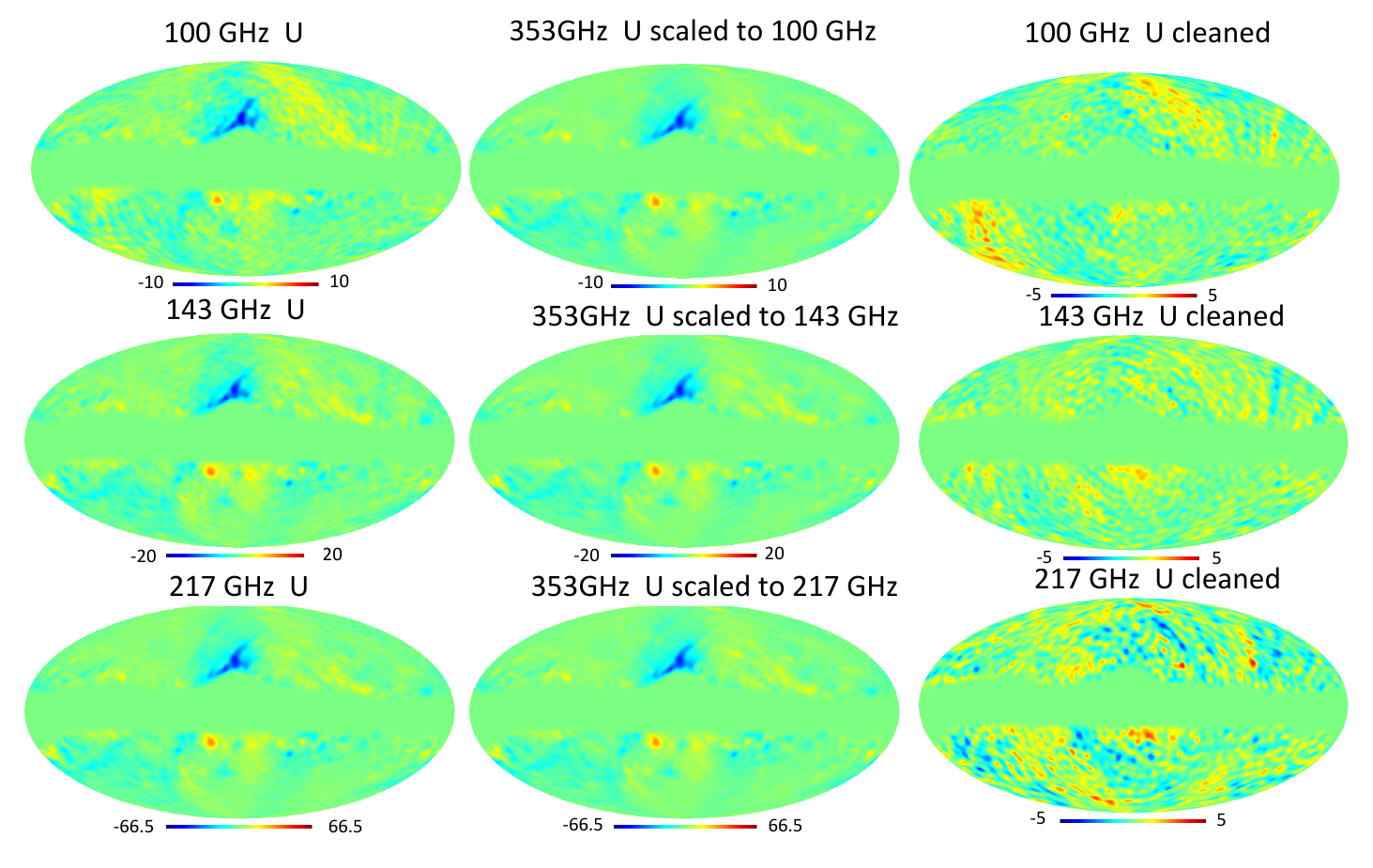} \\

\caption {Q and U polarization maps smoothed with a Gaussian beam of
  FWHM of 2 degrees.  The temperature mask80 has been applied to these
  maps. The top three rows show Q maps and the bottom three rows show
  U maps. In each row, the left-most map shows the polarization maps
  at each frequency on a scale such that Galactic dust emission is
  expected to have the same amplitude. The maps in the centre show
  353 GHz Q and U maps scaled to the dust amplitude for the appropriate
  frequency using the $\alpha_s^E$ template coefficients given in
  Table \ref{tab:pol_cleaning_coeffs}.  The figures on the right show
  the 353 GHz subtracted polarization maps at each frequency (i.e.\ the differences
between the left-most and middle figures in each row). The temperature scales are in units of
$\mu{\rm K}$.}
     
\label{fig:pol_cleaned_maps}
 \end{figure*}

We can get a visual feel for polarized dust emission from
Fig.\ \ref{fig:pol_cleaned_maps}. In this figure, we have smoothed the
full mission HFI \Planck\ polarization maps with a Gaussian of FWHM of
2 degrees. The plots to the left show the smoothed Q and U maps at
$100$, $143$ and $217$ GHz with a temperature scale based on the
$\alpha_s^E$ template coefficients given in Table
\ref{tab:pol_cleaning_coeffs} such that polarized Galactic dust
emission should look identical in each plot. The plots in the middle
panel show the 353 GHz Q and U maps rescaled using the $\alpha_s^E$
coefficients of Table \ref{tab:pol_cleaning_coeffs} to match the
respective lower frequency Q and U maps. The plots to the right show
the differences between the scaled 353 GHz and the lower frequency maps on
an expanded colour scale that is the same for all frequencies. As can
be seen, the scaled 353 GHz maps match extremely well with the lower
frequency maps, showing that polarized Galactic dust emission
is the dominant source of  the large scale features in all of the HFI polarized maps
(dominating over the CMB at multipoles $\ell \simlt 50$ at 100 GHz and
$\ell \simlt 250$ at 217 GHz, see Fig.\ \ref{fig:EEdust_residuals}).
The cleaned polarization maps are noise dominated, so it is not
possible to see the primordial CMB fluctuations in this figure. Large
scale systematic features are visible in the cleaned maps and are
largely caused by the distortion of the Solar dipole caused by
non-linearities in the analogue-to-digital conversion electronics, which are not
corrected accurately  by \SROLL\ (see Appendix B.4.2 of \citep{SROLL:2016}).

\begin{figure*}
 \centering
\includegraphics[width=75mm,angle=0]{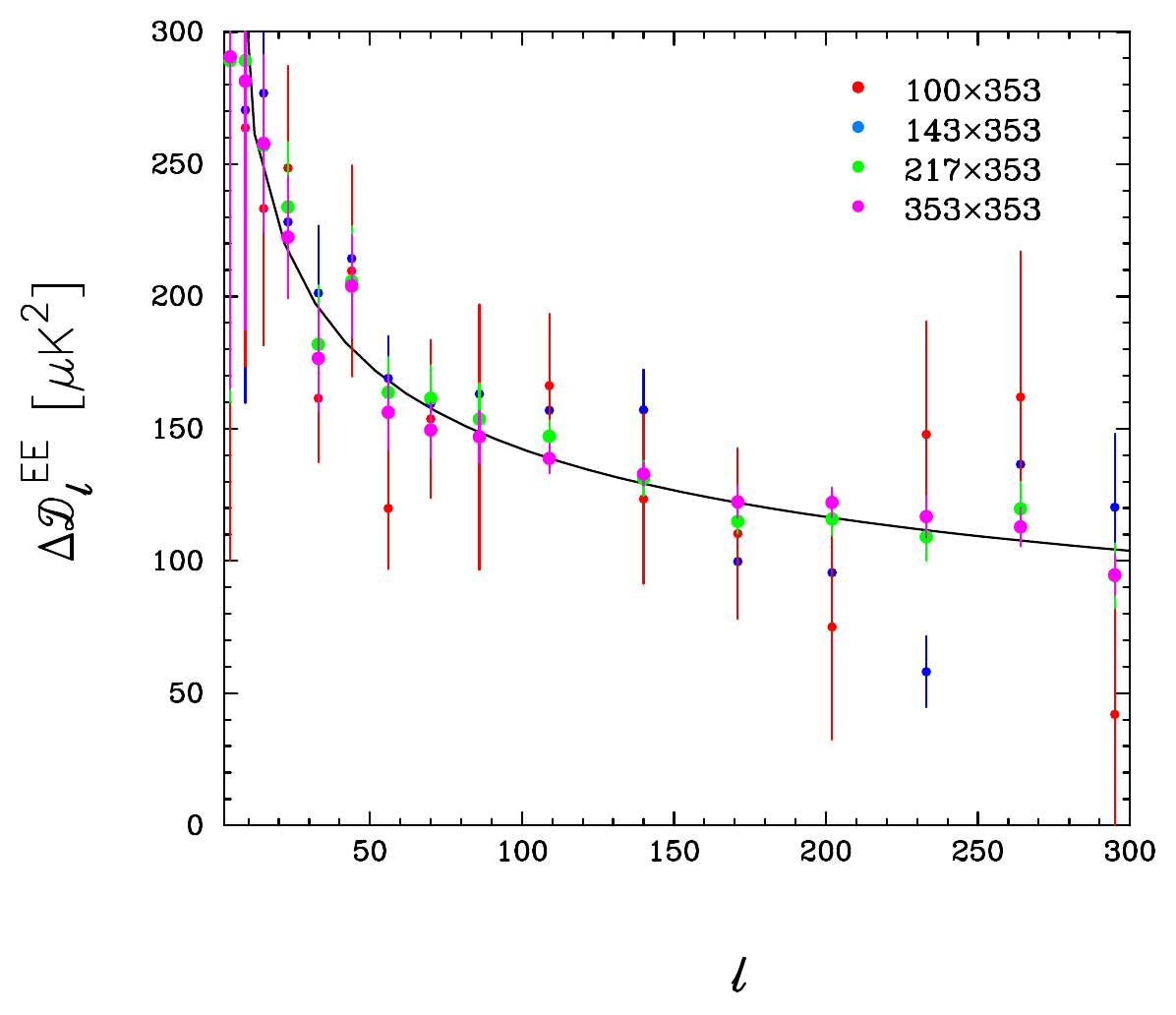} 
\includegraphics[width=75mm,angle=0]{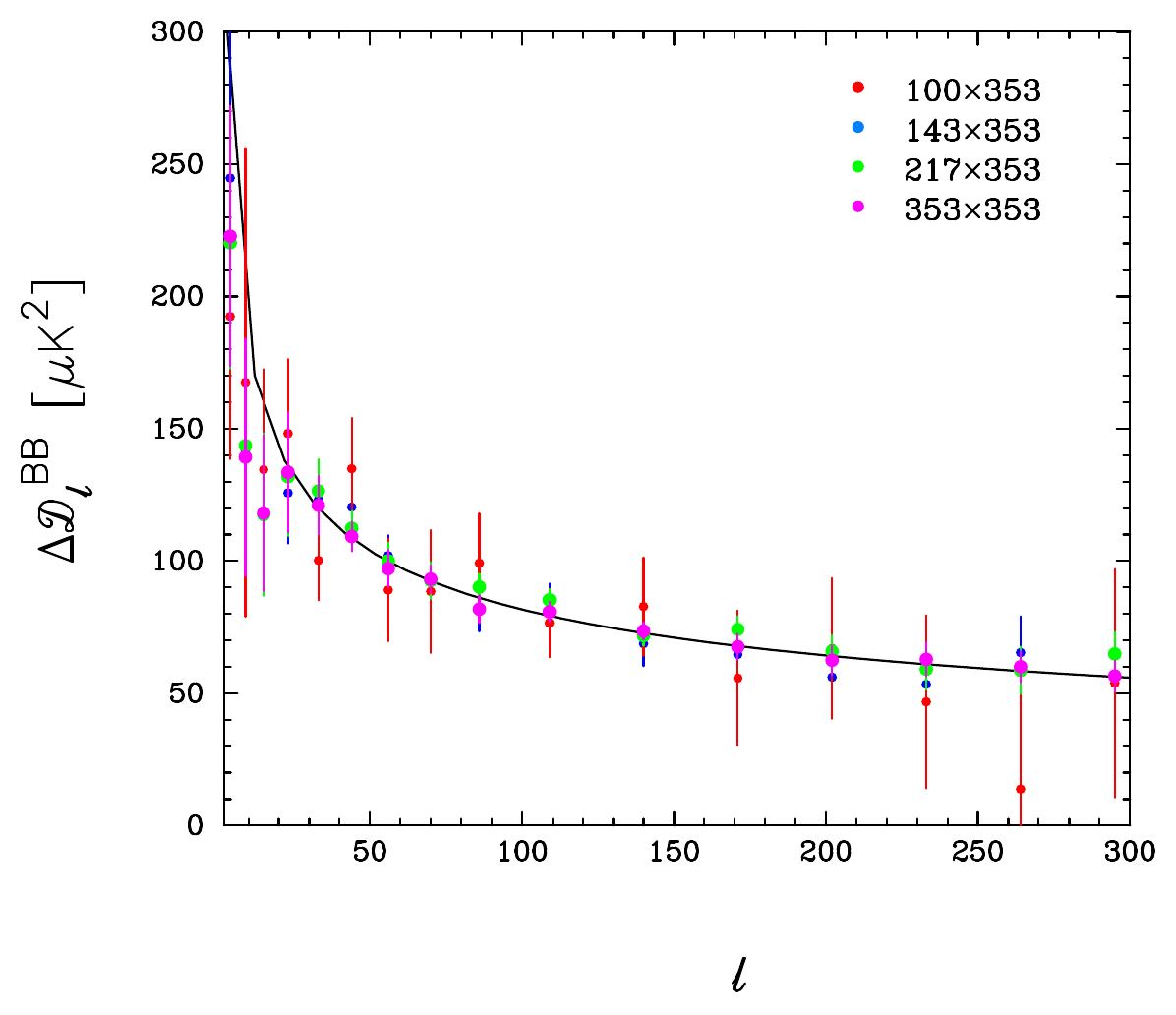} \\

\caption {Band averaged EE and BB cross-spectra involving $353$
  GHz computed for the 12.1HM \camspec\ polarization mask.  We have
  subtracted the fiducial base \LCDM\ E-mode spectrum from the EE
  spectra. The $100\times353$, $143\times 353$ and $217 \times 353$
  spectra have been renormalized to the amplitude of the $353\times
  353$ spectra by dividing by $\alpha^E_s/(1+\alpha^E_s)$,  using the
  cleaning coefficients listed in Table
  \ref{tab:pol_cleaning_coeffs}. (We use the same coefficients for the
  EE and BB spectra.) The solid lines show power-law fits to the
  $353\times 353$ spectrum (Eq.\ \ref{equ:PD1}) as described in the text. }
     
\label{pol_353}
 \end{figure*}

Figure \ref{pol_353} shows the half mission $353\times 353$, $217 \times 353$, $143 \times 353$
and $100 \times 353$ EE and BB spectra. Since there are no detectable extragalactic foreground
components in polarization, these spectra can be used to estimate the power spectrum of
Galactic polarized dust emission. The CMB E-modes contribute to the EE spectra,
so we have subtracted the E-mode spectrum of the fiducial base \LCDM\ model.  Each of the
spectra has been renormalized to match the $353 \times 353$ spectrum using the cleaning coefficients
$\alpha^E_S$ from Table \ref{tab:pol_cleaning_coeffs}. The error bars on the points are computed from the internal scatter
of the power spectra within each multipole band.  The solid lines show power-laws
\begin{equation}
 \hat D_\ell =  A \left ({\ell \over 200}  \right )^\epsilon,  \label{equ:PD1}  
\end{equation}
fitted over the multipole range $30 \le \ell \le 500$ to the $353\times 353$ $EE$ and $BB$ spectra.
We find:
\begin{subequations}
\begin{eqnarray}
   A^{EE} & = & 116.5 \pm 1.9 \  (\mu {\rm K})^2,  \quad  \epsilon^{EE} = -0.29 \pm 0.03, \label{equ:PD2a} \\
   A^{BB} & = & \ \ 64.2 \pm 1.6 \  (\mu {\rm K})^2,  \quad  \epsilon^{BB} = -0.35 \pm 0.03. \label{equ:PD2b} 
\end{eqnarray}
\end{subequations}
We see that $A^{BB}/A^{EE} \approx 0.55$, which is typical for polarized Galactic dust emission measured by 
\Planck\ over large areas of the sky \citep[e.g.][]{Planck_dust_pol:2016}. However, we find shallower spectral indices
than \cite{Planck_dust_pol:2016}, who concluded $\epsilon^{EE} = \epsilon^{BB} = -0.42 \pm 0.02$\footnote{\GE{The spectral indices in polarization are physically related to the power spectrum of the interstellar magnetic field (see e.g. \cite{Ghosh:2017}) and so are expected to differ from the steeper spectral index of Galactic dust emission in temperature, $D^{TT}_\ell \propto \ell^{-0.69}$}}. The reason for this small difference is not understood,
but may be related to the very different error model adopted in \cite{Planck_dust_pol:2016}\footnote{Which attempts to remove sample variance and so gives high weight to points at low multipoles.}.
 The amplitudes of the rescaled  frequency spectra in Fig.\ \ref{tab:pol_cleaning_coeffs}  match 
reasonably well for both EE and BB. However, the spectra are noisy and so do not provide a high precision test
of the accuracy of the cleaning coefficients. The effectiveness of template cleaning
at lower frequencies  provides a  more stringest test, as described in the next subsection.

Applying a similar analysis to half mission $353\times 353$ TE and ET spectra (using mask60 for the temperature maps)  we find:
\begin{subequations}
\begin{eqnarray}
   A^{TE} & = & 195.3 \pm 8.0 \  (\mu {\rm K})^2,  \quad  \epsilon^{TE} = -0.30 \pm 0.06, \label{equ:PD3a} \\
   A^{ET} & = & 189.0 \pm 8.0 \  (\mu {\rm K})^2,  \quad  \epsilon^{ET} = -0.27 \pm 0.06, \label{equ:PD3b} 
\end{eqnarray}
\end{subequations}
and these two estimates are consistent with each
other. Thus for all of the polarization spectra, we find shallower
slopes than for the Galactic dust temperature power spectrum (for
which  $\hat D^{TT}_\ell \propto \ell^{-0.69}$,  as shown in Sect.\ \ref{subsec:dust_power}).

\subsection{Cleaning EE, TE, ET spectra}
\label{subsec:pol_clean_spectra}

\begin{figure*}
 \centering
\includegraphics[width=52.0mm,angle=0]{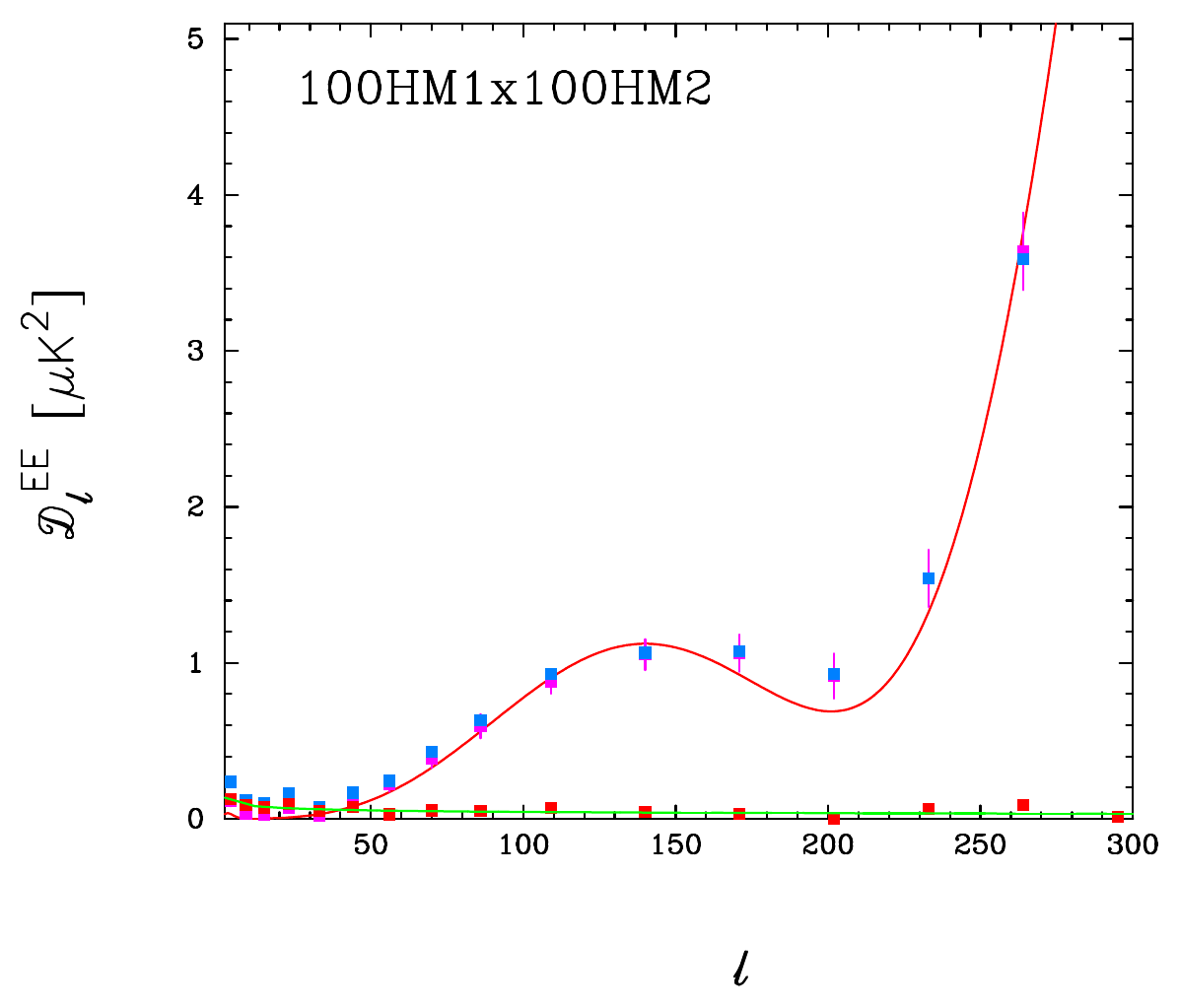} 
\includegraphics[width=52.0mm,angle=0]{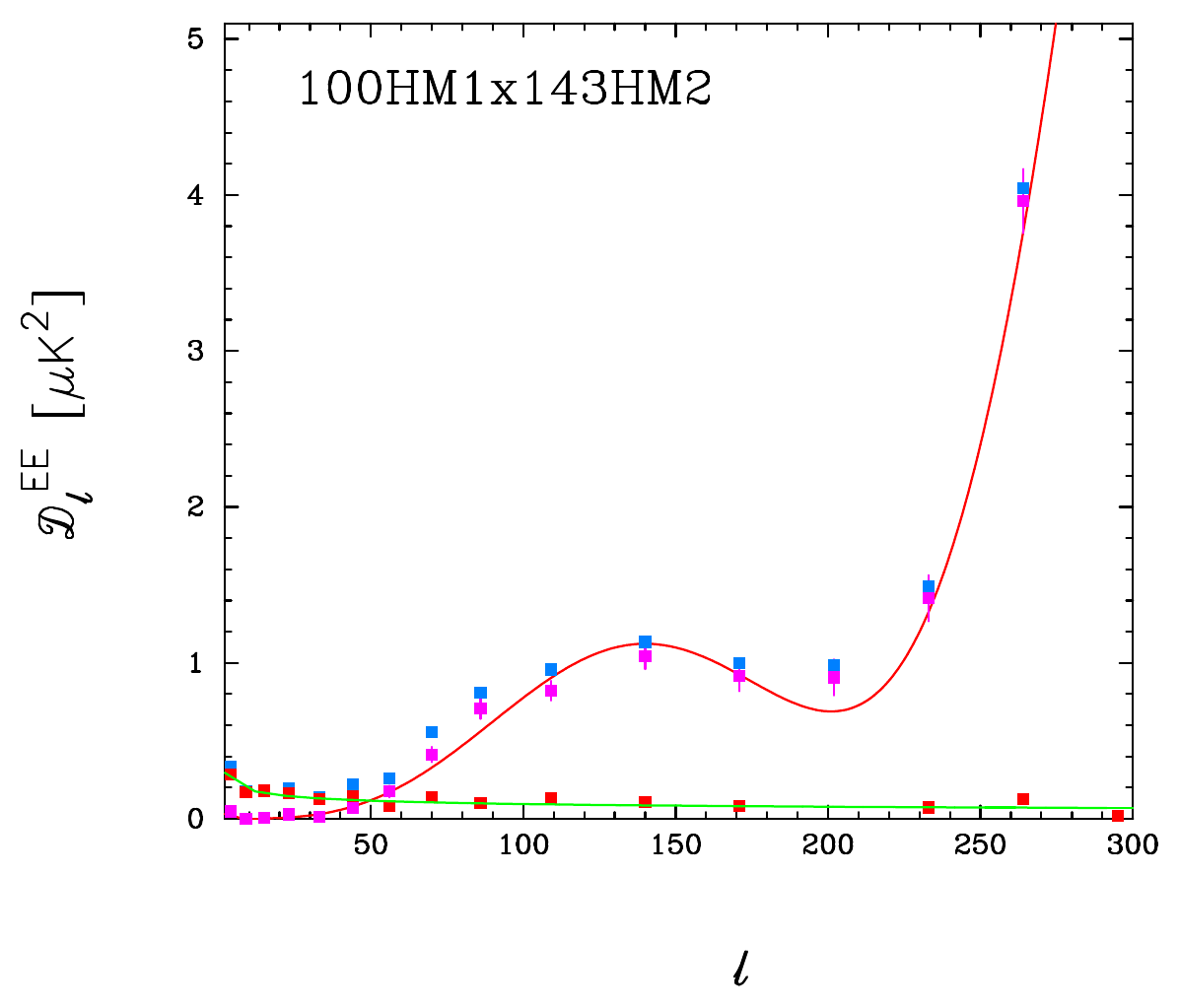} 
\includegraphics[width=52.0mm,angle=0]{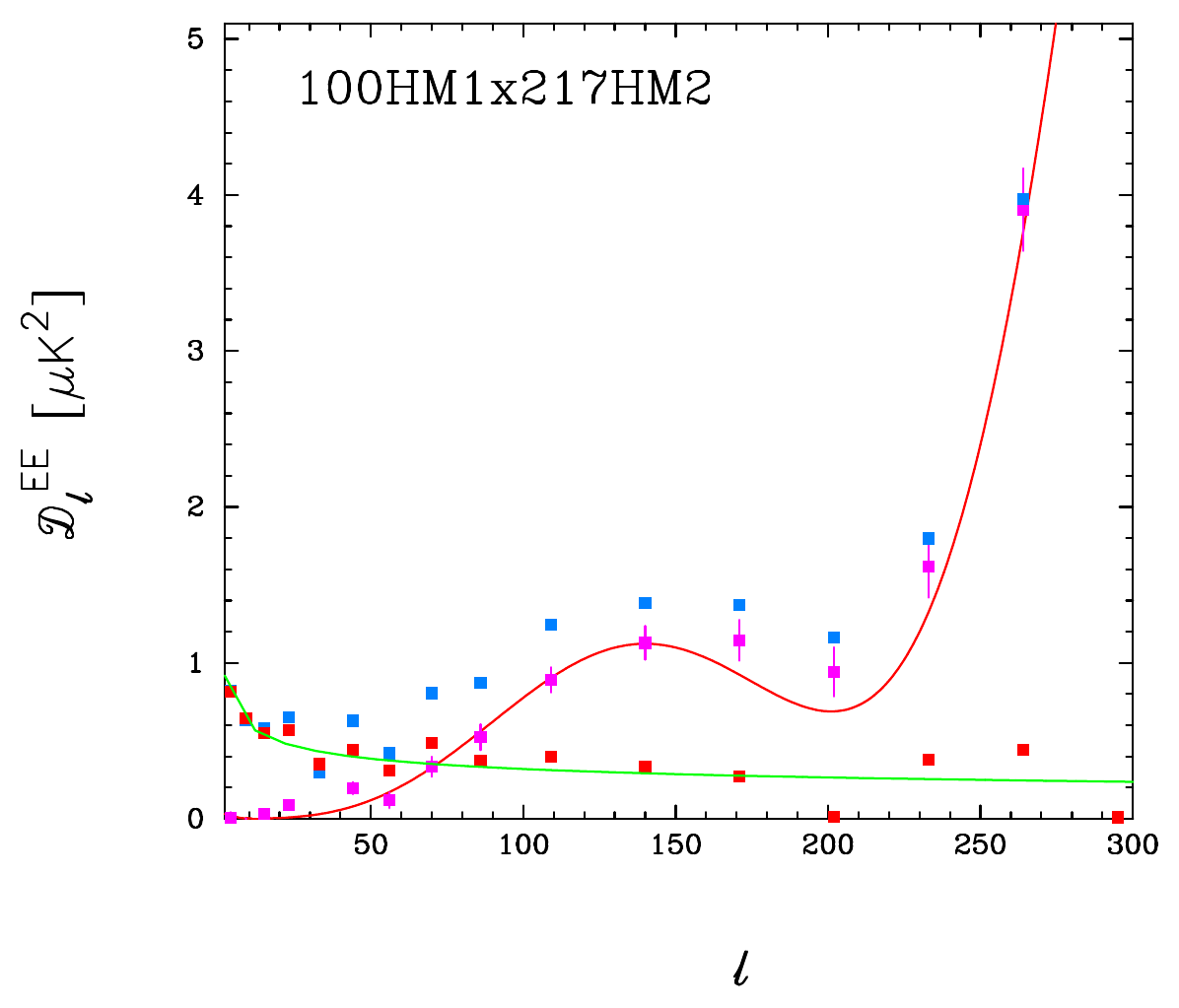} \\
\includegraphics[width=52.0mm,angle=0]{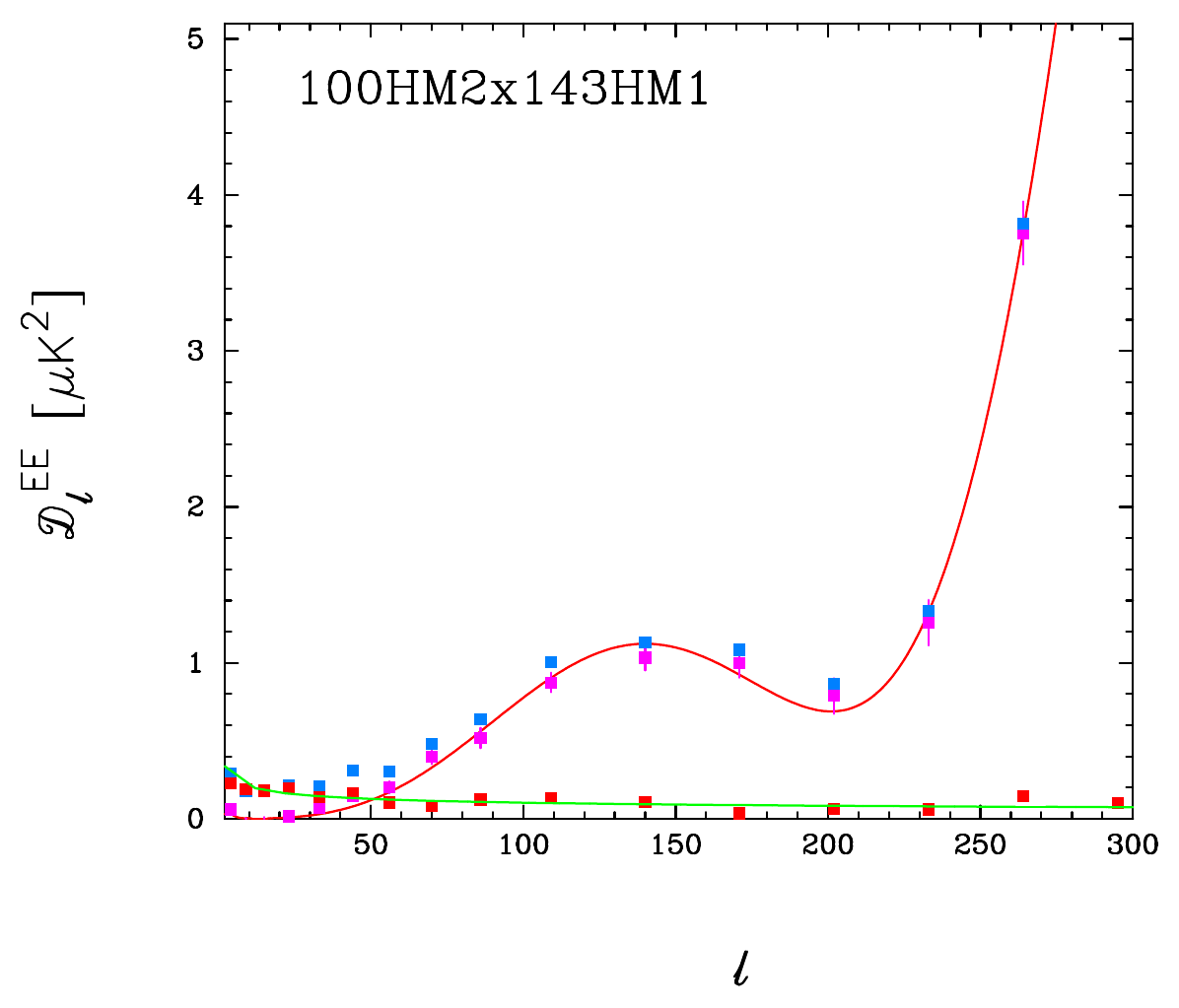} 
\includegraphics[width=52.0mm,angle=0]{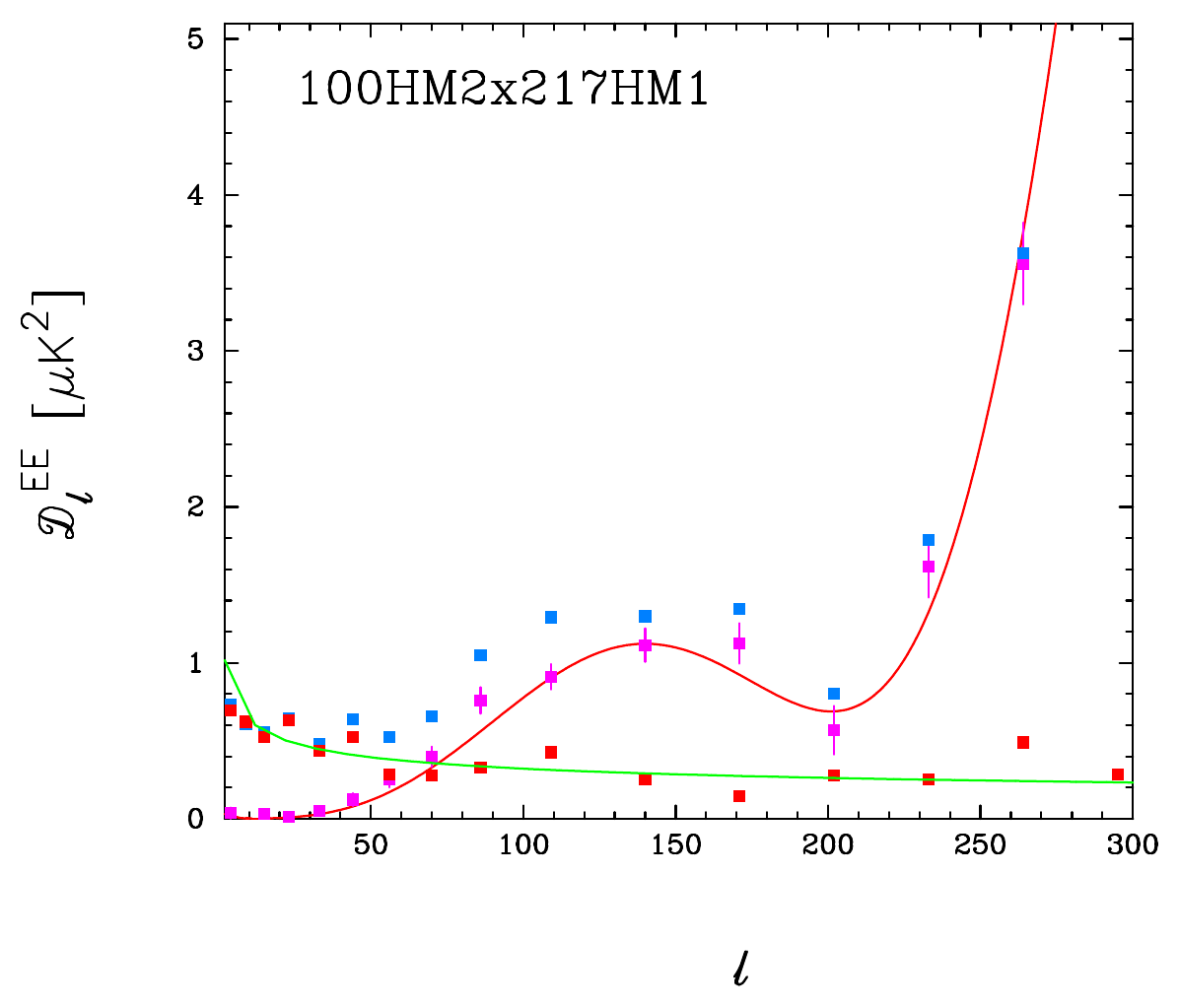} 
\includegraphics[width=52.0mm,angle=0]{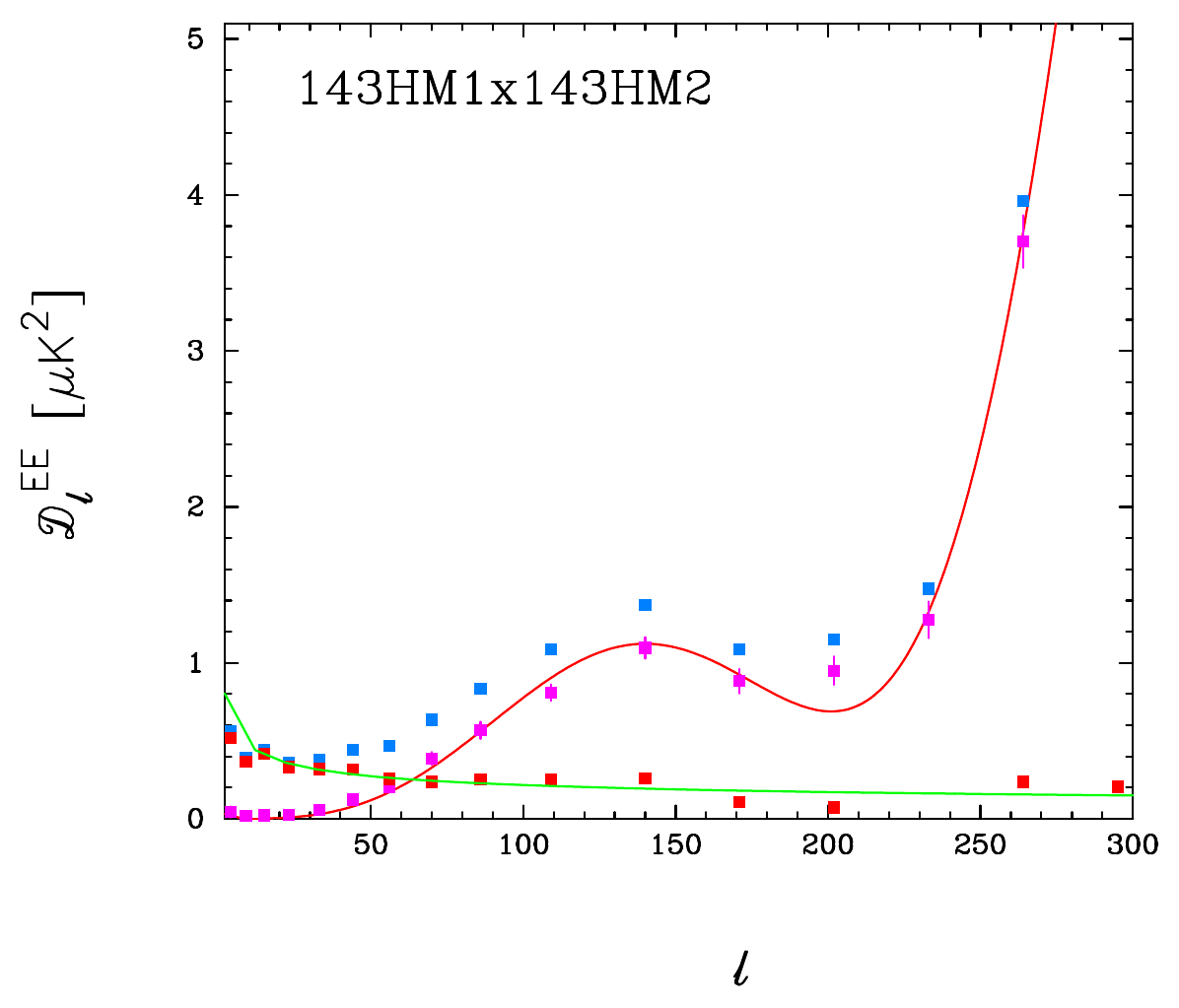} \\
\includegraphics[width=52.0mm,angle=0]{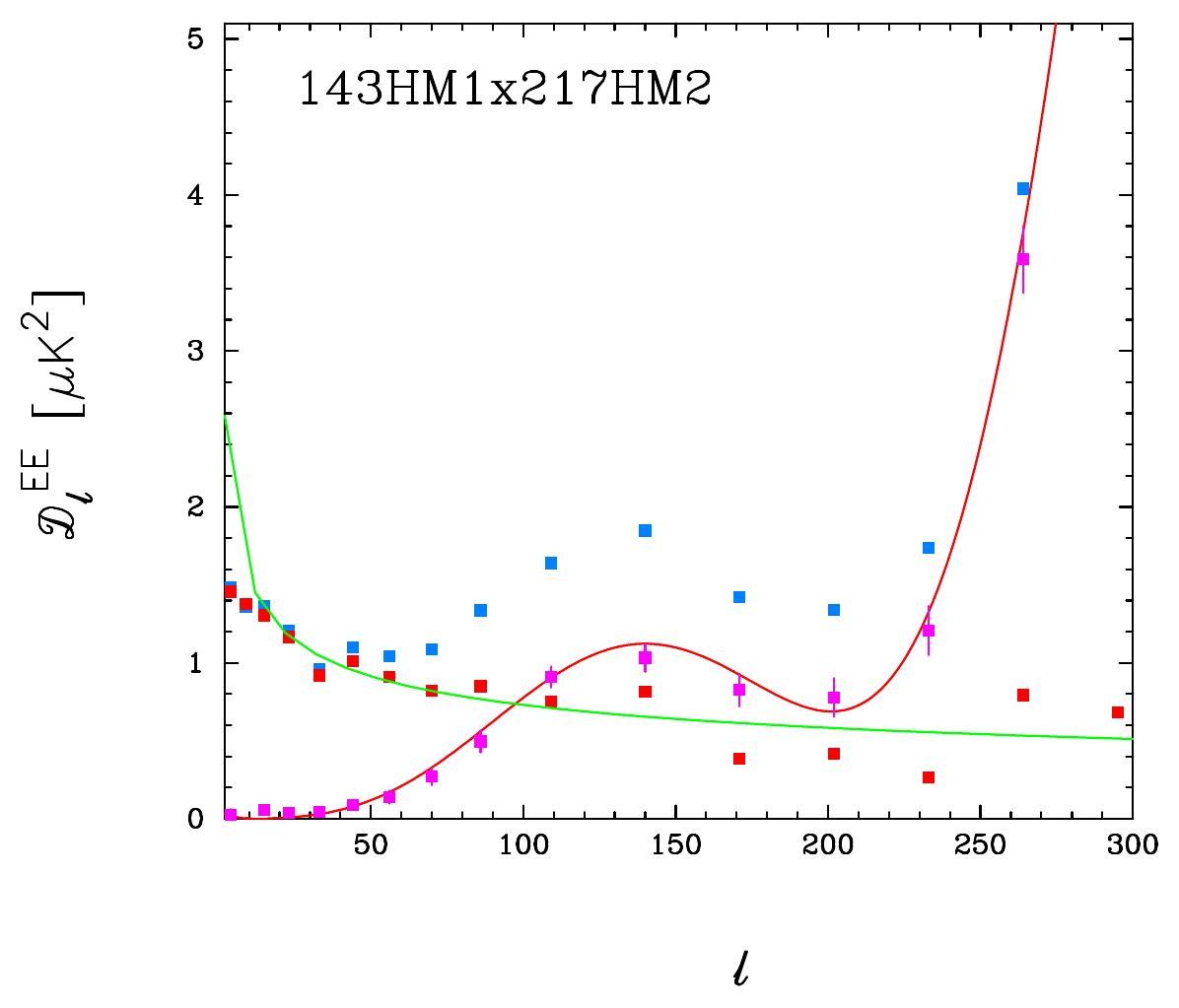} 
\includegraphics[width=52.0mm,angle=0]{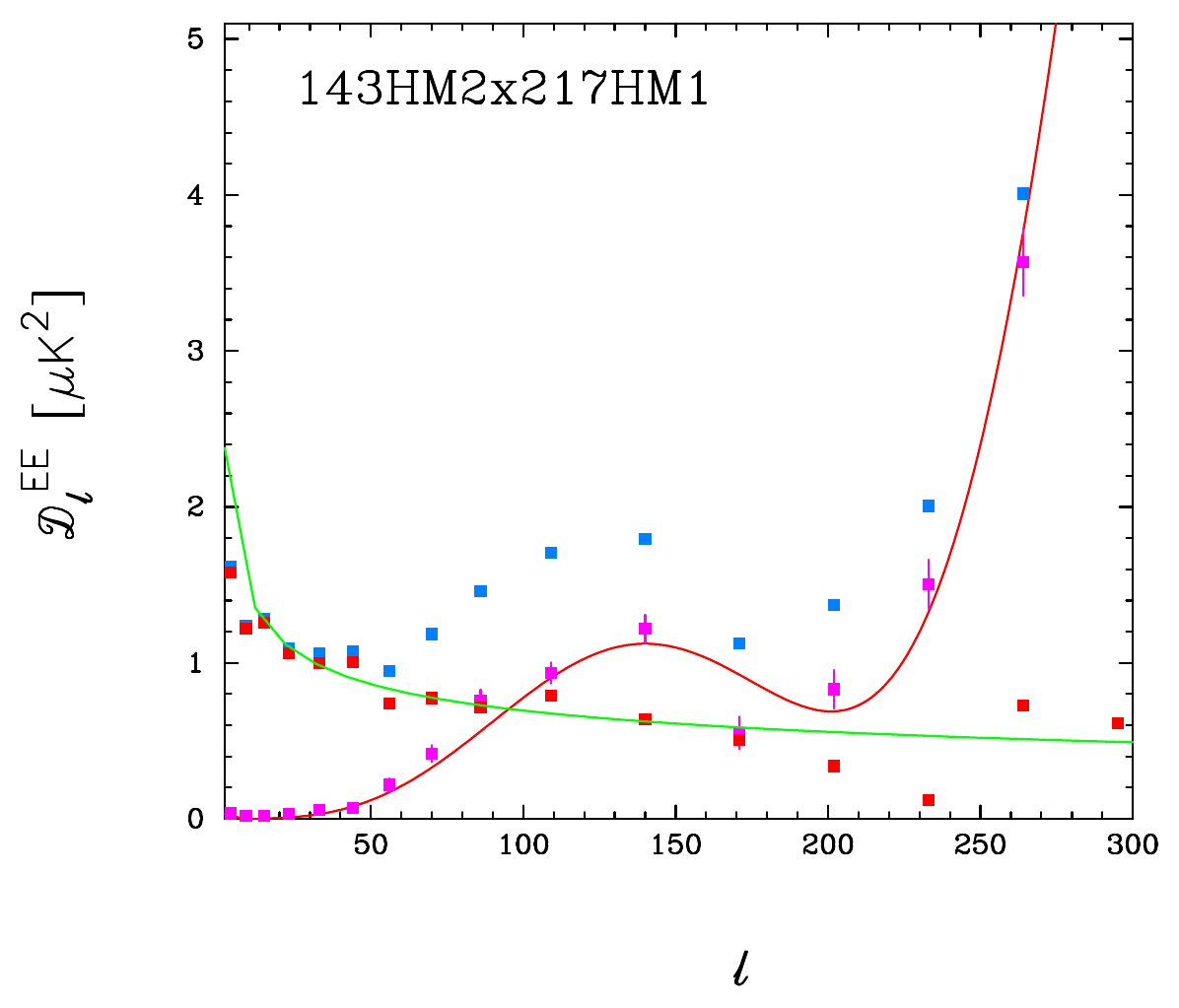} 
\includegraphics[width=52.0mm,angle=0]{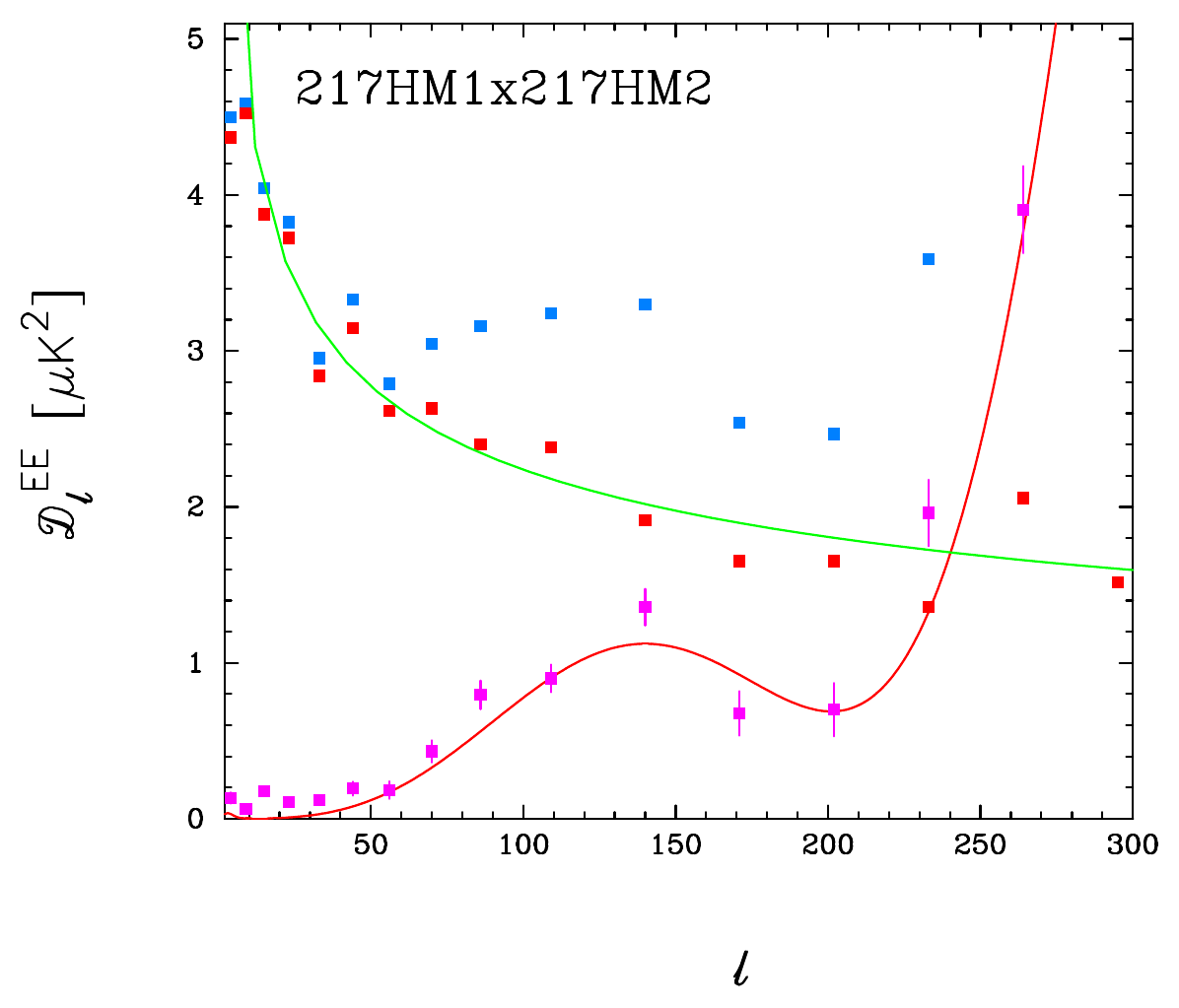}  \\
\caption {The nine EE spectra used to form the
  frequency averaged EE power spectrum in the 12.1HM
  \camspec\ likelihood plotted at low multipoles. The blue points show the uncleaned EE
  spectra. The red points show the difference between the $353$
  cleaned and uncleaned spectra. The green lines show the power law
  fits to the red points, with the parameters listed in Table
  \ref{tab:pol_dust_fits}. The pink points show the dust cleaned
  spectra. These are computed from the analogue of Eq.\ \ref{equ:MC2}
  for $\ell \le 150$; at higher multipoles we subtract the power-law
  fit from the blue points. The error bars are computed from the
  12.1HM \camspec\ covariance matrix. The red lines show the EE
  spectrum for the fiducial base \LCDM\ model.}

\label{fig:EEdust_residuals}

 \end{figure*}

\begin{figure*}
 \centering
\includegraphics[width=52.0mm,angle=0]{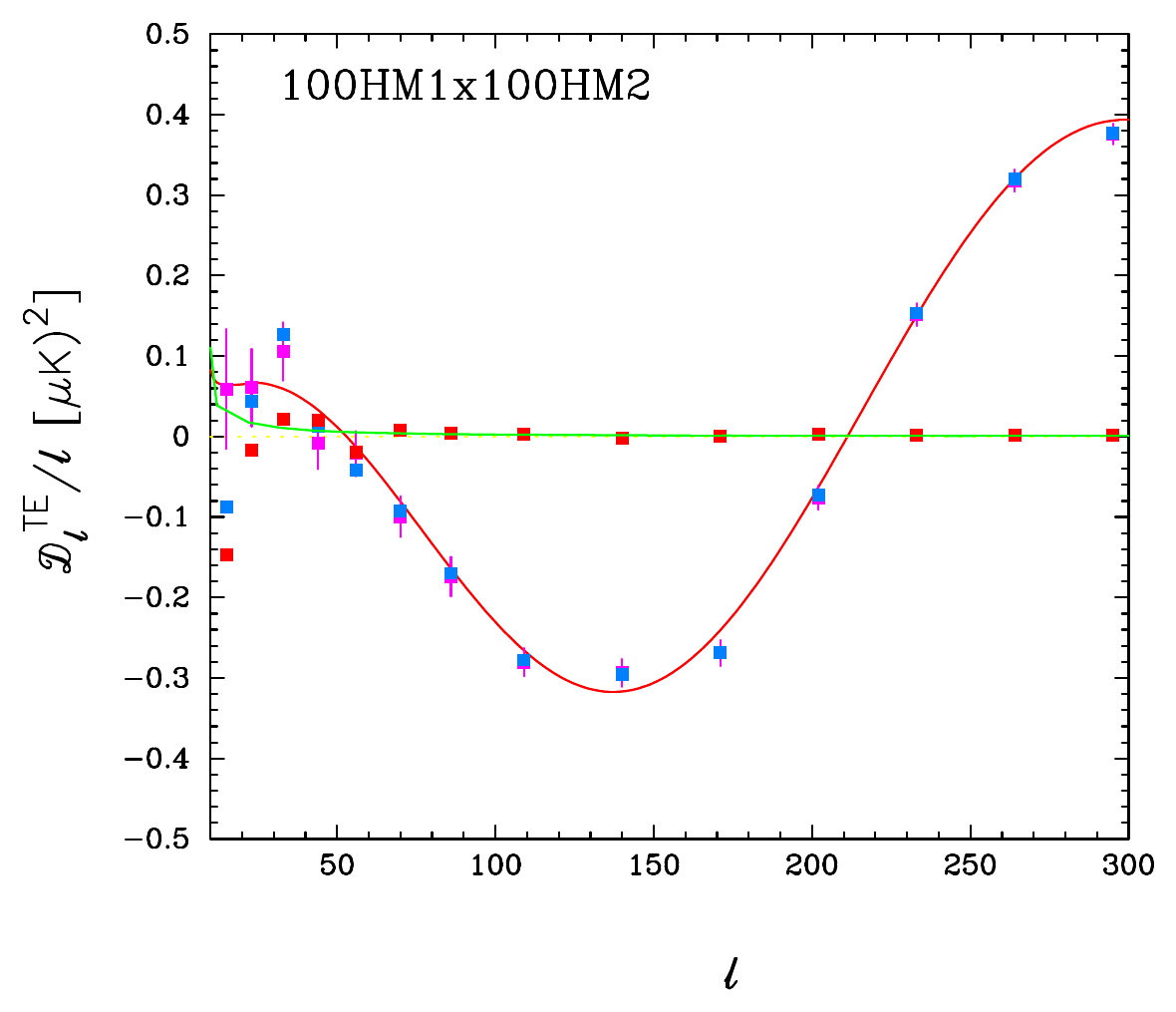} 
\includegraphics[width=52.0mm,angle=0]{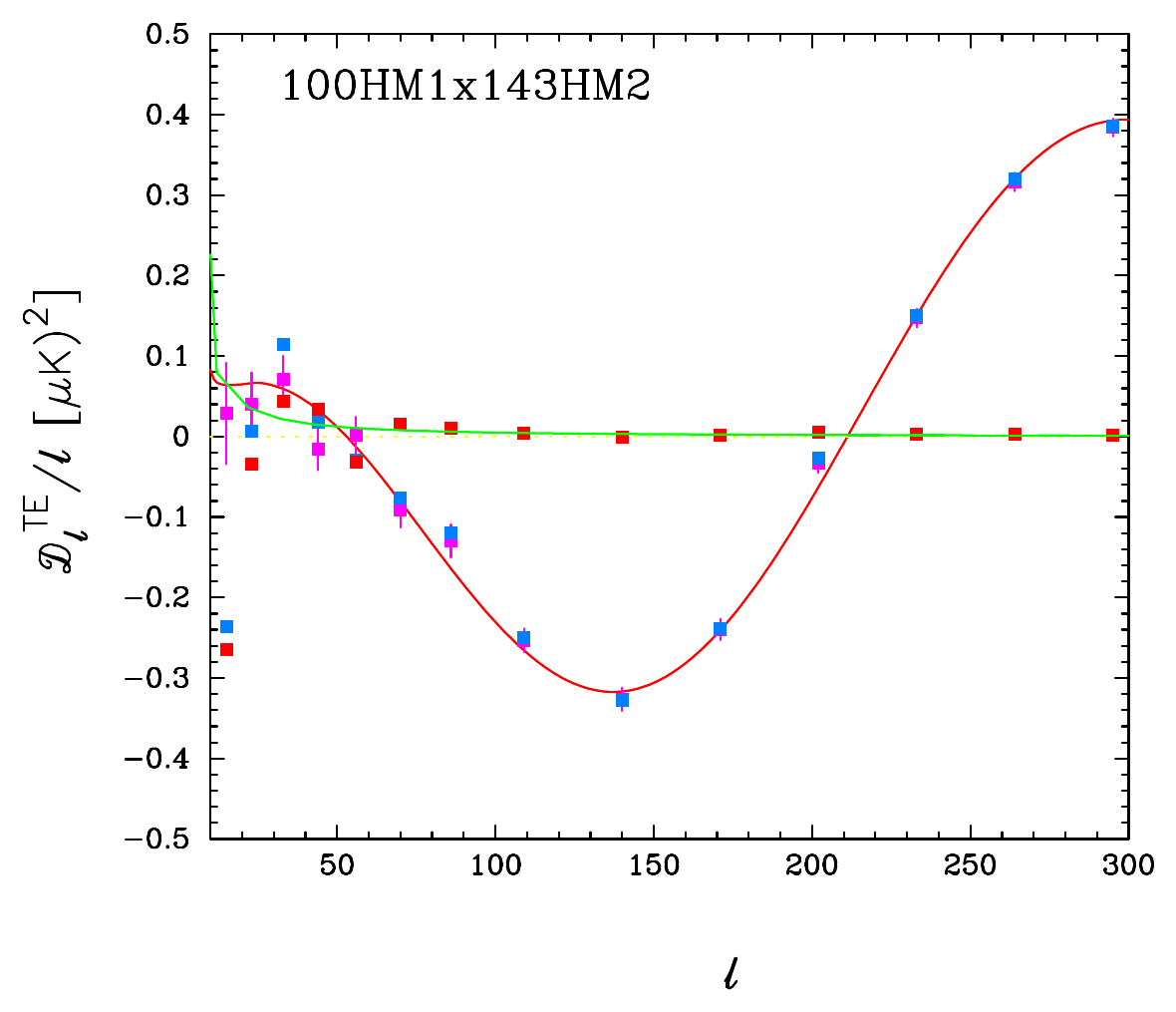} 
\includegraphics[width=52.0mm,angle=0]{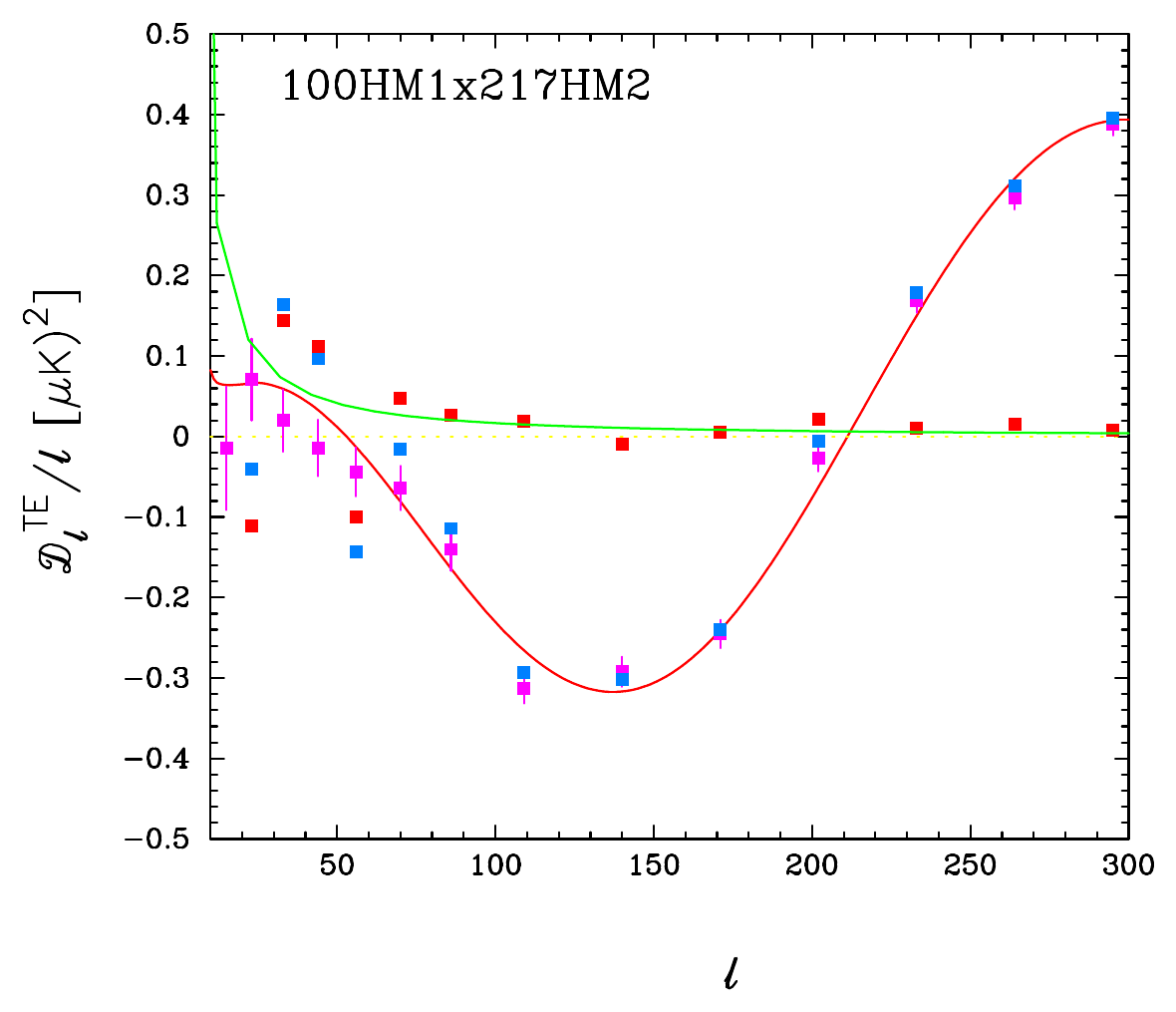} \\
\includegraphics[width=52.0mm,angle=0]{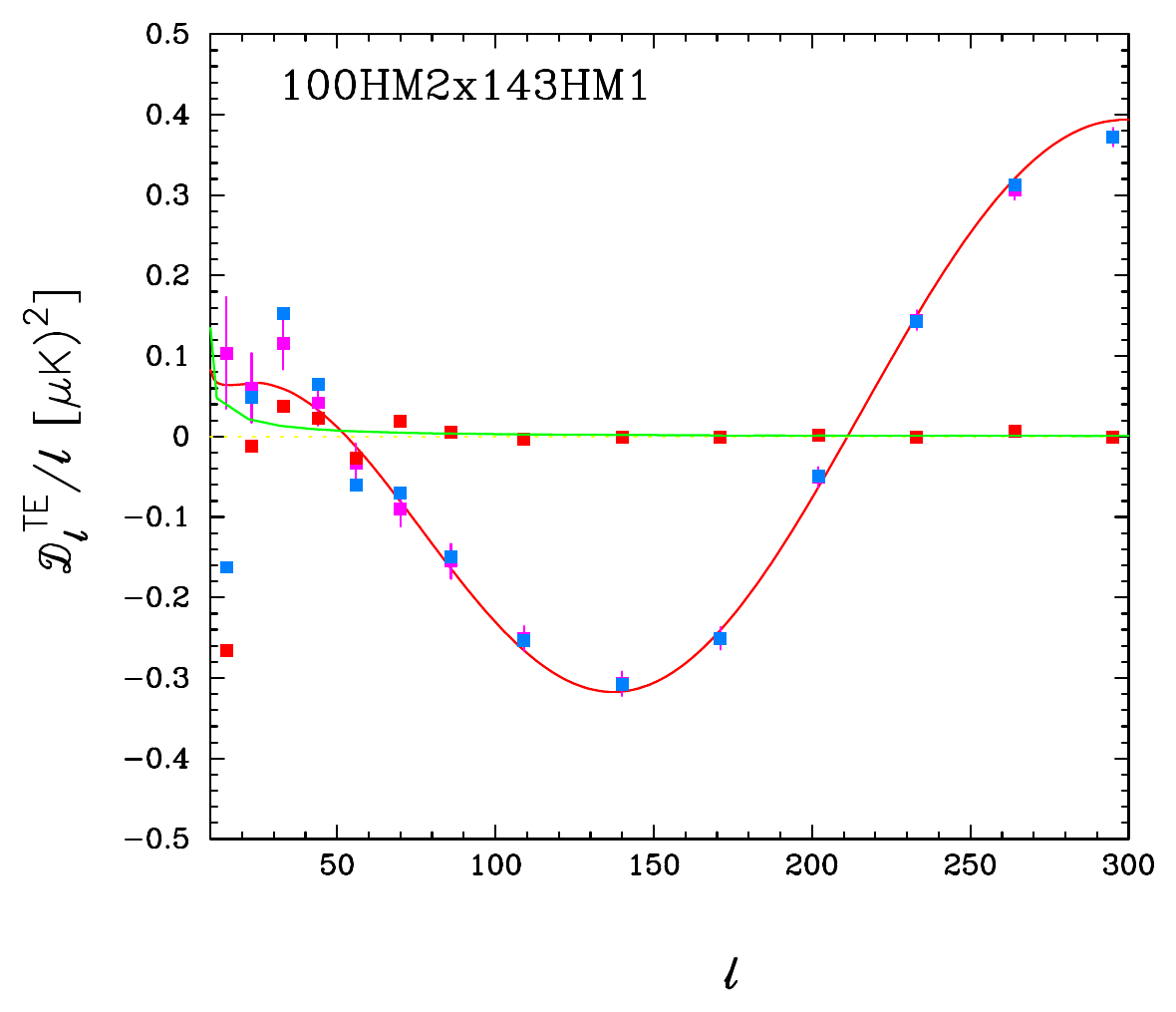} 
\includegraphics[width=52.0mm,angle=0]{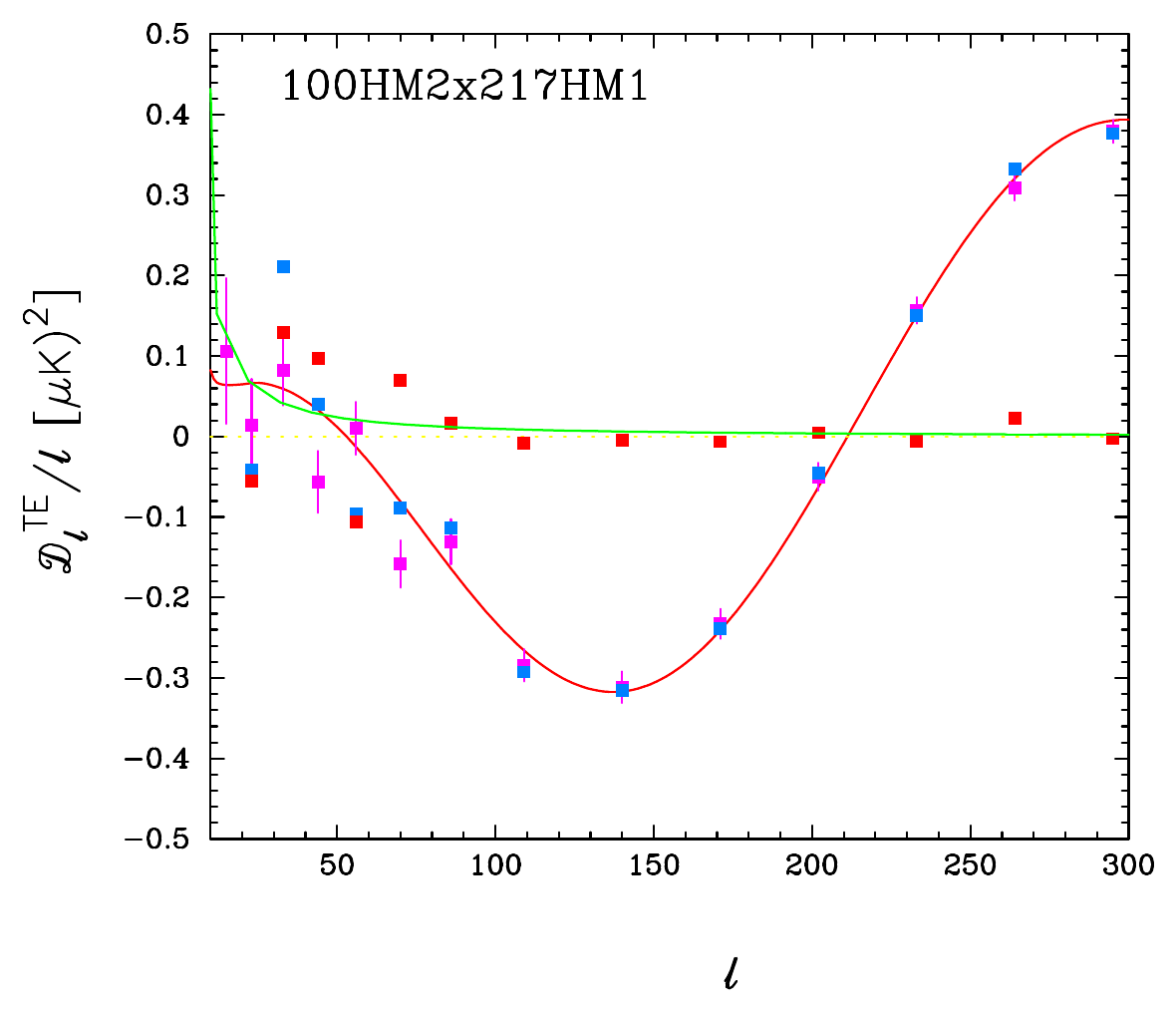} 
\includegraphics[width=52.0mm,angle=0]{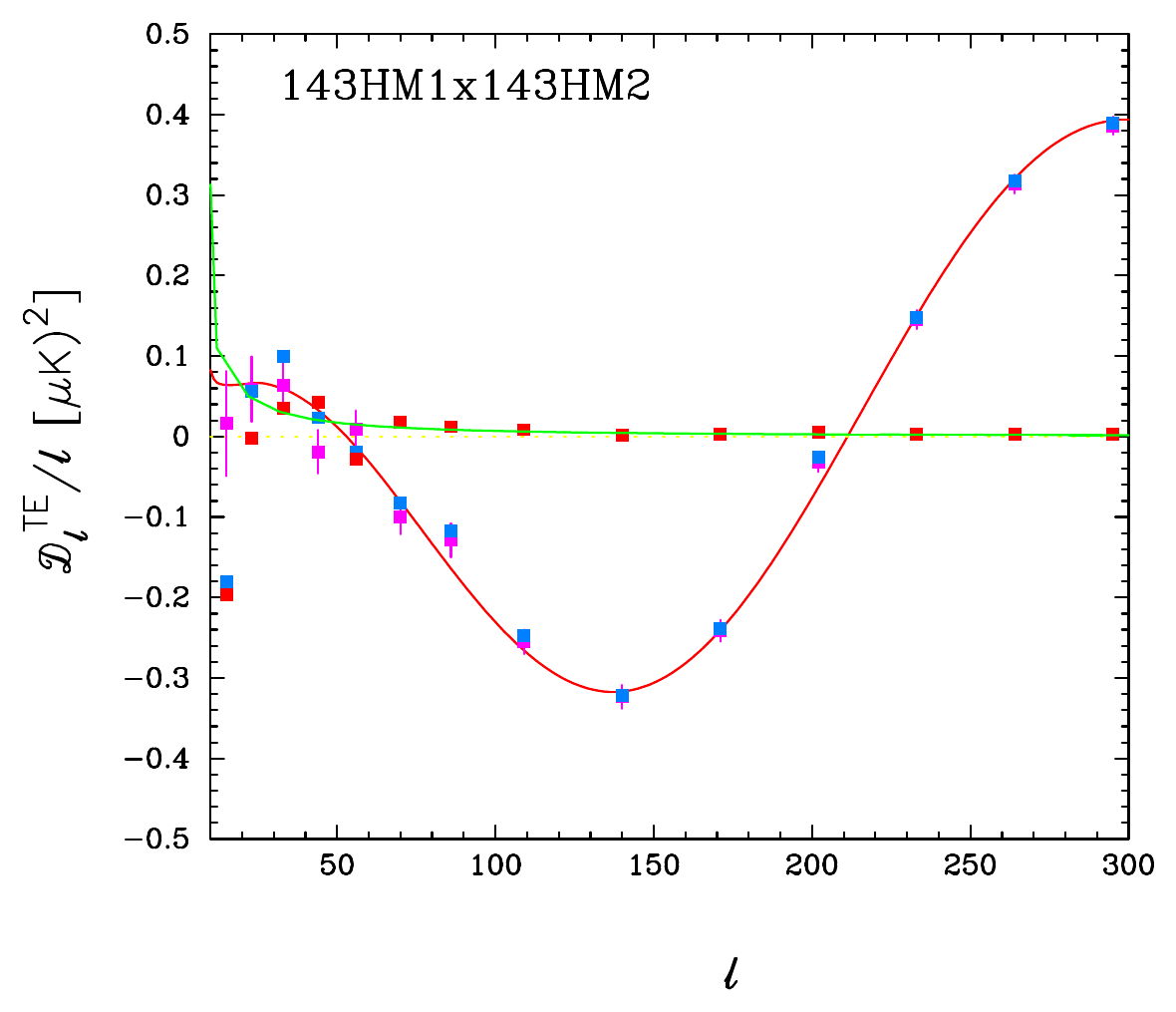} \\
\includegraphics[width=52.0mm,angle=0]{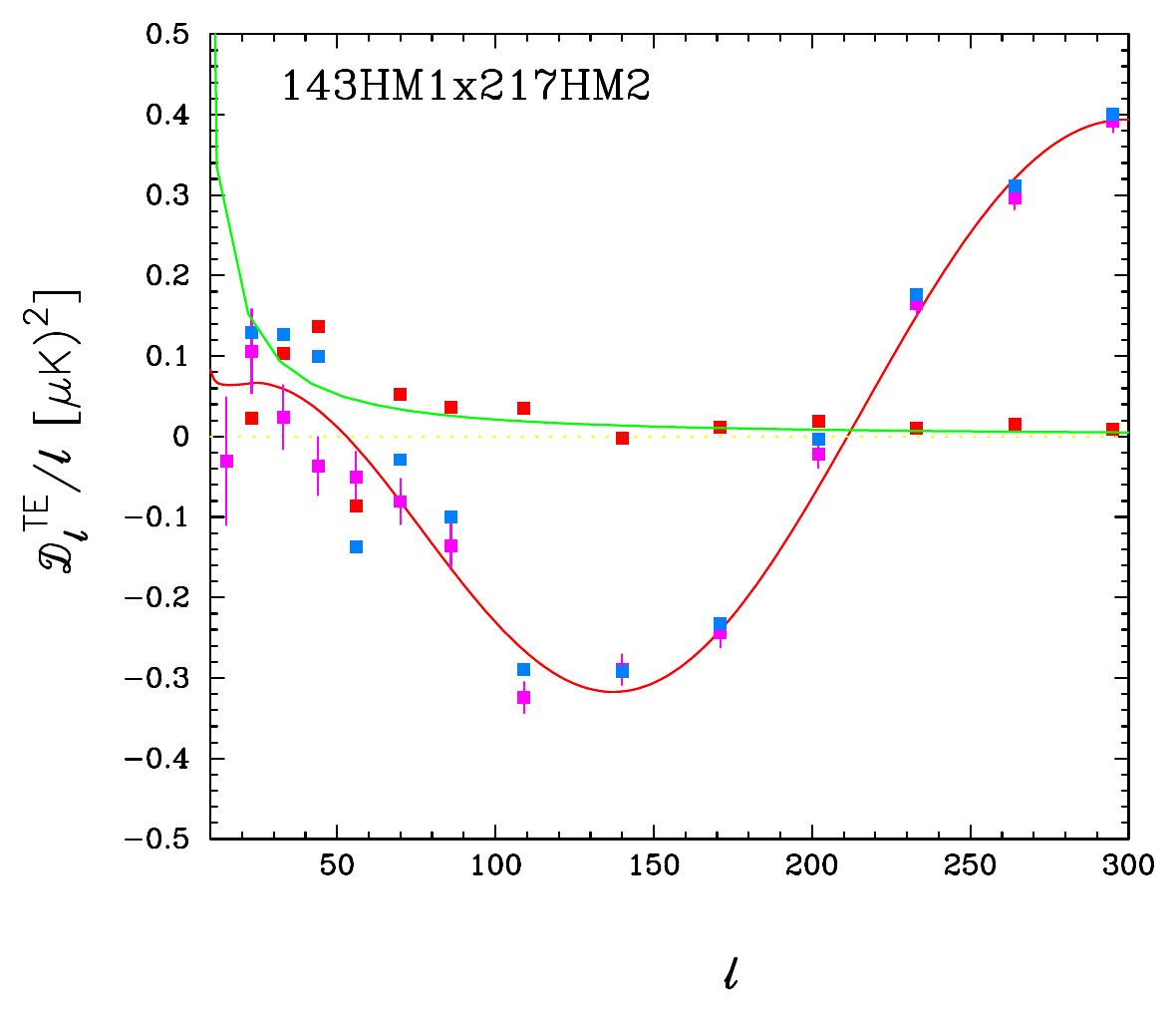} 
\includegraphics[width=52.0mm,angle=0]{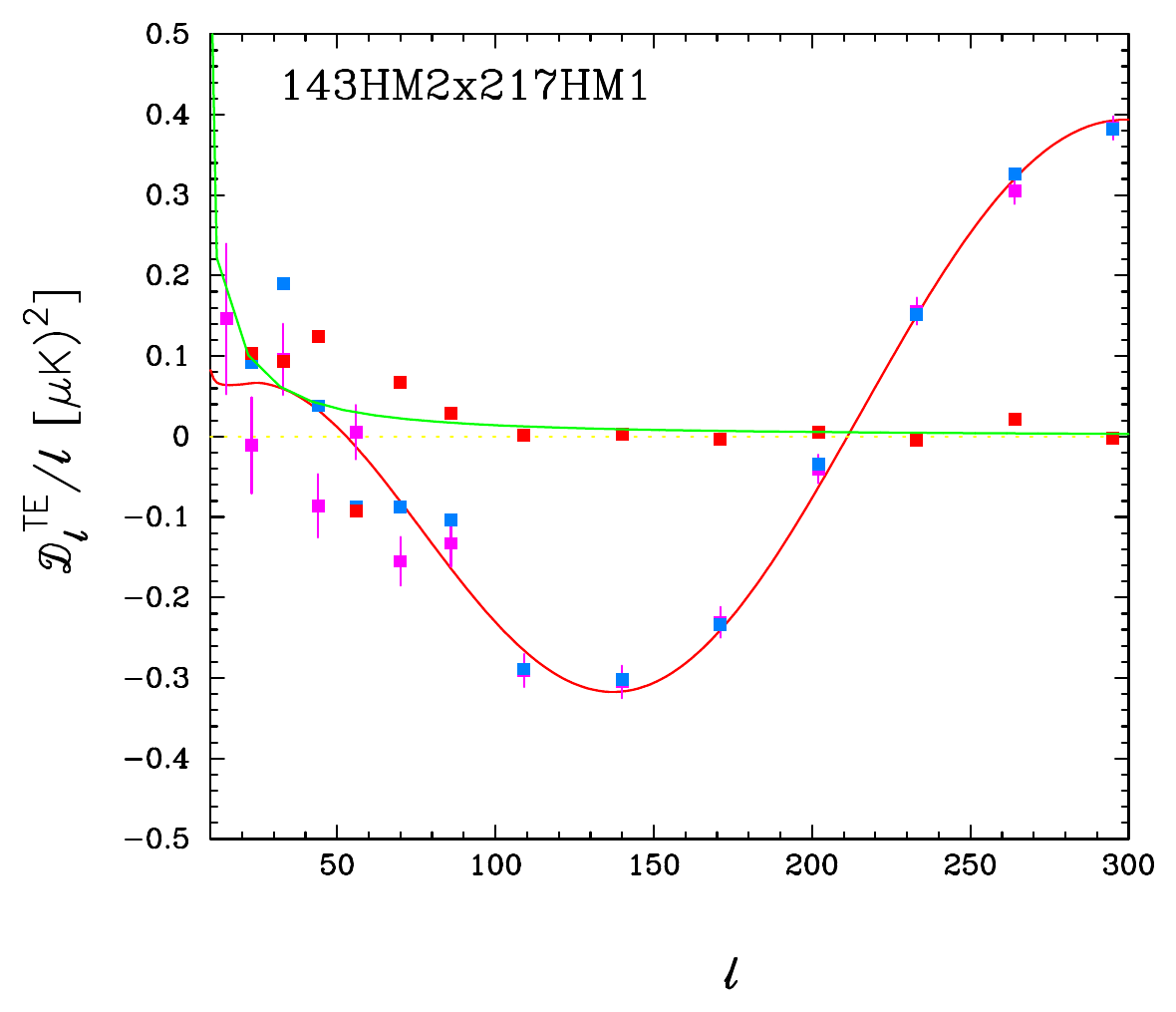} 
\includegraphics[width=52.0mm,angle=0]{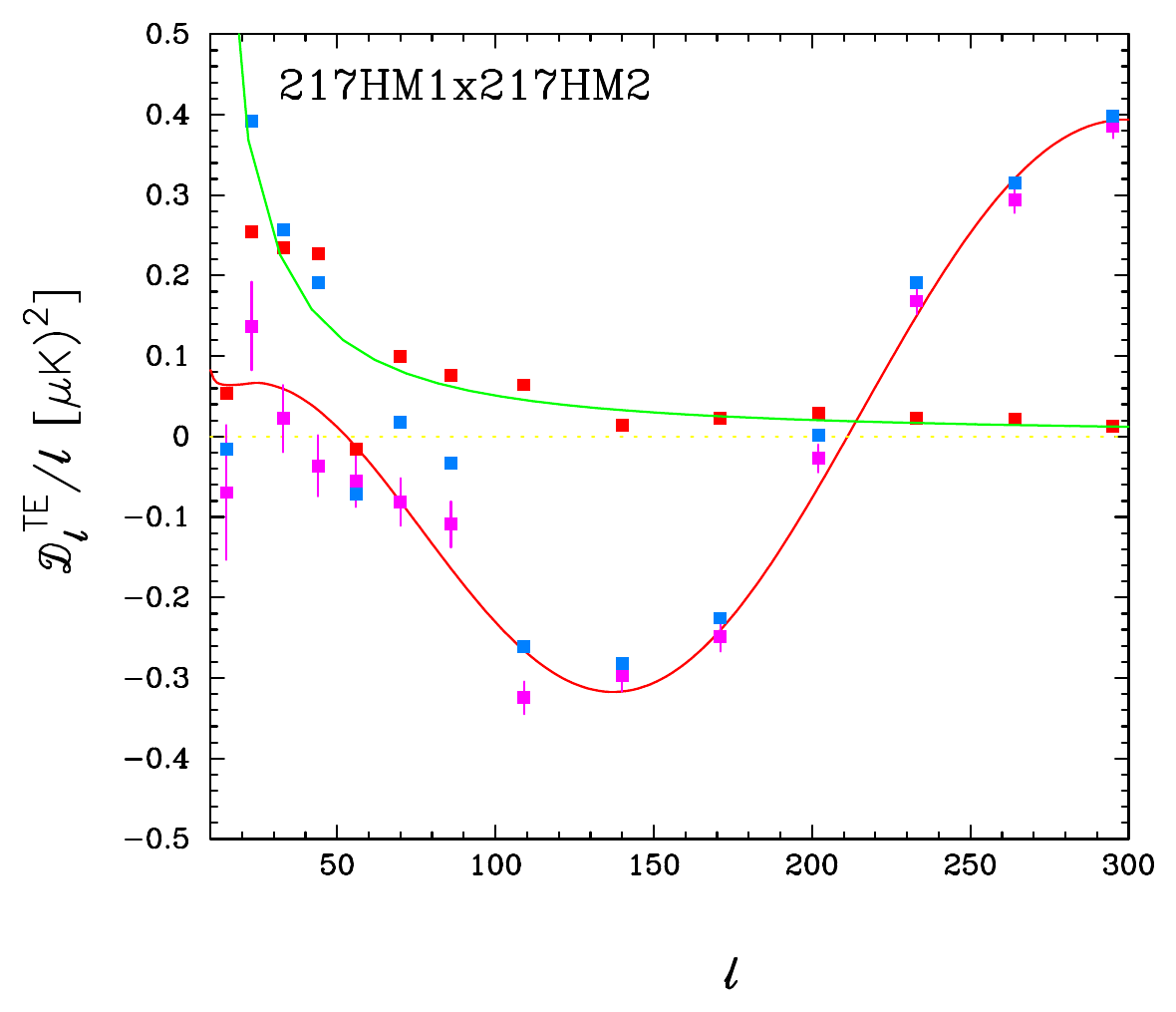} \\

\caption {As for Fig.\ \ref{fig:EEdust_residuals} illustrating the effects
  of $353$ GHz cleaning but for the TE spectra using the 12.1HM
  \camspec\ temperature and polarization masks.  The red lines show
  the TE spectrum for the fiducial base \LCDM\ model. Note that we
  plot ${\cal \hat D}^{TE}/\ell$. }

\label{fig:TEdust_residuals}

\vspace{0.1truein}     

 \end{figure*}

\begin{figure*}
 \centering
\includegraphics[width=52.0mm,angle=0]{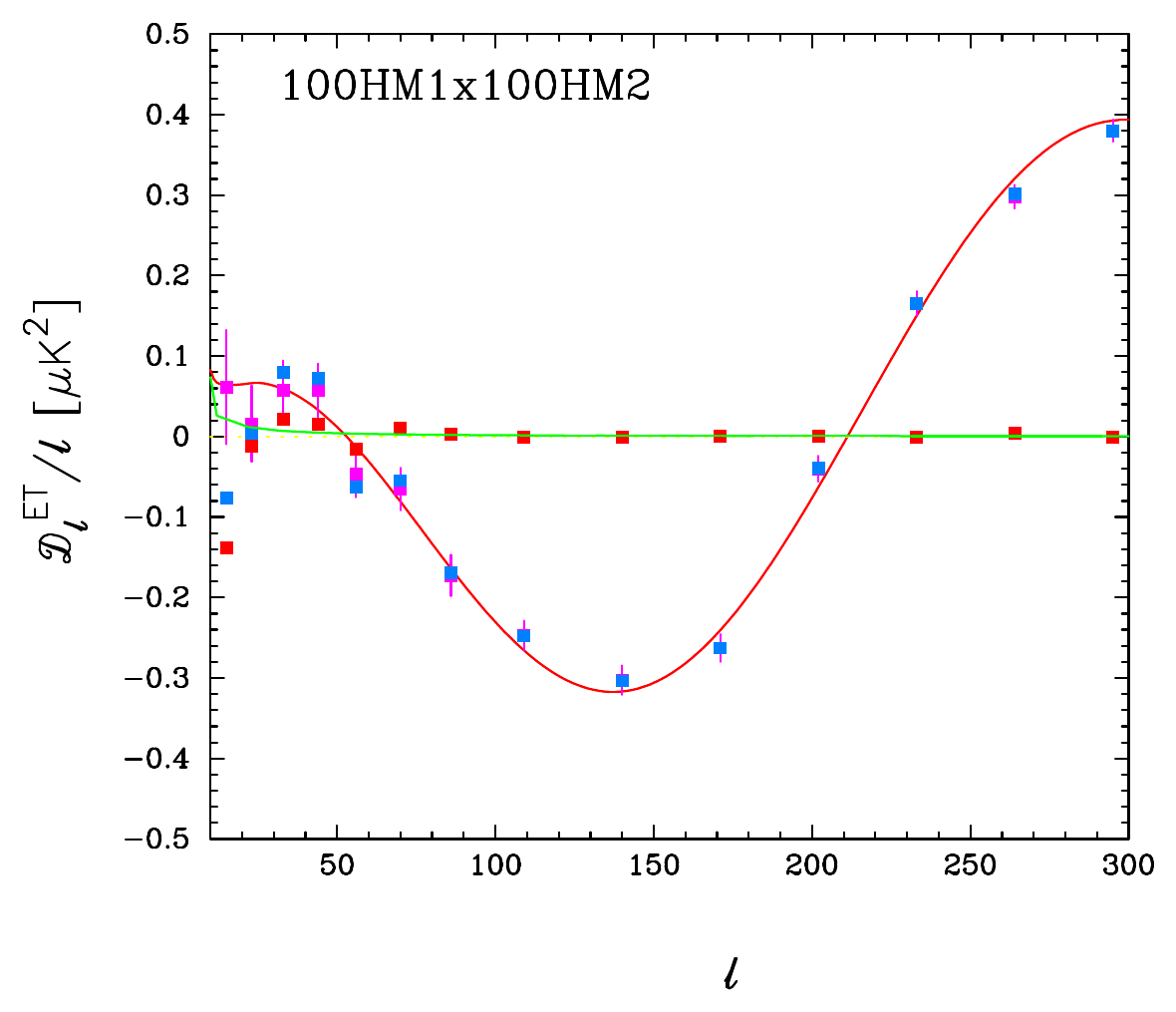} 
\includegraphics[width=52.0mm,angle=0]{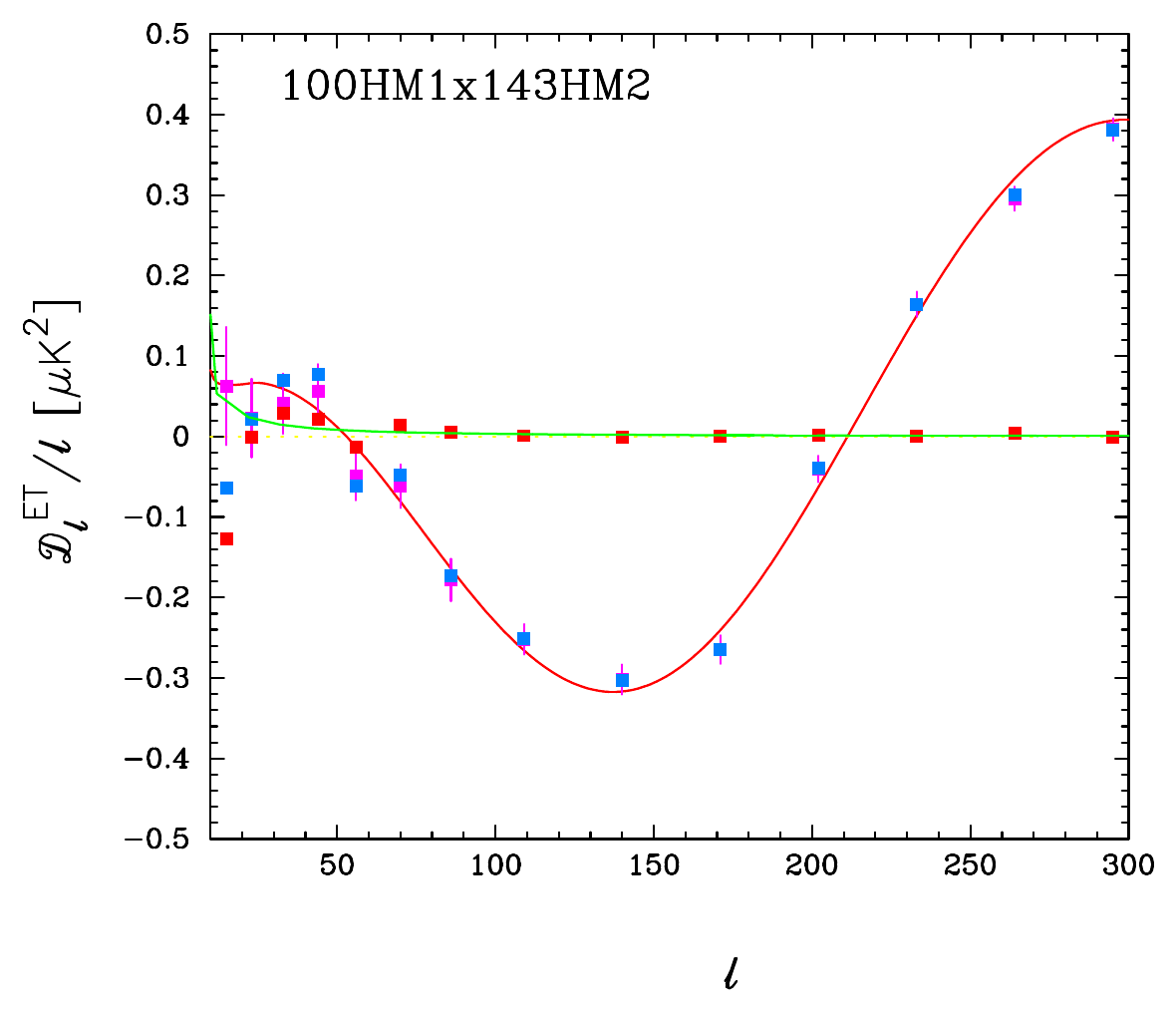} 
\includegraphics[width=52.0mm,angle=0]{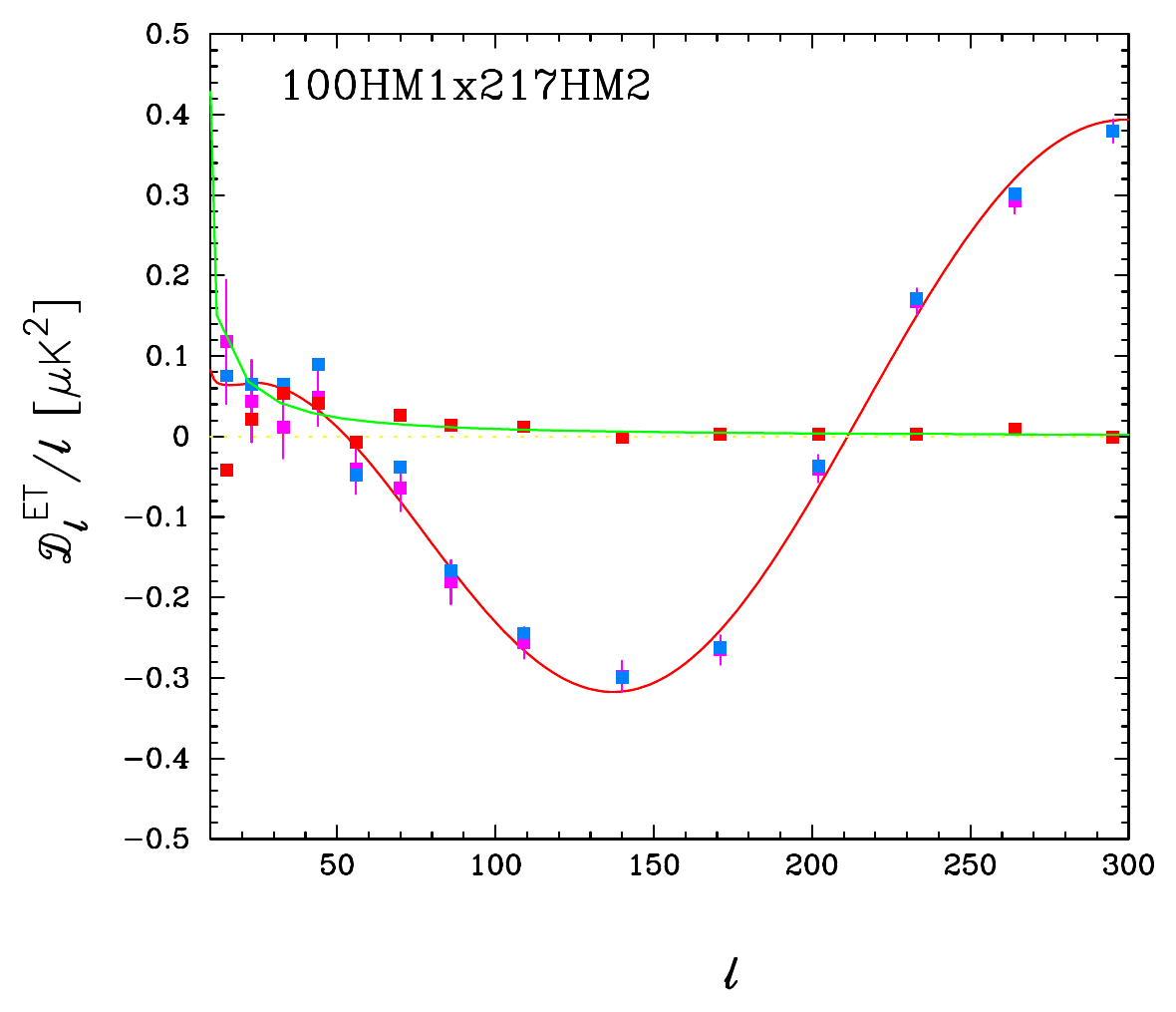} \\
\includegraphics[width=52.0mm,angle=0]{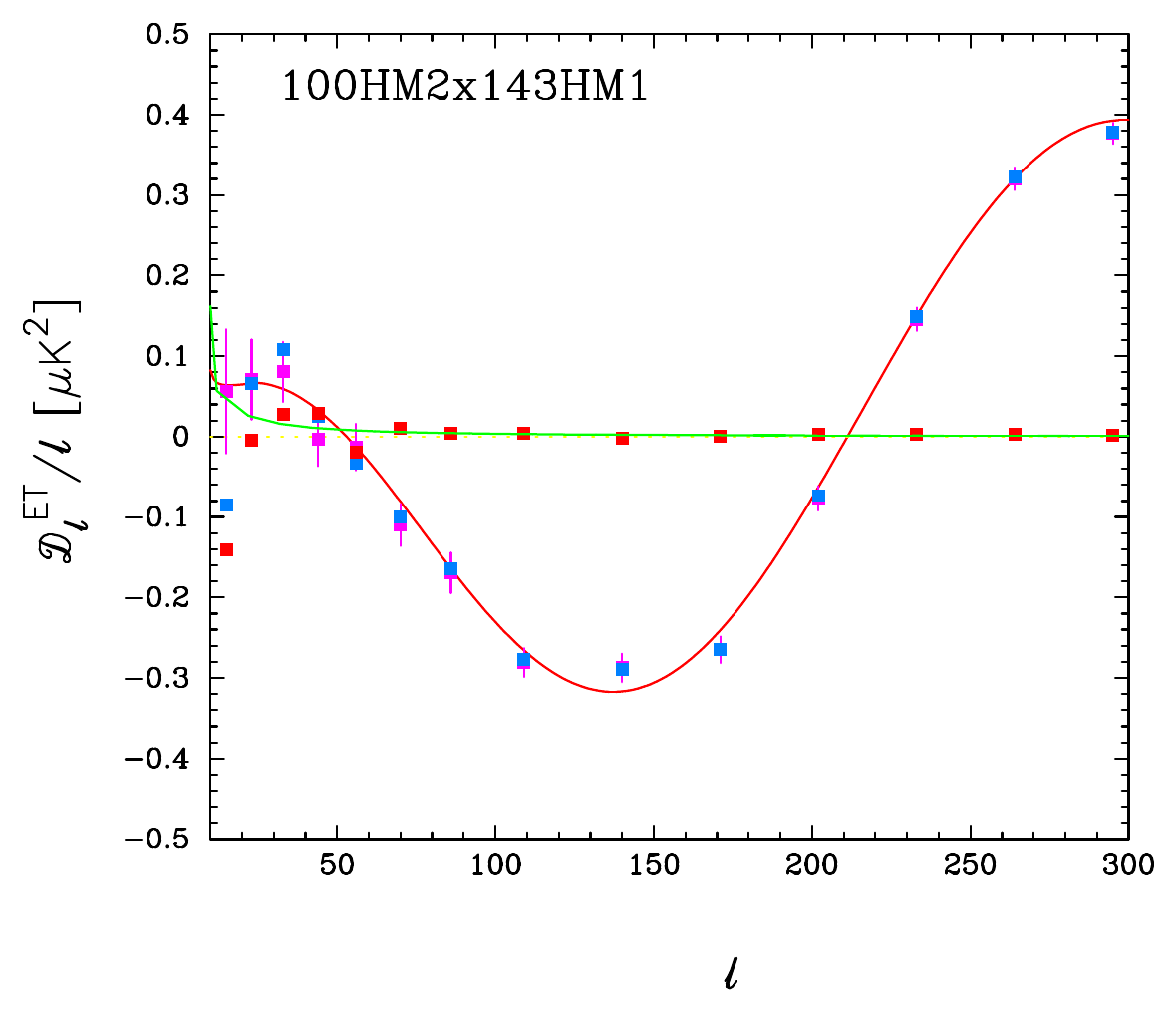} 
\includegraphics[width=52.0mm,angle=0]{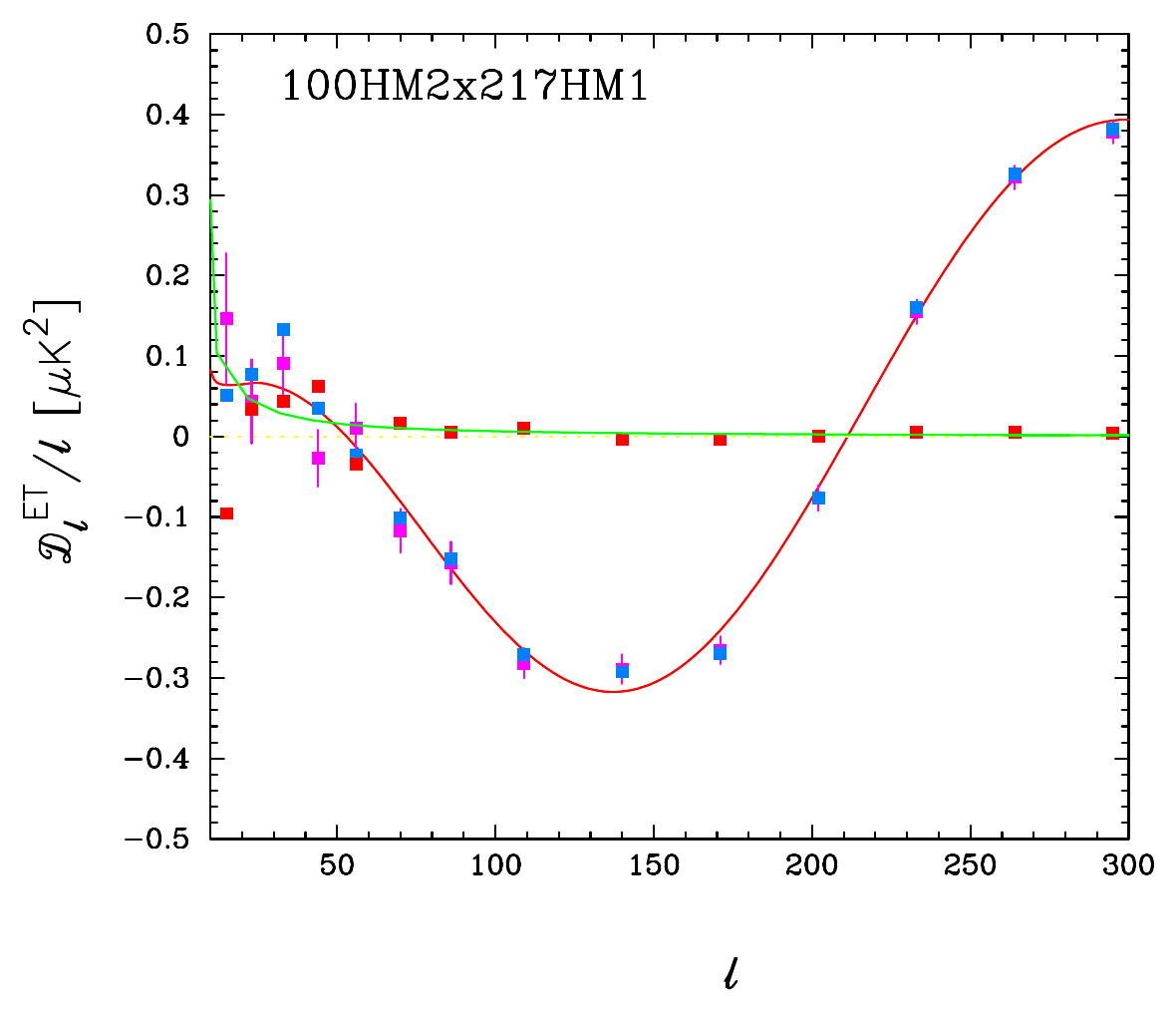} 
\includegraphics[width=52.0mm,angle=0]{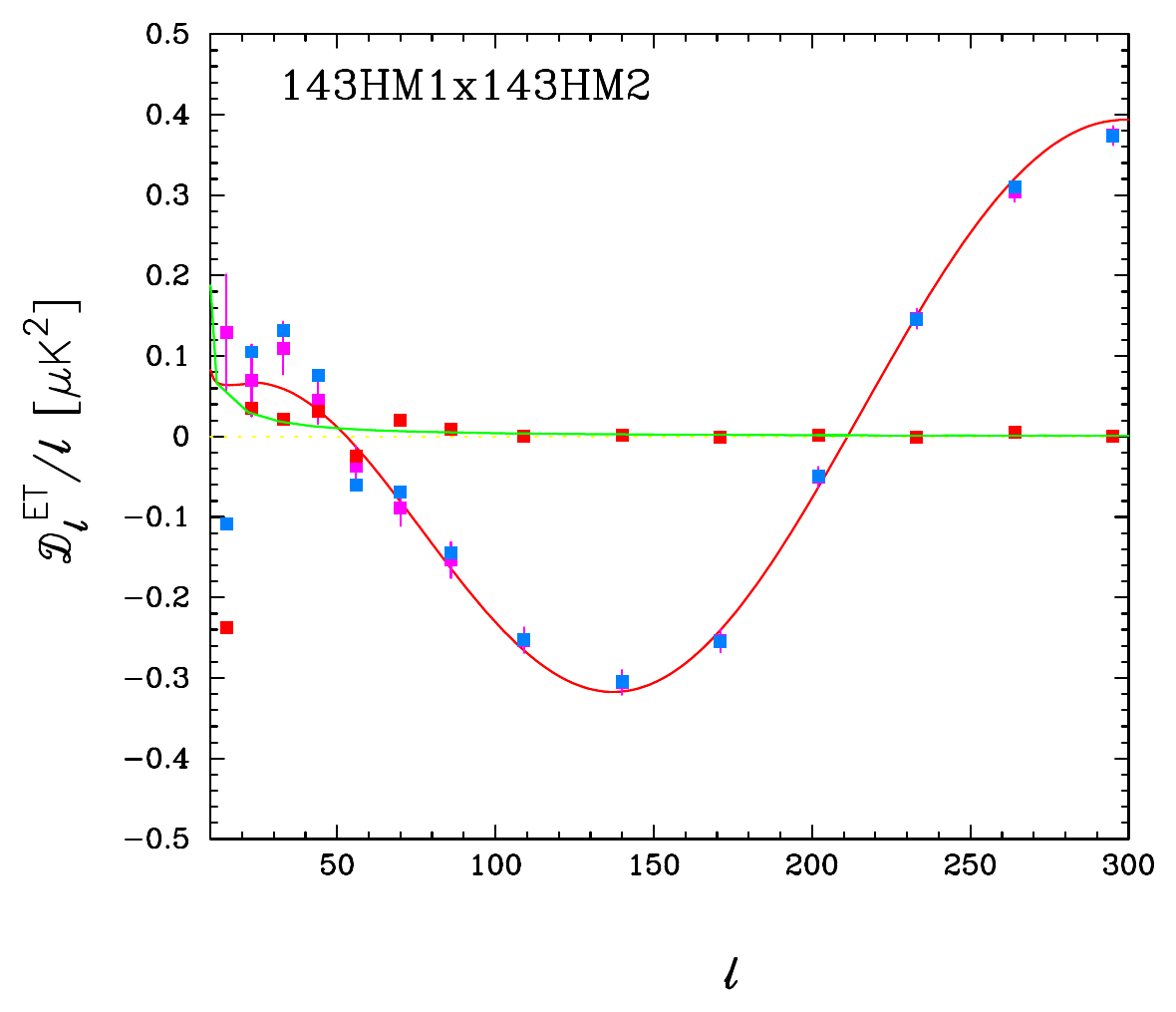} \\
\includegraphics[width=52.0mm,angle=0]{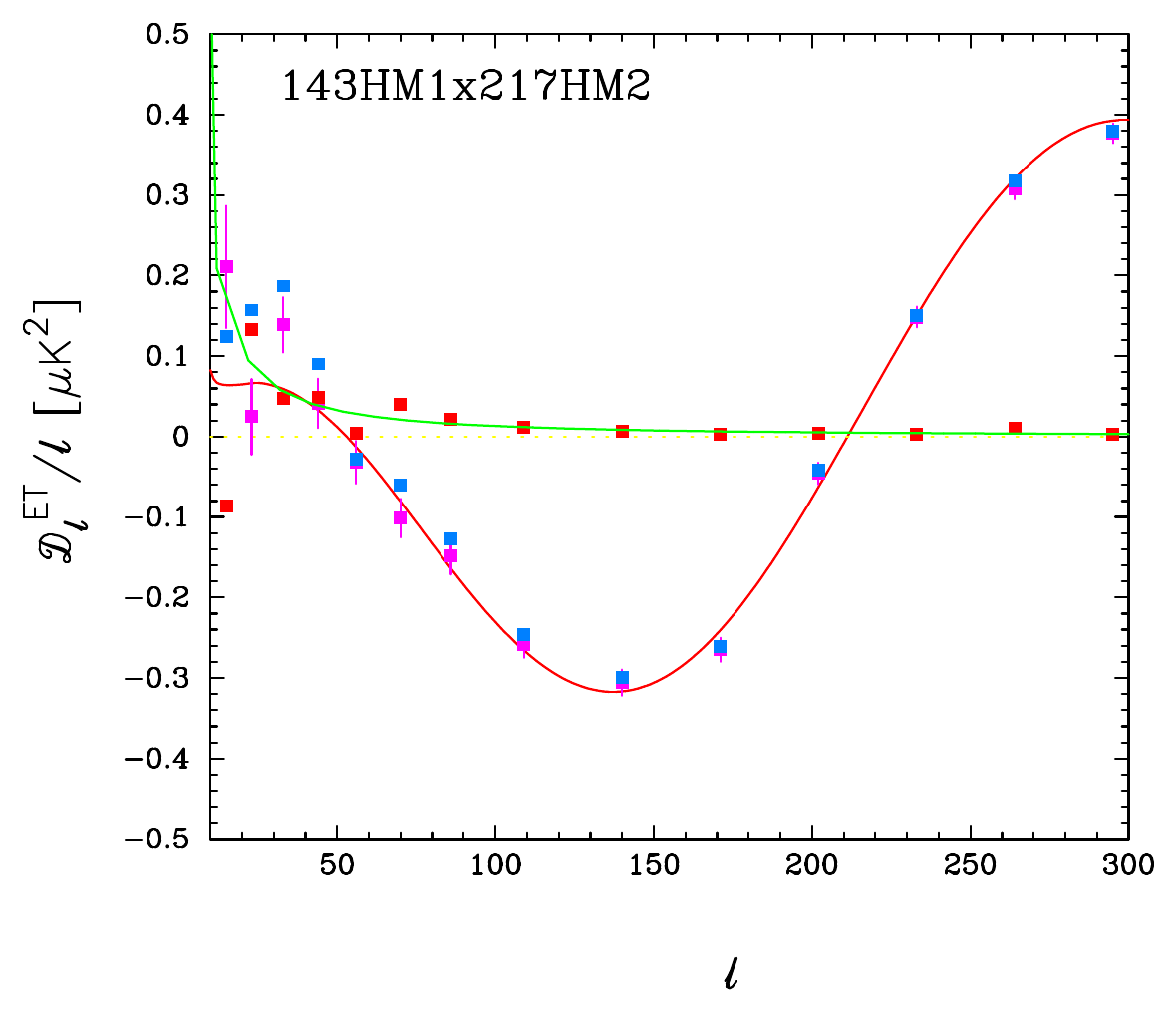} 
\includegraphics[width=52.0mm,angle=0]{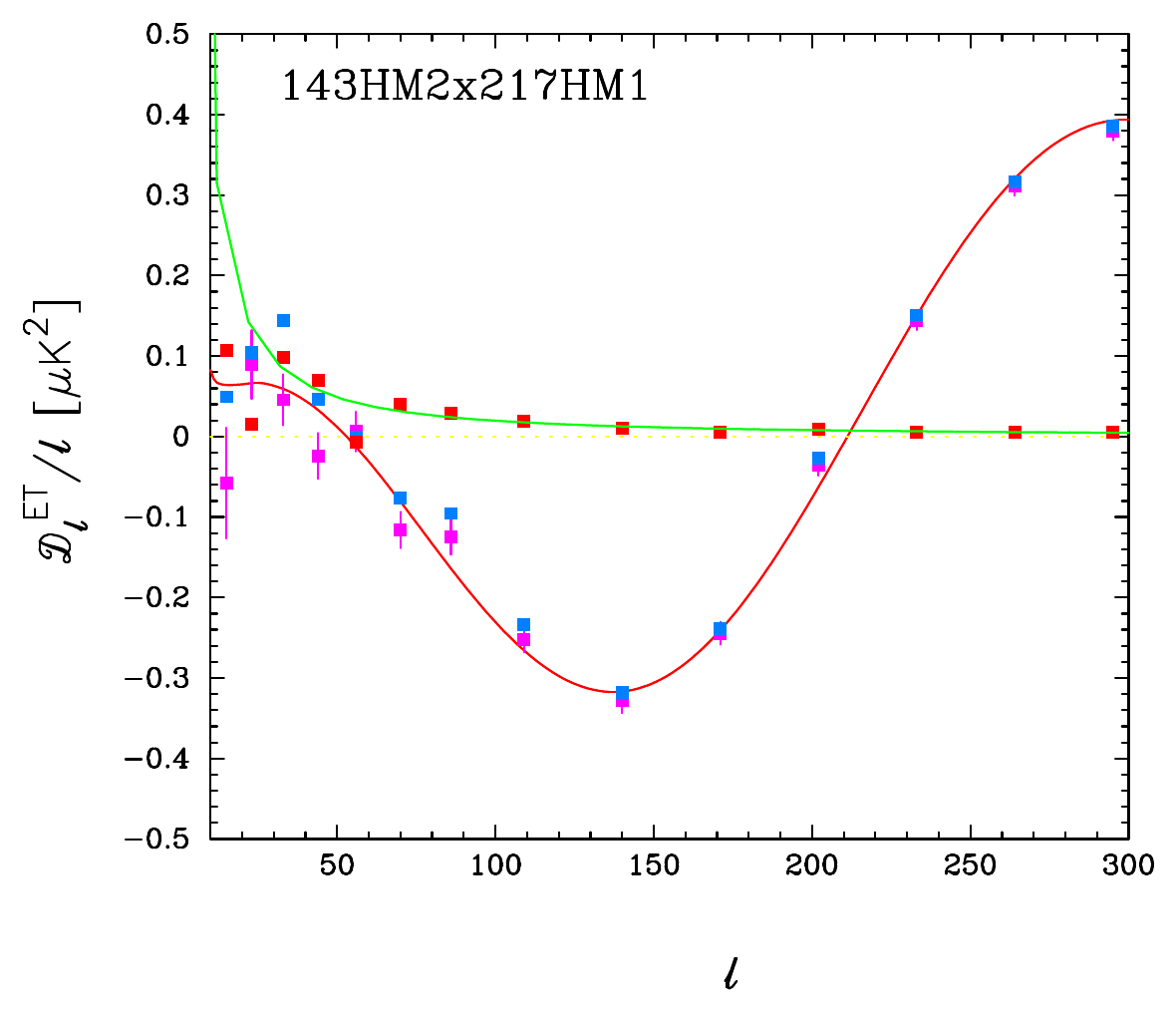} 
\includegraphics[width=52.0mm,angle=0]{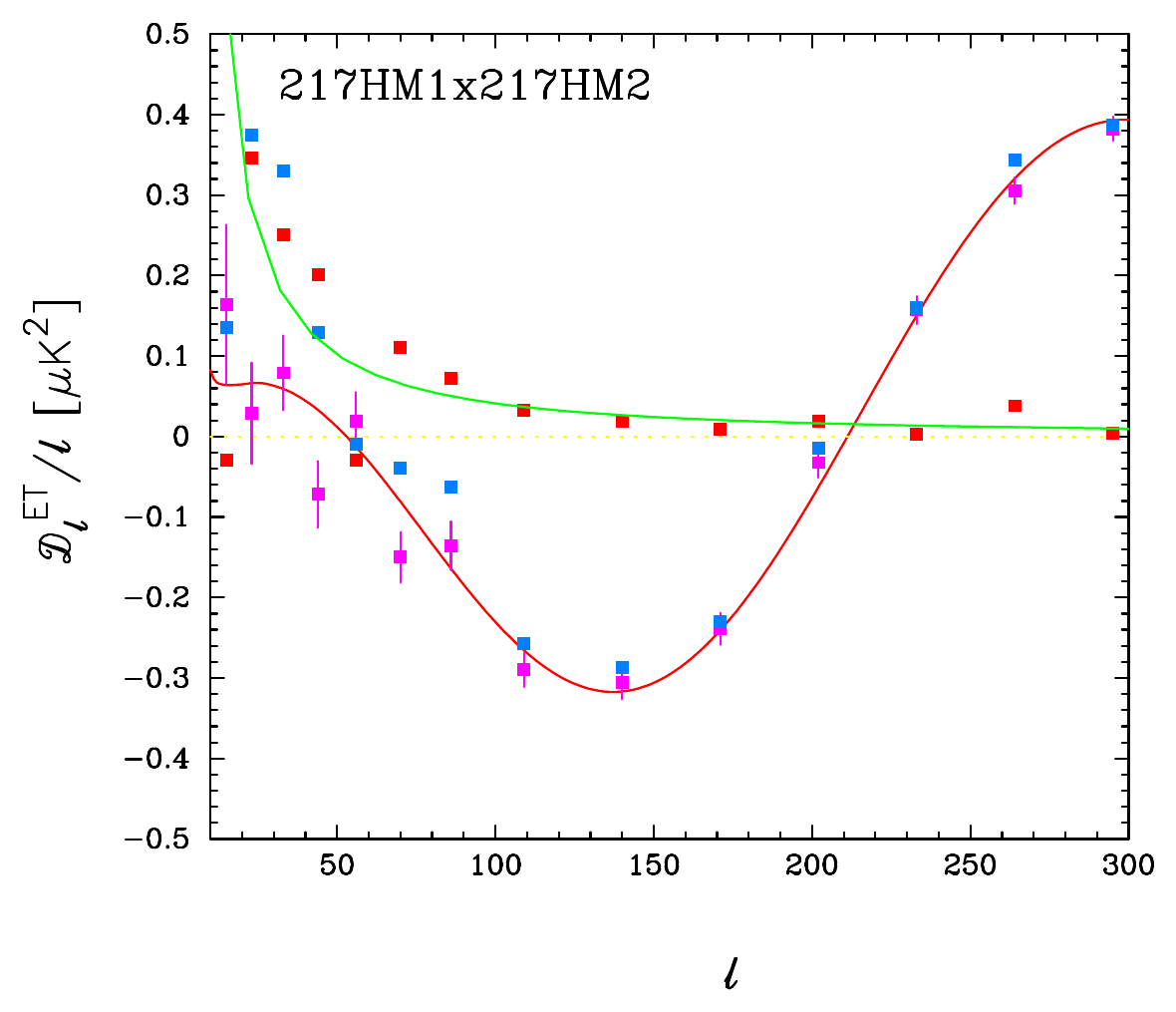} \\

\caption {As for Fig.\ \ref{fig:TEdust_residuals} but for the ET spectra.}

\label{fig:ETdust_residuals}

 \end{figure*}

Figs.\ \ref{fig:EEdust_residuals} - \ref{fig:ETdust_residuals} show the
impact of 353 GHz cleaning on the EE, TE and ET spectra for the 12.1HM
\camspec\ masks. The blue points in each figure show the deconvolved
half mission cross spectra (including corrections for
temperature-to-polarization leakage and polarization efficiencies
descibed in Sect.\ \ref{sec:beams}). The pink points with errors show
the 353 GHz cleaned spectra and the red solid lines show the
theoretical spectra for the 12.1HM TT fiducial base \LCDM\ cosmology. The red
points show the differences between the uncleaned and cleaned spectra,
which quantify the polarized Galactic dust contamination and CMB-dust
cross-correlations in each spectrum.  The green lines show power-laws
(Eq.\ \ref{equ:PD1}) fitted to the red points over the multipole
range $50 \le \ell \le 500$ with parameters listed in Table
\ref{tab:pol_dust_fits}. Since these dust spectra are noisy, we have
imposed Gaussian priors on the spectral indices $\epsilon^{EE},
\epsilon^{TE}$ and $\epsilon^{ET}$ of $\epsilon = -0.30 \pm 0.05$
based on the fits to the $353 \times 353$ spectra
(Eqs.\ \ref{equ:PD2a}, \ref{equ:PD3a} and \ref{equ:PD3b}). From Table
\ref{tab:pol_dust_fits} one can see that there is some sensitivity to
$\epsilon$ in the $217\times 217$ and $143 \times 217$ EE spectra,
but for the rest $\epsilon$ is fixed by the prior. These power law
fits are shown by the green lines in Figs \ref{fig:EEdust_residuals} -
\ref{fig:ETdust_residuals}.

\begin{table}[tp]
{\centering

\caption{\small{Power-law fits (Eq.\ \ref{equ:PD1}) to the polarized dust spectra plotted as the red points in Figs.\
\ref{fig:EEdust_residuals} - \ref{fig:ETdust_residuals}. The fits are performed over the multipole range
$50 \le \ell \le 500$.}}

\label{tab:pol_dust_fits}

\begin{small}

\smallskip

\smallskip

\hspace{14mm}\begin{tabular}{|c|c|c|c|c|c|c|} \hline
spectrum   & $A^{EE}$ [$\mu{\rm K}^2$] & $\epsilon^{EE}$ & $A^{TE}$ [$\mu{\rm K}^2$] & $\epsilon^{TE}$ & $A^{ET}$ [$\mu{\rm K}^2$] & $\epsilon^{ET}$  \\  \hline
217HM1x217HM2  &   $1.806 \pm 0.067$ & $-0.31 \pm 0.03$ &  $4.18 \pm 0.61$  & $-0.31 \pm 0.05$ & $3.41 \pm 0.61$ & $-0.31 \pm 0.05$  \\
143HM2x217HM1  &   $0.557 \pm 0.033$ & $-0.31 \pm 0.04$ &  $1.14 \pm 0.46$  & $-0.30 \pm 0.05$ & $1.52 \pm 0.22$ & $-0.31 \pm 0.05$  \\
143HM1x217HM2  &   $0.582 \pm 0.030$ & $-0.32 \pm 0.04$ &  $1.71 \pm 0.52$  & $-0.30 \pm 0.05$ & $1.00 \pm 0.24 $ & $-0.30 \pm 0.05$  \\
143HM1x143HM2  &   $0.172 \pm 0.010$ & $-0.33 \pm 0.04$ & $0.56 \pm 0.12$  & $-0.30 \pm 0.05$ & $0.34 \pm 0.15$ & $-0.30 \pm 0.05$  \\
100HM2x217HM1  &   $0.263 \pm 0.026$ & $-0.29 \pm 0.05$ & $0.79 \pm 0.31$  & $-0.30 \pm 0.05$ & $0.50 \pm 0.18$ & $-0.30 \pm 0.05$  \\ 
100HM2x143HM1  &   $0.084 \pm 0.008$ & $-0.30 \pm 0.05$ & $0.24 \pm  0.12$  & $-0.30 \pm 0.05$ & $0.29 \pm 0.08$ & $-0.30 \pm 0.05$  \\
100HM1x217HM2  &   $0.265 \pm 0.026$ & $-0.27 \pm 0.05$ & $1.35 \pm 0.22$  & $-0.30 \pm 0.05$ & $0.73 \pm 0.20$ & $-0.30 \pm 0.05$  \\
100HM1x143HM2  &   $0.078 \pm 0.008$ & $-0.29 \pm 0.05$ & $0.40 \pm 0.13$  & $-0.31 \pm 0.05$ & $0.27 \pm 0.08$ & $-0.30 \pm 0.05 $  \\
100HM1x100HM2  &   $0.036 \pm 0.005$ & $-0.28 \pm 0.05$ & $0.20 \pm 0.06$  & $-0.30 \pm 0.05$ & $0.13 \pm 0.05$ & $-0.30\pm 0.05$  \\ \hline
\end{tabular}
\end{small}}

\smallskip

\smallskip

\end{table}

For the EE spectra, polarized Galactic dust emission is a major
contaminant at low multipoles. However, polarized dust emission is
very accurately removed by $353$ GHz cleaning, even when the
foreground contamination exceeds the CMB signal by an order of
magnitude or more. {\it We see no evidence for any decorrelation of
polarized Galactic dust emission over the frequency range $100-353$
GHz} \citep[see also][]{Planck_dust_pol:2018}. The excess in the
cleaned $217\times 217$ spectrum at $\ell \simlt 50$ visible in
Fig.\ \ref{fig:EEdust_residuals} is caused primarily by systematic errors in the
$217$   GHz Q and U maps (see Fig.\ \ref{fig:pol_cleaned_maps}), not by errors in dust
subtraction. Additional evidence of systematic errors in the \SROLL\ maps
at low multipoles is presented in Sect.\ \ref{sec:inter_frequency_pol}.

The approach that we adopt in the \camspec\ likelihood is
to apply conservative multipole cuts, as listed in Table
\ref{tab:multipoleranges}. This avoids using polarization data at
multipoles where the spectra are heavily dominated by Galactic dust
emission. We use the 353 GHz cleaned spectra up to a multipole $\ell_{\rm  clean} = 150$ 
and subtract the power-law fits of Table \ref{tab:pol_dust_fits} 
from the uncleaned spectra at higher
multipoles.  We chose $\ell_{\rm clean}=150$ to limit the impact of
$353$ GHz noise on the cleaned spectra.  With these choices, polarized
Galactic dust emission and dust-CMB cross-correlations are removed
accurately from the polarization spectra. There is therefore no
need to model errors in  polarized dust subtraction in the \camspec\  likelihoods.

  Figs.\ \ref{fig:TEdust_residuals} and \ref{fig:ETdust_residuals}
  show that Galactic dust emission makes a very small contribution to
  the TE/ET spectra at $\ell \simgt 150$. At lower multipoles,
  CMB-dust cross-correlations become important and the dust corrections
  can be positive or negative. At these low multipoles, assuming a
  power-law dust spectrum (as is done in the \plik\ likelihood) is a
  very poor approximation and could potentially lead to biases in the
  TE likelihood.

In summary, by cleaning the polarization spectra at lower
frequencies using $353$ GHz maps, 
we produce spectra free of polarized dust emission. Since
there are no other frequency dependent foregrounds in the cleaned
polarization spectra, these are then coadded (after correction for
effective polarization efficiencies and small corrections for
temperature-to-polarization leakage) to produce a single EE and a
single TE spectrum which are used in the \camspec\ likelihoods. Thus
no nuisance parameters are required in the \camspec\ likelihoods to
describe foreground emission, though we include overall relative
calibration parameters $c^{TE}$ and $c^{EE}$, mainly as a consistency check of the likelihood,  as described in
Sect.\ \ref{sec:nuisance}. Comparisons of cosmological parameters
determined seperately from the TT, TE and EE likelihood blocks therefore provide
an important additional consistency check of systematics in the \Planck\
data and errors in  the TT foreground model.

\section{Nuisance parameters}
\label{sec:nuisance}

The \planck\ likelihoods fit parameters describing foreground power
spectrum templates and instrumental nuisance parameters at the
Monte-Carlo Markov Chain (MCMC) sampling stage\footnote{Throughout this paper we use the \COSMOMC\ sampler \citep{Lewis:2002} developed and maintained by Antony Lewis
({\tt https://cosmologist.info.cosmomc}). The version of \COSMOMC\ is almost identical to that used in PCP18.}. This approach has been adopted by ground-based
experiments \citep[e.g.][]{Dunkley:2011, Reichardt:2012}. The foreground/nuisance model used for the \Planck\ likelihoods has been described in 
described in previous \Planck\ papers PPL13, PPL15 and PPL18. For 
 completeness, in this section  we summarize the model adopted in this paper, 
detailing changes that we have made to the model since PCP18.

\subsection{Instrumental nuisance parameters}

\subsubsection{Inter-frequency relative calibrations in temperature}
\label{subsubsec:inter_frequency}

The discussion presented in Sect.\ \ref{subsec:intra_frequency} shows
that the relative calibrations of detset temperature spectra at fixed
frequency are consistent to better than about $0.1 \%$.  As described
in \citep{DataProcessing:2018}, the absolute calibrations of the
$100-217$ GHz frequency maps, based on the orbital dipole, are much
more accurate than this. Our interpretation of these small residual
`calibration' differences is that they reflect small transfer function
errors arising from the modelling of the beams beyond $100^\prime$ of
the beam axis. These errors can be absorbed to high accuracy by a
single multiplicative factor (i.e.  an effective calibration
factor). Given that we see such effects between detectors within a
frequency band, we would expect to find small effective calibration
differences between frequencies.  As demonstrated in
Sect.\ \ref{subsec:intra_frequency}, it is relatively straightforward
to measure small intra-frequency calibration factors accurately, since
each detector within a frequency band sees the same sky (apart from
small band-pass differences). It is more difficult to test effective
calibrations between frequency bands because the foregrounds are strongly
frequency dependent.

In previous versions of \camspec\ we have determined relative
calibrations between frequencies jointly with the foreground and
cosmological parameters.  We (arbitrarily) chose the $143\times 143$ spectrum as the
reference and solved for two relative calibration factors $c_{100}$,
$c_{217}$, setting $c_{143 \times 217} = \sqrt{c_{100}c_{217}}$. These
calibration factors multiply the data spectra (though as discussed in
PPL13, in the likelihood we divide the theory power spectra by these
factors when comparing to the \Planck\ spectra).

\begin{figure*}
\centering
\includegraphics[width=110mm,angle=0]{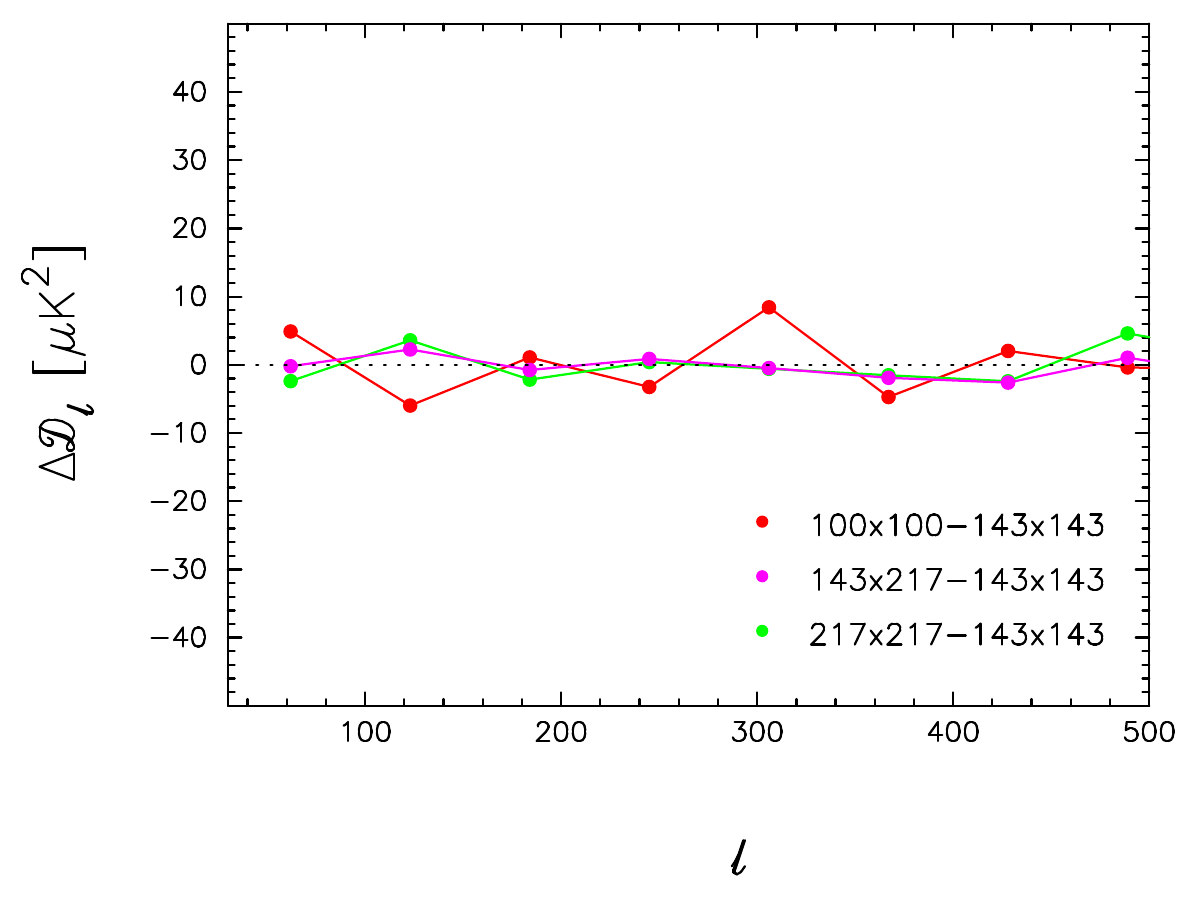}
\caption {Differences between recalibrated $545$ GHz cleaned
  half mission temperature cross spectra and the $143 \times 143$
  $545$ GHz cleaned spectrum. The relative calibration coefficients
  are given in Eqs. \ref{equ:Nu2a}- \ref{equ:Nu2c}.}

\label{fig:pgcalib}

\vspace{0.10 truein}
 \end{figure*}

However, since PCP18, we have developed a method to measure
inter-frequency relative calibrations to high accuracy.
Fig.\ \ref{fig:pgcalib} shows the differences between recalibrated
$100 \times 100$, $143 \times 217$ and $217 \times 217$ $545$ GHz
cleaned half mission cross spectra and the $143 \times 143$ $545$ GHz
cleaned cross spectra. We use mask30 together with the $217$ GHz point
source mask (and also eliminating missing pixels from all
frequencies). Subtraction of $143 \times 143$ eliminates cosmic
variance and by using mask30 and $545$ cleaning, we eliminate
contamination by Galactic dust and supress frequency-dependent
variations caused by foregrounds and CMB/foreground cross
correlations. We recalibrate the spectra by minimising
\begin{equation}
 \chi^2 =  \sum_k { (c_{xy} \hat D^{xy}_k - \hat D^{143 \times 143}_k )^2 \over \sigma^2_{xy} }, \label{equ:Nu1}
\end{equation}
for each spectrum $\hat D^{xy}$ over bandpowers $k$ within the range $50 \le l \le 500$. (For the $100 \times 100$
spectrum we also make a small correction for point sources and the thermal Sunyaev-Zeldovich effect, which are not
removed by $545$ GHz cleaning, using parameters from the base \LCDM\ fits
to the  12.1HM TT  likelihood). In Eq. \ref{equ:Nu1}, $\sigma_{xy}$ is the scatter of the bandpowers shown in
Fig.\ \ref{fig:pgcalib} over the fitted multipole range.
The results of recalibration give
\begin{subequations}
\begin{eqnarray}
 c_{100 \times 100} &=&   1.0022 \pm 0.0009,  \quad \sigma_{100 \times 100} = 4.5\ (\mu {\rm K})^2,  \label{equ:Nu2a}\\
 c_{143 \times 217} &=&   0.9989 \pm 0.0003,  \quad \sigma_{143 \times 217} = 1.5\ (\mu {\rm K})^2,  \label{equ:Nu2b}\\
 c_{217 \times 217} &=&   0.9972 \pm 0.0005,  \quad \sigma_{217 \times 217} = 2.6\ (\mu {\rm K})^2.  \label{equ:Nu2c}
\end{eqnarray}
\end{subequations}
 These relative calibration coefficients are determined very
precisely and show that the inter-frequency effective calibrations
display small $\sim 0.1 - 0.3 \%$ variations from unity (consistent with
our analysis of intra-frequency relative calibrations). These relative
calibrations are insensitive to multipole range and, provided one
restricts to high latitude areas of the sky, are insensitive to dust
cleaning\footnote{However, if instead of cleaning with 545 GHz, we
  remove dust using the smooth power spectrum dust templates discussed
  in Sect.\ \ref{subsec:dust_power}, the errors on the calibration
  coefficients are increased because of the additional scatter from
  CMB-foreground correlations.}. Note that at the position of the
first acoustic peak ($\ell \approx 220$), the four temperature spectra
in the \camspec\ likelihoods are consistent with each other to within
$\sim 0.08\%$ after recalibration.

 In the current version of \camspec\ we simply recalibrate the
 temperature spectra using the coefficients of \ref{equ:Nu2a} -
 \ref{equ:Nu2c} (or equivalents for the full mission detset
 likelihood) and no longer include relative calibrations as nuisance
 parameters.  This reduces degeneracies with other foreground
 components (principally dust amplitudes, if these are carried as nuisance parameters) but has little impact on
 cosmological parameters.

\subsubsection{Polarization calibrations}

Section \ref{subsec:pol_cal} described our procedure for recalibrating the individual 
polarization spectra, accounting for errors in polarization efficiencies and far-field beams.
These effective calibrations have typical accuracies of about $1\%$ (see Tables \ref{tab:EE_cal} and \ref{tab:TE_cal}). The corrections for
effective polarization efficiencies are applied to the \camspec\ TE, ET and EE spectra prior to coaddition. To account for possible residual calibration  differences with respect to the TT spectra, we include two nuisance parameters, $c_{\rm TE}$ and $c_{\rm EE}$, which multiply the coadded
TE and  EE spectra. We adopt Gaussian priors centred on unity with a standard deviation  of $0.01$.

\subsubsection{Absolute calibration}

  To account for possible errors in the absolute calibration of the
  \planck\ HFI spectra, we include an overall map-based calibration
  parameter $y_{\rm cal}$. All of the spectra are multiplied by
  $y^2_{\rm cal}$. As a result, this parameter has very little impact
  on cosmology, but adds an additional contribution to the errors on
  parameters related to the amplitude of the fluctuation spectrum. We
  adopt a Gaussian prior on $y_{\rm cal}$ centred on unity with a
  (very conservative) standard deviation of $0.25\%$. We emphasise
  that this calibration parameter describes an effective calibration
  error at high multipoles, and should not be confused with the
  absolute calibration error on the orbital dipole.  In reality, the
  dominant uncertainty in the {\it absolute} amplitude of the
  primordial fluctuation spectrum comes from systematic errors in the
  EE likelihood at $\ell < 30$, which is used to fix the optical depth
  to reionization. Systematic errors in the HFI EE spectrum at low multipoles  have been discussed
at length  \citep{SROLL:2016, Delouis:2019} and will not be revisited here.

\subsubsection{Beam errors}

  In PCP13, we modelled beam errors via a set of nuisance parameters multiplying beam error eigenmodes. For subsequent
\planck\ data releases, the errors on the $100-217$ HFI \planck\ main beams are so small that they have negligible impact
on cosmology. We therefore no longer include nuisance parameters describing beam errors.
Errors in the beams beyond the main beams are largely absorbed by the effective calibration factors.

\subsection{Galactic dust in temperature}
\label{subsec:galactic_dust_templates}

\begin{figure*}
\centering
\includegraphics[width=77mm,angle=0]{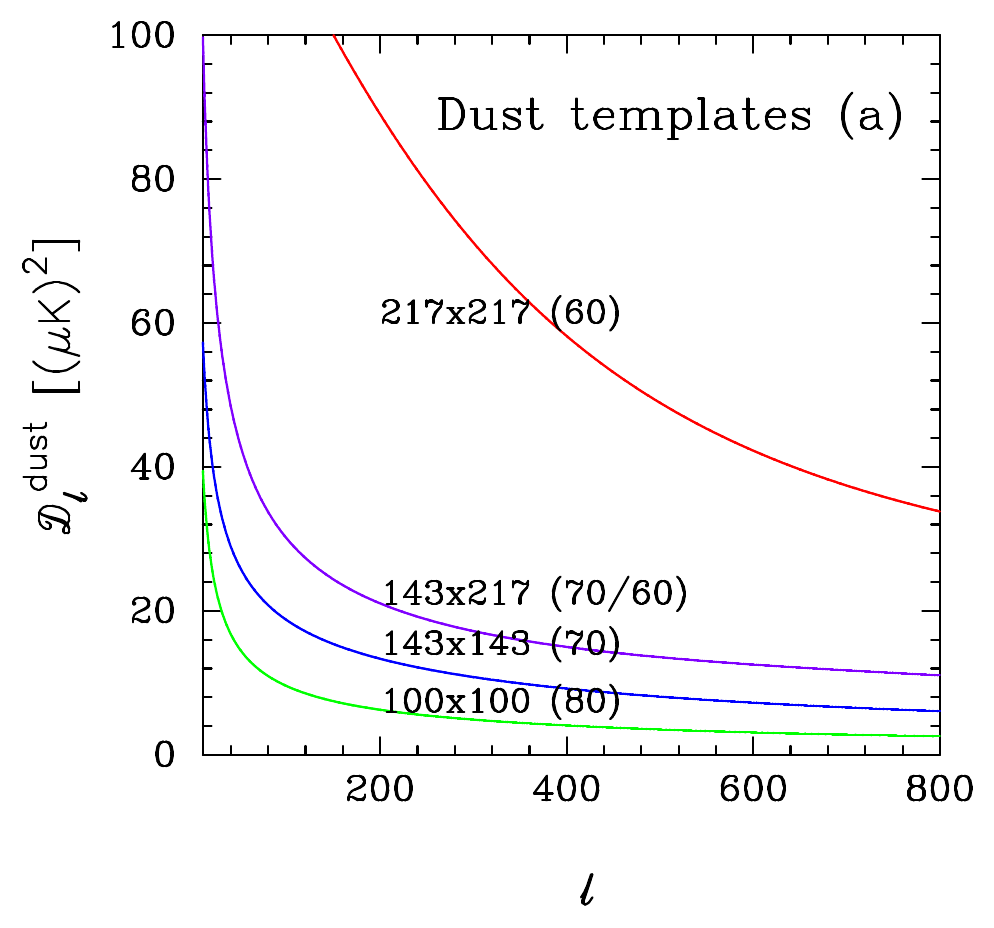} 
\includegraphics[width=77mm,angle=0]{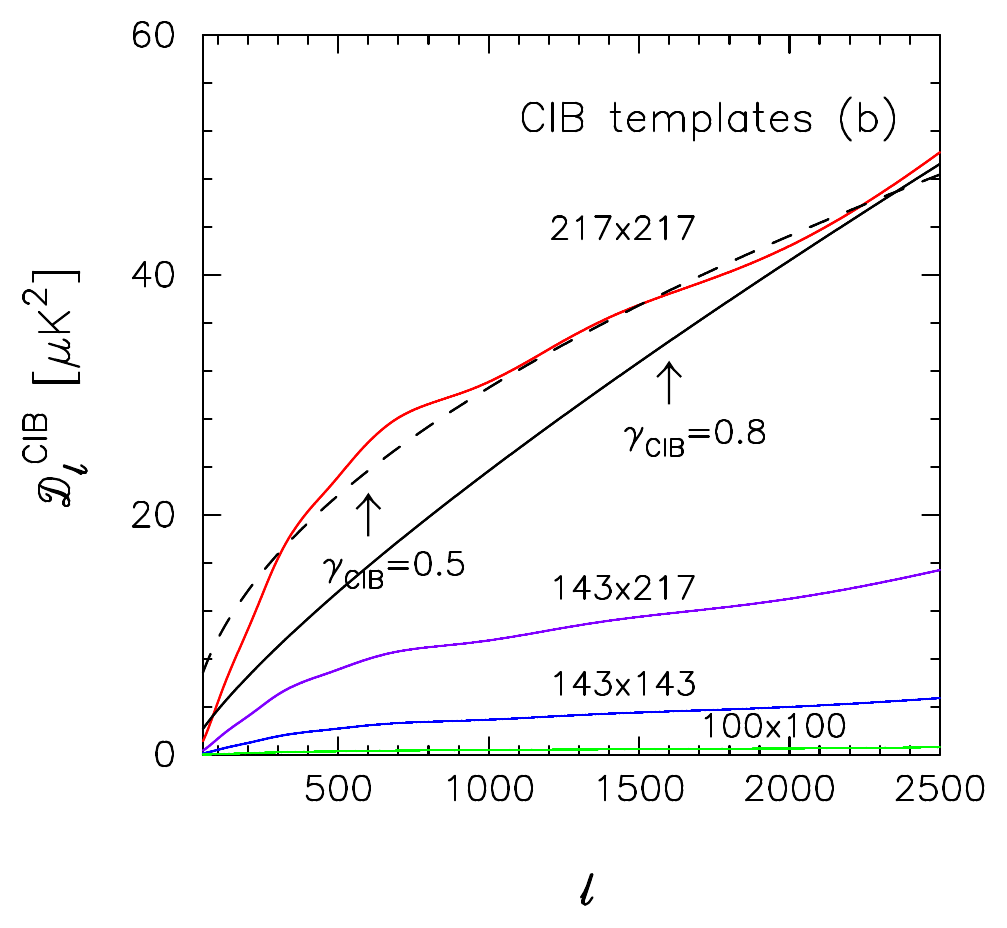} 
\caption {The figure to the left shows the $217\times 217$, $143\times 217$, $143 \times 143$ and $100 \times 100$ Galactic dust templates used in the  12.1HM likelihood (the sizes of the masks are given in brackets). The figure to the right shows the clustered CIB templates from the models of \cite{Planckdust:2014b}. The $217 \times  217$ CIB spectrum is normalized to the best fit value determined from the
12.1HM TT likelihood. The relative amplitudes of the other spectra are as given by the model of
\cite{Planckdust:2014b} (though we allow the amplitude of the $143 \times 217$ spectrum to float in the likelihood). The dashed and solid black lines show power laws, ${\hat D}^{\rm CIB}_\ell \propto \ell^{\gamma_{\rm CIB}}$ with 
${\gamma_{\rm CIB}} = 0.5$ and ${\gamma_{\rm CIB}} = 0.8$ respectively. }

\label{fig:pgdusttemplates}

\vspace{-0.03truein}

 \end{figure*}

As described in Sect.\ \ref{sec:dust_temp}, the power spectrum of
Galactic dust emission can be measured accurately, free from
extragalactic foregrounds, by differencing the high frequency power
spectra measured on different masks (see
Fig.\ \ref{fig:doublediff}). We therefore constructed a set of
template dust spectra from fits to the 545 GHz spectra using the
identical point source masks used in the \camspec\ likelihoods. This
accounts for the small differences in the shapes of the dust spectra
caused by differences in the point source holes (as discussed in
Sect.\ \ref{subsec:dust_power} and illustrated in
Fig.\ \ref{fig:dust545}). The template dust spectra for the 12.1HM
\camspec\ masks are plotted in Fig.\ \ref{fig:pgdusttemplates}. The
amplitudes of these templates depend on the cleaning coefficients
listed in bold face in Table \ref{tab:cleaning_coeffs} and so are not known
precisely \GE{(and also require a correction for the dust amplitude measured at high Galactic latitudes within mask25)}. For our `standard' likelihoods, we therefore sample over
four dust amplitude parameters, $A^{\rm dust}_{100 \times 100}$,
$A^{\rm dust}_{143 \times 143}$, $A^{\rm dust}_{143 \times 217}$, and
$A^{\rm dust}_{217 \times 217}$, with \GE{conservative} Gaussian priors centred on unity
and with a standard deviation of $0.2$.

\subsection{Extragalactic foregrounds}
\label{subsec:extragalactic foregrounds}

\subsubsection{Poisson point sources}

   The Poisson point source contributions are described by amplitude parameters
\begin{equation} 
              \hat D^{PS}_{\ell = 3000} =  A^{PS}   . \label{equ:Nu3}
\end{equation}
$A^{PS}_{100}$, $A^{PS}_{143}$ and $A^{PS}_{217}$ are the point source amplitudes of the 
$100\times 100$, $143 \times 143$ and $217 \times 217$ spectra respectively.
The point source amplitude of the $143 \times 217$ spectrum is described by a correlation
parameter $r^{PS}_{143 \times 217}$, so that $A^{PS}_{143 \times 217} = r^{PS}_{143 \times 217} \sqrt{A^{PS}_{143} A^{PS}_{217}}$.
We adopt uniform priors for these parameters with ranges as given in Table \ref{tab:priors}.

\begin{table}[tp]
{\centering

\caption{\small{Nuisance parameters: For parameters with Gaussian priors we list the mean and standard deviation.
For parameters with uniform priors, we give the range of the prior.}}

\label{tab:priors}
\begin{center}
\begin{small}
\begin{tabular}{|c|c|c|c|} \hline
Parameter  & Description  & Prior &   \\  \hline
$y_{\rm cal}$  &  absolute calibration & Gaussian &  $\mu =1$, $\sigma = 0.0025$ \\
$c_{\rm TE}$     &  relative  calibration of TE spectrum & Gaussian &  $\mu =1$, $\sigma = 0.01$ \\
$c_{\rm EE}$     &  relative  calibration of EE spectrum & Gaussian &  $\mu =1$, $\sigma = 0.01$ \\
$A^{\rm dust}_{100 \times 100}$     & dust amplitude $100 \times 100$ & Gaussian &  $\mu =1$, \GE{$\sigma = 0.2$} \\
$A^{\rm dust}_{143 \times 143}$     & dust amplitude $143 \times 143$ & Gaussian &  $\mu =1$, \GE{$\sigma = 0.2$} \\
$A^{\rm dust}_{143 \times 217}$     & dust amplitude $143 \times 217$ & Gaussian &  $\mu =1$, \GE{$\sigma = 0.2$} \\
$A^{\rm dust}_{217 \times 217}$     & dust amplitude $217 \times 217$ & Gaussian &  $\mu =1$, \GE{$\sigma = 0.2$} \\ 
$A^{\rm PS}_{100}$     & point source amplitude  $100 \times 100$ & uniform &  $0 - 360 \ (\mu{\rm K})^2$ \\ 
$A^{\rm PS}_{143}$     & point source amplitude  $143 \times 143$ & uniform &  $0 - 270 \ (\mu{\rm K})^2$ \\ 
$A^{\rm PS}_{217}$     & point source amplitude  $217 \times 217$ & uniform &  $0 - 450 \ (\mu{\rm K})^2$ \\ 
$r^{\rm PS}_{143 \times 217}$     & point source correlation coeff.  $143 \times 217$ & uniform &  $0-1$ \\ 
$A^{\rm CIB}_{217}$     & CIB amplitude  $217 \times 217$ & uniform &   $0 - 80 \ (\mu{\rm K})^2$ \\ 
$r^{\rm CIB}_{143 \times 217}$     & CIB correlation coeff.  $143 \times 217$ & uniform &  $0-1$ \\ 
$A^{\rm tSZ}_{143}$     & thermal SZ amplitude   $143 \times 143$ & see Sect.\ {\ref{subsubsec:SZpriors}} &   \\ 
$A^{\rm kSZ}$     & kinetic amplitude  &  see Sect.\ {\ref{subsubsec:SZpriors}}&  \\ 
$\xi^{\rm tSZ \times CIB}$     & tSZ-CIB correlation parameter  & uniform &  $0-1$ \\ 

\hline

\end{tabular}
\end{small}
\end{center}}
\end{table}

\subsubsection{Clustered CIB}

As in PCP15, we adopt the clustered CIB templates for $217 \times
217$, $143 \times 217$ and $143 \times 143$ from the halo model of
\cite{Planckdust:2014b}.  These templates are plotted in
Fig.\ \ref{fig:pgdusttemplates}.  Mak et al.\ \cite{Mak:2017} showed that at
multipoles $\ell \simgt 3000$, these templates become much steeper
than the power spectra of the CIB measured at $350 \ \mu {\rm m}$ and $500
\ \mu {\rm m}$ from Herschel \citep{Viero:2013}.  However, over the
multipole range accessible to \Planck, $\ell \simlt 2500$, the
templates of reference \cite{Planckdust:2014b}  are reasonably well approximated by
a power law $D^{\rm CIB}_{\ell} \propto \ell^{\gamma_{\rm CIB}}$ with
$\gamma_{\rm CIB} \approx 0.5$ (see Fig.\ \ref{fig:pgdusttemplates} and
Fig.\ 2 of \cite{Mak:2017}).  This is consistent with the results
presented in PCP13, where we found values of $\gamma_{\rm CIB} \approx
0.5$ from \Planck, with evidence of steepening to $\gamma_{\rm
  CIB}\approx 0.8$ when we combined \Planck\ with high multipole
ground-based CMB measurements.  Nevertheless, uncertainties in the
shapes of the CIB templates, particularly for the $217\times 217$
spectrum, are a potential source of systematic error in the foreground
model.

In PCP18, we varied a single amplitude
\begin{equation}
              \hat D^{\rm CIB}_{\ell = 3000} =  A^{\rm CIB}_{217}, \label{equ:Nu4}
\end{equation}
measuring the clustered CIB contribution to the $217 \times 217$
spectrum. The CIB contributions to the $143 \times 217$, $143 \times
143$ and $100 \times 100$ spectra were then fixed according to the
model of \cite{Planckdust:2014b}. Given the uncertainties in the
model of \cite{Planckdust:2014b},  we considered this to be too
restrictive and so have added a parameter $r^{\rm CIB}_{143 \times 217}$ to
adjust the amplitude of the CIB in the $143 \times 217$ spectrum,
$A^{\rm CIB}_{143 \times 217} = r^{\rm CIB}_{143 \times 217}
\sqrt{A^{\rm CIB}_{143}A^{\rm CIB}_{217}}$. Since the CIB makes a very
small contribution to the $143 \times 143$ and $100 \times 100$
spectra, we keep these amplitudes pinned to the amplitude of the $217
\times 217$ spectrum according to the model of
\cite{Planckdust:2014b}.

\subsubsection{Thermal and kinetic Sunyaev-Zeldovich effects}

As in PCP13 and later \Planck\ papers, we use the $\epsilon=0.5$ thermal Sunyaev-Zeldovich (tSZ) template from
\cite{Efstathiou:2012} normalized to a frequency of $143\,$GHz.
For cross-spectra between frequencies
$\nu_i$ and $\nu_j$, the tSZ template is normalized as
\begin{equation}
{\cal D}_\ell^{\mathrm{tSZ}_{\nu_i\times \nu_j}}  =
A^{\rm tSZ}_{143}\frac{f(\nu_i)f(\nu_j)}{f^2(\nu_0)}
{{\cal D}_{\ell}^{\rm{tSZ\, template}} \over {\cal D}_{3000}^{\rm{tSZ\, template}}}, \label{equ:Nu5}
\end{equation}
where $\nu_0$ is the reference frequency of $143\,\mathrm{GHz}$,
 ${\cal D}_{\ell}^{\rm{tSZ \ template}}$ is the template spectrum at $143\,\mathrm{GHz}$, and
\begin{equation}
 f(\nu) =  \left ( x{ e^x + 1 \over e^x -1 } - 4\right ),
 \quad {\rm with}\ x = {h\nu \over k_{\rm B} T_{\rm CMB}}. \label{equ:Nu6}
\end{equation}
We neglect the tSZ contribution for any spectra involving the \planck\
$217\,$GHz channels. The tSZ contribution is therefore characterized by the single 
amplitude parameter $A^{\rm tSZ}_{143}$.

Over the multipole range probed by \Planck, the tSZ template is a good
match to the tSZ power spectra measured from numerical simulations e.g.\
\citep{Battaglia:2010, McCarthy:2014}, though the amplitude and
shape of the tSZ spectrum  at multipoles $\ell \simgt 2000$ is sensitive to the details of
energy injection by active galactic nuclei into the intra-cluster
medium.

We adopt the kSZ template from
\cite{Trac:2011} and solve for the amplitude $A^{\rm kSZ}$:
\begin{equation}
    {\cal  D}^{\rm kSZ}_\ell = A^{\rm kSZ} {{\cal D}^{\rm kSZ\, template}_{\ell} \over
{\cal D}^{\rm kSZ\, template}_{3000}}. \label{equ:Nu7}
\end{equation}

\subsubsection{tSZ/CIB cross-correlation} 
The cross-correlation between dust emission from CIB galaxies and SZ
emission from clusters (tSZ$\times$CIB) is expected to be
non-zero. However, it is difficult to model this correlation reliably,
but fortunately over the multipole ranges probed by \Planck\ it only
makes a small contribution to the foregrounds (as confirmed by high
resolution ground-based experiments, e.g.\ \cite{Reichardt:2012}).  We
adopt the template spectrum computed by \cite{Addison:2012} in this
paper and model the frequency dependence of the power spectrum
according to:
\begin{eqnarray}
{\cal D}_\ell^{{\rm tSZ\times CIB}_{\nu_i \times \nu_j}} &=&- \xi^{\rm tSZ\times CIB}
{\cal D}_\ell^{\rm tSZ \times CIB\,\mathrm{template}} \nonumber \\
&&\hspace{-0.015\textwidth}\times \left ( \sqrt {{\cal D}^{{\rm CIB}_{\nu_i\times\nu_i}}_{3000} {\cal D}^{{\rm tSZ}_{\nu_j\times \nu_j}}_{3000}}
  +  \sqrt {{\cal D}^{{\rm CIB}_{\nu_j\times \nu_j}}_{3000} {\cal D}^{{\rm tSZ}_{\nu_i\times \nu_i}}_{3000}}
\right ) , \label{equ:Nu8}
\end{eqnarray}
where ${\cal D}_\ell^{\rm tSZ \times CIB\,\mathrm{template}}$ is the 
template spectrum  from \cite{Addison:2012} normalized to unity at $\ell=3000$ and
${\cal D}_\ell^{{\rm CIB}_{\nu_i\times \nu_i}}$ and ${\cal D}_\ell^{{\rm tSZ}_{\nu_i\times \nu_i}}$ are
given by Eqs.~\ref{equ:Nu5} and \ref{equ:Nu7}. The tSZ$\times$CIB contribution
is therefore characterized by the dimensionless cross-correlation coefficient
$\xi^{\rm tSZ \times CIB}$. With the definition of Eq.\ \ref{equ:Nu8}, a positive
value of $\xi^{\rm tSZ \times CIB}$ corresponds to an anti-correlation 
between the CIB and the tSZ signals.

\subsubsection{Priors on $A^{\rm tSZ}$, $A^{\rm kSZ}$, and $\xi^{\rm tSZ\times CIB}$}
\label{subsubsec:SZpriors}

The three parameters $A^{\rm tSZ}$, $A^{\rm kSZ}$,
and $\xi^{\rm tSZ\times CIB}$ are highly correlated with each other and
not well constrained by \Planck\ alone. Using high multipole data
from SPT,  Reichardt et al.\ \cite{Reichardt:2012} find strong constraints on  the linear combination
\begin{equation}
A^{\rm kSZ} + 1.55\,A^{\rm tSZ} = (9.2 \pm 1.3)\,\mu{\rm K}^2, \label{equ:NSZ1}
\end{equation}
after marginalizing over $\xi^{\rm tSZ\times CIB}$
 (where we have corrected the 
\cite{Reichardt:2012} constraints to the effective frequencies
used to define the \Planck\ amplitudes $A^{\rm kSZ}$
and $A^{\rm tSZ}$).

As in PCP15, in this paper we impose a conservative Gaussian prior
on $A^{\rm SZ}$, as defined by
\begin{equation}
A^{\rm SZ} = A^{\rm kSZ} + 1.6\,A^{\rm tSZ} = (9.5 \pm 3.0)\,\mu{\rm K}^2, \label{equ:NSZ1a}
\end{equation}
based on the PCP13 \Planck+highL solutions  (i.e., somewhat broader than the dispersion reported in 
\cite{Reichardt:2012}).  This condition prevents the individual SZ amplitudes from wandering
too far into unphysical regions of parameter space.   We apply a
uniform prior of [0,1] on $\xi^{\rm tSZ\times CIB}$. Results from the 
 complete 2540\,${\rm deg}^2$ SPT-SZ survey area
 \citep{George:2015} are consistent with Eq.\ \ref{equ:NSZ1}
and,  in addition,  constrain the correlation parameter to low values,
$\xi^{\rm tSZ\times CIB} = 0.113^{+0.057}_{-0.054}$. The parameter $\xi^{\rm tSZ\times CIB}$
is not well constrained by the \Planck\ data and so values are sampled by the \camspec\ likelihood
 that are excluded by 
 \citep{George:2015}.

\subsection{Cleaned likelihoods}
\label{subsection:cleaned_likelihood_nuisance}

We have also produced a set of `cleaned' likelihoods in which the $143
\times 143$, $143 \times 217$ and $217 \times 217$ TT spectra are
cleaned with 545 GHz as described in
Sect.\ \ref{subsec:spec_clean}. The TE and EE polarization spectra in these
likelihoods are cleaned using the 353 GHz maps, as described in
Sect.\ \ref{subsec:pol_clean_spectra}.  In the TT blocks of the
cleaned likelihoods, we discard the $100 \times 100$ spectra so that
we do not need to propagate foreground nuisance parameters for this
frequency. (Although 545 GHz cleaning removes Galactic dust emission,
it does very little to reduce the  extragalactic foregrounds, mainly point sources and tSZ, at high
multipoles in the $100 \times 100$ spectrum.) We adopt a heuristic
model for the foreground contributions to the remaining TT spectra.
For each of the $143\times 143$, $143 \times 217$ and $217 \times 217$
spectra, we adopt a power law foreground model:
\begin{equation}
         D_\ell^{\rm for}  =  A^{\rm for} \left( \ell \over 1500 \right )^{\epsilon^{\rm for}},  \label{equ:clean1}
\end{equation}
to capture the high multipole excesses seen in Fig.\ \ref{fig:545cleanedTT}. The temperature  foreground model
in the cleaned likelihoods  is therefore described by six parameters, instead of the thirteen parameters of the standard
foreground model. We adopt uniform priors within the ranges $0-50 \ (\mu{\rm K})^2$ for the amplitudes
  $A^{\rm for}$ and $0-5$ for the exponents $\epsilon^{\rm for}$.

The foreground corrected cleaned TT spectra are compared with those of
the standard foreground model in Sect.\ \ref{subsec:cleaned_TT} using
exactly the same sky masks. Furthermore, because Galactic dust
emission is subtracted accurately with 545 GHz cleaning, we have been
able to create a statistically powerful cleaned likelihood using
mask80 in both temperature and polarization at $143$ and $217$ GHz. Results from this
likelihood (12.5HMcl) are presented in Sects.\ \ref{sec:base_lcdm} and
\ref{sec:extensions_lcdm}.

\subsection{Summary}

   Extensive tests
described in PPL13 and PPL15 show that cosmological parameters for the base \LCDM\ and many simple extensions
to the base \LCDM\ model are remarkably insensitive to the nuisance parameters. We can gain further confidence
in the cosmological results by  comparing TE results with TT (since there are no nuisance parameters in the TE
likelihood apart from an overall calibration) and by comparing with results from
the  `cleaned' TT likelihoods which use a completely different  parameterization of the residual
extragalactic foregrounds involving fewer parameters. Such tests are described in Sects.\ \ref{sec:inter_frequency} and 
\ref{sec:base_lcdm}.

\section{Combined temperature and polarization likelihood}
\label{sec:Likelihood}

  Since  the extragalactic temperature foregrounds  depend strongly on
  frequency,  it is  necessary to apply the likelihoods 
  to  solve for
  nuisance and cosmological parameters in a full likelihood
  analysis
  {\it before} one can perform a detailed analysis of inter-frequency 
  residuals in the TT spectra. The   \camspec\ temperature-polarization likelihoods used in this paper
  are discussed in Sect.\ \ref{subsec:Camspec_likelihood}.
  Section \ref{subsec:low_multipole_likelihoods} provides a short
  summary of the TT and EE likelihoods used at $\ell < 30$.
  Section \ref{subsec:Base_LCDM} discusses fits to the base
  \LCDM\ cosmology using the 12.1HM \camspec\ half mission likelihood (which is similar to the
  \camspec\ likelihood used in PCP18) 
  and presents a number of  consistency tests of the spectra and
  likelihood. A detailed analysis of inter-frequency residuals is
  given in Sects.\ \ref{sec:inter_frequency} and  \ref{sec:inter_frequency_pol}.

\subsection{The \camspec\ likelihoods}
\label{subsec:Camspec_likelihood}

 The data vector of the combined temperature-polarization \camspec\ likelihoods consists of a set of spectra:
\begin{equation}
{\bf \hat C} = ({\bf \hat C^{TT}_1, \hat C^{TT}_2, \dots, \hat C^{TT}_N, \hat C^{TE}, \hat C^{EE}})^T, \label{equ:CL1}
\end{equation}
with covariance matrix: 
\begin{equation}
 {\pmb{$M$}} =  \left ( \begin{array} {ccccc} 
\langle \Delta \hat  C^{TT}_1 \Delta \hat C^{TT}_1 \rangle,  & \langle \Delta \hat C^{TT}_1 \Delta \hat C^{TT}_2 \rangle, &  \dots & \langle \Delta \hat C^{TT}_1 \Delta \hat C^{TE} \rangle,  & \langle \Delta \hat C^{TT}_1 \Delta  \hat C^{EE} \rangle  \\
\vdots &  & & & \vdots \\
\langle \Delta \hat C^{TE} \Delta \hat C^{TT}_1 \rangle,  & \langle \Delta \hat C^{TE} \Delta \hat C^{TT}_2 \rangle, &  \dots & \langle \Delta \hat C^{TE} \hat \Delta C^{TE} \rangle, & \langle \Delta \hat C^{TE}  \Delta \hat C^{EE} \rangle  \\
\langle \Delta \hat C^{EE} \Delta \hat C^{TT}_1 \rangle,  & \langle \Delta \hat C^{EE} \Delta \hat C^{TT}_2 \rangle, &  \dots & \langle \Delta \hat C^{EE} \Delta  \hat C^{TE} \rangle, & \langle \Delta \hat C^{EE} \Delta \hat C^{EE} \rangle  \\

              \end{array} \right ), \label{equ:CL2}
\end{equation}
which we can write as 
\begin{equation}
 {\pmb{$M$}} = \left ( \begin{array} {c|c} 
  {\bf M}_T &  {\bf M}_{TP}  \\  \hline
  {\bf M}_{TP}^T &   {\bf M}_P
              \end{array} \right ), \label{equ:CL3}
\end{equation}
where ${\bf M_P}$ is the polarization block:
\begin{equation}
 {\bf M}_P = \left ( \begin{array} {cc} 
\langle \Delta \hat C^{TE} \Delta \hat C^{TE} \rangle & \langle \Delta \hat C^{TE} \Delta \hat C^{EE} \rangle  \\
\langle \Delta \hat C^{EE} \Delta \hat C^{TE} \rangle &  \langle \Delta \hat C^{EE} \Delta \hat C^{EE} \rangle  \\ 
              \end{array} \right ). \label{equ:CL4}
\end{equation}
In Eq.\ \ref{equ:CL1}, $N=4$ for the `uncleaned'  \camspec\ temperature likelihoods,
corresponding to the $100 \times 100$, $143 \times 143$, $143 \times 217$ and $217 \times 217$
coadded TT spectra. For 545 GHz temperature cleaned likelihoods, $N=3$ since we discard the $100 \times
100$ TT spectrum. The polarization cross spectra $\bf \hat C^{TE}$, $\bf  \hat C^{EE}$, 
 are coadded over all frequency combinations, though some multipoles are excluded as discussed in Sect.\ \ref{subsec:multipole_ranges}.

We use a Gaussian approximation to the likelihood at multipoles $\ell \ge  30$:
\begin{equation}
-2 {\rm ln} {\cal L} = ({\bf \hat C} - {\bf \hat C}^{\rm model})^T \hat {\pmb{$M$}}^{-1}  ({\bf \hat C} - {\bf \hat C}^{\rm model}),    \label{equ:CL5}
\end{equation}
where ${\bf \hat C}^{\rm model}$ is the model prediction, including foreground and calibration parameters. 
The covariance matrix is computed as described in Sect.\ \ref{subsec:covariance_matrices},  assuming a 
fiducial theoretical model, which  is held fixed during MCMC sampling of the likelihood.

We have found it convenient to compute the inverse of ${\pmb{$M$}}$ as 
\begin{equation}
\hat {\pmb{$M$}}^{-1} = \left ( \begin{array} {cc} 
  {\bf M}^{-1}_T + {\bf M}^{-1}_T {\bf M}_{TP} {\bf M^\prime}^{-1}_P {\bf M}^T_{TP} {\bf M}^{-1}_{T},   &   - {\bf M}^{-1}_T {\bf M}_{TP}{\bf M^\prime}^{-1}_P   \\  
- {\bf M^\prime}^{-1}_P {\bf M}^T_{TP}{\bf M}^{-1}_T,  &   {\bf M^\prime}^{-1}_P
              \end{array} \right ),  \label{equ:CL5a}
\end{equation}
where ${\bf M^\prime}_P = ({\bf M}_P - {\bf M}^T_{TP} {\bf M}^{-1}_T{\bf M}_{TP})$. (This form
is useful for computing the temperature-polarization conditional spectra discussed in Sect. \ref{subsec:conditional}.)

\begin{table}
{\centering \caption{\small{The likelihoods constructed for this paper. The 12.1HM
    likelihoods are the most similar to the \camspec\ likelihoods used in
    PCP18. These likelihoods use the default choice of masks at each frequency in
    temperature  as illustrated in Fig.\ \ref{fig:tempmasks}. HM
    denotes half mission cross spectra and F denotes full mission detset
    cross spectra with corrections for correlated TT noise as
    described in Sect.\ \ref{subsec:correlatednoise}. The likelihoods use
    either the the standard temperature foreground model described in Sect.\
    \ref{sec:nuisance} or $545$ GHz cleaned spectra with the much simpler
    foreground model of Sect.\ \ref{subsection:cleaned_likelihood_nuisance}.
    All of the HM likelihoods use $353$ GHz
    cleaned TE and EE spectra as described in
    Sect.\ \ref{subsec:pol_clean_spectra}.}}

\label{tab:likelihood_summary}
\begin{center}

\smallskip

\begin{tabular}{|l|c|c|c|c|} \hline 
Likelihood  & TT foreground &Type & TT masks & Q/U masks   \\  \hline
12.1HM  & standard & half mission frequency maps & default & maskpol60\\ \hline
12.1HMcl & cleaned & half mission frequency maps & default & maskpol60 \\ \hline
12.1F &  standard &full mission detset maps & default & maskpol60 \\ \hline
12.2HM &  standard &half mission frequency maps & default & mask60 \\ \hline
12.3HM &  standard &half mission frequency maps & default & mask70 \\ \hline
12.4HM & standard  & half mission frequency maps & default & mask80 \\ \hline
12.5HMcl&  cleaned & half mission frequency maps & mask80 & mask80 \\ \hline
\end{tabular}
\end{center}}
\end{table}

The high-multipole likelihoods constructed for this paper are summarized in Table \ref{tab:likelihood_summary}. These likelihoods fall into three classes:

\smallskip

\noindent
$\bullet$ 12.1: The 12.1HM  likelihoods are similar to the
\camspec\ likelihoods used in PCP18 in that they use the same
temperature and polarization masks.  The main differences with the
PCP18 likelihoods are as follows: (a) in this paper we fix the temperature
inter-frequency calibrations as decribed in Section
\ref{subsubsec:inter_frequency}, rather than carrying them as nuisance
parameters; (b) we discard the $100\times 100$ spectrum from the
12.1HMcl likelihood, (c) we have reduced the multipole range over which we clean
the polarization spectra using 353 GHz template cleaning. 
The 12.1HM pair of likelihoods use revised calibrations of
effective polarization efficiencies as described in Sect.\
\ref{subsec:pol_cal}.  A comparison of results from the 12.1HM (which
uses the standard temperature foreground model of
Sect.\ \ref{sec:nuisance}) with those of the 12.1HMcl likelihood (which
uses $545$ GHz cleaned temperature spectra) is described in Sect.\
\ref{subsec:spec_clean}. This comparison tests the stability of the
cosmological parameters to the temperature foreground model.  The
12.1F likelihood uses the same temperature and polarization masks as in
the 12.1HM likelihood  but uses full mission detset spectra rather
than half mission spectra. The temperature full mission detset spectra
are corrected for correlated noise as described in
\ref{subsec:correlatednoise}. In the 12.1F likelihood, polarized dust
emission is subtracted from the TE and EE spectra using the power-law
fits from Table \ref{tab:pol_dust_fits} at all multipoles.
The 12.1F likelihood has higher signal-to-noise at high
multipoles in TT and EE compared to the 12.1HM likelihoods. Note that
because we use all non-cotemporal TE spectra in the half mission
likelihoods, there is very little improvement in the signal-to-noise
of the 12.1F TE spectrum compared to the 12.1HM TE spectrum.

\smallskip

\noindent
$\bullet$ 12.2HM-12.4HM: The TT component of these likelihoods is the
same as in 12.1HM. The sequence 12.2-12.4 explores changes in the TE and
EE components of the likelihood with variations in the Q/U sky
masks. Instead of using maskpol60, we apply the temperature masks (together with the 143 GHz
point-source mask) to the Q and U maps at each frequency. For 12.2HM, 12.3HM and 12.4HM we
apply mask60, mask70 and mask80 respectively to all polarization maps.
All of the polarization spectra in the HM likelihoods are cleaned using 353 GHz. 
It is worth noting here that we have found no evidence for any point source
contribution using maskpol60. However, we found some (albeit weak) evidence that a small
number of bright highly polarized point sources such as Centaurus A contribute to the
polarization spectra computed on extended polarization masks. To eliminate any possibility of bias, 
we decided to apply the $143$ GHz point source mask to the Q and U maps in the 12.2-12.5HM 
likelihoods.

\smallskip

\noindent
$\bullet$ 12.5HMcl: This is the most powerful likelihood that we have
produced, increasing the signal-to-noise over 12.1HM by using mask80
in both temperature and polarization\footnote{Applying  the 143 GHz
 point source mask to 143 GHz temperature maps, 143 and 217 GHz Q and U maps,
 and the $217$ GHz point source mask  to 217 GHz temperature maps.}.  The $143$ and $217$ GHz
temperature maps are cleaned using $545$ GHz maps as described in
Sect.\ \ref{subsec:spec_clean}. As discussed above, we discard 100 GHz maps.
  The TE and EE components of the 12.5HMcl
likelihood are identical to those of the 12.4HM likelihood.

\subsection{Low multipole likelihoods}
\label{subsec:low_multipole_likelihoods}

The \camspec\ likelihoods at $\ell \ge 30$ are patched on to low
multipole TT and EE likelihoods covering the multipole range
$2-29$. These low multipole likelihoods are identical to those used in
PCP18. In temperature we use the TT \commander\ likelihood, which is
a Gibbs-sample-based Blackwell-Rao likelihood based on the
\commander\ component separation algorithm. The \commander\ likelihood
is described in \cite{Planck_foregrounds_2018} and PPL18 and accounts accurately
for the non-Gaussian shape of the power spectrum posteriors at low
multipoles. To constrain the optical depth to reionization, $\tau$, we
use the \simall\ EE likelihood \GE{which is based on a quasi-QML estimate of the \Planck\ full mission $100
\times 143$} EE spectrum computed over the multipole range $2 \le \ell
\le 29$ \GE{(see \cite{deBelsunce:2021} for an exploration of different likelihood approximations
and frequency combinations}). The \simall\ likelihood is described in
\cite{DataProcessing:2018} and PPL18. We compare our TT and EE power spectra
with the \commander\ and \simall\ spectra in Figs.\
\ref{fig:commander} and \ref{fig:lowl_EE}.

\subsection{Notation}
\label{subsec:Likelihood_notation}

We adopt a simpler notation compared to that used in PCP18 to identify
results from different likelihoods. Unless otherwise stated (for
example, in Sect.\ \ref{subsec:base_ell_range}) cosmological parameter
results using a \camspec\ TT likelihood  include \commander\ and
\simall\ at low multipoles. The addition of \simall\ is necessary to
accurately constrain $\tau$.  Parameter constraints from \camspec\ TE, EE, or
combined TEEE likelihoods  include \simall\ at low multipoles but
not \commander.  Thus, for example:
\begin{eqnarray}
& &{\rm 12.1HM\ TT}  \equiv  {\rm 12.1HM \ TT \  likelihood+\commander+\simall}, \nonumber \\
& & {\rm 12.1HM\ TE}  \equiv  {\rm 12.1HM \ TE \  likelihood+\simall}, \nonumber \\
& &{\rm 12.1HM\ TEEE}  \equiv  {\rm 12.1HM \ TE+EE \  likelihood+\simall},    \nonumber\\
& & {\rm 12.5HMcl\ TTTEEE}  \equiv  {\rm 12.5HM \ cleaned \ TT+ TE+EE \  likelihood+\commander+\simall}. \nonumber 
\end{eqnarray}

\subsection{The fiducial base \LCDM\ cosmology}
\label{subsec:Base_LCDM}

In the next two sections, we will present a detailed analysis of
inter-frequency power spectrum residuals for both the temperature and
polarization spectra.  To do this, we need a fiducial cosmology and
foreground solution. In this section, we will discuss results for the
base \LCDM\ cosmology derived from the 12.1HM likelihood, since this
likelihood is the closest to the \camspec\  temperature
likelihood used in PCP18. We adopt the best-fit base \LCDM\ cosmology
derived from the 12.1HM TT likelihood as our fiducial theoretical
model. Subsequent sections will then discuss differences between the
power spectra estimated from different sky areas, between half mission
and full mission data and between different methods of temperature
foreground cleaning. Cosmological results from the likelihoods of
Table \ref{tab:likelihood_summary} are presented in
Sects.\ \ref{sec:base_lcdm} and \ref{sec:extensions_lcdm}.

\begin{table}

\small
\centering{
\caption{\small{Marginalized base \LCDM\ parameters with 68\% confidence intervals for the 12.1HM \camspec\ half mission likelihood.}}
\label{tab:LCDMcompare_camspec}

\begin{center}

\begin{tabular}{|l|c|c|c|c|c|c|}
\hline 
Parameter  & TT & TE \hfil&  EE & TTTEEE \cr \hline
$\Omega_{\mathrm{b}} h^2$ &$0.02218\pm 0.00022$&$0.02237\pm 0.00025$&$0.0235\pm 0.0012$&$0.02229\pm 0.00016$\cr
$\Omega_{\mathrm{c}} h^2$&$0.1202\pm 0.0021$&$0.1171\pm 0.0020$&$0.1177\pm 0.0048$&$0.1196\pm 0.0014$\cr
$100\theta_{\mathrm{MC}}$&$1.04081\pm 0.00047$&$1.04144\pm 0.00050$&$1.03945\pm 0.00087$&$1.04087\pm 0.00032$\cr
$\tau$&$0.0524\pm 0.0080$&$0.0504\pm 0.0082$&$0.0514\pm 0.0084$&$0.0529\pm 0.0078$\cr
${\rm ln}(10^{10} A_\mathrm{s})$&$3.043\pm 0.016$&$3.030\pm 0.022$&$3.054\pm 0.024$&$3.043\pm 0.016$\cr
$n_\mathrm{s}$&$0.9643\pm 0.0058$&$0.976\pm 0.011$&$0.974\pm 0.015$&$0.9664\pm 0.0044$\cr
\hline
$H_0\,[{\rm km}\,{\rm s}^{-1}\,{\rm Mpc}^{-1}]$&$67.09\pm 0.93$&$68.54\pm 0.92$&$68.6\pm 2.7$&$67.42\pm 0.63$\cr
$\Omega_\Lambda$&$0.682\pm 0.013$&$0.701\pm 0.012$&$0.697^{+0.035}_{-0.027}$&$0.6864\pm 0.0087$\cr
$\Omega_{\mathrm{m}}$&$0.318\pm 0.013$&$0.299\pm 0.012$&$0.303^{+0.027}_{-0.035}$&$0.3136\pm 0.0087$\cr
$\Omega_{\mathrm{m}} h^2$&$0.1431\pm 0.0020$&$0.1402\pm 0.0020$&$0.1418\pm 0.0039$&$0.1425\pm 0.0013$\cr
$\Omega_{\mathrm{m}} h^3$&$0.09596\pm 0.00044$&$0.09606\pm 0.00053$&$0.0973^{+0.0016}_{-0.0018}$&$0.09607\pm 0.00031$\cr
$\sigma_8$&$0.8117\pm 0.0090$&$0.799\pm 0.012$&$0.803 \pm 0.018$&$0.8097\pm 0.0076$\cr
$\sigma_8(\Omega_{\rm m}/0.3)^{0.5}$&$0.836\pm 0.024$&$0.798\pm 0.024$&$0.807^{+0.053}_{-0.062}$&$0.828\pm 0.017$\cr
$\sigma_8 \Omega_{\mathrm{m}}^{0.25}$&$0.610\pm 0.012$&$0.591\pm 0.013$&$0.595\pm 0.028$&$0.6059\pm 0.0086$\cr
$z_{\mathrm{re}}$&$7.50^{+0.83}_{-0.75}$&$7.21^{+0.91}_{-0.77}$&$7.09^{+0.87}_{-0.73}$&$7.53^{+0.81}_{-0.73}$\cr
$10^9 A_{\mathrm{s}}$&$2.097\pm 0.034$&$2.070\pm 0.046$&$2.121\pm 0.050$&$2.096\pm 0.033$\cr
$10^9 A_{\mathrm{s}} e^{-2\tau}$&$1.888\pm 0.014$&$1.871\pm 0.028$&$1.913\pm 0.032$&$1.885\pm 0.012$\cr
$\mathrm{Age}\,[\mathrm{Gyr}]$&$13.821\pm 0.037$&$13.769\pm 0.038$&$13.72\pm 0.14$&$13.805\pm 0.025$\cr
$z_\ast$&$1090.18\pm 0.41$&$1089.68\pm 0.42$&$1088.5^{+1.5}_{-1.8}$&$1089.99\pm 0.29$\cr
$r_\ast\,[\mathrm{Mpc}]$&$144.52\pm 0.47$&$145.17\pm 0.49$&$144.18\pm 0.71$&$144.61\pm 0.31$\cr 
$100\theta_\ast$&$1.04102\pm 0.00047$&$1.04162\pm 0.00049$&$1.03952\pm 0.00084$&$1.04106\pm 0.00031$ \cr
$z_{\mathrm{drag}}$&$1059.51\pm 0.45$&$1059.73\pm 0.55$&$1062.2\pm 2.3$&$1059.72\pm 0.33$ \cr
$r_{\mathrm{drag}}\,[\mathrm{Mpc}]$&$147.24\pm 0.47$&$147.85\pm 0.51$&$146.50\pm 0.75$&$147.30\pm 0.31$ \cr
$k_{\mathrm{D}}\,[\mathrm{Mpc}^{-1}]$&$0.14056\pm 0.00051$&$0.14006\pm 0.00058$&$0.1422\pm 0.0013$&$0.14059\pm 0.00034$ \cr
$z_{\mathrm{eq}}$&$3403\pm 48=7$&$3334\pm 47$&$3373\pm 90$&$3390\pm 32$ \cr
$k_{\mathrm{eq}}\,[\mathrm{Mpc}^{-1}]$&$0.01039\pm 0.00014$&$0.01018\pm 0.00014$&$0.01030\pm 0.00028$&$0.010347\pm 0.000097$ \cr
$100\theta_{\rm{s,eq}}$&$0.4492\pm 0.0046$&$0.4561\pm 0.0046$&$0.4524\pm 0.0094$&$0.4505\pm 0.0031$ \cr
$f_{2000}^{143}$&$29.8\pm 2.9$&&&$28.9\pm 2.7$ \cr
$f_{2000}^{217}$&$107.6\pm 2.0$&&&$107.1\pm 1.8$ \cr
$f_{2000}^{143\times217}$&$32.5\pm 2.1$&&&$31.8\pm 1.9$ \cr 
\hline

\end{tabular}

\end{center}}

\end{table}

Parameter constraints for base \LCDM\ derived from the 12.1HM
half mission likelihood are listed in Table
\ref{tab:LCDMcompare_camspec}, which can be compared to Table
1 of PCP18 (which compared base \LCDM\ parameters measured by \camspec\ and \plik).
Evidently, the
small changes that we have made to \camspec\ 12.1HM likelihood have
minor effects on the cosmological parameters of  base
\LCDM. For base \LCDM, the \camspec\ and \plik\ likelihoods
agree well (as they should since the input data, temperature masks and
methodology are similar).  However, as discussed in PCP18 the
agreement between \camspec\ and \plik\ is less good for extensions to
base \LCDM, particularly when polarization is added to the TT
likelihoods. This will be discussed in
Sect.\ \ref{sec:extensions_lcdm}.

\begin{figure}
\vspace{0.3truein}
\centering
\includegraphics[width=150mm,angle=0]{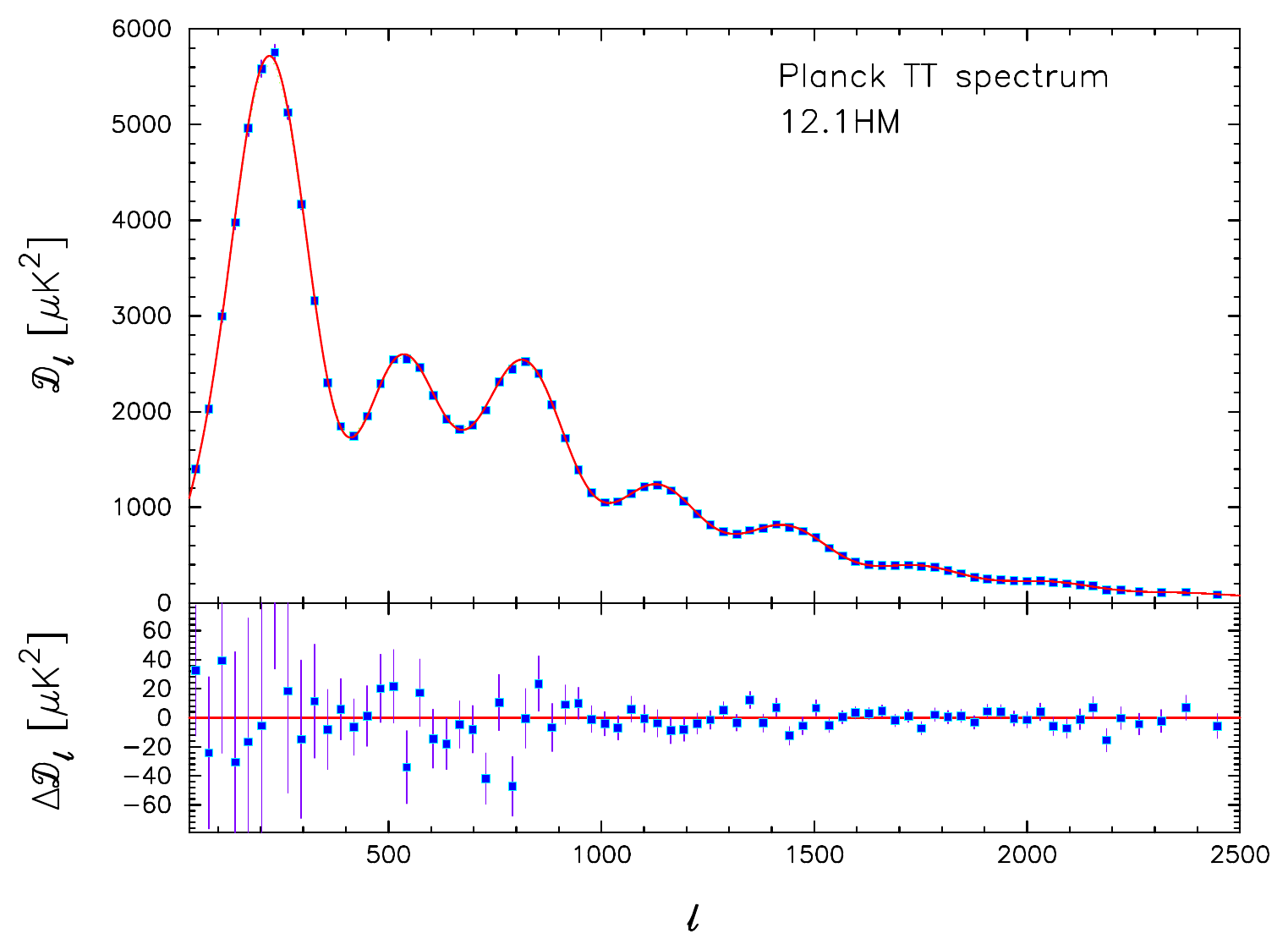} 
\caption {The maximum likelihood frequency averaged temperature power spectrum for the 12.1HM \camspec\ half mission
likelihood. The error bars on the band averages show $\pm 1\sigma$ ranges  computed from the covariance matrix of Eq.\ \ref{equ:CL7}. 
 The lower panel shows the residuals with respect to the fiducial base \LCDM\ cosmology (fitted to 12.1HM TT).}

\label{fig:12.1TT}

!\vspace{-0.2truein}
\end{figure}

\begin{figure}

\vspace{-0.2in}

\centering
\includegraphics[width=136mm,angle=0]{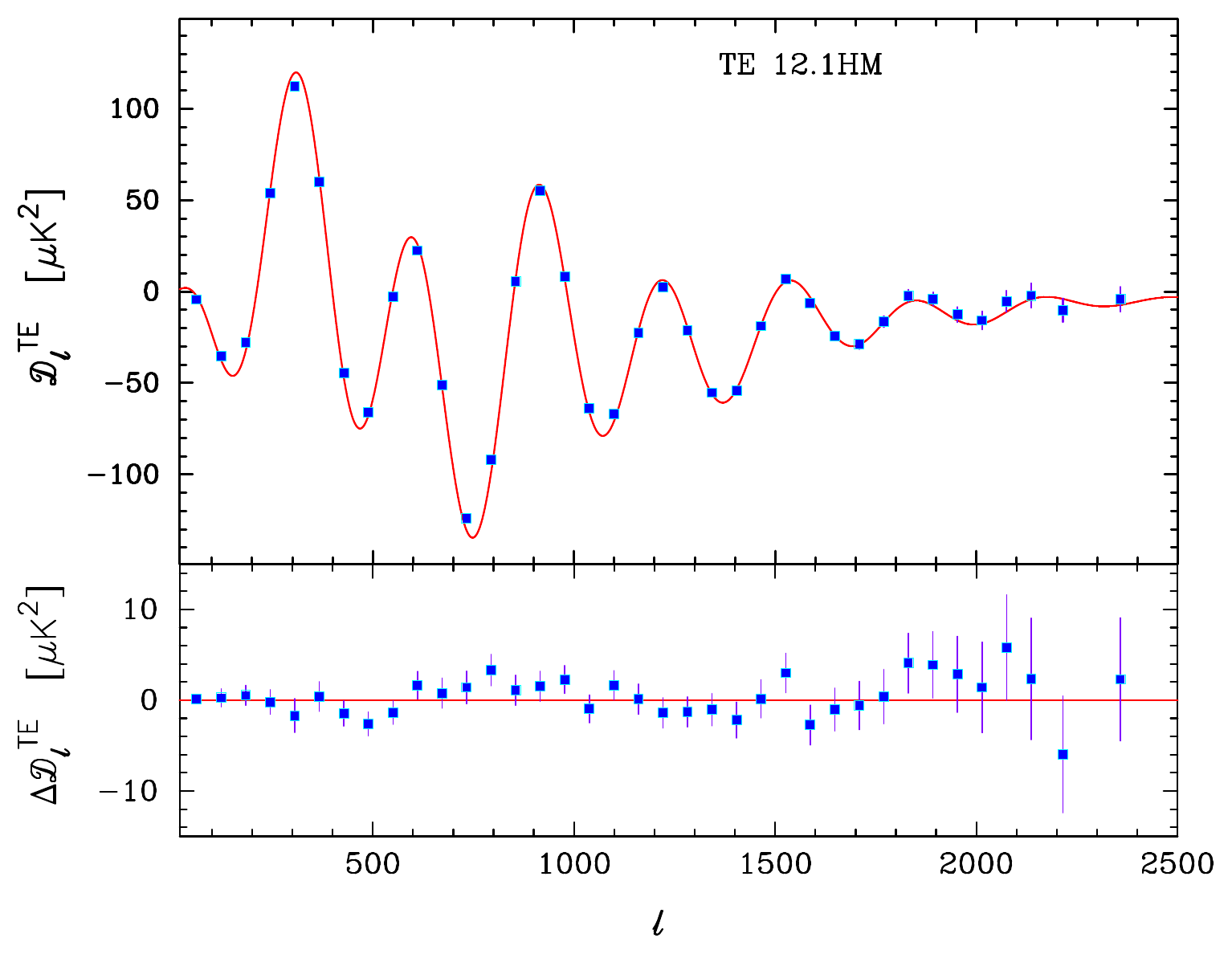} \\
\includegraphics[width=137mm,angle=0]{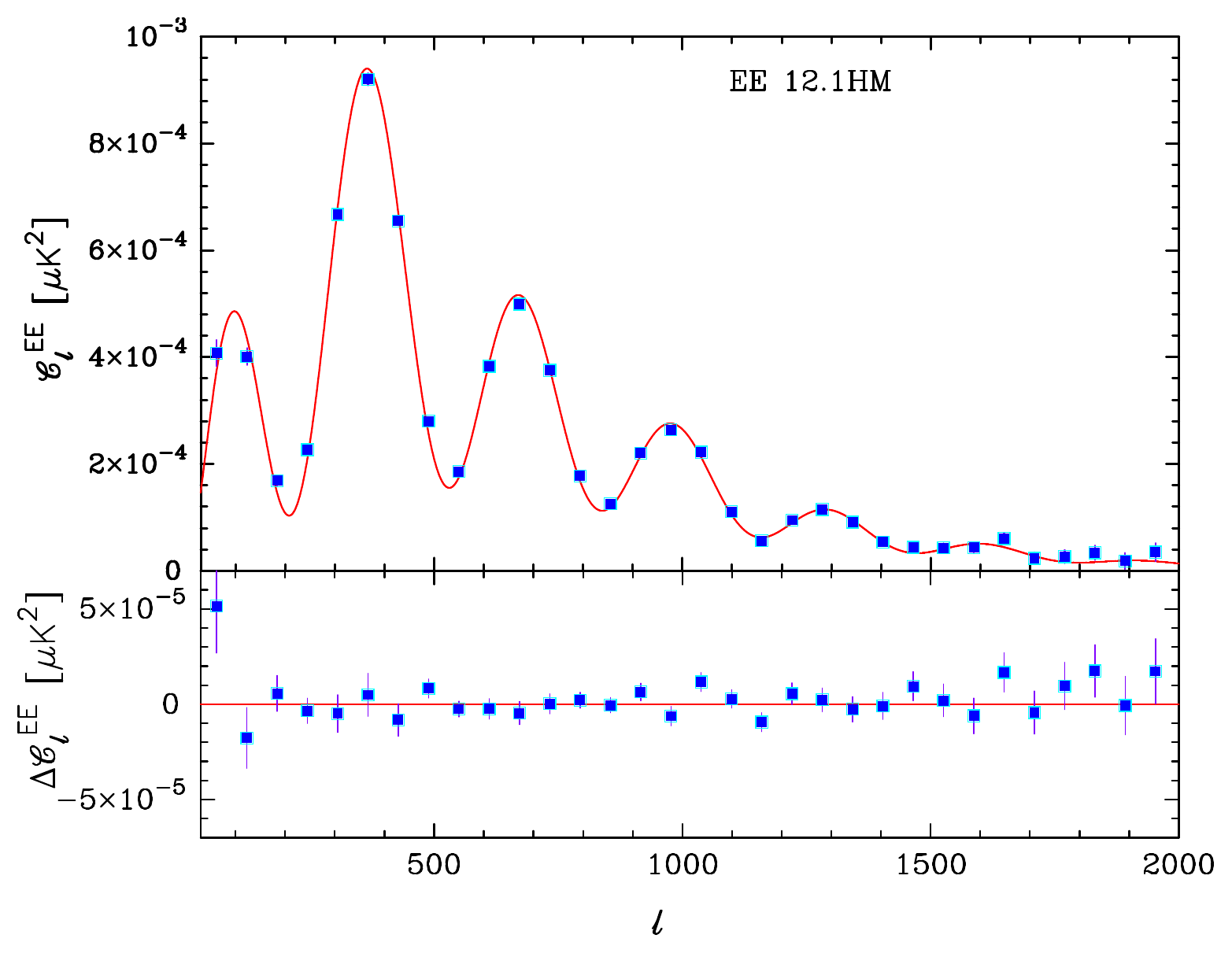} 

\caption {The coadded TE and EE power spectra of the 12.1HM half mission likelihood. The 
error bars on the band averages are computed from the \camspec\ covariance matrices.
The theoretical spectra show the fiducial base  \lcdm\ cosmology fitted to 12.1HM TT (i.e.\
they are not fits to the polarization spectra). We have applied (small)  corrections to the TE and EE
spectra using relative calibrations derived from fits to the 12.1HM TTTEEE likelihood.
Residuals with respect to the fiducial base \LCDM\  model are shown in the lower panels. Although the
\camspec\ likelihoods use TE only to $\ell_{\rm max} = 2000$, we have plotted the TE spectra out to 
$\ell_{\rm max} = 2500$.}

   \label{fig:12.1_TE_EE}

\end{figure}

Once we have solved for a best-fit cosmology and foreground model with
power spectrum $(C_\ell^{\rm for})^k$ for frequency combination $k$ we
can maximise the likelihood Eq.\ \ref{equ:CL5} to produce a `best-fit'
foreground-corrected spectrum, $\hat C^{TT}_\ell$, which is given by
the solution of:
\begin{equation}
\sum_{kk^{\prime}\ell^{\prime}}(\hat{{\cal M}}_{\ell\ell^{\prime}}^{-1})^{kk^{\prime}}\hat{C}_{\ell'}^{TT}=\sum_{kk^{\prime}\ell^{\prime}}(\hat{{\cal M}}_{\ell\ell^{\prime}}^{-1})^{kk^{\prime}}(\hat{C}_{\ell^{\prime}}^{k^{\prime}}-(\hat{C}_{\ell^{\prime}}^{\rm for})^{k^{\prime}}).\label{equ:CL6}
\end{equation}
 The covariance matrix of the estimates $\hat{C}_{\ell}^{TT}$ is given by the inverse of
the Fisher matrix:
\begin{equation}
\langle\Delta\hat{C}_{\ell}^{TT}\,\Delta\hat{C}_{\ell^{\prime}}^{TT}\rangle=\left(\sum_{kk^{\prime}}(\hat{{\cal M}}_{\ell\ell^{\prime}}^{-1})^{kk^{\prime}}\right)^{-1}.\label{equ:CL7}
\end{equation}
The spectrum $\hat C^{TT}_\ell$ is therefore simply a maximum likelihood coaddition of the individual TT spectra used in the
likelihood, corrected for foregrounds.

Fig.\ \ref{fig:12.1TT} shows the frequency averaged temperature power spectrum determined for the 12.1HM 
likelihood. Fig.\ \ref{fig:12.1_TE_EE} shows the coadded TE and EE power spectra compared to the best fit base \LCDM\
theory spectrum fitted to the TT likelihood. Note that there are  no foreground parameters in TE and EE; the only `nuisance'
parameters in the polarization spectra are overall calibration parameters which are very close to unity. In Fig.\ \ref{fig:12.1_TE_EE} we have multiplied the TE and EE spectra by factors of $0.9991$ and $0.9992$ respectively, which are the best fit values for the relative calibrations $c_{\rm TE}$ and $c_{\rm EE}$ determined from the base \LCDM\ fit to the
12.1HM TTTEEE likelihood.

\begin{table}[b]
\vspace{0.1truein}

{\centering \caption{\small{$\hat \chi^2$ values for the 12.1HM spectra and
    likelihood. For the first five rows, testing the TT spectra, we
    adopt the best fit base \LCDM\ model and nuisance parameters
    fitted to the 12.1HM TT likelihood.  For the remaining five rows,
    which test the components of the TTTEEE likelihood, we adopt the
    best fit model and nuisance parameters fitted to the 12.1HM TTTEEE
    likelihood. The second column gives the multipole range, $N_D$ is
    the size of the data vector (equal to the multipole range for
    single spectra). $\hat \chi^2 = \chi^2/N_D$ is the reduced
    $\chi^2$. \GE{The fifth column lists the number of standard
    deviations by which $\hat \chi^2$ differs from unity and the last column gives the
    probability to exceed (pte)}. `TT coadded'
    refers to the maximum likelihood frequency coadded spectrum
    plotted in Fig.\ \ref{fig:12.1TT}. The next four rows give $\hat
    \chi^2$ values for the individual foreground corrected TT spectra
    that enter the likelihood. `TT all' gives $\hat \chi^2$ for the
    complete TT likelihood and includes correlations between the
    frequency spectra. The next two lines give $\hat \chi^2$ for the
    TE and EE spectra plotted in Fig.\ \ref{fig:12.1_TE_EE}. The final
    two lines list $\hat \chi^2$ for the TEEE block and for the 12.1HM
    TTTEEE likelihood. }}
\label{tab:chi_squared}
\begin{center}

\smallskip

\begin{tabular}{|c|c|c|c|c|c|} \hline 
spectrum  & $\ell$ range & $N_D$ & $\hat \chi^2$ & $(\hat \chi^2-1)/\sqrt{2/N_D}$  & \GE{pte} \\  \hline
TT coadded &\;\;$30-2500$ & $2471$ & $1.01$ & $\;\;0.18$ & \GE{$0.43$}\\ 
TT $100\times 100$  & \;\;$30-1400$ & $1371$ & $1.04$ & $\;\;0.97$& \GE{$0.17$}\\ 
TT $143\times 143$  & \;\;$30-2000$ & $1971$ & $1.02$ & $\;\;0.56$& \GE{$0.29$}\\ 
TT $143\times 217$  &$500-2500$ & $2001$ & $0.98$ & $-0.57$ &\GE{$0.72$} \\ 
TT $217\times 217$  &$500-2500$ & $2001$ & $0.95$ & $-1.58$ &\GE{$0.94$}\\ 
TT All  &\;\;$30-2500$ & $7344$ & $0.99$ & $-0.38$&\GE{$0.64$}\\ 
TE & \;\;$30-2000$ & $1971$ & $1.01$ & \;\;$0.32$&\GE{$0.37$} \\
EE & \;\;$30-2000$ & $1971$ & $0.93$ & $-2.12$ &\GE{$0.98$}\\
TEEE & $\;\;30-2000$ & $3942$ & $1.02$ & \;\;$0.98$ &\GE{$0.16$} \\
TTTEEE & \;\;$30-2500$   & $11286$ &$0.97$ & $-2.20$& \GE{$0.99$} \\\hline
\end{tabular}
\end{center}}
\end{table}

Values of $\hat \chi^2$ (where the hat denotes the reduced $\chi^2$)
for the fits plotted in Figs.\ \ref{fig:12.1TT} and
\ref{fig:12.1_TE_EE} are listed in Table \ref{tab:chi_squared} for
various blocks of the 12.1HM likelihood.  In PCP13 we found acceptable
values of $\hat \chi^2$ for the individual TT spectra, but excess
$\hat \chi^2$ for the full TT likelihood.  The 2015 
\camspec\ likelihood
used in PCP15  had acceptable  $\hat \chi^2$
in  TT  but had excess $\hat \chi^2$ for the TE
and EE spectra. There are several reasons for the differences between the
2015 \camspec\ likelihood and the likelihoods produced for 
this paper. For
the TE and EE spectra, we correct for temperature-to-polarization
leakage and effective calibrations as described in
Sect.\ \ref{sec:beams}.  The temperature-to-polarization leakage
corrections are quite small for TE and are negligible for the EE
spectra. The effective calibrations in TE and EE, on the other hand,
have quite a large effect in reducing $\chi^2$ for individual
frequencies and for the coadded TE and EE spectra. The most significant
change, however, is in the noise model adopted in this paper, which is
now based on odd-even map differences instead of half-ring map
differences. As described in Sect.\ \ref{sec:noise}, the odd-even
differences lead to higher noise estimates than the half-ring
differences, particularly for the EE spectra. We also found evidence,
by comparing cross and auto spectra, that the odd-even differences
actually overestimate the noise in the Q and U maps, particularly at
100 GHz. A small overestimate of the noise in polarization is almost
certainly the explanation for the low $\hat \chi^2$ for the coadded EE
spectrum listed in Table \ref{tab:chi_squared} (though it is not unreasonably low). Nevertheless, the
$\hat \chi^2$ values are consistent with unity. Even for the full
TTTEEE likelihood, which has a large data vector length of $11286$,
$\hat \chi^2$ is consistent with unity to about $2\sigma$
(cf.\ Eq.\ \ref{equ:Noise6}).  In summary, the absolute values of
$\chi^2$ for the likelihoods used in this paper are acceptable, though
we have evidence from the coadded (and individual) EE spectra that the
Q and U noise power spectrum estimates used in this paper are too
high. Estimation of noise power spectra and noise correlations to a
precision of better than a percent remains a challenging problem for
\Planck\ analysis. End-to-end simulations have been used to
characterize the noise properties of polarized HFI maps at low
multipoles \cite{SROLL:2016}, but these fail to match the noise
properties of the real data at high multipoles because some important
aspects of the low-level data process (e.g.\ cosmic ray removal) are
not included self-consistently in the simulations.

\begin{figure}
\centering \includegraphics[width=75mm,angle=0]{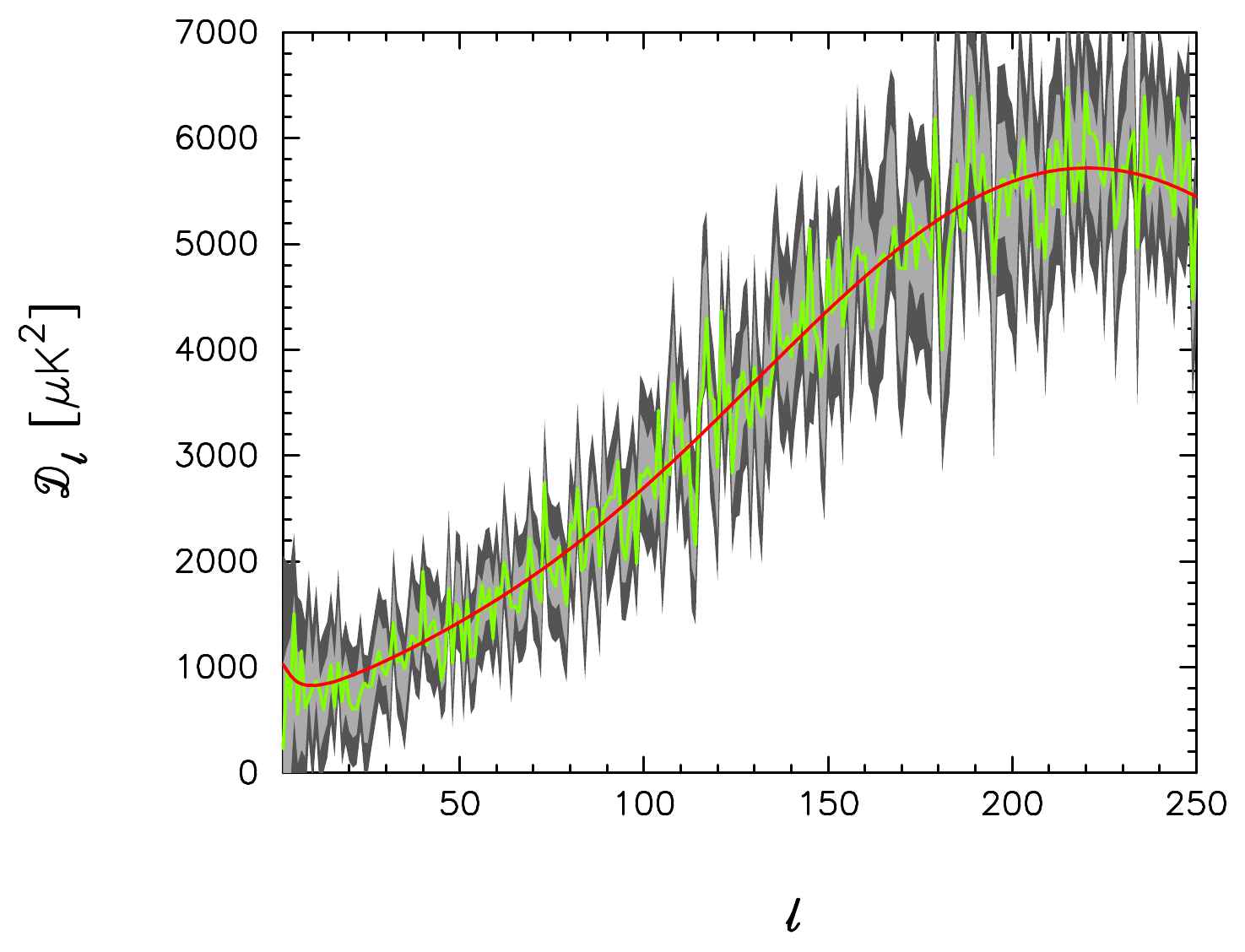}
\includegraphics[width=75mm,angle=0]{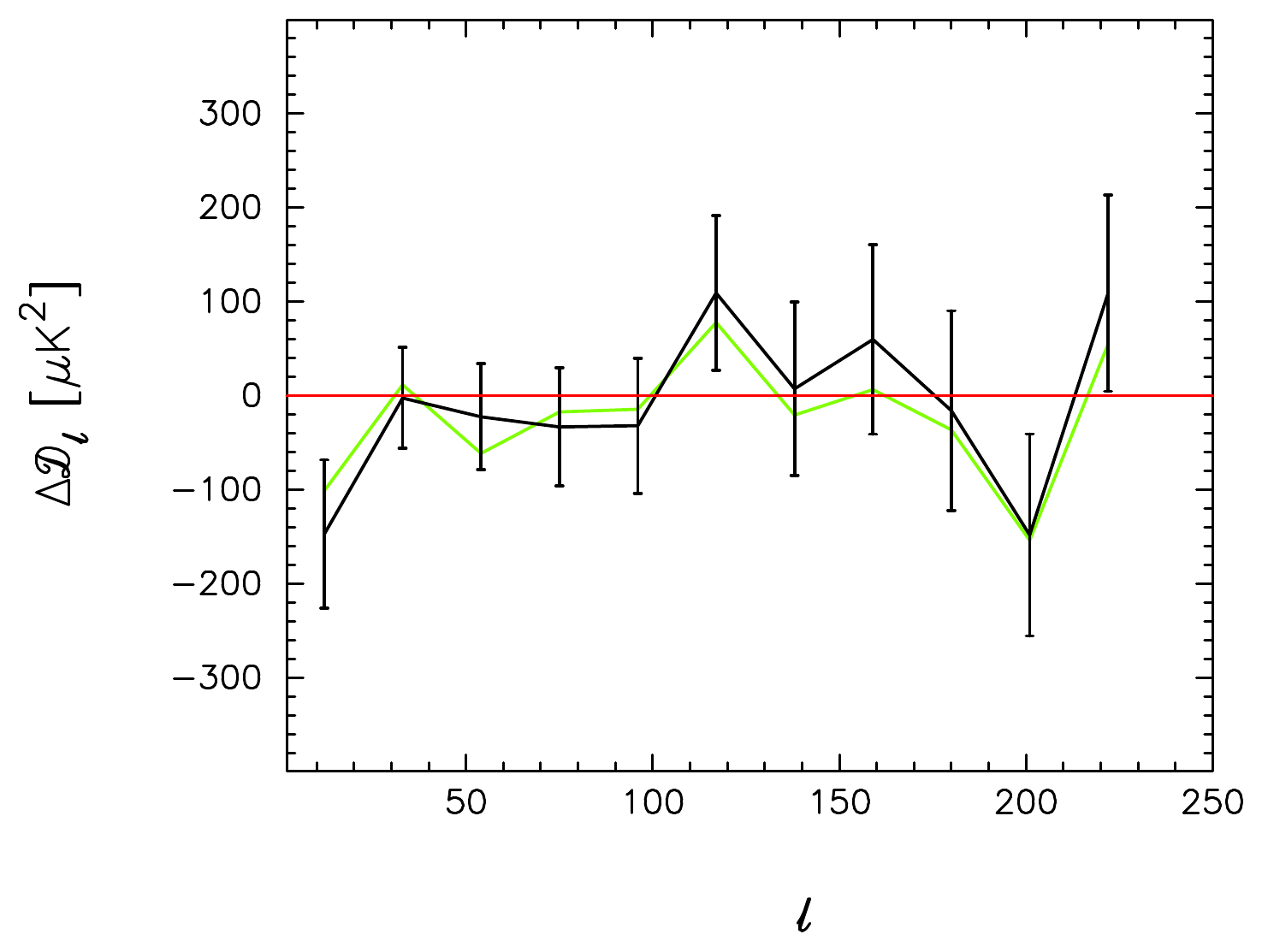}
\caption {Left hand figure: The green line shows the TT power spectrum
  plotted multipole-by-multipole from the \commander\ component
  separation algorithm. The \commander\ algorithm provides samples of component
 separated spectra which are 
  used to construct the TT likelihood at multipoles $\ell < 30$. The grey bands show the
  $1\sigma$ and $2\sigma$ ranges of the coadded foreground corrected
  \camspec\ 12.1HM  TT spectrum. The red line shows the fiducial base
  \LCDM\ model, as plotted in Fig.\ \ref{fig:12.1TT}. Right hand
  figure: Residuals of the power spectra with respect to the fiducial 
  base \LCDM\ model averaged in bands of width $\Delta \ell = 21$. The
  green line shows the \commander\ power spectrum residuals. The black
  line shows the \camspec\ residuals. \GE{The error bars show $1\sigma$ errors on the
  band-averaged \camspec\ points.} }

\label{fig:commander}

\end{figure}

Fig.\ \ref{fig:commander} compares the coadded foreground subtracted
spectrum from Fig.\ \ref{fig:12.1TT} to the \commander\ spectrum. The
\commander\ spectrum is computed using $86\%$ of the sky, which is
slightly larger than the (effective) $f^W_{\rm sky} = 70.1\%$ of sky
used at $100$ GHz in the 12.1HM likelihood (see Table
\ref{tab:sky_fractions}).  The left hand figure compares the power
spectra multipole-by-multipole up to the maximum of the first acoustic
peak. The figure to the right shows the residuals with respect to the
best-fit \LCDM\ model in band powers of width $\Delta \ell = 21$. We
have made no corrections for relative effective calibration
differences between the \commander\ and \camspec\ TT spectra. The
\camspec\ spectrum reproduces the features of the \commander\ spectrum
multipole-by-multipole even at multipoles $\le 30$. This demonstrates
that the choice of $\ell_{\rm min} = 30$ for the transition from the
\commander\ TT likelihood to \camspec\ is not particularly
critical. The \commander\ TT estimates at low multipoles have lower
variance than the PCL estimates used in \camspec. The 
reason for using the \commander\ likelihood is not primarily to improve on foreground removal,
but rather to model accurately the
non-Gaussian distributions of the power spectrum estimates at low
multipoles.

\subsection{Conditional spectra}
\label{subsec:conditional}

Having demonstrated a basic level of consistency of the TT component of the likelihood, in this subsection
we present additional tests of the coadded polarization spectra. 
Given the best fit cosmology and foreground parameters fitted to the four temperature spectra of the 12.1HM TT likelihood,
we can calculate the expected TE and EE spectra given the TT spectra.
Writing the data vector of Eq.\ \ref{equ:CL1}  as 
\begin{equation}
{\bf \hat C} = ({\bf \hat C}^{TT}_1, {\bf \hat C}^{TT}_2, \dots, {\bf \hat C}^{TT}_N, {\bf \hat C}^{TE}, {\bf \hat C}^{EE})^T = 
({\bf \hat X}_T, {\bf \hat X}_P)^T, \label{equ:CL8}
\end{equation}
(where all spectra are corrected for best-fit calibration factors) the expected value of the polarization vector given the temperature vector is
\begin{equation}
{\bf \hat X}_P = {\bf  X}^{\rm theory}_P + {\bf M}^T_{TP} {\bf M}^{-1}_T ({\bf\hat  X}_T - {\bf X}^{\rm theory}_T - {\bf X}^{\rm for}_T),  \label{equ:CL9}
\end{equation}
with covariance
\begin{equation}
{\bf \hat \Sigma}_P = {\bf M}_{P} -  {\bf M}^T_{TP} {\bf M}^{-1}_T {\bf M}_{TP}.  \label{equ:CL10}
\end{equation}
In  Eq.\ \ref{equ:CL9}, ${\bf X}^{\rm theory}_T$ and ${\bf X}^{\rm theory}_P$ are the theoretical temperature and polarization spectra
deduced from minimising the TT likelihood, and ${\bf X}^{\rm for}_T$ is the corresponding foreground/nuisance parameter solution.

\begin{figure}
\centering
\includegraphics[width=76mm,angle=0]{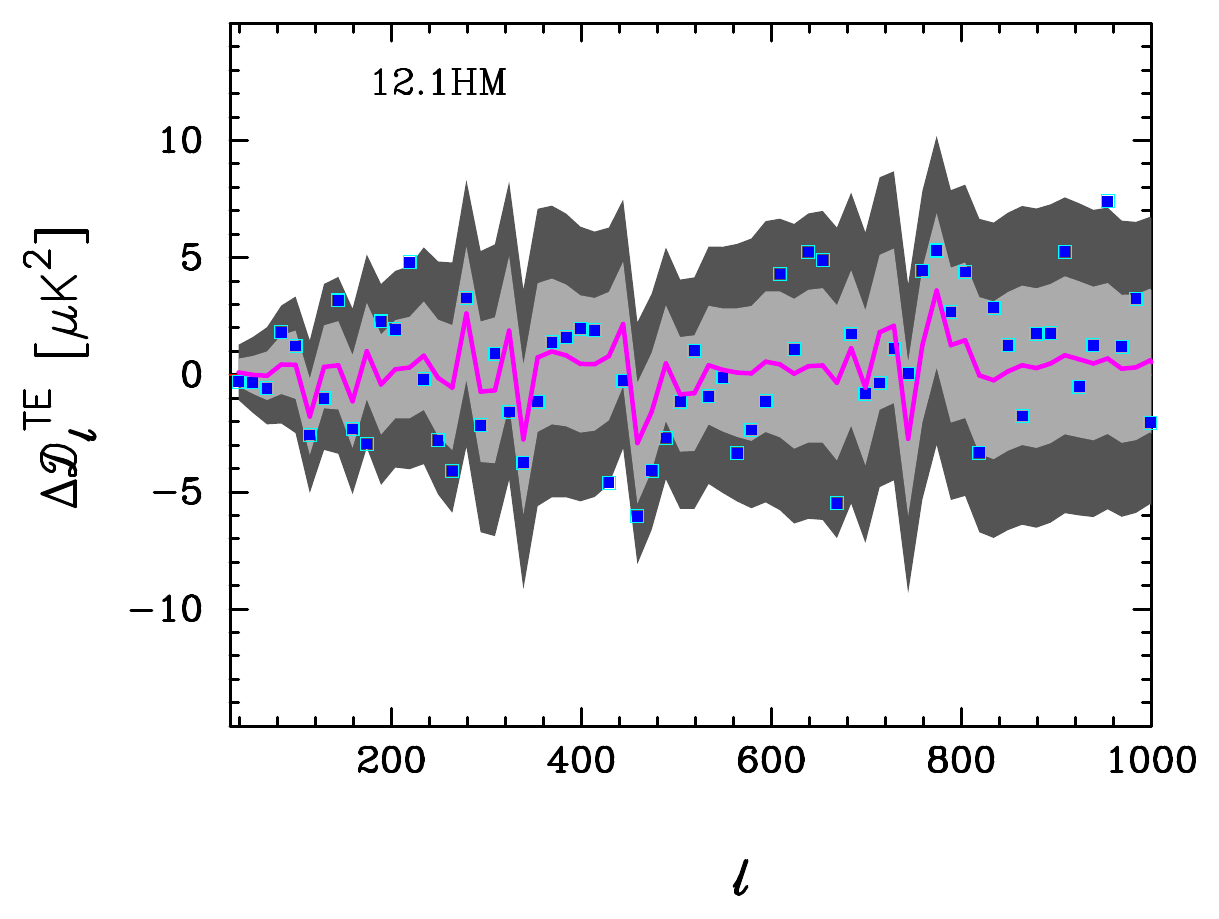} \includegraphics[width=76mm,angle=0]{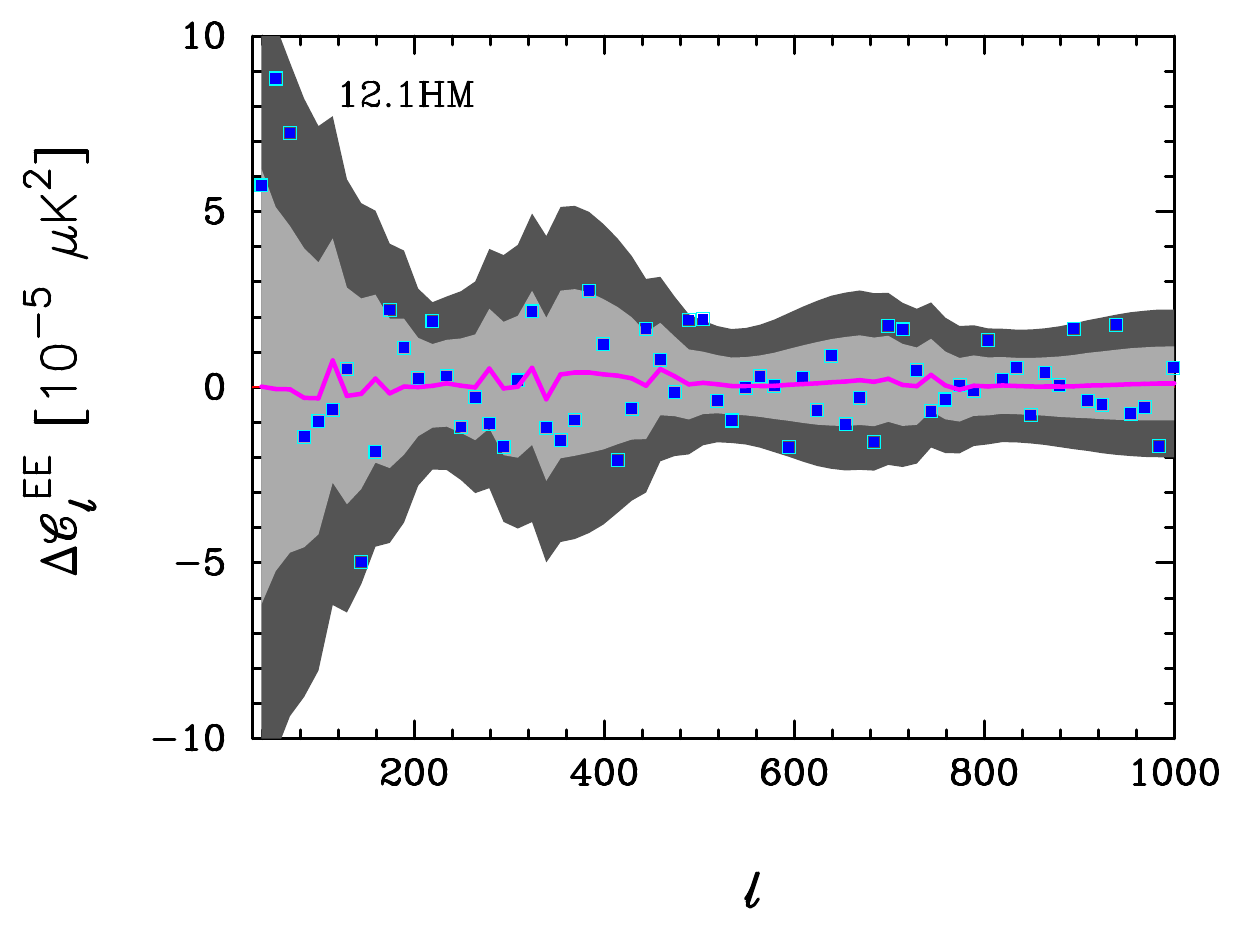} 
\caption {TE and EE  residuals with respect to the best-fit
base \lcdm\ cosmology fitted to the TT likelihood. The purple lines show the expected 
TE and EE spectra given the TT data (Eq.\ \ref{equ:CL9}). The shaded areas show the $1$ and $2\sigma$ ranges computed
from Eq.\ \ref{equ:CL10}.}

\label{fig:polconditional}

\end{figure}

Figure \ref{fig:polconditional} shows the results of applying
Eqs.\ \ref{equ:CL9} and \ref{equ:CL10}.  There is almost no correlation
between the TT, TE and EE spectra at multipoles $\ell \simgt 1000$
because the polarization spectra are dominated by noise. We therefore
plot the spectra in Fig.\ \ref{fig:polconditional} only up to $\ell =
1000$.  There is a correspondence between features in the TE spectrum
and the predicted spectrum; evidently some of the features in the TT spectrum,
for example at $\ell \approx 320$ and $\ell \approx 800$ have
correlated counterparts in TE.  In EE, however, the correlations with
the TT spectra are extremely weak. In both cases, the data points are
consistent with the error model with no obvious  outliers.
 These tests show that the polarization spectra are
statistically consistent with the TT spectra and with the base
\LCDM\ cosmology.

\vspace{0.1truein}

\section{Inter-frequency consistency in temperature}
\label{sec:inter_frequency}

  Given the high precision of the \Planck\ data, the possibility that
  unidentified systematics might be lurking within the dataset is an
  important concern.  We have already demonstrated the intra-frequency
  consistency of the detset spectra in
  Sect.\ \ref{subsec:intra_frequency}. The results of the previous
  section showed that after solving for a parameteric foreground
  model, the four TT spectra of the 12.1HM likelihood are 
  compatible with the best-fit base \LCDM\ cosmology as judged by
  $\chi^2$ statistics. In this section, we will discuss some more
  detailed consistency tests of the power spectra measured for
  different frequency combinations. This section deals exclusively
  with consistency of the temperature spectra. Inter-frequency
  consistency of the TE and EE spectra is discussed in the next
  section.

\subsection{Consistency of TT spectra in the 12.1HM half mission likelihood}
\label{subsec:inter_frequency_default}

\begin{figure}
\centering
\includegraphics[width=170mm,angle=0]{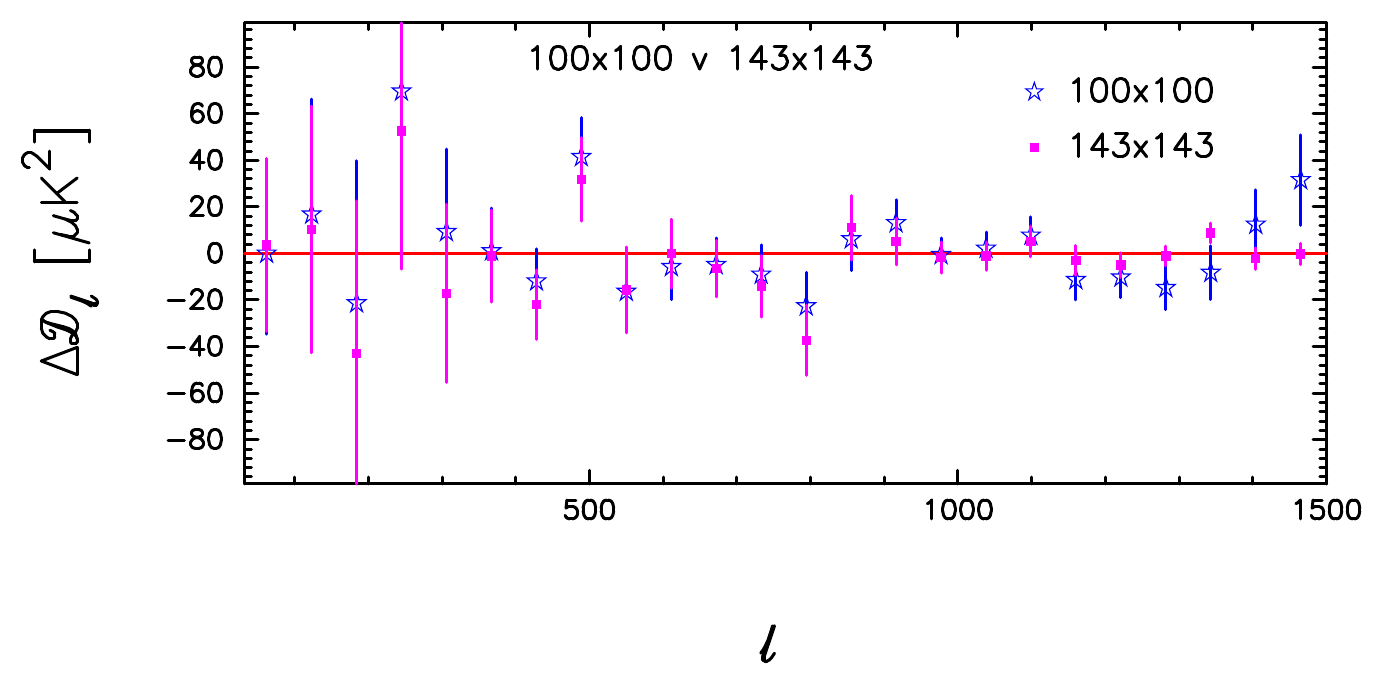} 
\caption {Residuals with respect to the best-fit base \LCDM\ cosmology and foreground model fitted to TT for the 
$100 \times 100$ and $143 \times 143$ half mission cross spectra used in the 12.1HM likelihood. 
Note that the $100 \times 100$ and $143 \times 143$ spectra  have been computed using different sky masks.}

\label{fig:inter_frequency100v143}

\end{figure}

\begin{figure}
\centering
\includegraphics[width=170mm,angle=0]{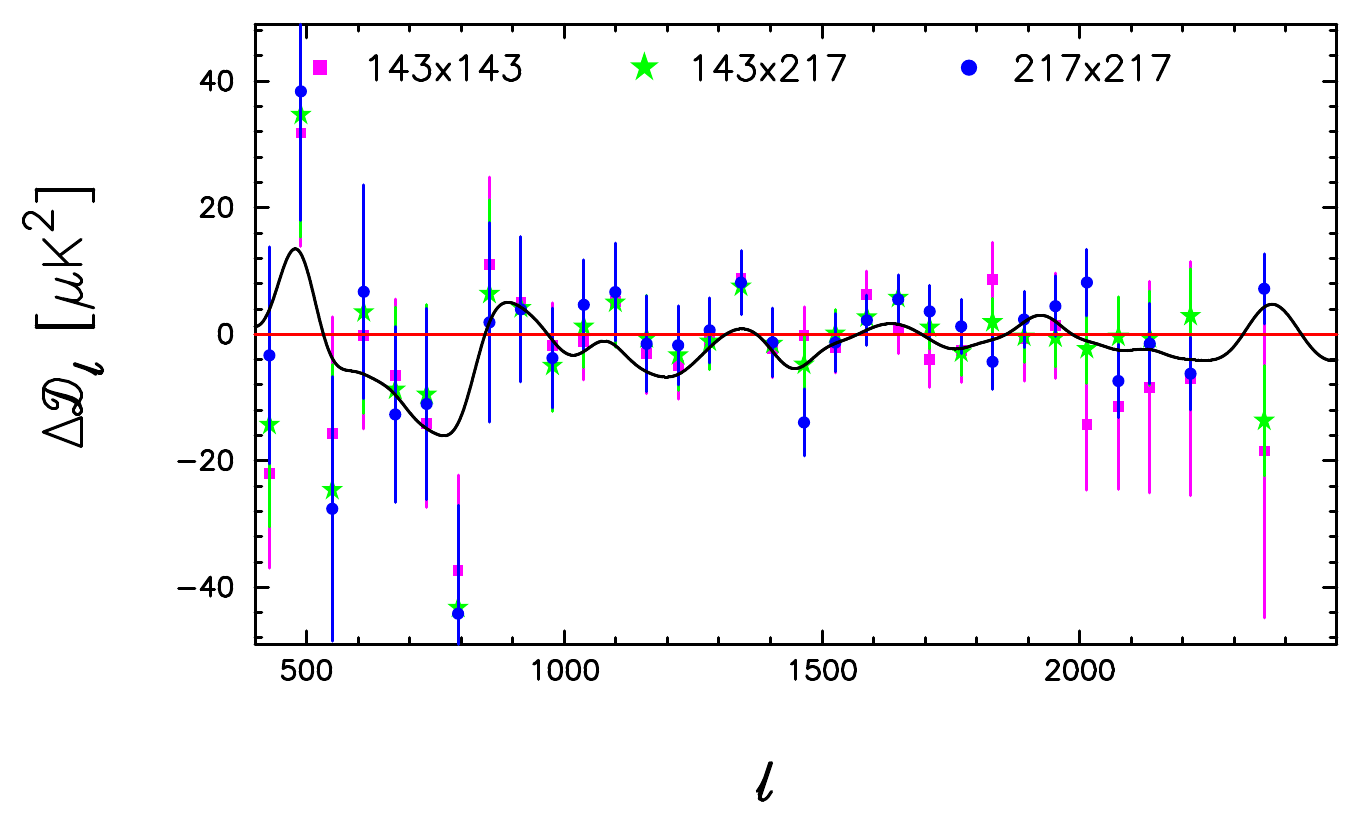} 
\caption {As Fig.\ \ref{fig:inter_frequency100v143}, but comparing residuals for the $143\times 143$, $143 \times 217$ and
$217 \times 217$ spectra used in the 12.1HM likelihood. The black line shows the residuals of the maximum likelihood coadded spectrum of Fig.
\ref{fig:12.1TT} smoothed with a Gaussian of width $\sigma_\ell = 40$. }

\label{fig:inter_frequency143v217}

\end{figure}

Figure \ref{fig:inter_frequency100v143} compares the foreground corrected $100 \times 100$ and $143 \times 143$
half mission spectra used in the 12.1HM likelihood. The residuals at multipoles $\ell \simlt 500$, where dust dominates
the foreground model, are small (differing by a maximum of $26\  \mu{\rm K}^2$ at $\ell =306$ for the band-powers shown in the figure, which have $\Delta \ell=31$),  demonstrating the consistency of the model for dust subtraction. These differences
are similar to those seen in Fig.\ \ref{fig:commander} comparing the \commander\ spectrum with the 
coadded foreground-corrected 12.1HM \camspec\ spectrum of Fig.\ \ref{fig:12.1TT}.  Typically, the consistency of the foreground corrected \camspec\
spectra is no better than $\sim 30 \ \mu{\rm K}^2$ at the maximum of the first acoustic peak (i.e.\ consistent to $\sim 0.5 \%$), 
though we see better consistency at these multipoles if we clean the spectra using 545 GHz and apply identical  masks at each frequency
(see Fig.\ \ref{fig:bands3_12_5}).  Having demonstrated the consistency of the
$100 \times 100$ and $143 \times 143$ spectra, in the remainder of
this section we concentrate on the consistency of the $143\times 143$, $143\times 217$ and $217 \times 217$ spectra.

\begin{table}

\small
\centering{
\caption{\small{Band-power residuals, $ \Delta \hat D_{\ell_b}$, with respect to the fiducial base \LCDM\ cosmology and foreground 
model for the $143 \times 143$, $143 \times 217$ and $217 \times 217$ spectra used in the 12.1HM \camspec\
likelihood. $\ell_b$ is the multipole at midpoint of the band. The columns labelled `error' give the $1\sigma$ uncertainties on 
$ \Delta \hat D_{\ell_b}$ assuming that the best-fit cosmology plus foreground model is exact.
 The columns labelled $N_\sigma$ give the number of standard deviations by which $\Delta \hat  D_{\ell_b}$ differs from zero. Entries which differ from zero by more than
$2 \sigma$ are coloured in red.}}
\label{tab:TT_residuals}

\begin{center}
  \begin{large}
    
\begin{tabular}{|c|c|c|c|c|c|c|c|c|c|}
\hline   
          & \multicolumn{3}{c|}{$143 \times 143$} & \multicolumn{3}{c|}{$143 \times 217$} &\multicolumn{3}{c|}{$217 \times 217$} \cr \hline
$\ell_b$  & $\Delta \hat D_{\ell_b}$  & error  &  $N_\sigma$   & $\Delta \hat D_{\ell_b}$  & error  &  $N_\sigma$   & $\Delta\hat D_{\ell_b}$  & error  &  $N_\sigma$   \cr \hline
     62&      3.66&     37.15&      0.10&      1.34&     40.37&      0.03&     36.46&     43.47&      0.84\cr
    123&     10.34&     53.01&      0.20&      8.97&     56.86&      0.16&      5.49&     60.12&      0.09\cr
    184&    -43.05&     65.60&     -0.66&    -33.25&     70.09&     -0.47&    -23.16&     73.70&     -0.31\cr
    245&     52.60&     59.33&      0.89&     85.82&     63.41&      1.35&    109.56&     66.65&      1.64\cr
    306&    -17.13&     38.26&     -0.45&    -37.51&     40.98&     -0.92&    -42.47&     43.29&     -0.98\cr
    367&     -1.13&     19.80&     -0.06&    -11.37&     21.32&     -0.53&    -13.52&     22.75&     -0.59\cr
    428&    -22.01&     14.94&     -1.47&    -14.34&     16.12&     -0.89&     -3.32&     17.16&     -0.19\cr
    489&     31.84&     17.96&      1.77&     34.68&     19.33&      1.79&     38.41&     20.40&      1.88\cr
    550&    -15.69&     18.46&     -0.85&    -24.61&     19.85&     -1.24&    -27.58&     20.91&     -1.32\cr
    611&     -0.12&     14.83&     -0.01&      3.48&     15.96&      0.22&      6.71&     16.89&      0.40\cr
    672&     -6.55&     12.12&     -0.54&     -8.77&     13.05&     -0.67&    -12.68&     13.89&     -0.91\cr
    733&    -14.08&     13.25&     -1.06&     -9.55&     14.25&     -0.67&    -11.00&     15.12&     -0.73\cr
    794&    \color{red} -37.33&  \color{red}   15.08&  \color{red}    -2.48&    \color{red}-43.22&  \color{red}   16.19&  \color{red}    -2.67&   \color{red} -44.15& \color{red}    17.12&  \color{red}   -2.58\cr
    855&     11.01&     13.85&      0.80&      6.37&     14.86&      0.43&      1.88&     15.76&      0.12\cr
    916&      5.09&      9.97&      0.51&      4.14&     10.70&      0.39&      3.95&     11.48&      0.34\cr
    977&     -1.73&      6.70&     -0.26&     -5.00&      7.20&     -0.69&     -3.75&      7.88&     -0.48\cr
   1038&     -1.19&      6.01&     -0.20&      1.21&      6.43&      0.19&      4.65&      7.12&      0.65\cr
   1099&      5.16&      6.59&      0.78&      4.99&      7.03&      0.71&      6.69&      7.75&      0.86\cr
   1160&     -2.98&      6.41&     -0.46&     -0.66&      6.80&     -0.10&     -1.49&      7.54&     -0.20\cr
   1221&     -4.98&      5.24&     -0.95&     -3.33&      5.51&     -0.60&     -1.74&      6.24&     -0.28\cr
   1282&     -1.12&      4.26&     -0.26&     -1.17&      4.42&     -0.27&      0.62&      5.14&      0.12\cr
   1343&    \color{red}  8.76&  \color{red}    4.23&  \color{red}    2.07&      7.53&      4.32&      1.74&      8.17&      5.07&      1.61\cr
   1404&     -2.24&      4.60&     -0.49&     -1.53&      4.63&     -0.33&     -1.28&      5.42&     -0.24\cr
   1465&     -0.25&      4.56&     -0.05&     -4.80&      4.47&     -1.07&  \color{red}  -13.92&   \color{red}   5.29& \color{red}    -2.63\cr
   1526&     -2.02&      4.06&     -0.50&      0.10&      3.79&      0.03&     -1.32&      4.60&     -0.29\cr
   1587&      6.27&      3.69&      1.70&      2.66&      3.19&      0.83&      2.22&      3.99&      0.56\cr
   1648&      0.80&      3.86&      0.21&      5.76&      3.11&      1.85&      5.51&      3.88&      1.42\cr
   1709&     -3.98&      4.42&     -0.90&      1.04&      3.36&      0.31&      3.62&      4.11&      0.88\cr
   1770&     -2.52&      5.07&     -0.50&     -2.88&      3.60&     -0.80&      1.26&      4.29&      0.29\cr
   1831&      8.72&      5.80&      1.50&      1.90&      3.78&      0.50&     -4.34&      4.35&     -1.00\cr
   1892&     -0.59&      6.82&     -0.09&     -0.37&      4.07&     -0.09&      2.34&      4.46&      0.53\cr
   1953&      1.35&      8.34&      0.16&     -0.53&      4.61&     -0.11&      4.46&      4.78&      0.93\cr
   2014&    -14.22&     10.42&     -1.36&     -2.34&      5.39&     -0.43&      8.18&      5.26&      1.55\cr
   2075&    -11.39&     13.15&     -0.87&     -0.40&      6.33&     -0.06&     -7.38&      5.79&     -1.27\cr
   2136&     -8.35&     16.70&     -0.50&     -0.63&      7.45&     -0.08&     -1.48&      6.35&     -0.23\cr
   2215&     -7.01&     18.50&     -0.38&      2.89&      7.45&      0.39&     -6.22&      5.73&     -1.09\cr
   2358&    -18.44&     26.38&     -0.70&    -13.67&      8.72&     -1.57&      7.19&      5.53&      1.30\cr

\hline
\end{tabular}

    \end{large}
\end{center}}

\end{table}

\begin{figure}
\vspace{-0.2in}
\centering
\includegraphics[width=100mm,angle=0]{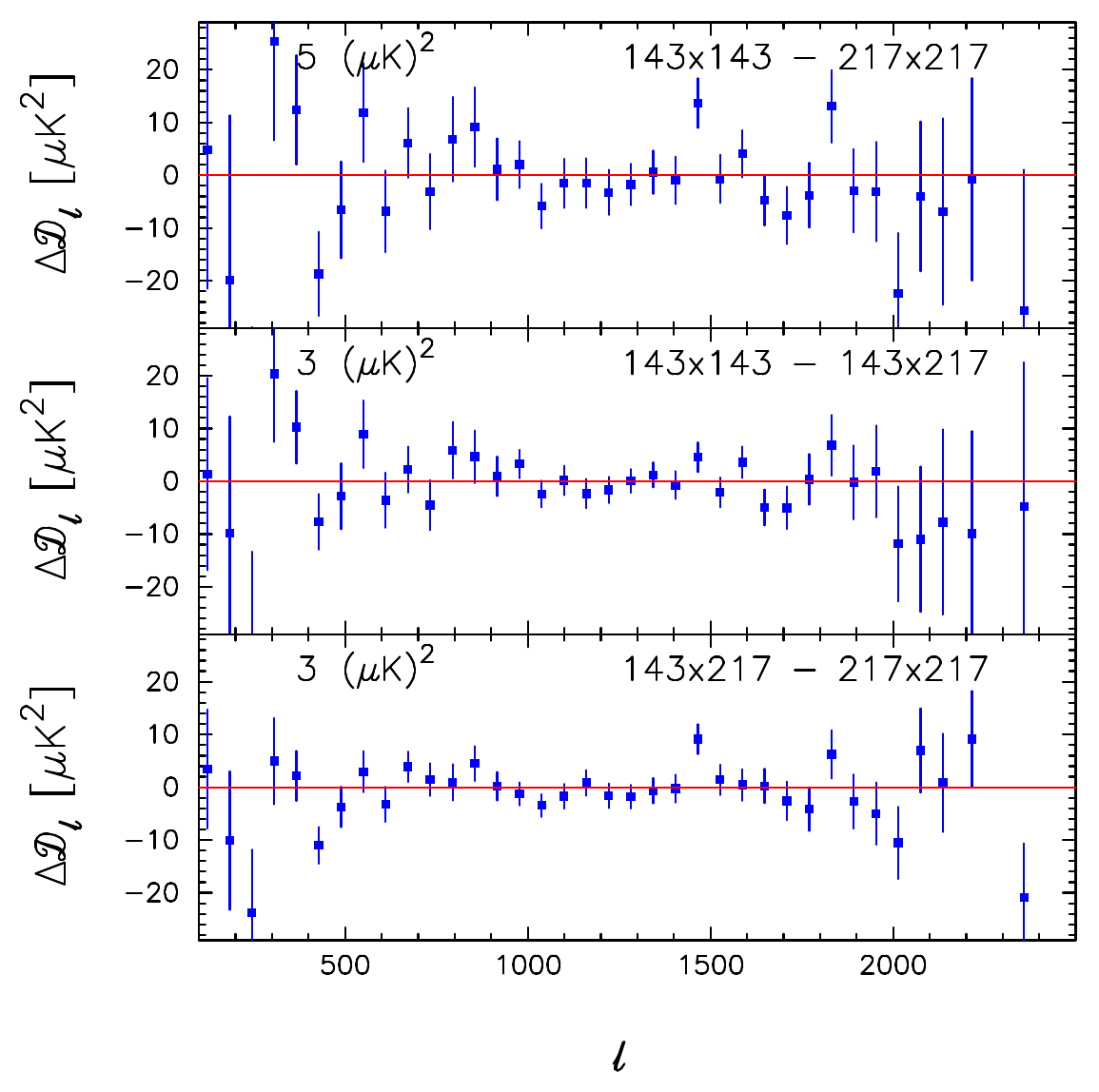} \\
\includegraphics[width=100mm,angle=0]{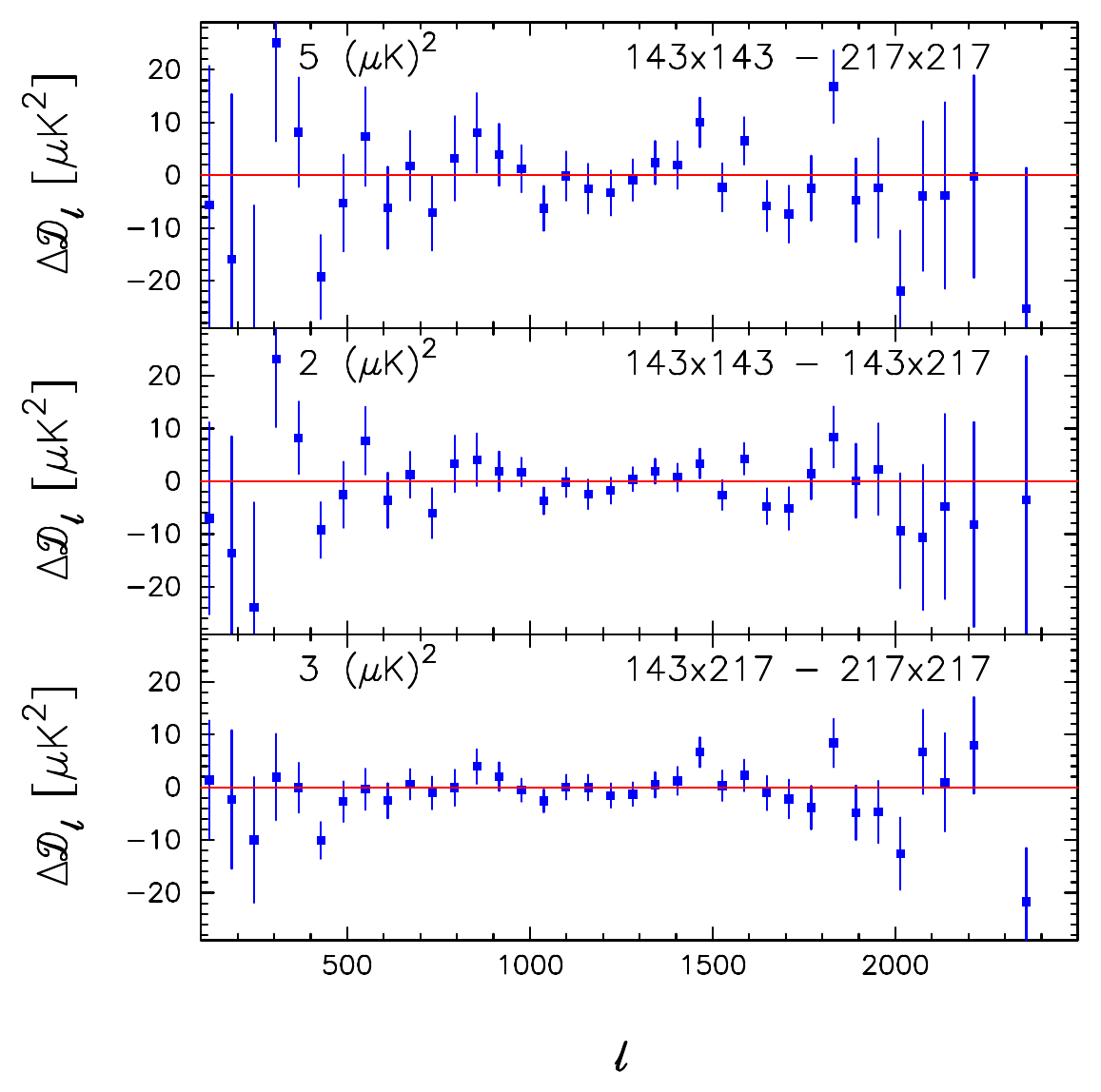} 
\caption {Differences in the foreground corrected  temperature power spectra used in  the 12.1HM likelihood  (upper figure) and for the $545$ GHz cleaned  spectra used in the 12.1HMcl likelihood (lower figure). We use the foreground solution fitting the base \LCDM\
model to the 12.1HM TT and 12.1HMcl TT likelihoods respectively.
Upper panels show the difference between the $143\times 143$ and $217 \times 217$ spectra,
middle panels show the difference between the $143 \times 217$ and $143 \times 143$ spectra, and the lower 
panels show the difference between the $143 \times 217$ and $217 \times 217$ spectra. The error bars for the
power spectrum differences are computed from linear combinations of the \camspec\ covariance matrices. The numbers
in each panel give the rms residuals of the bandpower differences over the multipole range $800 \le \ell \le 1500$.}

\label{fig:bands3}

\end{figure}

The residuals for the remaining three spectra are compared graphically
in Fig.\ \ref{fig:inter_frequency143v217} and listed numerically in
Table \ref{tab:TT_residuals}. Residuals that differ from zero by more
than $2 \sigma$ {\it assuming that the cosmology and foreground model
  are known exactly} are marked in red in Table
\ref{tab:TT_residuals}. The three spectra are in very good
agreement. The nearly $\sim 2\sigma$ upward fluctuation at $\ell_b
\approx 489$ and $\sim 2.5\sigma$ downward fluctuation at $\ell_b =794$ (which have been noted by some theorists e.g.\ \citep{Chen:2014})
are reproduced in all of the \Planck\ spectra and are clearly real
features of the primordial CMB spectrum. The general oscillatory
patterns in the residuals, which various authors (e.g.\ 
\citep{Henning:2018}) have claimed might be related to an
inconsistency in the lensing smoothing of the acoustic peaks (and
related to the high value of $A_L$ measured from \Planck\ TT, see
Fig.\ \ref{fig:inter_frequency143v217cleaned}) are also reproduced
across the three spectra (see also Fig.\ 12 of \cite{Galli:2017}). The most deviant point in
Fig.\ \ref{fig:inter_frequency143v217} is for the $217 \times 217$
spectrum in the band centred at $\ell = 1469$ (as noted in
PPL15 and PCP18). This band power deviates from zero by $2.63 \sigma$ on the
assumption that the cosmology and foreground model is known exactly
(which is, of course, not true).

The residuals and errors in Fig.\ \ref{fig:inter_frequency143v217} are
dominated by cosmic variance at multipoles $\ell \simlt 1500$. A more
sensitive consistency test is provided by differencing the power
spectra, so reducing cosmic variance and sensitivity to the
cosmological model. This is illustrated for the 12.1HM temperature
spectra in the upper plot in Fig.\ \ref{fig:bands3}. Note that: (a)
the errors on these spectral differences are constructed by forming
linear combinations of the \camspec\ covariance matrices; (b) cosmic
variance is not completely eliminated because we use different masks
at 143 and 217 GHz; (c) the errors do not accurately model
CMB-foreground cross-correlations, as discussed in
Sect.\ \ref{subsec:dust_power}.  The agreement between the spectra is
generally excellent. Nevertheless there are some outliers which might
appear to be statistically unlikely.  For example, in the panel
showing the $143 \times 143 - 217 \times 217$ difference, there are
outliers at $\ell_b = 428$ ($-2.35\sigma$) and  $\ell_b = 1465$
($2.92\sigma$). In the panel showing the $143 \times 217 -
217\times217$ difference, there are outliers in exactly the same
bands,  i.e.\  at $\ell_b = 428$ ($-3.16\sigma$) and $\ell_b = 1465$
($3.23\sigma$). In the central panel, showing the $143\times 143 - 143
\times 217$ spectrum differences, all of the bandpowers are consistent
with zero to within $2\sigma$. Our interpretation of these outliers is as
follows:

\smallskip
\noindent
(i) None of the outliers have high statistical significance. Statistically acceptable variations in the foreground model 
lead to tilts in the foreground corrected spectra at high multipoles ($\ell \simgt 1000$) and this can alter the statistical significance of a single bandpower residual by up to $\sim 1\sigma$.

\smallskip
\noindent
(ii) The amplitude of the CMB-foreground correlations is highest for
the $217\times 217$ spectra and adds to the variance of this spectrum. 

\smallskip
\noindent
(iii) At low multipoles ($\ell \simlt 1000$), the dust correction for the $217\times 217$ spectrum is large and inaccurate
at the $\sim 10-30 \ \mu{\rm K}^2$ level.
The residuals in the 143$\times$143 - 217$\times$217 and 143$\times$217 - 217$\times$217 spectra at $\ell_b = 428$ 
arise from inaccurate dust subtraction in the $217 \times 217$ spectrum. This is additional motivation to exclude the
$217 \times 217$ spectrum at $\ell < 500$ from the \camspec\ likelihood.

The clearest way of demonstrating these points is to repeat these inter-frequency comparisons using a completely different
model of dust cleaning and extragalactic foregrounds.

\subsection{Cleaned temperature  spectra}
\label{subsec:cleaned_TT}

Section \ref{subsec:spec_clean} introduced `cleaned' temperature
spectra using $353$, $545$ and $857$ GHz maps as Galactic dust
templates and demonstrated that high frequency cleaning accurately
removes Galactic dust and also much of the CIB.  In PCP15 and PCP18,
we formed `cleaned' likelihoods using $545$ GHz as a high frequency
template. As discussed in Sect.\  \ref{subsec:spec_clean} we focus on
545 GHz temperature cleaning in the rest of this paper, since there is a
significant signal-to-noise penalty if $353$ GHz is used as a
template.  (Note that using $353$ GHz temperature cleaning leads to almost
identical results to those presented here, though with additional
noise at $\ell \simgt 2000$.)

\begin{figure}
\centering
\vskip -12mm
\includegraphics[width=80mm,angle=0]{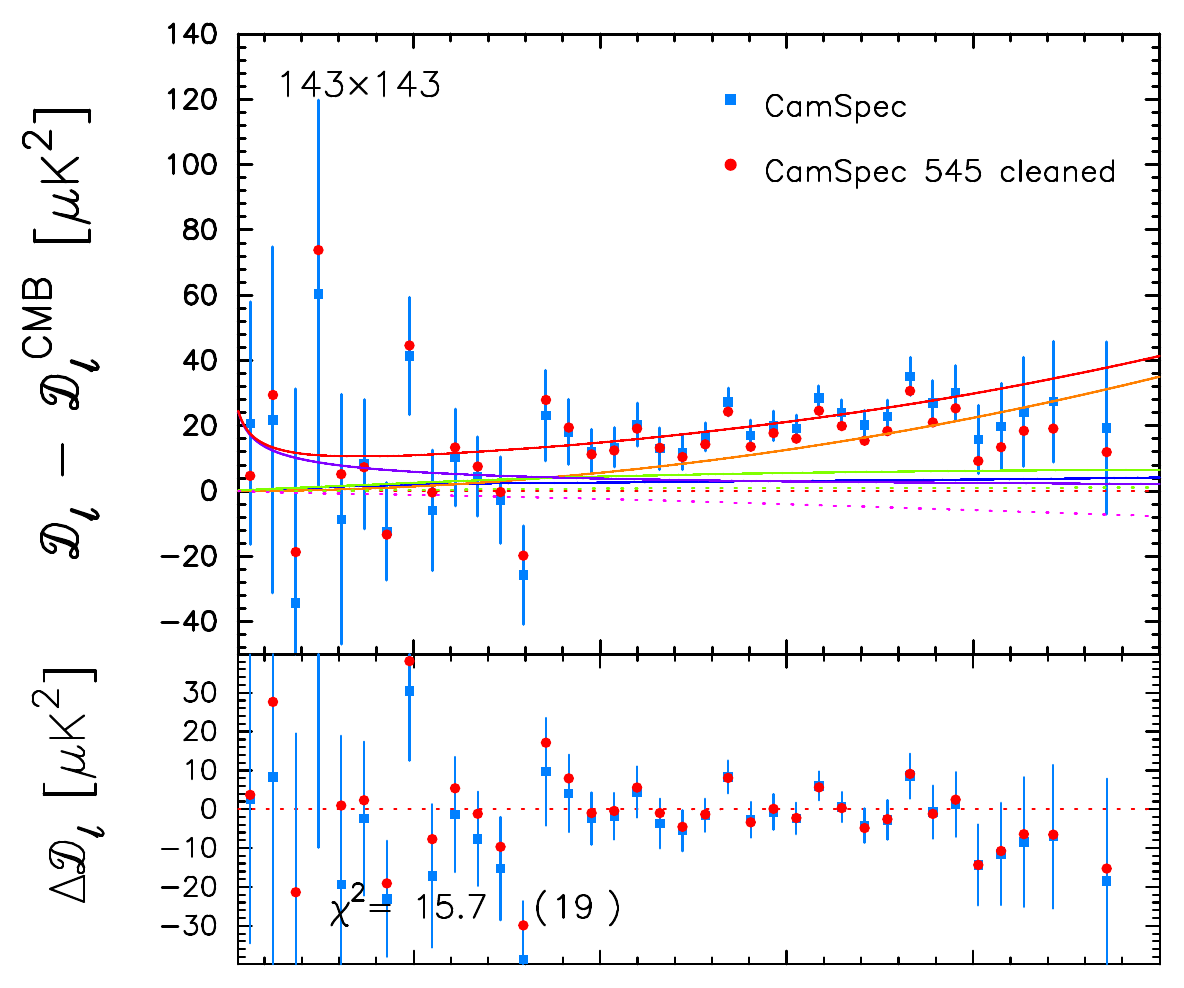} \\
\vskip -3mm
\includegraphics[width=80mm,angle=0]{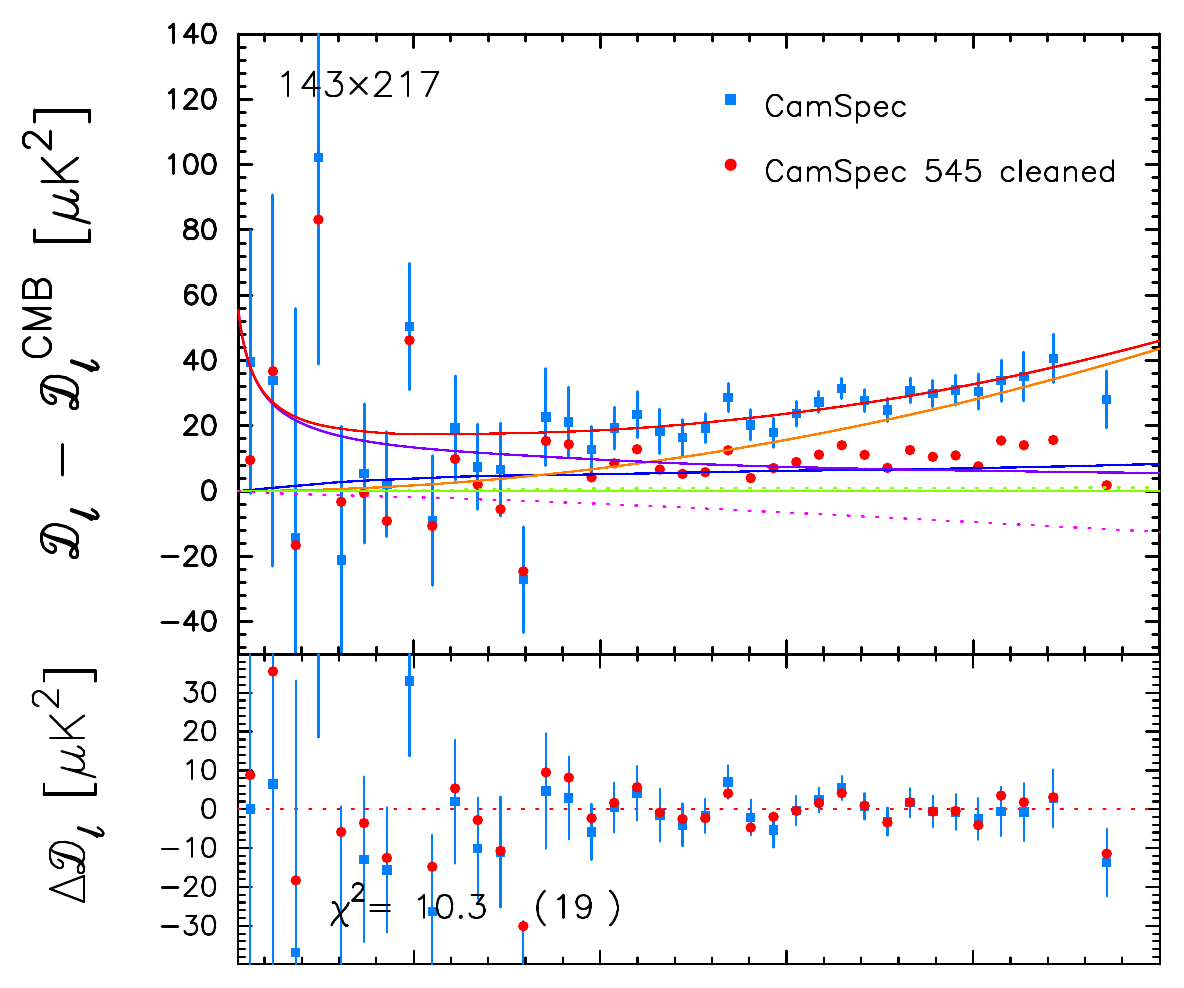} \\
\vskip - 3mm
\hskip 3mm \includegraphics[width=82mm,angle=0]{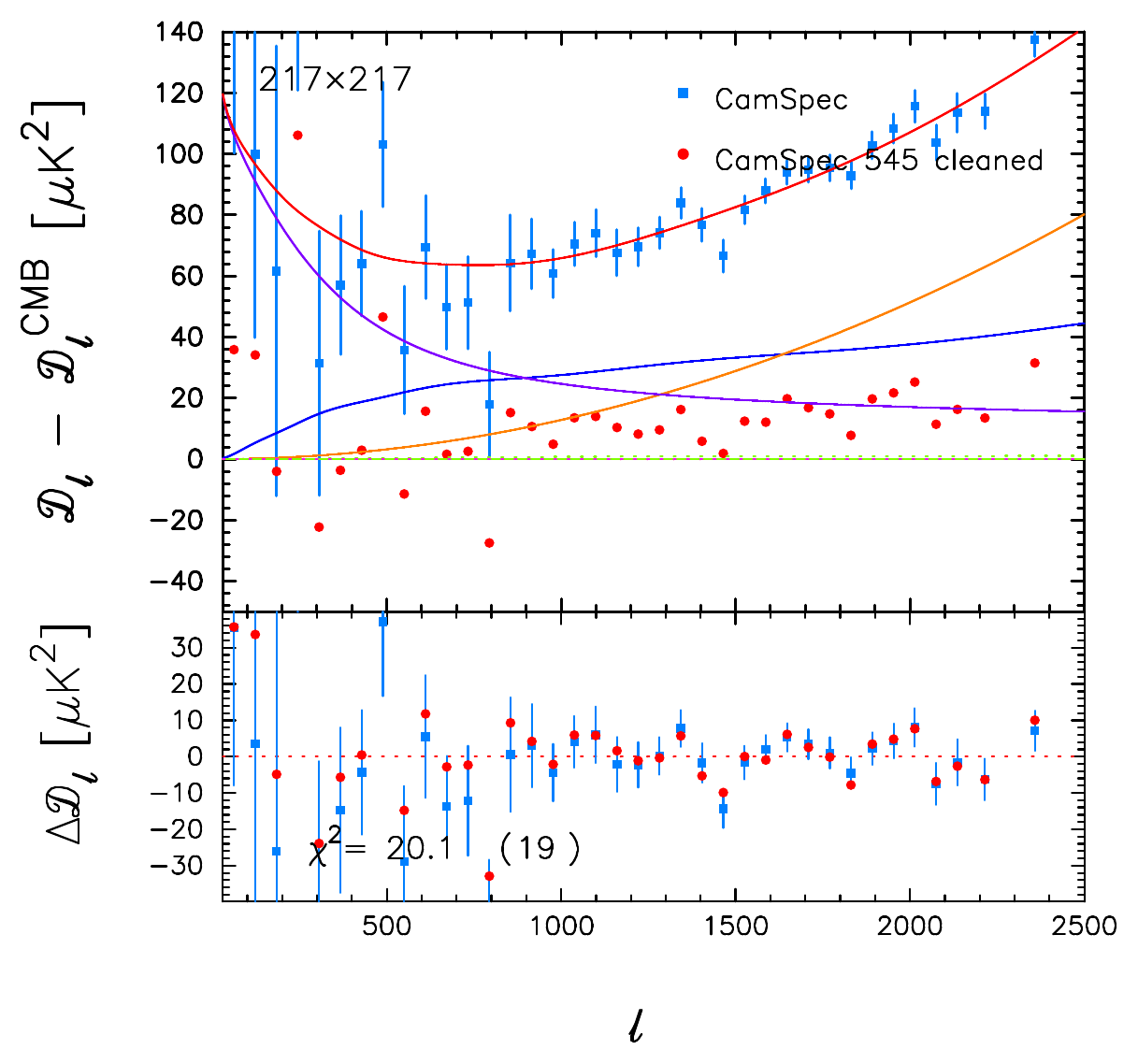}

   \caption {Comparison of the 12.1HM spectra (blue points) and $545$
     cleaned spectra (red points).  The upper plot in each figure
     shows the residuals with respect to the fiducial base
     \LCDM\ model. Major foregrounds are shown by the solid lines
     colour coded as follows: total foreground spectrum (red); Poisson
     point sources (orange), clustered CIB (blue); thermal SZ (green)
     and Galactic dust (purple). Minor foreground components are shown
     by the dotted lines colour coded as follows: kinetic SZ (green)
     and tSZxCIB (purple). The lower plots in each figure show the spectra after
     subtraction of the best-fit foreground model. For the cleaned
     spectra we adopt power-laws to model residual foregrounds, as
     described in Sect.\ \ref{subsection:cleaned_likelihood_nuisance}.  The $\chi^2$ values of the residuals of
     the blue band powers over the multipole range $1000 \le \ell \le 2200$, and the number of band powers, are listed in the
     lower panels.}

\label{fig:545cleanedTT}
\end{figure}

Figure \ref{fig:545cleanedTT} compares the 545 GHz cleaned spectra
with the 12.1HM temperature spectra. (As explained in
Sect.\ \ref{subsection:cleaned_likelihood_nuisance}, we exclude the
$100 \times 100$ spectrum from the `cleaned' likelihoods.) The upper
plots in each panel of Fig.  \ref{fig:545cleanedTT} show the difference of the spectra and the
fiducial base \LCDM\ model fitted to the 12.1HM TT likelihood. These
panels illustrate the effectiveness of 545 GHz cleaning. For all three
spectra, Galactic dust emission is accurately removed in the cleaned
spectra (as discussed in Sect.\ \ref {subsec:universality}).  545 GHz
cleaning also removes much of the extragalactic foreground at high
multipoles in the $217\times217$ and $143\times 217$ spectra (which are
dominated by the clustered and unclustered CIB). 545 GHz cleaning is
much less effective at removing extragalactic foregrounds in the $143
\times 143$ spectrum. Nevertheless, the most striking result in
Fig.\ \ref{fig:545cleanedTT} is the very close agreement between the
cleaned and uncleaned residuals of the foreground corrected spectra
plotted in the lower plots in each panel. (Note that for the 545 GHz cleaned
spectra, we have used the foreground solution determined from the
12.1HMcl TT likelihood fitted to base \LCDM.)

The 12.1HMcl and 12.1HM likelihoods give almost identical solutions
for the base \LCDM\ cosmology (see
Sect.\ \ref{subsubsec:base_consistency_TP}), even though the
foreground models used in the two likelihoods are very different. This
is perhaps the clearest demonstration that the cosmological results
from \Planck\ for \lcdm-like models are insensitive to unresolved foregrounds.

Although $545$ GHz temperature cleaning has no significant effect on
cosmology, it does have an impact on the inter-frequency
residuals. This can be seen from the lower figure in
Fig.\ \ref{fig:bands3} showing the foreground cleaned spectral
differences. At low multipoles, 545 GHz cleaning removes the
CMB-foreground correlations and so the spectra are more consistent at
$\ell \simlt 500$.  The $143\times 217 - 217 \times 217$ difference in 
the band centred at $\ell_b = 1465$ deviates from zero by $2.36\sigma$
for the cleaned spectra instead of $3.23 \sigma$ for the uncleaned spectra. 
Inter-frequency residuals are therefore sensitive to small differences in the foreground 
solution (and to the modelling of foreground errors in the covariance matrix). This needs
to be borne in mind when comparing inter-frequency residuals. For the 545 GHz cleaned
spectra, there is no evidence for unusual differences (i.e.\  $>2.5 \sigma$)
between the three spectra over the multipole ranges used in the 12.1HMcl likelihood.

\begin{figure}
\centering
\includegraphics[width=150mm,angle=0]{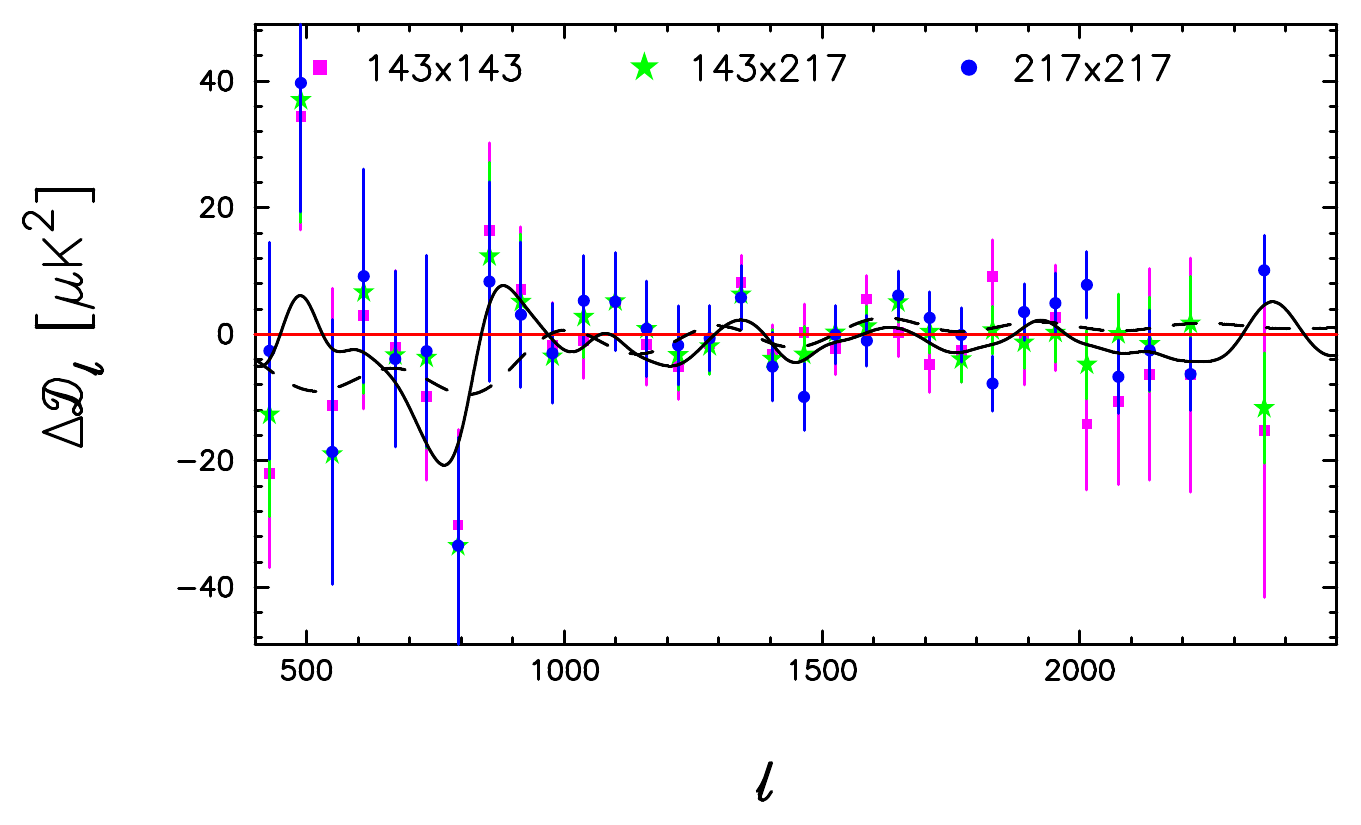} 
\caption {As Fig.\ \ref{fig:inter_frequency143v217}, but for spectra cleaned with 
 $545$ GHz as 
used in the 12.1HMcl likelihood. The residuals here are computed with respect to the best fit base \LCDM\
cosmology + foreground model derived from the 12.1HMcl TT likelihood.  
The black line shows the residuals of the maximum likelihood frequency coadded spectrum 
 smoothed with a Gaussian of width $\sigma_\ell =40$.  The dashed line shows the 
difference between the best-fit $A_L$ model (with $A_L = 1.14$) and the fiducial model 
smoothed with a Gaussian of width $\sigma_\ell =40$.}

\label{fig:inter_frequency143v217cleaned}

\end{figure}

Figure \ref{fig:inter_frequency143v217cleaned} shows the
inter-frequency residuals for the 12.1HMcl $545$ GHz cleaned
spectra. This figure can be compared with the equivalent plot
(Fig.\ \ref{fig:inter_frequency143v217}) for the uncleaned spectra. We
see a similar pattern of residuals and slightly (but barely
perceptible) improved consistency between the three spectra. The
general pattern of the residuals in Figs.\ \ref{fig:inter_frequency143v217} and
\ref{fig:inter_frequency143v217cleaned}  is consistent between frequencies,
despite the very different approaches to foreground modelling.
In particular, the oscillatory features seen in these plots
at multipoles $\simlt 2000$ are reproducible in all of these spectra
(and across detectors within a fixed frequency band,  cf. Fig.\ \ref{fig:ifresiduals}).

One of the unusual aspects of the \Planck\ TT data, evident since the
2013 data release, is the favouritism for high values of the
phenomenological lensing parameter, $A_L$. This issue has been
discussed at length in PCP15, \citep{Galli:2017} and PCP18 and will be
revisited in Sect.\ \ref{subsec:AL}.  The dashed line in Fig.\
\ref{fig:inter_frequency143v217cleaned} shows the best-fit $A_L$ model
fitted to the 12.1HMcl TT likelihood, for which $A_L=1.14$. Fig.\
\ref{fig:inter_frequency143v217cleaned} (which can be compared to
Fig.\ 24 in PCP18) shows that the oscillatory residuals of the $A_L$
model over the multipole range $800 - 1800$ correlate with the
residuals seen in both the cleaned and uncleaned likelihood. The
tendency for \Planck\ to favour high values of $A_L$ is driven by
features in this range of multipoles, which are consistent across the
frequency range, and not by features exclusive to the $217 \times 217$
spectrum.  As noted in PCP18, models with positive curvature show a
similar pattern of residuals to $A_L$ and so are also favoured by the
TT data. As far as we can see, the favouritism for high values of
$A_L$ is a modest statistical fluctuation.  We have found no evidence
to suggest that this result is driven by systematic errors in the
\Planck\ data. Sections \ref{subsec:AL} and \ref{subsec:curvature} will
discuss results for $A_L$ and $\omegak$ in further detail, including
constraints from the TE and EE spectra. First, we will discuss the
behaviour of the temperature spectra as a function of sky coverage.
Our aim is to create more powerful likelihoods than the 12.1HM pair by
using more sky and to quantify what happens to parameters such as
$A_L$ and $\omegak$.

\subsection{Residuals as a function of sky coverage}
\label{subsec:sky_coverage}

\begin{figure}
\vskip -10mm
\centering
\includegraphics[width=115mm,angle=0]{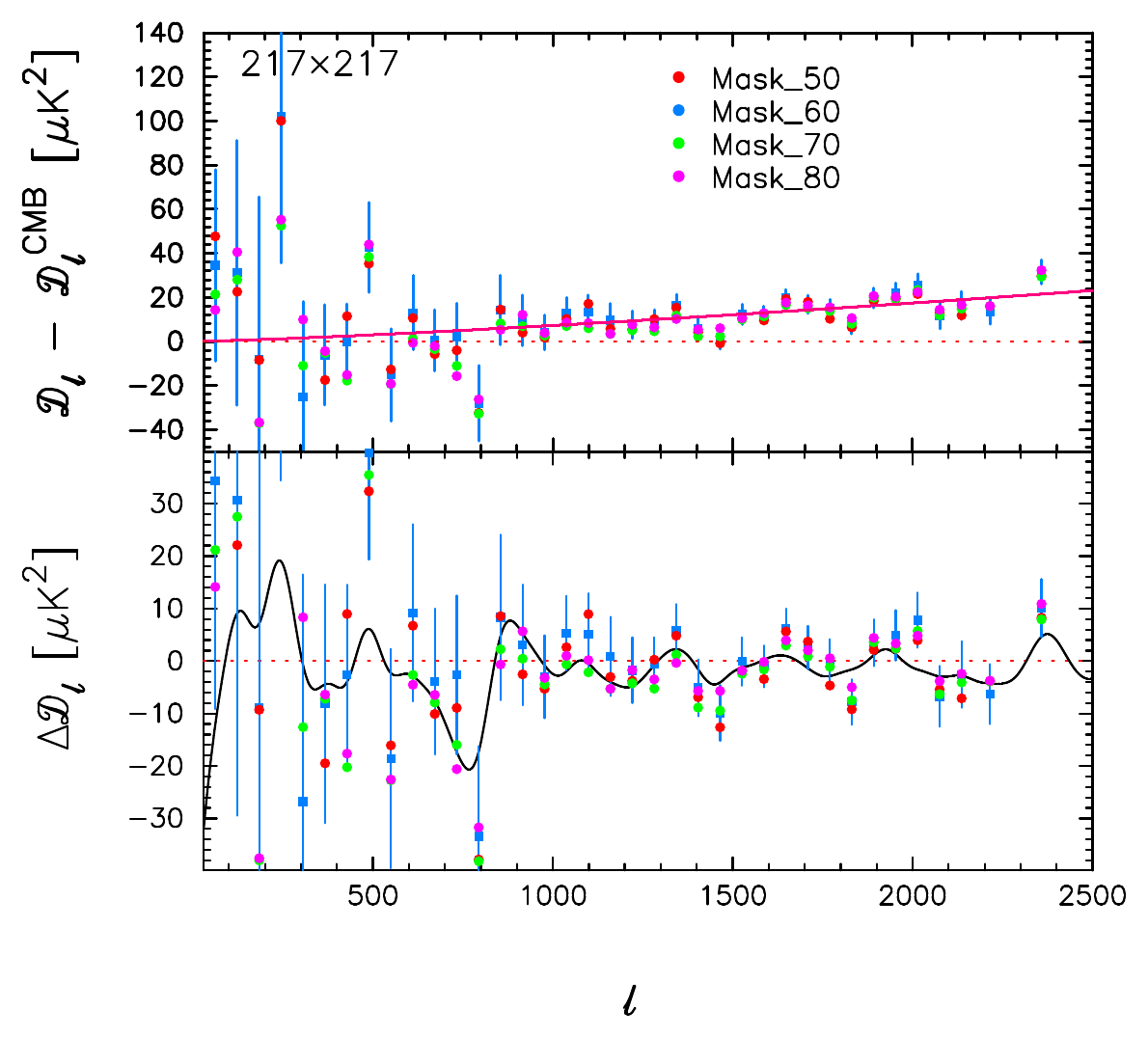}  \\
\includegraphics[width=115mm,angle=0]{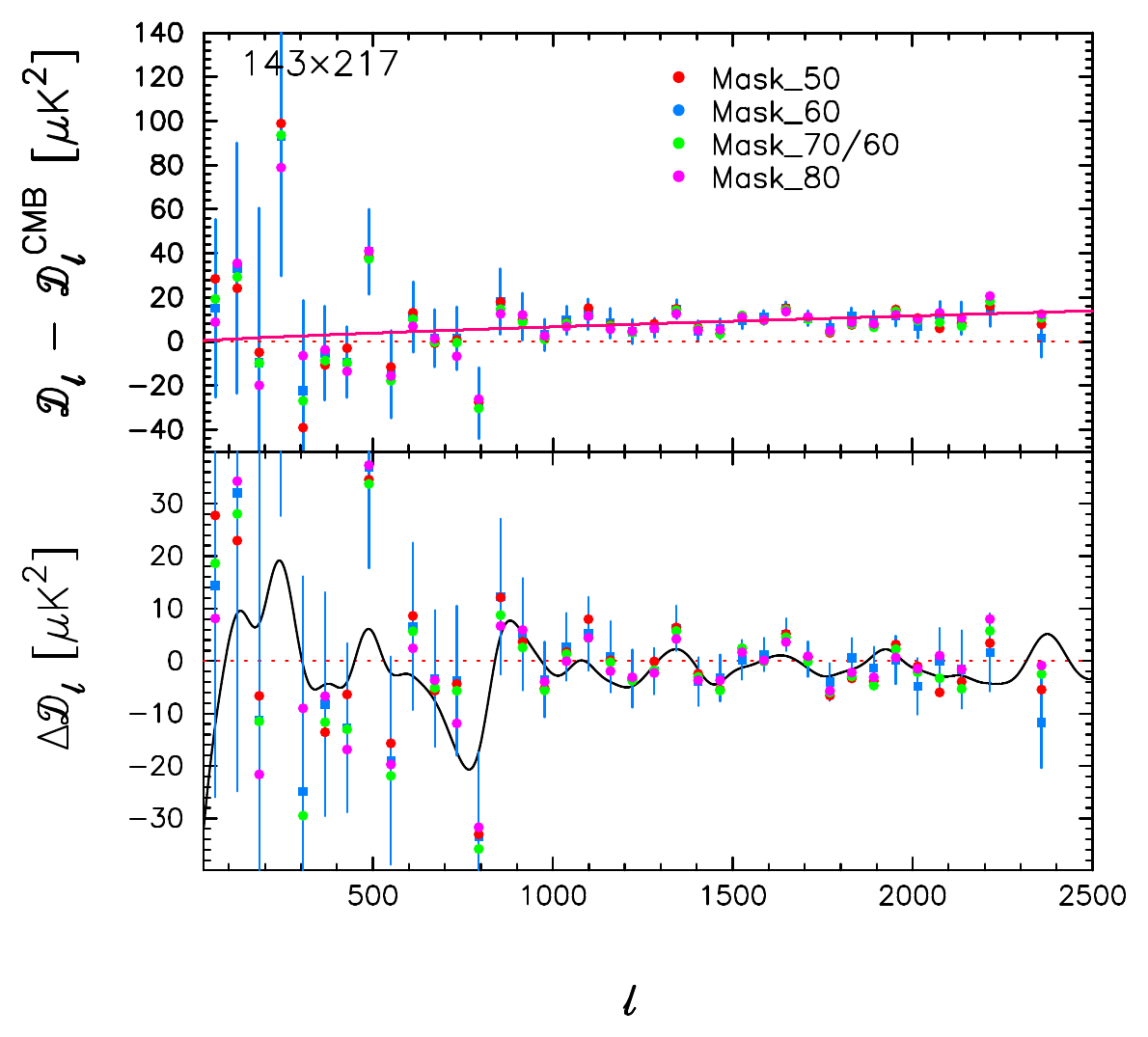}  
\caption {The upper panel in each plot shows the difference of the
  $545$ cleaned half mission spectra and the 12.1HMcl TT best-fit base
  \LCDM\ cosmology for masks of varying sizes. The upper figure shows the
  $217 \times 217$ spectrum and the lower figure shows the $143\times
  217$ spectrum. In the 12.1HMcl \camspec\ likelihood we use mask60
  for the 217 maps and mask70 for the 143 maps. Error bars are plotted
  on the spectra used in the 12.1HM likelihood.  The solid line in the
  upper panel in each figure shows the best-fit power-law foreground
  excess (Eq.\ \ref{equ:clean1}) derived from 12.1HMcl TT.  The lower
  panel in each figure shows the residuals after subtracting the
  foreground excesses. The solid line in the lower panel shows the 12.1HMcl TT maximum
  likelihood frequency coadded residuals, smoothed with a Gaussian of
  width $\sigma_\ell = 40$, as plotted in
  Fig.\ \ref{fig:inter_frequency143v217cleaned}.}

\label{fig:217_cleaned_sky}

\end{figure}

In this subsection, we analyse how the temperature spectra change with
increasing sky coverage. We focus on the $217 \times 217$ and $143
\times 217$ $545$ GHz cleaned spectra. Figure
\ref{fig:217_cleaned_sky} shows how these spectra change with
increasing sky coverage. We have shown in
Sect.\ \ref{subsec:universality} that $545$ cleaning accurately
removes Galactic dust emission leaving extragalactic residual
foregrounds. The remaining foreground excesses should therefore be
independent of the size of the mask.
 
This is what we see in Fig.\ \ref{fig:217_cleaned_sky}. The solid lines in the upper panel shows the power-law
fit to the excess foreground determined from the 12.1HMcl TT likelihood. The lower panels show the
residuals of the foreground subtracted spectra as a function of mask size. The solid lines in these panels show
the smoothed  maximum likelihood frequency coadded spectrum as plotted in Fig.\ \ref{fig:inter_frequency143v217cleaned}.
One can see that the oscillatory residuals are present for all mask sizes and in both spectra. Furthermore, the scatter around around the
black line {\it decreases} as more sky area is used. In fact, for the bandpowers plotted in Fig.\ \ref{fig:217_cleaned_sky}, the
rms scatter over the multipole range $1000 \le \ell \le 1800$ varies with mask as:
\begin{equation}
 \begin{cases}
      \sigma_{217 \times 217}  &=   3.9 \ (\mu {\rm K})^2,  \quad {\rm mask50},  \quad     \sigma_{143 \times 217}  =   2.9 \ (\mu {\rm K})^2,  \qquad {\rm mask50}, \\
      \sigma_{217 \times 217} &=   3.7 \ (\mu {\rm K})^2,  \quad {\rm mask60}, \quad \sigma_{143 \times 217}  =   2.3 \ (\mu {\rm K})^2,  \qquad {\rm mask60}, \\
      \sigma_{143 \times 217} &=   3.1 \ (\mu {\rm K})^2,  \quad {\rm mask70}, \quad \sigma_{143 \times 217}  =   2.3 \ (\mu {\rm K})^2,  \qquad {\rm mask70/60}, \\
      \sigma_{143 \times 217} &=   3.0 \ (\mu {\rm K})^2,  \quad {\rm mask80}, \quad \sigma_{143 \times 217}  =   2.0 \ (\mu {\rm K})^2,  \qquad {\rm mask80}, 
\end{cases} 
\end{equation}
where for the $143\times 217$ spectrum the notation mask70/60 denotes mask70 for $143$ GHz and mask60 for $217$ GHz
(as used in the 12.1HM likelihoods).
In other words, by using more sky the $217 \times 217$ and $143 \times 217$ spectra move
closer to the coadded spectrum (which averages all four spectra).
The residual in the $217 \times 217$ spectrum for the band centred
at $\ell_b = 1465$, which we have shown in the previous two
subsections is slightly anomalous on mask60, becomes progressively less anomalous
as the mask is extended to mask70 and mask80, as expected if
the mask60 residual is a statistical fluctuation.

\subsection{Comparison with full mission spectra}
\label{subsec:inter_frequency_full}

The noise levels in the \Planck\ spectra can be reduced by forming a
likelihood using the full mission detset spectra\footnote{PCP13 reported results based on a nominal mission detset likelihood, though 
we also performed an unpublished analysis of a full mission detset likelihood at that time.}.
However, as we have discussed in Sect.\ \ref{subsec:correlatednoise},
the noise between detsets is correlated at high multipoles. 
 The half mission temperature spectra
are signal dominated over most of the multipole range and as we have
demonstrated in PCP15 and PCP18, half mission likelihoods are
sufficiently powerful to determine the parameters of the base
\LCDM\ model, and most simple variants, to high precision. The main motivation
in constructing a full mission detset temperature likelihood is to test the consistency
of the data, rather than to improve constraints on cosmology.
In polarization, we use all non-cotemporal half mission cross spectra in TE
and so there is almost no gain in signal-to-noise in switching from half mission  to  full mission TE spectra.
There is, however, a gain in signal-to-noise in the full mission detset EE spectrum,
since the EE spectra are noise  dominated over most of the multipole range covered by
\Planck. In this section, we focus on a comparison of half mission and full mission temperature
spectra. A similar comparison of polarization spectra is presented in Sect.\ \ref{subsec:full_mission_pol}.

\begin{figure}
\centering
\includegraphics[width=150mm,angle=0]{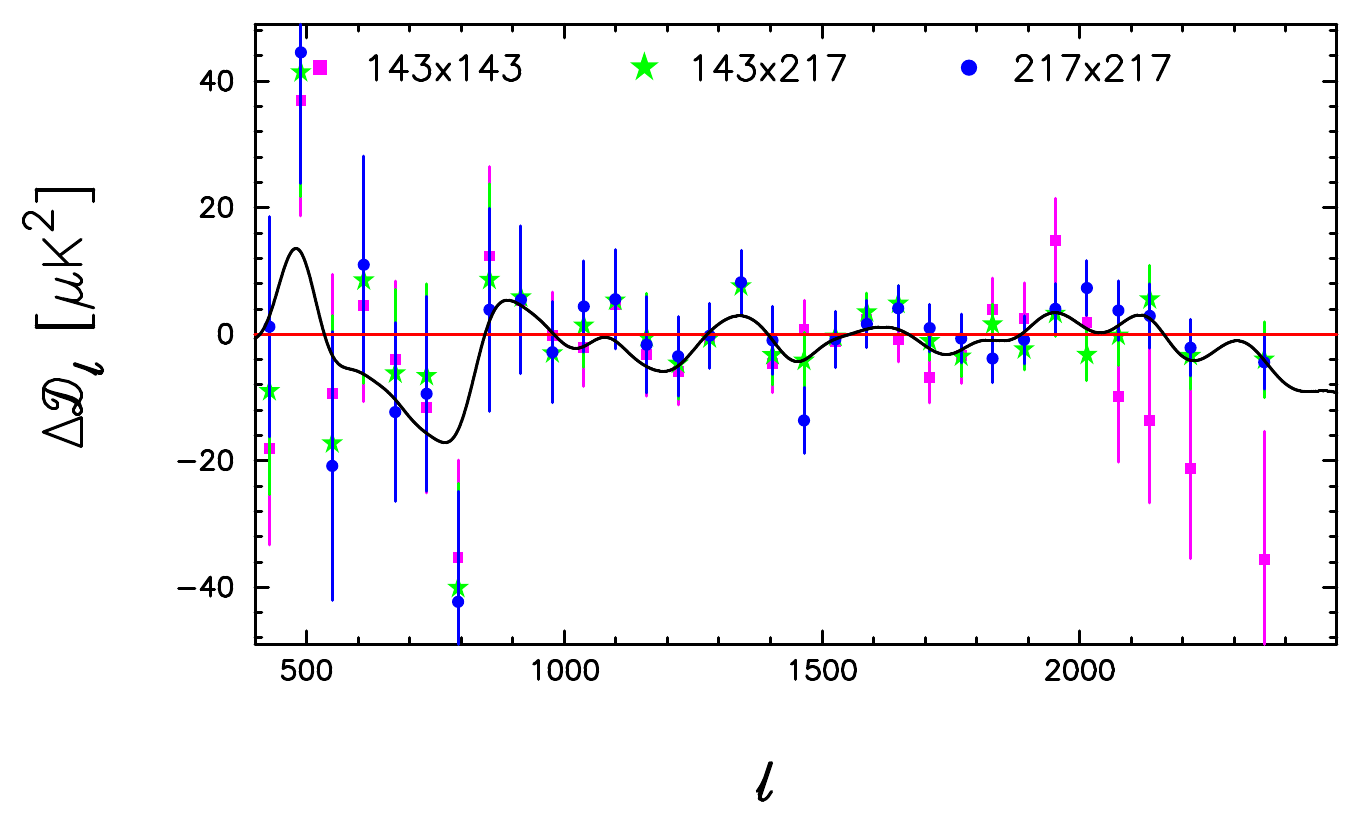} 
\caption {As Fig.\ \ref{fig:inter_frequency143v217}, but for the full mission (detset) 12.1F TT likelihood. 
The spectra
have been corrected for correlated noise between detsets, as discussed in the text, and foreground corrected
using the  base \LCDM\ best-fit solution to the 12.1F TT  likelihood. The residuals of these spectra are
computed with respect to the best-fit base \LCDM\ model fitted to this likelihood (which is very close to the 12.1HM TT best-fit
model).
The black line shows the residuals of the maximum likelihood coadded spectrum for the detset likelihood,
smoothed with a Gaussian of width $\sigma_\ell =40$. }

\medskip
\medskip

\label{fig:inter_frequency143v217full}

\end{figure}

Figure \ref{fig:inter_frequency143v217full} is the analogue of
Fig.\ \ref{fig:inter_frequency143v217} for the 12.1F full mission
detset likelihood. In forming this likelihood, we subtracted
correlated noise from each of the coadded $100 \times 100$, $143
\times 143$, $143 \times 217$ and $217 \times 217$ detset spectra
using smooth fits to the coadded odd-even differenced noise spectra
(these corrections are plotted as the red lines in
Fig.\ \ref{fig:corrnoise}). Note that these corrections have a
relatively small effect on the coadded spectra over the range of
multipoles included in the likelihood (for example, the $143\times
143$ spectrum is not included in the 12.1F likelihood for $\ell >
2000$, where the $143\times 143$ noise becomes large).  Note also,
that correlated noise is negligible in the $143 \times 217$ detset
spectra over the entire range of multipoles shown in
Fig.\ \ref{fig:inter_frequency143v217full}. In forming the
\camspec\ detset likelihood for PCP13, we chose to restrict the
maximum multipole ranges at each frequency so that none of the spectra
relied on beam calibrations at angular scales much smaller than the beam width.
This choice also protected the 2013 \camspec\ detset likelihood from biases
caused by correlated noise.

As in Figs. \ref{fig:inter_frequency143v217} and \ref{fig:inter_frequency143v217cleaned}, we see that
the detset spectra are consistent with each other and show a similar pattern of residuals to that seen in the 
half mission spectra, particularly in the multipole range
$800 \simlt \ell \simlt 1800$ which drives parameters such as $A_L$ and $\omegak$.

\subsection{Summary}

 In summary, the characteristic oscillatory pattern of residuals in
 the temperature spectra at multipoles $ \ell \simlt 2000$ are: (i)
 consistent across frequencies; (ii) insensitive to foreground
 modelling; (iii) insensitive to sky coverage (and actually decrease in amplitude
with increased sky coverage); (iv) are reproduced in
 the full mission detset spectra. For \LCDM-like models, almost all of
 the statistical power from \Planck\ comes from multipoles $\simlt
 1800$, so the tests described here add confidence that the
 cosmological results from \Planck\ are robust and not driven by systematic
errors.

\section{Inter-frequency consistency in polarization}
\label{sec:inter_frequency_pol}

\subsection{The 12.1HM likelihood}
\label{subsec:half_mission_inter_frequency_pol}

We process the half mission polarization spectra as follows:

\smallskip

\noindent
$\bullet$ The TE/ET and EE spectra are cleaned using $353$ GHz maps up
to a multipole $\ell=150$, as described in
Sect.\ \ref{subsec:pol_clean_spectra}. At higher multipoles, we
subtract power-law dust templates with parameters as given in Table
\ref{tab:pol_dust_fits}.

\smallskip

\noindent
$\bullet$ Each spectrum is corrected for temperature-to-polarization
(TP) leakage using the beam model described in
Sect.\ \ref{subsec:planck_beams}. TP leakage corrections are small but
non-negligible for TE and ET (see Fig.\ \ref{fig:leak}). The TP
corrections are negligible for EE. The consistency of the TP leakage
model for TE/ET spectra has been tested in
Sect.\ \ref{subsec:pol_leak}.

\smallskip

\noindent
$\bullet$ The beam-corrected, foreground subtracted spectra are then recalibrated against a fiducial cosmology
to determine effective polarization efficiencies as described in Sect.\ \ref{subsec:pol_cal}. The polarization efficiency
corrections are the most significant instrumental  corrections applied to the polarization spectra.

\begin{figure}
\centering
\includegraphics[width=50mm,angle=0]{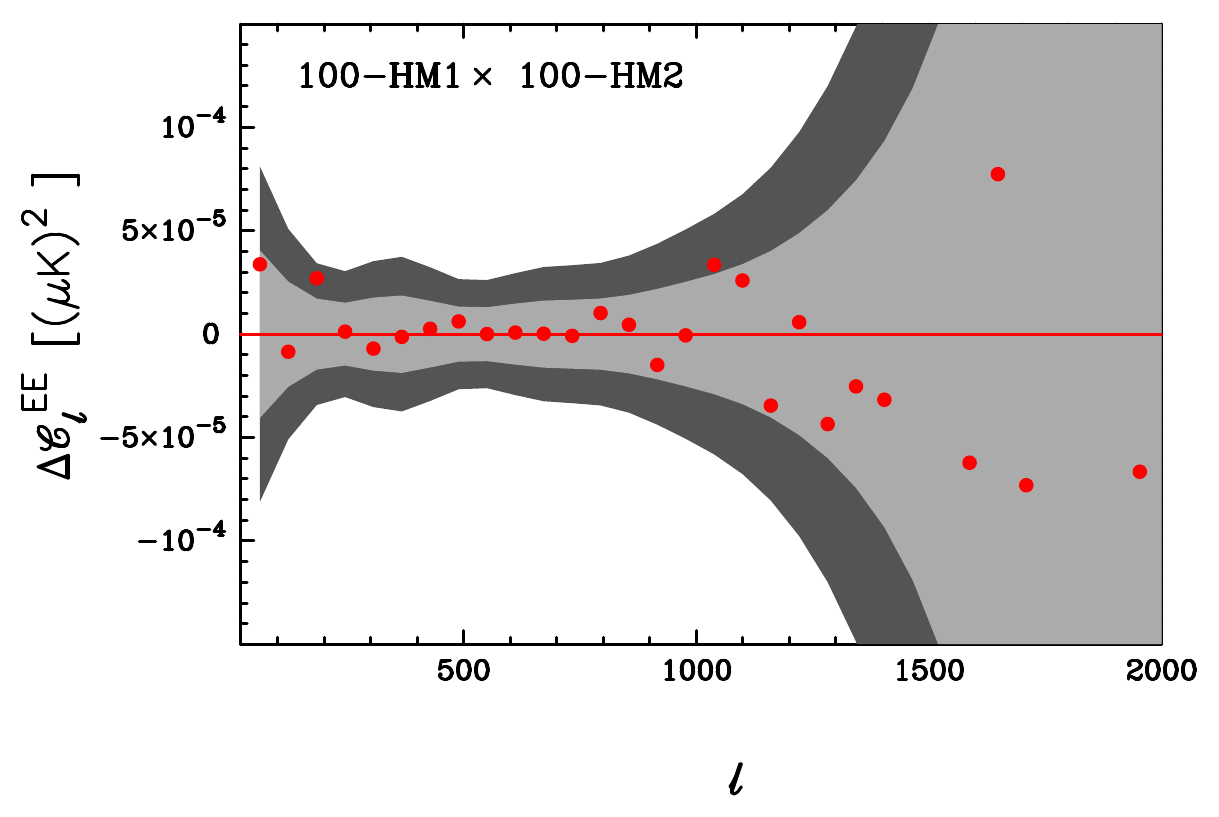} 
\includegraphics[width=50mm,angle=0]{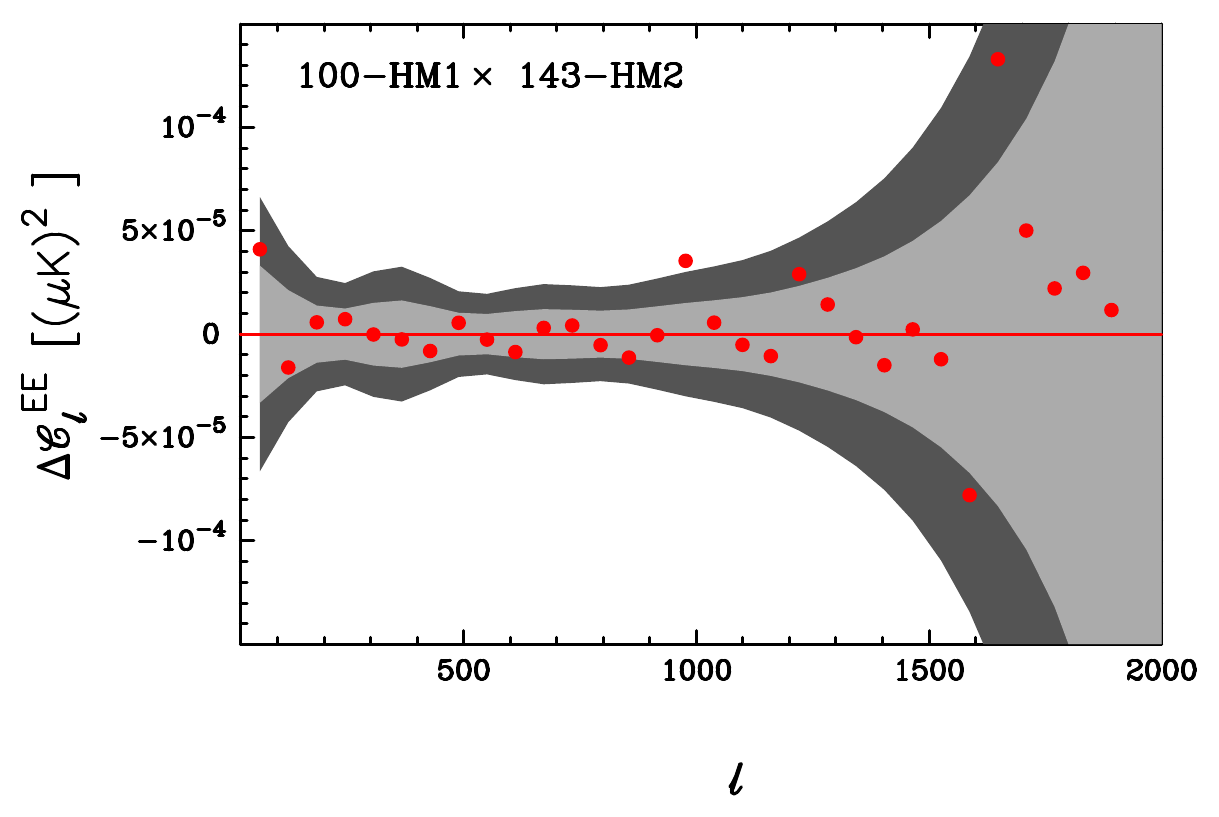} 
\includegraphics[width=50mm,angle=0]{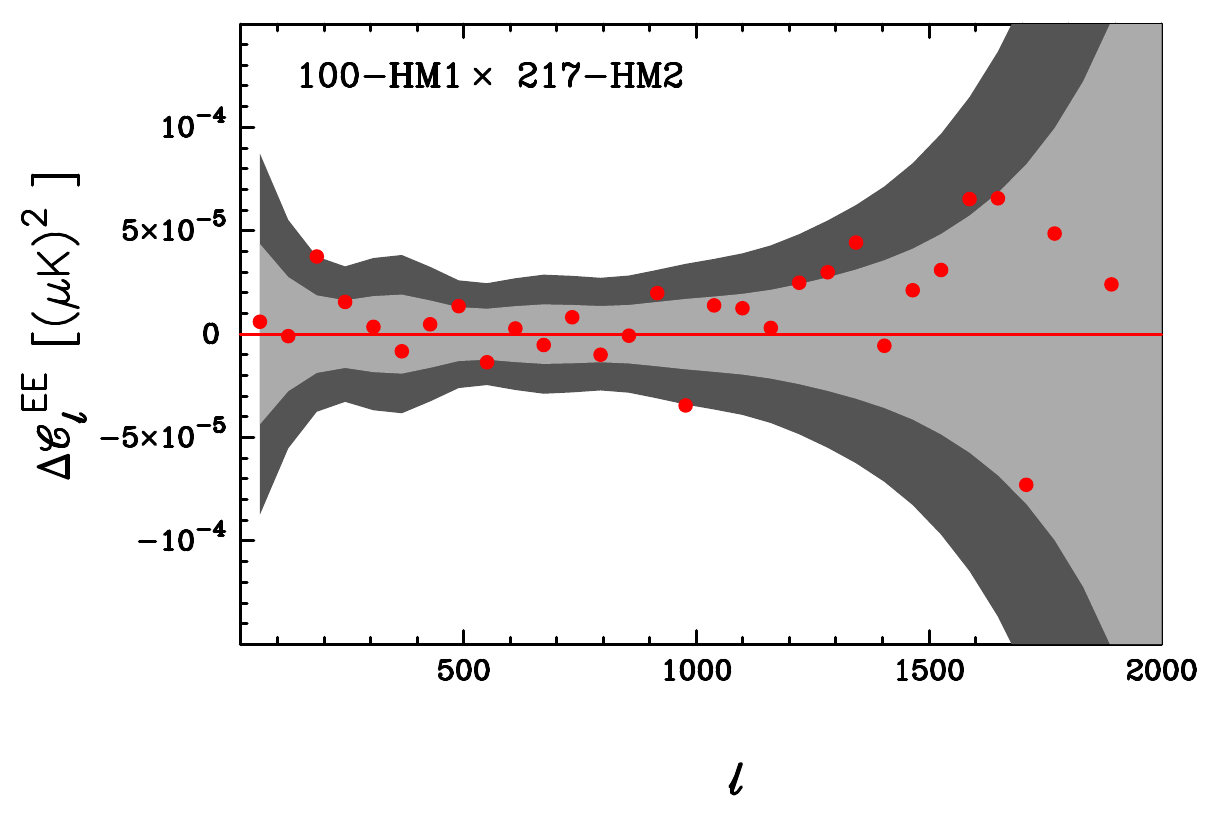}  \\

\includegraphics[width=50mm,angle=0]{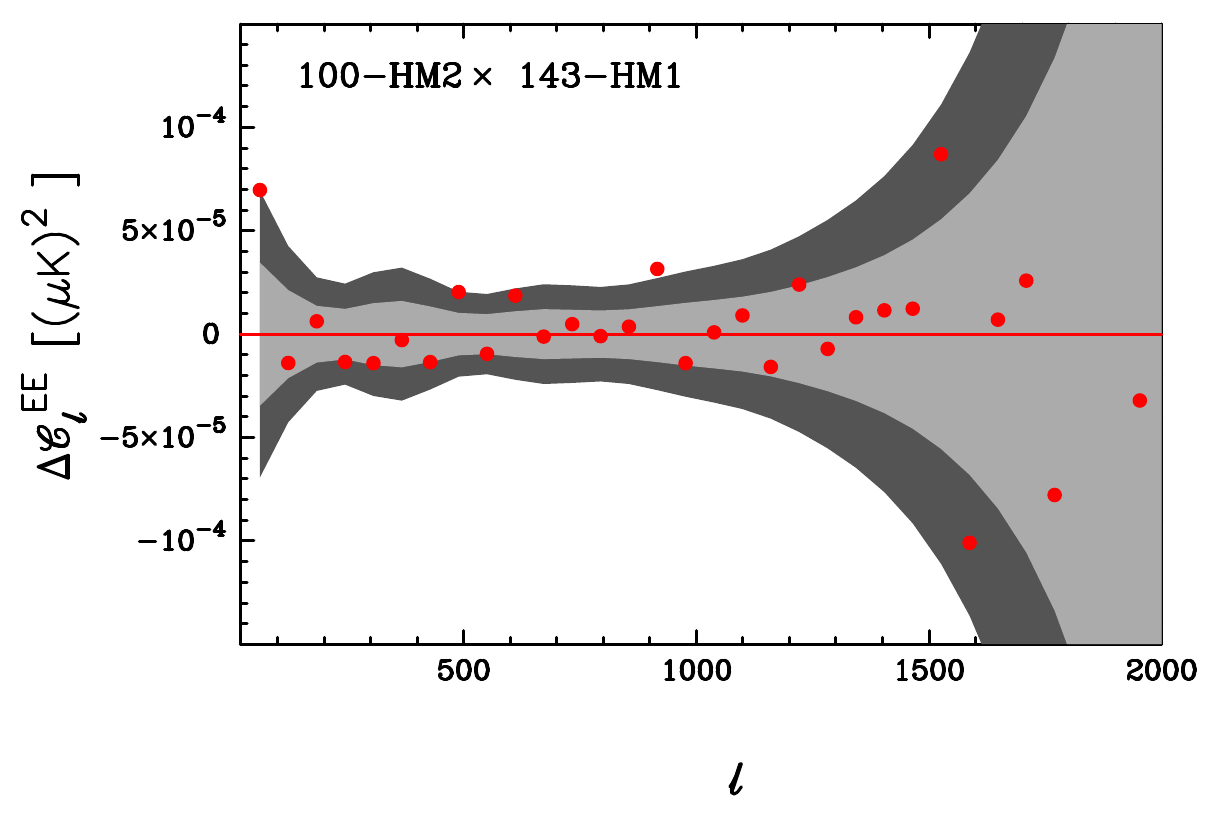} 
\includegraphics[width=50mm,angle=0]{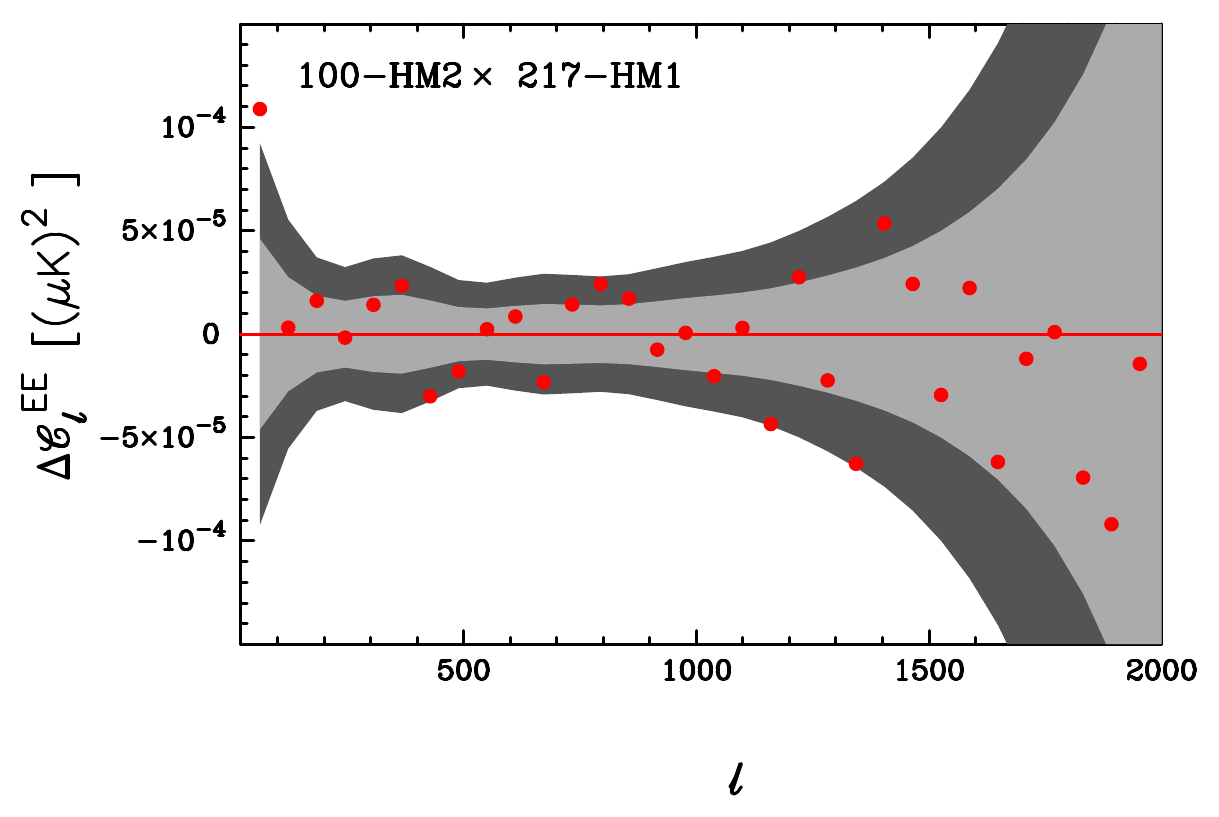} 
\includegraphics[width=50mm,angle=0]{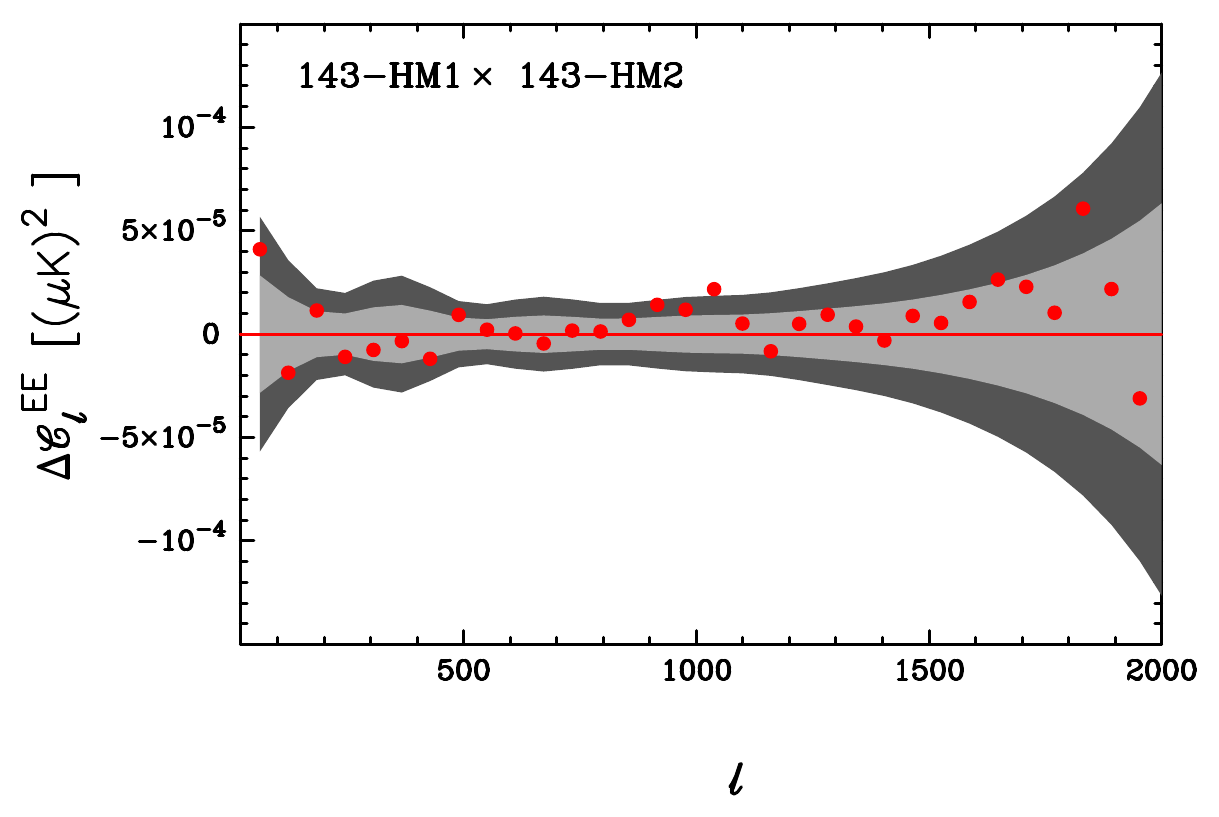}  \\

\includegraphics[width=50mm,angle=0]{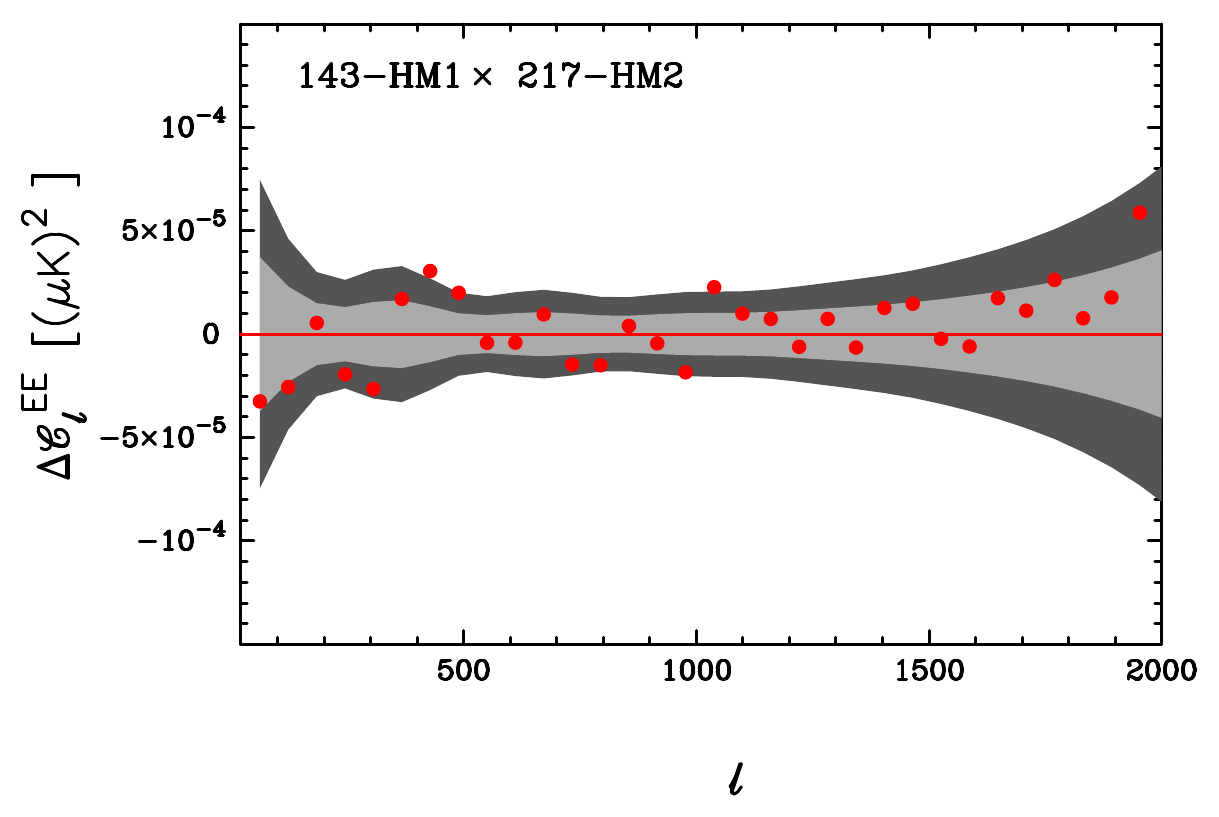} 
\includegraphics[width=50mm,angle=0]{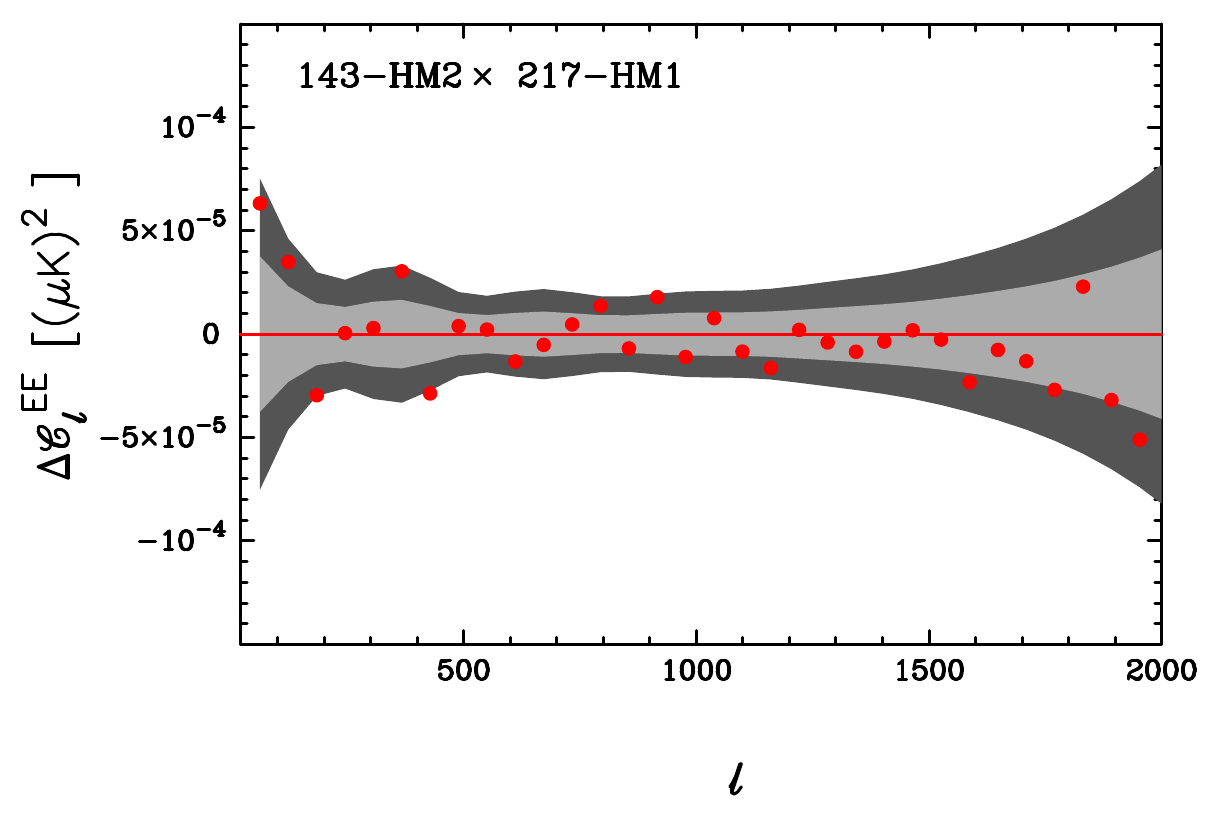} 
\includegraphics[width=50mm,angle=0]{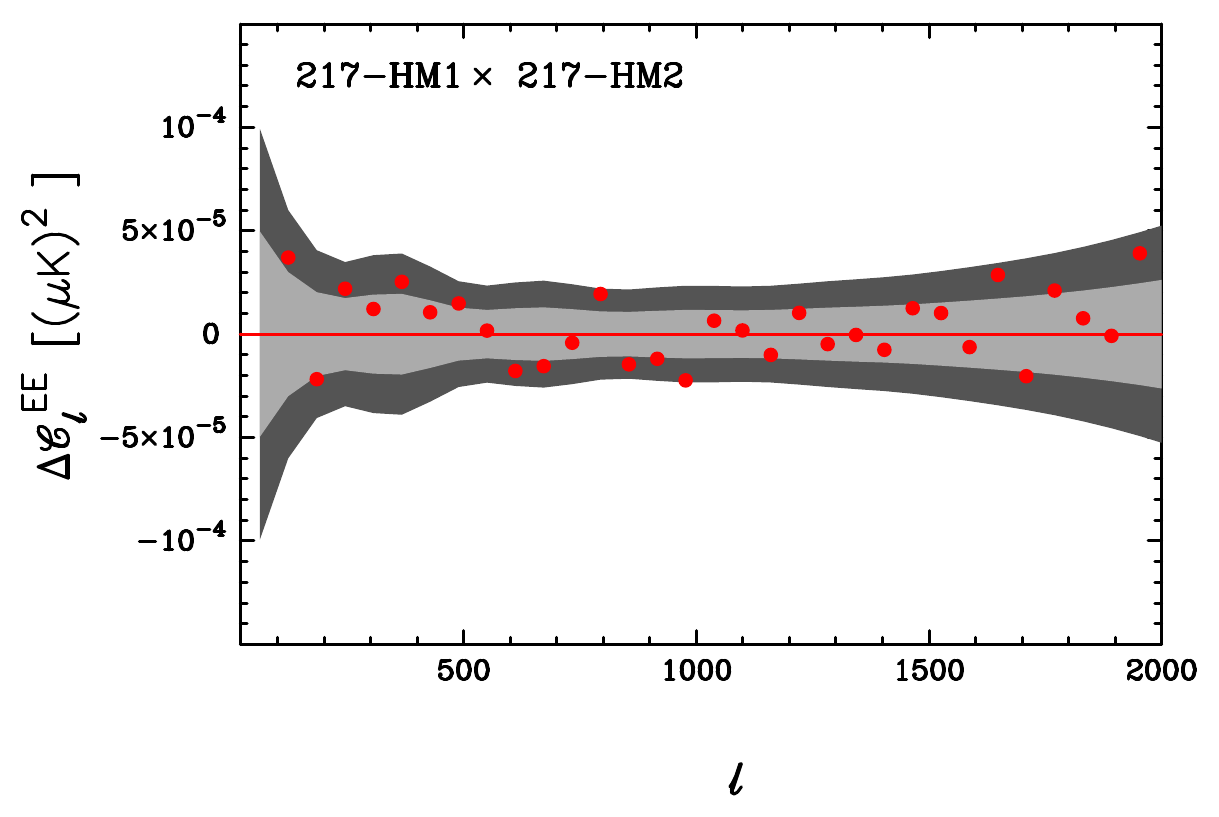}  \\

\caption {EE spectra used in the 12.1HM likelihood, corrected for dust emission, TP-leakage and effective polarization efficiencies.
Residuals are computed with respect to the fiducial base \LCDM\ cosmology fitted to 12.1HM TT.  The grey bands
show $1\sigma$ and $2\sigma$ error contours determined from the \camspec\ error model.}

\label{fig:inter_frequencyEE}

\end{figure}

\begin{figure}
\centering
\includegraphics[width=50mm,angle=0]{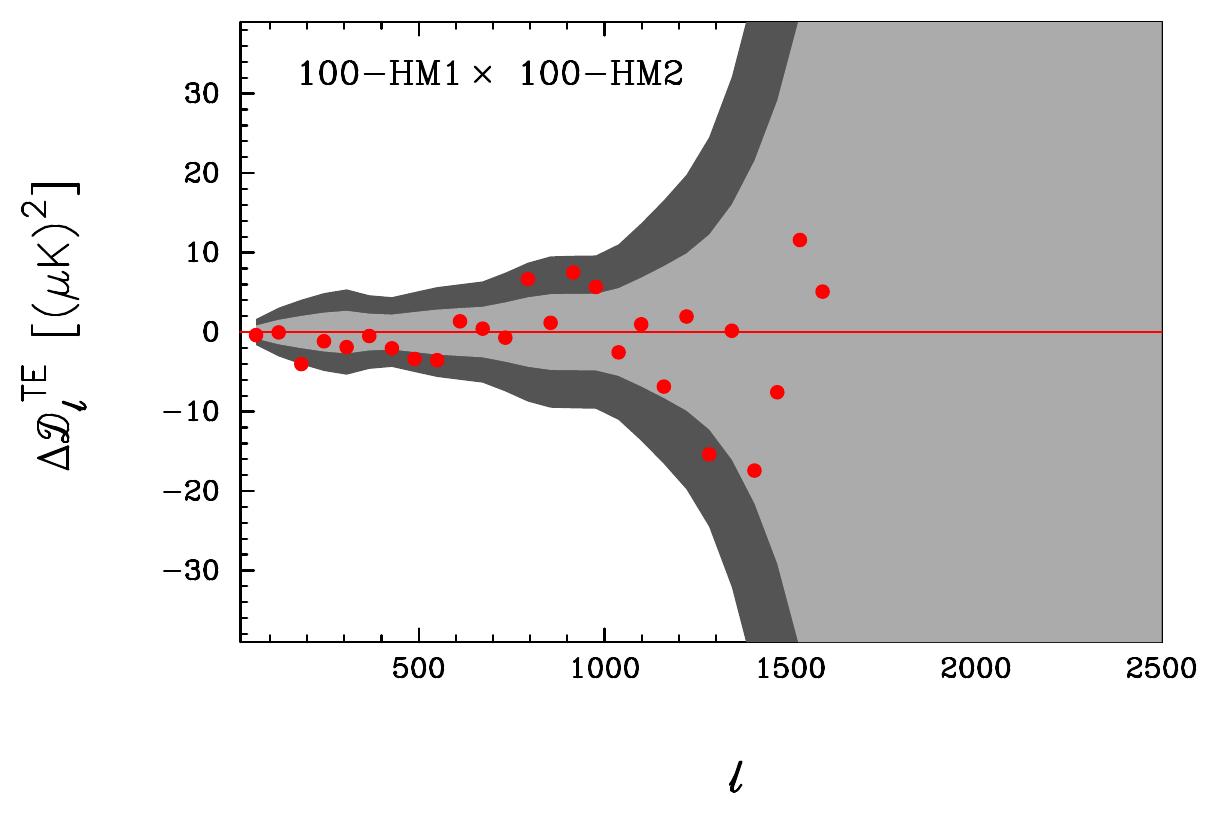} 
\includegraphics[width=50mm,angle=0]{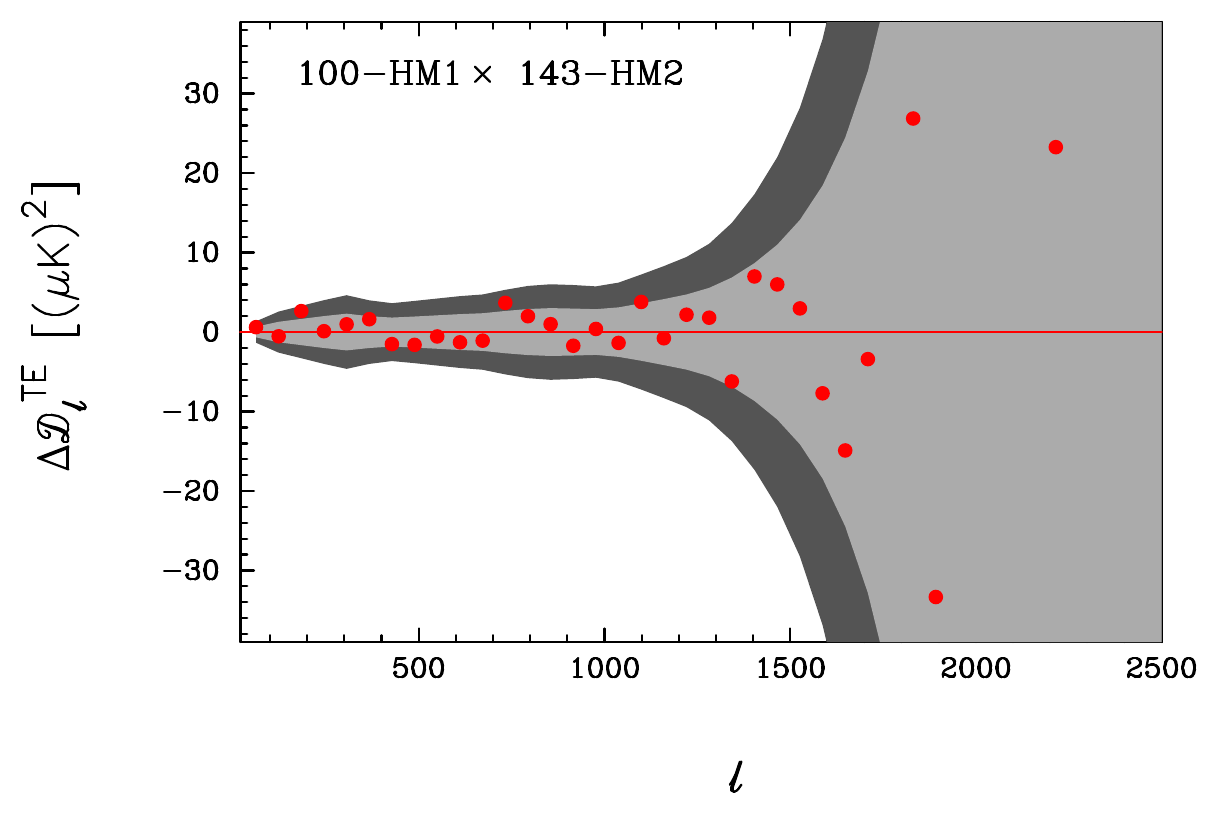} 
\includegraphics[width=50mm,angle=0]{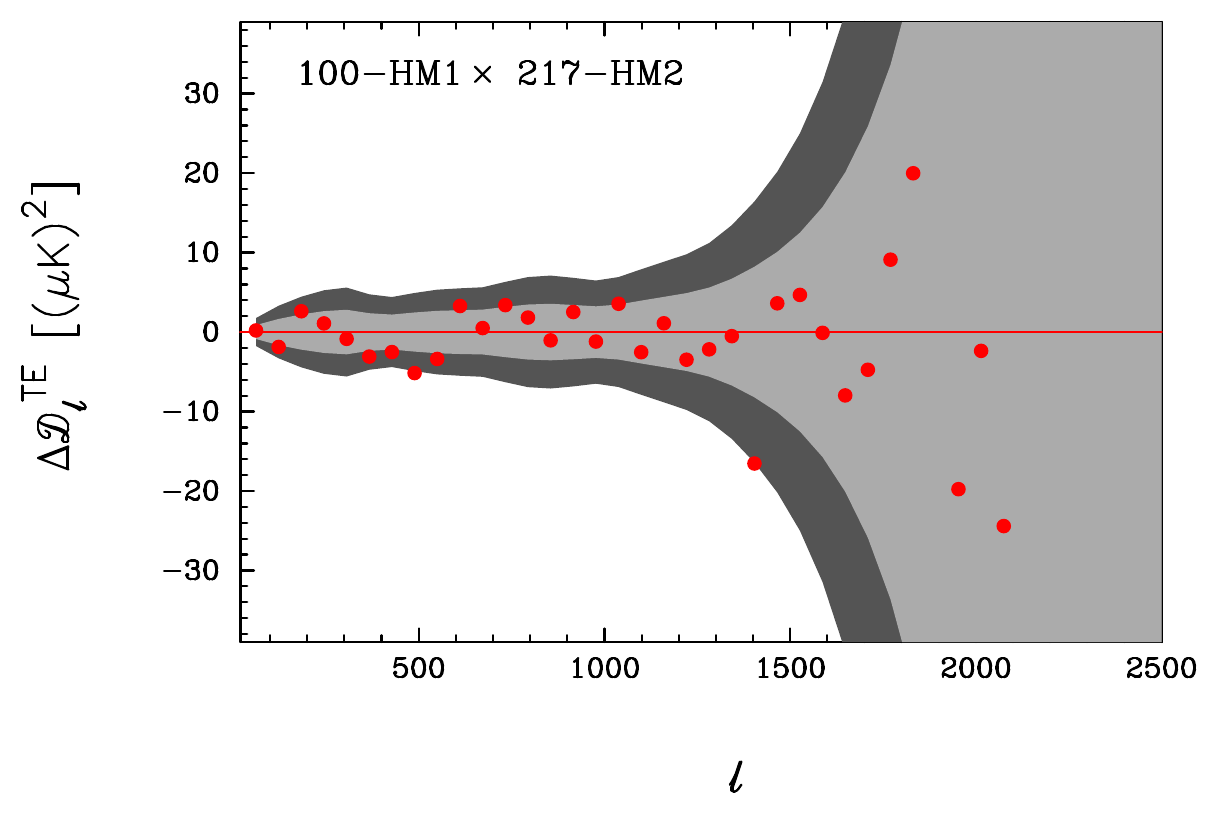}  \\

\includegraphics[width=50mm,angle=0]{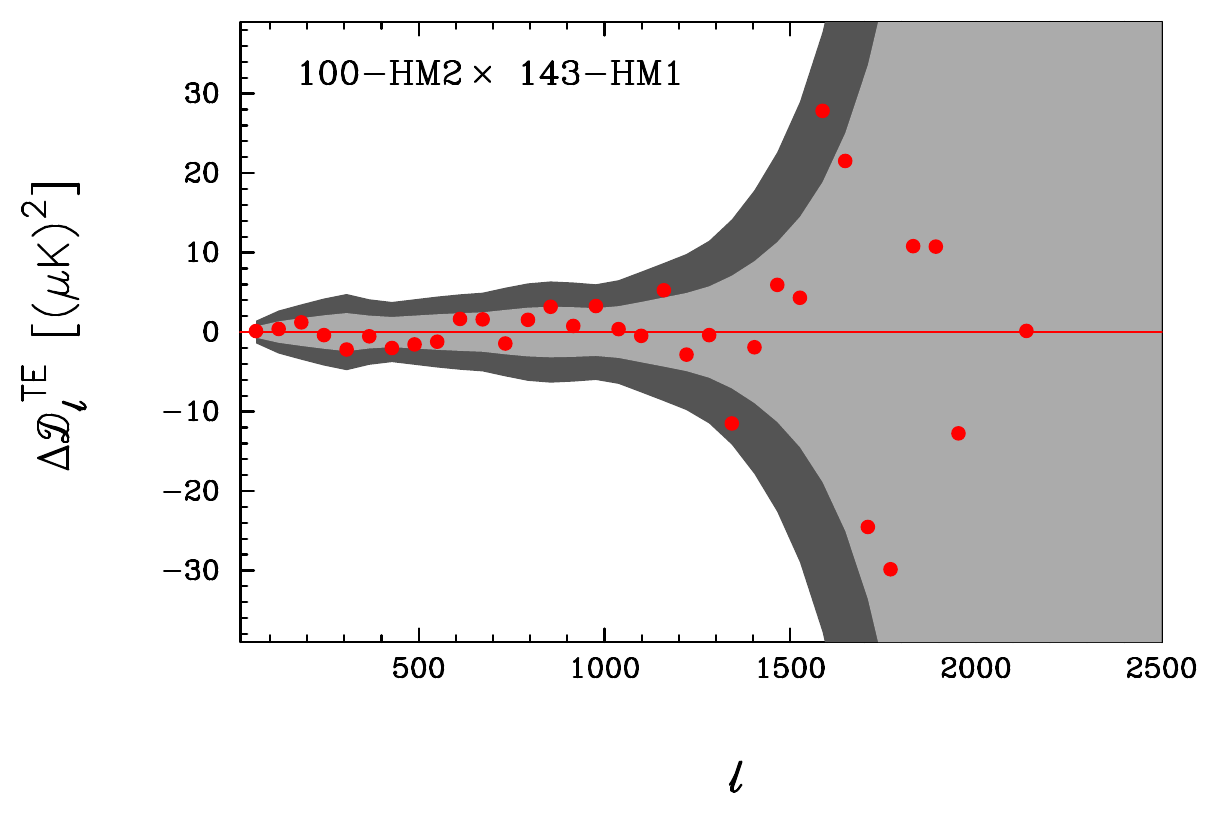} 
\includegraphics[width=50mm,angle=0]{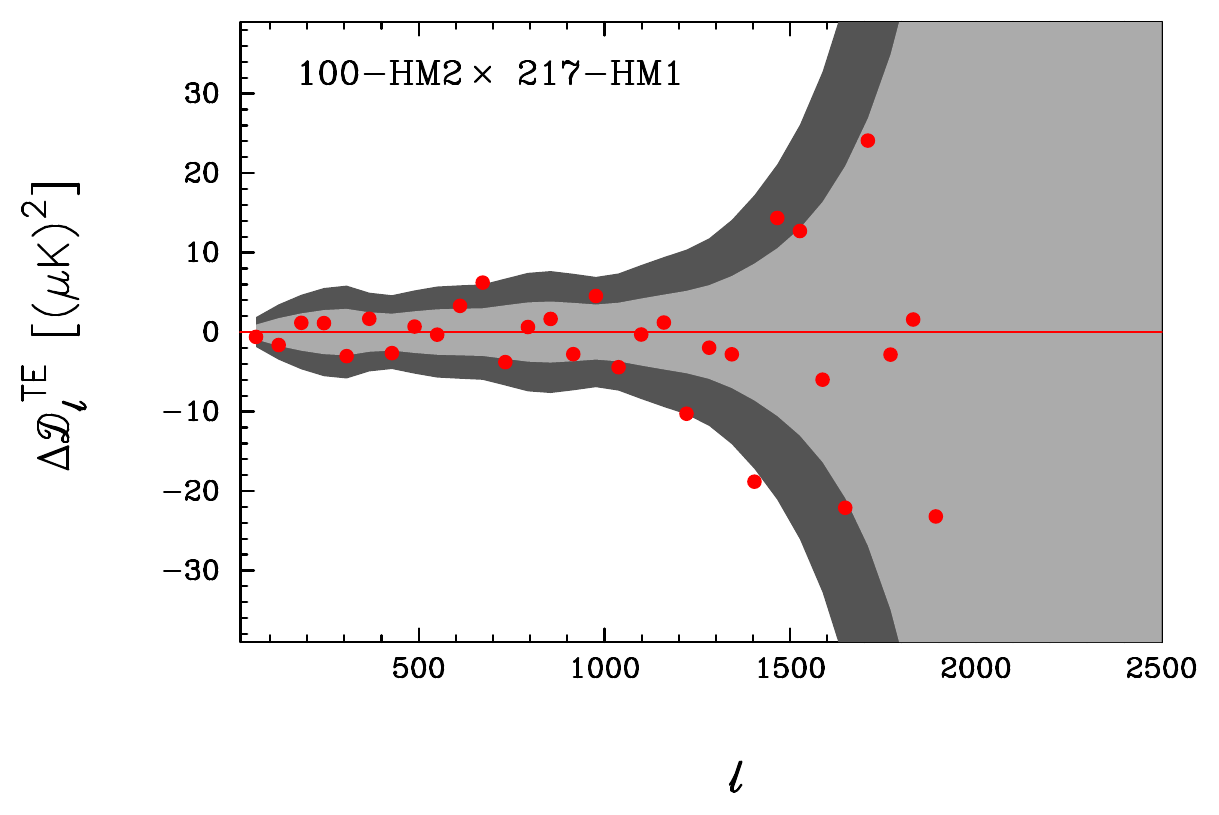} 
\includegraphics[width=50mm,angle=0]{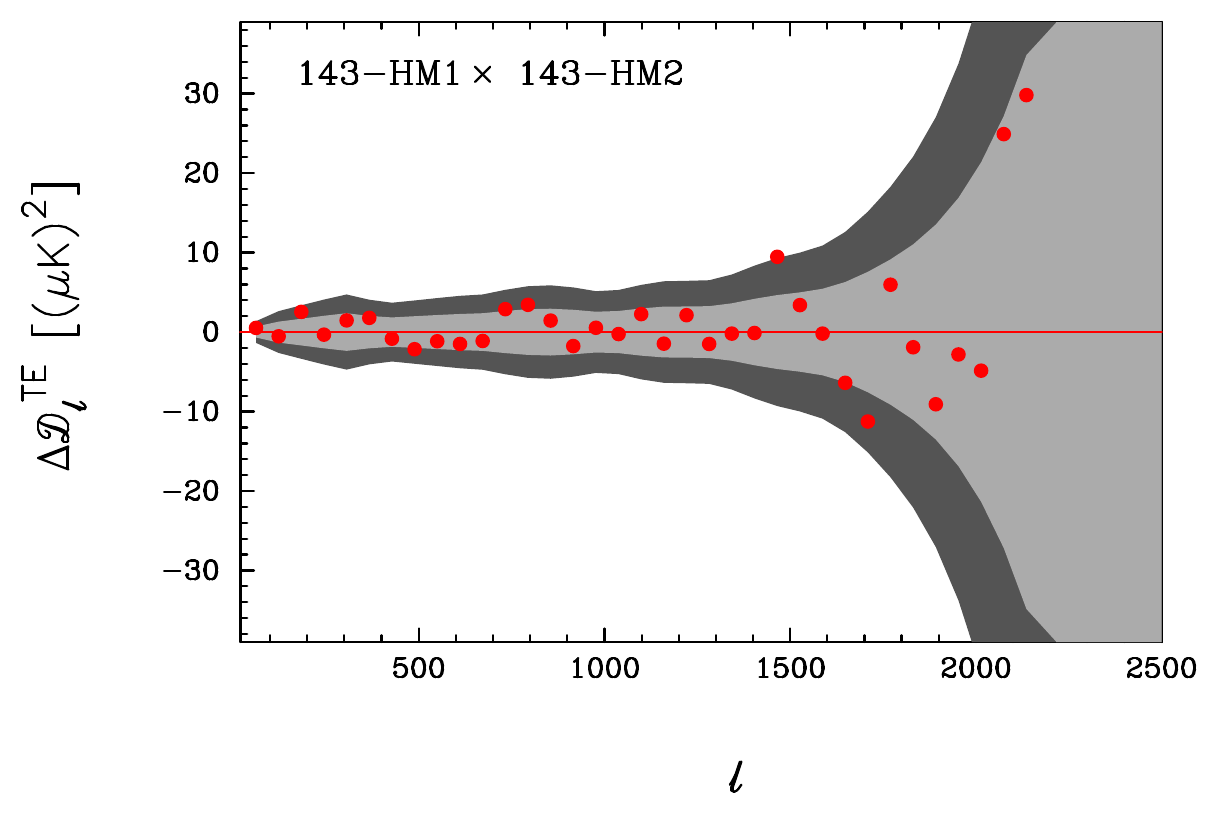}  \\

\includegraphics[width=50mm,angle=0]{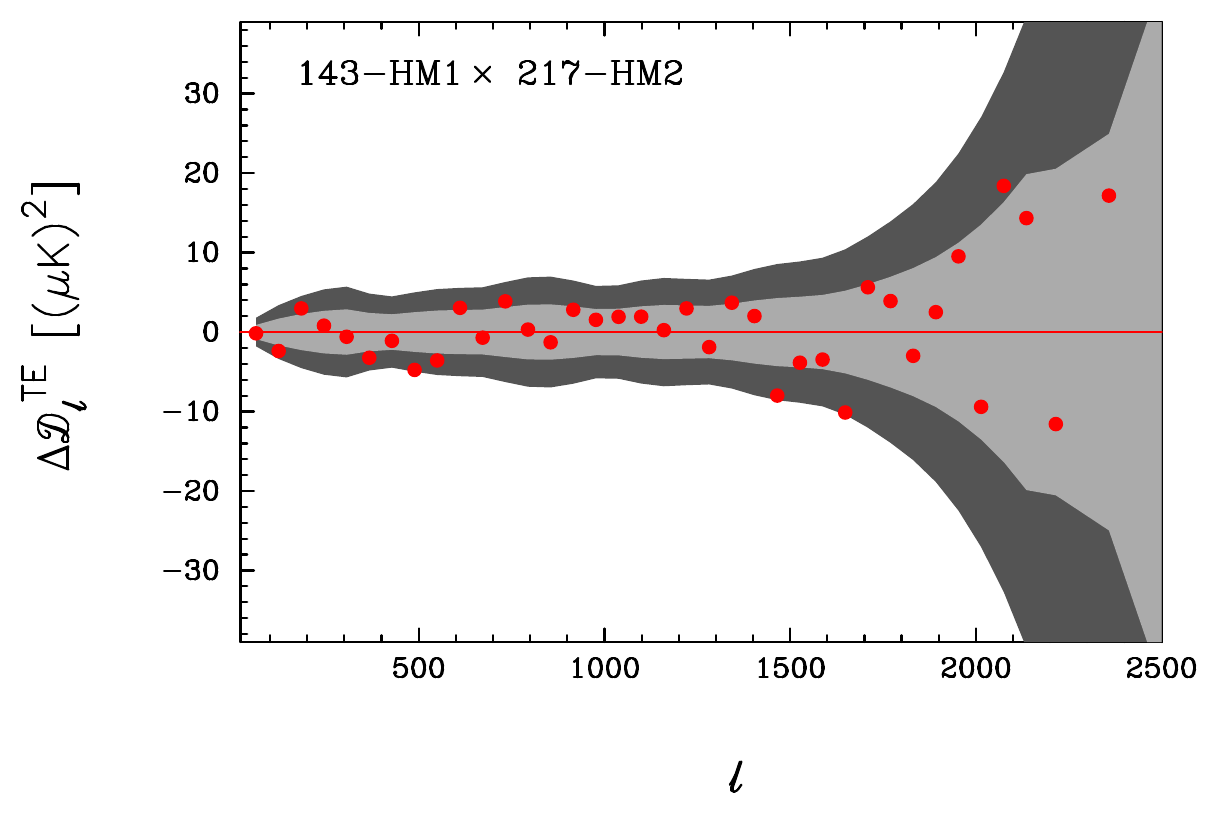} 
\includegraphics[width=50mm,angle=0]{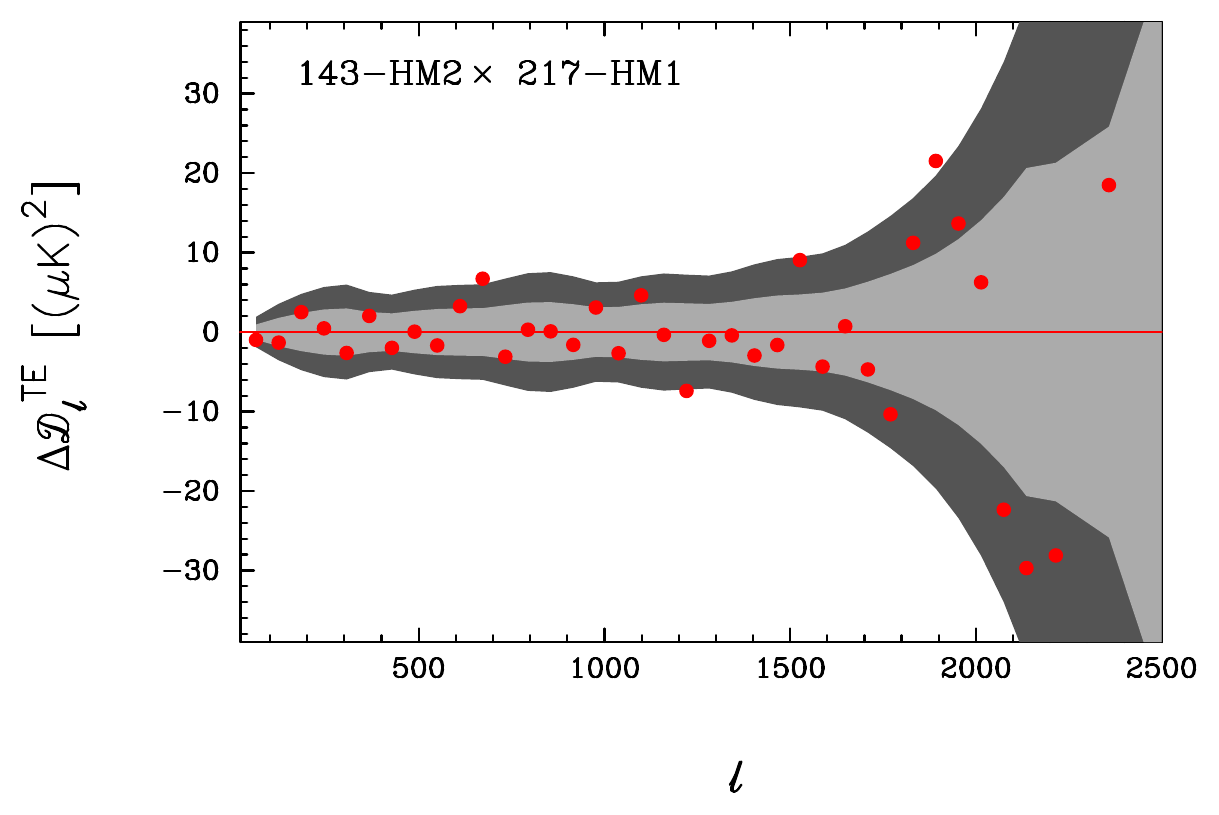} 
\includegraphics[width=50mm,angle=0]{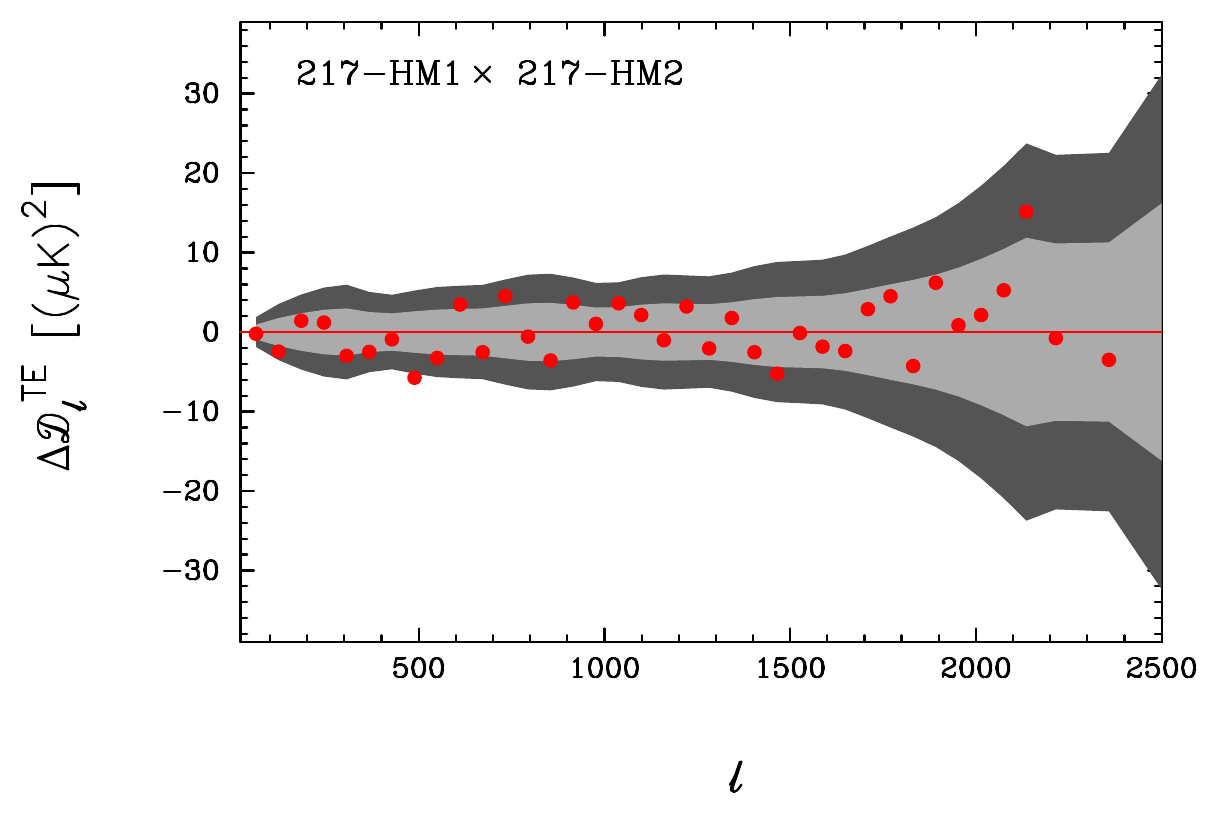}  \\

\caption {As for  Fig.\ \ref{fig:inter_frequencyEE} but for the TE spectra.}

\label{fig:inter_frequencyTE}

\end{figure}

\begin{figure}
\centering
\includegraphics[width=50mm,angle=0]{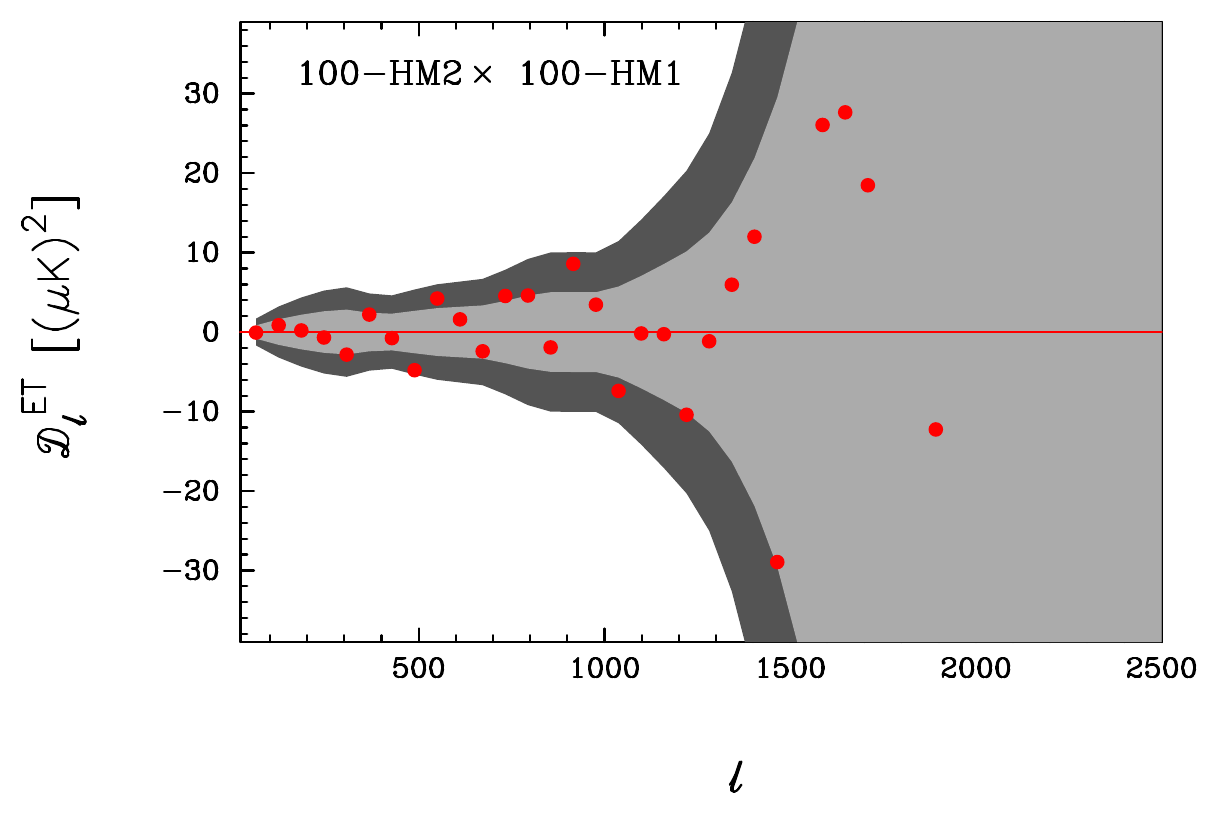} 
\includegraphics[width=50mm,angle=0]{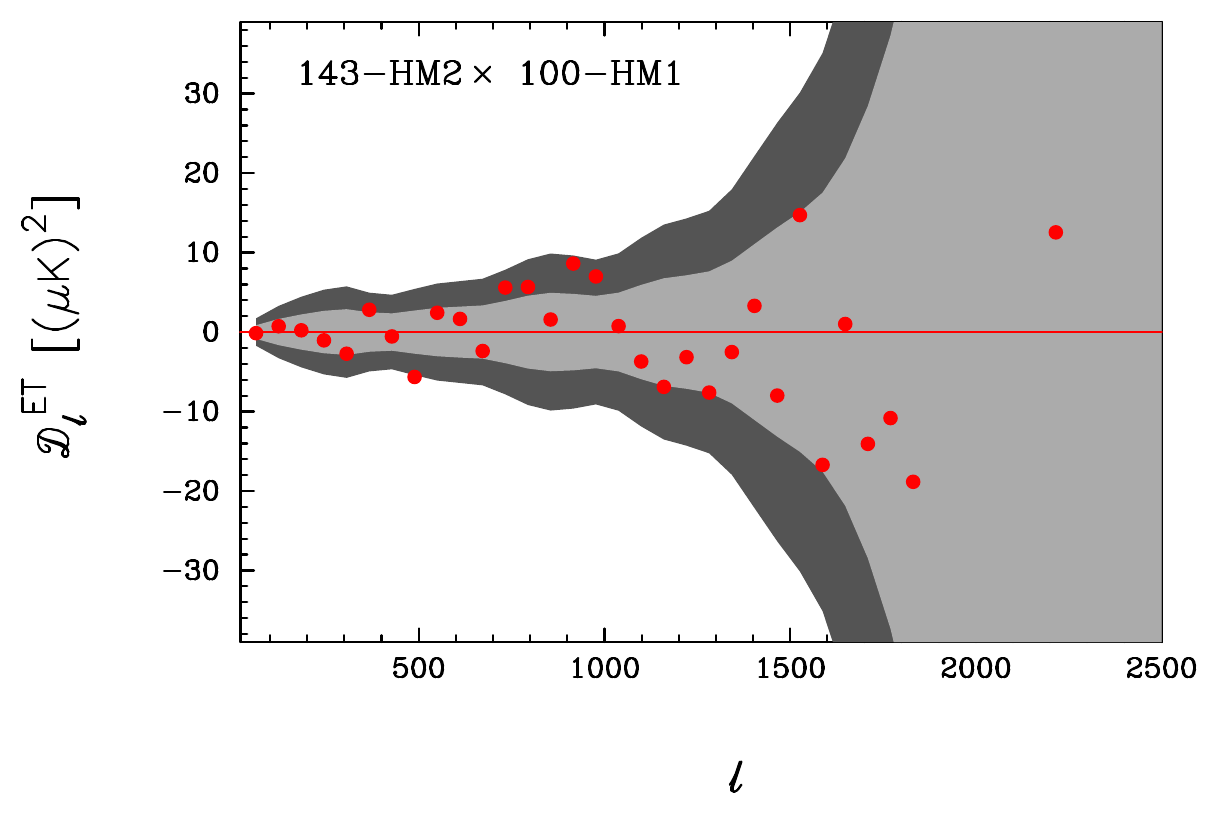} 
\includegraphics[width=50mm,angle=0]{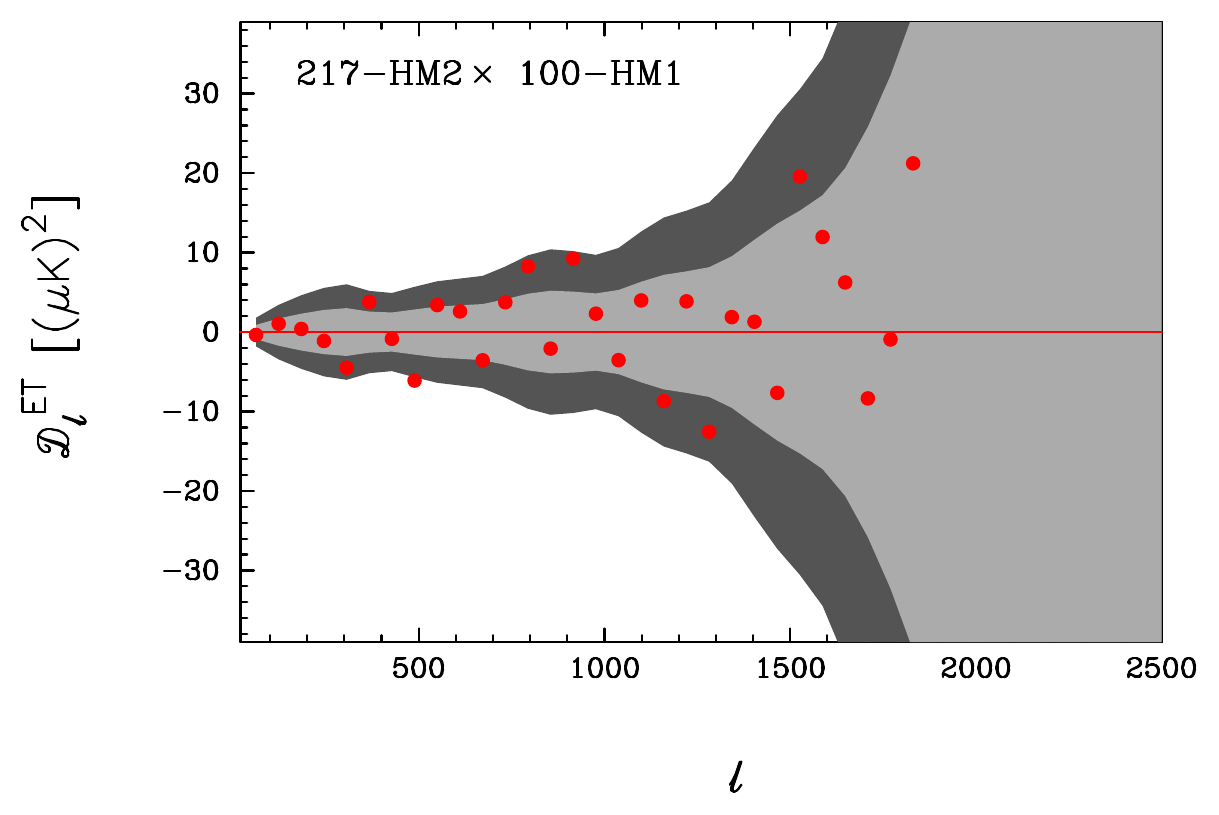}  \\

\includegraphics[width=50mm,angle=0]{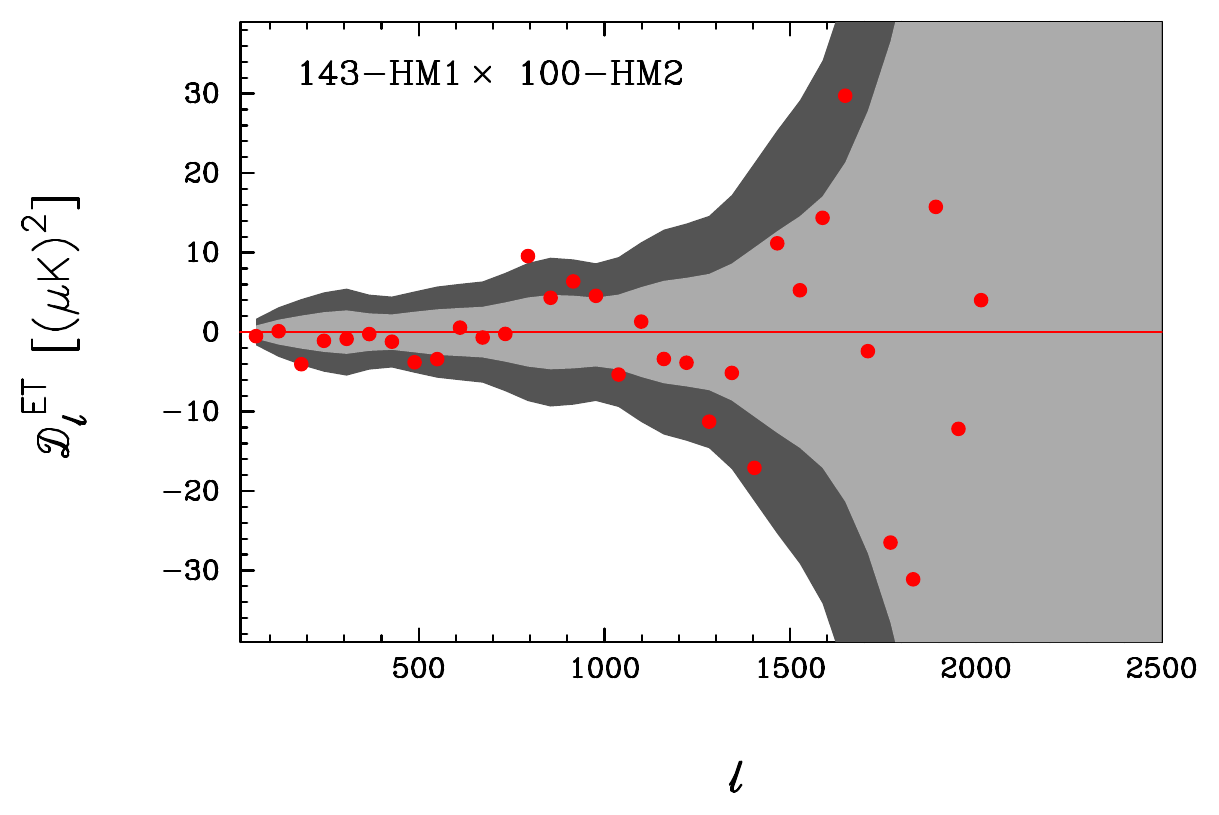} 
\includegraphics[width=50mm,angle=0]{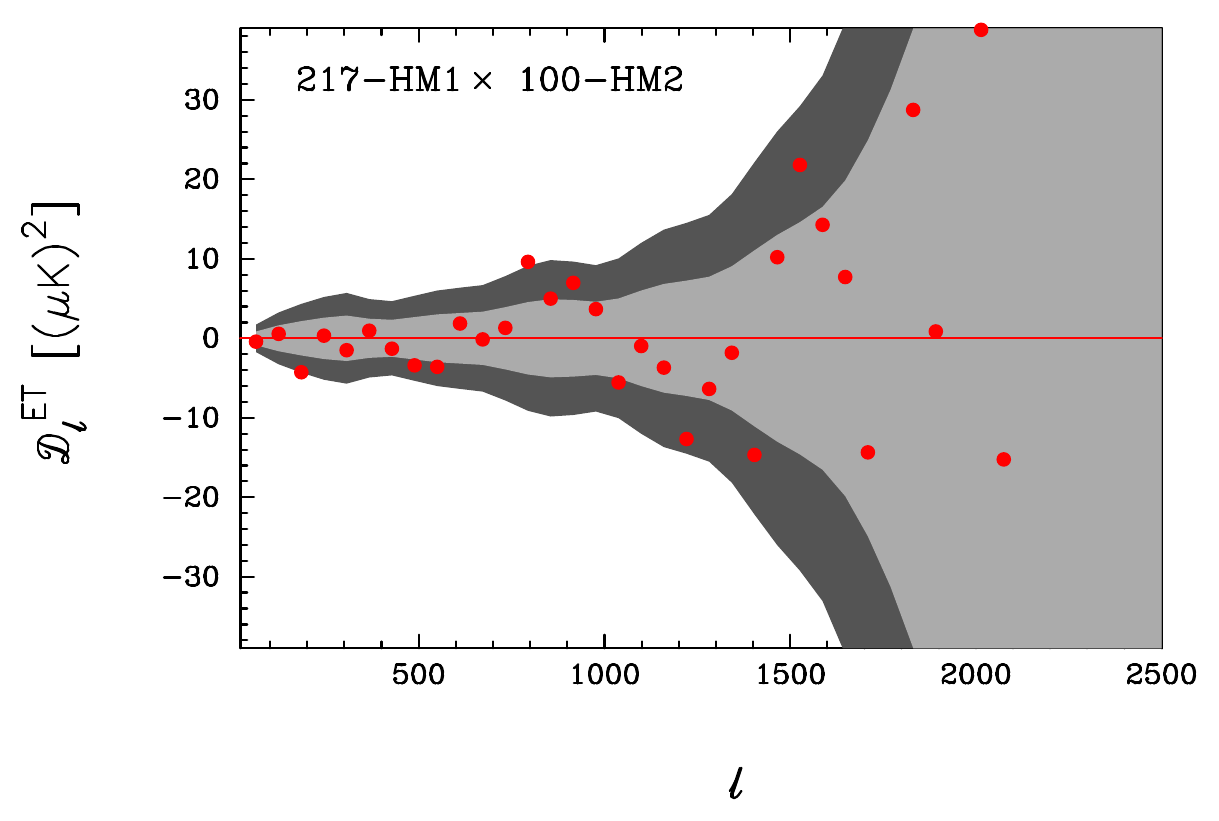} 
\includegraphics[width=50mm,angle=0]{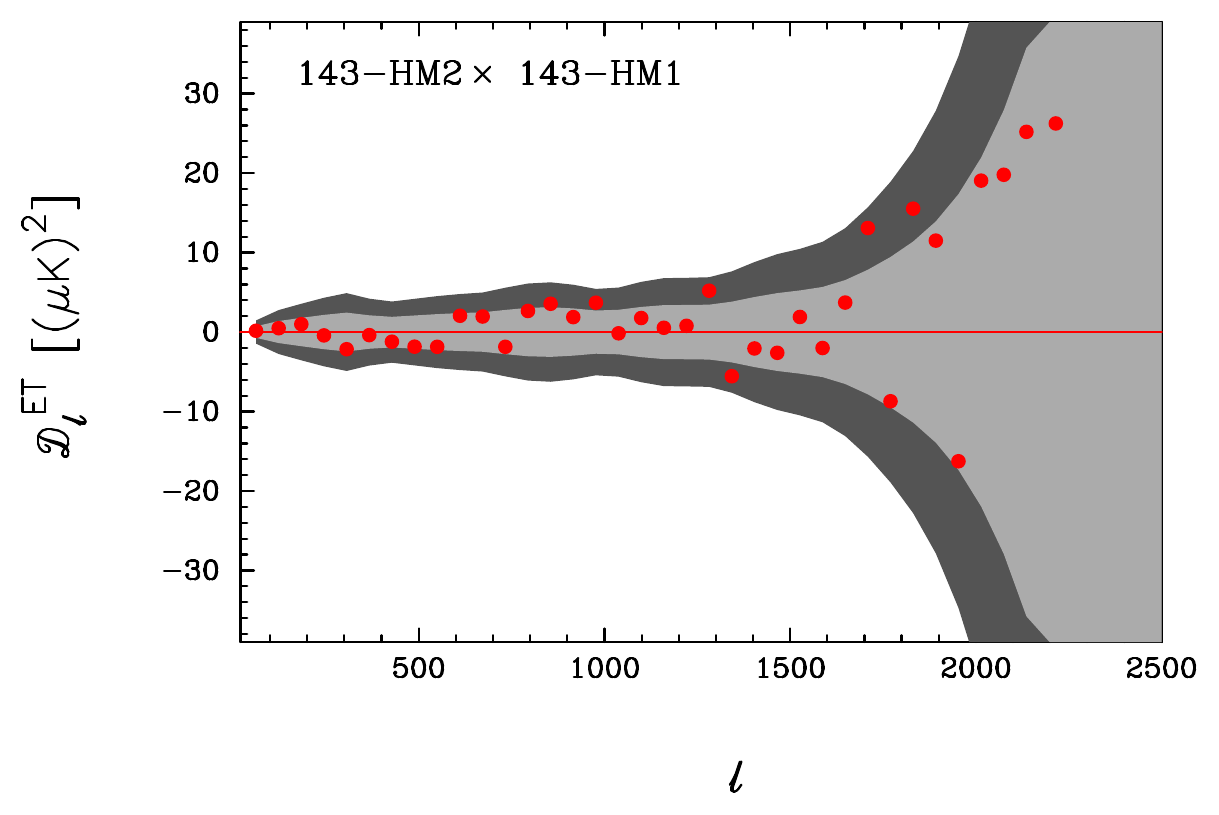}  \\

\includegraphics[width=50mm,angle=0]{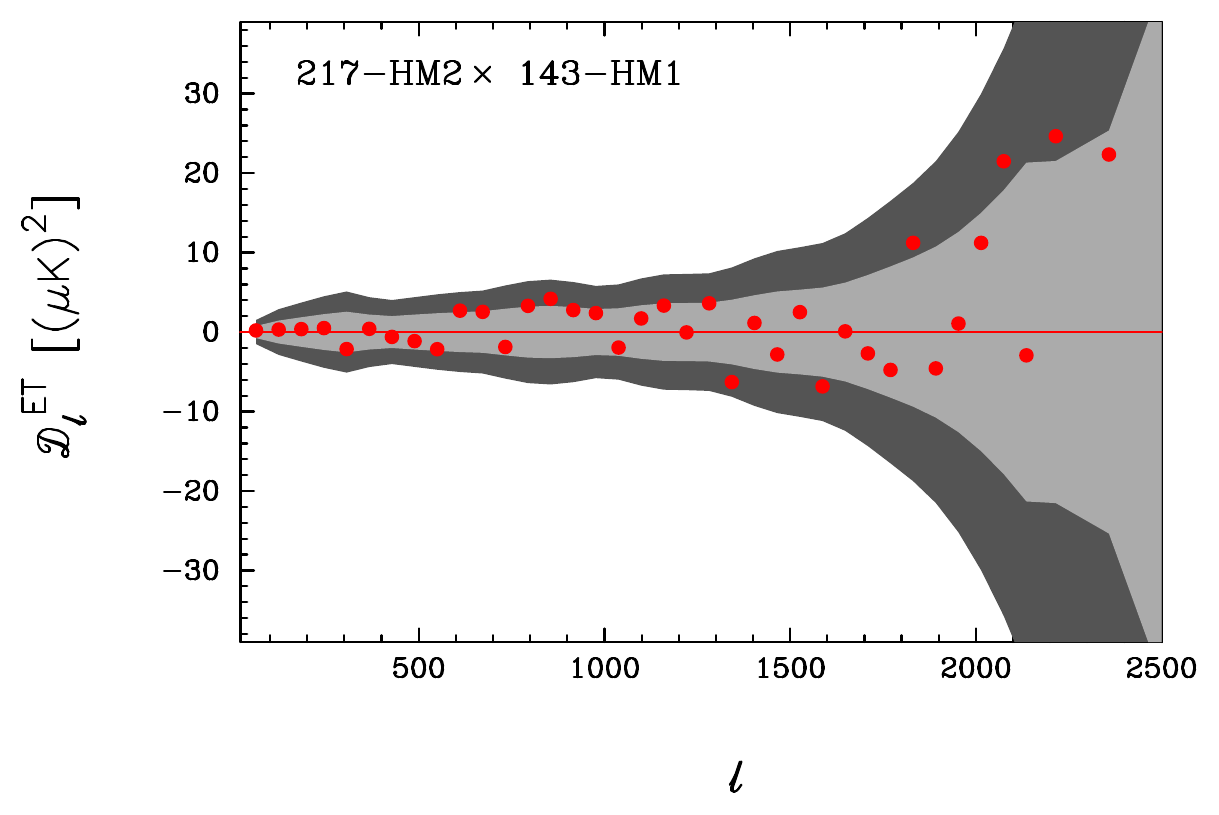} 
\includegraphics[width=50mm,angle=0]{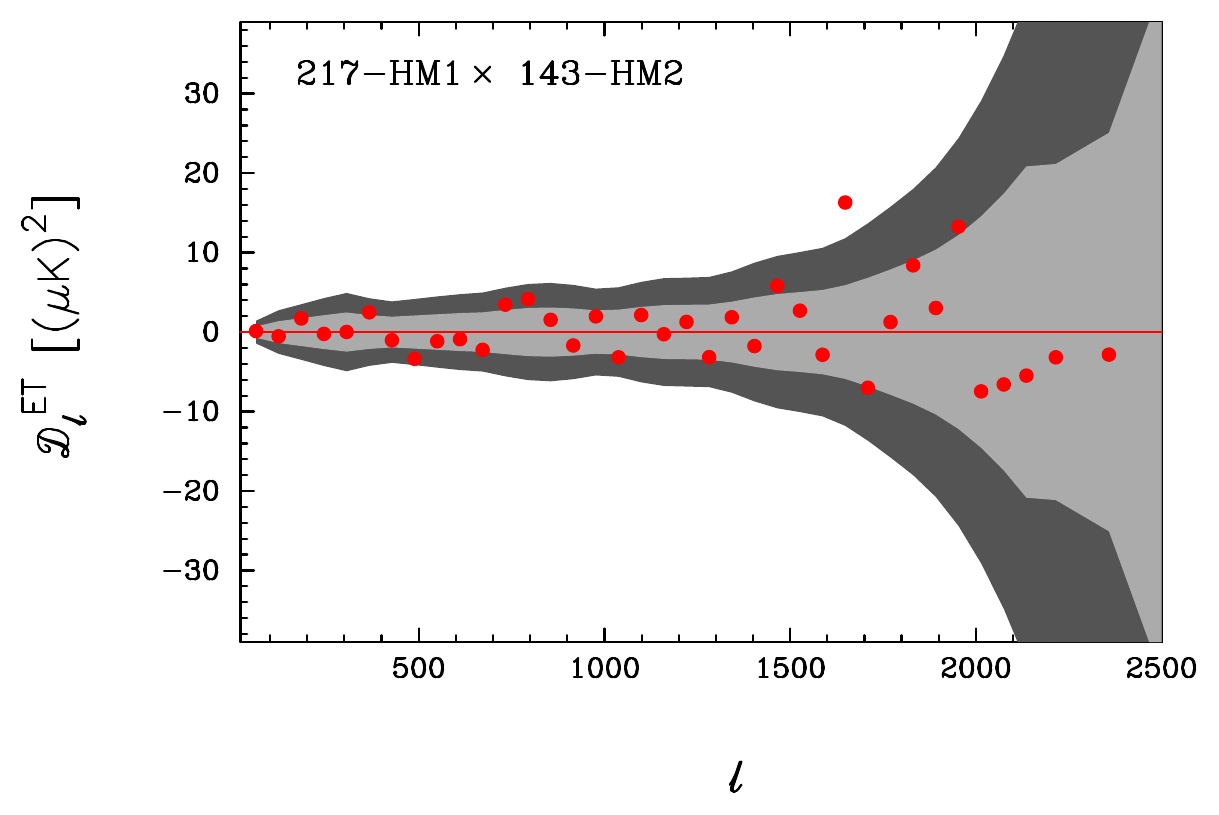} 
\includegraphics[width=50mm,angle=0]{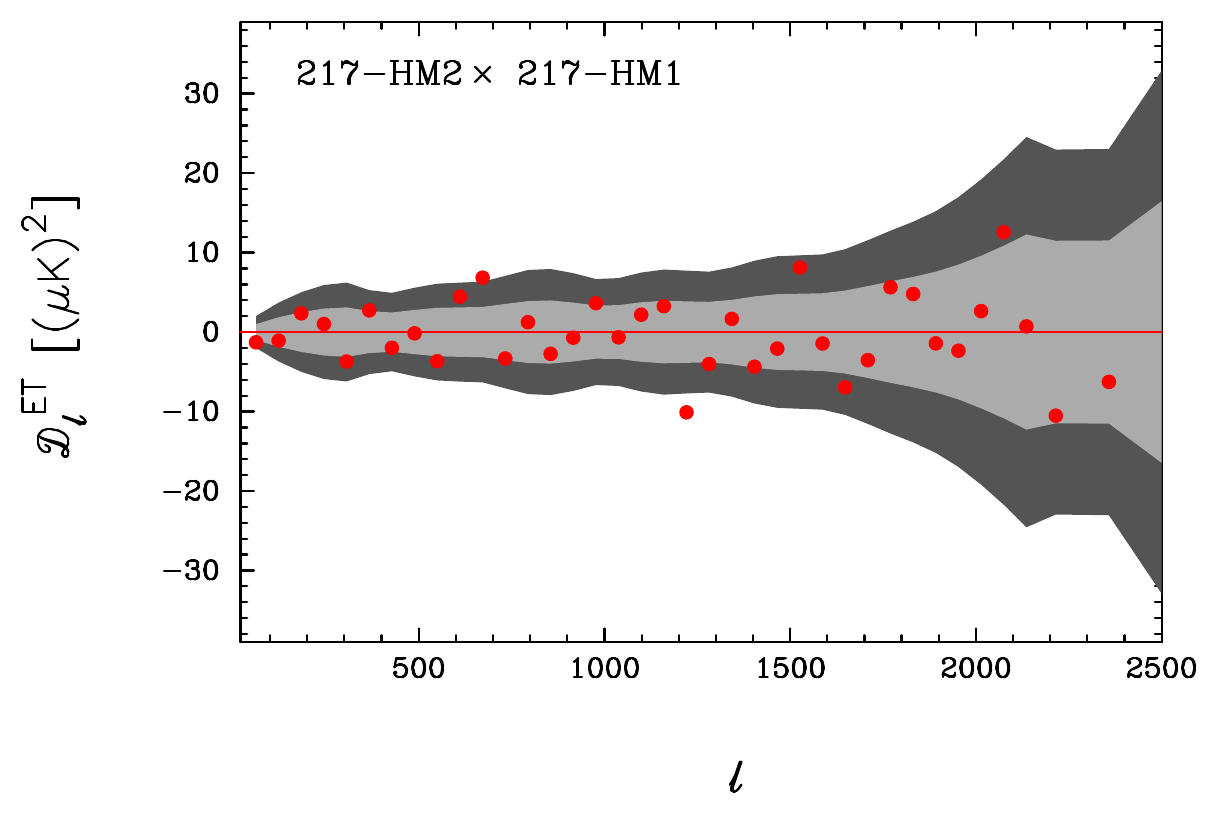}  \\

\caption {As for Fig.\ \ref{fig:inter_frequencyEE} but for the ET spectra.}

\label{fig:inter_frequencyET}

\vspace{0.1truein}
\end{figure}

With these corrections, all of the polarization spectra should be consistent with each other in the absence of
systematics. The residuals relative to the fiducial 12.1HM TT base \LCDM\  cosmology for the
individual polarization spectra of the 12.1HM  likelihood are plotted in 
Figs. \ref{fig:inter_frequencyEE}-\ref{fig:inter_frequencyET}. Table \ref{tab:chi_squared_pol} gives
reduced $\chi^2$  values for these spectra over the multipole ranges used in the 12.1HM likelihood.

\begin{table}[t]
{\centering \caption{\small{Reduced $\chi^2$ for the individual polarization spectra over the multipole ranges
used in the 12.1HM likelihood. $N_D$ lists the number of data points used to compute $\hat \chi^2$.}}
\label{tab:chi_squared_pol}
\begin{center}

\smallskip

\begin{tabular}{|c|c|c|c|c|c|c|c|} \hline 
spectrum  & $\ell$ range EE & $N_D$ & $\hat \chi^2_{EE}$ & $\ell$ range TE/ET& $N_D$ & $\hat \chi^2_{TE}$& $\hat \chi^2_{ET}$  \\  \hline
100HM1$\times$100HM2 &$200-1200$ & $1001$ & $0.85$ & \;\;$30-1200$ & $1171$ & $0.93$ & $0.96$ \\ 
100HM1$\times$143HM2 & \;\;$30-1500$ & $1471$ & $0.80$ & \;\;$30-1500$ & $1471$ & $0.93$ & $0.87$ \\ 
100HM1$\times$217HM2 & $200-1200$ & $1001$ & $0.92$ & $200-1500$ & $1301$ & $1.07$ & $1.00$ \\ 
100HM2$\times$143HM1 & \;\;$30-1500$ & $1471$ & $0.84$ & \;\;$30-1500$ & $1471$ & $0.94$ & $0.95$ \\ 
100HM2$\times$217HM1  &$200-1200$ & $1001$ & $0.92$ & $200-1500$ &   $1301$ & $1.01$ & $0.98$ \\ 
143HM1$\times$143HM2 &$200-2000$ & $1800$ & $0.83$ & \;\;$30-2000$ & $1971$ & $0.95$ & $0.95$  \\ 
143HM1$\times$217HM2 & $300-2000$ & $1701$ & $0.96$ & $200-2000$ & $1801$ & $1.00$ & $0.99$ \\
143HM2$\times$217HM1 & $300-2000$ & $1701$ & $0.94$ & $200-2000$ & $1801$ & $1.05$ & $0.97$ \\
217HM1$\times$217HM2 & $500-2000$ & $1501$ & $1.04$ & $500-2000$ & $1801$ & $1.03$ & $1.08$ \\ \hline
\end{tabular}
\end{center}}

\end{table}

One can see from both the figures and the table that there are no
obvious outliers. In fact, the $\hat \chi^2$ values for spectra
involving $100$ and $143$ GHz are low. This is a consequence of the
difficulties discussed in Sect.\  \ref{subsec:noise+power_spectrum} in
accurately determining the noise levels of \Planck\ in polarization,
particularly at $100$ GHz. The noise models used in this paper are
based on odd-even map differences and, as discussed in
Sect.\ \ref{subsec:noise+power_spectrum}, comparison with auto-spectra
suggests that the odd-even map differences overestimate the noise of
the $100$ GHz maps (see Fig.\ \ref{fig:noisespec}). The $\hat \chi^2$
listed in Table \ref{tab:chi_squared_pol} suggest that the noise
levels for several of the EE spectra are overestimated by a few
percent. Unfortunately, we have not been able to improve the accuracy
of the noise model.

The rationale for choosing the multipole ranges is as follows. The
values of $\ell_{\rm max}$ are chosen so that we do not use the
spectra when they become strongly noise dominated.  The choices of $\ell_{\rm
  max}$ are relatively unimportant since the coadded TE and EE spectra
are dominated by the highest signal-to-noise spectra. The values of
$\ell_{\rm min}$ are chosen so that we do not use spectra that are
heavily contaminated by dust emission. However, since we clean the
polarization spectra using $353$ GHz, polarized Galactic dust emission
is accurately characterized and so the choices of $\ell_{\rm min}$
should be  unimportant if instrumental systematics are negligible (see
Figs. \ref{fig:EEdust_residuals}-\ref{fig:ETdust_residuals}). This is
true for the TE/ET spectra but not for EE. At very low multipoles in
EE, we find clear evidence for residual systematics in the HFI maps.

\begin{figure*}
 \centering
\includegraphics[width=120mm,angle=0]{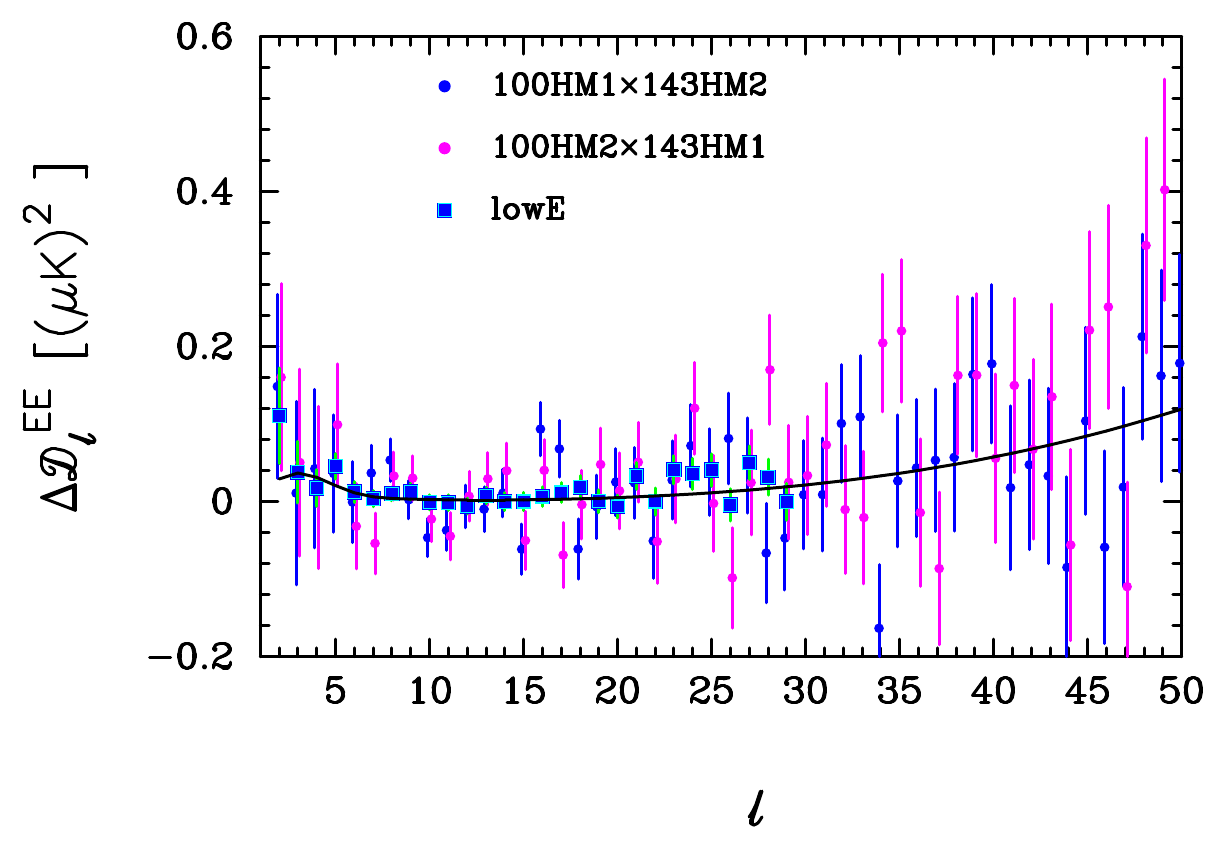} 
\includegraphics[width=120mm,angle=0]{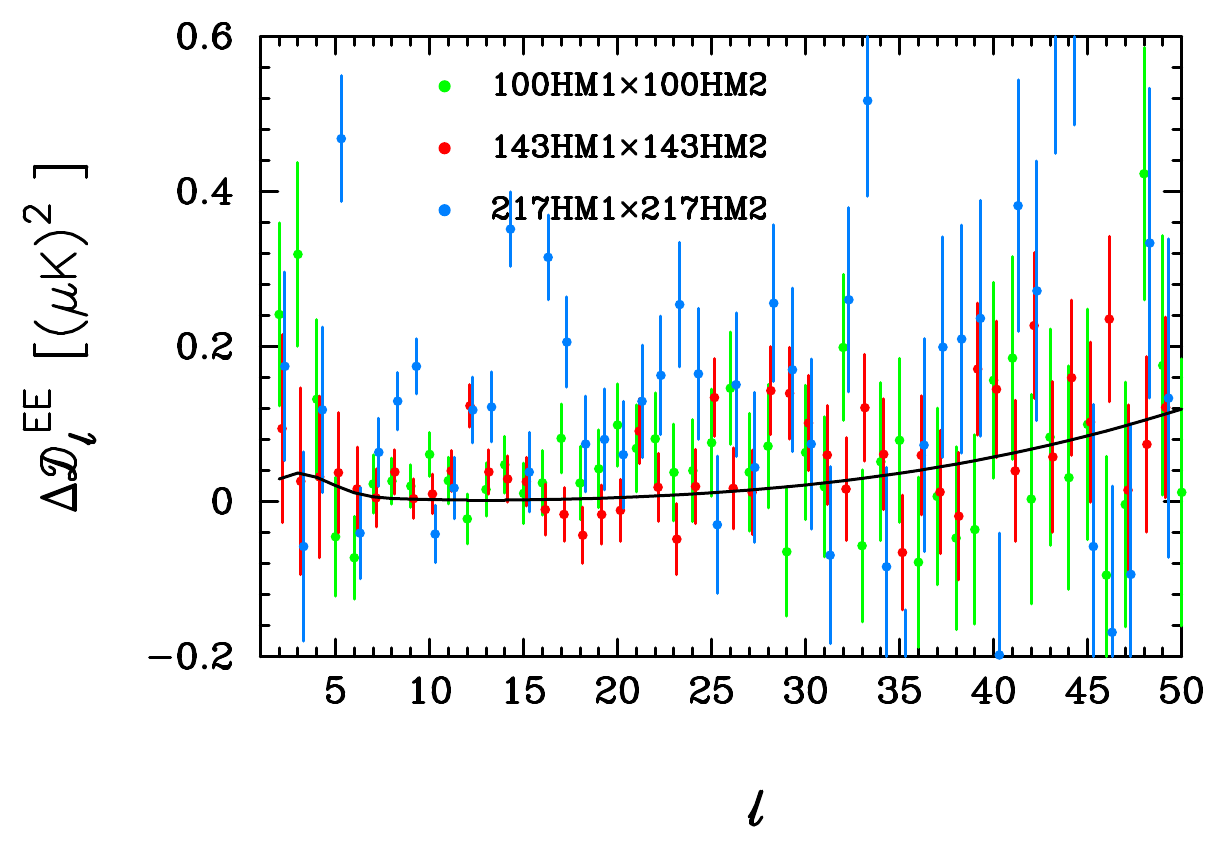}

\caption {EE half mission power spectra computed using maskpol60 at
  low multipoles.  The errors (which are highly correlated between
  multipoles) are computed from the diagonals of the
  \camspec\ covariance matrices. The lines show the EE spectrum for
  the base \LCDM\ cosmology fitted to the 12.1HM TT likelihood. The
  points labelled `lowE' in the upper plot show the $100 \times 143$
  EE quadratic maximum likelihood power spectrum used to form the
  \simall\ likelihood used to constrain $\tau$.}

\label{fig:lowl_EE}
     
\vspace{0.12truein}
 \end{figure*}

The latter point is illustrated by Fig.\ \ref{fig:lowl_EE} which shows selected
half mission EE power spectra at low multipoles.  The solid lines in
the figures show the EE power spectrum for the fiducial base
\LCDM\ cosmology. The optical depth to reionization, $\tau = 0.0524
\pm 0.0080$, in this model is constrained by the
\simall\ likelihood. The $100 \times 143$ EE power spectrum used in
\simall\ is plotted in the upper panel.  The upper figure shows the
two $100\times 143$ EE half mission power spectra. The lower figure
shows the $100\times 100$, $143 \times 143$ and $217 \times 217$ spectra. It
is clear from this figure that the $217 \times 217$ EE spectrum shows
excesses at low multipoles. The $100 \times 100$ and $143 \times 143$
spectra also have excess variance, though less pronounced than in the
$217\times 217$ spectrum.  The two $100 \times 143$ spectra fit well
with the theoretical model. All of the half mission EE cross-spectra
involving $217$ GHz maps show large excesses relative to the other
spectra and to the fiducial \LCDM\ cosmology extending to multipoles
$\ell \sim 100$.

The behaviour of the EE spectra at low multipoles is discussed in
detail in \citep{SROLL:2016} and \citep{DataProcessing:2018}. The main
systematics (visible in Fig.\ \ref{fig:pol_cleaned_maps}) are caused
by non-linearities in the bolometer analogue-to-digital converters
(ADC).  These non-linearities, together with other effects such as
long bolometer time constants and band-pass mismatches, introduce
systematic errors in the polariation maps.  The main aim of the
\SROLL\ map-making algorithm used in the 2018 \Planck\ data release is
to correct these systematic errors at low multipoles\footnote{The
  \SROLL\ maps produced for the $2018$ release give almost identical
  TT, TE and EE power spectra as the $2015$ \Planck\ maps at
  multipoles $\ell \simgt 200$. The changes in map making between the
  2015 and 2018 \Planck\ data releases (including the elimination of
  the last part of survey 5) have negligible impact on the power
  spectra at high multipoles.}.  As discussed in \citep{SROLL:2016} the
ADC non-linearities can be modelled and corrected to high accuracy for
$100$ and $143$ GHz bolometers, but the model is less accurate for
$217$ GHz. As a consequence, $217$ GHz polarization maps are strongly
affected by low multipole systematics. Even at $100$ and $143$ GHz,
there are small biases in $100\times100$ and $143\times 143$
spectra. The approach taken in \citep{SROLL:2016} and
\citep{DataProcessing:2018} is to construct end-to-end simulations of
the \SROLL\ pipeline, which are used to compute and remove biases and
to construct an empirical likelihood using the EE spectra. The lowE
likelihood used here is discussed in \citep{DataProcessing:2018} (and 
in abbreviated form in PPL18) and
uses the full mission $100 \times 143$ EE cross spectrum.

The end-to-end simulations show that biases are small in the $100 \times 143$
cross spectra. We therefore use only the $100\times143$  half mission spectra in the EE block of the \camspec\ 
likelihood at  multipoles $\ell < 200$. One can see from the upper panel of 
Fig.\ \ref{fig:lowl_EE} that the $100 \times 143$ half mission \camspec\ cross-spectrum matches smoothly with the
EE power spectrum used in the \simall\  likelihood at $\ell < 30$. One can also see that the \simall\ spectrum
has much smaller errors than the $100 \times 143$ half mission spectra. There are two reasons for this:
(i) the \simall\ errors are based on the end-to-end simulations, whereas the \camspec\ error model is heuristic
and unreliable at low multipoles; (ii) \camspec\ is based on pseudo-$C_\ell$ power spectrum estimates
which are sub-optimal at low multipoles. The \simall\ estimates use approximate quadratic maximum likelihood cross-spectrum
estimates  developed by us \citep{Efstathiou:2014} and  described in \citep{SROLL:2016}.

\subsection{Comparison with full mission  detset spectra}
\label{subsec:full_mission_pol}

The analysis of full mission detset polarization spectra is almost
identical to that of the half mission spectra.  We use the detset beams to
model TP leakage, and recalibrate each TE/ET and EE spectrum against
the fiducial 12.1HM TT cosmology. However, instead of cleaning the spectra using
$353$ GHz, we subtract the power-law dust model with the coefficients
given in Table \ref{tab:pol_dust_fits}\footnote{This is done because with half mission $353$ GHz it is
not possible to clean the detset spectra without introducing noise correlations in the cleaned
cross spectra.}.  The tests described in
Sect.\ \ref{subsec:correlatednoise} (see Fig.\ \ref {fig:corrnoise})
show that correlated noise between detsets is unimportant in the polarization
spectra  over the multipole ranges used in the \camspec\ likelihoods.

\begin{figure*}
 \centering
\includegraphics[width=140mm,angle=0]{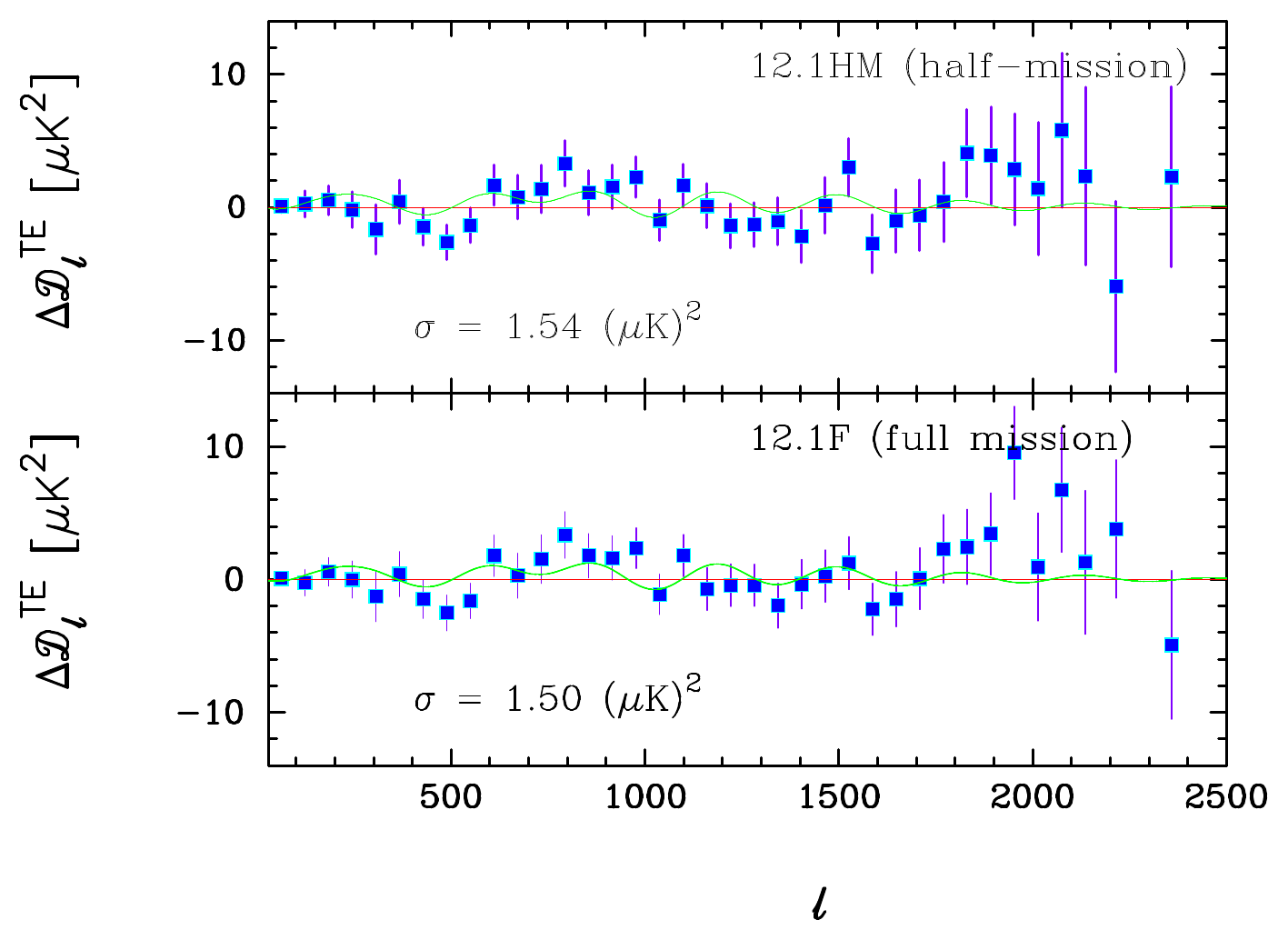} \\

\includegraphics[width=140mm,angle=0]{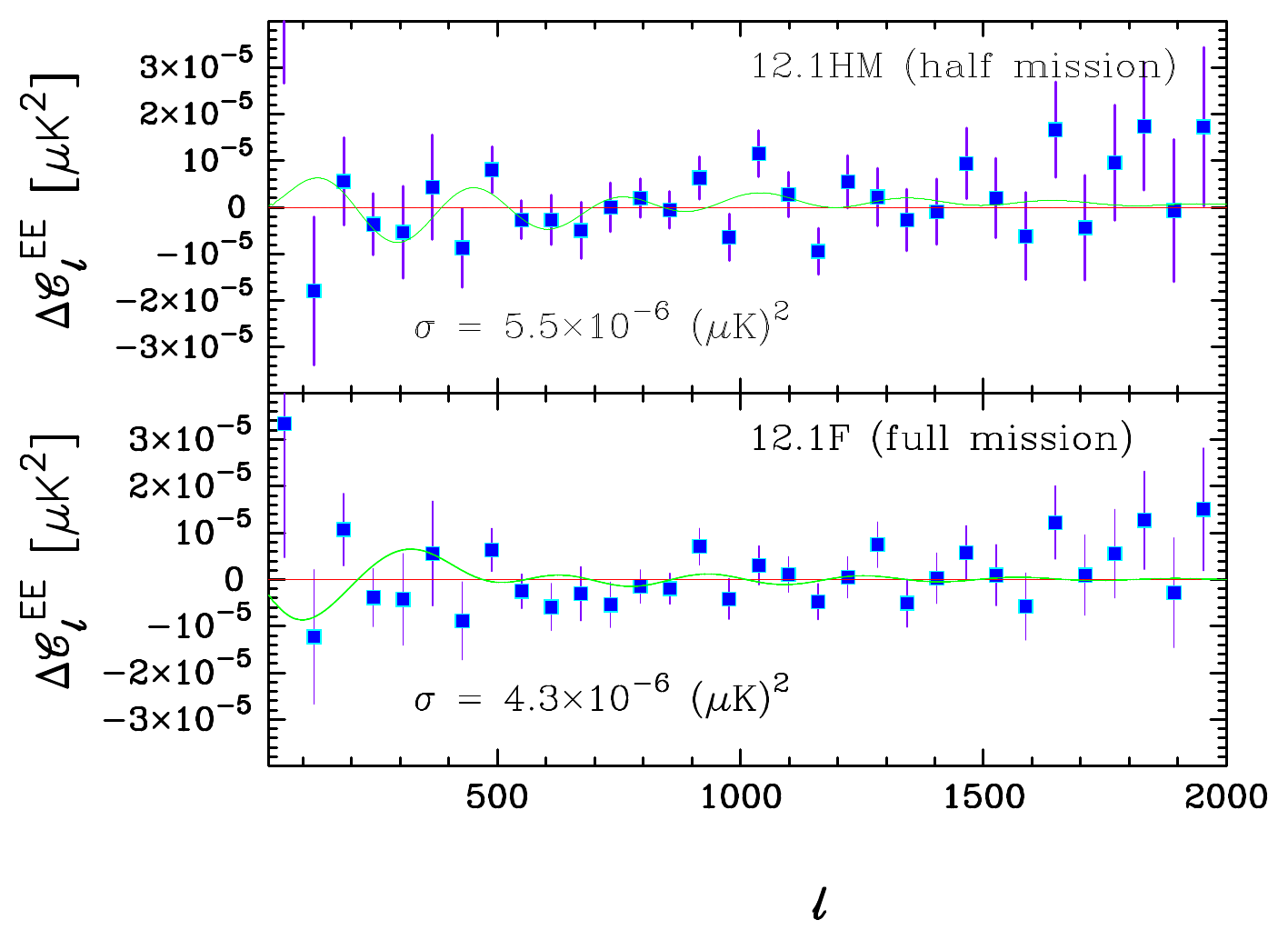} \\

\caption {Comparison of the 12.1HM half mission and 12.1F full mission
  detset TE spectra (upper figure) and EE spectra (lower figure).  The
  residuals are computed with respect to the fiducial 12.1HM TT base
  \LCDM\ cosmology. The numbers give the dispersion of the residuals
  over the multipole range $500 \le \ell \le 1500$. The green curves
  in each panel show the residuals relative to best-fit base
  \LCDM\ cosmology fitted to 12.1HM TE (upper panel, upper figure) and
  12.1F TE (lower panel, upper figure) and to 12.1HM EE (upper panel,
  lower figure) and 12.1F EE (lower panel, lower figure). }

\label{fig:fullvHM}

 \end{figure*}

Fig.\ \ref{fig:fullvHM} compares the coadded full mission detset TE and
EE spectra of the 12.1F likelihood to the half mission spectra used in
the 12.1HM likelihood.  We show the residuals with respect to the
fiducial base \LCDM\ cosmology fitted to 12.1HM TT\footnote{The
  calibration parameters $c_{\rm TE}$ and $c_{\rm EE}$ determined from
  the 12.1HM TTTEEE and 12.1F TTTEEE likelihoods are very close to unity. We have
  therefore not applied relative calibration factors in Fig.\
  \ref{fig:fullvHM}.}.  The green lines show the best fit base
\LCDM\ cosmology fitted to TE and EE blocks of the 12.1HM and 12.1F likelihoods.
The cosmological parameters determined from the polarization spectra
 differ slightly from (but are consistent with) those of the fiducial  base \LCDM\ model (see Table
\ref{tab:LCDMcompare_camspec} and  Fig.\ \ref{fig:base_lcdm_likelihoods}).

We note the following:

\smallskip

\noindent
$\bullet$ The full mission TE spectrum is almost identical to the half mission spectrum. As noted in Sect.\ \ref{subsec:Camspec_likelihood} the increase in signal-to-noise of the full mission TE spectrum compared to the half mission 
spectrum is marginal (and comes primarily from slight improvements in the signal-to-noise
of the temperature maps). Unsurprisingly, therefore, the 
12.1HM TE and 12.1F TE likelihoods lead to almost the same cosmologies.
The numbers in the figure give the dispersion in the band-powers over the
multipole range $500 \le \ell \le 1500$ and are almost identical for the two spectra.

\smallskip

\noindent
$\bullet$ The residuals of the full mission EE spectrum has noticeably
lower scatter compared to those of the half mission spectrum, particularly at
multipoles $\simgt 1000$. For EE there is a non-negligible improvement
in signal-to-noise in switching from half mission to the full mission
detset spectra. The green lines in this figure show the residuals
relative to the base \LCDM\ cosmology fitted to 12.1HM EE and 12.1F
EE. For the EE spectra, these two fits differ, with the 12.1F EE fit
lying closer to the 12.1HM TT cosmology. In other words, the improved
signal-to-noise of the 12.1F EE spectrum brings it closer to the TT
solution.

\smallskip

\noindent
$\bullet$ These results show that there is relatively little to be
gained in forming a full mission detset likelihood compared to a half
mission likelihood.  Although we see an improvement in the
signal-to-noise of the full mission EE spectrum, the EE spectra at
high multipoles are much less powerful than the TT and TE spectra in
constraining \LCDM-like models. The full mission EE spectra should
therefore be more appropriately considered a consistency check of the
\Planck\ polarization data.

\subsection{Variation of TE and EE with sky area}

\begin{figure*}
 \centering
\includegraphics[width=150mm,angle=0]{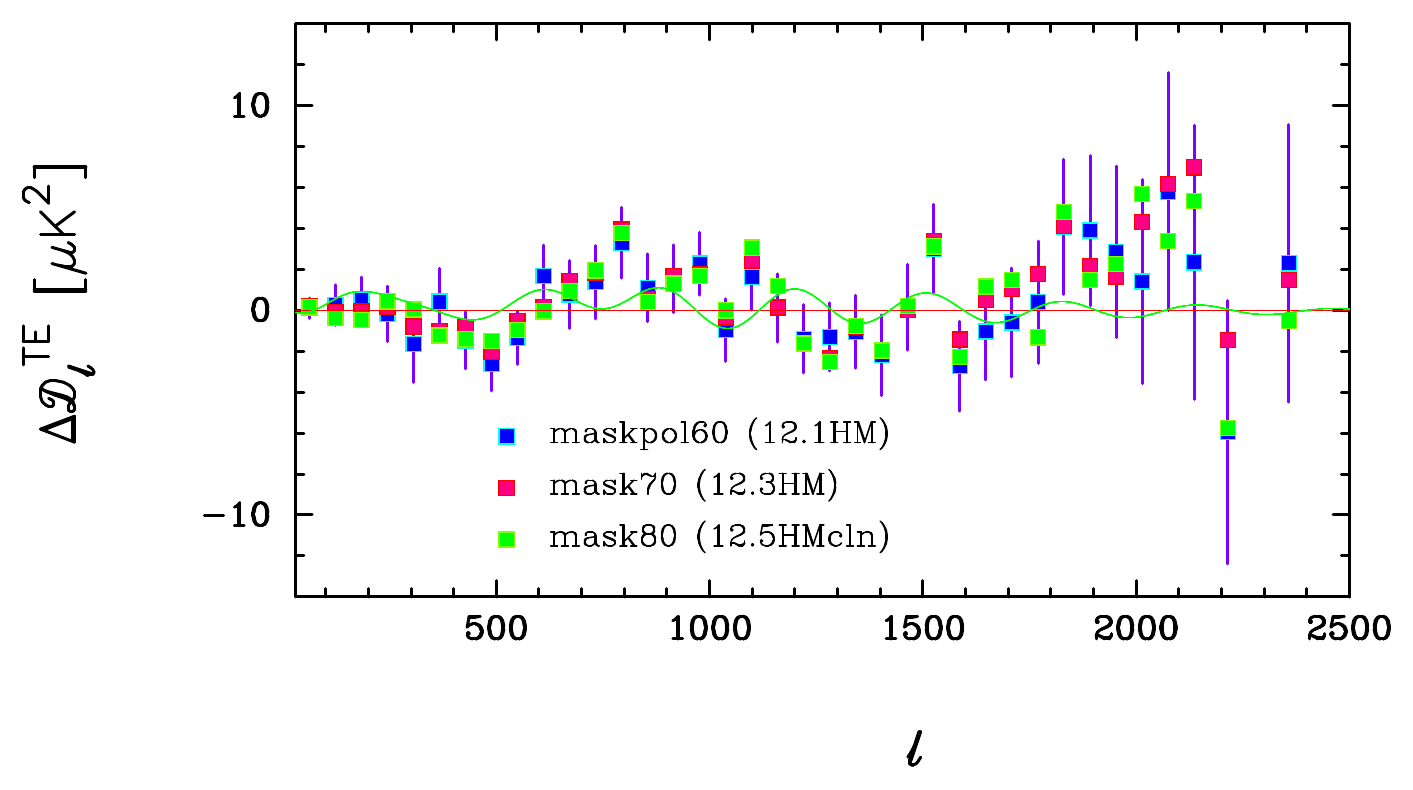} \\

\includegraphics[width=150mm,angle=0]{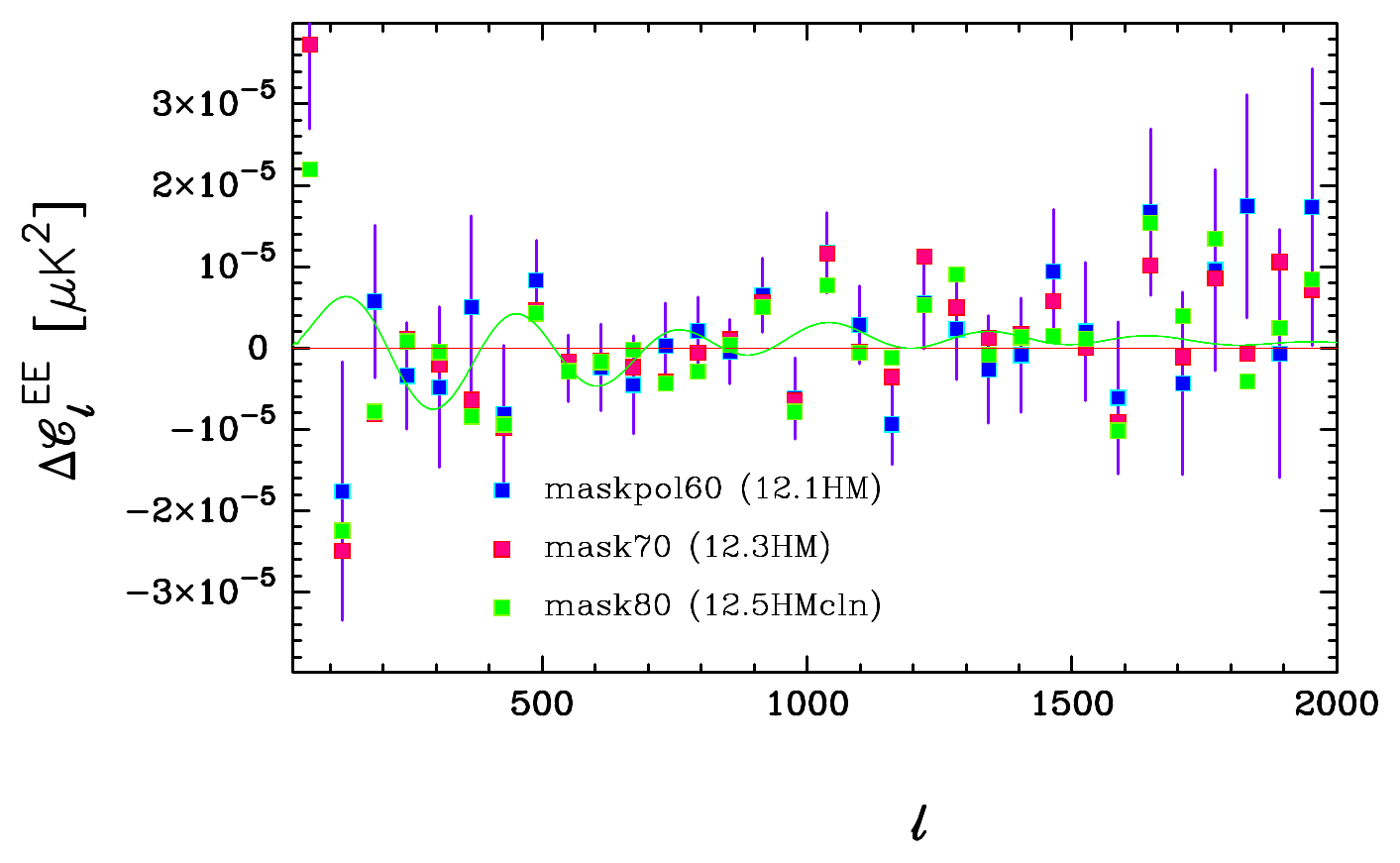} \\

\caption {Half mission TE spectra (upper figure) and EE spectra (lower
  figure) computed using different polarization masks.  The residuals
  are computed with respect to the 12.1HM TT fiducial \LCDM\ base
  cosmology. The blue points with error bars show the TE and EE
  spectra used in the 12.1HM likelihood. The red and green points show
  the spectra using the mask70 and mask80 temperature masks in
  polarization (including the 143 GHz point source+extended object
  mask) as used in the 12.3HM and 12.5HMcl likelihoods. Polarized
  dust emission is subtracted from  the polarization spectra using
  353 GHz, as described in Sect.\ \ref{subsec:pol_clean_spectra}.  The
  green curves in each panel show the residuals of the best-fit base
  \LCDM\ cosmologies fitted to 12.1HM TE (upper figure) and 12.1HM EE
  (lower figure).}

\label{fig:pol_skyarea}

 \end{figure*}

The only other way of improving the signal-to-noise of the
polarization spectra is to increase the sky area.
Fig.\ \ref{fig:pol_skyarea} shows the coadded TE and EE spectra of the
12.1HM likelihood compared to dust-subtracted coadded half mission
polarization spectra using larger areas of sky. The polarization
spectra computed on mask70 and mask80 show the same general features
as the 12.1HM spectra, showing that these spectra are stable with
respect to sky coverage. The 12.5HMcl uses mask80 in both temperature
and polarization. Because of the large sky coverage, 12.5HMcl is the most powerful likelihood
that we have produced from \Planck\ HFI data.

\section{The base \LCDM\ model}
\label{sec:base_lcdm}

\subsection{Supplementary Likelihoods}

The primary purpose of this paper is to explore the consistency of the
Planck power spectra rather than to perform an exhaustive analysis of
the consistency of \Planck\ with other types of data. We have
therefore limited the use of supplementary likelihoods to: (a)
\Planck\ lensing (described in detail in \citep{Planck_lensing:2018})
where we use the lensing likelihood as summarized in section 2.3 of
PCP18, and (b) baryon acoustic oscillation (BAO) measurements, where
we use the identical combination of BAO data as in PCP18.  Most of the
statistical weight in the BAO meaurements comes from the `consensus'
constraints on $D_M(z)$ and $H(z)$ (see below for definitions)
measured in three redshift bins ($z_{\rm eff}=0.38$, $0.51$ and
$0.61$) from the Baryon Oscillation Spectroscopic Survey (BOSS) DR12
analysis \citep{Alam:2017}. As in PCP18, we also include the
measurements of $D_V/r_{\rm drag}$ at lower redshift from the 6dFGS
\cite{Beutler:2011} and SDSS-MGS \cite{Ross:2015}.  We follow a
similar, but more compact, notation to that used in PCP18. For
example: 12.5HMcl TTTEEE+lensing+BAO denotes the TT+TE+EE 12.5HMcl
high multipole likelihood combined with \commander\ and \simall\ at
low multipoles together with the \Planck\ lensing and BAO likelihoods.

\subsection{Acoustic scale parameters}
\label{subsubsec:acoustic_scale}

\begin{table}
{\centering \caption{\small{Acoustic scale parameters in base \LCDM}}

\label{tab:acoustic_scale_summary}
\begin{center}

\smallskip

\begin{tabular}{|l|c|c|c|} \hline 
Likelihood     &   $100\theta_*$ &  $\Sigma$ &  $\Omega_mh^3$  \\ \hline
12.1HM \;  TT      &   $1.04103 \pm 0.00047$   & $101.089 \pm 0.052$   & $0.09707 \pm 0.00045$              \\
12.1HMcl TT    &   $1.04087 \pm 0.00048$   & $101.085 \pm 0.055$   & $0.09586 \pm 0.00047$              \\
12.1F \; \; \  TT       &   $1.04095 \pm 0.00047$   & $101.081 \pm 0.053$   & $0.09614 \pm 0.00045$              \\
12.5HMcl TT    &   $1.04095 \pm 0.00044$   & $101.083 \pm 0.052$   & $0.09576 \pm 0.00043$              \\
12.1HM \; TE      &   $1.04162 \pm 0.00050$   & $101.115 \pm 0.066$   & $0.09606 \pm 0.00053$ \\
12.1HMcl  TE      &   $1.04163 \pm 0.00050$   & $101.116 \pm 0.067$   & $0.09605 \pm 0.00053$ \\
12.1F \; \; \ TE      &   $1.04173 \pm 0.00050$   & $101.120 \pm 0.066$   & $0.09614 \pm 0.00054$ \\
12.2HM \; TE     &  $1.04155 \pm 0.00050$         & $101.112 \pm 0.067$   &      $0.09598 \pm 0.00054$ \\
12.3HM \; TE     &  $1.04160 \pm 0.00047$         & $101.139 \pm 0.065$   & $0.09612 \pm 0.00051$     \\
12.4HM \; TE     &  $1.04166 \pm 0.00048$         & $101.140 \pm 0.062$   &     $0.09612 \pm 0.00049$ \\
12.5HMcl  TE      &   $1.04165 \pm 0.00045$   & $101.148 \pm 0.060$   & $0.09616 \pm 0.00046$ \\
12.1HM \; EE      &   $1.03952 \pm 0.00085$   & $100.67 \; \pm 0.24 \;$     & $0.0973 \; \pm 0.0017$ \; \\
12.1F \; \; \ EE      &   $1.04031 \pm 0.00077$   & $100.58  \pm 0.23 \;$     &  $0.0987  \pm 0.0016$ \; \\
12.2HM \; EE      &   $1.04044 \pm 0.00091$   & $100.64 \; \pm 0.28 \;$     & $0.0978 \; \pm 0.0019$ \; \\
12.3HM \; EE      &   $1.04093 \pm 0.00084$   & $100.79 \; \pm 0.25 \;$     & $0.0977 \; \pm 0.0017$ \; \\
12.4HM \; EE      &   $1.04126 \pm 0.00079$   & $100.89 \; \pm 0.23 \; $     & $0.0973 \; \pm 0.0016$ \; \\
12.1HM \; TTTEEE  &   $1.04106 \pm 0.00032$   & $101.072 \pm 0.039$   & $0.09607 \pm 0.00031$ \\
12.1HMcl TTTEEE  &   $1.04100 \pm 0.00032$   & $101.067 \pm 0.039$   & $0.09604 \pm 0.00032$ \\ 
12.1F \;\; \ \  TTTEEE  &   $1.04123 \pm 0.00030$   & $101.076 \pm 0.039$   & $0.09622 \pm 0.00031$ \\ 
12.2HM \; TTTEEE  &   $1.04111 \pm 0.00033$   & $101.075 \pm 0.039$   & $0.09611 \pm 0.00032$ \\
12.3HM \; TTTEEE  &   $1.04103 \pm 0.00031$   & $101.085 \pm 0.038$   & $0.09610 \pm 0.00031$ \\
12.4HM \; TTTEEE  &   $1.04121 \pm 0.00030$   & $101.094 \pm 0.037$   & $0.09622 \pm 0.00031$ \\
12.5HMcl TTTEEE  &   $1.04124 \pm 0.00028$   & $101.096 \pm 0.036$   & $0.09613 \pm 0.00029$ \\ \hline
\end{tabular}
\end{center}}
\end{table}

The characteristic angular scale of CMB acoustic fluctuations, $\theta_*$,
is very accurately determined by the \Planck\ power
spectra. In base \LCDM, the CMB measurements of $\theta_*$ can be
approximated as a tight constraint on the parameter combination
\begin{equation}
     \Sigma = \left ( r_{\rm drag} h \over {\rm Mpc} \right ) \left (\Omega_m \over 0.3 \right )^{0.4}.  \label{param1}
\end{equation}
Typically, $\theta_*$ and $\Sigma$ are fixed to $\simlt 0.05\%$ by the
\Planck\ data (PCP18). Since the parameter combination $\Omega_bh^2$
is well determined by the relative heights of the CMB acoustic peaks,
the parameter combination $\Omega_m h^3$ \citep{Percival:2002} offers
a simpler proxy to the acoustic scale $\theta_*$, accurate to
typically $0.3\%$.

\begin{figure}[h]

\centering
\includegraphics[width=85mm,angle=0]{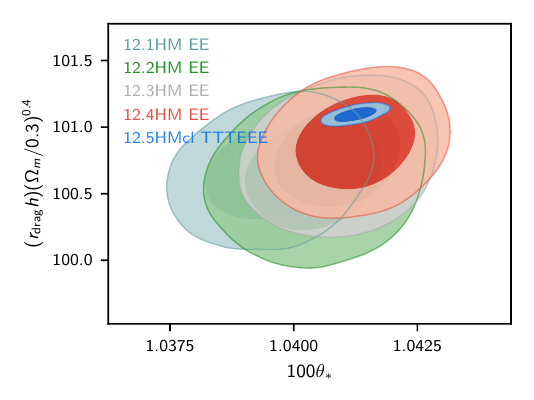} 
\includegraphics[width=85mm,angle=0]{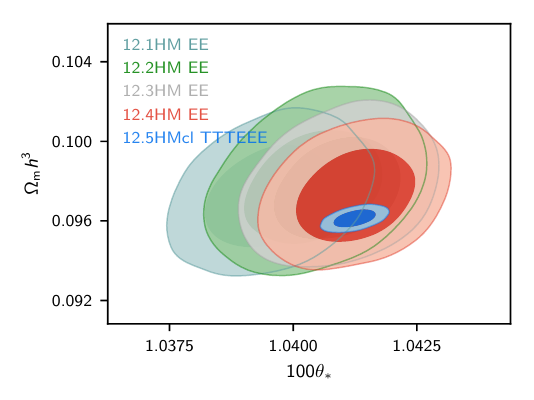}

\caption {Variation of the acoustic scale location parameters determined from the EE likelihoods. 
The sky area used in polarization increases from 12.1HM through to 12.4HM. (The EE block of the 12.5HMcl
likelihood is identical to that of the 12.4HM likelihood). We also show results for the
12.5HMcl TTTEEE likelihood.}

\label{fig:location_polEE}

\end{figure}

Table \ref{tab:acoustic_scale_summary} gives values for the acoustic
scale parameters determined from various likelihoods. Note that the
TT, TE and EE estimates determined from any given likelihood are
almost independent of each other. The agreement between these
estimates is excellent.  The EE spectra show an interesting feature,
however. Since the EE spectra are noisy, the acoustic scale parameters
from the EE half mission spectra are less accurate than those
determined from the TT and TE spectra.  As the sky area is increased
from the 12.1HM through to 12.4HM, the EE acoustic scale
parameters drift towards closer agreement with those determined from
the TT and TE spectra. This is illustrated in
Fig.\ \ref{fig:location_polEE}.   The acoustic scale parameters
determined from the TT, TE and TTTEEE likelihoods are remarkably
stable and show no trend with increasing sky area.

The final seven rows in Table \ref{tab:acoustic_scale_summary} give the acoustic scale
results for the TTTEEE likelihoods. The acoustic scale parameters determined from
 the 12.1HM and 12.5HMcl TTTEEE likelihoods are consistent to better than $0.03\%$, even though these
likelihoods use different sky areas in both temperature and polarization and very different foreground 
treatments in temperature.

\begin{figure}
\centering
\includegraphics[width=170mm,angle=0]{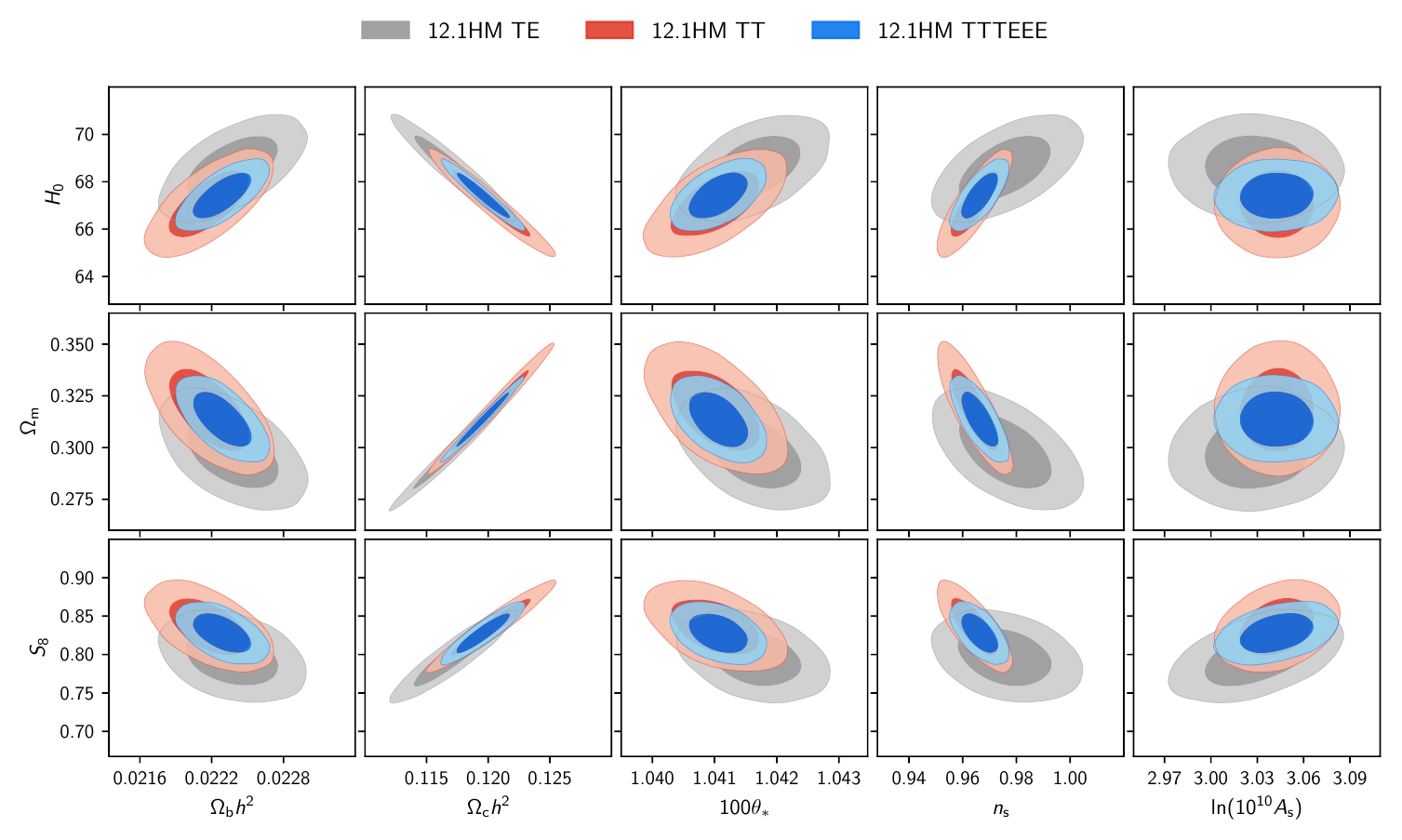} \\
\includegraphics[width=170mm,angle=0]{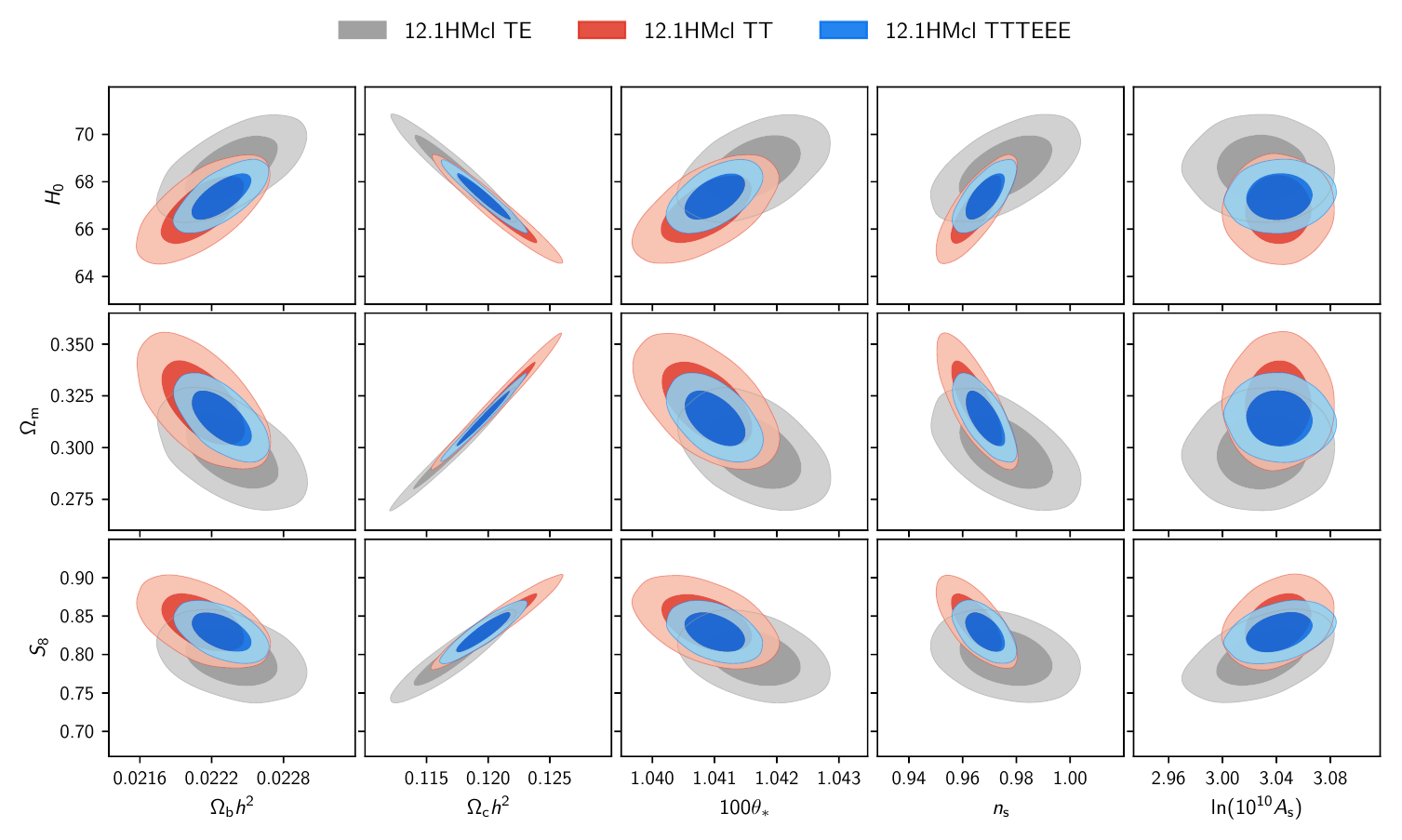}   
\caption {$68\%$ and $95\%$ confidence contours for base \lcdm\ cosmological parameters determined from the
TT, TE and TTTEEE likelihoods. The upper figure shows results for the 12.1HM likelihood, which is similar
to the \camspec\ likelihood discussed in PCP18. The lower figure shows results for the 545 GHz temperature cleaned
12.1HMcl likelihood. \GE{The  cosmological parameters are defined as in PCP15 and PCP18.}}

\label{fig:base_lcdm_likelihoods}

\end{figure}

\begin{figure}
\centering

\includegraphics[width=170mm,angle=0]{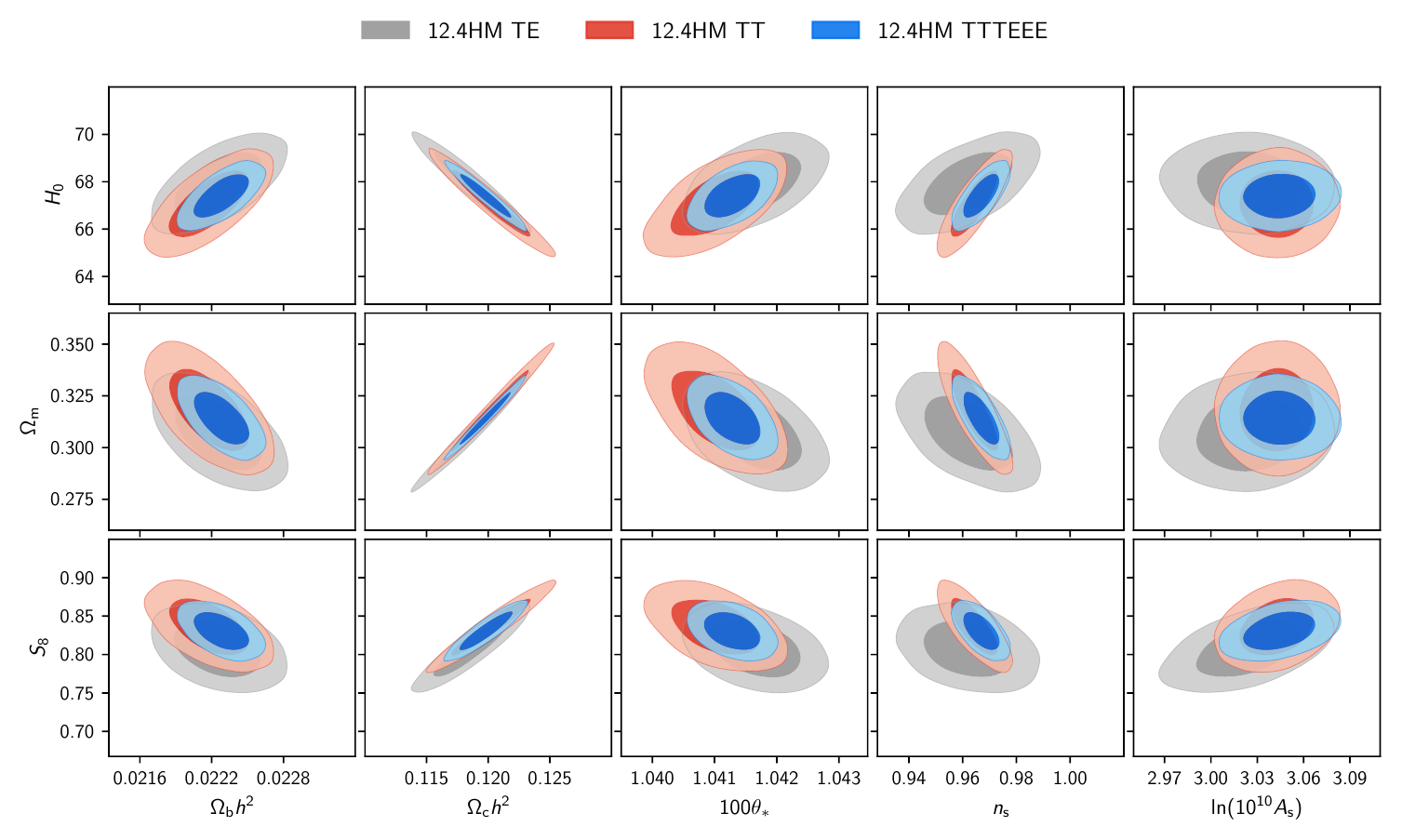}  \\
\includegraphics[width=170mm,angle=0]{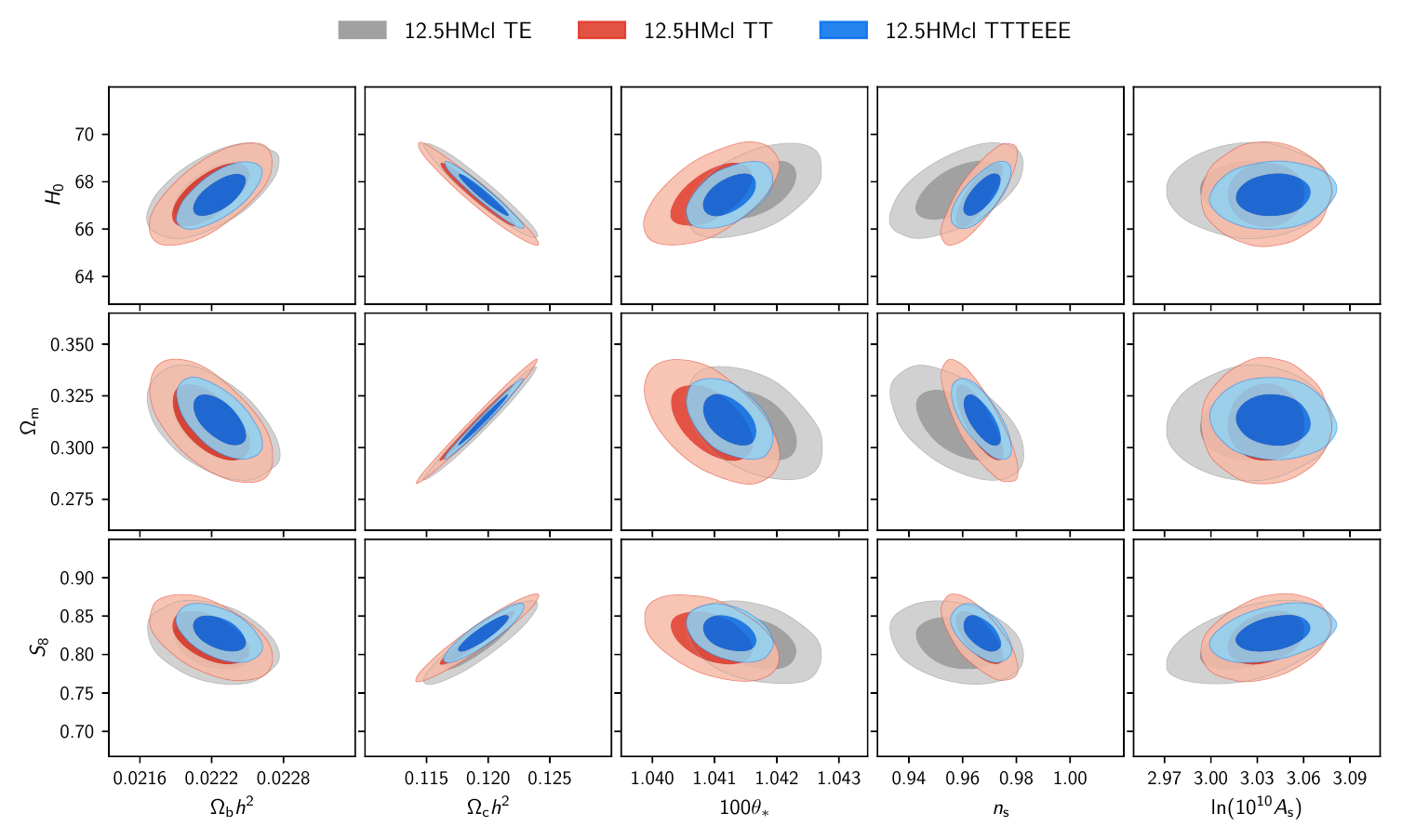}  

\caption*{{\bf  (Continued).}  The upper figure shows results for the 12.4HM likelihood, 
increasing the sky area in polarization compared to 12.1HM. The lower figure shows 
results for the 12.5HMcl likelihood which uses more sky area than 12.1HMcl in both 
temperature and polarization.}

\end{figure}

\subsection{Consistency between temperature and polarization}
\label{subsubsec:base_consistency_TP}

Figure  \ref{fig:base_lcdm_likelihoods} plots constraints on several
key cosmological parameters illustrating the consistency between the TT, TE and
the TTTEEE likelihoods. The 12.1HM and 12.1HMcl likelihoods are quite similar  to 
the \camspec\ likelihoods used in PCP18. The parameter constraints from these likelihoods
are in extremely close agreement with those reported in PCP18 (see e.g.\ Fig.\ A.1 of PCP18).

\begin{figure}
\centering
\includegraphics[width=150mm,angle=0]{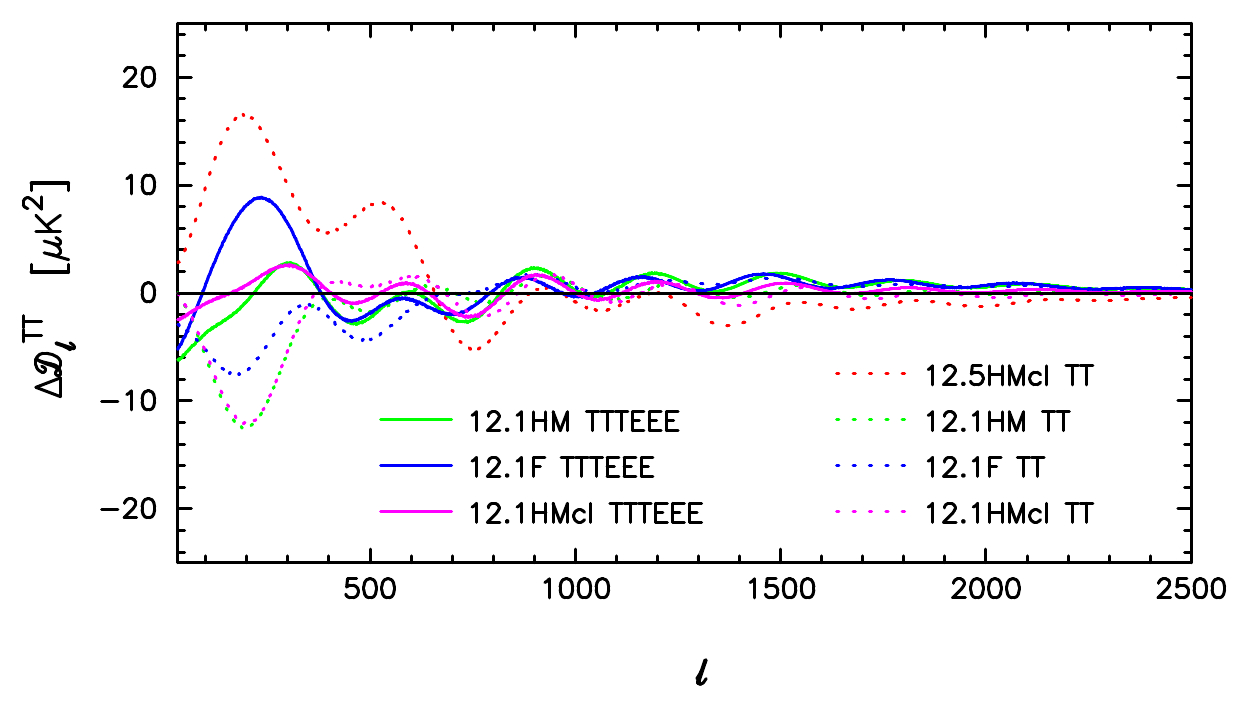}

\caption {Residuals of the best-fit base \LCDM\ TT theory  spectra determined
  from various likelihoods. The 12.5HMcl TTTEEE best-fit base
  \LCDM\ temperature spectrum is used as the reference.}

\label{fig:base_models}

\vspace{0.05truein}

\end{figure}

The TT, TE and TTTEEE parameter constraints are consistent with each
other in all of the \camspec\ likelihoods. The main trend apparent in
Fig.\ \ref{fig:base_lcdm_likelihoods} is for the TE measurement of
$n_s$ to drift to lower values as the sky area is increased, though
for all likelihoods, the TE measurements of $n_s$ are consistent with
those determined from TT and TTTEEE.  The most striking result from
Fig.\ \ref{fig:base_lcdm_likelihoods} is the stability of the TTTEEE
parameter constraints as we scan through the likelihoods. Comparing
the TTTEEE results from 12.1HM and 12.5HMcl, the results for most
parameters are consistent to better than $0.2\sigma$. The largest
deviation is found for $\theta_*$, which differs by $0.57\sigma$. This
is quite a large shift, but one must bear in mind that the formal
errors on $\theta_*$ from the TTTEEE likelihoods are very small (see
Table \ref{tab:acoustic_scale_summary}).  The values of $\theta_*$
determined from these two likelihoods are actually consistent to
better than $0.02 \%$.

\vspace{0.2truein}

\subsection{Best-fit models}

We can get an intuitive feel of the behaviour of these likelihoods by
looking at the  best-fit base
\lcdm\ temperature power spectra. We choose the 12.5HMcl TTTEEE best-fit model as a
reference and plot the residuals of the best-fit TT theory spectra for
various likelihoods in Fig.\ \ref{fig:base_models}. Since this type of 
plot is extremely sensitive to small absolute calibration differences 
we have rescaled the temperature spectra by minimising the rms scatter
of the spectral differences over the multipole range $500 \le \ell \le 1500$. This
rescaling largely removes multiplicative differences so that one can see differences
in the shapes of the best-fit models.

The residuals are well below $\sim 10 \ (\mu{\rm K})^2$ for all 
likelihoods  at $\ell \simgt 800$. At lower multipoles, we see
higher residuals of up to $\sim 16 \  (\mu{\rm K})^2$ at $\ell \sim 200$
(corresponding to the first acoustic peak).  The largest differences
are for the 12.5HMcl TT likelihood, for which we increased sky area to 80\%
for the 143 and 217 GHz maps. The inter-frequency residuals for the TT spectra
used in the 12.5HMcl TT likelihood (see Fig.\ \ref{fig:bands3_12_5}) are, however, 
significantly smaller than the differences seen in Fig.\ \ref{fig:base_models} which we
believe reflect differences in the response of the likelihood to cosmic variance
rather than inaccuracies in  dust subtraction. The differences between the 545 GHz cleaned and
uncleaned likelihoods may, however, be caused by errors in dust subtraction. (For the uncleaned
spectra, the foreground model is sufficiently flexible that dust subtraction errors of 
$\sim 20 \  (\mu{\rm K})^2$ at the first peak can be absorbed by the foreground model and
hence not show up in inter-frequency comparisons; see Figs.\ \ref{fig:inter_frequency100v143} and
\ref{fig:bands3}.) As expected from Fig.\ \ref{fig:base_lcdm_likelihoods}, the best-fit models
for the TTTEEE likelihoods agree extremely well over the entire multipole range.

\begin{figure}
\centering
\includegraphics[width=150mm,angle=0]{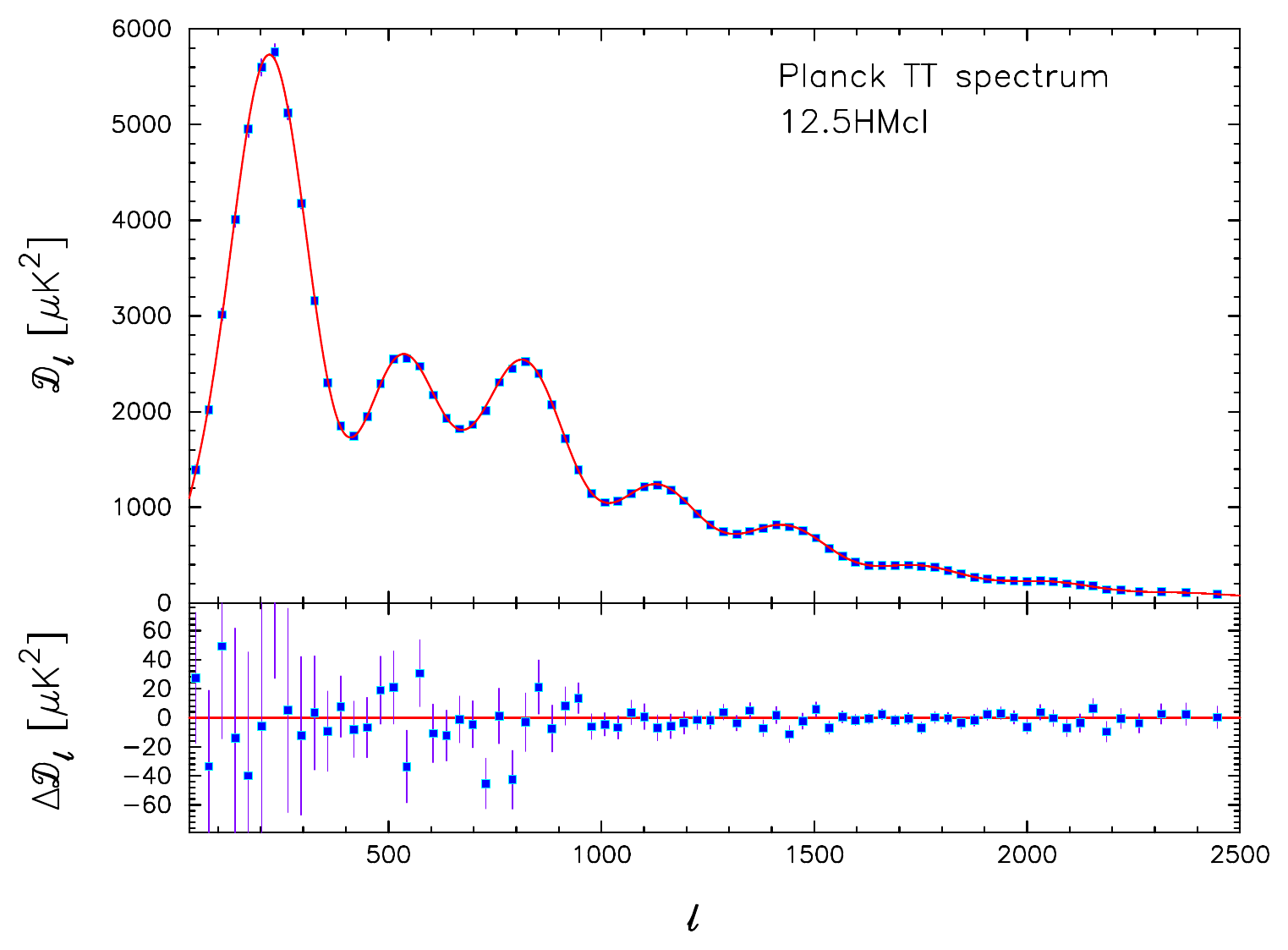} 
\caption {The maximum likelihood frequency averaged temperature power
  spectrum for the 12.5HMcl likelihood. This can be compared with the
  corresponding plot for the 12.1HM likelihood plotted in
  Fig.\ \ref{fig:12.1TT}. The best-fit base \LCDM\ cosmology fitted to the
  12.5HMcl TTTEEE likelihood is plotted in the upper panel and the
  residuals with respect to this theoretical model are plotted in the
  lower panel.}

\label{fig:12.5_TT}

\end{figure}

\begin{figure}

\centering
\hspace{3mm}\includegraphics[width=136mm,angle=0]{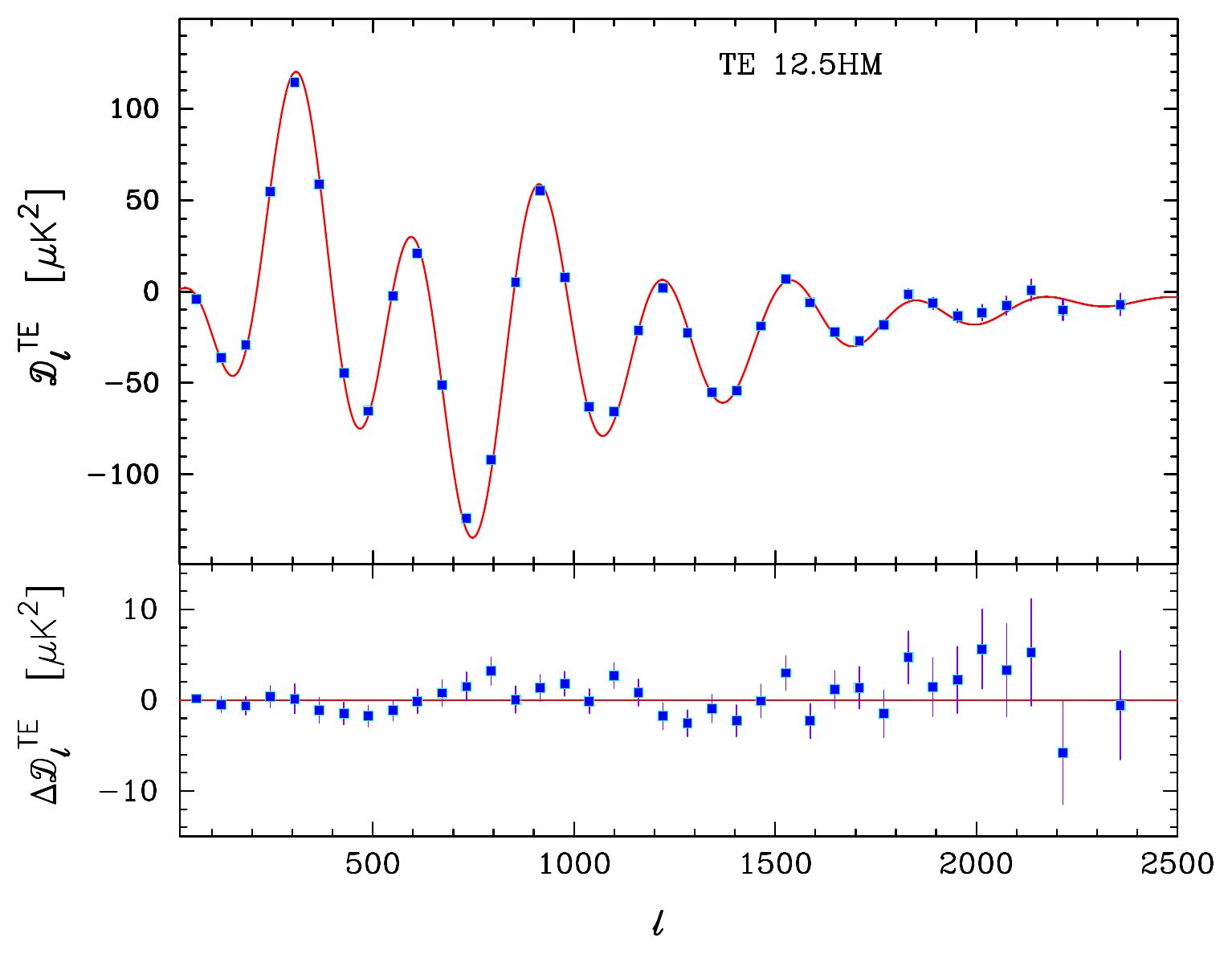} \\
\includegraphics[width=140mm,angle=0]{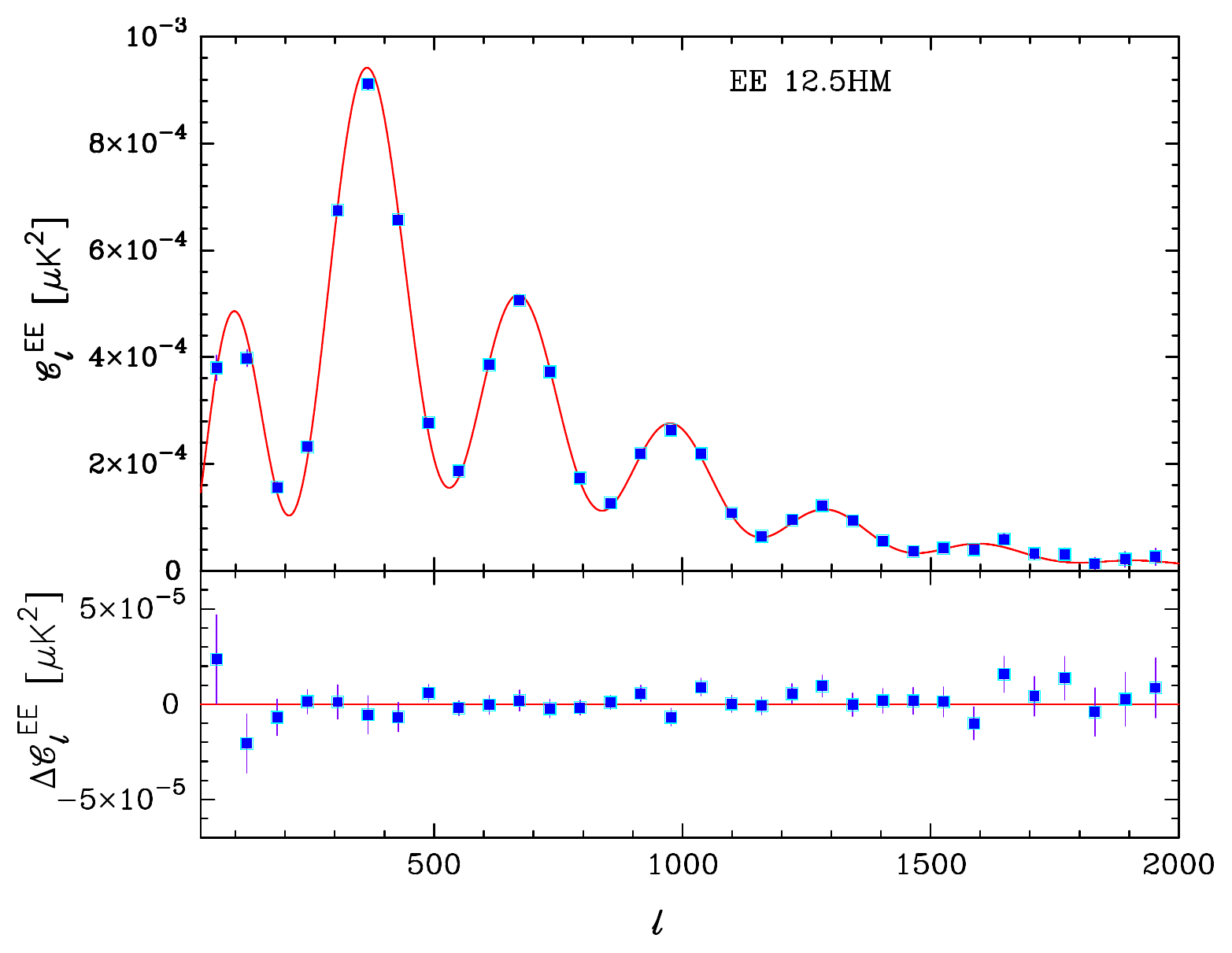} 

\caption {The coadded TE and EE power spectra for the 12.5HMcl
  likelihood. The best-fit base \LCDM\ cosmology fitted to
  12.5HMcl TTTEEE likelihood is plotted in the upper panels.  Residuals
  with respect to this theoretical model are shown in the lower
  panels.}

   \label{fig:12.5_TE_EE}

\end{figure}

The coadded 12.5HMcl TT, TE and EE spectra are plotted in
Figs.\ \ref{fig:12.5_TT} and \ref{fig:12.5_TE_EE}.  These plots can be
compared with the corresponding plots for the 12.1HM likelihood in
Figs.\ \ref{fig:12.1TT} and \ref{fig:12.1_TE_EE}. The polarization spectra of Fig. \ref{fig:12.5_TE_EE}
are compared with the polarization spectra measured by ACTpol and SPTpol in Appendix \ref{sec:appendix2}. The 12.5HMcl
likelihood is statistically more powerful than the 12.1HM and 12.1HMcl likelihoods
because of the larger sky area in both temperature and polarization.
For each of TT, TE and EE, the residuals plotted in the lower panels
of Figs.\ \ref{fig:12.5_TT} and \ref{fig:12.5_TE_EE} show less scatter
than seen in the 12.1HM spectra. In other words, the increase in sky
area leads to quieter spectra.  This is important evidence in favour
of the base \LCDM\ cosmological model.

\subsection{Comparison with Baryon Acoustic Oscillations and \Planck\ lensing}
\label{subsec:base_BOA}

\begin{figure}
\centering
\includegraphics[width=175mm,angle=0]{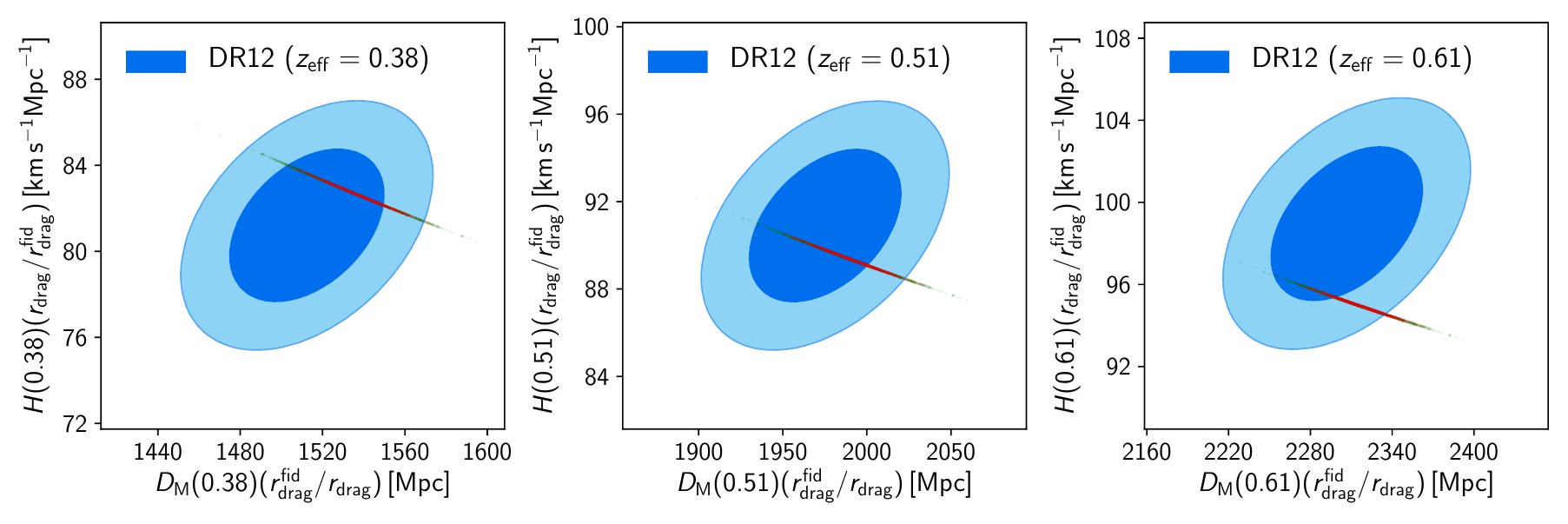}

\caption {The contours show 68\% and 95\% constraints on $D_M$ and
  $H(z)$ from the BOSS DR12 BAO analysis of \cite{Alam:2017} (which
  adopts a fiducial value of the sound horizon of $r^{\rm fid}_{\rm
    drag} = 147.78 \ {\rm Mpc}$). The green and red points show samples
  from the 12.5HMcl TT and 12.5HMcl TTTEEE chains respectively. For
  base \LCDM, the CMB constraints lie accurately on degeneracy lines
  specified by $\Omega_m h^2$ (see Eqs.\ \ref{equ:BAO1} -
  \ref{equ:BAO3}). Adding the TE and EE blocks to the 12.5HMcl TT
  likelihood narrows the range of allowed values of $\Omega_m h^2$,
  excluding high values  which are disfavoured by the
  BAO data.}

\label{fig:BAO_12_5}

\vspace{-0.4truein}
\end{figure}

 Fig.\ \ref{fig:BAO_12_5} shows the BOSS DR12 constraints on the
 comoving angular diameter distance $D_M(z)$ and $H(z)$ from the
 `consensus' results of \citep{Alam:2017} compared to the
 \Planck\ constraints from the 12.5HMcl likelihood.  Section
 \ref{subsubsec:acoustic_scale} shows that the acoustic scale
 parameters $\theta_*$, $\Sigma$ and $\Omega_mh^3$ are accurately
 determined by \Planck\ and are extremely stable. In base \LCDM,
fixing the acoustic scale forces the CMB constraints to lie on a 
degeneracy line in the $D_M(z)$-$H(z)$ plane depending on the value of
$\omega_m = \Omega_m h^2$ (or, equivalently $H_0$). In fact, to an accuracy of
about $0.4\%$, the \Planck\ results in Fig.\ \ref{fig:BAO_12_5} lie on the degeneracy lines: 
\begin{subequations}
\begin{equation}
\begin{cases}
H(z=0.38) ( r_d/r_d^F )   &= 57.07\;(\omega_m/0.1)^{-0.055}(1 + (\omega_m/0.1)^{-2.046}) \Hunit,     \\  \label{equ:BAO1} 
D_M(z=0.38) ( r_d^F/ r_d ) & =  2090\;\; (\omega_m/0.1)^{0.445}/(1 + (\omega_m/0.1)^{-1.451})  \ {\rm Mpc} ,
\end{cases}   
\end{equation}
\begin{equation}
\begin{cases}
H(z=0.51)(r_d/r_d^F) &= 59.14\; (\omega_m/0.1)^{0.097}(1 + (\omega_m/0.1)^{-2.149}) \Hunit, \\
D_M(z=0.51)(r_d^F/r_d)& = 2762\;\; (\omega_m/0.1)^{0.338}/(1 + (\omega_m/0.1)^{-1.586})  \ {\rm Mpc} ,
\end{cases}  \label{equ:BAO2}
\end{equation}
\begin{equation}
\begin{cases}
H(z=0.61)(r_d/r_d^F) &= 60.93\; (\omega_m/0.1)^{0.189}(1 + (\omega_m/0.1)^{-2.163}) \Hunit,  \\
D_M(z=0.61)(r_d^F/r_d)& = 3261\; \; (\omega_m/0.1)^{0.274}/(1 + (\omega_m/0.1)^{-1.649})  \ {\rm Mpc} .
\end{cases}  \label{equ:BAO3}
\end{equation}
\end{subequations}
The green points in Fig.\ \ref{fig:BAO_12_5} show samples from the
12.5HMcl TT chains, while the red points show samples from the
12.5HMcl TTTEEE chains. Adding the polarization data tightens the
constraints on $\Omega_m h^2$ (see
Fig.\ \ref{fig:base_lcdm_likelihoods}) bringing the \Planck\ data into
better agreement with the BOSS constraints (which disfavour the high
values of $\Omega_m h^2$ allowed by the 12.5HMcl TT chains). Adding
\Planck\ lensing to the \Planck\ temperature likelihoods produces a
similar effect (see Fig.\ 12 of PCP18); \Planck\ lensing combined with
\Planck\ temperature data disfavours high values of $\Omega_m h^2$
leading to better consistency with the BAO results.

\begin{figure}
\centering
\includegraphics[width=150mm,angle=0]{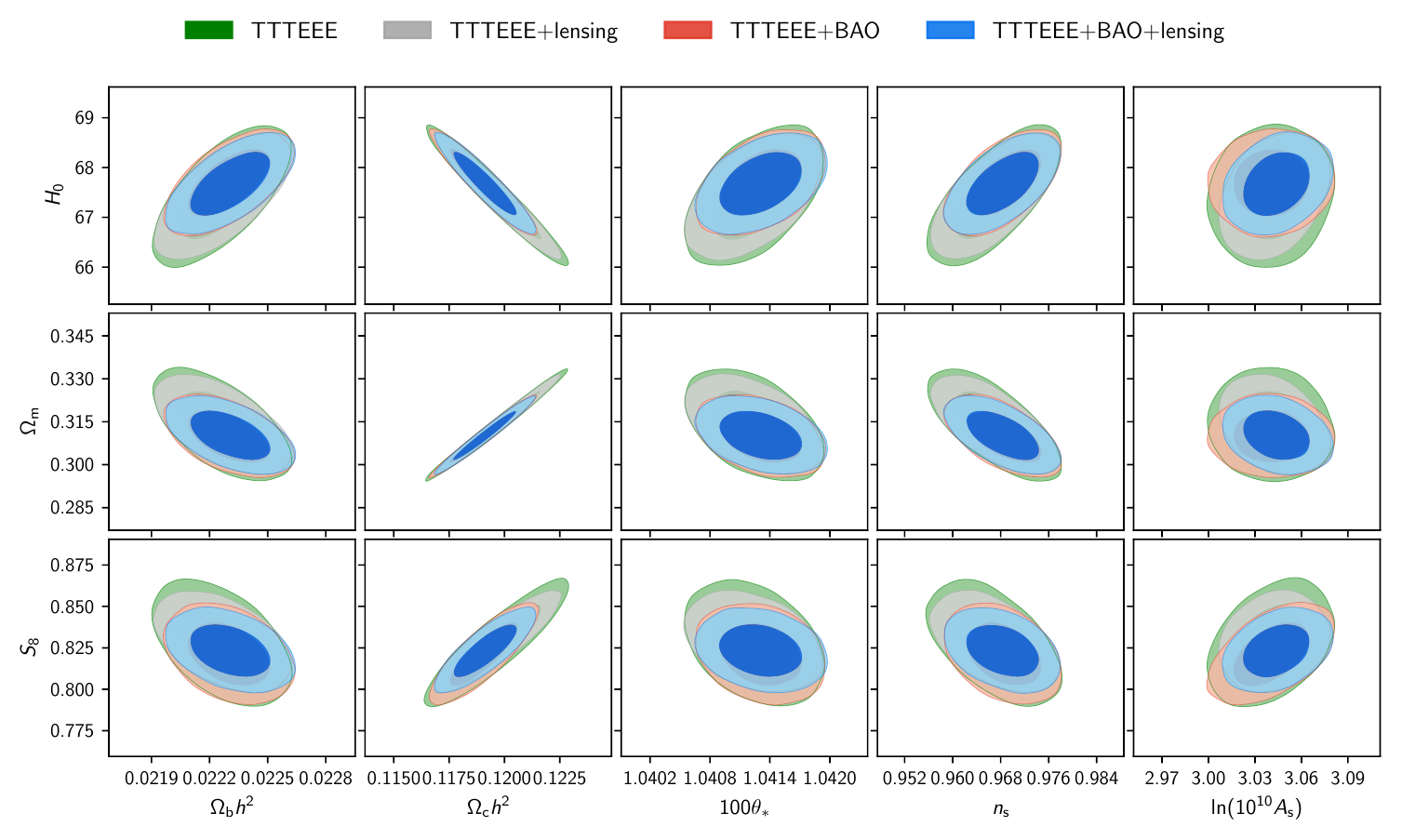}

\caption {$68\%$ and $95\%$ confidence contours for base \lcdm\ cosmological parameters determined from the
12.5HMcl TTTEEE likelihood combined with \Planck\ lensing and/or  BAO data. }

\label{fig:base_lcdm_BAO_lensing}

\vspace{0.2truein}

\end{figure}

The addition of BAO and/or lensing data to the 12.5HMcl TTTEEE
likelihood has relatively little effect on the cosmological parameters
of the base \LCDM\ model. This is illustrated in
Fig.\ \ref{fig:base_lcdm_BAO_lensing} and Table \ref{tab:LCDMcompare_12_5}. 
The \Planck\ lensing likelihood
is overwhelmed by the TTTEEE likelihood and so adding \Planck\ lensing
causes negligable shifts in cosmological parameters. As noted in
\citep{Planck_lensing:2018} and PCP18, the \Planck\ lensing likelihood constrains the
parameter combination 
\begin{subequations}
\begin{equation}
\sigma_8 \Omega_m^{0.25} = 0.589 \pm 0.020,  \quad {\rm Planck} \ {\rm lensing}, 
\end{equation}
which is compatible with the  constraint from the 12.5HMcl TTTEEE likelihood,
\begin{equation}
\sigma_8 \Omega_m^{0.25} = 0.6057 \pm 0.0081,  \quad {\rm 12.5HMcl \ TTTEEE}.
\end{equation}
\end{subequations}
Fig.\ \ref{fig:base_lcdm_BAO_lensing} shows that BAO measurements have 
a relatively little effect on the cosmological parameters of the base \LCDM\
model. The addition of the BAO data to the 12.5HMcl TTTEEE likelihood
causes a small shift towards lower values of $\Omega_ch^2$, lowering $\Omega_m$ and $S_8$ and 
raising $H_0$.

\begin{table}

\small
\centering{
\caption{\small{Marginalized base \LCDM\ parameters with 68\% confidence intervals determined from
the 12.5HMcl TT, TE and TTTEEE likelihoods. The last column combines
12.5HMcl TTTEEE with the \Planck\ lensing and BAO likelihoods.}}

\label{tab:LCDMcompare_12_5}

\medskip

\begin{center}

\hspace{-8mm}\begin{tabular}{|l|c|c|c|c|c|c|}
\hline 
Parameter  & TT & TE\hfil&  TTTEEE & TTTEEE+BAO+lensing \cr \hline
$\Omega_{\mathrm{b}} h^2$ &$0.02219\pm 0.00021$&$0.02221\pm 0.00022$&$0.02226\pm 0.0014$&$0.02231\pm 0.00013$\cr
$\Omega_{\mathrm{c}} h^2$&$0.1191\pm 0.0020$&$0.1193\pm 0.0019$&$0.1196\pm 0.0014$&$0.11914\pm 0.00094$\cr
$100\,\theta_{\mathrm{MC}}$&$1.04075\pm 0.00044$&$1.04145\pm 0.00046$&$1.04105\pm 0.00029$&$1.04112\pm 0.00028$\cr
$\tau$&$0.0521\pm 0.0080$&$0.0486^{+0.0087}_{-0.0073}$&$0.0533\pm 0.0077$&$0.0554\pm 0.0073$\cr
${\rm ln}(10^{10} A_\mathrm{s})$&$3.036\pm 0.016$&$3.025\pm 0.021$&$3.040\pm 0.016$&$3.044\pm 0.014$\cr
$n_\mathrm{s}$&$0.9661\pm 0.0058$&$0.958\pm 0.010$&$0.9671\pm 0.0046$&$0.9683\pm 0.0040$\cr
\hline
$H_0\,[{\rm km}\,{\rm s}^{-1}\,{\rm Mpc}^{-1}]$&$67.47\pm 0.88$&$67.64\pm 0.82$&$67.44\pm 0.58$&$67.68\pm 0.42$\cr
$\Omega_\Lambda$&$0.688\pm 0.012$&$0.689\pm 0.011$&$0.6865 \pm 0.0081$&$0.6897\pm 0.0056$\cr
$\Omega_{\mathrm{m}}$&$0.312\pm 0.012$&$0.311\pm 0.011$&$0.3135 \pm 0.0081$&$0.3103\pm 0.0056$\cr
$\Omega_{\mathrm{m}} h^2$&$0.1420\pm 0.0019$&$0.1422\pm 0.0018$&$0.1425\pm 0.0013$&$0.14209\pm 0.00089$\cr
$\Omega_{\mathrm{m}} h^3$&$0.09576\pm 0.00043$&$0.09616\pm 0.00046$&$0.09612 \pm 0.00029$&$0.09616\pm 0.00029$\cr
$\sigma_8$&$0.8057\pm 0.0091$&$0.800\pm 0.011$&$0.8095 \pm 0.0074$&$0.8097\pm 0.0060$\cr
$\sigma_8(\Omega_{\rm m}/0.3)^{0.5}$&$0.822\pm 0.023$&$0.815\pm 0.022$&$0.828\pm 0.016$&$0.823\pm 0.010$\cr
$\sigma_8 \Omega_{\mathrm{m}}^{0.25}$&$0.602\pm 0.011$&$0.598\pm 0.012$&$0.6057\pm 0.0081$&$0.6043\pm 0.0056$\cr
$z_{\mathrm{re}}$&$7.45^{+0.81}_{-0.79}$&$7.08^{+0.96}_{-0.72}$&$7.58^{+0.81}_{-0.74}$&$7.78^{+0.71}_{-0.70}$\cr
$10^9 A_{\mathrm{s}}$&$2.082\pm 0.034$&$2.060\pm 0.044$&$2.091\pm 0.033$&$2.100\pm 0.030$\cr
$10^9 A_{\mathrm{s}} e^{-2\tau}$&$1.876\pm 0.013$&$1.869\pm 0.027$&$1.880\pm 0.012$&$1.879\pm 0.010$\cr
$\mathrm{Age}\,[\mathrm{Gyr}]$&$13.817\pm 0.035$&$13.795\pm 0.034$&$13.803\pm 0.023$&$13.794\pm 0.020$\cr
$z_\ast$&$1090.07\pm 0.38$&$1090.07\pm 0.38$&$1090.02 \pm 0.26$&$1089.92\pm 0.21$\cr
$r_\ast\,[\mathrm{Mpc}]$&$144.80\pm 0.46$&$144.73\pm 0.45$&$144.61\pm 0.30$&$144.70\pm 0.22$\cr 
$100\,\theta_\ast$&$1.04095\pm 0.00043$&$1.04165\pm 0.00045$&$1.04124\pm 0.00028$&$1.04131\pm 0.00027$ \cr
$z_{\mathrm{drag}}$&$1059.46\pm 0.44$&$1059.51\pm 0.46$&$1059.66\pm 0.30$&$1059.73\pm 0.29$ \cr
$r_{\mathrm{drag}}\,[\mathrm{Mpc}]$&$147.53\pm 0.47$&$147.45\pm 0.46$&$147.31\pm 0.31$&$147.39\pm 0.24$ \cr
$k_{\mathrm{D}}\,[\mathrm{Mpc}^{-1}]$&$0.14027\pm 0.00051$&$0.14036\pm 0.00053$&$0.14056\pm 0.00034$&$0.14050\pm 0.00029$ \cr
$z_{\mathrm{eq}}$&$3377\pm 46$&$3382\pm 43$&$3391\pm 30$&$3380\pm 21$ \cr
$k_{\mathrm{eq}}\,[\mathrm{Mpc}^{-1}]$&$0.01031\pm 0.00014$&$0.01032\pm 0.00013$&$0.010349\pm 0.000092$&$0.010316\pm 0.000065$ \cr
$100\,\theta_{\rm{s,eq}}$&$0.4516\pm 0.0044$&$0.4514\pm 0.0042$&$0.4505\pm 0.0030$&$0.4515\pm 0.0020$ \cr
\hline

\end{tabular}

\end{center}

}

\end{table}

Cosmological parameters derived from the 12.5HMcl likelihood are
summarized in Table \ref{tab:LCDMcompare_12_5}. We consider the
12.5HMcl TTTEEE results to be the most reliable set of cosmological
parameters derived from \Planck\ power spectra, based on the
consistency of the TT, TE and EE likelihoods and the small residuals
in Figs.\ \ref{fig:12.5_TT} and \ref{fig:12.5_TE_EE}.

\vspace{0.2truein}

\subsection{Dependence on  multipole range}
\label{subsec:base_ell_range}

Over the multipole range probed by \WMAP\ (which we assume to be
approximately $2$ -- $800$), there is excellent agreement between \WMAP\ and
\Planck\ temperature data at both the power spectrum and map level\footnote{After
correcting for a $1.3\%$ calibration error in the 2013 \Planck\ HFI maps.} 
(see e.g. Appendix A of PCP13 and \citep{Planck_consistency:2014,
  Huang:2018}). As a consequence, if we restrict the
\Planck\ temperature likelihood to a maximum multipole $\ell_{\rm max}
= 800$, the base \LCDM\ cosmological parameters are very close to
those determined from \WMAP\ \citep{Bennett:2013}. The question then
arises as to whether the shifts in cosmological parameters
measured by \Planck\  are statistically consistent with the expectations of the
base \LCDM\ cosmology as $\ell_{\rm max}$ is increased to higher
multipoles.  This issue has been addressed in PCP13, PPL15, \citep{Galli:2017} and
PCP18. In particular, the analysis presented in
\cite{Galli:2017} concluded that the parameter shifts seen in the
\Planck\ temperature data were broadly consistent with those expected
in base \LCDM, with no compelling evidence for any anomalies.

However, this conclusion has been questioned by \citep{Addison:2016}
who raised the possibility that systematic errors in the \Planck\ data
at high multipoles may be driving cosmological parameter shifts. We
revisit this issue in this section using the statistically more powerful 12.5HMcl
likelihood. Addison et al.\ \citep{Addison:2016} applied a multipole cut
of $1000 \le \ell \le 2500$ to the 2015 \plik\ likelihood. \GE{With this
choice of multipole range, the spectral index $n_s$ is extremely
poorly constrained leading to large degeneracies with cosmological
parameters of interest. In addition, the standard template-based
foreground model contains a large number of parameters. Foreground
model parameters, in particular the Galactic dust amplitudes, 
become poorly constrained if low multipoles are excluded.  Small
inconsistencies between the foreground model and reality can then affect
the cosmological parameters. While we agree with \citep{Addison:2016}
on the general trends of cosmological parameter shifts, quantifying their statistical
significance to sub-$\sigma$ accuracy (which is necessary to
interpret this exercise)  depends sensitively  on the accuracy of the foreground model.}

\begin{figure}
\centering
\includegraphics[width=170mm,angle=0]{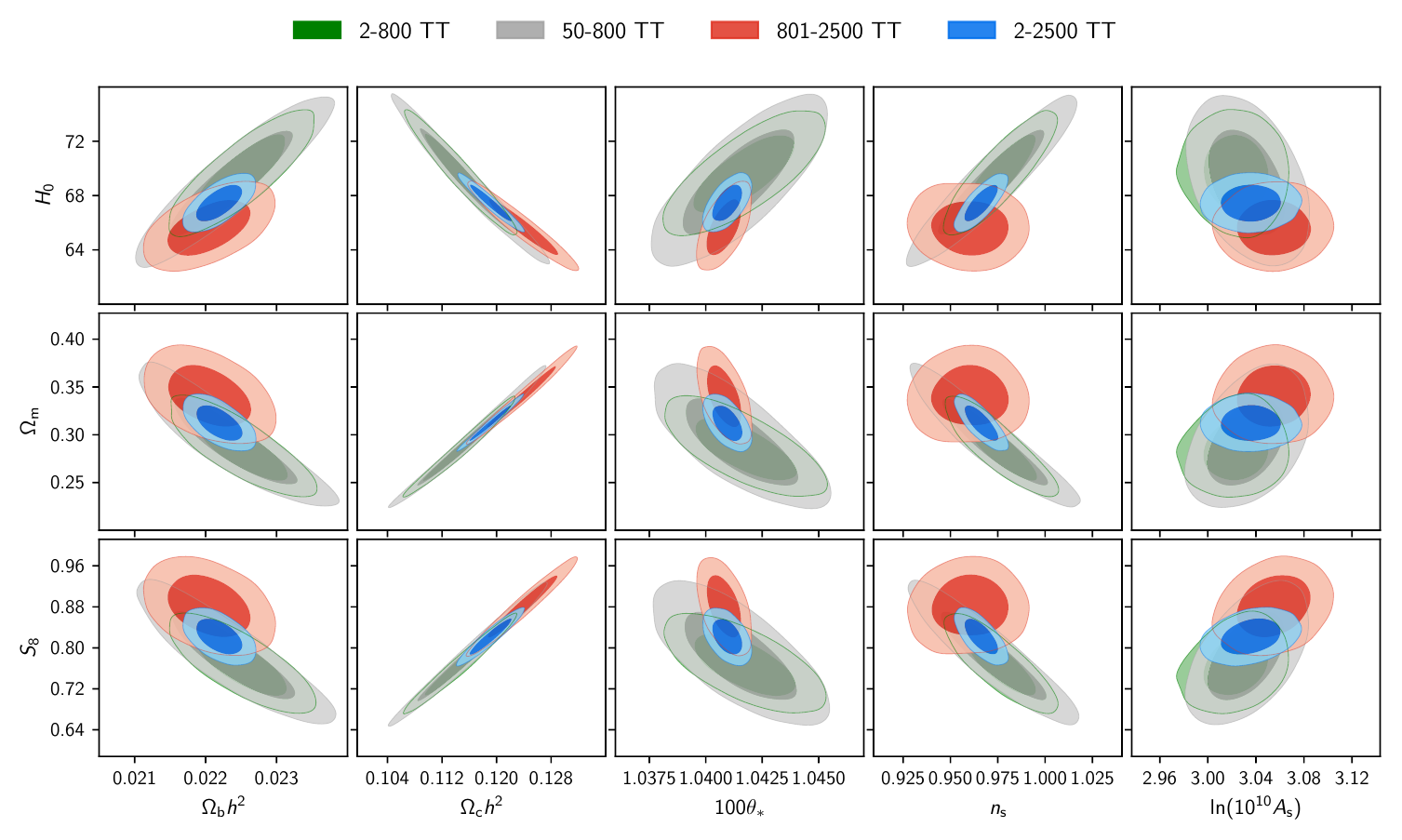} \\
\includegraphics[width=170mm,angle=0]{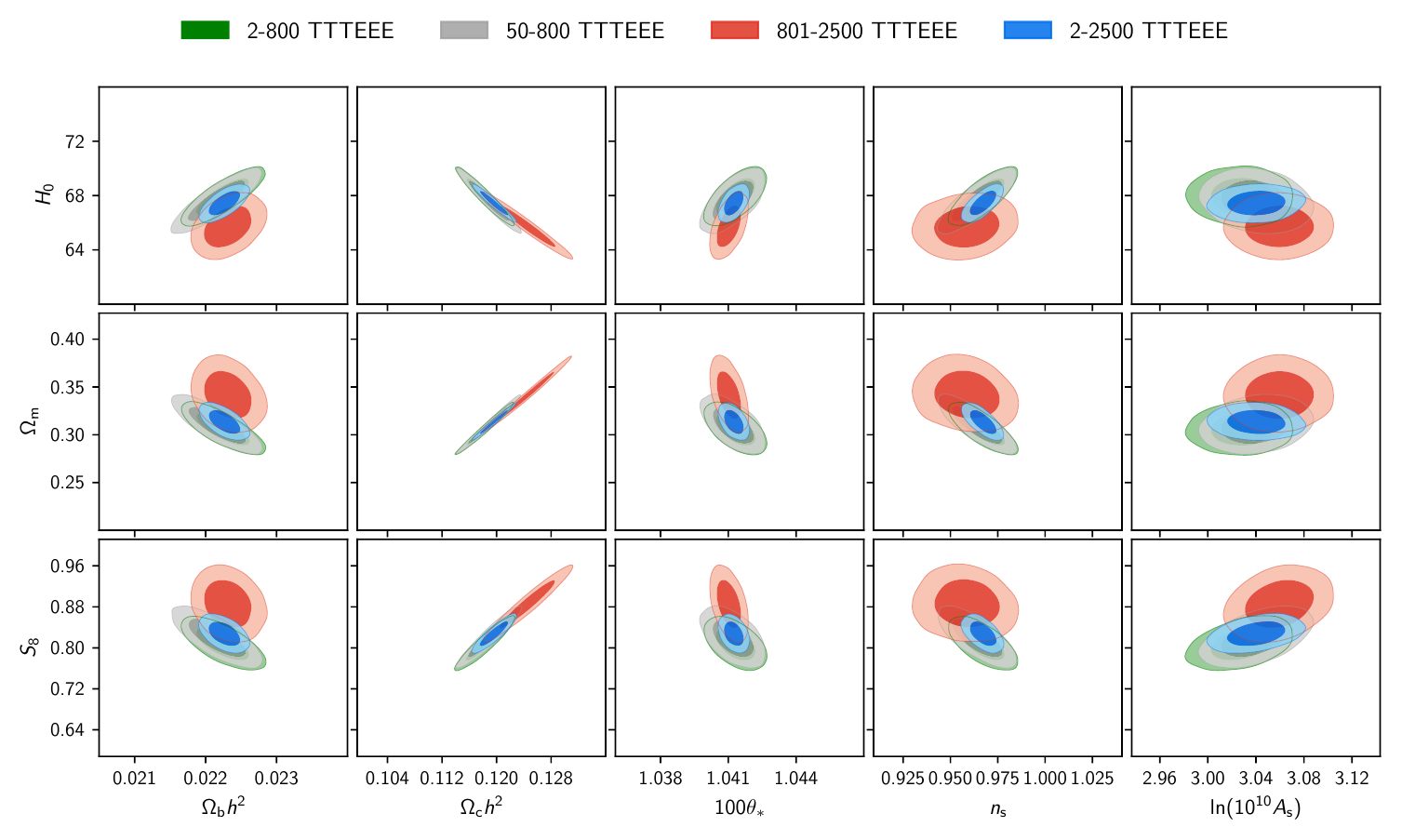}

\caption {\small{Base \LCDM\ parameter constraints determined from the 12.5HMcl likelihood
for various multipole ranges. The upper panels show results from the  TT likelihood and 
lower panels show results from the TTTEEE likelihood.}}

\label{fig:base_ell_range}

\end{figure}

\begin{table}
{\centering 

\caption{\small{Base \LCDM\ cosmological parameters for different multipole ranges 
determined from the 12.5HMcl likelihood. 
The numbers in the fourth and seventh columns list the parameter shifts in standard deviations
assuming that the low and high multipole parameters are independent. $H_0$ is given in units of
$\Hunit$.}}

\label{tab:param_shifts_table}

\begin{center}
\small
 \begin{tabular}{|l|c|c|c|c|c|c|}
\hline 
Param.  & [1] 2-800 TT  & [2]  801-2500 TT & [1]-[2] & [3] 2-800 TTTEEE & [4] 801-2500 TTTEEE & [3]-[4] \cr \hline
$\Omega_{\mathrm{b}} h^2$ &$0.02249\pm 0.00041$&$0.02205\pm 0.00038$& $\; \; 1.16 \sigma$ &$0.02243\pm 0.00024$ & $0.02232 \pm 0.00022$  & $\; 0.30 \sigma$ \cr
$\Omega_{\mathrm{c}} h^2$&$0.1147\pm 0.0032$&$0.1238\pm 0.0033$&$-1.86 \sigma$ &$0.1183\pm 0.0018$ & $0.1246 \pm 0.0027$ & $-1.96 \sigma$ \cr
$100\, \theta_*$&$1.0417\pm 0.0014$&$1.04081\pm 0.00050$&$\; \; 0.09 \sigma$&$1.04133\pm 0.00057$ & $1.04102 \pm 0.00035$& $\;\; 0.46 \sigma$ \cr
$n_\mathrm{s}$&$0.976\pm 0.012$&$0.960\pm 0.013$&$\;\; 0.90 \sigma$&$0.9672\pm 0.0074$ & $0.958 \pm 0.011$  &  $\;\;0.66 \sigma$  \cr
$H_0$&$69.58\pm 1.80$&$65.67\pm 1.30$&$\;\; 1.76\sigma$&$67.93\pm 0.89$ & $65.72 \pm 1.00$& $\;\; 1.65 \sigma$  \cr
$\Omega_{\mathrm{m}}$&$0.286\pm 0.021$&$0.340\pm 0.021$&$-1.82 \sigma$&$0.306\pm 0.0011$ & $0.342 \pm 0.016$ & $-1.85 \sigma$ \cr
$\sigma_8$&$0.790\pm 0.0040$&$0.827\pm 0.014$&$-1.92 \sigma$&$0.8002\pm 0.0091$ & $0.830 \pm 0.011$ & $-1.94 \sigma$ \cr
$S_8$ &$0.771\pm 0.040$&$0.881\pm 0.039$&$-1.98\sigma$&$0.808\pm 0.021$ & $0.886 \pm 0.031$ & $-2.09 \sigma$ \cr
$\sigma_8 \Omega_{\mathrm{m}}^{0.25}$&$0.577\pm 0.020$&$0.631\pm 0.019$&$-1.98\sigma$&$0.595\pm 0.011$ & $0.635 \pm 0.015$ & $-2.15 \sigma$ \cr
\hline

\end{tabular}
\end{center}}

\medskip
\medskip

\end{table}

 In PCP18, parameter shifts were analyzed using the {\tt plik$\_$lite}
\Planck\ likelihood (described in PPL18) which marginalizes over
foreground and nuisance parameters. With this approach, the foreground parameters are therefore
constrained by the full \Planck\ multipole range. 
We adopt a different approach in this paper by using the
12.5HMcl likelihood. Cleaning the temperature maps with 545 GHz lowers
the foreground levels and it is then possible to constrain the
residual foreground levels accurately using either low or high
multipole cuts. Likelihood analyses using disjoint multipole ranges
are then strictly independent. \GE{The analysis of parameter shifts 
presented in this Section is therefore much less sensitive to foreground modelling 
than the analysis presented in \citep{Addison:2016}. As we will see, in our analysis we find parameter shifts 
that are consistent with modest statistical fluctuations.  }

\begin{figure*}
 \centering
\includegraphics[width=170mm,angle=0]{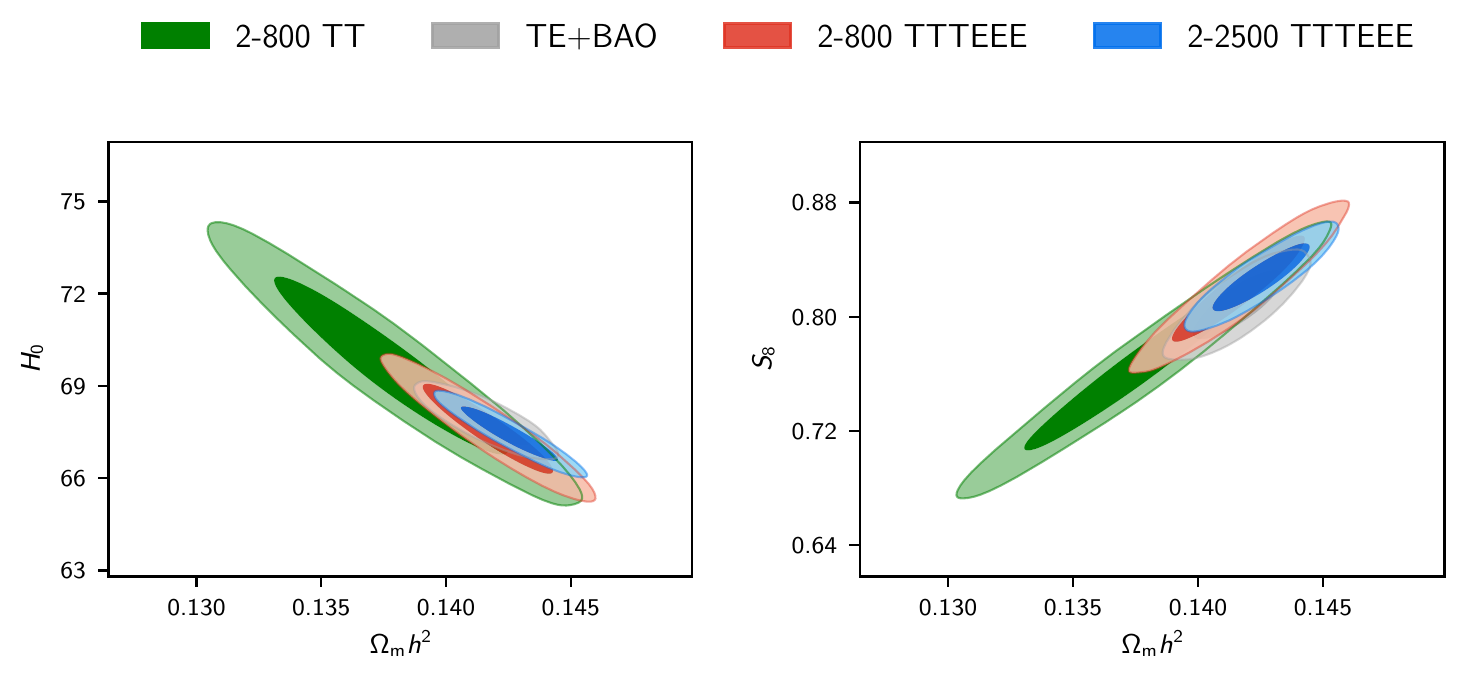}

\caption {Parameter shifts showing the progression to high $\Omega_m h^2$, low $H_0$ and high $S_8$
for the 12.5HMcl TE+BAO likelihood, for the 12.5HMcl TTTEEE likelihood limited to $2 \le \ell \le 800$
and for the full 12.5HMcl TTTEEE covering $2 \le \ell \le 2500$.}

\label{fig:params_base_ellrangeTE}

 \end{figure*}

Results are shown in Fig.\ \ref{fig:base_ell_range} for multipole
ranges $2 \le \ell \le 800$, $50 \le \ell \le 800$, $801 \le \ell \le
2500$ and for the full multipole range $2 \le \ell \le 2500$. To
simplify the subsequent discussion, we will refer to the multipole
splits as follows \LOW\ ($2 \le \ell \le 800$), \HIGH\ ($801 \le \ell \le
2500$) and \FULL\ ($2 \le \ell \le 2500$). We have included the
 multipole split $50 \le \ell \le 800$ in Fig.\ \ref{fig:base_ell_range} 
so that it is possible to assess the impact of the temperature power
spectrum in the low multipole range $ 2 \le \ell \le 50$, which has a
slightly lower amplitude than expected from the best-fit \Planck\ base
\LCDM\ model (see PCP13, \cite{Galli:2017}).

Table
\ref{tab:param_shifts_table} gives numerical values for selected
parameters and quantifies the shifts assuming the low and high
multipole cuts are independent. The \LOW\ TT parameters
shown in Fig.\ \ref{fig:base_ell_range} are close to those measured
by \WMAP, whereas the \HIGH\ TT parameters
prefer higher values of $\Omega_c h^2$ (qualitatively similar to the
results found by \citep{Addison:2016}). Since the acoustic peak scale
is insensitive to multipole range, this shift to higher values of
$\Omega_ch^2$ leads to lower values of $H_0$ for the \HIGH\ likelihood 
(cf.\ Eqs.\ \ref{equ:BAO1}-\ref{equ:BAO3}).
The \HIGH\ likelihood
also favours higher values of the amplitude parameters, as measured by
$\sigma_8$, $S_8$ and the CMB lensing combination $\sigma_8
(\Omega_m)^{0.25}$. The parameter shifts are not particularly
anomalous and both the \LOW\ and \HIGH\ multipole contours overlap with
the \FULL\  TT contours. We note that the very low TT multipoles,
$2 \le \ell \le 50$, have a relatively small effect on the parameter
shifts.  The lower panels in Fig.\ \ref{fig:base_ell_range} show the
equivalent results for the TTTEEE likelihoods. Interestingly, the addition of the TE and
EE likelihoods over the multipole range $2 \le \ell \le 800$
reduces the parameter errors substantially, driving the cosmological
parameters close to those of the \FULL\ TTTEEE likelihood.
In contrast, since TE and EE from \Planck\ are noisy at $\ell \simgt 800$,
 the addition of the polarization data has a relatively small effect 
on the \HIGH\ TTTEEE likelihood.
The \LOW\ and \HIGH\ TTTEEE parameters listed in Table \ref{tab:param_shifts_table} 
are consistent to better than $2.2 \sigma$.

From these results, we conclude that the base \LCDM\ cosmological
parameters from the \Planck\ high multipoles are displaced towards
higher values of $\Omega_c h^2$, $S_8$ and lower values of $H_0$
compared to the \FULL\ TTTEEE solution. The \LOW\ TT likelihood is
displaced towards lower values of $\Omega_c h^2$, $S_8$ and higher
values of $H_0$. Both of these results are consistent with statistical
fluctuations. As shown in Fig.\ \ref{fig:base_ell_range}, adding TE
and EE to TT over the multipole range $2 \le \ell \le 800$ drives the
parameters close to those of the full TTTEEE solution. Adding BAO to
$\ell \le 800$ TT also excludes low $\Omega_m h^2$, as does
\Planck\ TE+BAO (see Fig.\ \ref{fig:params_base_ellrangeTE}). We
therefore strongly disagree with the conclusions of
\citep{Addison:2016}.  The parameter shifts seen in \Planck\ do not
suggest systematics in the \Planck\ data at high multipoles,  instead
there is considerable evidence that the $\ell=2-800$ and $\ell =
801-2500$ TT parameters both differ from the truth as a result of
modest statistical fluctuations.

\begin{figure*}
 \centering
\includegraphics[width=130mm,angle=0]{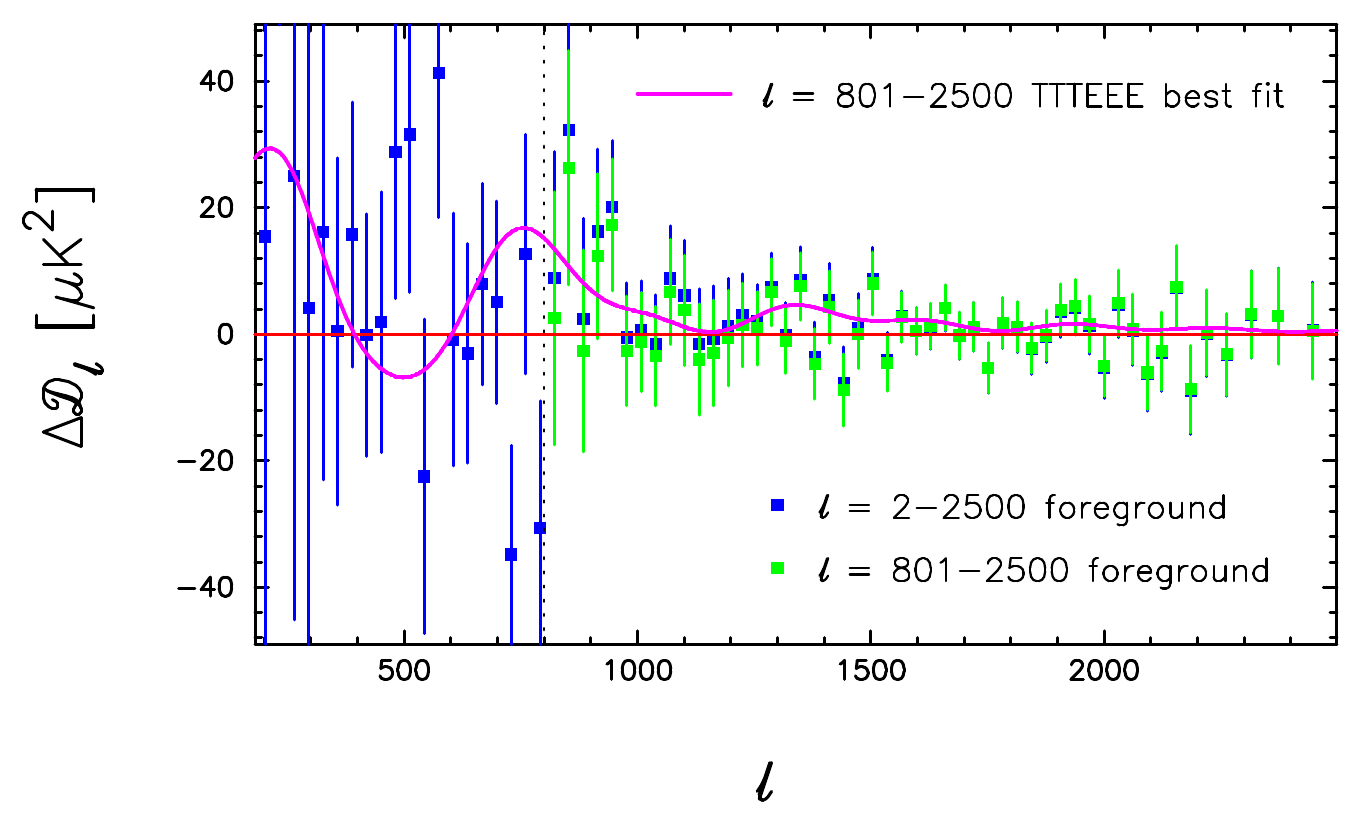}

\caption {The blue points show the residuals of the coadded 12.5HMcl TT spectrum with respect to the best fit
base \LCDM\ model fitted to the 12.5HMcl full TTTEEE likelihood. The green points show the residuals of the coadded
12.5HMcl TT spectrum at $\ell > 800$ using the foreground solution of the 12.5HMcl TTTEEE likelihood fitted over the multipole
range $801 \le \ell \le 2500$. The best fit base \LCDM\ cosmology fitted to the $801 \le \ell \le 2500$ 12.5HMcl TTTEEE likelihood
is shown as the purple line.}

\label{fig:cell_range}

 \end{figure*}

\begin{figure}
\centering
\includegraphics[width=150mm,angle=0]{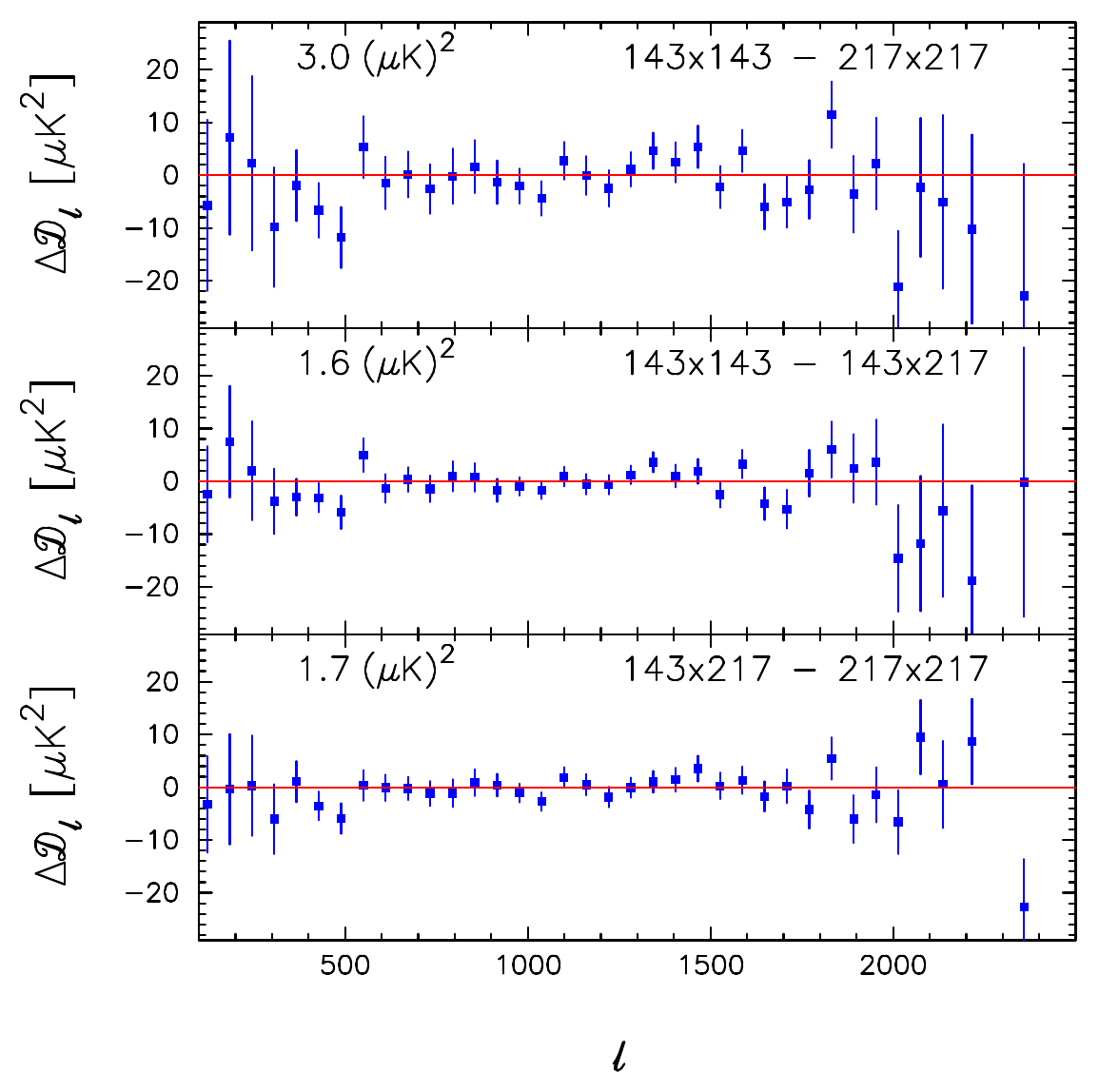} 
\caption {Inter-frequency differences for the 12.5HMcl TT spectra (as in Fig.\ \ref{fig:bands3}). This figure uses the 12.5HMcl TTTEEE foreground solution.  The error bars for the
power spectrum differences are computed from appropriate linear combinations of the \camspec\ covariance matrices. The numbers
in each panel give the rms residuals of the bandpower differences over the multipole range $800 \le \ell \le 1500$.}

\label{fig:bands3_12_5}

\end{figure}

The cause of these parameter shifts is apparent in Fig.\ \ref{fig:cell_range}. The blue points in this figure 
show the residuals of the coadded TT spectrum with respect to the $2 \le \ell \le 2500$ 12.5HMcl TTTEEE best fit cosmology. The green points show
the residuals of the  coadded spectrum at $\ell > 800$ using the foreground solution determined from the $800 \le \ell \le 2500$ TTTEEE likelihood. The differences between the blue and green points show the impact of the different foreground solution,
which is small but non-negligible. The purple line shows the best-fit base \LCDM\ TTTEEE cosmology fitted to $800 \le \ell \le 2500$ (which is disfavoured by the points at $\ell \le 800$). The purple line responds to the oscillatory features in the multipole
range $800 \simlt \ell \simlt 1500$ (which is also apparent in Figs.\ \ref{fig:inter_frequency143v217} and \ref{fig:inter_frequency143v217cleaned}) and is reproducible to high precision across frequencies (see Fig.\ \ref{fig:bands3_12_5}).

 In summary, our analysis is consistent with the conclusions of
 \citep{Galli:2017} and PCP18, namely that parameter shifts between
 low and high multipoles are consistent with statistical
 fluctuations. The features which drive the parameters from the
 $801-2500$ TT likelihood are mainly located in the multipole range
 $801-1500$ and are reproducible to high accuracy in the $143\times
 143$, $143 \times 217$ and $217 \times 217$ spectra. The base
 \LCDM\ parameters derived from the \LOW\ TTTEEE likelihood are very
 close to those derived from the \FULL\ TT and
 TTTEEE likelihoods. This is a very important consistency check of the
 \Planck\ data. Most of the \Planck\ TT results for base \LCDM\ can be
 recovered to comparable accuracy from $\ell \le 800$ {\it if one
   includes the polarization spectra.}

\subsection{Tensions with other astrophysical data}
\label{subsec:tensions}

We will not make an extensive comparison of these results with other astrophysical data in this paper, since this
topic has been reviewed in detail in PCP18. However we make the following observations:

\smallskip

\noindent
{\it $H_0$ tension:} As noted in PCP13, the best-fit \Planck\ value of
$H_0$ differs from direct measurements based on the Cepheid-Type Ia
supernova distance ladder \citep{Riess:2011, Riess:2016,
  Riess:2018}. The latest value from the SH0ES\footnote{Supernovae,
  $H_0$, for the Equation of State of Dark Energy.}  collaboration
\citep{Riess:2019} is $H_0 = 74.03 \pm 1.42 \Hunit$, differing by
$6.59 \Hunit$ from the 12.5HMcl TTTEEE value of $H_0$ listed in Table
\ref{tab:LCDMcompare_12_5}.  Interestingly, the statistical significance of the discrepancy
between \Planck\ and SH0ES has grown from $2.5 \sigma$ in PCP13 to  $4.3\sigma$ today. There
are also hints, for example, from strong gravitational lensing time delays \citep{Birrer:2019}
 that the late time value of $H_0$ may differ from the \Planck\ value. The $H_0$ discrepancy is
perhaps the most intriguing tension with the base \LCDM\ model at this time. There is a general consensus, from
application of the inverse distance ladder to BAO and Type Ia supernovae data, that this discrepancy,
if real, requires physics that reduces the value of the sound horizon $r_{\rm drag}$ {\it irrespective of the
nature of dark energy} \citep{Aubourg:2015, Cuesta:2015, Bernal:2016, Lemos:2018b, Macaulay:2018, Efstathiou:2021}. \GE{For reviews of possible theoretical explanations of this
tension see \cite{Knox:2019, diValentino:2021, Schloneberg:2021}.}

\smallskip
\noindent
{\it Weak Gravitational Lensing:} \GE{ Recently, three large cosmic shear surveys have reported constraints on the
parameter combination $S_8 = \sigma_8 (\Omega_m/0.3)^{0.5}$. 
The Subaru Hyper Suprime-Cam
(HSC) first year data give \citep{Hikage:2018}
\begin{subequations}
\begin{equation}
\qquad \ \  \ \  S_8 = 0.780^{+0.030}_{-0.033}, \quad {\rm HSC}. \qquad \qquad  \label{WL3}
\end{equation}
 The Kilo-Degree Survey (KiDS)-1000 3x2pt function analysis (shear-shear, galaxy-galaxy lensing, galaxy-galaxy) \citep{Heymans:2021} gives
\begin{equation}
 \quad \ \ S_8 = 0.766^{+0.020}_{-0.014}, \quad \text{KiDs--1000}, \label{WL2}
\end{equation}
updating earlier results from  \citep{Hildebrandt:2018}.
The  Dark Energy Survey (DES) Year 3 3x2pt function analysis of 
\cite{DES3:2021} gives
\begin{equation}
\qquad \qquad \ \  S_8 = 0.776\pm 0.017, \quad {\rm DESY3}, \qquad \quad \label{WL1}
\end{equation}
where the neutrino mass is allowed to vary. (Allowing the heaviest
neutrino mass to vary rather than fixing it to $0.06\ {\rm eV}$ has a
small effect on $S_8$, see Fig.\ 28 of \cite{DES3:2021}, which we will
ignore for the qualitative comparison presented in this
Section.) Eq.\ \ref{WL1}  updates the DES Year-1 results reported in
\cite{DES1:2018}.
\end{subequations}

For base \LCDM\ the 12.5HMcl likelihoods give\footnote{\GE{The \plik\ TTTEEE likelihood gives a somewhat higher value of 
$S_8 = 0.834 \pm 0.016$.}}
\begin{subequations}
\begin{eqnarray}
       S_8 & =&  0.822 \pm 0.023, \qquad {\rm TT},     \label{equ:S8a} \\
       S_8 & =&  0.828 \pm 0.016, \qquad {\rm TTTEEE},  \label{equ:S8b} \\
       S_8 & =&  0.829 \pm 0.012, \qquad {\rm TTTEEE+lensing}.  \label{equ:S8c}
\end{eqnarray}
\end{subequations}
As noted by many authors, the cosmic shear analysis   consistently give lower values of $S_8$ 
than the \Planck\ results of Eqs. \ref{equ:S8a}-\ref{equ:S8c}. 
 If we crudely combine these
estimates,  assuming that they are statistically independent,  we find
$S_8 = 0.773 \pm 0.012$ which is  about $3.3\sigma$ lower than the
\Planck\ result of Eq. \ref{equ:S8c}\footnote{\GE{For more rigorous combinations of variants of the KiDs and DES Year1 shear surveys see \cite{Joudaki:2020, Asgari:2020, Garcia-Garcia:2021}. The analysis of \cite{Garcia-Garcia:2021}, in particular, combines cosmic shear data from DES Year 1 and KiDs-1000, together with the \Planck\ weak lensing map and clustering measurements of galaxies and quasars, to derive the very tight limit of
  $S_8 = 0.7769 \pm 0.0095$.}}.   As the cosmic shear surveys have increased in size (and with the addition of
galaxy-galaxy and galaxy-galaxy lensing two-point correlation functions), the statistical significance of the
$S_8$ tension with the \Planck\ base \LCDM\ expectation has increased. It is now unlikely that the
$S_8$ discrepancy is simply a statistical fluctuation.

There are, of course, a number of potential sources of systematic
error in both the forward distance scale and weak lensing
measurements, which we will not discuss here. The key point that we
wish to make in this paper is  that the \Planck\ results are remarkably robust
between frequencies, between temperature and polarization and between  sky
fractions. The \Planck\ results are therefore unlikely to be affected
by systematic errors to any significant degree. If the tensions with
the distance scale and weak lensing measurements persist,  and the measurements can be shown to
be free of systematic errors,  new physics will be required beyond that assumed in the  base \LCDM\ model. }

\section{Extensions to  \LCDM\ }
\label{sec:extensions_lcdm}

PCP18 reported two unusual results related to extensions to the base
\LCDM\ cosmology involving the phenomenological $A_L$ parameter and
spatial curvature $\omegak$. For both parameters, the TTTEEE
\camspec\ and \plik\ likelihoods behaved differently. As noted in PCP18, the primary
reason for these differences is that \plik\ used polarization efficiency corrections
derived from the EE spectra. As discussed in Sect. \ref{sec:beams} in \camspec\ we use polarization efficiency
corrections derived from TE and EE spectra, which are clearly more accurate for the TE spectra. In this section,
we investigate how these parameters vary using the statistically more
powerful 12.5HMcl likelihood. For completeness, we also investigate
constraints on the neutrino mass $\sum m_\nu$, on the number of
relativistic species $N_{\rm eff}$ and on tensor modes.

\subsection{The $A_L$ parameter}
\label{subsec:AL}

 It has been noted since PCP13 that the \Planck\ temperature data
 favour values of $A_L > 1$.  In PCP18, the \plik\ likelihood gave
 $A_L = 1.243 \pm 0.096$ (TT+lowE) and $1.180\pm 0.065$ (TTTEEE+lowE),
 favouring $A_L > 1$ at $2.5 \sigma$ and $2.8 \sigma$
 respectively. The \camspec\ likelihood used in PCP18 (which is
 similar to the 12.1HM likelihood produced for this paper) gave $A_L =
 1.246 ^{+0.092}_{-0.100}$ (TT+lowE) and $A_L = 1.149 \pm 0.072$
 (TTTEEE+lowE), favouring $A_L > 1$ at $2.5 \sigma$ and $2.1 \sigma$
 respectively. For the \plik\ likelihood, adding TE and EE made the
 discrepancy with $A_L=1$ worse, whereas for \camspec\ the addition of
 polarization reduced the discrepancy.  (Though, importantly, for both
 likelihoods the addition of polarization data caused the best fit
 value of $A_L$ to fall.)

\begin{figure}
\centering \includegraphics[width=57mm,angle=0]{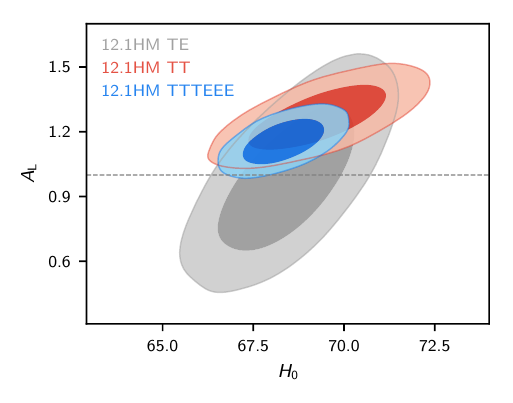}
\includegraphics[width=57mm,angle=0]{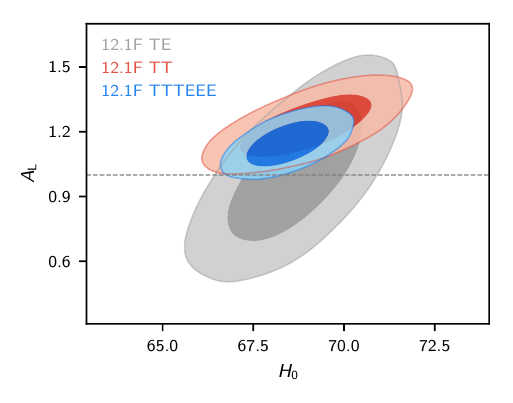}
\includegraphics[width=57mm,angle=0]{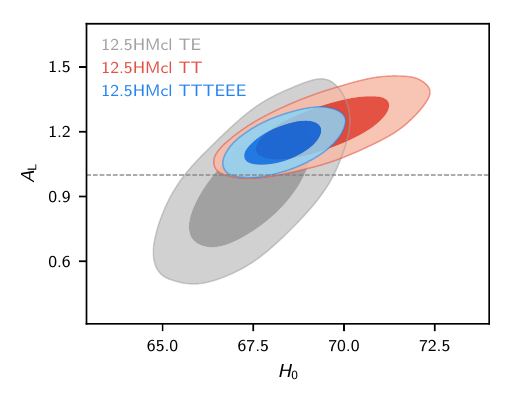}
\caption {68\% and 95\% contours on the parameters $A_L$ and $H_0$ (in units of
$\Hunit$) for various likelihood combinations:  12.1HM likelihood
(left),  12.1F (middle)  and 12.5HMcl (right). The horizontal dotted line shows 
$A_L = 1$.}

\label{fig:Alens}

\end{figure}

It is important to note that the $A_L$ parameter is very poorly
constrained by the power spectra at low multipoles. For example, over
the multipole range $2 \le \ell \le 800$, the 12.5HMcl TT likelihood
gives $A_L = 1.32 \pm 0.48$. The $A_L$ parameter is therefore
extremely sensitive to the \Planck\ data at high multipoles. Results
for $A_L$ for the 12.1HM and 12.5HMcl likelihoods are given in Table
\ref{tab:extensions_LCDM}.  Compared to the \camspec\ TT likelihood
used in PCP18, $A_L$ from the 12.1HM TT likelihood is slightly higher,
differing from unity by about $2.6 \sigma$. The only significant
difference between these likelihoods is that we fix the relative
calibrations between frequencies in the 12.1HM likelihood, as
described in Sect.\ \ref{subsubsec:inter_frequency}, rather than
allowing them to vary as nuisance parameters. This illustrates the
extreme sensitivity of the $A_L$ parameter to the nuisance
parameter/foreground model. The 12.5HMcl TT likelihood covers more sky
area at 143 and 217 GHz compared to 12.1HMcl and we see that the
amplitude of $A_L$ goes down, differing from unity by $2.2
\sigma$. This is what we would have expected to see given that the
residuals of the $217 \times 217$ and $143 \times 217$ spectra with
respect to the base \LCDM\ best fit (see
Fig.\ \ref{fig:inter_frequency143v217cleaned}) decrease in amplitude
as sky area is increased. We find similar results for the full mission
12.1F likelihood, which improves the signal-to-noise of the
temperature spectra.  The behaviour of $A_L$ is consistent with a
moderate statistical fluctuation, driven by a chance match of the TT
power spectrum residuals in the multipole range $\sim 800 - 1500$
which are reproducible over the frequency range $143-217$ GHz. These
residuals decline in amplitude with increasing sky area and also  by
switching to the full mission spectra. The TE and EE spectra do not
provide strong constraints on $A_L$ because they are noisy at high
multipoles. As can be seen from Fig.\ \ref{fig:Alens}, which shows
constraints in the $A_L - H_0$ plane, the TE spectrum disfavours high
values of $H_0$, so when the polarization blocks of the likelihood are
added to the TT blocks, the value of $A_L$ goes down\footnote{Note that none of
  our likelihoods reproduce the $2.8\sigma$ deviation from $A_L=1$
  reported for  the \plik\ TTTEEE likelihood in PCP18.}. The 12.5HMcl
TTTEEE constraints give $A_L = 1.149 \pm 0.067$, i.e.\ a $2.2\sigma$
deviation from unity. Adding \Planck\ lensing, the value of $A_L$ goes
down further reducing the discrepancy to $1.6 \sigma$. The best-fit
12.5HMcl TTTEEE $A_L$ model is plotted against the temperature data in
Fig.\ \ref{fig:extensions}.

\begin{table}
{\centering \caption{\small{Parameters for extensions to \LCDM.}}

\label{tab:extensions_LCDM}
\begin{center}

\small

\begin{tabular}{|l|c|c|c|c|c|} \hline 
12.1HM Likelihood     &   $A_L$ &  $\omegak$ &  $N_{\rm eff}$ & $m_\nu$ (eV)  & $r_{0.002}$ (+BK15)   \\ \hline
 TT   &$1.267^{+0.095}_{-0.102}$      & $-0.061^{+0.027 }_{-0.018}$   & $2.89^{+0.27}_{-0.30}$  &  $<0.47$  & $< 0.052$        \\
 TE    &  $0.98^{+0.21}_{-0.24}$     & $-0.032^{+0.056}_{-0.022}$   & $2.82^{+0.43}_{-0.53}$  & $<1.47$   &  $< 0.070$       \\
 TE+BAO & $0.93 \pm 0.19$          &  $-0.0008 \pm 0.0024$  & $2.78^{+0.34}_{-0.38}$  &  $<0.35$      &     $<0.071$ \\ 
 TTTEEE  &$1.156 \pm 0.070$       & $-0.037^{+0.019}_{-0.013}$ &  $2.89 \pm 0.21$       &  $<0.36$  & $<0.061$      \\
 TTTEEE+lensing   & $1.061 \pm 0.042$        & $0.0092^{+0.0065}_{-0.0064}$   & $2.84 \pm 0.21$ & $< 0.30$ & $< 0.059$  \\  
 TTTEEE+lensing+BAO & $1.058 \pm 0.038$       & $0.0004 \pm 0.0020$   & $2.95 \pm 0.19$ & $<0.14$  & $<0.060$ \\  \hline
12.5HMcl Likelihood     &   $A_L$ &  $\omegak$ &  $N_{\rm eff}$ & $m_\nu$ (eV)  & $r_{0.002}$ (+BK15)   \\ \hline
  TT      &   $1.218 \pm 0.097$    & $-0.048^{+0.025}_{-0.018}$   & $2.92^{+0.30}_{-0.34}$  &  $<0.67$ &$<0.058$           \\
 TE    &   $0.96 \pm 0.19$   & $-0.020^{+0.044}_{-0.020}$     & $2.95^{+0.40}_{-0.47}$ &     $<1.32$   & $<0.065$      \\
 TE+BAO       &   $1.00 \pm 0.17$   & $0.0010 \pm 0.0023$   & $3.05^{+0.32}_{-0.37}$  & $<0.25$   &$<0.069$          \\
 TTTEEE    &   $1.149 \pm 0.067$   & $-0.035^{+0.018}_{-0.013}$   & $2.96^{+0.21}_{-0.24}$   &  $<0.34$  & $<0.056$    \\
 TTTEEE+lensing      &   $1.064 \pm 0.040$   & $-0.0101^{+0.0066}_{-0.0065}$   & $2.91^{+0.20}_{-0.24}$ & $<0.26$ &$<0.057$ \\  
 TTTEEE+lensing+BAO  &   $1.062 \pm 0.036$   & $0.0004 \pm 0.0019$   & $3.01 \pm 0.19$ & $<0.13$ &$<0.058$ \\  \hline
\end{tabular}
\end{center}}
\end{table}

\newpage

\subsection{Spatial Curvature }
\label{subsec:curvature}

 As discussed  in PCP18, the fluctuations in the \Planck\ spectra that cause $A_L$ to be higher than unity
also couple with  spatial curvature,  driving the best fit \planck\ cosmology towards 
closed universes. This tendency has been noted by a number of authors \cite{Park:2019b, 
DiValentino:2019, DiValentino:2020, Handley:2019}. 

\begin{figure}
\centering
\includegraphics[width=76mm,angle=0]{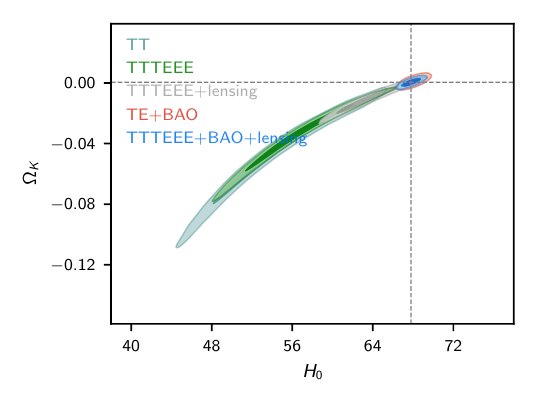} \includegraphics[width=76mm,angle=0]{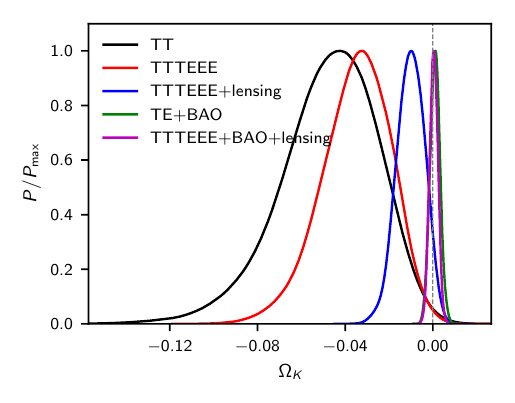} 
\caption {The figure to the left shows 68\% and 95\% contours in the $H_0$-$\omegak$ plane for various likelihood
combinations using the 12.5HMcl likelihood. The dashed lines show the best-fit values of $\omegak$ and $H_0$ for the TTTEEE
12.5HMcl+BAO+lensing likelihood. The figure to the right shows posteriors for $\omegak$ illustrating
that $\omegak=0$ is consistent
with the 12.5HMcl TT and TTTEEE likelihoods to within about $2 \sigma$.}

\label{fig:omegak}

\end{figure}

Constraints on $\omegak$ for various likelihood combinations\footnote{Note that
these results assume a uniform prior for $\omegak$  over the range $-0.3 \le \omegak \le 0.3$.} 
are plotted in Fig.\ \ref{fig:omegak} and listed in Table
\ref{tab:extensions_LCDM}.  From TT alone, the pull towards closed
Universes is at about the $2 \sigma$ level. The polarization spectra
are relatively neutral towards $\omegak$, so for the 12.5HMcl TTTEEE
likelihood, the significance level for $\omegak < 0$ drops slightly.
 From the CMB power spectra alone, it is difficult to
constrain $\omegak$ because of the near-exact geometrical degeneracy
\citep{Bond:1997}, which is broken only by the lensing of the CMB
\citep{Stompor:1999}. As a consequence of the geometrical degeneracy,
the parameter $\omegak$ is highly degenerate with the value of the
Hubble constant (see Fig.\ \ref{fig:omegak}) with much of the
parameter range allowed by the CMB corresponding to low values of
$H_0$ which are strongly disfavoured by direct measurements. The
addition of BAO and/or \Planck\ CMB lensing breaks this degeneracy
very effectively. This is illustrated in Fig.\ \ref{fig:omegak}. For
example, the addition of the BAO data to the TE likelihood constrains
the Universe to be nearly spatially flat to an accuracy that is almost
as good as the TTTEEE+BAO+lensing likelihood (see also Table
\ref{tab:extensions_LCDM}).

Posteriors for $\omegak$ are shown in the right hand panel of
Fig.\ \ref{fig:omegak} for various likelihood combinations.  As with
$A_L$, the \Planck\ results are consistent with a moderate statistical
fluctuation in the temperature spectra that favours closed universes
at about the $2 \sigma$ level.  With the addition of BAO and/or
\Planck\ lensing data, we find strong evidence to support a spatially
flat Universe. Given the recent interest in models with spatial
curvature, we have given a more detailed statistical analysis of the
results from the 12.5HMcln likelihood in \cite{Efstathiou:2020},
including the role of assuming a uniform prior in $\omegak$.

\subsection{Relativistic species and massive neutrinos }
\label{subsec:neutrinos}

\begin{figure}
\centering
\includegraphics[width=76mm,angle=0]{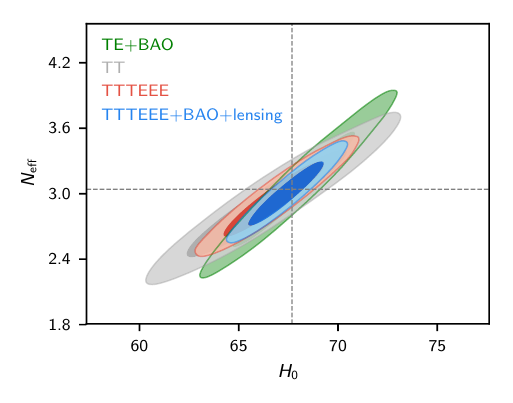} \includegraphics[width=76mm,angle=0]{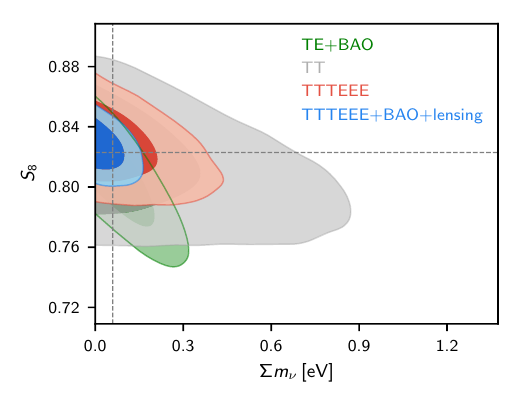} 
\caption {68\% and 95\% contours derived using the 12.5HMcl likelihood
  in combination with the BAO and \Planck\ lensing likelihoods. The
  figure to the left shows constraints in the $H_0$-$N_{\rm eff}$
  plane (with $H_0$ in units of $\Hunit$) if $N_{\rm eff}$ is added as
  an additional parameter to base \LCDM. The figure to the right shows
  constraints in the $\sum m_\nu$-$S_8$ plane if the sum of neutrino
  masses is allowed to vary as an additional parameter to the base
  \LCDM\ cosmology.  The dashed lines show best fit base
  \LCDM\ parameters of $H_0$ and $S_8$ determined from the 12.5HMcl TTTEEE likelihood.  }

\label{fig:neutrinos}

\end{figure}

For completeness, Table \ref{tab:extensions_LCDM} gives results for the number of relativistic species and the sum of neutrino masses for the 12.1HM and 12.5HMcl likelihoods. These results are consistent with those reported in PCP18. Increasing the number 
of relativistic species above the value $N_{\rm eff} = 3.046$  of the base \LCDM\ model  has been proposed as a 
possible solution of the tension between direct measurements of $H_0$ and the value inferred from the CMB  \citep{Riess:2016, Bernal:2016}. However, it clear from Fig.\ \ref{fig:neutrinos}  that this solution is quite strongly disfavoured by \Planck. Allowing $N_{\rm eff}$ to vary, the 12.5HMcl TTTEEE+BAO+lensing likelihood gives $N_{\rm eff}=3.01 \pm 0.19$, $H_0 = 67.42 \pm 1.25 \Hunit$, i.e. very close to the best fit parameters of the base \LCDM\ model, and discrepant by $3.1 \sigma$ from the $H_0$ determination reported in \citep{Riess:2019}.

The plot to the right of Fig.\ \ref{fig:neutrinos} shows constraints on
the sum of neutrino masses. As explained in PCP18, the \Planck\ power
spectra constrain neutrino masses through the effects of lensing. The
fluctuations in the TT power spectrum that favour $A_L>1$ could shift
the neutrino mass constraints towards lower values. Since the 12.5HMcl
likelihood favours lower values of $A_L$ than reported in PCP18, it is
interesting to investigate the constraints on neutrino masses derived
from the 12.5HMcl likelihood. In fact, we find almost no difference in
the 95\% upper limits, with $\sum m_\nu < 0.36\ {\rm eV}$ from the
12.1HM TTTEEE likelihood and $\sum m_\nu < 0.34 \ {\rm eV}$ for the
12.5HMcl TTTEEE likelihood.  The addition of BAO and \Planck\ lensing
to the 12.5HMcln TTTEEE likelihood lowers this limit to $\sum m_\nu < 0.13 \ {\rm eV}$, 
almost identical to the constraint reported in PCP18.
It has been  argued \cite{McCarthy:2018} that  massive neutrinos with $\sum m_\nu$
in the range $0.2 - 0.4 \ {\rm eV}$  may explain the tension between low redshift
measurements of the amplitude of the mass fluctuations (including the
weak galaxy lensing measurements summarized in  Sect.\ \ref{subsec:tensions})
and the base \LCDM\ results from \Planck. However, Fig.\ \ref{fig:neutrinos} 
shows that this solution is quite strongly disfavoured by the \Planck\ and BAO data. 

\begin{figure}
\centering
\includegraphics[width=135mm,angle=0]{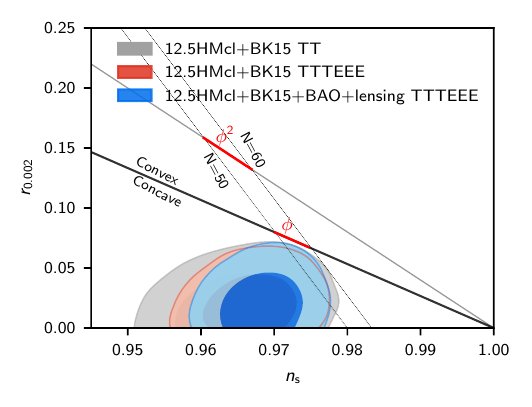}
\caption {68\% and 95\% constraints in the $n_s$-$r_{0.002}$ plane for
  various likelihood combinations combined with the BK15 B-mode
  measurements of \cite{BICEP:2018}. The thick black line shows the $n_s$-$r$
  relation for a linear potential, $V(\phi) \propto \phi$,  to first
  order in slow roll parameters. The red portion shows the range of
  parameters allowed in inflationary models with e-folding numbers in
  the range $50$-$60$. The thin black line shows the $n_s$-$r$ relation for a
quadratic potential $V(\phi) \propto \phi^2$. }
\label{fig:tensors}

\vspace{0.4 truein}
\end{figure}

\subsection{Tensor amplitude}
\label{subsec:tensors}

The \Planck\ results, combined with BAO measurements, show that our
Universe is almost spatially flat with a spectrum of nearly scale
invariant adiabatic fluctuations. In addition, the \Planck\ data show
no evidence for primordial non-Gaussianity \citep{PlanckNG:2014,
  PlanckNG:2016}. These results are consistent with single field
models of inflation (see \cite{PlanckInflation:2014,
  PlanckInflation:2016,PlanckInflation:2018} and references
therein). If the Universe did indeed experience an inflationary phase,
there should exist a nearly scale-invariant spectrum of tensor
fluctuations \citep{Starobinski:1979} with a highly uncertain
amplitude that depends on the energy scale of inflation.  Allowing
tensor modes is therefore one of the best motivated extensions of the
base \LCDM\ model and their detection would provide an important clue
towards understanding the physics of inflation.  The
\Planck\ temperature spectra provide relatively poor constraints on
tensor modes because of cosmic variance on large scales
\cite{Knox:1994}. However, tensor modes can be detected via B-mode
polarization anisotropies \citep{Kamionkowski:1997, Seljak:1997}.  At
present, the strongest constraints on primordial B-mode anistropies
come from the BICEP2-Keck Array collaboration \citep{BICEP:2015,
  BICEP:2016, BICEP:2018}, developing on earlier work by the BICEP2
collaboration \citep{BICEP:2014}\footnote{\GE{We do not consider the analysis of \cite{Tristram:2021} which claims
a $95\%$ upper limit of $r<0.069$ from \Planck\ polarization spectra covering the multipole range $2-150$. This result 
disagrees  with the $95\%$ upper limit of $r< 0.41$ reported in PPL18 from an analysis of SROLL1  polarization  maps at low multipoles (covering the reionization bump) and with a detailed analysis by one of us (GE) of BB spectra covering the 
`recombination bump', which gave a $68\%$ upper limit of $r<0.31$.}}. Joint constraints using our
\camspec\ likelihoods combined with the BICEP2-Keck Array measurements
of \citep{BICEP:2018} (denoted BK15) are plotted in
Fig.\ \ref{fig:tensors}\footnote{As in PCP13-PCP18, we report
  constraints on the tensor-scalar ratio $r_{0.002}$, which is the
  relative amplitude of the tensor and scalar primordial fluctuation
  spectra at a pivot scale of $k = 0.002 \ {\rm Mpc}^{-1}$. The tensor
  spectral index is set to the value expected in single-field inflation
  models, $n_t = -r_{0.05}/8$, where $r_{0.05}$ is the
  tensor-to-scalar ratio at a pivot scale $k = 0.05 \ {\rm
    Mpc}^{-1}$. The scalar spectral index $n_s$ is defined at a pivot
  scale of $k = 0.05 \ {\rm Mpc}^{-1}$ (see Eq.\ 2 of
  PCP18).}. 
 The 95\% upper limits on the tensor-scalar ratio given by the 12.5HMcl
TTTEEE+BK15+BAO+lensing likelihood are
\begin{equation}
  r_{0.002} < 0.058, \qquad r_{0.05} < 0.062, \label{equ:tensor1}
\end{equation}
determined mainly by the BK15 measurements. Excluding the BK15 likelihood,
the 95\% upper limit is $r_{0.002} < 0.11$, so the main contribution of the 
\Planck\ data to Fig.\ \ref{fig:tensors} is to constrain the spectral index 
$n_s$.

As discussed in detail in PCP18 and \citep{PlanckInflation:2018}, the most striking result to emerge
from \Planck\ and the BICEP2-Keck Array results is the requirement of unusually flat inflationary
potentials and therefore a hierarchy in the magnitudes of the inflationary slow-roll parameters. 
Inflationary model building has therefore shifted towards ideas on how to explain this hierarchy
(see e.g.\ \citep{Kallosh:2018, Achucarro:2018, Akrami:2018} and references therein).

\subsection{Summary}
\label{subsec:extensions_summary}

\begin{figure}
\centering
\includegraphics[width=150mm,angle=0]{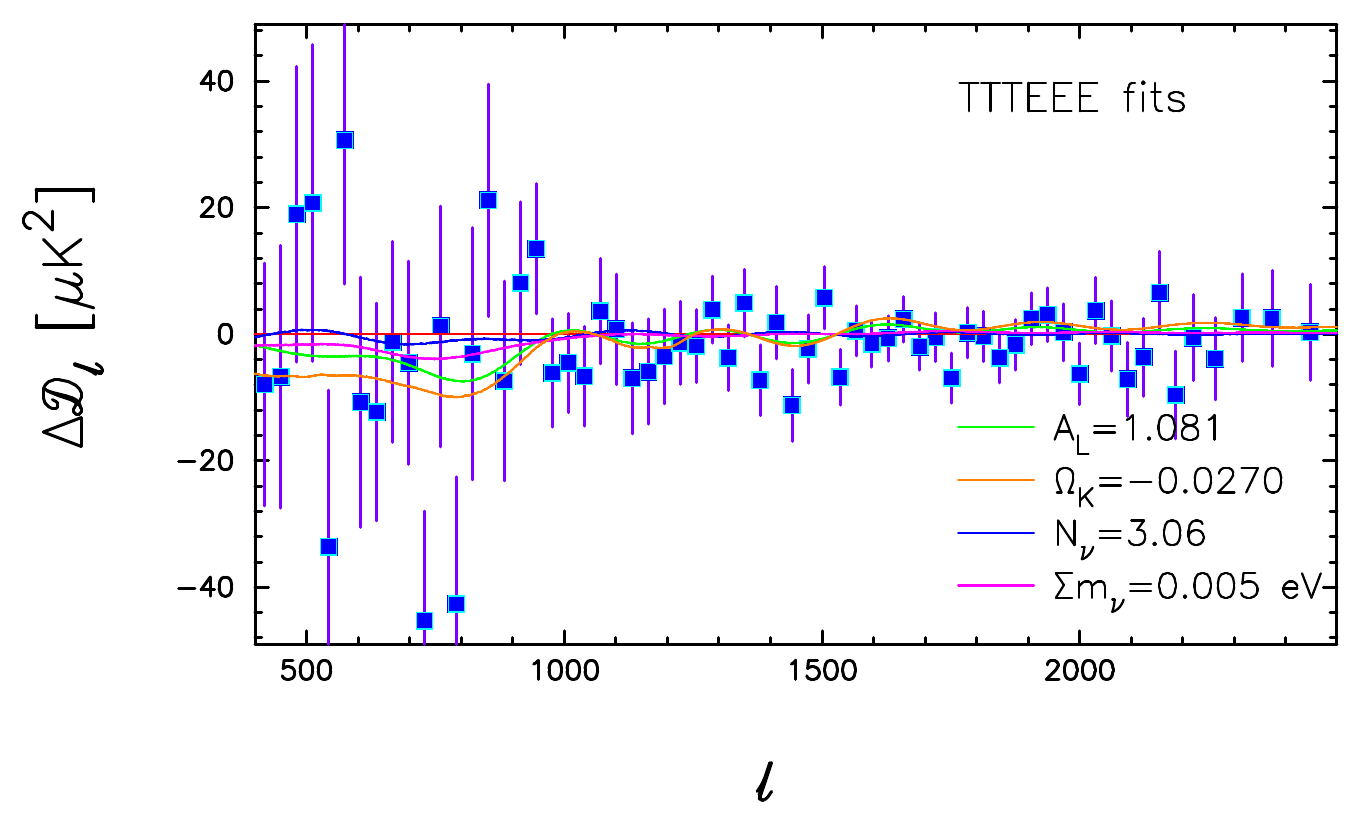} 
\caption {The blue points show the residuals of coadded 12.5HMcl TT spectrum 
relative to the best fit base \LCDM\ cosmology fitted to the 12.5HMcl TTTEEE likelihood.
The lines show the residuals of the best-fit TT spectra for one-parameter extensions to base \LCDM\
fitted to the 12.5HMcl TTTEEE likelihood, with best-fit parameters as listed in the figure.}

\label{fig:extensions}

\vspace{0.2truein}
\end{figure}

This section has investigated some simple one-parameter extensions to
the base \LCDM\ cosmology.  None of these extensions are strongly
favoured by \Planck\ data. Figure \ref{fig:extensions} shows the
best-fit temperature power spectra for extended models fitted to the
12.5HMcl TTTEEE likelihood. These models have nearly identical
temperature power spectra as a consequence of degeneracies with other
cosmological parameters. Note that for models such as $\omegak$, external data is
necessary to break the strong internal degeneracies inherent in the
CMB data (for example, as shown in Fig.\ \ref{fig:omegak}, the
addition of BAO data breaks the geometrical degeneracy leading to
tight constraints on $\omegak$). Models with strong internal
degeneracies are extremely sensitive to systematics in the
\Planck\ data.  However, for the extended models considered here, we
find consistency between the \Planck\ temperature and polarization
spectra and consistency between the \Planck\ power spectra,
\Planck\ lensing and BAO data. There is no evidence of systematics in
the \Planck\ power spectra, even for strongly degenerate models such
as $A_L$ and $\omegak$.

\section{Conclusions}
\label{sec:conclusions}

 Our main aim in this paper has been to present a description of
 \camspec\ in sufficient detail that an independent researcher should
 be able to reproduce our half-mission likelihoods  working with 
 \Planck\ 2018 data available from the PLA. \Planck\ is a complex data set and
 requires an appreciation of the intricacies of the data if one is to
 build an accurate power spectrum-based likelihood. We hope that this paper will give readers such an appreciation. 

A further aim has been to demonstrate the remarkable consistency of the \Planck\ power spectra between individual
detectors, between frequencies and with varying sky areas, reinforcing the conclusions of PCP18, PPL18
and earlier \Planck\ collaboration papers. The main data systematics that we have
investigated  are as follows:

\noindent
$\bullet$ correlated noise between detsets;

\smallskip

\noindent
$\bullet$ small effective calibration differences in the temperature maps;

\smallskip

\noindent
$\bullet$  effective polarization efficiencies;

\smallskip

\noindent
$\bullet$ temperature-to-polarization leakage.

\smallskip

Throughout this paper, we have developed internal consistency checks to correct  these data systematics
or, as in the case of temperature-to-polarization leakage,  to check our models for these corrections.
With the exception of low multipole polarization,  we find no evidence for any
further instrumental systematics in the \Planck\ power spectra that could impact on the fidelity of our
high multipole likelihoods.

We have analysed the properties of Galactic dust emission in
temperature and polarization, separating out isotropic extragalactic
foregrounds such as the CIB from anisotropic dust emission. We find
that over large areas of sky ($\sim 80\%$), Galactic emission is
remarkably universal and can be subtracted to high accuracy using high
frequency \Planck\ maps as templates. In temperature, we have
  demonstrated that subtraction of Galactic dust and CIB emission at
  $143$ and $217$ GHz using $545$ GHz maps leaves low amplitude
  statistically isotropic (i.e. extragalactic) foregrounds with power
  spectra that can be described accurately by power-laws. These
  findings have allowed us to create `cleaned' temperature likelihoods
  using 80\% of the sky at $143$ and $217$ GHz therefore extending the sky
  coverage compared to the likelihoods used in the \Planck\ 2018
  papers. Similarly, polarized dust emission is found to be remarkably
  universal over the frequency range $100-353$ GHz., with no evidence
  to high accuracy for any decorrelation with frequency.

PCP18 reported a number of `unusual' results, though not at high
statistical significance. These included large residuals in the
$217\times217$ spectrum at $\ell \sim 1460$ and a tendency for the
\Planck\ temperature power spectra to favour high values of $A_L$ and
 $\Omega_k<0$. PCP18 also showed that the \plik\ and
\camspec\ likelihoods behaved differently for some models if the
polarization spectra were added to the temperature spectra. We have
investigated these issues further by creating a set of likelihoods
using different sky coverage in temperature and polarization and
  very different models for temperature foregrounds. For base \LCDM,
the cosmological parameters are extremely stable as we scan through
these likelihoods as shown in
Fig.\ \ref{fig:base_lcdm_likelihoods}. The main trend that we have found is
for the TE parameters to come into better agreement with the TT
parameters as the sky area is increased. The TTTEEE parameters from
these likelihoods are highly consistent and in excellent agreement
with the base \LCDM\ cosmological parameters reported in PCP18. Given
the internal consistency of the \Planck\ spectra, the agreement
between temperature and polarization spectra, and insensitivity of the
likelihoods to sky area and foreground removal, we believe that the
\Planck\ results, as presented in the \Plancks collaboration papers,
are secure.

We disagree strongly with the conclusions of \cite{Addison:2016}
concerning the statistical significance of parameter shifts inferred
from \Planck\ spectra at low and high multipoles. We find no evidence
for any anomalous parameter shifts. It is straightforward to isolate
the features in the high multipole region of the temperature spectrum
responsible for the cosmological parameters determined from the
multipole range $801 \le \ell \le 2500$. These features are consistent
across the 143$\times$143, 143$\times$217 and 217$\times$217
spectra. Significantly, if we restrict the combined TTTEEE likelihood to
the multipole range $2\le \ell \le 800$, we infer cosmological parameters for
base \LCDM\ that are in very good agreement, and have comparable
accuracy, to the parameters determined from the $2 \le \ell \le 2500$ TT
likelihoods. Our analysis reveals no evidence that systematics at high
multipoles have any significant impact on the parameters of the base
\LCDM\ model derived from \Planck.

As we scan through the likelihoods in sequence of increasing
statistical power, the internal tensions reported in PCP18 decrease in
statistical significance. As noted in the introduction, the temperature and polarization spectra
also become `quieter' and come into closer agreement with the best-fit
base \LCDM\ cosmology. The results on $A_L$ and $\omegak$ reported in
PCP18 are consistent with modest statistical fluctuations which
decline in statistical significance as we increase the statistical
power of the likelihoods. For all extensions of base \LCDM\ considered
here, the addition of the \Planck\ polarization spectra to the
temperature spectra {\it always} drives parameters closer to those of
base \LCDM.  As far as we can see, the base \LCDM\ model is a perfect
fit to the \Planck\ spectra within the statistical errors.

What does this mean for the future of CMB research? If external data,
such as direct determinations of $H_0$, or cosmic shear surveys,
are shown to be discrepant with \Planck\ then it is unlikely that the
\Planck\ data are at fault. This would require us to search for new
physics that mimics,  to high accuracy,  the primordial CMB and lensing
power spectra measured by \Planck. It may be possible to find
extensions to \LCDM\ with strong internal parameter degeneracies that
achieve this (see for example, \citep{Kreisch:2019}). 
For such models, the \Planck\ data will
be essentially neutral and the evidence in favour of new physics will rely
entirely on the fidelity of external data. The \Planck\ data are,
however, limited. They have  little statistical power at
multipoles $\ell \simgt 2000$ and so cannot strongly constrain
theoretical models that modify the damping tail of the CMB
fluctuations.  There have been some claims, at low statistical
significance, of an inconsistency between base \LCDM\ and CMB
polarization power spectra at high multipoles
\citep{Henning:2018}. Fortunately, an ambitious programme of ground
based CMB polarization measurements should provide a strong test of 
\LCDM\ via high resolution observations of the CMB damping tail\footnote{Since the submission of this paper, the ACT and SPT 
teams have released  new results on the temperature and polarization power
spectra extending to $\ell \approx 4000$ 
 \cite{Choi:2020, Aiola:2020, Dutcher:2021}. The new ACT and SPT results are in very good agreement with the \Planck\ best fit \LCDM\ cosmology as reported in PCP18 and in this paper. A qualitative
comparison of the new ground based  TE and EE spectra with the \Planck\ 12.5HMcl 
polarization spectra is given in 
Appendix \ref{sec:appendix2}.}   and CMB lensing
\citep{Ade:2019}. The detection of tensor modes would, of
course, have profound implications for inflationary cosmology and
early Universe physics. The continuing search for primordial B-modes, from ground and
space experiments, e.g.\ \citep{BICEP:2018, Ade:2019, Sekimoto:2018},
is therefore of paramount importance. The next decade will see a new generation of
large-scale structure, weak lensing and low frequency radio surveys. We can only hope
that we are lucky, and that we will learn more about early Universe physics, dark matter
and dark energy -- all of which remain mysterious at this time.

\vspace{0.1 truein}

\section*{Acknowledgments} GPE would like to thank Caltech (and especially
Anthony Readhead) for supporting a visiting fellowship where part of
this work was done.  SG acknowledges financial support from STFC and
the award of a Kavli Institute Fellowship at KICC.  We are grateful to
our Cambridge colleagues, particularly Anthony Challinor, Marina
Migliaccio, Mark Ashdown and Anthony Lasenby for many useful comments
on (and contributions to) \Planck\ power spectrum analysis over the
years. \GE{We thank the referee for  providing
 a conscientious and comprehensive report which has helped us improve this paper.}
 We are indebted to Antony Lewis, for his incredible
contributions in developing the \COSMOMC\ Monte-Carlo Markov Chain
software, which we used for all of the MCMC results reported in this
paper and for his graphics scripts which we have used for the
production of many of the figures showing parameter constraints.  We
also thank the \plik\ team, Karim Benabed, Francois Bouchet, Silvia
Galli, Eric Hivon, Marius Millea, Simon Prunet, and many other members
of the \planck\ collaboration, with whom we have had constructive
discussions over the years. We thank Erminia Calabrese for discussions
concerning the \Planck-ACTpol comparison and we are grateful to Zack Li for 
pointing out \GE{a typographical}  error in Appendix A.  We are indebted to the
\Planck\ Collaboration for their enormous efforts in producing such a
wonderful set of data and we thank the \Plancks Science team for
permission to use the 2018 detset maps in this paper. We are also
grateful to the \Plancks Editorial Board,  and particularly to Antony
Lewis, for comments on a draft of this paper.  We wish to make it
clear, however, that the conclusions presented here represent the
views of the authors.

\newpage

\appendix

\section{Mathematical Details}
\label{sec:appendix}

\subsection{Coupling Matrices}
\label{subsec:coupling_matrices}

Assuming statistical isotropy, the expectation value of (2) is related to the theoretical spectra 
$C^{TT}, C^{TE},  \dots$, via a set of coupling matrices:
\begin{equation}  
     \begin{array}{l}
        \langle \tilde C^{T_iT_j} \rangle = K^{T_iT_j}  C^{TT},  \\
        \langle \tilde C^{T_iE_j} \rangle  = K^{T_iE_j} C^{TE},  \\
        \langle \tilde C^{E_iT_j} \rangle  = K^{E_iT_j} C^{TE},  \\
        \langle \tilde C^{E_iE_j} \rangle  = K^{E_iE_j} C^{EE} + K^{E_iB_j} C^{BB},  \\
        \langle \tilde C^{B_iB_j} \rangle  = K^{E_iB_j}C^{EE} + K^{E_iE_j}  C^{BB},  \\
        \langle \tilde C^{E_iB_j} \rangle  = [K^{E_iE_j}  -  K^{E_iB_j}]  C^{EB},  \\
        \langle \tilde C^{B_iE_j} \rangle  = [K^{E_iE_j}  -  K^{B_iE_j}]  C^{EB},  \\
        \langle \tilde C^{T_iB_j} \rangle  = K^{T_iE_j} C^{TB},   \\
        \langle \tilde C^{B_iT_j} \rangle  = K^{E_iT_j} C^{TB}.   \\
       \end{array}   \label{C2}
\end{equation}

The various blocks of the coupling matrix appearing in equation (\ref{C2})  are given by the 
following expressions \citep{Hivon:2002, Kogut:2003}:
\begin{subequations}
\begin{eqnarray}
  K^{T_iT_j}_{\ell_1 \ell_2}  & = &  {(2 \ell_2 + 1) \over 4 \pi}
\sum_{\ell_3 }  (2 \ell_3 + 1)\tilde W^{T_iT_j}_{\ell_3}
{\left ( \begin{array}{ccc}
        \ell_1 & \ell_2 & \ell_3  \\
        0  & 0 & 0
       \end{array} \right )^2} \equiv (2 \ell_2 + 1) \Xi_{TT}(\ell_1, \ell_2, \tilde W^{T_iT_j}), \ \ \qquad \label{A1a} \\
  K^{T_iE_j}_{\ell_1 \ell_2}  & = &  
{(2 \ell_2 + 1) \over 8 \pi}
\sum_{\ell_3 } (2 \ell_3 + 1) \tilde W^{T_iP_j} _{\ell_3}
(1 + (-1)^L)
{\left ( \begin{array}{ccc}
        \ell_1 & \ell_2 & \ell_3  \\
        0  & 0 & 0
       \end{array} \right ) } {\left ( \begin{array}{ccc}
        \ell_1 & \ell_2 & \ell_3  \\
        -2  & 2 & 0
       \end{array} \right ) } \nonumber \\  
 & &  \equiv (2 \ell_2 + 1) \Xi_{TE}(\ell_1, \ell_2, \tilde W^{T_iP_j}), \qquad L = \ell_1 + \ell_2 + \ell_3, \label{A1b} \\
  K^{E_iE_j}_{\ell_1 \ell_2}  & = &    
{(2 \ell_2 + 1) \over 16 \pi}
\sum_{\ell_3 }  (2 \ell_3 + 1) \tilde W^{P_iP_j} _{\ell_3}
(1 + (-1)^L)^2
 {\left ( \begin{array}{ccc}
        \ell_1 & \ell_2 & \ell_3  \\
        -2  & 2 & 0
       \end{array} \right )^2 } \nonumber \\
   & &  \equiv (2 \ell_2 + 1) \Xi_{EE}(\ell_1, \ell_2, \tilde W^{P_iP_j}),  \qquad L = \ell_1 + \ell_2 + \ell_3,  \label{A1c} \\
  K^{E_iB_j}_{\ell_1 \ell_2}  & = &   K^{B_iE_j}_{\ell_1 \ell_2}  = 
{(2 \ell_2 + 1) \over 16 \pi}
\sum_{\ell_3 }  (2 \ell_3 + 1) \tilde W^{P_iP_j} _{\ell_3}
(1 - (-1)^L)^2
 {\left ( \begin{array}{ccc}
        \ell_1 & \ell_2 & \ell_3  \\
        -2  & 2 & 0
       \end{array} \right )^2 } \\ \nonumber
 & & \equiv (2 \ell_2 + 1) \Xi_{EB}(\ell_1, \ell_2, \tilde W^{P_i P_j}), \qquad L = \ell_1 + \ell_2 + \ell_3, \label{A1d}
\end{eqnarray}
\end{subequations}

 where for the cross spectrum $(i,j)$, $\tilde W^{X_iX_j}_\ell$ is the power spectrum of the ``window'' function
defined by the mask and weighting scheme
\begin{equation}
   \tilde W^{X_iX_j}_\ell = {1 \over (2 \ell + 1)} \sum_m \tilde w^{X_i}_{\ell m} \tilde w^{X_j*}_{\ell m} ,  \label{A2}
\end{equation}
and $X$ denotes the mode (either temperature $T$ or polarization $P$).

\subsection{Covariance matrices}
\label{subsec:covariance_matrices}

\camspec\ uses analytic approximations to the covariance matrices of the pseudo-spectra 
derived under the assumptions of narrow window functions and uncorrelated (but anisotropic) noise pixel noise ($(\sigma^T_i)^2$,
$(\sigma^Q_i)^2$, $(\sigma^U_i)^2$) \citep{Efstathiou:2004, Challinor:2005, Efstathiou:2006, Hamimeche:2008}.
 The expressions are quite cumbersome, and so we give the
expresssions for only those covariance matrices used to form the TTTEEE likelihoods. The following equations
are  based on the analytic formulae developed in \cite{Efstathiou:2006, Hamimeche:2008}:

\hspace{1truein}

\begin{subequations}
\begin{eqnarray}
\langle \Delta \tilde C^{T_iT_j}_\ell \Delta \tilde C^{T_pT_q}_{\ell^\prime} \rangle
&\approx& ({\bar C}^{T_iT_p}_\ell {\bar C}^{T_iT_p}_{\ell^\prime} {\bar C}^{T_jT_q}_{\ell} {\bar C}^{T_jT_q}_{\ell^\prime})^{1/2} \Xi_{TT}(\ell, \ell^\prime, \tilde W^{(ip)(jq)} )  \nonumber \\
 & +&  ({\bar C}^{T_iT_q}_\ell {\bar C}^{T_iT_q}_{\ell^\prime} {\bar C}^{T_jT_p}_{\ell} {\bar C}^{T_jT_p}_{\ell^\prime})^{1/2} \Xi_{TT}(\ell, \ell^\prime, \tilde W^{(iq)(jp)} )  \nonumber \\
   &+ & ({\bar C}^{T_jT_q}_\ell {\bar C}^{T_jT_q}_{\ell^\prime})^{1/2}  \Xi_{TT}(\ell, \ell^\prime, \tilde W^{2T(ip)(jq)}) \delta_{ip} \nonumber \\
 &+&  ({\bar C}^{T_jT_p}_\ell {\bar C}^{T_jT_p}_{\ell^\prime})^{1/2}\Xi_{TT}(\ell, \ell^\prime, \tilde W^{2T(iq)(jp)} ) \delta_{iq} \nonumber \\
  &+ & ({\bar C}^{T_iT_p}_\ell {\bar C}^{T_iT_p}_{\ell^\prime})^{1/2}\Xi_{TT}(\ell, \ell^\prime, \tilde W^{2T(jq)(ip)}) \delta_{jq}  \nonumber \\
  &+ & ({\bar C}^{T_iT_q}_\ell {\bar C}^{T_iT_q}_{\ell^\prime})^{1/2} \Xi_{TT}(\ell, \ell^\prime, \tilde W^{2T(jp)(iq)}) \delta_{jp}   \nonumber \\
& + & \Xi_{TT}(\ell, \ell^\prime, \tilde W^{TT(ip)(jq)})\delta_{ip}\delta_{jq} +  
\Xi_{TT}(\ell, \ell^\prime, \tilde W^{TT(iq)(jp)}) \delta_{iq}\delta_{jp},  \label{CV1a} 
\end{eqnarray}

\newpage

\begin{eqnarray}
\langle \Delta \tilde C^{T_iE_j}_\ell \Delta \tilde C^{T_pE_q}_{\ell^\prime} \rangle
&\approx& ( \bar C^{T_iT_p}_\ell \bar C^{T_iT_p}_{\ell^\prime} \bar C^{E_jE_q}_\ell \bar C^{E_jE_q}_{\ell^\prime})^{1/2}  \Xi_{TE}(\ell, \ell^\prime, \tilde W^{(ip)(jq)} ) \nonumber \\
&+& \bar C^{T_iE_q}_\ell \bar C^{T_pE_j}_{\ell^\prime} \Xi_{TT}(\ell, \ell^\prime, \tilde W^{(iq)(jp)} )  \nonumber  \\
  & + &  (\bar C^{T_iT_p}_\ell \bar C^{T_iT_p}_{\ell^\prime})^{1/2} \Xi_{TE}(\ell, \ell^\prime, \tilde W^{2P(ip)(jq)} )\delta_{jq} \nonumber \\
  & + &
   (\bar C^{E_jE_q}_\ell \bar C^{E_jE_q}_{\ell^\prime})^{1/2} \Xi_{TE}(\ell, \ell^\prime, \tilde W^{2T(jq)(ip)} )\delta_{ip}  \nonumber \\
&+ &  \Xi_{TE}(\ell, \ell^\prime, \tilde W^{TP(ip)(jq)})\delta_{ip}\delta_{jq},  \label{CV1b} \\
\langle \Delta \tilde C^{E_iE_j}_\ell \Delta \tilde C^{E_pE_q}_{\ell^\prime} \rangle
&\approx& (\bar C^{E_iE_p}_\ell \bar C^{E_iE_p}_{\ell^\prime} \bar C^{E_jE_q}_\ell \bar C^{E_jE_q}_{\ell^\prime})^{1/2}   \Xi_{EE}(\ell, \ell^\prime, \tilde W^{(ip)(jq)})  \nonumber \\ 
&+ &  (\bar C^{E_iE_q}_\ell \bar C^{E_iE_q}_{\ell^\prime} \bar C^{E_jE_p}_\ell \bar C^{E_jE_p}_{\ell^\prime})^{1/2} \Xi_{EE}(\ell, \ell^\prime, \tilde W^{(iq)(jp)} )     \nonumber \\
& +&  (\bar C^{E_jE_q}_\ell \bar C^{E_jE_q}_{\ell^\prime})^{1/2}  \Xi_{EE}(\ell, \ell^\prime, \tilde W^{2P(ip)(jq)})\delta_{ip} \nonumber \\
& +&  (\bar C^{E_jE_p}_\ell \bar C^{E_jE_p}_{\ell^\prime})^{1/2}\Xi_{EE}(\ell, \ell^\prime, \tilde W^{2P(iq)(jp)})\delta_{iq} \nonumber \\
& +&  (\bar C^{E_iE_p}_\ell \bar C^{E_iE_p}_{\ell^\prime})^{1/2}\Xi_{EE}(\ell, \ell^\prime, \tilde W^{2P(jq)(ip)})\delta_{jq} \nonumber \\
& +&  (\bar C^{E_iE_q}_\ell \bar C^{E_iE_q}_{\ell^\prime})^{1/2}\Xi_{EE}(\ell, \ell^\prime, \tilde W^{2P(jp)(ip)})\delta_{jp} \nonumber \\
&+& \Xi_{EE}(\ell, \ell^\prime, \tilde W^{PP(ip)(jq)})\delta_{ip}\delta_{jq} + \Xi_{EE}(\ell, \ell^\prime, \tilde W^{PP(iq)(jp)}) \delta_{iq}\delta_{jp},   \qquad  \label{CV1c} \\
\langle \Delta \tilde C^{T_iT_j}_\ell \Delta \tilde C^{T_pE_{q}}_{\ell^\prime} \rangle
&\approx& {1 \over 2} (\bar C^{T_iT_p}_\ell \bar C^{T_iT_p}_{\ell^\prime})^{1/2}  (\bar C^{T_jE_q}_\ell + \bar C^{T_jE_q}_{\ell^\prime}) \Xi_{TT}(\ell, \ell^\prime, \tilde W^{(ip)(jq)})  \nonumber \\
&+&  {1 \over 2} (\bar C^{T_jT_p}_\ell \bar C^{T_jT_p}_{\ell^\prime})^{1/2}  (\bar C^{T_iE_q}_\ell + \bar C^{T_iE_q}_{\ell^\prime})  \Xi_{TT}(\ell, \ell^\prime, \tilde W^{(jp)(iq)} )     \nonumber \\
& +&  {1 \over 2}(\bar C^{T_jE_q}_\ell + \bar C^{T_jE_q}_{\ell^\prime})  \Xi_{TT}(\ell, \ell^\prime, \tilde W^{2T(ip)(jq)}) \delta_{ip} \nonumber \\
& +& {1 \over 2}(\bar C^{T_iE_q}_\ell + \bar C^{T_iE_q}_{\ell^\prime}) \Xi_{TE}(\ell, \ell^\prime, \tilde W^{2T(jp)(iq)}) \delta_{jp},  \\
\langle \Delta \tilde C^{E_iE_j}_\ell \Delta \tilde C^{T_pE_q}_{\ell^\prime} \rangle
&\approx& {1 \over 2} (\bar C^{E_jE_q}_\ell \bar C^{E_jE_q}{\ell^\prime})^{1/2}  (C^{T_pE_i}_\ell + C^{T_pE_i}_{\ell^\prime}) 
\Xi_{EE}(\ell, \ell^\prime, \tilde W^{(ip)(jq)} ) \nonumber \\
&+ & {1 \over 2} (\bar C^{E_iE_q}_\ell \bar C^{E_iE_q}{\ell^\prime})^{1/2}  (\bar C^{T_pE_j}_\ell + \bar C^{T_pE_j}_{\ell^\prime})   \Xi_{EE}(\ell, \ell^\prime, \tilde W^{(iq)(jp)} )     \nonumber \\
& +&  {1 \over 2}(\bar C^{T_pE_i}_\ell + \bar C^{T_pE_i}_{\ell^\prime})  \Xi_{EE}(\ell, \ell^\prime, \tilde W^{2P(ip)(jq)})\delta_{jq}  \nonumber \\
& + & {1 \over 2}(\bar C^{T_pE_j}_\ell + \bar C^{T_pE_j}_{\ell^\prime})\Xi_{EE}(\ell, \ell^\prime, \tilde W^{2P(jp)(iq)})\delta_{iq}, \\
\langle \Delta \tilde C^{T_iT_j}_\ell \Delta \tilde C^{E_pE_q}_{\ell^\prime} \rangle
&\approx&  (\bar C^{T_iE_p}_\ell \bar C^{T_jE_q}_{\ell^\prime}) \Xi_{TT}(\ell, \ell^\prime, \tilde W^{(ip)(jq)}) \nonumber \\
&+&  (\bar C^{T_iE_q}_\ell \bar C^{T_jE_p}_{\ell^\prime})  \Xi_{TT}(\ell, \ell^\prime, \tilde W^{(iq)(jp)} ),   \label{CV1f}   
\end{eqnarray}
\end{subequations}
where the matrices $\Xi$ are defined in Eqs. (\ref{A1a})-(\ref{A1d}), and the window
functions are given by
\begin{subequations}
\begin{eqnarray}
   \tilde W^{(ij)(pq)}_\ell &=& {1 \over (2 \ell + 1)} \sum_m \tilde w^{(ij)}_{\ell m} \tilde 
w^{(pq)*}_{\ell m} ,  \label{B2a} \\
   \tilde W^{TT(ij)(pq)}_\ell &=& {1 \over (2 \ell + 1)} \sum_m \tilde w^{T(ij)}_{\ell m} \tilde 
w^{T(pq)*}_{\ell m} ,  \label{B2b} \\
   \tilde W^{TP(ij)(pq)}_\ell &=& {1 \over (2 \ell + 1)} \sum_m {1 \over 2} [\tilde w^{T(ij)}_{\ell m} \tilde 
w^{Q(pq)*}_{\ell m}  +  \tilde w^{T_(ij)}_{\ell m} \tilde 
w^{U(pq)*}_{\ell m}] ,  \label{B2c} \\
   \tilde W^{2T(ij)(pq)}_\ell &=& {1 \over (2 \ell + 1)} \sum_m \tilde w^{(ij)}_{\ell m} \tilde 
w^{T(pq)*}_{\ell m} ,  \label{B2d} \\
   \tilde W^{2P(ij)(pq)}_\ell &=& {1 \over (2 \ell + 1)} \sum_m {1 \over 2}
[\tilde w^{(ij)}_{\ell m} \tilde w^{Q(pq)*}_{\ell m} + \tilde w^{(ij)}_{\ell m} \tilde w^{U(pq)*}_{\ell m}] ,  \label{B2e} \\ 
   \tilde W^{PP(ij)(pq)}_\ell &=& {1 \over (2 \ell + 1)} \sum_m {1 \over 2}
[\tilde w^{Q(ij)}_{\ell m} \tilde w^{Q(pq)*}_{\ell m} + \tilde w^{U(ij)}_{\ell m}
 \tilde w^{U(pq)*}_{\ell m} ] ,  \label{B2f} 
\end{eqnarray}
\end{subequations}
and
\begin{subequations}
\begin{eqnarray}
w^{(ij)}_{\ell m} & = & \sum_s w^i_s w^j_s \Omega_s Y^*_{\ell m}({\pmb{$\theta$}}_i), \label{equ:A3a} \\
w^{T(ij)}_{\ell m} & = & \sum_s (\sigma^T_s)^2w^i_s w^j_s \Omega^2_s Y^*_{\ell m}({\pmb{$\theta$}}_i), \label{equ:A3b}\\
w^{Q(ij)}_{\ell m} & = & \sum_s (\sigma^Q_s)^2w^i_s w^j_s \Omega^2_s Y^*_{\ell m}({\pmb{$\theta$}}_i), \label{equ:A3c}\\
w^{U(ij)}_{\ell m} & = & \sum_s (\sigma^U_s)^2w^i_s w^j_s \Omega^2_s Y^*_{\ell m}({\pmb{$\theta$}}_i)  \label{equ:A3d}.
\end{eqnarray}
\end{subequations}
\adjust where the sums in Eqs.\ ref{equ:A3a}-{equ:A3d} run over the number of pixels.

\adjust As in PPL13, we adopt a heuristic approach to model non-white noise. We compute the noise spectra from odd-even difference
maps as described in Sect. \ref{subsec:noise+power_spectrum} which we
fit to the functional form given in Eq. \ref{equ:Noise5}.  This
defines a set of weight factors:
\begin{equation}
   \psi^X_\ell = {N^{{\rm fit}X} \over {\rtensor N}^{X}}, \label{equ:A4}
\end{equation}
where $X=(T,Q,U)$ (each treated as a scalar map) and ${\rtensor N}^X$
is the white noise spectrum given by a summation over the weighted
pixels (Eq. \ref{equ:Noise1}). The pixel noise estimates
$(\sigma^{T})^2_i$, $(\sigma^{Q})^2_i$ and $(\sigma^{U})^2_i$ in
Eqs.\ \ref{equ:A3b}- \ref{equ:A3d} are then replaced by
$\sqrt{\psi^T_\ell\psi^T_{\ell^\prime}}(\sigma^{T})^2_i$, 
$\sqrt{\psi^Q_\ell\psi^Q_{\ell^\prime}}(\sigma^{Q})^2_i$ and
$\sqrt{\psi^U_\ell\psi^U_{\ell^\prime}}psi^U_\ell (\sigma^{U})^2_i$, where $(\sigma^{T})^2_i$,
$(\sigma^{Q})^2_i$ and $(\sigma^{U})^2_i$ are the diagonal components
of the pixel noise covariance matrix returned by the map making code.
To keep data storage to manageable levels, we  only store 
 covariance
matrix elements  for $\Delta \ell = \vert \ell - \ell^\prime \vert
 \le  200$.

\section{Comparison with ACT and SPT polarization spectra}
\label{sec:appendix2}

The TE and EE spectra measured by \Planck\ are noisy and so have
little statistical power at multipoles $\ell \simgt 1000$. At higher
multipoles, the ground based polarization measurements by ACTpol \citep{Louis:2017}
and SPTpol \citep{Henning:2018} have much higher sensitivity.  These experiments have
measured the TE and EE spectra with high precision up to multipoles of
a few thousand, i.e. well into the CMB damping tail. The \Planck\ base
\LCDM\ best-fit model is very well determined and makes accurate predictions
of the polarization spectra at high multipoles. It is therefore worth 
asking whether this model is consistent with the results from ACT
and SPT. Tests of the damping tail provide a particularly important 
check of extended \LCDM\ models, including variants that could lead to 
a change in the sound horizon (e.g.\ \citep{Lancaster:2017,Kreisch:2019,Park:2019}).
At the time of the submission of this
paper, the ACT data release 4 (DR4) results  \cite{Choi:2020, Aiola:2020} and SPT-3G results \cite{Dutcher:2021}
 had not yet appeared.
As in the submitted version of this paper,  we  discuss  the polarization measurements from SPTpol and ACTpol, 
but have added a brief comparison with ACT DR4 and SPT-3G polarization spectra.

\begin{figure}
\centering
\includegraphics[width=145mm,angle=0]{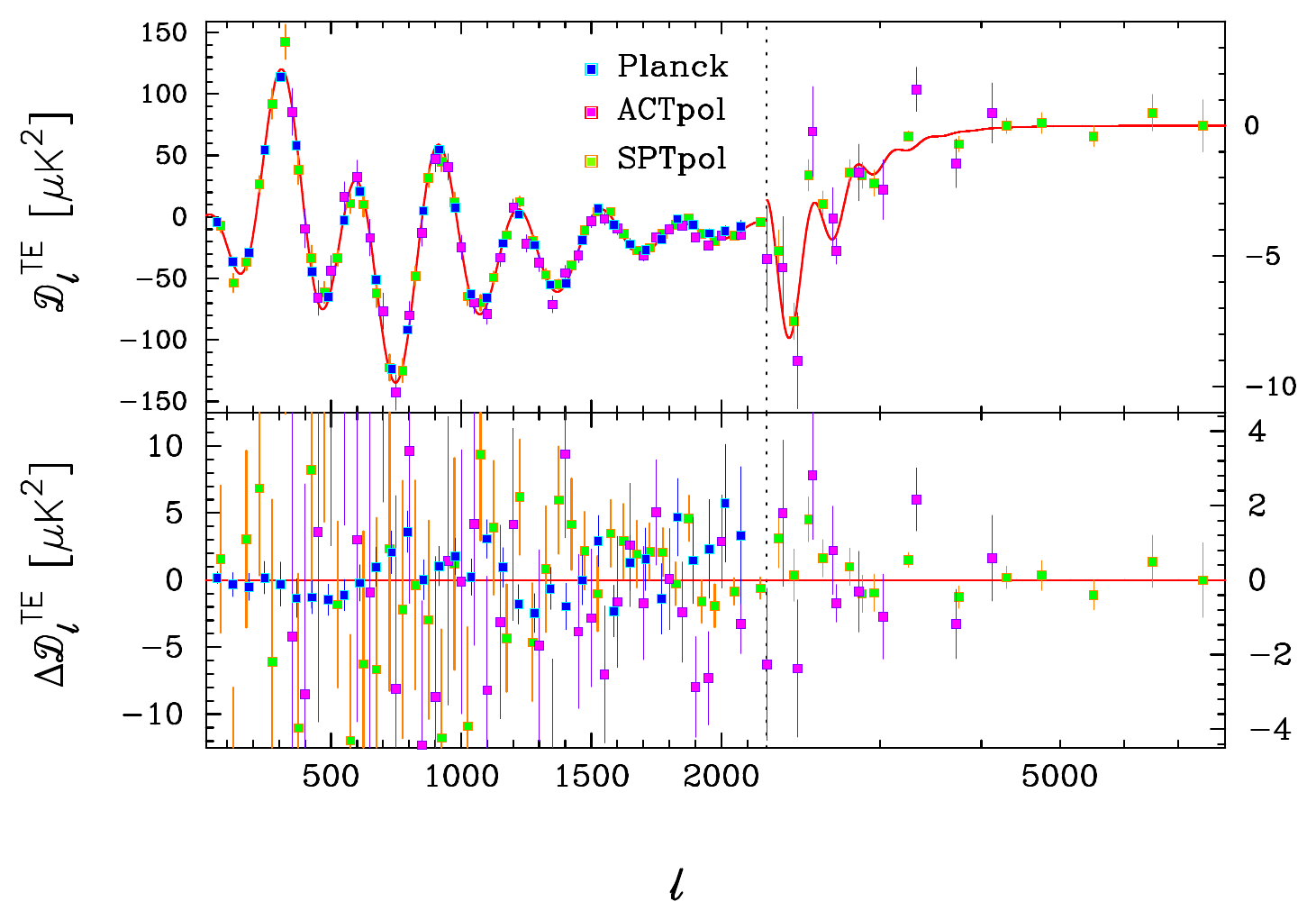}  \\
\hspace{8mm}\includegraphics[width=155mm,angle=0]{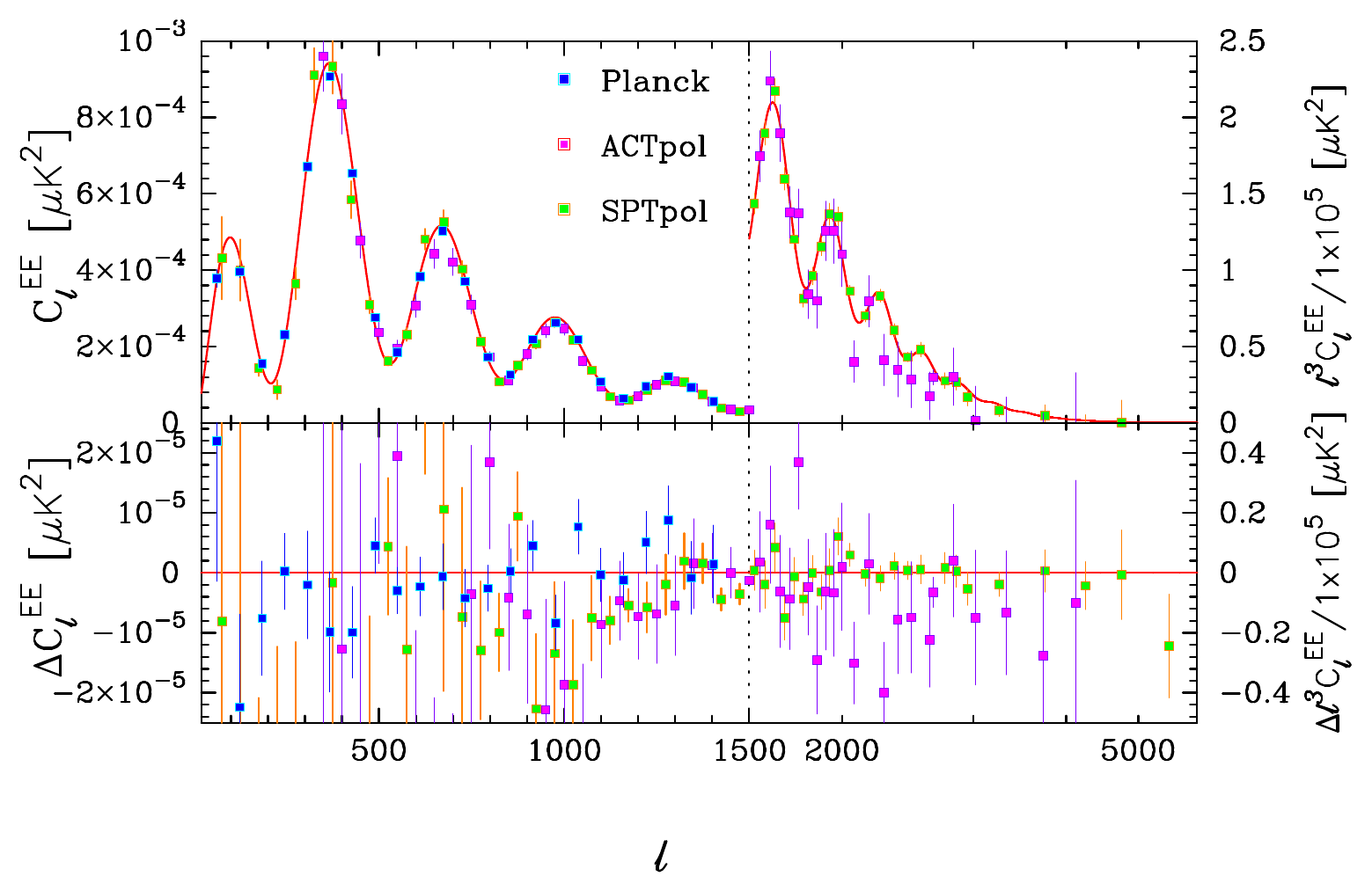}

\caption {The 12.5HMcl TE and EE spectra (blue points) compared to the
  ACTpol and SPTpol spectra. The red line shows the 12.5HMcl TTTEEE
  best-fit base \LCDM\ cosmology. Each plot is split into four panels,
  with the left-hand panels showing the low multipoles on a linear
  multipole scale, and the right-hand panels showing the high
  multipoles on a logarithmic scale. The lower panels in each plot
  show the residuals relative to the best fit-base \LCDM\ model.  In
  the lower figure, showing the EE spectra, we have plotted $\ell^3
  C^{EE}_\ell$, and its residuals, at high multipoles to emphasise the
  the shape of the damping tail. }

\label{fig:actspt}

\end{figure}

Fig.\ \ref{fig:actspt} compares the \Planck, ACTpol and SPTpol
TE and EE spectra. We make the following observations:

\smallskip

\noindent
[1] The overall agreement between \Planck, ACTpol and SPTpol is
extremely good, delineating the shapes of the acoustic peaks in both
the TE and EE spectra out to  multipoles of a few thousand. The fact that three
independent investigations agree so well is a remarkable experimental achievement
and provides strong support that the primordial fluctuations were dominated by adiabatic
modes.

\smallskip

\noindent
[2] In detail, however, we see that it is difficult to quantitatively test
consistency of the spectra to high accuracy. Over the multipole range where
\Planck\ errors are small, the errors on the ACTpol and SPTpol spectra are
large (and vice versa). This means that it is not possible to determine accurate
polarization  calibrations for ACTpol and SPTpol relative to \Planck\ by comparing the power spectra.
Without accurate relative calibrations in polarization, one cannot easily stitch
together \Planck, ACTpol and SPTpol polarization spectra to test theoretical predictions 
well into the CMB damping tail. If one tries to do this for base \LCDM\  by allowing relative calibrations
to vary as nuisance parameters in, say, a joint \Planck+SPTpol likelihood analysis, 
\Planck\ overwhelms SPTpol and the recovered cosmology is almost identical to that derived
from \Planck\  alone \cite{Lemos:2018}.

\smallskip

\noindent
[3] Since the ACTpol and SPTpol TE and EE errors are large compared to those from \Planck\
at multipoles $\ell \simlt 1500$, the base \LCDM\ cosmological parameter constraints determined 
from ACTpol and SPTpol are much weaker than those determined from \Planck. The ground based experiments
do not cover a wide enough range of multipoles to strongly constrain critical parameters such as $n_s$.
 The base \LCDM\ parameters from
both ACTpol and SPTpol are consistent with those determined by   \Planck, though 
\citep{Henning:2018} report hints of parameter shifts at high multipoles at low statistical 
significance. The analysis discussed in PCP18 (based on work reported in \cite{Lemos:2018}) 
shows that the SPTpol base \LCDM\ TEEE parameters converge
by $\ell_{\rm max} = 2500$, so any parameter shifts are driven by the SPTpol spectra in the multipole
range $\sim 1000-2500$, not by higher multipoles.

\smallskip

\begin{figure}
\centering
\includegraphics[width=145mm,angle=0]{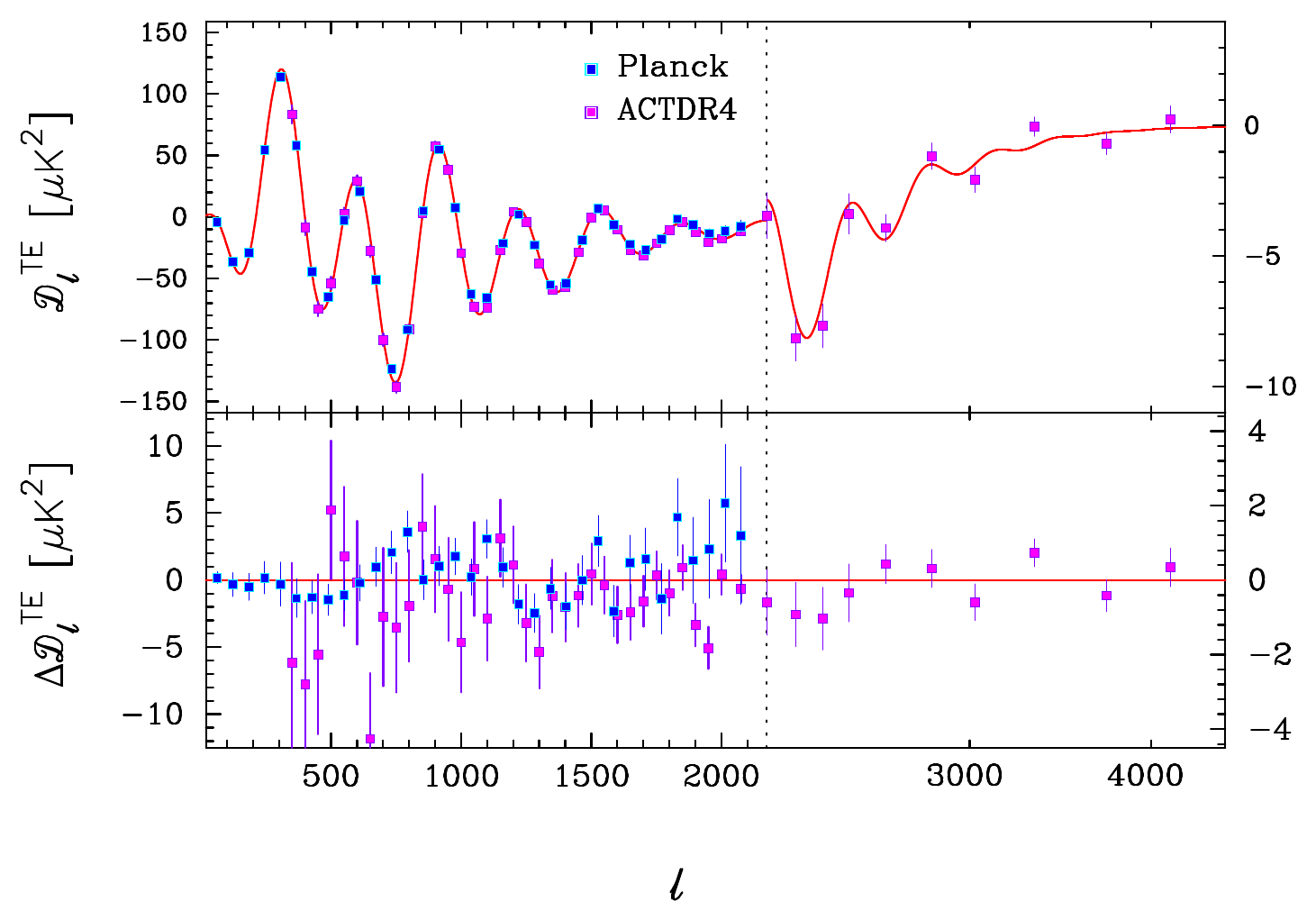}  \\
\hspace{8mm}\includegraphics[width=155mm,angle=0]{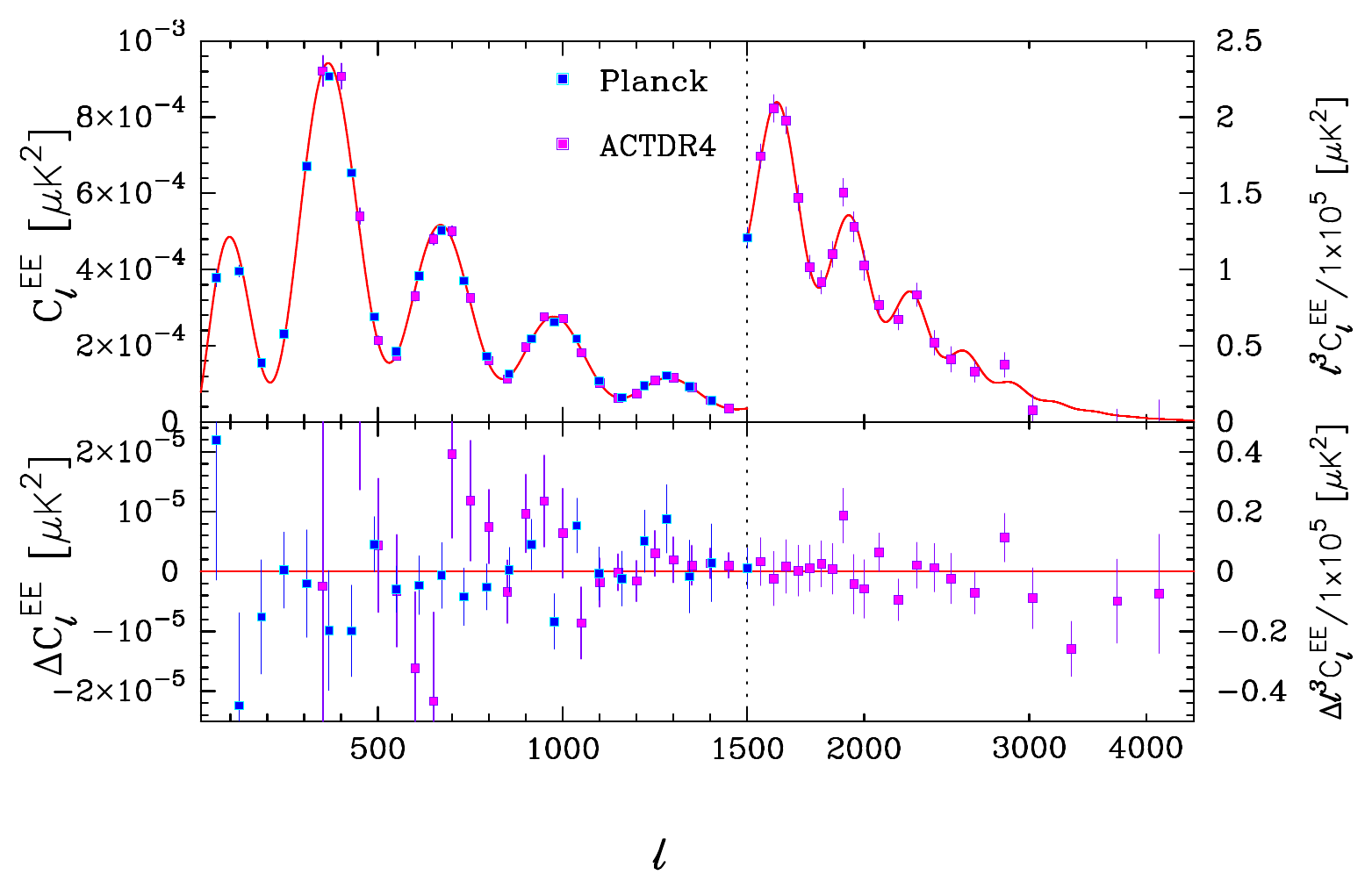}

\caption {As Fig. \ref{fig:actspt} but for the ACT DR4 spectra as reported by \cite{Choi:2020}.}

\label{fig:actDR4}

\end{figure}

\begin{figure}
\centering
\includegraphics[width=145mm,angle=0]{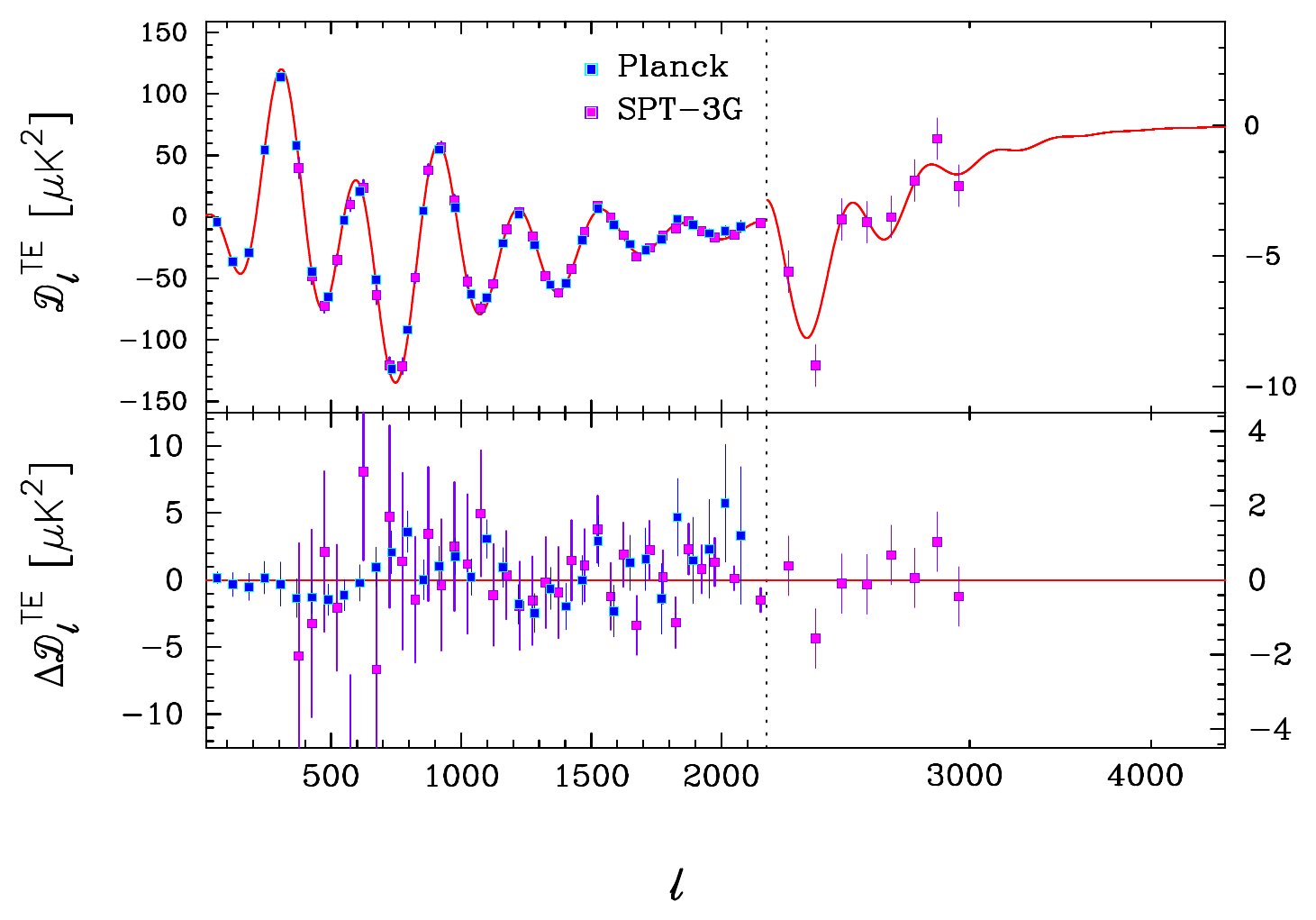}  \\
\hspace{8mm}\includegraphics[width=155mm,angle=0]{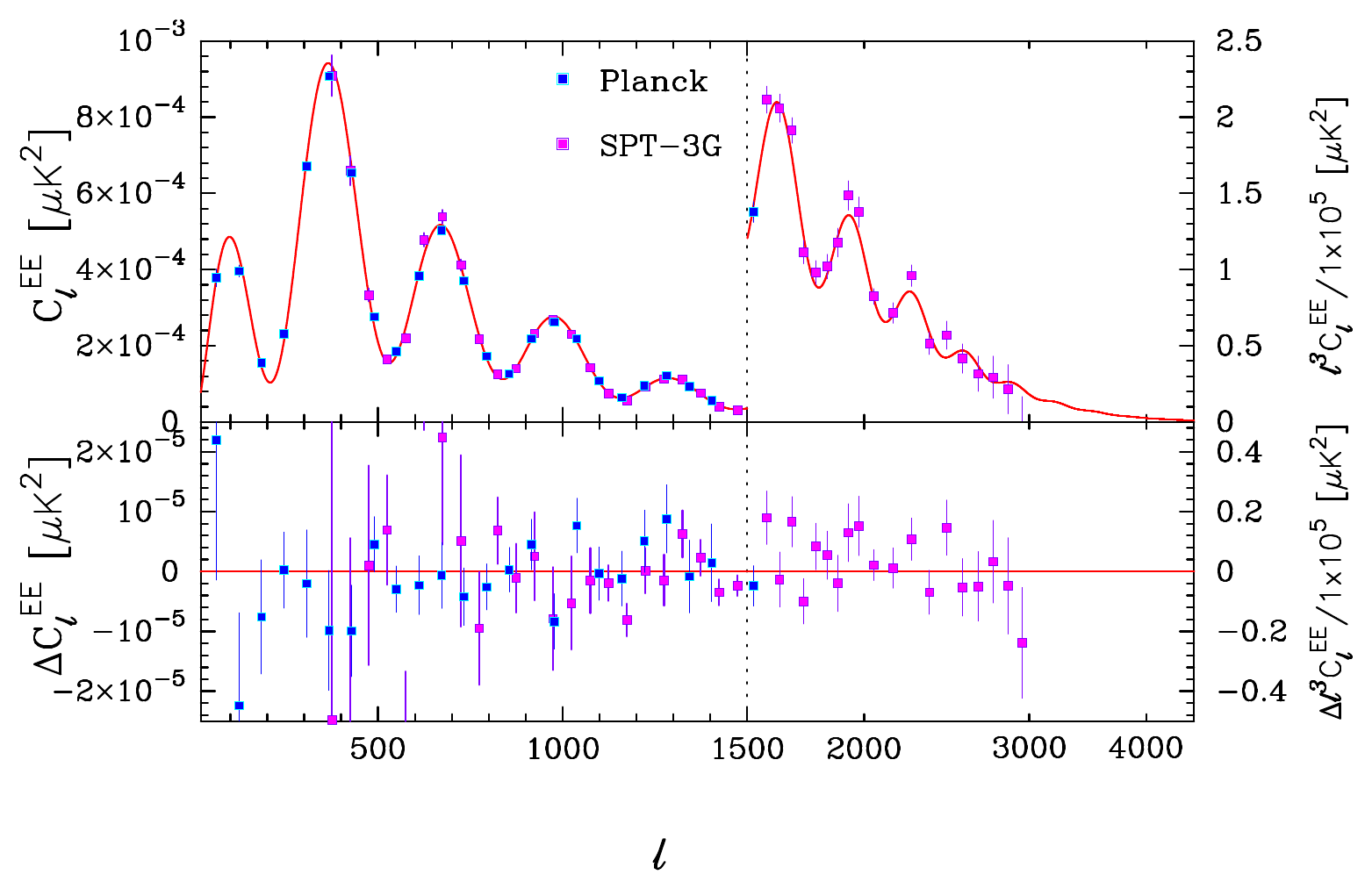}

\caption {As Fig. \ref{fig:actspt} but for the SPT-3G bandpower spectra reported in Table IV of
 \cite{Dutcher:2021}.}

\label{fig:SPT3G}

\end{figure}

\noindent
[4] ACTpol and SPTpol calibrate temperature at the map level by
cross-correlating against \Planck\ 143 GHz temperature maps. ACTpol
then cross-correlate with the 2015 143 GHz \Planck\ Q and U
maps\footnote{Note that the inferred effective polarization
  efficiencies of the 2015 \Planck\ maps are very close to those of
  the 2018 maps discussed in Sect.\ \ref{subsec:pol_cal}.} and infer a
polarization efficiency of $0.990 \pm 0.025$ (which is not corrected
for in the ACTpol spectra shown in Fig.\ \ref{fig:actspt}).  The
laboratory study of \citep{Rosset:2010} measured polarization
efficiences for the 143 GHz detectors in the range $84 - 94 \%$,
significantly lower than the efficiencies of detectors at other
frequencies. These laboratory values are assumed in the \Planck\ HFI
map making.  However, from the discussion in
Sect.\ \ref{subsec:pol_cal} we infer an effective polarization
efficiency for the 143 GHz Q and U maps of $1.015$ based on the EE
spectra and $1.007$ based on the TE spectra. Averaging these estimates
gives an effective polarization efficiency of \Planck\ 143 GHz Q and U
maps of $1.011$ with an uncertainty of about $0.005$. This would bring
the ACTpol polarization efficiency closer to unity.  Polarization
efficiencies  $1.011$ cannot, however, explain the run of low
ACTpol EE points at $\ell \simgt 2000$ seen in Fig.\ \ref{fig:actspt}
(which may indicate that the contribution from polarized point sources
has been over subtracted in the analysis of \cite{Louis:2017}). Note that this discrepancy 
with the ACTpol EE spectrum is not seen in the ACT DR4 EE spectrum (see  Fig. \ref{fig:actDR4}).
SPTpol
cross-correlated their polarization maps with the 2015 \Planck\ 143
GHz polarization maps and inferred an effective polarization
efficiency of $1.06\pm 0.01$. This is higher than expected from their
polarization calibration measurements using an external polarized
source \citep{Crites:2015, Keisler:2015} for reasons that are not yet
understood. The polarization efficiency of $1.06$ was applied to the
published SPTpol TE and EE bandpowers plotted in
Fig. \ref{fig:actspt}. With this factor applied, the SPTpol spectra
provide a very good match to the damping tail of our best fit base
\LCDM\ model.

The key point that we wish to make is that the effective polarization
efficiencies of the \GE{publicly} available \Planck\ polarization maps
differ from unity. This needs to be borne in mind if these maps are
used to calibrate other experiments. Uncertainties in polarization
efficiencies therefore limit the precision with which the
\Planck\ polarization power spectra can be `stitched' to ground-based
polarization power spectra extending to higher multipoles. The best
way of achieving high precision tests of cosmology with ground-based
polarization experiments is to make them self-contained by increasing the
multipole range, i.e.\ to improve the accuracy of the
polarization spectra at multipoles $\simlt 2000$ as well as at higher
multipoles.

As noted above, the ACTpol and SPTpol results shown in Fig. \ref{fig:actspt} have been superseded recently by the ACT DR4 
and SPT-3G releases. Figures \ref{fig:actDR4} and \ref{fig:SPT3G} shows a comparison of the new ACT and SPT TE and EE spectra with the \Planck\ 12.5HMcl spectra. The ACT spectra have been downloaded from NASA lambda web site\footnote
{\tt{https://lambda.gsfc.nasa.gov/product/act/act$\_$dr4$\_$spectra$\_$get.cfm}} 
and we have plotted TE and EE spectra from the files
{\tt{act$\_$dr4$\_$D$\_$ell$\_$TE$\_$cmbonly.txt}} and
{\tt{act$\_$dr4$\_$D$\_$ell$\_$EE$\_$cmbonly.txt}} with no additional
calibration corrections. The SPT-3G spectra are minimum variance coaadded spectra
over the $95$, $150$ and $220$ GHz from Table IV of \cite{Dutcher:2021} with polarization efficiencies
calibrated by comparing with \Plancks\ spectra (inferring high values of effective polarization efficiencies
of $1/1.028$, $1/1.057$ and $1/1.13$ for $95$, $150$ and $220$ GHz respectively).
Both ACT DR4 and SPT-3G  polarizations spectra agree extremely well with
the \Planck\ base \LCDM\ cosmology well into the damping tail. The
latest ground based  polarization spectra are therefore consistent with the base
\LCDM\ model out to multipoles of $\sim 4000$. A more quantitative
comparison requires a detailed map-based relative calibration of the
ACT and SPT  maps relative to \Planck\ accounting (as discussed above) for  the
variation of the \Planck\ polarization efficiencies with frequency.

\newpage

\bibliographystyle{h-physrev}

\bibliography{oja_paper}

\begin{thebibliography}{100}

\bibitem{PCP18}
{\ Planck Collaboration} {\em et~al.},
\newblock \aap\ {\bf 641}, A6 (2020), astro-ph/1807.06209.

\bibitem{Penzias:1965}
{\ A}.~A. {Penzias} and R.~W. {Wilson},
\newblock \apj\ {\bf 142}, 419 (1965).

\bibitem{Planck_mission:2014}
{\ Planck Collaboration} {\em et~al.},
\newblock \aap\ {\bf 571}, A1 (2014), astro-ph/1303.5062.

\bibitem{Planck_mission:2015}
{\ Planck Collaboration} {\em et~al.},
\newblock \aap\ {\bf 594}, A1 (2016), astro-ph/1502.01582.

\bibitem{Smoot:1992}
{\ G}.~F. {Smoot} {\em et~al.},
\newblock \apjl\ {\bf 396}, L1 (1992).

\bibitem{Bennett:2003}
{\ C}.~L. {Bennett} {\em et~al.},
\newblock \apjs\ {\bf 148}, 1 (2003), astro-ph/0302207.

\bibitem{Bennett:2013}
{\ C}.~L. {Bennett} {\em et~al.},
\newblock \apjs\ {\bf 208}, 20 (2013), astro-ph/1212.5225.

\bibitem{PCP13}
{\ Planck Collaboration} {\em et~al.},
\newblock \aap\ {\bf 571}, A16 (2014), astro-ph/1303.5076.

\bibitem{PCP15}
{\ Planck Collaboration} {\em et~al.},
\newblock \aap\ {\bf 594}, A13 (2016), astro-ph/1502.01589.

\bibitem{PPL13}
{\ Planck Collaboration} {\em et~al.},
\newblock \aap\ {\bf 571}, A15 (2014), astro-ph/1303.5075.

\bibitem{PPL15}
{\ Planck Collaboration} {\em et~al.},
\newblock \aap\ {\bf 594}, A11 (2016), astro-ph/1507.02704.

\bibitem{PPL18}
{\ Planck Collaboration} {\em et~al.},
\newblock \aap\ {\bf 641}, A5 (2020), astro-ph/1907.12875.

\bibitem{Lueker:2010}
{\ M}.~{Lueker} {\em et~al.},
\newblock \apj\ {\bf 719}, 1045 (2010), astro-ph/0912.4317.

\bibitem{Spergel:2015}
{\ D}.~N. {Spergel}, R.~{Flauger}, and R.~{Hlo{\v z}ek},
\newblock \prd\ {\bf 91}, 023518 (2015), astro-ph/1312.3313.

\bibitem{Louis:2017}
{\ T}.~{Louis} {\em et~al.},
\newblock \jcap\ {\bf 6}, 031 (2017), astro-ph/1610.02360.

\bibitem{Henning:2018}
{\ J}.~W. {Henning} {\em et~al.},
\newblock \apj\ {\bf 852}, 97 (2018), astro-ph/1707.09353.

\bibitem{Choi:2020}
{\ S}.~K. {Choi} {\em et~al.},
\newblock \jcap\ {\bf 2020}, 045 (2020), astro-ph/2007.07289.

\bibitem{Aiola:2020}
{\ S}.~{Aiola} {\em et~al.},
\newblock \jcap\ {\bf 2020}, 047 (2020), astro-ph/2007.07288.

\bibitem{Dutcher:2021}
{\ D}.~{Dutcher} {\em et~al.},
\newblock \prd\ {\bf 104}, 022003 (2021), astro-ph/2101.01684.

\bibitem{Efstathiou:2004}
{\ G}.~{Efstathiou},
\newblock \mnras\ {\bf 348}, 885 (2004), astro-ph/0310207.

\bibitem{Challinor:2005}
{\ A}.~{Challinor} and G.~{Chon},
\newblock \mnras\ {\bf 360}, 509 (2005), astro-ph/0410097.

\bibitem{Efstathiou:2006}
{\ G}.~{Efstathiou},
\newblock \mnras\ {\bf 370}, 343 (2006), astro-ph/0601107.

\bibitem{Hamimeche:2008}
{\ S}.~{Hamimeche} and A.~{Lewis},
\newblock \prd\ {\bf 77}, 103013 (2008), astro-ph/0801.0554.

\bibitem{Gorski:2005}
{\ K}.~M. {G{\'o}rski} {\em et~al.},
\newblock \apj\ {\bf 622}, 759 (2005), astro-ph/0409513.

\bibitem{Hivon:2002}
{\ E}.~{Hivon} {\em et~al.},
\newblock \apj\ {\bf 567}, 2 (2002), astro-ph/0105302.

\bibitem{Kogut:2003}
{\ A}.~{Kogut} {\em et~al.},
\newblock \apjs\ {\bf 148}, 161 (2003), astro-ph/0302213.

\bibitem{Mak:2017}
{\ D}.~S.~Y. {Mak}, A.~{Challinor}, G.~{Efstathiou}, and G.~{Lagache},
\newblock \mnras\ {\bf 466}, 286 (2017), astro-ph/1609.08942.

\bibitem{DataProcessing:2018}
{\ Planck Collaboration} {\em et~al.},
\newblock \aap\ {\bf 641}, A3 (2020), astro-ph/1807.06207.

\bibitem{Flauger:2017}
{\ R.}.~{Flauger}, L.~{McAllister}, E.~{Silverstein}, and A.~{Westphal},
\newblock \jcap\ {\bf 2017}, 055 (2017), hep-th/1412.1814.

\bibitem{Planck_CO13}
{\ Planck Collaboration} {\em et~al.},
\newblock \aap\ {\bf 571}, A13 (2014), astro-ph/1303.5073.

\bibitem{Lemos:2018}
{\ P}.~{Lemos},
\newblock PhD Thesis, University of Cambridge  (2018).

\bibitem{deBelsunce:2021}
{\ R}.~{de Belsunce}, S.~{Gratton}, W.~{Coulton}, and G.~{Efstathiou},
\newblock arXiv e-prints  (2021), astro-ph/2103.14378.

\bibitem{Rosset:2010}
{\ C}.~{Rosset} {\em et~al.},
\newblock \aap\ {\bf 520}, A13 (2010), astro-ph/1004.2595.

\bibitem{Planck_parity:2016}
{\ Planck Collaboration} {\em et~al.},
\newblock \aap\ {\bf 596}, A110 (2016), astro-ph/1605.08633.

\bibitem{SROLL:2016}
{\ Planck Collaboration} {\em et~al.},
\newblock \aap\ {\bf 596}, A107 (2016), astro-ph/1605.02985.

\bibitem{Planck_beams:2014}
{\ Planck Collaboration} {\em et~al.},
\newblock \aap\ {\bf 571}, A7 (2014), astro-ph/1303.5068.

\bibitem{Mitra:2011}
{\ S}.~{Mitra} {\em et~al.},
\newblock \apjs\ {\bf 193}, 5 (2011), astro-ph/1005.1929.

\bibitem{Hivon:2017}
{\ E}.~{Hivon}, S.~{Mottet}, and N.~{Ponthieu},
\newblock \aap\ {\bf 598}, A25 (2017), astro-ph/1608.08833.

\bibitem{Hivon:2015}
{\ E}.~{Hivon}, A.~{Ducout}, S.~{Mottet}, and G.~{Roudier},
\newblock Planck Internal Communication  (2015).

\bibitem{Planck_components:2014}
{\ Planck Collaboration} {\em et~al.},
\newblock \aap\ {\bf 571}, A12 (2014), astro-ph/1303.5072.

\bibitem{Planck_component2015}
{\ Planck Collaboration} {\em et~al.},
\newblock \aap\ {\bf 594}, A9 (2016), astro-ph/1502.05956.

\bibitem{Planck_foregrounds_2018}
{\ Planck Collaboration} {\em et~al.},
\newblock \aap\ {\bf 641}, A4 (2020), astro-ph/1807.06208.

\bibitem{Eriksen:2006}
{\ H}.~K. {Eriksen} {\em et~al.},
\newblock \apj\ {\bf 641}, 665 (2006), astro-ph/0508268.

\bibitem{Eriksen:2008}
{\ H}.~K. {Eriksen} {\em et~al.},
\newblock \apj\ {\bf 676}, 10 (2008), astro-ph/0709.1058.

\bibitem{Cardoso:2008}
{\ J}.-F. {Cardoso}, M.~{Martin}, J.~{Delabrouille}, M.~{Betoule}, and
  G.~{Patanchon},
\newblock arXiv e-prints  (2008), astro-ph/0803.1814.

\bibitem{Delabrouille:2009}
{\ J}.~{Delabrouille} {\em et~al.},
\newblock \aap\ {\bf 493}, 835 (2009), astro-ph/0807.0773.

\bibitem{Leach:2008}
{\ S}.~M. {Leach} {\em et~al.},
\newblock \aap\ {\bf 491}, 597 (2008), astro-ph/0805.0269.

\bibitem{SEVEM}
{\ R}.~{Fern{\'a}ndez-Cobos}, P.~{Vielva}, R.~B. {Barreiro}, and
  E.~{Mart{\'{\i}}nez-Gonz{\'a}lez},
\newblock \mnras\ {\bf 420}, 2162 (2012), astro-ph/1106.2016.

\bibitem{Planckdust:2014b}
{\ Planck Collaboration} {\em et~al.},
\newblock \aap\ {\bf 571}, A30 (2014), astro-ph/1309.0382.

\bibitem{Planck_dust}
{\ Planck Collaboration} {\em et~al.},
\newblock \aap\ {\bf 571}, A11 (2014), astro-ph/1312.1300.

\bibitem{Larsen:2016}
{\ P}.~{Larsen}, A.~{Challinor}, B.~D. {Sherwin}, and D.~{Mak},
\newblock Physical Review Letters {\bf 117}, 151102 (2016),
  astro-ph/1607.05733.

\bibitem{PlanckIX:2014}
{\ Planck Collaboration} {\em et~al.},
\newblock \aap\ {\bf 571}, A9 (2014), astro-ph/1303.5070.

\bibitem{Planck_calib:2016}
{\ Planck Collaboration} {\em et~al.},
\newblock \aap\ {\bf 594}, A8 (2016), astro-ph/1502.01587.

\bibitem{Finkbeiner1999}
{\ D}.~P. {Finkbeiner}, M.~{Davis}, and D.~J. {Schlegel},
\newblock \apj\ {\bf 524}, 867 (1999), astro-ph/9905128.

\bibitem{Meisner2015}
{\ A}.~M. {Meisner} and D.~P. {Finkbeiner},
\newblock \apj\ {\bf 798}, 88 (2015), astro-ph/1410.7523.

\bibitem{Planckdust:2015}
{\ Planck Collaboration} {\em et~al.},
\newblock \aap\ {\bf 576}, A107 (2015), astro-ph/1405.0874.

\bibitem{Planck_dust_pol:2016}
{\ Planck Collaboration} {\em et~al.},
\newblock \aap\ {\bf 586}, A133 (2016), astro-ph/1409.5738.

\bibitem{Ghosh:2017}
{\ T}.~{Ghosh} {\em et~al.},
\newblock \aap\ {\bf 601}, A71 (2017), astro-ph/1611.02418.

\bibitem{Planck_dust_pol:2018}
{\ Planck Collaboration} {\em et~al.},
\newblock \aap\ {\bf 641}, A11 (2020), astro-ph/1801.04945.

\bibitem{Lewis:2002}
{\ A}.~{Lewis} and S.~{Bridle},
\newblock \prd\ {\bf 66}, 103511 (2002), astro-ph/0205436.

\bibitem{Dunkley:2011}
{\ J}.~{Dunkley} {\em et~al.},
\newblock \apj\ {\bf 739}, 52 (2011), astro-ph/1009.0866.

\bibitem{Reichardt:2012}
{\ C}.~L. {Reichardt} {\em et~al.},
\newblock \apj\ {\bf 755}, 70 (2012), astro-ph/1111.0932.

\bibitem{Delouis:2019}
{\ J}.~M. {Delouis}, L.~{Pagano}, S.~{Mottet}, J.~L. {Puget}, and L.~{Vibert},
\newblock \aap\ {\bf 629}, A38 (2019), astro-ph/1901.11386.

\bibitem{Viero:2013}
{\ M}.~P. {Viero} {\em et~al.},
\newblock \apj\ {\bf 772}, 77 (2013), astro-ph/1208.5049.

\bibitem{Efstathiou:2012}
{\ G}.~{Efstathiou} and M.~{Migliaccio},
\newblock \mnras\ {\bf 423}, 2492 (2012), astro-ph/1106.3208.

\bibitem{Battaglia:2010}
{\ N}.~{Battaglia}, J.~R. {Bond}, C.~{Pfrommer}, J.~L. {Sievers}, and
  D.~{Sijacki},
\newblock \apj\ {\bf 725}, 91 (2010), astro-ph/1003.4256.

\bibitem{McCarthy:2014}
{\ I}.~G. {McCarthy}, A.~M.~C. {Le Brun}, J.~{Schaye}, and G.~P. {Holder},
\newblock \mnras\ {\bf 440}, 3645 (2014), astro-ph/1312.5341.

\bibitem{Trac:2011}
{\ H}.~{Trac}, P.~{Bode}, and J.~P. {Ostriker},
\newblock \apj\ {\bf 727}, 94 (2011), astro-ph/1006.2828.

\bibitem{Addison:2012}
{\ G}.~E. {Addison}, J.~{Dunkley}, and D.~N. {Spergel},
\newblock \mnras\ {\bf 427}, 1741 (2012), astro-ph/1204.5927.

\bibitem{George:2015}
{\ E}.~M. {George} {\em et~al.},
\newblock \apj\ {\bf 799}, 177 (2015), astro-ph/1408.3161.

\bibitem{Chen:2014}
{\ X}.~{Chen} and M.~H. {Namjoo},
\newblock Physics Letters B {\bf 739}, 285 (2014), astro-ph/1404.1536.

\bibitem{Galli:2017}
{\ Planck Collaboration} {\em et~al.},
\newblock \aap\ {\bf 607}, A95 (2017), astro-ph/1608.02487.

\bibitem{Efstathiou:2014}
{\ G}.~{Efstathiou} and S.~{Gratton},
\newblock Planck Internal Communication  (2014).

\bibitem{Planck_lensing:2018}
{\ Planck Collaboration} {\em et~al.},
\newblock \aap\ {\bf 641}, A8 (2020), astro-ph/1807.06210.

\bibitem{Alam:2017}
{\ S}.~{Alam} {\em et~al.},
\newblock \mnras\ {\bf 470}, 2617 (2017), astro-ph/1607.03155.

\bibitem{Beutler:2011}
{\ F}.~{Beutler} {\em et~al.},
\newblock \mnras\ {\bf 416}, 3017 (2011), astro-ph/1106.3366.

\bibitem{Ross:2015}
{\ A}.~J. {Ross} {\em et~al.},
\newblock \mnras {\bf 449}, 835 (2015), astro-ph/1409.3242.

\bibitem{Percival:2002}
{\ W}.~J. {Percival} {\em et~al.},
\newblock \mnras\ {\bf 337}, 1068 (2002), astro-ph/0206256.

\bibitem{Planck_consistency:2014}
{\ Planck Collaboration} {\em et~al.},
\newblock \aap\ {\bf 571}, A31 (2014), astro-ph/1508.03375.

\bibitem{Huang:2018}
{\ Y}.~{Huang}, G.~E. {Addison}, J.~L. {Weiland}, and C.~L. {Bennett},
\newblock \apj\ {\bf 869}, 38 (2018), astro-ph/1804.05428.

\bibitem{Addison:2016}
{\ G}.~E. {Addison} {\em et~al.},
\newblock \apj\ {\bf 818}, 132 (2016), astro-ph/1511.00055.

\bibitem{Riess:2011}
{\ A}.~G. {Riess} {\em et~al.},
\newblock \apj\ {\bf 730}, 119 (2011), astro-ph/1103.2976.

\bibitem{Riess:2016}
{\ A}.~G. {Riess} {\em et~al.},
\newblock \apj\ {\bf 826}, 56 (2016), astro-ph/1604.01424.

\bibitem{Riess:2018}
{\ A}.~G. {Riess} {\em et~al.},
\newblock \apj\ {\bf 861}, 126 (2018), astro-ph/1804.10655.

\bibitem{Riess:2019}
{\ A}.~G. {Riess}, S.~{Casertano}, W.~{Yuan}, L.~M. {Macri}, and D.~{Scolnic},
\newblock \apj\ {\bf 876}, 85 (2019), astro-ph/1903.07603.

\bibitem{Birrer:2019}
{\ S}.~{Birrer} {\em et~al.},
\newblock \mnras\ {\bf 484}, 4726 (2019), astro-ph/1809.01274.

\bibitem{Aubourg:2015}
{\ \'E}.~{Aubourg} {\em et~al.},
\newblock \prd\ {\bf 92}, 123516 (2015), astro-ph/1411.1074.

\bibitem{Cuesta:2015}
{\ A}.~J. {Cuesta}, L.~{Verde}, A.~{Riess}, and R.~{Jimenez},
\newblock \mnras\ {\bf 448}, 3463 (2015), astro-ph/1411.1094.

\bibitem{Bernal:2016}
{\ J}.~L. {Bernal}, L.~{Verde}, and A.~G. {Riess},
\newblock \jcap\ {\bf 10}, 019 (2016), astro-ph/1607.05617.

\bibitem{Lemos:2018b}
{\ P}.~{Lemos}, E.~{Lee}, G.~{Efstathiou}, and S.~{Gratton},
\newblock \mnras\  (2018), astro-ph/1806.06781.

\bibitem{Macaulay:2018}
{\ E}.~{Macaulay} {\em et~al.},
\newblock \mnras\ {\bf 486}, 2184 (2019), astro-ph/1811.02376.

\bibitem{Efstathiou:2021}
{\ G}.~{Efstathiou},
\newblock \mnras\ {\bf 505}, 3866 (2021), astro-ph/2103.08723.

\bibitem{Knox:2019}
{\ L}.~{Knox} and M.~{Millea},
\newblock \prd\ {\bf 101}, 043533 (2020), astro-ph/1908.03663.

\bibitem{diValentino:2021}
{\ E}.~{Di Valentino} {\em et~al.},
\newblock Classical and Quantum Gravity {\bf 38}, 153001 (2021),
  astro-ph/2103.01183.

\bibitem{Schloneberg:2021}
{\ N}.~{Sch{\"o}neberg} {\em et~al.},
\newblock arXiv e-prints  (2021), astro-ph/2107.10291.

\bibitem{Hikage:2018}
{\ C}.~{Hikage} {\em et~al.},
\newblock \pasj\ {\bf 71}, 43 (2019), astro-ph/1809.09148.

\bibitem{Heymans:2021}
{\ C}.~{Heymans} {\em et~al.},
\newblock \aap\ {\bf 646}, A140 (2021), astro-ph/2007.15632.

\bibitem{Hildebrandt:2018}
{\ H}.~{Hildebrandt} {\em et~al.},
\newblock \aap\ {\bf 633}, A69 (2020), astro-ph/1812.06076.

\bibitem{DES3:2021}
{\ DES Collaboration} {\em et~al.},
\newblock arXiv e-prints  (2021), astro-ph/2105.13549.

\bibitem{DES1:2018}
{\ T}.~M.~C. {Abbott} {\em et~al.},
\newblock \prd\ {\bf 98}, 043526 (2018), astro-ph/1708.01530.

\bibitem{Joudaki:2020}
{\ S}.~{Joudaki} {\em et~al.},
\newblock \aap\ {\bf 638}, L1 (2020), astro-ph/1906.09262.

\bibitem{Asgari:2020}
{\ M}.~{Asgari} {\em et~al.},
\newblock \aap\ {\bf 634}, A127 (2020), astro-ph/1910.05336.

\bibitem{Garcia-Garcia:2021}
{\ C}.~{Garc{\'\i}a-Garc{\'\i}a} {\em et~al.},
\newblock arXiv e-prints  (2021), astro-ph/2105.12108.

\bibitem{Park:2019b}
{\ C}.~{Park} and B.~{Ratra},
\newblock \apj\ {\bf 882}, 158 (2019), astro-ph/1801.00213.

\bibitem{DiValentino:2019}
{\ E}.~{Di Valentino}, A.~{Melchiorri}, and J.~{Silk},
\newblock Nature Astronomy , 484 (2019), astro-ph/1911.02087.

\bibitem{DiValentino:2020}
{\ E}.~{Di Valentino}, A.~{Melchiorri}, and J.~{Silk},
\newblock \apjl\ {\bf 908}, L9 (2021), astro-ph/2003.04935.

\bibitem{Handley:2019}
{\ W}.~{Handley},
\newblock arXiv e-prints  (2019), astro-ph/1908.09139.

\bibitem{Bond:1997}
{\ J}.~R. {Bond}, G.~{Efstathiou}, and M.~{Tegmark},
\newblock \mnras\ {\bf 291}, L33 (1997), astro-ph/9702100.

\bibitem{Stompor:1999}
{\ R}.~{Stompor} and G.~{Efstathiou},
\newblock \mnras\ {\bf 302}, 735 (1999), astro-ph/9805294.

\bibitem{Efstathiou:2020}
{\ G}.~{Efstathiou} and S.~{Gratton},
\newblock \mnras\ {\bf 496}, L91 (2020), astri-ph/2002.06892.

\bibitem{McCarthy:2018}
{\ I}.~G. {McCarthy} {\em et~al.},
\newblock \mnras\ {\bf 476}, 2999 (2018), astro-ph/1712.02411.

\bibitem{BICEP:2018}
{\ BICEP2 Collaboration} {\em et~al.},
\newblock Physical Review Letters {\bf 121}, 221301 (2018),
  astro-ph/1810.05216.

\bibitem{PlanckNG:2014}
{\ Planck Collaboration} {\em et~al.},
\newblock \aap\ {\bf 571}, A24 (2014), astro-ph/1303.5084.

\bibitem{PlanckNG:2016}
{\ Planck Collaboration} {\em et~al.},
\newblock \aap\ {\bf 594}, A17 (2016), astro-ph/1502.01592.

\bibitem{PlanckInflation:2014}
{\ Planck Collaboration} {\em et~al.},
\newblock \aap\ {\bf 571}, A22 (2014), astro-ph/1303.5082.

\bibitem{PlanckInflation:2016}
{\ Planck Collaboration} {\em et~al.},
\newblock \aap\ {\bf 594}, A20 (2016), astro-ph/1502.02114.

\bibitem{PlanckInflation:2018}
{\ Planck Collaboration} {\em et~al.},
\newblock \aap\ {\bf 641}, A10 (2020), astro-ph/1807.06211.

\bibitem{Starobinski:1979}
{\ A}.~A. {Starobinskij},
\newblock Pisma v Zhurnal Eksperimentalnoi i Teoreticheskoi Fiziki {\bf 30},
  719 (1979).

\bibitem{Knox:1994}
{\ L}.~{Knox} and M.~S. {Turner},
\newblock Physical Review Letters {\bf 73}, 3347 (1994), astro-ph/9407037.

\bibitem{Kamionkowski:1997}
{\ M}.~{Kamionkowski}, A.~{Kosowsky}, and A.~{Stebbins},
\newblock \prd\ {\bf 55}, 7368 (1997), astro-ph/9611125.

\bibitem{Seljak:1997}
{\ U}.~{Seljak} and M.~{Zaldarriaga},
\newblock Physical Review Letters {\bf 78}, 2054 (1997), astro-ph/9609169.

\bibitem{BICEP:2015}
{\ BICEP2/Keck Collaboration} {\em et~al.},
\newblock Physical Review Letters {\bf 114}, 101301 (2015),
  astro-ph/1502.00612.

\bibitem{BICEP:2016}
{\ BICEP2 Collaboration} {\em et~al.},
\newblock Physical Review Letters {\bf 116}, 031302 (2016),
  astro-ph/1510.09217.

\bibitem{BICEP:2014}
{\ BICEP2 Collaboration} {\em et~al.},
\newblock Physical Review Letters {\bf 112}, 241101 (2014), astro-ph/1403.3985.

\bibitem{Tristram:2021}
{\ M}.~{Tristram} {\em et~al.},
\newblock \aap\ {\bf 647}, A128 (2021), astro-ph/2010.01139.

\bibitem{Kallosh:2018}
{\ R}.~{Kallosh}, A.~{Linde}, D.~{Roest}, A.~{Westphal}, and Y.~{Yamada},
\newblock Journal of High Energy Physics {\bf 2}, 117 (2018),
  hep-th/1707.05830.

\bibitem{Achucarro:2018}
{\ A}.~{Ach{\'u}carro}, R.~{Kallosh}, A.~{Linde}, D.-G. {Wang}, and
  Y.~{Welling},
\newblock \jcap\ {\bf 4}, 028 (2018), hep-th/1711.09478.

\bibitem{Akrami:2018}
{\ Y}.~{Akrami}, R.~{Kallosh}, A.~{Linde}, and V.~{Vardanyan},
\newblock \jcap\ {\bf 6}, 041 (2018), hep-th/1712.09693.

\bibitem{Kreisch:2019}
{\ C}.~D. {Kreisch}, F.-Y. {Cyr-Racine}, and O.~{Dor{\'e}},
\newblock \prd\ {\bf 101}, 123505 (2020), astro-ph/1902.00534.

\bibitem{Ade:2019}
{\ P}.~{Ade} {\em et~al.},
\newblock \jcap\ {\bf 2}, 056 (2019), astro-ph/1808.07445.

\bibitem{Sekimoto:2018}
{\ Y}.~{Sekimoto} {\em et~al.},
\newblock {Concept design of the LiteBIRD satellite for CMB B-mode
  polarization},
\newblock in {\em Space Telescopes and Instrumentation 2018: Optical, Infrared,
  and Millimeter Wave}, , Society of Photo-Optical Instrumentation Engineers
  (SPIE) Conference Series Vol. 10698, p. 106981Y, 2018.

\bibitem{Lancaster:2017}
{\ L}.~{Lancaster}, F.-Y. {Cyr-Racine}, L.~{Knox}, and Z.~{Pan},
\newblock \jcap\ {\bf 7}, 033 (2017), astro-ph/1704.06657.

\bibitem{Park:2019}
{\ M}.~{Park}, C.~D. {Kreisch}, J.~{Dunkley}, B.~{Hadzhiyska}, and F.-Y.
  {Cyr-Racine},
\newblock \prd\ {\bf 100}, 063524 (2019), astro-ph/1904.02625.

\bibitem{Crites:2015}
{\ A}.~T. {Crites} {\em et~al.},
\newblock \apj\ {\bf 805}, 36 (2015), astro-ph/1411.1042.

\bibitem{Keisler:2015}
{\ R}.~{Keisler} {\em et~al.},
\newblock \apj\ {\bf 807}, 151 (2015), astro-ph/1503.02315.

\end{thebibliography}

\end{document}